\documentclass[30pt,english]{book}
\usepackage[T1]{fontenc}
\usepackage[latin1]{inputenc}
\usepackage{geometry}
\usepackage{fancyhdr}
\usepackage{rotating}
\usepackage{subfigure}
\usepackage{graphicx}
\usepackage{setspace}
\usepackage{amssymb}
\usepackage{epsf}
\usepackage{rotating}
\makeatletter

\newcommand{\noun}[1]{\textsc{#1}}

\providecommand{\tabularnewline}{\\}

 \newcommand{\lyxaddress}[1]{
   \par {\raggedright #1 
   \vspace{1.4em}
   \noindent\par}
 }
 \newcommand{\lyxrightaddress}[1]{
   \par {\raggedleft \begin{tabular}{l}\ignorespaces
   #1
   \end{tabular}
   \vspace{1.4em}
   \par}
 }
 \usepackage{verbatim}

\usepackage{framed}

\usepackage{babel}

\def    \be             {\begin{equation}}
\def    \ee             {\end{equation}}
\def    \ba             {\begin{eqnarray}}
\def    \ea             {\end{eqnarray}}
\def    \nn             {\nonumber}

\def\beq{\begin{equation}}
\def\eeq{\end{equation}}
\def\beqn{\begin{eqnarray}}
\def\ba{\begin{eqnarray}}
\def\eeqn{\end{eqnarray}}
\def\ea{\end{eqnarray}}
\def\slash#1{#1\hskip-6pt/\hskip6pt}

\newcommand{\beqa}{\begin{eqnarray}}
\newcommand{\eeqa}{\end{eqnarray}}

\def\slashed{\ds}

\def\b{\beta}

\def\g{\gamma}

\def\tr{\mathop{\rm Tr}}
\def\ds#1{#1\kern-1ex\hbox{/}}
\def\dsh{h\kern-1.2ex /}

\newcommand{\bea}{\begin{eqnarray}}
\newcommand{\eea}{\end{eqnarray}}
\def\ul{\underline}
\def\nn{\nonumber}

\def\beq{\begin{equation}}
\def\eeq{\end{equation}}
\def\beqn{\begin{eqnarray}}
\def\eeqn{\end{eqnarray}}
\def\ba{\begin{eqnarray}}
\def\ea{\end{eqnarray}}
\newcommand{\lbl}[1]{\label{eq:#1}}
\newcommand{\eps}{\epsilon}
\newcommand{\veps}{\varepsilon}
\newcommand{\la}{\lambda}
\newcommand{\ro}{\rho}
\newcommand{\si}{\sigma}
\newcommand{\Li}[2]{{\mbox{Li}}_{#1}\left(#2\right)}

\newcommand{\ksls}{\not \! k}

\newcommand{\qsls}{\not \! q}
\newcommand{\pslsh}{\not \! p}
\newcommand{\dd}{\!\cdot\!}

\makeatother
\begin{document}
\begin{center}{\huge \thispagestyle{empty}}\end{center}{\huge \par}

\begin{center}{\large UNIVERSITÀ DEL SALENTO}\end{center}{\large \par}

\begin{center}{\large DIPARTIMENTO DI FISICA}\end{center}{\large \par}

\begin{center}\vspace{1cm}\end{center}

\begin{center}\includegraphics[%
  width=4cm]{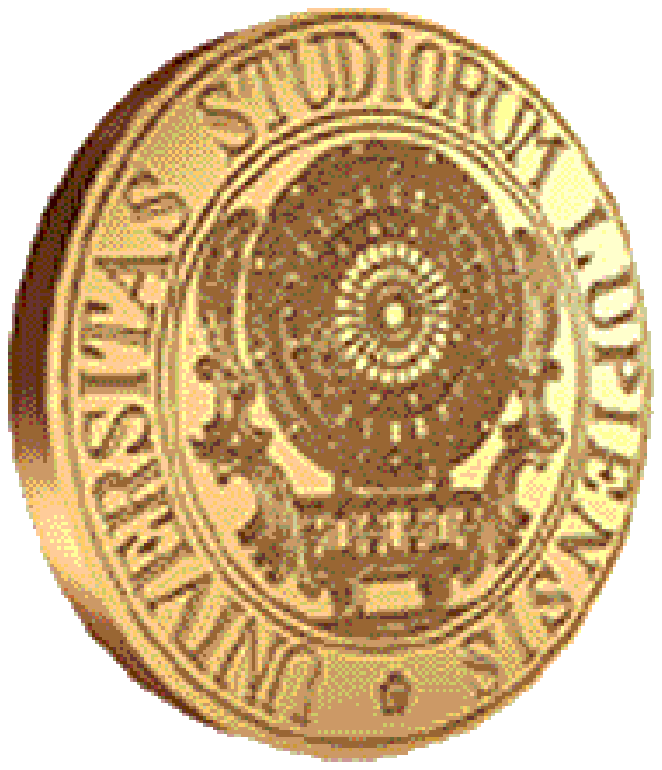}\vspace{2cm}\end{center}

\begin{center}
\textbf{\huge St\"uckelberg Axions}\\
\vspace{.3cm}
\textbf{\huge and Anomalous Abelian Extensions}\\
\vspace{.3cm}
\textbf{\huge of the Standard Model}  \\
\vspace{3cm}
\end{center}

\lyxaddress{\textbf{Advisor}\\
Claudio Corianò}

\lyxrightaddress{\textbf{Candidate}\\
Simone Morelli}

\lyxrightaddress{\vspace{0.5cm}}

\begin{center}\noindent

\rule{16cm}{0.1mm}\renewcommand{\headrulewidth}{0pt}\end{center}

\begin{center}\noun{\large Tesi di Dottorato di Ricerca in Fisica
-- XXI Ciclo}\fancyhf{}

\renewcommand{\headrulewidth}{0pt}

\fancyhead[LE,RO]{\thepage}

\fancyhead[RE]{\nouppercase\rightmark}

\fancyhead[LO]{\nouppercase\leftmark}\end{center}

\newpage
\begin{flushright}
{\em Truth is one, \\but the wise call it by different names.\\ (Rig Veda; section 1, hymn 164, verse 64.) }
\end{flushright}

\tableofcontents{}

\chapter*{Acknowledgements}

\addcontentsline{toc}{chapter}{Acknowledgements}\markboth{}{Acknowledgements}

I wish to express an extreme gratitude to my advisor Claudio Corian\`o. His patience and firmness of purposes together with an exceptional passion for Physics have always been my guide during the last years and will be reason of real inspiration to me during all my life.  

A special thank is due to my friend and collaborator Marco Guzzi for all the nice time we spent working together. I am grateful to him for his constant help and for pushing me to do better and better every time.       

I wish to thank Nikos Irges for giving me the precious opportunity to work with him and for his support. Every discussion with him has always been illuminating and of great importance for my work. 

A great thank is addressed to Cosmas K. Zachos for his kind hospitality and for the opportunity to work with him. I am particularly grateful to him for the three months we spent together in Chicago, for all the time he dedicated to me for discussions and last, but not least, for his "English lessons".  

Many thanks are due to Marco Roncadelli, Francesco Fucito and Theodore Tomaras for their support and hospitality.  

Finally, I cannot miss to thank all my friends and colleagues of the Department of Physics in Lecce for their sympathy and all the fun we had together. Their friendship has always been a safe shelter for me.    

This work is dedicated to my family.
\chapter*{List of publications}

\addcontentsline{toc}{chapter}{List of publications}\markboth{}{List of publications}

The chapters presented in this thesis are based on the following research papers.

\begin{enumerate}
\item \emph{Stueckelberg axions and the effective action of anomalous Abelian models. 1. A Unitarity analysis of the Higgs-axion mixing.} \\
C. Corian\`o, N. Irges and S. Morelli, JHEP 07 (2007) 008, hep-ph/0701010. 
\item \emph{Stueckelberg axions and the effective action of anomalous Abelian models. 2. A $SU(3)_C \times SU(2)_W \times U(1)_Y \times U(1)_B$ model and its signature at the LHC.}  \\
 C. Corian\`o, N. Irges and S. Morelli, Nucl. Phys. B789 (2008) 133, hep-ph/0703127. 
\item \emph{Unitarity Bounds for Gauged Axionic Interactions and the Green-Schwarz Mechanism.}\\
C. Corian\`o, M. Guzzi and S. Morelli, Eur. Phys. J. C55 (2008) 629, arXiv:0801.2949 [hep-ph].   
\item \emph{Axions and Anomaly-Mediated Interactions: The Green-Schwarz and Wess-Zumino Vertices at Higher Orders and g-2 of the muon.} \\ 
R. Armillis, C. Corian\`o, M. Guzzi and S. Morelli, JHEP 10 (2008) 034, arXiv:0808.1882 [hep-ph].  
\item  \emph{An Anomalous Extra Z Prime from Intersecting Branes with Drell-Yan and Direct Photons at the LHC.}\\
R. Armillis, C. Corian\`o, M. Guzzi and S. Morelli, Nucl. Phys. B814 (2009) 15679, arXiv:0809.3772 [hep-ph].  
 \end{enumerate}

\chapter*{Introduction\addcontentsline{toc}{chapter}{Introduction}\markboth{}{Introduction}}
\fancyhead[LO]{\nouppercase{Introduction}}
One of the challenges of the next several years in the area of particle theory will be the search for confirmation of the 
Standard Model (SM) at the Large Hadron Collider, which has provided so far the fundamental scheme within which to explain a large part of the electroweak phenomenology and the physics of the strong interactions.  Based on a Yang-Mills theory and incorporating the mechanism of spontaneous symmetry breaking in order to give mass to the electroweak gauge bosons, this theory remains, however, still to be proven to be correct in all of its sectors. For instance, the mechanism of spontaneous symmetry breaking itself, which 
is a structural part of the theory, while elegant and 
so far unchallenged as a theoretical model for the breaking of the gauge symmetry  -while preserving the renormalizability
 and the unitarity of the theory-  involves a scalar field, the Higgs field, which has not been found yet. Whether the mechanism is 
 correct or not, at least experimentally, remains to be seen, although there are reasonable hopes to believe that the Higgs field 
 will be found. Naturally, even the discovery of the Higgs and the confirmation of the correctness of the scalar sector 
 of the model does not stop the hopes that in the next decade we will be able to enlarge quite substantially our knowledge 
 of the physics of the fundamental interactions at the energy scale of several TeV's. One among the many possibilities is to look for minimal extensions of the SM which are characterized by modest departures from the 
 $SU(3) \times SU(2) \times U(1)_Y$ gauge structure and which can be more easily tested at the Large Hadron Collider (LHC).  Naturally, there are plenty of options and roads that one can take, from  supersymmetry to extra dimensions to brane 
 constructions. 
 
 Many scenarios have been put forward in the last few years in which many 
 of these ideas have been blended and synthesized in new ways, providing a fertile ground for further elaborations and modeling of the fundamental interactions with a substantial enrichment of the spectrum of the physical possibilities. 
 
 It is clear, at this point, that, given all these new developements, the selection of the most promising directions to follow and 
 on which contribute is not an issue with a straight "yes or not" answer,
  since there are personal tastes and motivations of different nature, some of them sociological, that motivate the pattern to follow, the theory 
  to develope and to believe in, and the answer to give to these questions. If one assumes that unification is for sure an important 
  issue in current and future elementary particle theory, then for sure, the simplest answer is provided within the context 
  of Grand Unified theories (GUT), the most impressive result in favour of it being the unification of the gauge coupling 
in a supersymmetric  extension of the SM. In all these scenarios, string theory has played a considerable role, especially after the discovery that special vacua of 
string/brane theory may show up in some way at future collider experiments. 
While string theory has always been connected with physics at the Planck scale, with the advent of brane theory and extra dimensional models, strings are expected to play a role also at rather modest energies, such as at the TeV scale. The study of this energy range will be at the center of the 
two major experiments planned at the LHC, that is ATLAS and CMS. In fact, a large effort has gone into the classification of these vacua of string theory, 
especially in the theoretical analysis of models of intersecting branes which are at the root of the class of theories that we are going to investigate 
in this thesis work. 

In our study we will start with an analysis of these models in a rather simplified context, focusing on their basic structure and studying the class of effective actions wich are at the origins of these scenarios. Being the gauge structure of these gauge theories rather simple, 
we will be able to go quite far in the developement of the corresponding field theory and we will provide accurate predictions at the LHC for these models, 
having identified the key parameters which characterize them and which are the objects of experimental verification. 

Our work is not a work in string theory, but is a study of what string theory may imply at field theory level in a special class of its vacua. 
One of the key discoveries of these class of vacua is the presence of a physical axion in the spectrum of this class
of theories. The axion, termed the "axi-Higgs" in \cite{Coriano:2005js}, is a natural by-product of the presence of anomalous symmetries in the gauge structure of these theories. The effective actions of these models, which contain several extra anomalous $U(1)$'s, have their anomaly cancelled by a Green-Schwarz mechanism, which involves an axion. At the same time, the axion appears in the form of a combination of St\"uckelberg fields and $CP$-odd contributions of an extended Higgs sector, characterized by two Higgs doublets, with the possibility of some "phases" (Peccei-Quinn breaking terms) added to its canonical structure. 

In our analysis we will characterize these models from the ground up, working out at a very fine level of detail the effective action of these models, starting 
from the simplest ones and moving towards more general and complex analysis of them. The objective of these studies is to provide a firm characterization of the 
field theory and the phenomenology of these models, stressing on the various mechanisms of cancellation of the anomalies which are allowed in these types of constructions. A key role, in these scenarios, is played by the anomaly, and several of its properties are analyzed under a new light. 
One important point of our study concerns the discovery of a unitarity bound for this class of theories, which appears when a mechanism of anomaly cancellation involves a local counterterm of Wess-Zumino type. Other mechanisms are proposed and analyzed as well, one of them being the subtraction of the anomaly pole and its formulation in terms of an axion and a ghost. 

As we have mentioned, our analysis stresses mostly the role of the physical (St\"uckelberg) axion, that survives after electroweak symmetry breaking as a physical field, and its potential significance as a new weakly interacting pseudoscalar which might play a role as a component of dark matter. 
  
\chapter{St\"{u}ckelberg Axions and the Effective Action of Anomalous Abelian Models 1. 
\\ A unitarity analysis of the Higgs-axion mixing  \label{chap:AbelianModels1}}
\fancyhead[LO]{\nouppercase{Chapter 1. A unitarity analysis of the Higgs-axion mixing}}
\section{Introduction to the chapter}
The search for the identification of possible extensions of the Standard Model (SM) is a challenging area both from the theoretical and the experimental perspectives. It is even more so with the upcoming experiments at the 
LHC, where the hopes are that at least some among the many phenomenological scenarios that have been formulated in the last three decades 
can finally be tested.  The presence of so many wide and diverse possibilities certainly render these studies very 
challenging. Surely, among these, the choice 
of simple Abelian extension of the basic gauge structure of the SM is one of 
the simplest to take into consideration. These extensions will probably 
be the easiest to test and be also the 
first to be confirmed or ruled out. Though $U(1)$ extensions are ubiquitous, they are far from being trivial. These theories predict new gauge bosons, the extra $Z^\prime$, with masses that are likely to be detected if they are 
up to $4 \div 5$ TeV (see for instance \cite{Langacker:1991pg, DelAguila:1995fa, Cvetic:1995zs, Capstick:1987uc, Leike:1998wr} for an overview and topical studies). 
These extensions are formulated, with a variety of motivations,  
within a well-defined theoretical framework and involve phenomenological studies which are far 
simpler than those required, for instance, in the case of supersymmetry, where a large 
set of parameters and soft-breaking terms clearly render the theoretical description much more involved.

On the other hand, simple Abelian extensions are also quite numerous, since new neutral currents are predicted both by Grand Unified Theories 
(GUT's) and/or by superstring inspired models based on $E_6$ and $SO(10)$ (see \cite{Faraggi:2002nq, Faraggi:2001dw, Coriano:2002fr} for instance). 
One of the common features of these models is the absence of an anomalous fermion spectrum,  as for the 
SM, with the anomaly cancellation mechanism playing a key role in fixing the couplings of the fermions to the gauge fields and in guaranteeing their inner consistency. 
In this respect, unitarity and renormalizability, tenets of the effective theory, are preserved. 

When we move to enlarge the gauge symmetry of the SM,
the unitarity has to be preserved, but not necessarily the renormalizability of the model.
In fact, operators of dimension-5 and higher which may appear at higher energies 
have been studied and classified under quite general assumptions \cite{Hagiwara:1986vm}.

Anomalous Abelian models, differently from the non-anomalous ones, show some striking features, which have been 
exploited in various ways, for example in the generation of realistic 
hierarchies among the Yukawa couplings \cite{Binetruy:1996cb, Irges:1998ax} and to analyze neutrino mixing. There are obvious reasons that justify these studies: the mechanism of anomaly cancellation that Nature selects may not just be based on an anomaly-free spectrum, but may require 
a more complex pattern, similar to the Green-Schwarz (GS) 
anomaly cancellation mechanism 
of string theory, that invokes an axion.  
Interestingly enough, the same pattern appears if, for a completely different and purely dynamical reason, part of the fermion spectrum of an anomaly free theory is integrated out, together with  part of the Higgs 
sector \cite{Coriano:2006xh}. In both cases, the result is a theory that shows the features 
discussed in this work, though some differences between the two different realizations may remain in the effective theory. For instance, it has been suggested that the PVLAS result can be easily 
explained within this class of models incorporating a single anomalous $U(1)$. The anomaly 
can be real (due to anomaly inflow from extra dimensions, (see \cite{Hill:2004uc, Hill:2006ei} as an example), 
or effective, due to the partial decoupling of a heavy Higgs, and the St\"uckelberg field is the remnant phase of this partial decoupling. The result is a ``gauging'' of the $PQ$ axion \cite{Coriano:2006xh}.
\subsection{The quantization of anomalous Abelian models and the axion}
The interest on the quantization of anomalous models and their proper field theoretical description has beeen a key topic 
for a long period, in an attempt to clarify under which conditions an anomalous gauge theory may be improved by the introduction of suitable interactions so to become unitary and renormalizable. 
The introduction of the Wess-Zumino term (WZ), a $\theta F\wedge F$ term, which involves a pseudoscalar $\theta$ times the divergence of a topological current, has been proposed as a common cure in order to restore the gauge invariance 
of the theory \cite{Faddeev:1986pc,Krasnikov:1985bn}. Issues related to the unitarity of models incorporating Chern-Simons (CS) and anomalous interactions 
in lower dimensions have also been analyzed in the past \cite{Jackiw:1984zi, Mitra:1990sf}.

Along the same lines of thought, also non-local counterterms have been proposed as a way to achieve 
the same objective \cite{Adam:1997gj}.
The gauge dependence of the WZ term and its introduction into the spectrum so 
to improve the power counting in the loop expansion of the theory has also been a matter of 
debate \cite{Preskill:1990fr}. Either with or without a WZ term, renormalizability is clearly lost, 
while unitarity, in principle, can be maintained. As we are going to illustrate in specific and realistic 
examples, gauge invariance and anomaly cancellation play a subtle role in guaranteeing the 
gauge independence of matrix elements in the presence of symmetry breaking.

So far, the most interesting application of this line of reasoning in which the Wess-Zumino term acquires a physical meaning 
is in the Peccei-Quinn solution of the 
strong $CP$-problem of QCD \cite{Peccei:1977ur}, where the SM Lagrangian is augmented by a global anomalous U(1) and involves an axion. 

The $PQ$ symmetry, in its original form, is a global symmetry broken only by instanton effects. 
The corresponding axion, which in the absence of non perturbative 
effects would be the massless Nambu-Goldstone boson of the global 
(chiral) symmetry, acquires a tiny mass. In the $PQ$ case 
the mass of the axion and its coupling to the gauge field are correlated,  
since both quantities are defined in terms of the same factor $1/f_a $, 
with $f_a $ being the $PQ$-breaking scale, which is currently bounded, by 
terrestrial and astrophysical searches, to be very large ($\approx 10^9$ GeV) \cite{Abbott:1982af, Kim:1999ia}. 

This tight relation between the axion mass and the coupling is a specific feature of models of $PQ$ type where a global symmetry is invoked and, as we are going to see, it can be relaxed if the anomalous interaction is gauged.
The axion discussed in the thesis and its effective action has some special features that render it an interesting 
physical state, quite distinct from the $PQ$ axion. The term ``gauged axion'' or ``St\"{u}ckelberg axion'' or ``axi-Higgs'' all 
capture some of its main properties. Depending on the size of the $PQ$-breaking potential, 
the value of the axion mass gets corrected in the form of additional factors which are absent in the standard 
$PQ$ axion. 

 Although some of the motivations to investigate this class of models come from the interest toward special vacua of string theory \cite{Coriano:2005js}, the study of anomalous Abelian interactions, in the particular construction that we are going to discuss in this work, are applicable to a wide variety of models which share the typical features of those studied here. 
\subsection{The Case of String/Branes Inspired Models}
As we have mentioned, we work under quite general assumptions that apply 
to Abelian anomalous models that combine both the Higgs and the St\"{u}ckelberg mechanisms 
\cite{ Stueckelberg:1900zz} in order to give mass to the extra (anomalous) gauge bosons.  
There are various low-energy effective theories which can be included into this framework, one example being 
low energy orientifold models, but we will try to stress on the generality of the construction rather 
than on its stringy motivations, which, from this perspective, are truly just optional.

These models have been proposed as a possible scenario for physics beyond the Standard Model, with 
motivations that have been presented in \cite{Coriano:2005js}. Certain features of these models have been studied in some generality \cite{Kiritsis:2003mc}, and their formulation relies on the Green-Schwarz mechanism of anomaly cancellation that incorporates axionic and Chern-Simons interactions. At low energy, the Green-Schwarz term is nothing but the long known Wess-Zumino term. 
In particular, the mechanism of spontaneous symmetry breaking, that now involves 
both the St\"{u}ckelberg field (the axion) and the Standard Model Higgs, has been elucidated \cite{Coriano:2005js}. 
While the general features of the theory have been presented before, the selection of a specific gauge structure (the number of anomalous $U(1)$'s in 
\cite{Coriano:2005js} was generic), in our case of a single additional anomalous $U(1)$, allows us to specify the model in much 
more detail and discuss the structure of the effective action to a larger extent. This study is needed and provides a new 
step toward a phenomenological analysis for the LHC that we will 
present in the following chapters.

This requires the choice of a specific 
and simplified gauge structure which can be amenable to experimental testing. While in the minimal formulation of the models derived from intersecting branes the number of anomalous Abelian gauge groups is larger or equal to three, 
the simplest case (and the one for which a quantitative phenomenological analysis is possible) is the one in which a single anomalous 
$U(1)$ is present. This simplified structure appears once we make the assumption that the masses of the new Abelian gauge interactions are widely (but not too widely) 
separated so to guarantee an effective decoupling of the heavier $Z^\prime$s. 
Clearly, in this simplified setting, the analysis of 
\cite{Coriano:2005js} can be further specialized and, even more interestingly, 
one can try to formulate possible experimental predictions. 
\subsection{The content of this chapter} 
This chapter and the following address the construction of anomalous 
Abelian models in the presence of an extra anomalous $U(1)$, called $U(1)^{}_B$. 
This extra gauge boson becomes massive via a combined Higgs-St\"uckelberg 
mechanism and is accompanied by one axion, $b$. We illustrate the physical 
role played by the axion when both the Higgs and the St\"uckelberg mechanisms are present. The physical axion, that emerges in the scalar sector when $b$ is 
rotated into its physical component (the axi-Higgs, denoted by $\chi$) 
interacts with the gauge bosons with dimension-5 operators (the WZ terms). 
The presence of these interactions renders the theory non-renormalizable 
and one needs a serious study of its unitarity in order to make sense of it,  
which is the objective of these two chapters. Here the analysis 
is exemplified in the case of two simple models (the A-B and Y-B Models) where 
the non-Abelian sector is removed. A complete model will be studied in Chap.~\ref{chap:AbelianModels2}. Beside the WZ term the theory clearly shows that additional Chern-Simons interactions become integral part of the effective action. 
\subsection{The role of the Chern-Simons interactions} 
There are some very interesting features of these models which deserve a careful study, and which differ from the case of the Standard Model (SM). 
In this last case the cancellation of the anomalies is enforced by charge assignments. As a result of 
this, before electroweak symmetry breaking, all the anomalous trilinear gauge interactions vanish.  This cancellation continues to hold also after symmetry breaking if all the fermions of each generation are mass 
degenerate. Therefore, trilinear gauge interactions containing axial couplings are only sensitive to the mass differences among the 
fermions. In the case of extensions of the SM which include an anomalous $U(1)$ this pattern changes considerably, since the massless contributions in anomalous diagrams do not vanish. In fact, these theories become consistent only if a suitable set of axions and Chern-Simons (CS) interactions are included as counterterms in the defining Lagrangian. The role of the CS interactions is to re-distribute the partial anomalies 
among the vertices of a triangle in order to restaure the gauge invariance of the one-loop effective action before symmetry breaking. For instance, a hypercharge current involving a generator $Y$,would be anomalous at one-loop level in a trilinear interaction of the form $YBB$ or $YYB$, if $B$ is an anomalous gauge boson. In fact, while anomalous diagrams of the form $YYY$ are automatically vanishing by charge assignment, the former ones are not. The theory requires that in these anomalous interactions the CS counterterm moves the partial anomaly from the $Y$ vertex to the $B$ vertex, rendering in these diagrams the hypercharge current effectively vector-like. The $B$ vertex then carries all the anomaly of the trilinear interaction, but $B$ is accompanied by a Green-Schwarz axion $b$ and its anomalous gauge variation is canceled by the 
GS counterterm. It is then obvious that these theories show some new features which have never fully discussed in the past and require a very careful study. In particular, one is naturally forced to develope 
a regularization scheme that allows to keep track correctly of the distribution of the anomalies 
on the various vertices of the theory. This problem is absent in the case of the SM since the vanishing of the anomalous vertices in the massless phase renders any momentum parameterization of the diagrams acceptable. We describe in detail some of these more technical points in the appendices, 
where we illustrate how these theories can be treated consistently in dimensional regularization but with the addition of suitable shifts that take the form of CS counterterms.
\section{Massive $U(1)$'s {\it \`a la} St\"{u}ckelberg} 
One of the ways to render an Abelian $U(1)$ gauge theory massive is by the mechanism proposed 
by St\"{u}ckelberg \cite{Stueckelberg:1900zz} and extensively studied in the past. Before that another mechanism, 
the Higgs mechanism, was proposed as a viable and renormalizable method to give mass both to Abelian and to non-Abelian gauge theories. There are various ways in which, nowadays, this 
mechanism is implemented, and St\"{u}ckelberg fields appear quite naturally 
in the form of {\em compensator} fields in many supergravity and string models. On the phenomenological side, one of the first successful investigations of this mechanism for model building has been presented in \cite{Kors:2004dx}, while, rather recently, supersymmetric extensions of this mechanism have been investigated \cite{Kors:2004ri, Feldman:2006wb, Feldman:2006ce}. In other recent work some of its perturbative aspects have also been addressed, 
in the case of non anomalous Abelian models.

In the seventies, 
the St\"{u}ckelberg field (also called the ``St\"{u}ckelberg ghost'') re-appared in the analysis of the 
properties of renormalization of Abelian massive Yang-Mills theory by Salam and Strathdee \cite{Salam:1971sp}, 
Delbourgo \cite{Delbourgo:1987np} and others \cite{Ruegg:2003ps}, while Gross and Jackiw \cite{Gross:1972pv} introduced it in their analysis of the role of the 
anomaly in the same theory. According to these analysis the perturbative properties of a massive Yang-Mills 
theory, which is not renormalizable in its direct formulation, can be ameliorated by the introduction 
of this field. Effective actions in massive Yang-Mills theory 
have been also investigated in the past, and shown to have 
some predictivity also without the use of the St\"{u}ckelberg variables \cite{Coriano:1989pv, Coriano:1992bh}, but clearly 
the advantages of the Higgs mechanism and its elegance remains a firm result of the current formulation of the Standard Model. We briefly review these points to make our treatment self-contained but also to show 
that the role of this field completely changes in the presence of an anomalous fermion spectrum, when 
the need to render the theory unitary requires the introduction of an interaction $b F\tilde F$, 
spoiling renormalizability, but leaving the resulting theory well defined as an 
{\em effective} theory. For this to happen one needs to check explicitly the unitarity 
of the theory, which is not obvious, especially if the Higgs and St\"{u}ckelberg mechanisms are combined. 
This study is the main objective of the first part of this investigation, which is focused on the issues of unitarity of simple models which include 
both mechanisms. Various technical aspects of this analysis are important 
for the study of realistic models, as discussed in Chap.~\ref{chap:AbelianModels2}, where we move toward the study of an extension of the SM with two Abelian factors $U(1)^{}_Y \times U(1)^{}_B$, one of them being the standard 
hypercharge $Y$. The charge assignments for the anomalous diagrams 
involving a combinations of both gauge bosons are such that additional Ward Identities are needed to render the theory unitary, starting from gauge 
invariance. We study most of the features of this model in depth, and show 
how the neutral vertices of the model are affected by the new anomaly cancellation mechanism. We will work out an application of the theory in the process 
$Z\to \gamma \gamma $, which can be tested at forthcoming experiments at the 
LHC.
\subsection{The St\"{u}ckelberg action from a field-enlarging transformation}
We start with a brief introduction on the derivation of an action 
of St\"{u}ckelberg type to set the stage for further elaborations. 

A massive Yang-Mills theory can be viewed as a gauge-fixed version of a more general action involving the 
St\"{u}ckelberg scalar. A way to recognize this is to start from the standard Lagrangian 
\beq
\mathcal{L}= -\frac{1}{4} F_{\mu\nu}F^{\mu\nu} + \frac{1}{2} M_1^2 (B_\mu)^2
\label{fixed}
\eeq
with $F_{\mu\nu}=\partial_\mu B_\nu - \partial_\nu B_\mu$ and perform a field-enlarging transformation 
(see the general discussion presented in \cite{Alfaro:1989rx})
\beq
B_\mu = B_\mu' - \frac{1}{M_1}\partial_\mu b,
\eeq
that brings the original (gauge-fixed) theory (\ref{fixed}) into the new form 
\beq
\mathcal{L}=-\frac{1}{4} F_{\mu\nu}F^{\mu\nu} + \frac{1}{2} M_1^2 (B_\mu)^2 +
\frac{1}{2}(\partial_\mu b)^2 - M^{}_1 B_\mu \partial^\mu b 
\label{enlarged}
\eeq
which now reveals a peculiar gauge symmetry. It is invariant under the transformation 
\beqa
b&\to&b'= b - M^{}_1 \theta \nonumber \\
B_\mu &\to& B_\mu'=B_\mu + \partial_\mu \theta. 
\eeqa    
We can trace back our steps and gauge-fix this Lagrangian in order to obtain a new version of the original Lagrangian that now contains a scalar. One can choose to remove the mixing between $B_\mu$ and $b$ by the gauge-fixing condition 
\beq
\mathcal{L}_{gf} =- \xi \left( \partial\cdot B + \frac{M^{}_1}{2 \xi}b\right)^2 
\eeq
giving the gauge-fixed Lagrangian 
\beq
\mathcal{L}=-\frac{1}{4} F_{\mu\nu}F^{\mu\nu} + \frac{1}{2} M_1^2 (B_\mu)^2 +
\frac{1}{2}(\partial_\mu b)^2 - \xi (\partial B)^2 - \frac{M_1^2}{4 \xi} b^2.
\eeq
It is easy to show that the BRST charge of this model generates exactly the St\"{u}ckelberg condition on the 
physical subspace, decoupling the unphysical Faddeev-Popov ghosts from the physical spectrum. 

Different gauge choices are possible. 
The choice of a unitary gauge $(b=0)$ in the Lagrangian (\ref{enlarged}) brings us back to the original 
massive Yang-Mills model (\ref{fixed}). In the presence of a chiral fermion, the same field-enlarging transformation trick goes 
through, though this time we have to take into account the contribution of the anomaly 
\beq
\mathcal{L}=-\frac{1}{4}F_B^2 + \frac{M_1^2}{2}(B_\mu + \frac{1}{M_1} \partial_\mu b)^2 + 
i \overline{\psi}_L \gamma^\mu(\partial_\mu + i g B_\mu + i g \partial_\mu b)\psi_L,
\eeq
where $\psi_L=\frac{1}{2}(1- \gamma^5)$ is the left-handed anomalous fermion.
The Fujikawa method can be used to derive from the anomalous variation 
of the measure the relation 
\beq
g \overline{\psi}_L \gamma^\mu \partial_\mu b\psi_L = \frac{g^3}{32 \pi^2} 
\epsilon^{\mu\nu\rho\sigma}F_{\mu\nu}F_{\rho\sigma}
\eeq
thereby obtaining the final anomalous action 
\beq
\mathcal{L}=-\frac{1}{4}F_B^2 + \frac{M_1^2}{2}(B_\mu + \frac{1}{M_1}\partial_\mu b)^2 + 
i \overline{\psi}_L \gamma^\mu(\partial_\mu + i g B_\mu)\psi_L
- \frac{g^3}{32 \pi^2}b\,\epsilon^{\mu\nu\rho\sigma}F_{\mu\nu}F_{\rho\sigma}.
\eeq

Notice that the $b$ field can be integrated out \cite{Gross:1972pv}.
In this case one obtains an alternative effective action of the form
\beqa
\mathcal{L} &=& -\frac{1}{4}F_B^2 + \frac{M_1^2}{2}(B_\mu)^2 + 
i \overline{\psi}_L \gamma^\mu(\partial_\mu + i g B_\mu)\psi_L \nonumber \\
&& -
\frac{g^3}{96 \pi^2}\int d^4 y F_B^{\alpha\beta}\tilde{F}_{B\alpha \beta}(x)D(x-y|M_1^2 \xi)
F^{\mu\nu}(y)\tilde{F}_{\mu\nu}(y) 
\eeqa
with $(\square + M_1^2 \xi^2)D(x|M_1^2)=-\delta^4(x)$. The locality of the description is clearly 
lost. It is also obvious that the role of the axion, in this case, is to be an unphysical field. However, in the case 
of a model incorporating both spontaneous symmetry breaking and the St\"{u}ckelberg mechanism, the axion plays a physical 
role and can be massless or massive depending whether it is part of the scalar potential or not. 
Our interest, in this chapter, is to analyze in detail the contribution to the 
one-loop effective action of anomalous Abelian models, here defined as the classical Lagrangian plus its 
anomalous trilinear fermionic interactions. Anomalous Ward Identities in these 
effective actions are eliminated once the divergences from the triangles are removed either by 1) suitable charge assignments for some of generators, or by 2) shifting axions or by 3) a distribution of the partial anomalies on each vertex. 

Since this approach of anomaly cancellations is more involved than in the SM case, we have decided to analyze it using some simple 
(purely Abelian) models as working examples, before considering a realistic extension of the Standard Model which will be developed in Chap.~\ref{chap:AbelianModels2}. There, all the methodology developed in this chapter will 
be widely applied to the analysis of a string-inspired model derived from the orientifold construction \cite{Coriano:2005js}.
In fact, this analysis tries to clarify some unobvious issues that naturally appear once an effective anomaly-free gauge theory 
is generated at lower energies from an underlying renormalizable theory at a higher energy. For this purpose we will use a simple approach based on $s$-channel unitarity, inspired by the classic work of Bouchiat, Iliopoulos and Meyer \cite{Bouchiat:1972iq}.
\subsection{Implications at the LHC}
A second comment concerns the possible prospects for the discovery of a $Z^\prime$ of anomalous origin. Clearly $Z^\prime$ bosons being ubiquitous in GUT's and other SM extensions, discerning an anomalous $Z^\prime$ from a non-anomalous one is subtle, but possible. 
In Chap.~\ref{chap:AbelianModels2} we propose the Drell-Yan mechanism as a possible way to make this distinction, since some new effects related to the treatment of the anomalies are already apparent near the 
$Z$ resonance in this process. The numerical results for the anomalous corrections to Drell-Yan will be presented in Chap.~\ref{chap:LHC}. Anomalous vertices involving the $Z$ gauge boson 
appear both in the production mechanism and in its
decay into two gluons or two photons. In the usual Drell-Yan 
process of the SM, these contributions, because of anomaly cancellations, 
are sensitive only to the mass difference between the fermion of a given 
generation and are usually omitted in NNLO computations. 
If these resonances, predicted by theories with extra Abelian gauge structures, are very weakly coupled, then a precise determination of the QCD background is necessary to detect them (see \cite{Armillis:2008vp} for details).
\section{The Effective Action in the A-B Model}
As we have already mentioned, we will focus our analysis on the anomalous effective actions of simple 
Abelian theories. We will analyze two models: a first one called ``A-B'', with a $A$ vector-like 
(anomaly-free) and $B$ 
axial-vector-like (anomalous) and a second model called the ``Y-B'' Model where $B$ is anomalous and 
$Y$ is anomaly-free with both vector and axial-vector interactions. 
We start defining the so called ``A-B'' Model by the Lagrangian 
\beqa
\mathcal{L}_0 &=& |(\partial_{\mu} + i g^{}_B q^{}_B B_{\mu} ) \phi | ^{2} -\frac{1}{4} F_{A}^{2}
-\frac{1}{4} F_{B}^{2}   + \frac{1}{2}( \partial_{\mu} b + 
M_1\ B_{\mu})^{2} -\lambda( |\phi|^{2} - \frac{v^{2}}{2})^{2}   \nonumber\\
&& + \overline{\psi} i \gamma^{\mu} ( \partial_{\mu} +i e A_{\mu}
+ i g^{}_{B} \gamma^{5} B_{\mu}  ) \psi - \lambda_1 \overline{\psi}_L \phi \psi_R 
- \lambda_1 \overline{\psi}^{}_R \phi^{*} \psi^{}_L
\label{llagrangeBC}
\eeqa
which exhibits a non anomalous ($A$) and an anomalous ($B$) gauge interaction. 
 \begin{figure}[t]
{\centering \resizebox*{10cm}{!}{\rotatebox{0}
{\includegraphics{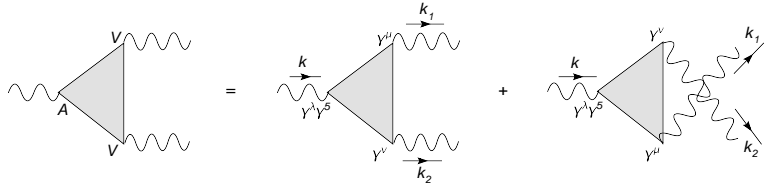}}}\par}
\caption{\small The ${\bf AVV}$ diagrams.}
\label{VAA1}
\end{figure}
\begin{figure}[t]
{\centering \resizebox*{10cm}{!}{\rotatebox{0}
{\includegraphics{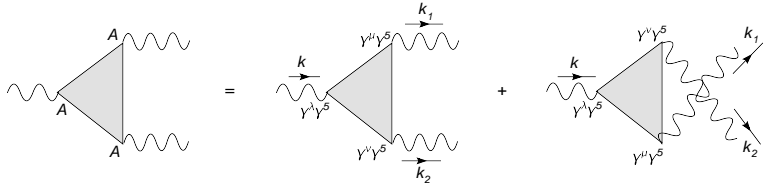}}}\par}
\caption{\small The ${\bf AAA}$ diagrams.}
\label{AAA1}
\end{figure}
Its couplings are summarized in Tabs.~\ref{tab1_AB} and \ref{tab2_AB}, where ``S'' refers to the presence of a St\"{u}ckelberg mass term for the corresponding gauge boson, if present. We have indicated with a small lowercase $b$ the corresponding axion. The $U(1)_A$ symmetry is unbroken 
while $B$ gets its mass by the combined Higgs-St\"{u}ckelberg mechanism. 
Another feature of the model, as we are going to see, is the presence of 
an Higgs-axion mixing generated not by a scalar potential (such as $V(\phi,b)$), 
as we will show in Sec.~\ref{sec:massaxion}, but by the fact that both mechanisms communicate their mass 
to the same gauge boson $B$. The axion remains a massless field in this case.  
\begin{table}[tbh]
\begin{center}
\begin{tabular}{|c|c|c|} \hline  
                         &     A                   &             B                    \\
                 \hline
 $\psi$               & $q^{A}_{L}=q^{A}_{R}=1$   &   $q^{B}_{R}= -q^{B}_{L}=1$       \\
                 \hline
\end{tabular}
\end{center}
\caption{\small Fermion assignments, A-B Model.}
\label{tab1_AB}

\begin{center}
\begin{tabular}{|c|c|c|} \hline  
                   &     $\phi$         &    $S$        \\
                 \hline
 $A$               &   $q^{A}= 0$    &     $0$     \\
                 \hline
  $B$              & $q^{B}= -2 $     &    $b$    \\
                 \hline
\end{tabular}
\end{center}
\caption{\small Gauge structure, A-B Model.}
\label{tab2_AB}
\end{table}
Our discussion relies on the formalism of the one-loop effective action, which is the generating functional of the one-particle irreducible (1PI) correlation functions of a given model. The correlators are multiplied by external classical fields and the formalism allows to derive quite directly the anomalous Ward Identities of the theory. The reader can find a discussion of the formalism in the appendix, where we 
study the properties of the Chern-Simons and Wess-Zumino vertices of the model and their gauge variations.

\begin{figure}[tbh]
{\centering \resizebox*{14cm}{!}{\rotatebox{0}
{\includegraphics{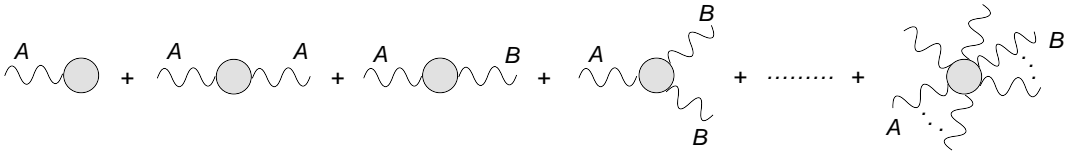}}}\par}
\caption{\small Expansion of the effective action.}
\label{effective11}
\end{figure}

\begin{figure}[tbh]
{\centering \resizebox*{8.5cm}{!}{\rotatebox{0}
{\includegraphics{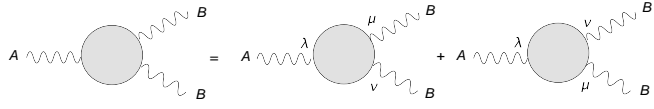}}}\par}
\caption{\small Triangle diagrams with permutations.
\label{effective22}}
\end{figure}

 In the A-B Model, this will involve the classical defining action plus the anomaly 
diagrams with fermionic loops and we will require its invariance under gauge transformations.   
The structure of the (total) effective action is summarized, in the case of
one vector ($A$) and one axial-vector ($B$) interaction by an expansion of the form
\beqn
W[A,B] &=& 
 \sum^{\infty}_{n_{1} = 1}  \sum^{\infty}_{n_{2} = 1} \frac{i^{\,n_1 + n_2}}{n_{1}! n_{2}!} \int dx_{1}...dx_{n_{1}}
dy_{1}...dy_{n_{2}} T^{\lambda_{1}...\lambda_{n_{1}} \mu_{1}...\mu_{n_{2}}}(x_{1}...x_{n_1}, y_1...y_{n_2})   \nonumber\\
&& \hspace{3.5cm}  B^{\lambda_1}(x_1)...B^{\lambda_{n_1}}(x_{n_1}) 
A_{\mu_1}(y_1)...A_{\mu_{n_2}}(y_{n_2}),
\eeqn
corresponding to the diagrams in Fig.~\ref{effective11}
where we sum, for each diagram, over the symmetric exchanges of all the indices (including the momentum) of the identical gauge 
bosons (see also Fig.~\ref{effective22}). 
As we are going to discuss next, also higher-order diagrams of the form, 
for instance, AVVV will be affected by the presence of an undetermined shift in the triangle 
amplitudes, amounting to Chern-Simons interactions (CS). 
They turn to be well-defined once the distribution of the anomaly on three-point functions is performed according 
to the correct Bose symmetries of these correlators of lower order. 
\begin{figure}[tbh]
{\centering \resizebox*{7.5cm}{!}{\rotatebox{0}
{\includegraphics{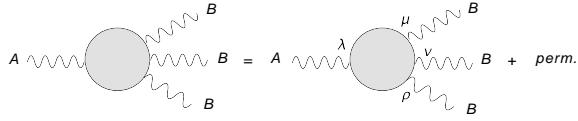}}}\par}
\caption{\small Symmetric expansion.
\label{effective33}}
\end{figure}
Computing the variation of the generating functional we obtain 
\beqn
\delta_{B} W^{an} [A,B] &=& \frac{1}{2!} \int dx \, dy \, dz \,T^{\lambda \mu \nu}(z,x,y)\, \delta B_{\lambda}(z)
A_{\mu}(x)A_{\nu}(y) \nonumber\\ 
&&+ \frac{1}{3!} \int dx \, dy \, dz\,  dw \,T^{\lambda \mu \nu \rho}(z,x,y,w) B_{\lambda}(z)A_{\mu}(x)A_{\nu}(y)A_{\rho}(w),
\eeqn
using $\delta B_{\lambda}(z) = \partial_{\mu} \theta_{B}(z)$ and integrating 
by parts we get 
\beqn
\delta_{B} W^{an} [A,B] &=& -\frac{1}{2!} \int dx \, dy \, dz \,\frac{\partial}{\partial z^{\lambda}}T^{\lambda \mu \nu}(z,x,y) 
\,\theta_{B}(z) A_{\mu}(x) A_{\nu}(y) \nonumber\\ 
&&- \frac{1}{3!} \int dx \, dy \, dz \, dw \,\frac{\partial}{\partial z^{\lambda}} T^{\lambda \mu \nu \rho}(z,x,y,w) \,\theta_{B}(z) 
A_{\mu}(x) A_{\nu}(y) A_{\rho}(w). \nonumber \\
\eeqn
Notice that in configuration space the four- and the three- point function correlators 
are related by 
\beqn
\frac{\partial}{\partial z^{\lambda}} \overline{T}^{\, \lambda \mu \nu \rho}(z,x,y,w) = \delta(z-w) \overline{T}^{\, \rho \mu \nu}(w,x,y) 
- \delta(z- x) \overline{T}^{\, \mu \nu \rho}(x,y,w) \mbox{+ perm.}
\eeqn
For $T$ we will be using the same conventions as for $\Delta$, with 
$\overline{T}^{\,\lambda \mu \nu}$ indicating a single diagram with non-permuted external gauge lines, while 
$T^{\lambda \mu \nu}$ will denote the symmetrized one (in $\mu \nu$).
Clearly, the one-loop effective theory of this model contains anomalous interactions that need to be cured by the introduction 
of suitable compensator fields. The role of the axion $b$ is to remove the anomalies associated to the triangle 
diagrams which are correlators of one and three chiral currents respectively 
\beq
T^{(\bf AVV)}_{\la\mu\nu}(x,y,z)= \langle 0| T\left( J_\mu(x) J_\nu(y) J^5_\la(z)\right) |0\rangle
\eeq
and 
\beq
T^{(\bf AAA)}_{\la\mu\nu}(x,y,z)= \langle 0| T\left( J^5_\mu(x) J^5_\nu(y) J^5_\la(z)\right) |0\rangle,
\eeq
where
\beq 
J_\mu=-\overline{\psi}\gamma_\mu \psi \qquad J^5_\mu=-\overline{\psi}\gamma_\mu \gamma_5 \psi.
\eeq
We denote by $\Delta(k_1,k_2)$ and $\Delta_3(k_1,k_2)$ their corresponding expressions 
in momentum space 
\beq
(2 \pi)^4 \delta(k - k_1 - k_2) 
\Delta^{\la\mu\nu}(k_1,k_2)=\int dx \,dy \,dz \,e^{ i k_1\cdot x + i k_2\cdot y - i k\cdot z} \, T^{\la\mu\nu}_{(\bf AVV)}(x,y,z)
\label{dd1}
\eeq
\beq
(2 \pi)^4 \delta(k - k_1 - k_2) 
\Delta_3^{\la\mu\nu}(k_1,k_2)=\int dx \,dy \,dz \, e^{ i k_1\cdot x + i k_2\cdot y - i k\cdot z} \, T^{\la\mu\nu}_{(\bf AAA)}(x,y,z).
\label{dd2}
\eeq
Another point to remark is that the invariant amplitudes linear in momenta 
in the definition of the {\bf AVV } triangle diagram correspond, in configuration space, 
to Chern-Simons interactions. In momentum space these are proportional to
\beq
V_{CS}^{\la\mu\nu}(k_1,k_2)=- i \epsilon^{\la\mu\nu\sigma}(k_1 - k_2)_\sigma
\eeq
and we denote with $T_{CS}^{\la\mu\nu}$ the corresponding contribution to the effective Lagrangian 
in Minkowski space (see App.~\ref{app:momentumCS_GS} for an explicit computation) 
\beq
\mathcal{L}_{CS, ABA}=\int dx \, A^{\mu}(x)B^{\nu}(x) F_A^{\rho\sigma}(x) \epsilon_{\mu\nu\rho\sigma}.
\eeq
Moving to the anomalous part of the effective action, this takes the form, for generic gauge bosons $A^{}_i $
\beq
\mathcal{S}^{}_{eff}=\mathcal{S}^{}_0 + \mathcal{S}^{}_1,
\eeq
where 
\beq
\mathcal{S}_1= \sum_{i\, j\, k}\frac{1}{n_i! n_j! n_k!} \, g_{i j k}\int dx \, dy \, dz \,A_i^\la(x)A_j^\mu(y)A_k^\nu(z)  T^{\la\mu\nu}_{A_i A_j A_k}(x,y,z)
\label{Seff}
\eeq
and where $A^{}_i$ indicates an $A$ or a $B$ gauge boson, while 
$g^{}_{ijk}$ is the product of the three coupling constants $g^{}_{A_i} g^{}_{A_j} g^{}_{A_k}$, with an additional normalization due to a counting of identical external gauge bosons $(n_i!)$. All the anomalous contributions are included in the definition.
In order to derive its explicit structure in our simplified cases, we consider the case of the $BAA$ vertex, the other examples being similar. We have, for instance, the partial contribution 
\beq
\mathcal{S}^{BAA}= g^{}_B \int dx \,dy \, dz \, B_{\la}(z)\langle J_5^\la (z) e^{i g_A \int d^4 x J^\mu (x) A_\mu (x)}\rangle, 
\eeq
where the gauge fields are treated as classical fields and $\langle,\rangle$ indicate the vacuum expectation value. 
Expanding to second order, we keep only the connected contributions obtaining for instance
\beq
\mathcal{S}^{BAA}= \frac{i^2}{2!} g^{}_B g_A^2\int dx \, dy \, dz \, B_{\la}(z)A_\mu(x) A_\nu(y) \langle J_5^\la (z) J^\mu(x) J^\nu(y)\rangle.
\eeq
This expression is our starting point for all the 
further analysis. Most of the manipulations concerning the proof of gauge-invariance of the effective 
action are more easily worked out in this formalism. Moving from momentum space to configuration space and back, may also be quite useful in order to detail 
the Ward Identities of a given anomalous effective action. 

In the presence of 
spontaneous symmetry breaking and of St\"{u}ckelberg mass terms one has to 
decide whether the linear mixing between the St\"{u}ckelberg field and the 
gauge boson is kept or not. 
One can keep the mixing and derive ordinary Ward Identities for a given model. 
This is a possibility which is clearly at hand and can be useful. The 
disadvantage of this approach is that there is no gauge-fixing parameter that can be used to analyze the gauge dependence of a given set of 
amplitudes and their cancellation. When the mixing is removed by going to a 
$R^{}_\xi $ gauge, one can identify the set of gauge-invariant 
contributions to a given amplitude and identify more easily the conditions under which a given model becomes unitary. We follow this second approach. We are then able to combine gauge-dependent contributions in such a way that the unphysical poles of 
a given amplitude cancel. The analysis that we perform is limited to the $s$-channel, but the results are easily generalizable to the $t$ and $u$ channels as well. From this simple analysis one can easily extract information on the 
perturbative expansion of the effective action. 
\subsection{The anomalous effective action of the A-B Model }
In the A-B Model, defined in  Eq.~(\ref{llagrangeBC}), the contribution to the anomalous effective action is given by 
\beqa
\mathcal{S}_{an} &=& \mathcal{S}_1 + \mathcal{S}_3 \nonumber \\
\mathcal{S}_{1} &=&  \int d x \, d y \, d z \, \left( \frac{g^{}_{B} \, g^{2}_{A}}{2!} \, 
T_{\bf AVV}^{\la\mu\nu}(x,y,z) B_{\la}(z) A_{\mu}(x)A_{\nu}(y)
 \right),  \nonumber \\
\mathcal{S}_3 &=&  \int d x \, d y \, d z \, \left( \frac{ g^{3}_{B}}{3!} \, 
 T_{\bf AAA}^{\la\mu\nu}(x,y,z) B_\la(z) B_\mu(x) B_\nu(y) \right), 
\eeqa
where we have collected all the anomalous diagrams of the form {\bf AVV} and {\bf AAA}. We can easily express the 
gauge transformations of $A$ and $B$ in the form
\beqa
 \frac{1}{2!} \delta_B  \langle T_{\bf AVV} BAA \rangle &=&  \frac{i}{2!} a^{}_3(\beta) \frac{1}{4} \langle F_A \wedge F_A  
\theta_B \rangle, \nonumber\\  
 \frac{1}{3!} \delta_B \langle T_{\bf AAA} BBB \rangle &=&  \frac{i}{3!} \frac{a^{}_n}{3} \frac{3}{4} \langle  F_B \wedge F_B  
\theta_B \rangle,
\eeqa
where we have left open the choice over the parameterization of the loop momentum, 
denoted by the presence of the arbitrary parameter $\beta$ with
\beq
a_3(\beta)=- \frac{i}{4 \pi^2} + \frac{i}{2 \pi^2} \beta\qquad a_3\equiv \frac{a_n}{3}=-\frac{i}{6 \pi^2},
\eeq
while
\beqa
\frac{1}{2!}\delta_A\langle T_{\bf AVV} BAA\rangle =  \frac{i}{2!}  a_1(\beta) \frac{2}{4} \langle F_B \wedge F_A 
\theta_A \rangle. 
\eeqa
We have the following equations for the anomalous variations of the effective Lagrangian 
\beqa
\delta_B \mathcal{L}_{an} &=&  \frac{i g^{}_{B} g^{\,2}_{A}}{2!} \, a_3(\beta) \frac{1}{4}  F_A \wedge F_A  \theta_B 
+  \frac{ i g^{\,3}_{B} }{3!}\, \frac{a_n}{3}  \frac{3}{4} F_B \wedge F_B \theta_B 
\nonumber \\
\delta_A \mathcal{L}_{an} &=&  \frac{i g^{}_{B} g^{\,2}_{A}}{2! } \, a_1(\beta) \frac{2}{4} F_B \wedge F_A \theta_A, 
\eeqa
while the axionic contributions $\mathcal{L}_{b}$ (Wess-Zumino counterterms), needed to 
restore the gauge symmetry violated at one-loop level, are given by 
\beq
\mathcal{L}_{b} =  \frac{C^{}_{AA}}{M} b \, F_A \wedge F_A  + \frac{C^{}_{BB}}{M} b \, F_B \wedge F_B.  
\eeq
Notice that since the axion shifts only under a gauge variation of the anomalous $U(1)$ gauge field $B$ (and not under $A$), gauge invariance of the effective 
action under a gauge transformation of the gauge field $A$ requires that
\beq
\delta_A \mathcal{L}_{an}=0.
\eeq
Clearly, this condition fixes $\beta= -1/2\equiv \beta_0$ and is equivalent to the CVC condition on $A$ that we had relaxed at 
the beginning. 
Imposing gauge invariance under $B$ gauge transformations, on the other hand, we obtain 
\beqa
\delta_B \left(  \mathcal{L}_b + \mathcal{L}_{an}  \right) = 0
\eeqa
which implies
\beq
C^{}_{AA} =  \frac{i \,g^{}_{B}  g^{\,2}_{A} }{2!} \frac{1}{4} \, a_3(\beta_0) \, \frac{M}{M_1}, 
\qquad C^{}_{BB} =  \frac{i g^{\,3}_{B}}{3!} \frac{1}{4} \, a_n \, \frac{M}{M_1}.
\eeq
These conditions on the coefficients $C$ are sufficient to render gauge-invariant the total Lagrangian. We observe that the presence of an Abelian symmetry which has 
to remain exact and is not accompanied by a shifting axion has important 
implications on the consistency of the theory. We have brought up this example 
because in more complex situations in which a given gauge symmetry is broken 
and the pattern of breakings is such to preserve a final symmetry (for instance 
$U(1)_{em}$), the structure of the anomalous correlators, in some case, is drastically 
constrained to assume the CVC form. However this is not a general result. 

Under a more general assumption, we could have allowed some Chern-Simons contributions in the counterterm Lagrangian. This is an interesting variation 
that can be worked out at a diagrammatic level in order to identify the role 
played by the CS interactions. We will get back to this point once we start our diagrammatic analysis of these simple models.
\section{Higgs-Axion Mixing in $U(1)$ Models: massless axi-Higgs}
Having discussed the consistency to all orders of the effective 
action, we need to discuss the role played by the shifting axions in the spectrum of the theory. We have already pointed out that the axion will mix with the 
remaining scalars of the model. In the presence of a Higgs sector such a mixing can take place at the level of the 
scalar potential, with implications on the mass and the coupling of the axion to the remaining particles of 
the model. In order to understand how this mixing occurs, we consider the case of the A-B Model.

This model has two scalars: the Higgs and the St\"uckelberg fields. We assume that the Higgs field takes a non-zero v.e.v.
and, as usual, the scalar field is expanded around the minimum $< \phi > =v$
\beq
\phi = \frac{1}{\sqrt{2}}\left( v + \phi^{}_{1} + i \phi^{}_{2} \right), 
\eeq
while from the quadratic part of the Lagrangian we can easily read out the mass terms and the 
Goldstone mode present in the spectrum in the broken phase. 
This is given by 
\beqa
\mathcal{L}_q &=& \frac{1}{2} \left(\partial_\mu \phi_{1}\right)^2 +
\frac{1}{2} \left(\partial_\mu \phi_{2}\right)^2 + \frac{1}{2}\left(\partial_\mu b\right)^2  
+ \frac{1}{2}\left(M_1^2 + (q^{}_{B} g^{}_{B} v)^2\right) B_\mu B^{\mu} 
 - \frac{1}{2} m_{1}^2 \phi_{1}^2   \nonumber\\
&& + B_\mu \partial^\mu \left(M_1  b +  v g^{}_{B} q^{}_{B} \phi^{}_{2} \right),   
\eeqa
from which, after diagonalization of the mass terms, we obtain 
\beqa
\mathcal{L}_q &=&\frac{1}{2} \left(\partial^{}_\mu \chi  \right)^2 +
\frac{1}{2} \left(\partial_\mu G^{}_{B}\right)^2  + \frac{1}{2}\left(\partial_\mu h^{}_{1} \right)^2  
 + \frac{1}{2} M_B^2 B_\mu B^{\mu} - \frac{1}{2} m_{1}^2 h_{1}^2     \nonumber\\
&& + M_B B^\mu\partial_\mu G^{}_{B} 
\eeqa
where we have redefined $\phi^{}_{1}(x) = h^{}_{1}(x)$ and $m^{}_{1}=v \sqrt{2 \lambda}$, for the Higgs field and its 
mass. We have identified the linear combinations 
\beqa
\chi &=& \frac{1}{M_B} \left(- M^{}_1 \, \phi^{}_{2} + q^{}_{B} g^{}_{B} v \, b\right),   \nonumber\\
G^{}_{B} &=& \frac{1}{M_B}\left(q^{}_{B} g^{}_{B} v \, \phi^{}_{2} + M^{}_1 \, b\right), 
\eeqa
corresponding to a massless particle, the axi-Higgs $\chi$, and a massless Goldstone mode $G^{}_{B}$. 
The rotation matrix that allows the change of fields $(\phi^{}_{2},b) \to (\chi,G^{}_{B})$ is given by 
\beq
U=\left(
\begin{array}{ll}
 -\cos \theta^{}_{B} & \sin \theta^{}_{B} \\
 \sin  \theta^{}_{B} & \cos \theta^{}_{B}
\end{array}
\right)
\eeq
with $\theta^{}_{B}={\arccos} ({M_1/M_B})={\arcsin}(q^{}_{B} g^{}_{B} v/ M_B)$. 
The axion $b$ can be expressed as linear combination of the rotated fields $\chi, G^{}_{B}$ as 
\beqa
b = \alpha_1 \chi + \alpha_2 G^{}_{B} = \frac{q^{}_{B} g^{}_{B} v}{M_B} \chi + \frac{M_1}{M_B} G^{}_{B}, 
\label{projection}
\eeqa
while the gauge field $B^{}_\mu$ gets its mass $M^{}_B$ through the combined Higgs-St\"{u}ckelberg 
mechanism  
\beq
M_B=\sqrt{M_1^2 + (q^{}_{B} g^{}_{B} v)^2}.
\eeq
To remove the mixing between the gauge field and the Goldstone mode we work in the $R^{}_\xi$ gauge. 
The gauge-fixing Lagrangian is given by 
\beq
\mathcal{L}_{gf}= -\frac{1}{2}\mathcal{G}_B^2 
\eeq
where
\beq
\mathcal{G}_B=\frac{1}{\sqrt{\xi_B}}\left( \partial \cdot B - \xi_B M_B G^{}_{B}\right), 
\eeq
and the corresponding ghost Lagrangian 
\beqa
\mathcal{L}_{B\,gh}&=& {\bar{c}^{}_B}\left( - \square - \xi^{}_B  v (h^{}_{1} + v ) - \xi^{}_B M_1^2\right) c^{}_B. 
\eeqa
The full Lagrangian in the physical basis after diagonalization of the mass matrix becomes 
\beqa
\mathcal{L}&=& -\frac{1}{4} F_A^2 -\frac{1}{4} F_B^2  + 
\mathcal{L}_{B gh} + \mathcal{L}_{f} + \mathcal{L}_{B}
\eeqa
where $\mathcal{L}_f$ denotes the fermion contribution. Its expression can be found in \cite{Coriano:2007fw}.

At this stage there are some observations to be made. 
In the St\"uckelberg phase the axion $b$ is a Goldstone mode, since it can be set to vanish by a gauge transformation on the $B$ gauge boson, while $B$ is massive (with a mass $M^{}_1$) and has 3 degrees of freedom 
(d.o.f.). Therefore in this phase the number of physical d.o.f.'s is  
3 for $B$, 2 for $A$, 2 for the complex scalar Higgs $\phi$, for a total of 7.
After electroweak symmetry breaking we have 3 d.o.f.'s for $B$, 2 for $A$ which remains massless, 1 real Higgs field $h^{}_{1}$ and 1 physical axion $\chi$, for a total of 7.
The axion, in this case, on the contrary of what happens in the case of ordinary 
symmetry breaking is a {\em massless physical } scalar, being not part of the scalar potential. 
Not much surprise so far. Let's now move to the analysis of the case when the axion is part of the 
scalar potential. In this second case the physical axion (the axi-Higgs) gets its mass by the combined 
Higgs-St\"uckelberg mechanism.  
\subsection{Higgs-Axion Mixing in $U(1)$ Models: massive  axi-Higgs \label{sec:massaxion}}
We now illustrate the mechanism of mass generation for the physical axion $\chi$. We focus on the breaking of 
the $U(1)^{}_{B}$ gauge symmetry of the A-B Model. We have a gauge-invariant Higgs potential given by 
\beqn
V_{PQ}(\phi) = \mu^{2} \phi^{*} \phi + \lambda \left( \phi^{*} \phi \right)^2
\eeqn
plus the new $PQ$-breaking terms, allowed by the symmetry \cite{Coriano:2005js}
\beqn
V_{\ds{P}\ds{Q}}(\phi, b) = b^{}_{1} \left( \phi \, e^{- i q^{}_{B} g^{}_{B} \frac{b}{M^{}_{1}}}   \right) 
+ \lambda^{}_1 \left( \phi \, e^{- i q^{}_{B} g^{}_{B} \frac{b}{M^{}_1} }   \right)^{2} 
+2 \lambda^{}_{2} \left(  \phi^{*} \phi  \right)   \left( \phi  \, e^{- i q^{}_B g^{}_{B} \frac{b}{M^{}_1} } \right) 
\eeqn
so that the complete potential considered is given by
\beqn
V(\phi, b) = V_{PQ} +  V_{\slash P \slash Q} + V^{*}_{\slash P \slash Q}.
\label{ppqq}
\eeqn 
We require that the minima of the potential are located at 
\beq
\langle b \rangle=0 \qquad \langle \phi \rangle =v,
\eeq
which imply that the mass parameter satisfies
\beqn
\mu^2 =   - \frac{b_1}{v}  -  2 v^2  \lambda - 2  \lambda_1 - 6 v \lambda_2.
\eeqn
We are interested in the matrix describing the mixing of the $CP$-odd Higgs sector with the axion field $b$, given by
\beqn
\left( \,\,\, \phi_2, \,\,\,  b \,\,\,  \right)  {\mathcal M_{2} }
\left(   \begin{array}{c}
 \phi_2\\
b
 \end{array}   \right)
\eeqn
where $\mathcal M_{2}$ is a symmetric matrix
 \beqn
 \mathcal M_{2}  = -\frac{1}{2} c_{\chi} v^2 \pmatrix{ 1    &  - v \frac{q^{}_B g^{}_{B}}{M^{}_1}  \cr
                                                     - v \frac{q^{}_B g^{}_{B}}{M^{}_1}  &  v^2 \frac{q^2_B g^{2}_{B}}{M^2_1}  \cr}
 \eeqn
and where the dimensionless coefficient multiplied in front is given by
\beqn
c^{}_{\chi} = 4 \left( \frac{b^{}_1}{v^3} + \frac{4 \lambda^{}_1}{v^2} + \frac{2 \lambda^{}_2}{v} \right).
\label{chimass}
\eeqn
Notice that this parameter plays an important role in establishing the size of the mass of the physical axion, after diagonalization. It encloses all the dependence of the mass from the 
$PQ$ corrections to the standard Higgs potential. They can be regarded as corrections 
of order $p/v$, with $p$ being any parameter of the $PQ$ potential. If $p$ is very small, which is the case if 
the $V_{\ds{P} \ds{Q}}$ term of the potential is generated non-perturbatively
(for instance by instanton effects in the case of QCD), the mass of the axi-Higgs 
can be pushed far below the typical mass of the electroweak 
breaking scenario (the Higgs mass), as discussed in \cite{Coriano:2006xh}.   \newline
The mass matrix has 1 zero eigenvalue corresponding to the Goldstone boson $G^{}_B$ and 1 non-zero eigenvalue corresponding to a physical 
axion field $- \chi -$ with mass
\beqn
m^2_{\chi} =  - \frac{1}{2} c_{\chi} v^2 \left[  1 + \frac{q^2_B g^{2}_{B} v^2}{M^2_1} \right] = - \frac{1}{2} c_{\chi} \, v^2 \, 
\frac{M^{2}_{B}}{M^{2}_{1}}. 
\eeqn 
The mass of this state is positive if $c_{\chi} < 0$. 
The rotation matrix that takes from the interaction eigenstates to the mass eigenstates is denoted by $O^{\chi}$
\beqn
\left(   \begin{array}{c}
    \chi     \\
   G^{}_B
 \end{array}   \right) =   O^{\chi}  \left(   \begin{array}{c}
    \phi_{2}     \\
      b
 \end{array}   \right) 
\eeqn
so that we obtain the rotations
\beqn
\phi^{}_2  &=& \frac{1}{M^{}_B}  (- M^{}_1 \, \chi + q^{}_B g^{}_{B} \, v \, G^{}_B   )   \\
b &=& \frac{1}{M^{}_B} (  q^{}_B g^{}_{B} \, v \,\chi  +  M^{}_1 \, G^{}_B   ).   
\eeqn
The mass squared matrix can be diagonalized as
\beqn
(\,\, \chi , \,G^{}_B \,\,)  O^{\chi} \, {\mathcal M_2} (O^{\chi})^T  \left(   \begin{array}{c}
    \chi     \\
   G^{}_B
 \end{array}   \right)   = (\,\, \chi,\, G^{}_B\,\,)       \pmatrix{   m^2_\chi   &   0     \cr
                                                       0   &   0  \cr}   \left(   \begin{array}{c}
    \chi     \\
   G^{}_B
 \end{array}   \right) 
\eeqn
so that $G^{}_B$ is a massless Goldstone mode and $m^{}_\chi$ is the mass of the physical axion. In 
\cite{Coriano:2006xh} one can find a discussion of some physical implications of this field when its mass 
is driven to be small in the instanton vacuum, similarly to the Peccei-Quinn axion 
of a global symmetry. However, given the presence of both mechanisms, the St\"uckelberg and the Higgs, 
 it is not possible to decide whether this axion can be a valid dark-matter candidate. In the same work 
it is shown that the entire St\"uckelberg mechanism can be the result of a partial decoupling of 
a chiral fermion.
\section{Unitarity  issues in the A-B Model in the exact phase} 
In this section we start discussing the issue of unitarity of the model that we have introduced. This is a rather involved 
topic that can be addressed by a diagrammatic analysis of those amplitudes with $s$-channel exchanges of the gauge bosons, 
the axi-Higgs and the 
NG modes generated in the various phases of the theory 
(before and after symmetry breaking, with/without Yukawa couplings). The analysis could, of course, be repeated in the other channels ($t, u$) as well, but no further condition would be 
obtained. 
We will gather all the information coming from the study of the $S$-matrix amplitudes  
to set constraints on the parameters of the model. We have organized our analysis 
as a case-by-case study verifying the cancellation of the unphysical singularities in the amplitude 
in all the phases of the theory, establishing also their gauge independence. This is worked out in the 
$R^{}_\xi$ gauge so to make evident the disappearance of the gauge-fixing parameter in each amplitude. 
The scattering amplitudes are built out of two anomalous diagrams with $s$-channel exchanges of gauge dependent propagators, and in all the cases we are brought back to the analysis of a set of anomalous 
Ward Identities to establish our results.
\subsection{Unitarity and CS interactions in the A-B Model}
The first point that we address in this section concerns the role played by the CS interactions 
in the unitarity analysis of simple $s$-channel amplitudes. This analysis clarifies that CS interactions can be 
included or kept separately from the anomalous vertices with no consequence. 
To show this,  we consider the following modification on the A-B Model, where the CS interactions are generically introduced as possible counterterms in the one-loop effective action, which is given by 
\beqn
{\mathcal L} = {\mathcal L_0} + {\mathcal L_{GS}} + {\mathcal L_{CS}},
\eeqn
where ${\mathcal L_0}$ is already known from the previous sections, but in particular we focus on the components
\beqn
{\mathcal L_{GS}} =  \frac{C^{}_{AA}}{M} b F^{}_{A} \wedge F^{}_{A} +\frac{C^{}_{BB}}{M} b F^{}_{B} \wedge F^{}_{B} 
\eeqn
and
\beqn
{\mathcal L_{CS}} =  d^{}_{1}B^{\mu} A^{\nu} F^{\rho \sigma}_{A} \epsilon_{\mu \nu \rho \sigma}
\equiv d^{}_{1}BA\wedge F^{A}.
\eeqn
Under an $A$-gauge transformation we have \footnote{In the language of the effective action the multiplicity factors are proportional to the number $(n!)$ of external gauge lines of a given type. We keep these factors explicitly to backtrack their origin.}
\beqn
\delta^{}_{A} {\mathcal L} = \frac{d^{}_{1}}{2}\, \theta^{}_{A}  F^{}_{B} \wedge F^{}_{A} 
+ \frac{i}{2!} a^{}_{1}(\beta) \frac{2}{4} \, \theta^{}_{A}  F^{}_{B} \wedge F^{}_{A}, 
\eeqn
so that we obtain
\beqn
\frac{d^{}_{1}}{2} + \frac{i a^{}_{1}(\beta)}{4} = 0 \,\,\,\,\,\leftrightarrow \,\,\,\,\, d^{}_{1}= -\frac{i}{2} a^{}_{1}(\beta).
\eeqn
Analogously, under a $B$-gauge transformation we have 
\beqn
&&\delta^{}_{B} {\mathcal L} = -\frac{d^{}_{1}}{2} \theta^{}_{B}  F^{}_{A} \wedge F^{}_{A} 
+ \frac{i}{2!} a^{}_{3}(\beta) \frac{1}{4}  F^{}_{A} \wedge F^{}_{A} \theta^{}_{B} 
- C^{}_{AA} \frac{ M^{}_{1} \theta^{}_{B} }{M}\,  F^{}_{A} \wedge F^{}_{A}  \nonumber\\
&&\,\,\,\,\,\,\,\,\,\,\,\,\,\,\,\,\,\,- C^{}_{BB} \frac{ M^{}_{1} \theta^{}_{B} }{M} F^{}_{B} \wedge F^{}_{B} 
+ \frac{i}{3!} \frac{a^{}_{n}}{3} \frac{3}{4} \theta^{}_{B} F^{}_{B} \wedge F^{}_{B}, 
\eeqn
to obtain
\beqn
-\frac{d^{}_{1}}{2} - C^{}_{AA} \frac{M^{}_{1}}{M} + \frac{i}{2!} a^{}_{3}(\beta) \frac{1}{4} = 0 \,\,\,\,\,
\leftrightarrow  \,\,\,\,\,
C^{}_{AA}= \left( - \frac{d^{}_{1}}{2} + \frac{i}{2!}a^{}_{3}(\beta)\frac{1}{4}  \right)\frac{M}{M^{}_{1}}.
\label{GSexplicit}
\eeqn
\begin{figure}[t]
{\centering \resizebox*{10cm}{!}{\rotatebox{0}
{\includegraphics{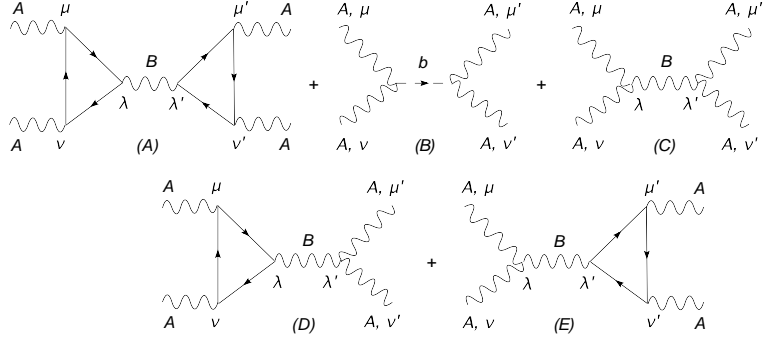}}}\par}
\caption{\small Diagrams involved in the unitarity analysis with external CS interactions.}
\label{chern}
\end{figure}
We refer the reader to one of the appendices where the computation is performed in detail. 
We have shown that the presence of external Ward Identities forcing the 
invariant amplitudes of a given anomalous triangle diagram to assume a specific form, allow to 
re-absorb the CS coefficients inside the triangle, thereby simplifyig the computations. In this specific case the CVC condition for $A$ is a property of the theory. In other cases this does 
not take place. For instance, instead of the condition $a_1(\beta)=0$, a less familiar condition such as $a_3(\beta)=0$ (conserved axial current, CAC) may be needed. In this sense, 
if we define the CVC condition to be the ``standard case'', the CAC condition points toward a {\em new} anomalous interaction. We remark once more that $\beta$ remains ``free'' in the 
SM, since the anomaly traces cancel for all the generators, differently 
from this new situation. The theory allows new CS interactions, with the understanding 
that, at least in these cases, these interactions can be absorbed into a redefinition 
of the vertex. However, the presence of a WI, that allows us to re-express $a_1$ and $a_2$ in terms of $a_3\dots a_6$ in {\em different} 
ways (see Eq.~(\ref{RRos}) for the structure of the triangle diagram), at the same time allows us to come up with different gauge invariant 
expressions of the same vertex function (fixed by a CVC or a CAC condition, depending on the case). These different versions of these 
{\bf AVV} three-point functions are characterized by different (gauge-variant) contact 
interactions since $a_1$ and $a_2$ in Minkowski space contain, indeed, CS 
interactions. We will elaborate on this point in a following section where we discuss the structure of the effective action in Minkowski space.
 
\begin{figure}[t]
{\centering \resizebox*{10.5cm}{!}{\rotatebox{0}
{\includegraphics{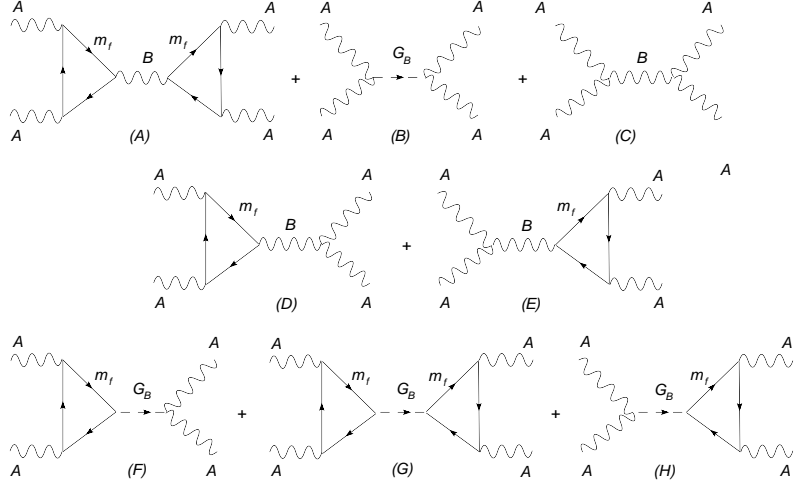}}}\par}
\caption{\small Unitarity issue for the A-B Model in the broken phase.}
\label{chernmassive}
\end{figure}

The extension of this pattern to the broken Higgs phase can be understood 
from Fig.~\ref{chernmassive} where the additional contributions have been 
explicitly included. We have 
depicted the CS terms as separate contributions and shown perfect cancellation also in this case. 
The complete set of diagrams is 
\beq
S^{}_{\xi} = A^{}_{\xi} + B^{}_{\xi} + C^{}_{\xi} + D^{}_{\xi} + E^{}_{\xi} + F^{}_{\xi} + 
G^{}_{\xi} + H^{}_{\xi} \eeq
and the total gauge dependence vanishes. Details on the calculation can be found in the App.~\ref{app:brokenHiggs}.
\section{Gauge independence of the $A$ self-energy to higher orders and the loop counting}
In this section we move to an analysis performed to higher orders that illustrates how the loop expansion and the counterterms get organized so to have a consistent gauge independent theory.

\begin{figure}[t]
{\centering \resizebox*{9cm}{!}{\rotatebox{0}
{\includegraphics{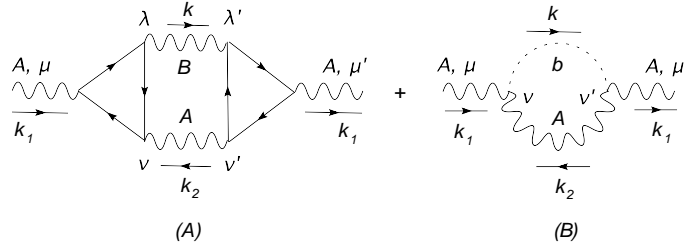}}}\par}
\caption{\small Three-loop cancellations of the gauge dependence.}
\label{bbox}
\end{figure}

For this purpose, let's consider the diagrams in Fig.~\ref{bbox}, which are relevant in order to verify this 
cancellation in the massless fermion case. It shows the self-energy of the $A$ gauge boson. From now on we are dropping all the coupling constants to simplify the notation, which can be 
re-inserted at the end. We have omitted diagrams which are 
symmetric with respect to the 
two  intermediate lines of the $B$ and $A$ gauge bosons, for simplicity. This symmetrization is responsible for the 
cancellation of the gauge dependence of the propagator of $A$ and the vector interaction of $B$, while the gauge 
dependence of the axial-vector contribution of $B$ is canceled by the corresponding Goldstone (shown).
Diagram 
(A) involves three loops and therefore we need to look for cancellations induced by a diagram involving 
the $s$-channel exchange both of an $A$ and of a $B$ gauge boson plus the one-loop 
interactions involving the relevant counterterms. In this case one easily identifies diagram (B) as the 
only possible additional contribution. 

\begin{figure}[h]
{\centering \resizebox*{14cm}{!}{\rotatebox{0}
{\includegraphics{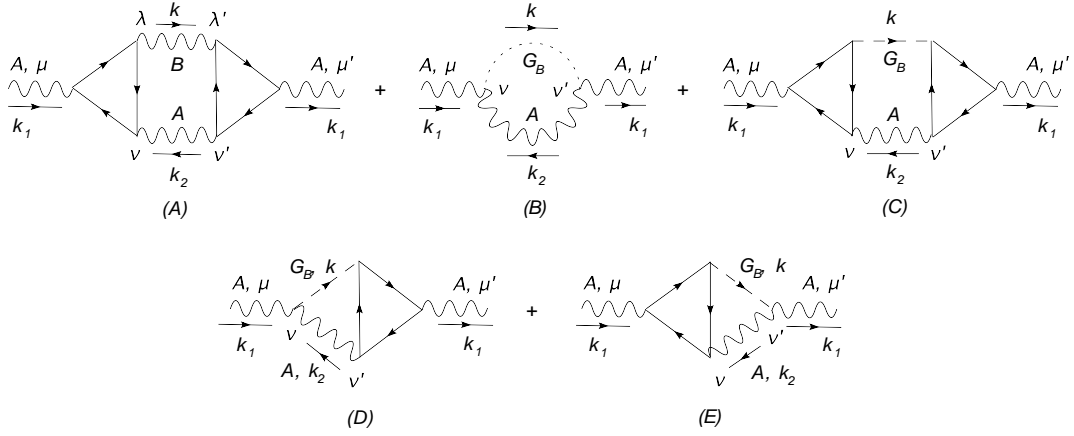}}}\par}
\caption{\small The complete set of diagrams in the broken phase.}
\label{bboxmass}
\end{figure}

To proceed with the demonstration we first isolate the gauge dependence in the propagator for the gauge boson exchanged in the $s$-channel 
\beqa
\frac{- i}{k^2 - M_1^2}\left[  g^{\, \lambda\, \lambda^{\prime}} - \frac{k^\lambda \, k^{\lambda^\prime}}{k^2 - \xi^{}_B M_1^2}
(1 - \xi_B) \right]  &=& 
\frac{- \,i}{ k^2 - M_1^2} \left( g^{\,\lambda\, \lambda^\prime} - \frac{k^\lambda \, 
k^{\lambda^\prime}}{M_1^2} \right) 
+  \frac{- \,i}{ k^2 - \xi_B \,M_1^2}  \left( \frac{ k^{\lambda} k^{\lambda^\prime}}{M_1^2}    \right)  \nonumber\\
&=& P_0^{\lambda \, \lambda^\prime} +  P_{\xi}^{\lambda \, \lambda^\prime}.
\eeqa
and, using this separation, the sum involving the two diagrams gives
\beqn
S = \int \frac{d^{4} k^{}_{2}}{(2 \pi)^{4}} ( {\mathcal A} + {\mathcal B} ),
\eeqn
with the gauge-dependent contributions being given by
\beqn
{\mathcal A}^{}_{ \xi 0} &=&  \, \Delta^{\lambda \mu \nu} (-k^{}_{1}, -k^{}_{2})  P^{\lambda \lambda'}_{\xi} 
\Delta^{\lambda' \mu' \nu'} (k^{}_{1}, k^{}_{2})    P^{\nu \nu'}_{o}    \nonumber\\
&=&  \, \Delta^{\lambda \mu \nu} (-k^{}_{1}, -k^{}_{2})
\left[ \frac{- i }{k^{2} - \xi^{}_{B} M^{2}_{1}} \left( \frac{k^{\lambda} k^{\lambda'}}{ M^{2}_{1}} \right)  \right] 
\Delta^{\lambda' \mu' \nu'} (k^{}_{1}, k^{}_{2}) \left[  \frac{-i g^{\nu \nu'}}{k^{2}_{2}}  \right]   \nonumber\\
{\mathcal B}_{\xi 0} &=& 4 \times  \left( \frac{4}{M} C^{}_{AA} \right)^{2} \epsilon^{\, \mu \nu \rho \sigma} 
k^{}_{1\rho} k^{}_{2\sigma} \frac{i}{k^{2} - \xi^{}_{B} M^{2}_{1}} \epsilon^{\, \mu' \nu' \rho' \sigma'} 
k^{}_{1\rho'} k^{}_{2\sigma'}   P^{\nu \nu'}_{o}. 
\eeqn
Using the anomaly equations and substituting the appropriate value already obtained for the WZ-coefficient, we obtain a vanishing expression
\beqn
{\mathcal A}^{}_{ \xi 0} + {\mathcal B}^{}_{ \xi 0}  
&=&   ( - a^{}_{3}(\beta) \epsilon^{\, \mu \nu \rho \sigma} 
k^{}_{1\rho} k^{}_{2\sigma})
\left[ \frac{- i }{k^{2} - \xi^{}_{B} M^{2}_{1}}  \frac{1}{ M^{2}_{1} }  \right] 
 ( a^{}_{3}(\beta) \epsilon^{\, \mu' \nu' \rho' \sigma'} k^{}_{1\rho'} k^{}_{2\sigma'} )  P^{\nu \nu'}_{o}  \nonumber\\
&+& 4 \, \frac{16}{ M^{2} } \left( \frac{i}{2!} \frac{1}{4} a^{}_{3}(\beta)  \frac{M}{M^{}_{1}} \right)^{2}
 \epsilon^{\, \mu \nu \rho \sigma} 
k^{}_{1\rho} k^{}_{2\sigma} \frac{i}{k^{2} - \xi^{}_{B} M^{2}_{1}} \epsilon^{\, \mu' \nu' \rho' \sigma'} 
k^{}_{1\rho'} k^{}_{2\sigma'}P^{\nu \nu'}_{o}= 0.
\eeqn
After symmetry breaking, with massive fermions, the pattern gets far more involved and is described in 
Fig.~\ref{bboxmass}. In this case we have the sum  
\beqn
S = \int \frac{d^{4} k^{}_{2}}{(2 \pi)^{4}} ( {\mathcal A} + {\mathcal B}+ {\mathcal C} + {\mathcal D} + {\mathcal E} )
\eeqn
and it can be shown by direct computation that the gauge dependences cancel in this combination. 
The details about this cancellation are given in App.~\ref{app:self_brokenHiggs}. 
Notice that the pattern to follow in order to identify the relevant diagrams is quite clear, 
and it is not difficult to identify at each order of the loop expansion
contributions with the appropriate $s$-channel exchanges that combine into gauge invariant amplitudes. These are built identifying direct contributions and counterterms in an appropriate fashion, counting the counterterms at their appropriate order in Planck's constant $(h)$.
\section{Unitarity Analysis of the A-B Model in the broken phase}
\begin{figure}[t]
\begin{center}
\includegraphics[scale=0.7,angle=0]{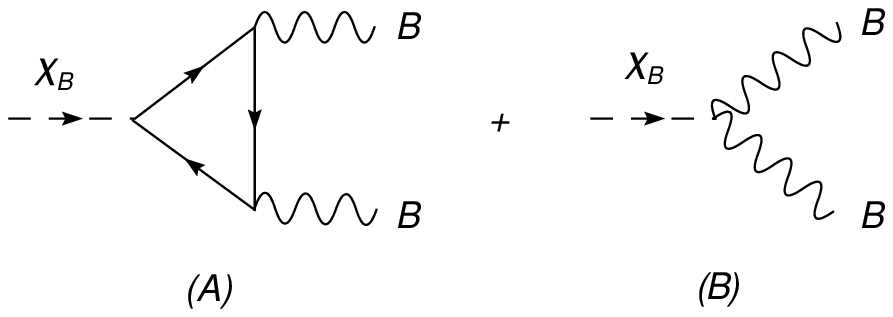}
\caption{\small Decay amplitude for $\chi\to B B $.}
\label{chi-decay}
\end{center}
\end{figure}

In the broken phase and in the presence of Yukawa couplings there are some modifications that take place, since the $s$-channel 
exchange of the $b$ axion is rotated on the two components (the Goldstone and physical axion $\chi$) as shown later in Fig.~\ref{gs}. 
The introduction of the Yukawa interaction and the presence of a symmetry breaking phase determines an interaction of the axion with the 
fermions. Therefore, let's consider the Yukawa Lagrangian    
\beqa
{\cal{L}}^{}_{Yuk} =- \lambda^{}_1 \, \overline{\psi}_{L} \phi \, \psi^{}_{R} 
- \lambda^{}_1 \, \overline{\psi}_{R} \phi^{*} \psi^{}_{L},
\eeqa
which is needed to extract the coupling between the axi-Higgs and the 
fermions. We focus on the term
\beqa
{\cal{L}}^{}_{Yuk}(\phi^{}_{2}) =- \frac{ \lambda^{}_1 }{2} \left[ \overline{\psi} ( 1 + \gamma^5) \psi 
\, \frac{ i \phi_{2} }{\sqrt 2} 
-  \overline{\psi} (1 - \gamma^5) \psi\, \frac{i \phi_{2} }{\sqrt 2}   \right],
\eeqa
having expanded around the Higgs vacuum. Performing a rotation to express the pseudoscalar 
Higgs phase $\phi^{}_{2}$ in  terms of the physical axion and the NG boson 
$$\phi^{}_{2} = - \frac{M_1}{M_B} \chi +  \frac{ q^{}_{B} g^{}_{B} v}{M_B} G^{}_{B},$$ one extracts a 
$\chi\overline{\psi} \psi$ coupling of the kind
\beqa 
{\cal{L}}^{}_{Yuk}(\chi) =  \frac{\lambda^{}_1}{\sqrt 2} \frac{M_1}{M_B} i \, \overline{\psi}  \chi \gamma^5 \psi 
=  \frac{m^{}_f}{v} \frac{M^{}_1}{M^{}_B}  i \, \overline{\psi}  \chi \gamma^5 \psi ,
\eeqa
and a coupling $G\overline{\psi} \psi$ for the Goldstone mode
\beqa
{\cal{L}}^{}_{Yuk}(G^{}_{B}) = -
\frac{\lambda^{}_1}{\sqrt 2} \frac{q^{}_{B} g^{}_{B} v}{M_B} i  \,  \overline{\psi} \, G^{}_{B}  \gamma^5 
\psi =  2 g^{}_{B} \frac{ m^{}_{f} }{M_B} 
i \overline{\psi} \gamma^5 \psi G^{}_{B}.  
\eeqa
Having fixed the Yukawa couplings of the model, we move to the 
analysis of the same diagrams of the previous section in the broken phase. Preliminarily, we need to identify the 
structure of the anomaly equation for the fermionic three-point functions with their complete mass dependence.  
In the case of massive fermions the anomalous Ward Identities for an {\bf AVV} triangle are of the form 
\beqn
k_{1\mu}\Delta^{\lambda\mu\nu}(\beta,k_1,k_2)&=& a_1(\beta) \epsilon^{\lambda\nu\alpha\beta}
k_1^\alpha k_2^\beta,\nonumber\\
k_{2\nu}\Delta^{\lambda\mu\nu}(\beta,k_1,k_2)&=& a_1(\beta) \epsilon^{\lambda\mu\alpha\beta}
k_2^\alpha k_1^\beta,\nonumber\\
k_\lambda\Delta^{\lambda\mu\nu}(\beta,k_1,k_2)&=& a_3(\beta) \epsilon^{\mu\nu\alpha\beta}
k_1^\alpha k_2^\beta +  2 m_{f} \Delta^{\mu \nu},
\label{bbshift}
\eeqn
and in the case of {\bf AAA} triangle $\Delta_{3}^{\lambda \mu \nu}(\beta, k_1, k_2)= \Delta_{3}^{\lambda \mu \nu}( k_1, k_2)$, with  Bose symmetry providing a factor 1/3 for the distribution of the anomalies among the 3 vertices
\beqn
k_{1\mu}\Delta_{3}^{\lambda\mu\nu}(k_1,k_2)&=& \frac{a_n}{3} \epsilon^{\lambda\nu\alpha\beta}
k_1^\alpha k_2^\beta +   2 m_{f} \Delta^{\lambda \nu}       ,\nonumber\\
k_{2\nu}\Delta_{3}^{\lambda\mu\nu}(k_1,k_2)&=& \frac{a_n}{3} \epsilon^{\lambda\mu\alpha\beta}
k_2^\alpha k_1^\beta +   2 m_{f} \Delta^{\lambda \mu}          ,\nonumber\\
k_\lambda \Delta_{3}^{\lambda\mu\nu}(k_1,k_2)&=& \frac{a_n}{3} \epsilon^{\mu\nu\alpha\beta}
k_1^\alpha k_2^\beta +   2 m_{f} \Delta^{\mu \nu},
\label{bbshift}
\eeqn
where we have dropped the appropriate coupling constants common to both sides. The amplitude $\Delta^{\mu\nu}$ is given by 
\beqn 
\Delta^{\mu \nu} =  \int \frac{ d^4 q }{ ( 2 \pi )^4 } \frac{ Tr \left[ \gamma^{5} ( \slash{q} - \slash{k} + m_{f}) 
\gamma^{\nu} \gamma^{5} ( \slash{q} - \ds{k}_1 + m_{f})  \gamma^{\mu} \gamma^5 (\slash{q} + m_{f}) \right] }{
[ q^2 - m_{f}^2] [ ( q - k )^2 - m_{f}^2 ] [ (q - k_1)^2 - m_{f}^2 ] } \mbox{+ exch.}
\label{chidec}
\eeqn
and can be expressed as a two-dimensional integral using the Feynman parameterization. We find
\beqn 
\Delta^{\mu\nu}&=&  \epsilon^{\alpha \beta \mu \nu }  k_1^{\alpha} 
k_2^{\beta} m_f \left( \frac{1}{ 2 \pi^2}  \right) I, 
\label{chigg}
\eeqn
with $I$ denoting the formal expression of the integral
\beq
 I\equiv \int^{1}_{0} dx \int^{1-x}_{0} dy  \frac{1}{\Delta(x,y,m_f,m_\chi,M_B)}. 
\eeq
We have dropped the charge dependence since we have normalized the charges to unity and we have defined 
\beq
\Delta(x,y,m_f,m_\chi,M_B) = \Sigma^2 - D =  m_f^2 - x\,y \,m_{\chi}^2 + M_B^2 (x+y)^2 - x M_B^2 - y M_B^2 \equiv \Delta(x,y)
\label{delta}.
\eeq
We can use this amplitude to compute the one-loop decay of the axi-Higgs in this simple model, 
shown in Fig.~\ref{chi-decay}, which is given by
\beqn
{\mathcal M}^{\mu \nu}(\chi \rightarrow BB) &=&  {\mathcal A} + {\mathcal B}   \nonumber\\
&=& i \frac{m^{}_f}{v} \frac{M^{}_{1}}{M^{}_{B}} \Delta^{\mu \nu} (k^{}_{1}, k^{}_{2})
+ \alpha^{}_{1} \frac{4}{M} C^{}_{BB} \epsilon^{\alpha \beta \mu \nu }  k_1^{\alpha} 
k_2^{\beta}    \nonumber\\
&=& i \frac{m^{}_f}{v} \frac{M^{}_{1}}{M^{}_{B}} \Delta^{\mu \nu} (k^{}_{1}, k^{}_{2})
 -    \frac{2 g^{}_{B} v}{M^{}_{B}} \left(  \frac{4}{M} \frac{i}{3!} \frac{1}{4} a^{}_{n} \frac{M}{M^{}_{1}}  \right) 
 \epsilon^{\alpha \beta \mu \nu }  k_1^{\alpha} 
k_2^{\beta},
\eeqn 
where $\alpha^{}_{1}$ is the coefficient that rotates the axion $b$ on the axi-Higgs particle $\chi$.
\subsection{A-B Model: $BB \rightarrow BB$ mediated by a $B$ gauge boson in the broken phase}
\begin{figure}[t]
\begin{center}
\includegraphics[scale=0.7,angle=0]{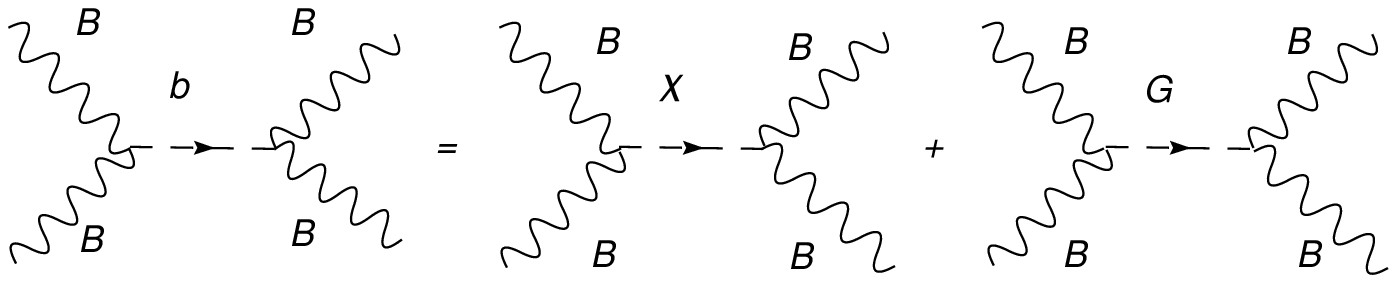}
\caption{\small Diagrams from the Green-Schwarz coupling after symmetry breaking.}
\label{gs}
\end{center}
\end{figure}
\begin{figure}[tbh]
\begin{center}
\includegraphics[scale=0.6,angle=0]{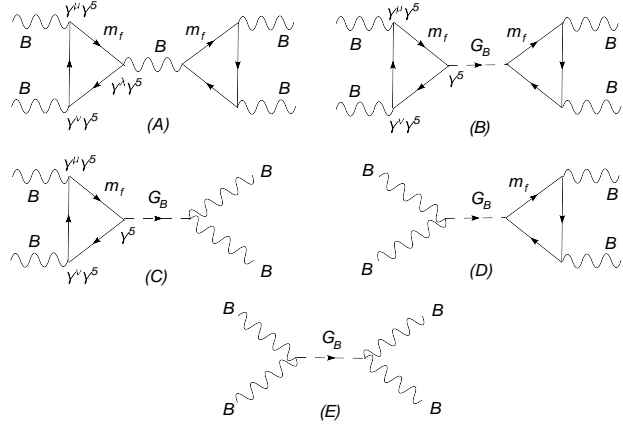}
\caption{\small Cancellation of the gauge dependence after spontaneous symmetry breaking.}
\label{brokenphase}
\end{center}
\end{figure}

A second class of contributions that require a different distribution of the partial anomalies are those involving $BBB$ diagrams. They appear in the $BB\to BB$ amplitude, mediated by the exchange of a $B$ gauge boson of mass 
$M^{}_B = \sqrt{M_1^2 + (2 g^{}_{B} v)^2}$. Notice that $M^{}_B$ gets its mass both from the Higgs and the St\"{u}ckelberg sectors. 
 The relevant diagrams for this check are shown in Fig.~\ref{brokenphase}. 
We have not included the exchange of the physical axion, since this is not gauge-dependent. We have only 
left the gauge-dependent contributions. 
Notice that the expansion is valid at 
two-loop level and involves two-loop diagrams built as combinations of 
the original diagrams and of the one-loop counterterms. There 
are some comments that are due in order to appreciate the way the 
cancellations take place. If we neglect the Yukawa couplings the diagrams 
(B), (C) and (D) are absent, since the Goldstone does not couple to a massless fermion. In this case, the axion $b$ is rotated, as in the previous sections, into a Goldstone mode $G_B$ and a physical axion $\chi$ (see Fig.~\ref{gs}). On the other hand, if we include the Yukawa couplings then the entire set of diagrams is needed. 
From diagram (E) we obtain the partial contribution 
\beqn
\mathcal{E}^{}_{\xi} = 
4 \times \left( \frac{4}{M} \alpha^{}_2 C^{}_{BB} \right)^2 \epsilon^{\mu\nu\rho\sigma} k_{1}^\rho k_{2}^\sigma  
 \left( \frac{i}{k^2 - \xi_B M_{B}^2} \right)  \epsilon^{\mu^\prime  \nu^\prime   \rho^\prime \sigma^\prime} 
k_{1}^{\rho^\prime} k_{2}^{\sigma^\prime},
\eeqn
where the overall factor of 4 in front is a symmetry factor, 
the coefficient $\alpha^{}_2$ comes from the rotation of the $b$ axion 
over the Goldstone boson ($\alpha_2 = \frac{M_1}{M_B}$), and the coefficient $C^{}_{BB}$ has already been determined from the condition of 
gauge invariance of the anomalous effective action before symmetry breaking.
Similarly, from diagram (B) we get the term 
\beqn
\mathcal{B}_{\xi} = 
( g^{}_{B} )^3 \Delta^{\mu \nu}(-k^{}_1,- k^{}_2) \left( 2i \, \frac{ m_{f} }{M_B} \right)  \frac{i}{k^2 - \xi_B M_B^2} 
\left( 2 i \frac{ m_{f} }{M_B} \right)
( g^{}_{B})^3 \Delta^{\mu^\prime \nu^\prime}(k^{}_1, k^{}_2),
\eeqn 
while diagram (C) gives
\beqn
\mathcal{C}_{\xi} = 
2 \times ( g^{}_{B} )^3 \Delta^{\mu \nu}(-k^{}_1,- k^{}_2) \left( 2\,i\frac{ m_{f} }{M_B} \right)  \frac{i}{k^2 - \xi_B M_B^2}  
\left( \frac{4}{M} \alpha^{}_2 C^{}_{BB} \, 
\epsilon^{\mu^\prime \nu^\prime \rho^\prime \sigma^\prime}  k_1^{\rho^\prime}  k_2^{\sigma^\prime} \right),
\eeqn
having introduced also in this case a symmetry factor. 
Finally (D) gives 
\beqn
\mathcal{D}^{}_{\xi} = 
2\times \left( \frac{4}{M} \alpha_2 C^{}_{BB} \epsilon^{\mu \nu \rho \sigma}  k_1^{\rho}  k_2^{\sigma} \right)
\frac{i}{k^2 - \xi_B M_B^2} \left(2\,i \frac{ m_{f} }{M_B} \right) \Delta^{\mu^\prime \nu^\prime}(k^{}_{1}, k^{}_{2}) ( g^{}_{B} )^3.
\eeqn
We will be using the anomaly equation in the massive fermion case, having distributed the anomaly equally among the three $B$ vertices 
\beqn
k^\lambda \Delta^{\lambda \mu \nu}=  \frac{a_n}{3} \epsilon^{\mu \nu \alpha \beta} k_1^{\alpha} k_2^{\beta} 
+ 2 m_{f} \Delta^{\mu \nu}. 
\eeqn 
Separating in diagram (A) the gauge dependent part of the propagator of the 
boson $B$  from the rest we obtain
\beqa
\mathcal{A}^{}_{\xi}&=& \Delta^{\lambda \mu \nu } \frac{- i }{k^2 - \xi_B M_B^2} \left( \frac{k^\lambda k^{\lambda^\prime}}{M^{2}_{B}} \right) 
\Delta^{\lambda^\prime \mu^\prime \nu^\prime}  \nonumber \\ 
&=&  \left( \frac{a_n}{3} \epsilon^{\mu \nu \alpha \beta} k_1^{\alpha} k_2^{\beta} 
+ 2 m^{}_{ f} \Delta^{\mu \nu} \right) ( g^{}_{B})^{3} \frac{ i }{k^2 - \xi^{}_B M_B^2} \frac{1}{M_B^2} \left( \frac{a_n}{3} 
\epsilon^{\mu^\prime \nu^\prime \alpha^\prime \beta^\prime} 
k_1^{\alpha^\prime} k_2^{\beta^\prime} +  2 m^{}_{ f} \Delta^{\mu^\prime \nu^\prime}   \right)  ( g^{}_{B})^{3}  \nonumber\\ 
&=& \frac{ i }{k^2 - \xi_B M_B^2} \frac{g^{\,6}_{B}}{M_B^2}  \left[  \left( \frac{a_n}{3} \right)^2 \epsilon^{\mu \nu \alpha \beta } 
\epsilon^{\mu^\prime \nu^\prime \alpha^\prime \beta^\prime } k_1^\alpha k_2^\beta k_1^{\alpha^\prime} 
k_2^{\beta^\prime} \right. \nonumber\\ 
&&\left. +\frac{a_n}{3} \epsilon^{\mu \nu \alpha \beta}  k_1^\alpha k_2^\beta \, 2 m^{}_{ f} \Delta^{\mu^\prime \nu^\prime} 
 + 2 m^{}_{ f} \Delta^{\mu \nu} \frac{a_n}{3} \epsilon^{\mu^\prime \nu^\prime \alpha^\prime \beta^\prime } k_1^{\alpha^\prime}
k_2^{ \beta^\prime} \right.  \nonumber\\
&&\left.+ \left( 2 m^{}_{ f} \Delta^{\mu \nu} \right) ( 2 m^{}_{ f} \Delta^{\mu^\prime \nu^\prime}) \right].  
\label{brokenABC}
\eeqa
The first term in (\ref{brokenABC}) is exactly canceled by the contribution 
from diagram (E). The last contribution cancels by the contribution from diagram (B). Finally diagrams (C) 
and (D) cancel against the second and third contributions from diagram (A).
\section{The Effective Action in the Y-B Model}
\begin{figure}[h]
\begin{center}
\includegraphics[scale=0.7,angle=0]{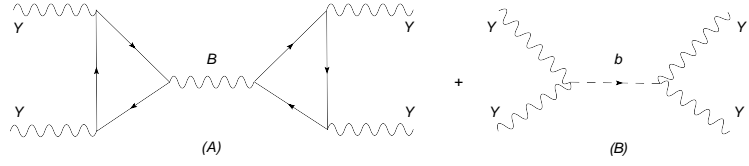}
\caption{\small Unitarity diagrams in the Y-B Model.}
\label{yb-unitarity}
\end{center}
\end{figure}

We anticipate in this section some of the methods that will be used 
in the next chapter for the analysis 
of a realistic model. In the A-B Model, in order to simplify the 
analysis, we have assumed that the coupling of the $B$ gauge boson was 
purely axial-vector-like while $A$ was purely vector-like. Here we discuss 
a gauge structure which allows both gauge bosons to have combined vector and axial-vector 
couplings. We will show that the external Ward Identities of the model involve a specific 
definition of the shift parameter in one of the triangle diagrams that forces the axial-vector current to be conserved ($a^{}_3(\beta)=0$). This result, new compared to the case of the SM, shows the presence of an effective CS term in some amplitudes. 
The Lagrangian that we choose to exemplify this new situation is given by 
\beqn
{\mathcal L}_{0} &=& 
| (\partial^{}_{\mu} + i g^{}_{Y} q^{Y}_{u} Y^{}_{\mu} + i g^{}_{B} q^{B}_{u} B^{}_{\mu}) \phi_{u} |^{2} + 
| (\partial^{}_{\mu} + i g^{}_{Y} q^{Y}_{d} Y^{}_{\mu} + i g^{}_{B} q^{B}_{d} B^{}_{\mu}) \phi_{d} |^{2}
- \frac{1}{4} F^{2}_{Y} - \frac{1}{4} F^{2}_{B}   \nonumber\\
&+& \frac{1}{2} (\partial^{}_{\mu} b + M^{}_{1}B^{}_{\mu})^{2} - \lambda^{}_{u} \left( |\phi^{}_{u}|^{2} 
- \frac{v^{}_{u}}{2} \right)^{2} - \lambda^{}_{d} \left( |\phi^{}_{d}|^{2} 
- \frac{v^{}_{d}}{2} \right)^{2} + {\mathcal L}_{f} + {\mathcal L}_{Yuk},
\eeqn
where the Yukawa couplings are given by
\beqn
{\mathcal L}_{Yuk} =- \lambda^{}_{1} \overline{\psi}_{1L} \phi^{}_{u} \psi^{}_{1R} - \lambda^{}_{1} \overline{\psi}_{1R}
 \phi^{*}_{u} \psi^{}_{1L}
-  \lambda^{}_{2} \overline{\psi}_{2L} \phi^{}_{d} \psi^{}_{2R} - \lambda^{}_{2} \overline{\psi}_{2R} \phi^{*}_{d} \psi^{}_{2L},
\eeqn
with $L$ and $R$ denoting left- and right- handed fermions. The fermion currents are
\beqn
{\mathcal L}_{f} &=& \overline{\psi}_{1L} i \gamma^{\mu} \left[ \partial^{}_{\mu} + i g^{}_{Y} q^{Y}_{1L} Y^{}_{\mu} 
+ i g^{}_{B} q^{B}_{1L}B^{}_{\mu}  \right] \psi^{}_{1L} + 
 \overline{\psi}_{1R} i \gamma^{\mu} \left[ \partial^{}_{\mu} + i g^{}_{Y} q^{Y}_{1R} Y^{}_{\mu} 
+ i g^{}_{B} q^{B}_{1R}B^{}_{\mu}  \right] \psi^{}_{1R} \nonumber\\ 
&+& \overline{\psi}_{2L} i \gamma^{\mu} \left[ \partial^{}_{\mu} + i g^{}_{Y} q^{Y}_{2L} Y^{}_{\mu} 
+ i g^{}_{B} q^{B}_{2L}B^{}_{\mu}  \right] \psi^{}_{2L}  + 
 \overline{\psi}_{2R} i \gamma^{\mu} \left[ \partial^{}_{\mu} + i g^{}_{Y} q^{Y}_{2R} Y^{}_{\mu} 
+ i g^{}_{B} q^{B}_{2R}B^{}_{\mu}  \right] \psi^{}_{2R} \nonumber\\ 
\eeqn
so that, in general, without any particular charge assignment, both gauge bosons show vector and axial-vector couplings.
In this case we realize an anomaly-free charge assignment for the 
hypercharge by requiring that $q^{Y}_{2L} = - q^{Y}_{1L}, q^{Y}_{2R} = - q^{Y}_{1R}$,
which cancels the anomaly for a $YYY$ triangle since 
\beqn
\sum^{}_{f=1,2} (q^{Y}_{f})^{3} = (q^{Y}_{1R})^{3} -  (q^{Y}_{1L})^{3} + (q^{Y}_{2R})^{3} - (q^{Y}_{2L})^{3} 
=  (q^{Y}_{1R})^{3} -  (q^{Y}_{1L})^{3} - (q^{Y}_{1R})^{3} 
+ (q^{Y}_{1L})^{3} = 0.
\eeqn
This condition is similar to the vanishing of the $YYY$ anomaly for the hypercharge in the 
SM, and for this reason we will assume that it holds also in our simplified model.
Before symmetry breaking the $B$ gauge boson has a Goldstone coupling coming from the St\"{u}ckelberg mass term due 
to the presence of a Higgs field.
The effective action for this model is given by 
\beqn
{\mathcal S}^{}_{eff} =  {\mathcal S}^{}_{an} + {\mathcal S}^{}_{WZ},  
\eeqn
which reads
\beqn
{\mathcal S}^{}_{an} 
&=&  \frac{1}{3!}\langle T^{\lambda \mu \nu}_{BBB}(z,x,y) B^{\lambda} B^{\mu} B^{\nu}\rangle + 
\frac{1}{2!}\langle  T^{\lambda \mu \nu}_{BYY}(z,x,y) B^{\lambda} Y^{\mu} Y^{\nu} \rangle
+ \frac{1}{2!} \langle T^{\lambda \mu \nu}_{YBB}(z,x,y) Y^{\lambda} B^{\mu} B^{\nu}  \rangle \nonumber \\ 
\eeqn
with
\beqn
{\mathcal S}^{}_{WZ} =  \frac{C^{}_{YY}}{M}\,\langle b\, F^{Y} \wedge F^{Y}\rangle + 
 \frac{C^{}_{BB}}{M}\, \langle b\, F^{B} \wedge F^{B}\rangle + \frac{C^{}_{BY}}{M}\,\langle b\, F^{B} \wedge F^{Y} \rangle
\eeqn
denoting the WZ counterterms.
Only for the triangle $BBB$ we have assumed an anomaly symmetrically distributed, all the other anomalous diagrams having an {\bf AVV} anomalous 
structure, given in momentum space by
\beqn
\Delta^{\lambda \mu \nu}_{ijk} = \frac{i^{3}}{2} \sum_{f}\left[ q^{i}_{f} q^{j}_{f} q^{k}_{f} \right] 
\int \frac{d^{4} p}{(2 \pi)^{4}} 
\frac{ Tr[ \gamma^{\lambda} \gamma^{5} \slash{p} \gamma^{\mu} 
(\slash{p} - \ds{k}_{1}) \gamma^{\nu} (\slash{p} - \ds{k}) ]}{p^{2} 
(p-k^{}_{1})^{2} (p-k)^{2} } 
+ \{ \mu, k^{}_{1} \,\, \leftrightarrow \,\, \nu, k^{}_{2} \}, 
\eeqn
with indices running over i, j, k =Y, B. The sum over the fermionic spectrum involves the charge operators in the chiral basis
\beqn
D^{}_{ijk} = \frac{1}{2} \sum_{f=1,2} \left[ q^{i}_{f} q^{j}_{f} q^{k}_{f} \right] \equiv
\frac{1}{2} \sum_f\left( q^{i}_{fR} q^{j}_{fR} q^{k}_{fR} - q^{i}_{fL} q^{j}_{fL} q^{k}_{fL}\right).  
\eeqn
Computing the $Y$-gauge variation for the effective one-loop anomalous action under the trasformations 
$Y^{}_{\mu} \rightarrow Y^{}_{\mu} + \partial^{}_{\mu} \theta^{}_{Y}$ we obtain
\beqn
\delta_{Y} {\mathcal S}_{an} = \frac{i}{2!} a^{}_{1}(\beta^{}_{1}) \frac{2}{4} \theta^{}_{Y} F^{}_{B} \wedge F^{}_{Y} D_{BYY} +  
\frac{i}{2!} a^{}_{3}(\beta^{}_{2}) \frac{1}{4} \theta^{}_{Y} F^{}_{B} \wedge F^{}_{B} D^{}_{YBB},
\eeqn
and, similarly,  for $B$-gauge transformations $B^{}_{\mu} \rightarrow B^{}_{\mu} + \partial^{}_{\mu} \theta^{}_{B}$ we obtain
\beqn
\delta_{B} {\mathcal S}_{an} &=&  \frac{i}{3!} \frac{a^{}_{n}}{3} \frac{3}{4}  \theta^{}_{B} F^{}_{B} \wedge F^{}_{B} D^{}_{BBB}  + 
 \frac{i}{2!} a^{}_{3}(\beta^{}_{1}) \frac{1}{4} \theta^{}_{B} F^{}_{Y} \wedge F^{}_{Y} D^{}_{BYY}   \nonumber\\
&&+ \frac{i}{2!} a^{}_{1}(\beta^{}_{2}) \frac{2}{4} \theta^{}_{B} F^{}_{B} \wedge F^{}_{Y} D^{}_{YBB},  
\eeqn
so that to get rid of the anomalous contributions due to gauge variance we have to fix the parameterization of the loop momenta with parameters
\beqn
\beta^{}_{1} = \overline{\beta^{}}_{1}  = - \frac{1}{2},  \qquad  
\beta^{}_{2} = \overline{\beta^{}}_{2} = + \frac{1}{2}.
\eeqn
Notice that while $\beta^{}_{1}$ corresponds to a canonical choice (CVC condition), the second amounts to a 
condition for a conserved axial-vector current, which can be interpreted as a condition that forces a CS counterterm in the parameterization of the triangle amplitude. 
Having imposed these conditions to cancel the anomalous variations for the $Y$ gauge boson, we can determine the WZ coefficients as
\beqn
C^{}_{BB}=  \frac{M}{M^{}_{1}} \frac{i}{3!} a^{}_{n} \frac{1}{4} D^{}_{BBB}, \,\,\,\,\,
C^{}_{YY}=  \frac{M}{M^{}_{1}} \frac{i}{2!} a^{}_{3}(\overline{\beta}_{1}) \frac{1}{4} D^{}_{BYY},   
\,\,\,\,\,  C^{}_{BY}=  \frac{M}{M^{}_{1}} \frac{i}{2!} a^{}_{1}(\overline{\beta}_{2}) \frac{2}{4} D^{}_{YBB}.
\eeqn 
Having determined all the parameters in front of the counterterms we can test the unitarity of the model. Consider the process $YY \rightarrow YY$ 
mediated by an $B$ gauge boson depicted in Fig.~\ref{yb-unitarity}, one can easily check that the gauge dependence vanishes. 
In fact we obtain
\beqn
{\mathcal S}^{}_{\xi} &=& 
{\mathcal A}_{\xi} + {\mathcal B}_{\xi}  \nonumber\\ 
&=&\Delta^{\lambda \mu \nu}(-k^{}_{1}, -k^{}_{2}) 
\left[  \frac{- i}{ k^2 - \xi_{B} M^{2}_{1}} \left( \frac{k^{\lambda} k^{\lambda'}}{M^{2}_{1}}
\right) \right]  \Delta^{\lambda' \mu' \nu'}(k^{}_{1}, k^{}_{2})  (D^{}_{BYY})^{2}  \nonumber\\
&&+ 4 \left(  \frac{4}{M} C_{YY}  \right)^{2}  \epsilon^{\mu \nu \rho \sigma} 
k^{\rho}_{1} k^{\sigma}_{2}  \left( \frac{i}{k^2 - \xi_{B} M^{2}_{1}} \right)  
 \epsilon^{\mu' \nu' \rho' \sigma'} k^{\rho'}_{1} k^{\sigma'}_{2}  
  \nonumber\\
&=&   \frac{- i}{ k^2 - \xi_{B} M^{2}_{1}} \frac{1}{M^{2}_{1}} 
 \left(  - a^{}_{3}( \overline{\beta^{}}_{1} ) \epsilon^{\mu \nu \rho \sigma} k^{\rho}_{1} k^{\sigma}_{2} \right) 
  \left( a^{}_{3}(\overline{\beta^{}}_{1} )    \epsilon^{\mu' \nu' \rho' \sigma'} k^{\rho'}_{1} k^{\sigma'}_{2} \right) 
 (D^{}_{BYY})^{2}    \nonumber\\
 &+& 4 \frac{16}{M^{2}}\left( \frac{M}{M^{}_{1}} \frac{i}{2!} a^{}_{3}( \overline{\beta}_{1}) \frac{1}{4} D^{}_{BYY}\right)^{2}  
 \epsilon^{\mu \nu \rho \sigma} 
k^{\rho}_{1} k^{\sigma}_{2}  \left( \frac{i}{k^2 - \xi_{B} M^{2}_{1}} \right) 
  \epsilon^{\mu' \nu' \rho' \sigma'} k^{\rho '}_{1} k^{\sigma '}_{2}    = 0
\eeqn 
where we have included the corresponding symmetry factors. There are some comments that are in order. 
In the basis of the interaction eigenstates, characterized by $Y$ and $B$ before symmetry breaking, the CS counterterms can be absorbed into the diagrams, thereby obtaining a re-distribution of the partial anomalies on each anomalous gauge interaction. As we have already mentioned, the role of the CS terms is to render vector-like an axial-vector current at one-loop level in an anomalous trilinear coupling. The anomaly is moved 
from the $Y$ vertex to the $B$ vertex, and then canceled by a WZ counterterm. However, after symmetry 
breaking, in which $Y$ and $B$ undergo mixing, the best way to treat these anomalous interactions is to 
keep the CS term, rotated into the physical basis, separate from the triangular contribution. This separation is scheme dependent, being the CS term gauge variant. These theories are clearly characterized 
by direct interactions which are absent in the SM which can be eventually 
tested in suitable processes at the LHC (as suggested in Chap.~\ref{chap:AbelianModels2} and developed in Chap.~\ref{chap:LHC}).
\section{The fermion sector} 
Moving to analyze the gauge consistency of the fermion sector, 
we summarize some of the features of the organization of some typical fermionic 
amplitudes. These considerations, naturally, can also be 
generalized to more complex cases. For our discussion we work directly in the A-B Model for simplicity. Applications of 
this analysis can be found in the next chapter.

\begin{figure}[tbh]
{\centering \resizebox*{7.5cm}{!}{\rotatebox{0}
{\includegraphics{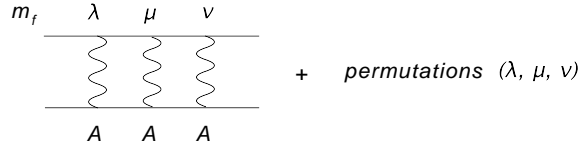}}}\par}
\caption{\small The massive fermion sector with massive vector exchanges 
in the $t$-channel.}
\label{fermion1}
\end{figure}

\begin{figure}[tbh]
{\centering \resizebox*{8cm}{!}{\rotatebox{0}
{\includegraphics{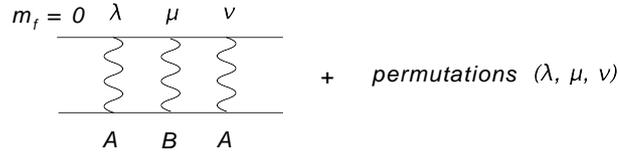}}}\par}
\caption{\small $t$-channel exchanges with vector and axial-vector interactions of 
massive gauge bosons. Being the fermion massless, permutation of the exchanges is sufficient to 
generate a gauge invariant result.}
\label{fermion2}
\end{figure}

\begin{figure}[tbh]
{\centering \resizebox*{14cm}{!}{\rotatebox{0}
{\includegraphics{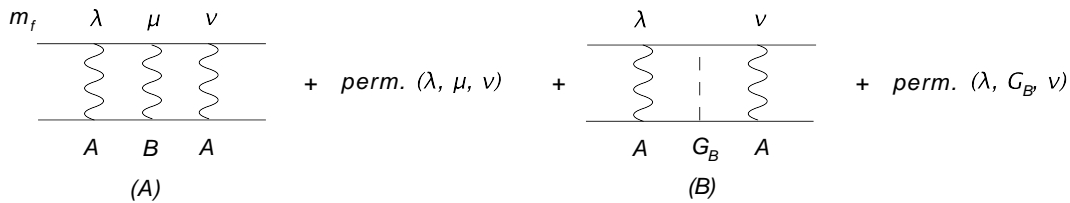}}}\par}
\caption{\small As in Fig.~\ref{fermion2} but in the massive case and with a Goldstone.}
\label{fermion3}
\end{figure}

 We start from Fig.~\ref{fermion1} that describes the $t$-channel exchange of $A$-gauge bosons. We have 
explicitly shown the indices $(\lambda\mu\nu)$ over which we perform permutations. In the absence of axial-vector interactions the gauge independence of diagrams of these types is obtained just 
with the symmetrization of the $A$-lines, both in the massive and in the massless fermion ($m^{}_f$=0) case. When, 
instead, we allow for a $B$ exchange in diagrams of the same topology, 
the cases $m^{}_f=0$ and $m^{}_f\neq 0$ involve a different (see Fig.~\ref{fermion2}) organization of the expansion. In the first case, the derivation of the gauge independence in this class of diagrams 
is obtained again just by a permutation of the attachments of the gauge boson lines. In the massive fermion case, instead, we need to add to this class of diagrams also the corresponding Goldstone exchanges together with their similar symmetrizations (Fig.~\ref{fermion3}). 
\begin{figure}[tbh]
\begin{center}
\includegraphics[scale=0.65,angle=0]{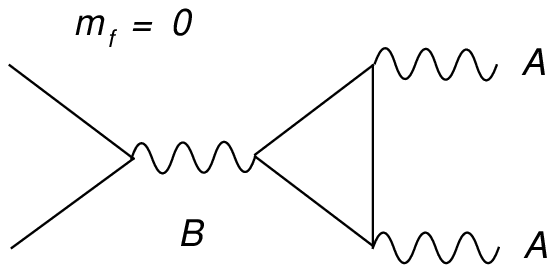}
\caption{\small $f$$\overline{f}$ annihilation in the massless case.}
\label{fermion4}
\end{center}
\end{figure}
\begin{figure}[tbh]
\begin{center}
\includegraphics[scale=0.6,angle=0]{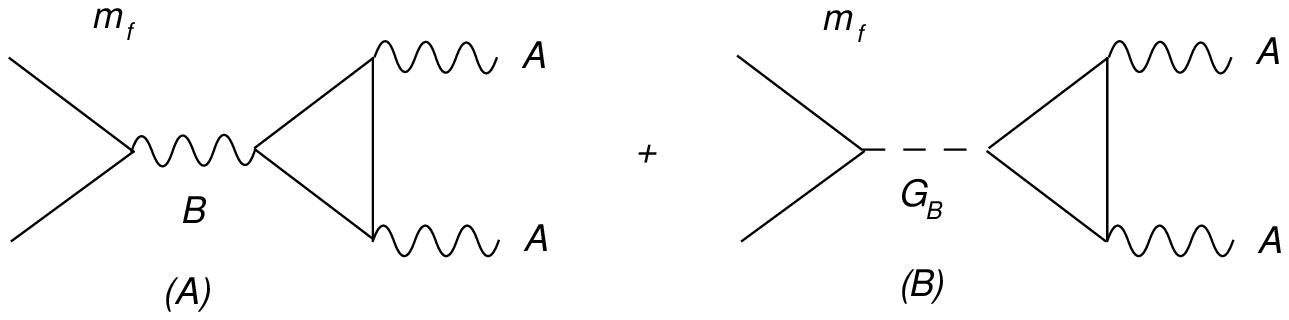}
\caption{\small $f \overline{f}$ in the massive case.}
\label{fermion5}
\end{center}
\end{figure}
\begin{figure}[tbh]
\begin{center}
\includegraphics[scale=0.7,angle=0]{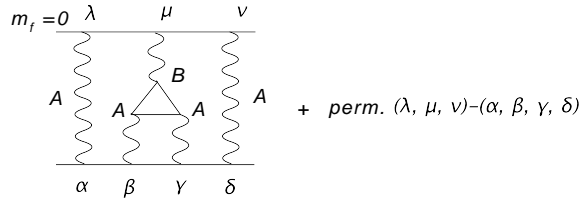}
\caption{\small Anomaly in the $t$-channel.}
\label{fermion7}
\end{center}
\end{figure}
\begin{figure}[tbh]
\begin{center}
\includegraphics[scale=0.8,angle=0]{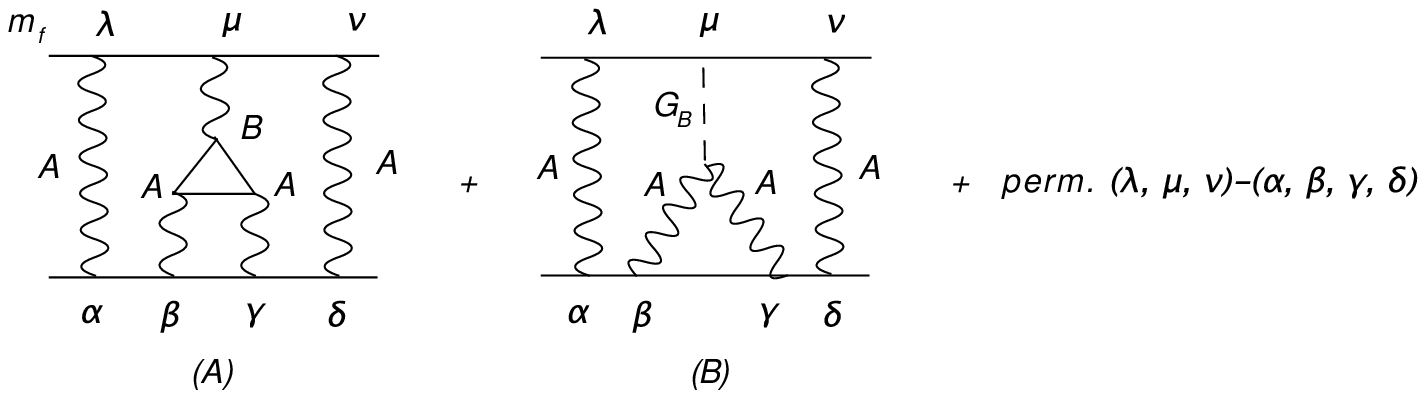}
\caption{\small The WZ counterterm for the restauration of gauge invariance.}
\label{fermion8}
\end{center}
\end{figure}
We illustrate in Figs.~\ref{fermion4} and \ref{fermion5} 
the organization of the expansion to lowest order for a process of the type $f\bar{f} \to A A$ which is the analogous of $q\bar{q} \to \gamma \gamma$ in this simple model. The presence of a 
Goldstone exchange takes place, obviously, only in the massive case. 
Finally, we have included 
the set of gauge-invariant diagrams describing the exchange of $A$ and $B$ gauge bosons in the 
$t$-channel and with an intermediate triangle anomaly diagram ($BAA$) (Figs. \ref{fermion7} and \ref{fermion8}). 
In the massless fermion case gauge invariance is 
obtained simply by adding to the basic diagram all the similar ones obtained by permuting the attachments of the gauge 
lines (this involves both the lines at the top 
and at the bottom) and summing only over the topologically independent configurations. 
In the massive case one needs to add to this set of diagrams 2 additional sets: those containing a Goldstone exchange and those involving a WZ interaction. The contributions of these additional diagrams have 
to be symmetrized as well, by moving the attachments of the gauge 
boson/scalar lines.
\section{ The Effective Action in configuration space}
Hidden inside the anomalous three-point functions are some Chern-Simons interactions. 
Their ``extraction'' can be done quite easily if we try to integrate out 
completely a given diagram and look at the structure of the effective action that is so generated directly in configuration space. The resulting action is non-local but contains a contact term that is present independently from the type of external Ward Identities that need to be imposed on the external vertices. 
This contact term is a dimension-4 contribution that is identified with a CS 
interaction, while the higher dimensional contributions have 
a non-trivial structure. The variation of both the local and the non-local effective vertex is still a local operator, proportional to $F\wedge F$. The coefficient in front of the local CS interaction changes, 
depending on the external conditions imposed on the diagram (the external Ward Identities). In this sense, different vertices may carry different 
CS terms. 

To illustrate this issue in a simple way, we proceed as follows. 
Consider the special case in which the two lines $\mu \nu  $ are on-shell, so that $k_1^2=k_2^2=0$. This simplifies our derivation, though 
a more general analysis can also be considered. We work in the specific parameterization in which the vertex satisfies the vector WI on the 
$\mu \nu$ lines, with the anomaly brought entirely on the $\lambda$ line. 

In this case we have 
\beqa
A_1(k_1,k_2)&=&k_1\cdot k_2 A_3(k_1,k_2) \nonumber \\ 
A_2(k_1,k_2) &=& -A_1(k_2,k_1) \nonumber \\
A_5(k_1,k_2) &=& -A_4(k_2,k_1) \nonumber \\
A_6(k_1,k_2) &=& -A_3(k_2,k_1), \nonumber \\
\eeqa 
and defining 
$s=2 k_1\cdot k_2= k^2$, the explicit expressions of $A_1$ and $A_2$ are summarized in the form 
\beq
A_1(k^2)= -i \frac{1}{4 \pi^2} + i C_0(m_f^2,s)
\eeq
with $C_0$ a given function of the ratio $m_f^2/s $ that we redefine as $R(m_f^2/s)$. The typical expression of these functions can be found in the appendix. Here we assume that $s > 4 m_f^2  $, but other regions can be reached by suitable analytic continuations. The important point to be appreciated is the presence of a constant term in this 
invariant amplitude. Notice that the remaining amplitudes do not share this property. If we denote as $T_{c}$ the vertex in configuration space, the contribution to the effective action becomes 
\beqa
&&\langle T_c^{\lambda\mu\nu}(z,x,y)B_{\lambda}(z) A_\mu(x)A_\nu(y) \rangle   \nonumber\\
&&= \frac{1}{4 \pi^2} \langle \epsilon_{\mu\nu\rho\sigma}B^\lambda A^\mu 
F_A^{\rho\sigma} \rangle + \langle R\left(-m_f^2/\square_z)\right)
\left[ \delta(x-z)\delta(y-z)\right]B^{\lambda}(z)A^\mu(x) A^\nu(y) \rangle
\eeqa
where the $1/\square_z$ operator acts only on the distributions inside the 
squared brackets ($\left[\,\, \right]$). It is not difficult to show that if we perform a gauge variation, say under $B$, of the vertex written in this form, 
then the first term trivially gives the $F_A\wedge F_A$ contribution, while 
the second (non-local) expression summarized in $R$, vanishes identically after an integration by parts. For this one needs to use the Bianchi identities 
of the $A, B$ gauge bosons. 

A similar computation on $A_3, A_4$ etc, can be carried out, but this time 
these contributions do not have a contact interaction as $A_1$ and $A_2$, 
but they are, exactly as the $R()$ term, purely non-local. Their gauge variations are also vanishing. When we impose a parameterization of the triangle 
diagram that redistributes the anomaly in such a way that some axial 
interactions are conserved or, for that reason, any other distribution of the 
partial anomalies on the single vertices, than we are actually introducing 
into the theory some specific CS interactions. We should think of 
these vertices as new effective vertices, fixed by the Ward Identities 
imposed on them. Their form is dictated by the conditions 
of gauge invariance. These conditions may appear with an axion term if 
the corresponding gauge boson, such as $B$ in this case, is anomalous. 
Instead, if the gauge boson is not paired to a shifting axion, such as for 
$Y$, then gauge invariance under $Y$ is restaured by suitable CS terms.  
The discussion of the phenomenological relevance of these vertices will be 
addressed in the next chapters.
\section{Conclusions} 
We have analyzed in some detail unitarity issues that emerge in the context of anomalous Abelian models 
when the anomaly cancellation mechanism involves a WZ term, CS interactions and traceless conditions on some 
of the generators. We have investigated the 
features of these types of theories both in their exact and in their broken phases, and we have used $s$-channel 
unitarity as a simple strategy to achieve this. 
We have illustrated in a simple model (the ``A-B Model'') how the axion $b$ is decomposed into a physical 
field $\chi$ and a Goldstone mode $G^{}_B$ (see Eq.~(\ref{projection})), and how the cancellation of the gauge dependences 
in the $S$-matrix involves either $b$ or $G_B$, in the St\"uckelberg or the Higgs-St\"uckelberg phase respectively. In 
the St\"uckelberg phase the axion is a Goldstone mode. The physical component of the field, $\chi$, 
appears after spontaneous symmetry breaking, and becomes massive via a combination of both the St\"uckelberg and the Higgs mechanism. Its mass can be driven 
to be light if the Peccei-Quinn breaking contributions in the scalar potential (see Eq.~(\ref{ppqq})) appear with small parameters $(b^{}_1,\lambda^{}_1, \lambda^{}_2)$ compared to the Higgs v.e.v. (see Eq.~(\ref{chimass})). Then we have performed a unitarity analysis of this model first in the St\"uckelberg phase and then in the Higgs-St\"uckelberg phase, summarized 
in the set of diagrams collected in Figs.~\ref{chern} and~\ref{chernmassive}. A similar analysis has been presented for self-energy like diagrams, and is summarized in Figs.~\ref{bbox} and \ref{bboxmass} respectively. In the broken phase, the most demanding pattern of cancellation is the 
one involving several anomalous interactions ($BBB$), and the analysis is summarized 
in Fig.~\ref{brokenphase}. We have also shown that in the simple models discussed in this chapter, Chern-Simons interactions can be absorbed into the triangle diagrams 
by a re-definition of the momentum parameterization, if one rewrites a given amplitude in the basis of the interaction eigenstates. Isolation of the Chern-Simons terms may 
however help in the computation of trilinear gauge interactions in realistic extensions of the SM 
and can be kept separate from the fermionic triangles. 
Their presence is the indication that the theory requires external Ward Identities to be correctly defined at one-loop. 
Our results will be generalized in the following chapter
and applied to the analysis of effective string models derived from the orientifold construction in the following chapters.
\newpage
\section{Appendix. The triangle diagrams and their ambiguities}
We have collected in this and in the following appendices some of the more technical material 
which is summarized in the main sections. We present also a 
rather general analysis of the main features of anomalous diagrams, some of 
which are not available in the similar literature on the Standard Model, for instance
due to the different pattern of cancellations of the anomalies required in our case study.

The consistency of these models, in fact,
requires specific realizations of the vector WI for gauge trasformations 
involving the vector currents, which implies a specific parameterization of the fermionic triangle diagrams. 
While the analysis of these triangles is well known in the massless fermion 
case, for massive fermions it is slightly more involved. We have gathered here some 
results concerning these diagrams. 

The typical ${\bf AVV}$ diagram with two vectors and one 
axial-vector current (see Fig.~\ref{VAA1}) is  described here using a 
specific parameterization of the loop momenta given by
\beq 
\Delta_{\bf AVV}^{\lambda\mu\nu} = \Delta^{\lambda\mu\nu} =  i^3\int \frac{d^4 q}{(2\pi)^4} 
\frac{Tr\left[\gamma^\mu(\ds{q}+m) \gamma^\lambda\gamma^5(\ds{q}-\ds{k}+m)\gamma^\nu(\ds{q}-\ds{k}_1+m)\right]}
{(q^2-m^2)[(q-k_1)^2-m^2][(q-k)^2-m^2]} + \makebox{exch}.
\eeq
Similarly, for the ${\bf AAA}$ diagram we will use the parameterization
\beq 
\Delta_{\bf AAA}^{\lambda\mu\nu}=\Delta_3^{\lambda\mu\nu}= i^3 \int \frac{d^4 q}{(2\pi)^4} 
\frac{Tr\left[\gamma^\mu\gamma^5 (\ds{q}+m) \gamma^\lambda\gamma^5(\ds{q}-\ds{k}+m)\gamma^\nu\gamma^5(\ds{q}-\ds{k}_1+m)\right]}
{(q^2-m^2)[(q-k_1)^2-m^2][(q-k)^2-m^2]} + \makebox{exch}.
\eeq
In both cases we have included both the direct and the exchanged contributions 
\footnote{Our conventions differ from \cite{Zee:2003mt} by an overall (-1) since our currents 
are defined as $j_\mu^B=-q^{}_B g^{}_B \overline{\psi}\gamma_\mu \psi$}.
 
In our notation $\overline{\Delta}^{\lambda \mu \nu}$ denotes a single diagram while we will use the 
symbol $\Delta$ to denote the Bose symmetric expression
\beqn
\Delta^{\lambda \mu \nu} = \overline{\Delta}^{\, \lambda \mu \nu}(k_1, k_2) + \mbox{exchange of}\,\, \{(k_1, \mu),(k_2, \nu)\}.
\eeqn
To be noticed that the exchanged diagram is equally described by a diagram equal to the first diagram but with a reversed fermion flow. Reversing the fermion flow is sufficient to guarantee Bose symmetry of the two {\bf V} lines. Similarly, for an ${\bf AAA}$ diagram, the exchange of any two {\bf A} lines is sufficient to render the entire diagram completely symmetric under cyclic permutations of the three ${\bf AAA}$ lines. 

Let's now consider the ${\bf AVV}$ contribution and work out some preliminaries.
It is a simple exercise to show that the parameterization that we have used 
above indeed violates the 
vector WI on the $\mu\nu $ vector lines giving 
\beqa
k_{1\mu} \Delta^{\la\mu\nu}(k_1,k_2) = a_1 \epsilon^{\lambda\nu\alpha\beta} 
k_1^\alpha k_2^\beta \nonumber \\
k_{2\nu} \Delta^{\la\mu\nu}(k_1,k_2) = a_2 \epsilon^{\lambda\mu\alpha\beta} 
k_2^\alpha k_1^\beta \nonumber \\
k_{\la} \Delta^{\la\mu\nu}(k_1,k_2) = a_3 \epsilon^{\mu\nu\alpha\beta} 
k_1^\alpha k_2^\beta, \nonumber \\
\eeqa
where
\beq
 a_1=-\frac{i}{8 \pi^2} \qquad a_2=-\frac{i}{8 \pi^2} \qquad a_3=-\frac{i}{4 \pi^2}.
\label{basic}
\eeq
Notice that $a_1=a_2 $, as expected from the Bose symmetry of the two {\bf V} lines. It is also well known that the total anomaly 
$a_1+a_2 + a_3 \equiv  a_n$ is regularization scheme independent 
($a_n=-\frac{i}{2 \pi^2}$). We do not impose any WI on the {\bf V} lines, 
conditions which would bring the anomaly only to the axial vertex, as done for the SM case,   
but we will determine consistently the value of the three anomalies at a later stage from the requirement of gauge invariance of the effective action, with the inclusion of 
the axion terms. To render our discussion self-contained, and define our notations, we briefly review the issue of the shift dependence of these diagrams. 

We recall that a shift of the momentum in the integrand $(p\rightarrow p + a)$ where $a$ is the most general momentum written in terms of the two independent external momenta of the triangle diagram $(a=\alpha (k_1 + k_2) + \beta(k_1 - k_2))$ induces on  $\Delta$ changes that 
appear only through a dependence on one of the two parameters characterizing $a$, that is 
\beq
\Delta^{\la\mu\nu}(\beta,k_1,k_2)= \Delta^{\la\mu\nu}(k_1,k_2) - \frac{i}{4 \pi^2}\beta \epsilon^{\lambda\mu\nu\sigma}\left( k_{1\sigma} - 
k_{2\sigma}\right).
\eeq
We have introduced the notation $\Delta^{\la\mu\nu} (\beta,k_1,k_2)$ to denote the shifted three-point function, while 
$\Delta^{\la\mu\nu}(k_1,k_2)$ denotes the original one, with a 
vanishing shift.
In our parameterization, the choice $\beta=-\frac{1}{2}$ corresponds to conservation of the vector current 
and brings the anomaly to the axial vertex
\beqn
k_{1\mu}\Delta^{\lambda\mu\nu}(a,k_1,k_2)&=& 0,\nonumber\\
k_{2\nu}\Delta^{\lambda\mu\nu}(a,k_1,k_2)&=&0,\nonumber\\
k_\lambda\Delta^{\lambda\mu\nu}(a,k_1,k_2)&=&-\frac{i}{2 \pi^2}\epsilon^{\mu\nu\alpha\beta}k_1^\alpha k_2^\beta
\label{bshift}
\eeqn
with $a_n=a_1+a_2+a_3=-\frac{i}{2 \pi^2}$ still equal to the total anomaly. Therefore, starting from generic values of $(a_1=a_2,a_3)$, for instance from the values deduced from the 
basic parameterization (\ref{basic}), an additional shift with parameter $\beta' $ gives
\beq
\Delta^{\lambda\mu\nu}(\beta',k_1,k_2)= \Delta^{\lambda\mu\nu}(\beta,k_1,k_2)-\frac{i \beta'}{4 \pi^2}
\epsilon^{\lambda\mu\nu\sigma}(k_1-k_2)_\sigma
\eeq
and will change the Ward Identities into the form 
\beqn
k_{1\mu}\Delta^{\lambda\mu\nu}(\beta',k_1,k_2)&=& (a_1 -\frac{i \beta'}{4 \pi^2})
\epsilon^{\lambda\nu\alpha\beta}k_1^\alpha k_2^\beta,\nonumber\\
k_{2\nu}\Delta^{\lambda\mu\nu}(\beta',k_1,k_2)&=&(a_2-\frac{i \beta'}{4 \pi^2})
\epsilon^{\lambda\mu\alpha\beta}k_2^\alpha k_1^\beta,\nonumber\\
k_\lambda\Delta^{\lambda\mu\nu}(\beta',k_1,k_2)&=&(a_3+\frac{i \beta'}{2 \pi^2})
\epsilon^{\mu\nu\alpha\beta}k_1^\alpha k_2^\beta,
\label{bbshift}
\eeqn
where $a_2=a_1$. There is an intrinsic ambiguity in the definition of the amplitude, which 
can be removed by imposing CVC on the vector vertices, as done, for instance, in Rosenberg's 
original paper \cite{Rosenberg:1962pp} and discussed later. We remark 
once more that, in our case, this condition is not automatically required. 
The distribution of the anomaly may, in general, be different and we are 
defining, 
in this way, new effective parameterizations of the three-point anomalous vertices.

It is therefore convenient to introduce a notation that makes explicit this dependence and for this reason we define 
\beqa
a_1(\beta)=a_2(\beta)=-\frac{i}{8\pi^2} - \frac{i}{4 \pi^2}\beta, \,\, \,\,\qquad \,\,\,\, 
a_3(\beta) = -\frac{i}{4\pi^2} + \frac{i}{2 \pi^2}\beta,
\label{a12}
\eeqa
with
\beq
a_1(\beta) + a_2(\beta) + a_3(\beta)=a_n=-\frac{i}{2 \pi^2}.
\eeq
Notice that the additional $\beta$-dependent contribution amounts 
to a Chern-Simons interaction. Clearly, this contribution can be moved around at will and is related to the presence of two 
divergent terms in the general triangle diagram that need to be fixed appropriately using the underlying Bose symmetries 
of the three-point functions. 
\begin{figure}[t]
{\centering \resizebox*{10cm}{!}{\rotatebox{0}
{\includegraphics{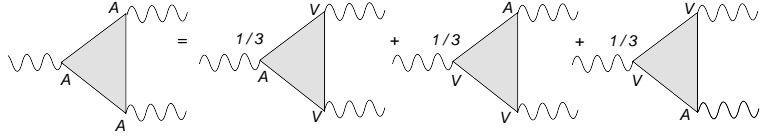}}}\par}
\caption{\small Distribution of the axial anomaly for the ${\bf AAA}$ diagram.}
\label{distribution}
\end{figure}
The regularization of the $\bf AAA$ vertex, instead, has to respect the complete Bose symmetry of the diagram and this can be achieved with the symmetric expression 
\beq
\Delta_3^{\lambda\mu\nu}(k_1,k_2)=\frac{1}{3}[\Delta^{\lambda\mu\nu}(k_1,k_2)+\Delta^{\mu\nu\lambda}(k_2,-k)+\Delta^{\nu\lambda\mu}
(-k,k_1)].
\label{delta3}
\eeq
It is an easy exercise to show that this symmetric choice is independent from the momentum shift 
\beq
\Delta_3^{\lambda\mu\nu}(\beta,k_1,k_2)=
\Delta_3^{\lambda\mu\nu}(k_1,k_2)
\eeq
and that the anomaly is equally distributed among the 3 vertices, $a_1=a_2=a_3=a_n/3$, as shown in 
Fig.~\ref{distribution}.
We conclude this section with some comments regarding the kind of invariant amplitudes appearing in the definition 
of $\Delta$ which help to clarify the role of the CS terms in the parameterization of these diagrams in momentum space. 
For $\overline{\Delta}_{{\la}\mu\nu}^{({\bf AVV})}$, expressed using Rosenberg's parametrization, one obtains 
\beqa
\overline{\Delta}_{{\la}\mu\nu} &=& \hat{a}_1 \epsilon[k_1,\mu,\nu,\la] 
+ \hat{a}_2 \epsilon[k_2,\mu,\nu,\la] +
\hat{a}_3 \epsilon[k_1,k_2,\mu,\la]{k_1}^{\nu}\nonumber \\
&& + \hat{a}_4\epsilon[k_1,k_2,\mu,\la]k_2^{\nu}
 + \hat{a}_5 \epsilon[k_1,k_2,\nu,\la]k_1^\mu 
+ \hat{a}_6 \epsilon[k_1,k_2,\nu,\la]k_2^\mu,
\label{RRos}
\eeqa
originally given in \cite{Rosenberg:1962pp}, with $\lambda$ being the axial-vector vertex. 
By power-counting, two invariant amplitudes are divergent, $a_1$ and $a_2$, while the $a_i$ with $i\geq 3$ are finite\footnote{
We will be using the notation $\epsilon[a,b,\mu,\nu]\equiv \epsilon_{\alpha\beta\mu\nu}
a^\alpha b^\beta$ to denote the structures in the expansion of the anomalous triangle diagrams.}. 

Instead, for the direct plus the exchanged diagrams we use the expression
\beqa
\Delta_{\underline{\la}\mu\nu} + \Delta_{\underline{\la}\nu\mu} &=& 
(\hat{a}_1 - \hat{a}_2) \epsilon[k_1,\mu,\nu,\la] + (a_2- a_1)\epsilon[k_2,\mu,\nu,\la]\nonumber \\
&& +(\hat{a}_3- \hat{a}_6) \epsilon[k_1,k_2,\mu,\la]{k_1}^{\nu}
 + (\hat{a}_4 - \hat{a}_5)\epsilon[k_1,k_2,\mu,\la]k_2^{\nu}\nonumber \\
&& +(\hat{a}_5 - \hat{a}_4 ) \epsilon[k_1,k_2,\nu,\la]k_1^\mu 
+ (\hat{a}_6 - \hat{a}_3)\epsilon[k_1,k_2,\nu,\la]k_2^\mu \nonumber \\
&=& \underline{a}_1 \epsilon[k_1,\mu,\nu,\la] + 
\underline{a}_2\epsilon[k_2,\mu,\nu,\la] 
+\underline{a}_3 \epsilon[k_1,k_2,\mu,\la]{k_1}^{\nu}\nonumber \\
&& +\underline{a}_4 \epsilon[k_1,k_2,\mu,\la]{k_2}^{\nu} +
\underline{a}_5 \epsilon[k_1,k_2,\nu,\la]k_1^\mu +
\underline{a}_6\epsilon[k_1,k_2,\nu,\la]k_2^\mu
\label{Ros1}
\eeqa
where clearly $\ul{a}_2=-\ul{a}_1, \ul{a}_3=-\ul{a}_6$ and $\ul{a}_4=-\ul{a}_5$. 
The CS contributions are those proportional to the two terms linear in the external momenta. 
We recall that, in Rosenberg, these linear terms are re-expressed in terms of the remaining ones by imposing the vector Ward Identities on the {\bf V}-lines. 
As already explained, we will instead assume, in our case, that the distribution of the anomaly among the 3 vertices of all the anomalous diagrams of the theory respects 
the requirement of Bose symmetry, with no additional constraint. 
A discussion of some 
technical points concerning the regularization of this and other diagrams both in 4 dimensions and in other schemes, such as Dimensional Regularization (DR) can be found below. For instance, one can find there the proof of the identical vanishing 
of $\Delta_{\bf VVV}$ worked out in both schemes. 
In this last case 
this result is obtained after removing the so called 
hat-momenta of the t'Hooft-Veltman scheme on the external lines. 
In this scheme this is possible since one can choose the external momenta to lay on a four-dimensional subspace (see \cite{Coriano:1996us, Chang:1997ik} for a discussion of these methods). 
We remark also that it is also quite useful to be able to switch from momentum space to configuration 
space with ease, and for this purpose we introduce the 
Fourier transforms of (\ref{dd1}) and (\ref{dd2}) in the anomaly equations, obtaining their 
expressions in configuration space
\beqa
\frac{\partial}{\partial x^\mu}T^{\la\mu\nu}_{\bf AVV}(x,y,z)&=& 
i a_1(\beta) \epsilon^{\la\nu\alpha\beta}\frac{\partial}{\partial x^\alpha}
\frac{\partial}{\partial y^\beta}\left(\delta^4(x-z)\delta^4(y-z)\right),
\nonumber \\
\frac{\partial}{\partial y^\nu}T^{\la\mu\nu}_{\bf AVV}(x,y,z)&=& 
i a_2(\beta) \epsilon^{\la\mu\alpha\beta} \frac{\partial}{\partial y^\alpha}
\frac{\partial}{\partial x^\beta}\left(\delta^4(x-z)\delta^4(y-z)\right),
\nonumber \\
\frac{\partial}{\partial z^\la}T^{\la\mu\nu}_{\bf AVV}(x,y,z)&=& 
i a_3(\beta) \epsilon^{\mu\nu\alpha\beta}\frac{\partial}{\partial x^\alpha}
\frac{\partial}{\partial y^\beta}\left(\delta^4(x-z)\delta^4(y-z)\right),
\eeqa
with $a_1, a_2$ and $a_3$ as in (\ref{a12}), for the $\bf AVV$ case
and 
\beqa
\frac{\partial}{\partial x^\mu}T^{\la\mu\nu}_{\bf AAA}(x,y,z)&=& 
i \frac{a_n}{3}\epsilon^{\la\nu\alpha\beta}\frac{\partial}{\partial x^\alpha}
\frac{\partial}{\partial y^\beta}\left(\delta^4(x-z)\delta^4(y-z)\right),
\nonumber \\
\frac{\partial}{\partial y^\nu}T^{\la\mu\nu}_{\bf AAA}(x,y,z)&=& 
i \frac{a_n}{3} \epsilon^{\la\mu\alpha\beta} \frac{\partial}{\partial y^\alpha}\frac{\partial}{\partial x^\beta}\left(\delta^4(x-z)\delta^4(y-z)\right),
\nonumber \\
\frac{\partial}{\partial z^\la}T^{\la\mu\nu}_{\bf AAA}(x,y,z)&=& 
i \frac{a_n}{3} \epsilon^{\mu\nu\alpha\beta}\frac{\partial}{\partial x^\alpha}\frac{\partial}{\partial y^\beta}\left(\delta^4(x-z)\delta^4(y-z)\right),
\eeqa
for the $\bf AAA$ case. Notice that in this last case we have distributed 
the anomaly equally among the three vertices. These relations will be needed when we derive the anomalous variation of the effective action directly in configuration space. 
\section{Appendix. Chern-Simons cancellations}
Having isolated the CS contributions, as shown in 
Fig.~\ref{chern}, the cancellation of the gauge dependence can be obtained 
combining all these terms so to obtain\footnote{The symmetry factor of each configuration 
is easily identified as the first factor in each separate contribution.} 
\beqn
S_{\xi} &=& \Delta^{\lambda \mu \nu}(-k^{}_{1}, -k^{}_{2}) 
\left[  \frac{-i}{k^{2} - \xi^{}_{B} M^{2}_{1}} \frac{k^{\lambda} k^{\lambda'}}{ M^{2}_{1} } \right]
 \Delta^{\lambda' \mu' \nu'}(k^{}_{1}, k^{}_{2})   \nonumber\\
&+& 4 \times \left( \frac{4}{M} C^{}_{AA} \right)^{2} 
\epsilon^{\mu\nu\rho\sigma} k^{}_{1\rho} k^{}_{2\sigma} \left[ \frac{i}{k^{2} - \xi^{}_{B} M^{2}_{1}} \right] 
\epsilon^{\mu'\nu'\rho'\sigma'} k^{}_{1\rho'} k^{}_{2\sigma'} \nonumber\\
&+&  4 \times ( i d^{}_{1} \epsilon^{\mu \nu \lambda \sigma} (k^{}_{1}-k^{}_{2})_{\sigma} ) 
\left[ \frac{-i}{k^{2} - \xi^{}_{B} M^{2}_{1}} \frac{k^{\lambda} k^{\lambda'}}{ M^{2}_{1} } \right] 
(- i d^{}_{1} \epsilon^{\mu' \nu' \lambda' \sigma'} (k^{}_{1}-k^{}_{2})_{\sigma'} )  \nonumber\\
&+&2 \times \Delta^{\lambda \mu \nu}(-k^{}_{1}, -k^{}_{2}) 
\left[  \frac{-i}{k^{2} - \xi^{}_{B} M^{2}_{1}} \frac{k^{\lambda} k^{\lambda'}}{ M^{2}_{1} } \right] 
(- i d^{}_{1} \epsilon^{\mu' \nu' \lambda' \sigma'} (k^{}_{1}-k^{}_{2})_{\sigma'} )   \nonumber\\
&+&2 \times ( i d^{}_{1} \epsilon^{\mu \nu \lambda \sigma } (k^{}_{1}-k^{}_{2})_{\sigma} ) 
 \left[  \frac{-i}{k^{2} - \xi^{}_{B} M^{2}_{1}} \frac{k^{\lambda} k^{\lambda'}}{ M^{2}_{1} } \right] 
 \Delta^{\lambda' \mu' \nu'}(k^{}_{1}, k^{}_{2}),
\eeqn
and using the relevant Ward Identities these simply to 
\beqn
S_{\xi}&=&(- a^{}_{3}(\beta) \epsilon^{\mu \nu \rho \sigma} k^{}_{1\rho} k^{}_{2\sigma}  )  
\left[  \frac{-i}{k^{2} - \xi^{}_{B} M^{2}_{1}} \frac{1}{ M^{2}_{1} } \right] 
 ( a^{}_{3}(\beta) \epsilon^{\mu' \nu' \rho' \sigma'} k^{}_{1\rho'} k^{}_{2\sigma'}  )  \nonumber\\
&+&  4 \times \frac{16}{M^{2}} \left[ \left( -\frac{ d^{}_{1} }{2} + \frac{i}{2} a^{}_{3}(\beta)\frac{1}{4} \right)^{2} 
\frac{M^{2}}{M^{2}_{1}} \right] \epsilon^{\mu \nu \rho \sigma} k^{}_{1\rho} k^{}_{2\sigma} 
\left[ \frac{i}{k^{2} - \xi^{}_{B} M^{2}_{1}} \right] \epsilon^{\mu' \nu' \rho' \sigma'} k^{}_{1\rho'} k^{}_{2\sigma'} \nonumber\\
&+&  4 \times d^{2}_{1} \left[  \frac{-i}{k^{2} - \xi^{}_{B} M^{2}_{1}} \frac{1}{ M^{2}_{1} } \right] 
4 \epsilon^{\mu \nu \rho \sigma} k^{}_{1\rho} k^{}_{2\sigma} 
\epsilon^{\mu' \nu' \rho' \sigma'} k^{}_{1\rho'} k^{}_{2\sigma'}   \nonumber\\
&+&  2 \times (- a^{}_{3}(\beta) \epsilon^{\mu \nu \rho \sigma} k^{}_{1\rho} k^{}_{2\sigma}  )  
\left[  \frac{-i}{k^{2} - \xi^{}_{B} M^{2}_{1}} \frac{1}{ M^{2}_{1} } \right] 
(+ i d^{}_{1} 2 \epsilon^{\mu' \nu' \lambda' \sigma'} k^{\lambda'}_{1} k^{\sigma'}_{2} ) \nonumber\\
&+& 2 \times (- i d^{}_{1} 2 \epsilon^{\mu \nu \lambda \sigma } k^{\lambda}_{1} k^{\sigma}_{2} ) 
\left[  \frac{-i}{k^{2} - \xi^{}_{B} M^{2}_{1}} \frac{1}{ M^{2}_{1} } \right]  
( a^{}_{3}(\beta) \epsilon^{\mu' \nu' \rho' \sigma'} k^{}_{1\rho'} k^{}_{2\sigma'}  ) = 0.
\eeqn
Having shown the cancellation of the gauge-dependent terms, the gauge independent contribution becomes 
\beqn
S_{0} &=& \Delta^{\lambda \mu \nu}(-k^{}_{1}, -k^{}_{2}) 
\left[  \frac{-i}{k^{2} - M^{2}_{1}} \left( g^{\lambda \lambda'} - \frac{k^{\lambda} k^{\lambda'}}{ M^{2}_{1} }  \right)\right]
 \Delta^{\lambda' \mu' \nu'}(k^{}_{1}, k^{}_{2})   \nonumber\\
&+& 4 \times ( i d^{}_{1} \epsilon^{\,\mu \nu \lambda \sigma} (k^{}_{1}-k^{}_{2})_{\sigma} ) 
\left[    \frac{-i}{k^{2} - M^{2}_{1}} \left( g^{\lambda \lambda'} - \frac{k^{\lambda} k^{\lambda'}}{ M^{2}_{1} } \right) \right] 
(- i d^{}_{1} \epsilon^{\,\mu' \nu' \lambda' \sigma'} (k^{}_{1}-k^{}_{2})_{\sigma'} )  \nonumber\\
&+& 2 \times \Delta^{\lambda \mu \nu}(-k^{}_{1}, -k^{}_{2}) 
\left[  \frac{-i}{k^{2} - M^{2}_{1}} \left( g^{\lambda \lambda'} - \frac{k^{\lambda} k^{\lambda'}}{ M^{2}_{1} } \right) \right] 
(- i d^{}_{1} \epsilon^{\,\mu' \nu' \lambda' \sigma'} (k^{}_{1}-k^{}_{2})_{\sigma'} )   \nonumber\\
&+&2 \times ( i d^{}_{1} \epsilon^{\,\mu \nu \lambda \sigma } (k^{}_{1}-k^{}_{2})_{\sigma} ) 
 \left[  \frac{-i}{k^{2} - M^{2}_{1}} \left( g^{\lambda \lambda'} - \frac{k^{\lambda} k^{\lambda'}}{ M^{2}_{1}} \right)  \right] 
 \Delta^{\lambda' \mu' \nu'}(k^{}_{1}, k^{}_{2}). 
\eeqn  
At this point we need to express the triangle diagrams in terms of their shifting parameter $\beta$ using the shift-relations
\beqn
&&\Delta^{\lambda \mu \nu}(\beta, k^{}_{1}, k^{}_{2}) = \Delta^{\lambda \mu \nu}(k^{}_{1}, k^{}_{2}) 
- \frac{i}{4 \pi^{2}} \beta \epsilon^{\lambda \mu \nu \sigma} (k^{}_{1}- k^{}_{2})_{\sigma},  \\
&& \Delta^{\lambda \mu \nu}(\beta,- k^{}_{1},- k^{}_{2}) = \Delta^{\lambda \mu \nu}(- k^{}_{1},- k^{}_{2}) 
+ \frac{i}{4 \pi^{2}} \beta \epsilon^{\lambda \mu \nu \sigma} (k^{}_{1}- k^{}_{2})_{\sigma}, 
\eeqn
and with the substitution $d^{}_{1} = -i a^{}_{1}(\beta)/2$ we obtain 
\beqn
S^{}_{0} &=& \left( \Delta^{\lambda \mu \nu}(- k^{}_{1},- k^{}_{2}) 
+ \frac{i}{4 \pi^{2}} \beta \epsilon^{\,\lambda \mu \nu \sigma} (k^{}_{1}- k^{}_{2})_{\sigma}  \right) P^{\lambda \lambda'}_{0}
\left( \Delta^{\lambda' \mu' \nu'}( k^{}_{1}, k^{}_{2}) 
- \frac{i}{4 \pi^{2}} \beta \epsilon^{\,\lambda' \mu' \nu' \sigma'} (k^{}_{1}- k^{}_{2})_{\sigma'} \right) \nonumber\\
&+&4 \times \left( \frac{1}{2} a^{}_{1}(\beta)
\epsilon^{\,\mu \nu \lambda \sigma} (k^{}_{1}- k^{}_{2})_{\sigma} \right) P^{\lambda \lambda'}_{0} 
\left( - \frac{1}{2} a^{}_{1}(\beta)
\epsilon^{\,\mu' \nu' \lambda' \sigma'} (k^{}_{1}- k^{}_{2})_{\sigma'} \right)  \nonumber\\
&+&2 \times  \left( \Delta^{\lambda \mu \nu}(- k^{}_{1},- k^{}_{2}) 
+ \frac{i}{4 \pi^{2}} \beta \epsilon^{\,\lambda \mu \nu \sigma} (k^{}_{1}- k^{}_{2})_{\sigma}  \right) P^{\lambda \lambda'}_{0}
\left( - \frac{1}{2} a^{}_{1}(\beta)
\epsilon^{\, \mu' \nu' \lambda' \sigma'} (k^{}_{1}- k^{}_{2})_{\sigma'} \right) \nonumber\\
&+& 2 \times  \left( \frac{1}{2} a^{}_{1}(\beta)
\epsilon^{\, \mu \nu \lambda \sigma} (k^{}_{1}- k^{}_{2})_{\sigma} \right) P^{\lambda \lambda'}_{0} 
\left( \Delta^{\lambda' \mu' \nu'}(k^{}_{1}, k^{}_{2}) 
- \frac{i}{4 \pi^{2}} \beta \epsilon^{\,\lambda' \mu' \nu' \sigma'} (k^{}_{1}- k^{}_{2})_{\sigma'}  \right).
\eeqn
Introducing the explicit expression for $a^{}_{1}(\beta)$, it is an easy 
exercise to show the equivalence between $S^{}_{0}$ and diagram (A) of Fig.~\ref{chern}, with a choice of the shifting parameter that corresponds 
to the CVC condition ($\beta=-1/2$) 
\beqn
S^{}_{0} \equiv \left( \Delta^{\lambda \mu \nu}(- k^{}_{1},- k^{}_{2}) 
- \frac{i}{8 \pi^{2}} \epsilon^{\,\lambda \mu \nu \sigma} (k^{}_{1}- k^{}_{2})_{\sigma}  \right) P^{\lambda \lambda'}_{0}
\left( \Delta^{\lambda' \mu' \nu'}( k^{}_{1}, k^{}_{2}) 
+ \frac{i}{8 \pi^{2}} \beta \epsilon^{\,\lambda' \mu' \nu' \sigma'} (k^{}_{1}- k^{}_{2})_{\sigma'} \right). \nonumber \\
\eeqn 
\subsection{Cancellation of gauge dependences in the broken Higgs phase \label{app:brokenHiggs}}
In this case we have (see Fig.~\ref{chernmassive})
\beqn
S^{}_{\xi} &=& A^{}_{\xi} + B^{}_{\xi} + C^{}_{\xi} + D^{}_{\xi} + E^{}_{\xi} + F^{}_{\xi} + G^{}_{\xi} + H^{}_{\xi} \nonumber \\
&=& \Delta^{\lambda \mu \nu}(-k^{}_{1}, -k^{}_{2}) 
\left[  \frac{-i}{k^{2} - \xi^{}_{B} M^{2}_{B}} \frac{k^{\lambda} k^{\lambda'}}{ M^{2}_{B} } \right]
 \Delta^{\lambda' \mu' \nu'}(k^{}_{1}, k^{}_{2})   \nonumber\\
&+& 4 \times \left( \frac{4}{M} \alpha^{}_{2} C^{}_{AA} \right)^{2} 
\epsilon^{\mu\nu\rho\sigma} k^{}_{1\rho} k^{}_{2\sigma} \left[ \frac{i}{k^{2} - \xi^{}_{B} M^{2}_{B}} \right] 
\epsilon^{\mu'\nu'\rho'\sigma'} k^{}_{1\rho'} k^{}_{2\sigma'} \nonumber\\
&+&  4 \times ( i d^{}_{1} \epsilon^{\mu \nu \lambda \sigma} (k^{}_{1}-k^{}_{2})_{\sigma} ) 
\left[ \frac{-i}{k^{2} - \xi^{}_{B} M^{2}_{B}} \frac{k^{\lambda} k^{\lambda'}}{ M^{2}_{B} } \right] 
(- i d^{}_{1} \epsilon^{\mu' \nu' \lambda' \sigma'} (k^{}_{1}-k^{}_{2})_{\sigma'} )  \nonumber\\
&+&2 \times \Delta^{\lambda \mu \nu}(-k^{}_{1}, -k^{}_{2}) 
\left[  \frac{-i}{k^{2} - \xi^{}_{B} M^{2}_{B}} \frac{k^{\lambda} k^{\lambda'}}{ M^{2}_{B} } \right] 
(- i d^{}_{1} \epsilon^{\mu' \nu' \lambda' \sigma'} (k^{}_{1}-k^{}_{2})_{\sigma'} )   \nonumber\\
&+&2 \times ( i d^{}_{1} \epsilon^{\mu \nu \lambda \sigma } (k^{}_{1}-k^{}_{2})_{\sigma} ) 
 \left[  \frac{-i}{k^{2} - \xi^{}_{B} M^{2}_{B}} \frac{k^{\lambda} k^{\lambda'}}{ M^{2}_{B} } \right] 
 \Delta^{\lambda' \mu' \nu'}(k^{}_{1}, k^{}_{2})   \nonumber\\
&+&  2 \times \Delta^{\mu \nu}(-k^{}_{1},- k^{}_{2}) \left( 2i \frac{m^{}_{f}}{M^{}_{B}} \right) 
\left[ \frac{i}{k^{2} - \xi^{}_{B} M^{2}_{B} } \right] \left( \frac{4}{M} \alpha^{}_{2} C^{}_{AA} 
\epsilon^{\mu' \nu' \rho' \sigma'} k^{}_{1\rho'}   k^{}_{2\sigma'}\right)  \nonumber\\
&+& \Delta^{\mu \nu}(-k^{}_{1},- k^{}_{2}) \left( 2i \frac{m^{}_{f}}{M^{}_{B}} \right) 
\left[ \frac{i}{k^{2} - \xi^{}_{B} M^{2}_{B} } \right] \left( 2i \frac{m^{}_{f}}{M^{}_{B}} \right) 
\Delta^{\mu' \nu'}(k^{}_{1}, k^{}_{2})    \nonumber\\
&+&  2 \times \left( \frac{4}{M} \alpha^{}_{2} C^{}_{AA} 
\epsilon^{\mu \nu \rho \sigma} k^{}_{1\rho}   k^{}_{2\sigma}\right)  
\left[ \frac{i}{k^{2} - \xi^{}_{B} M^{2}_{B} } \right]   \left( 2i \frac{m^{}_{f}}{M^{}_{B}} \right) 
  \Delta^{\mu \nu}(k^{}_{1}, k^{}_{2}).
\eeqn
The vanishing of this expression can be checked as in the previous case, using the massive version of 
the anomalous Ward Identities in the triangular graphs involving $\Delta$.
\subsection{Cancellations in the A-B Model: $BB \rightarrow BB$ mediated by a $B$ gauge boson}
\begin{figure}[t]
{\centering \resizebox*{9.5cm}{!}{\rotatebox{0}
{\includegraphics{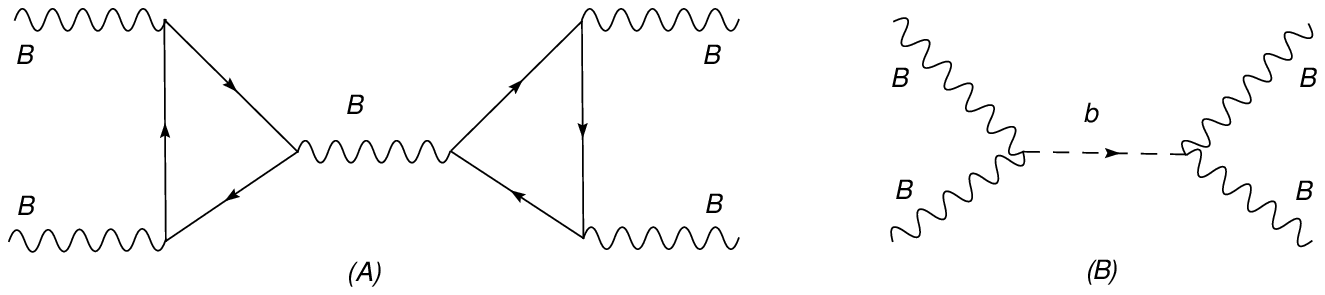}}}\par}
\caption{\small Relevant diagrams for the unitarity check before symmetry breaking.
\label{unitaritycheck}}
\end{figure}

Let's now discuss the exchange of a $B$ gauge boson in the $s$-channel before spontaneous symmetry breaking. 
The relevant diagrams are shown in Fig.~\ref{unitaritycheck}. We remark, obviously, that each diagram has to be inserted with the correct multiplicity factor in order to obtain the cancellation of the unphysical poles. 
In this case, from Bose-symmetry, the anomaly is equally distributed among the 3 vertices, $a_1=a_2=a_3=a_n/3$, as we have discussed 
above. We recall that from the variations $\delta_{B} \mathcal{L}_{an}$ and $\delta_{B} \mathcal{L}_{b}$ the relevant terms are
\beqa
\frac{1}{3!} \delta_{B} \langle T_{BBB}^{\lambda \mu \nu} B^{\lambda}(z) 
B^{\mu}(x) B^{\nu}(y) \rangle &=&  \,  \frac{i g^{\,3}_{B} }{ 3! } \, a_n \langle   \theta_{B} \, \, \frac{F^{B}\wedge F^{B}}{4}  \rangle      \nonumber\\
\delta_{B} \langle \frac{ C^{}_{BB} }{M} \, b F^{B}\wedge F^{B}  \rangle &=& - \, C^{}_{BB} \frac{M_1}{M} 
\langle \theta^{}_B F^{B}\wedge F^{B}  \rangle
\; \;\; \mbox{from}  \;\; \;\delta_{B} b = - M_1 \theta_{B}, \nonumber \\
\eeqa
so that from the condition of anomaly cancellation we obtain 
\beqa
- C^{}_{BB} \frac{M_1}{M} + \frac{ i g^{\,3}_{B} }{ 3! }\, \frac{1}{4} \, a^{}_{n} = 0 \,\, \Longleftrightarrow \,\,
  C^{}_{BB} =  \,\frac{i g^{\,3}_{B} }{ 3! }   \, \frac{1}{4}  \,a^{}_n \frac{M}{M_1},
\eeqa
which fixes the appropriate value of the coefficient of the WZ term. 
One can easily show the correspondence between a Green-Schwarz term $\frac{C^{}_{BB}}{M}\, b \, F^B \wedge F^B$ and a vertex 
$4 \frac{ C^{}_{BB}}{M} \, \epsilon^{\mu \nu \rho \sigma} k_1^{\rho} k_2^{\sigma}$ in momentum representation, as will be shown in App.~\ref{app:momentumCS_GS}. Taking into account only the gauge-dependent parts of the two diagrams, we have that the diagram with the exchange of the gauge boson 
$B$ can be written as 
\beqa
\mathcal{A}^{}_{\xi} =
\Delta^{\lambda \, \mu \, \nu} (- k^{}_{1}, -k^{}_{2}) \left[  \frac{- \, i}{  k^2  - \xi_{B} M_1^2} \left( 
\frac{k^\lambda \, k^{\lambda^\prime}}{M_1^2} \right)  \right] \Delta^{\lambda^\prime \, \mu^\prime \, \nu^\prime}(k^{}_{1}, k^{}_{2})
\eeqa
and the diagram with the exchange of the axion $b$ is
\beqa
\mathcal{B}^{}_{\xi} = 
4 \times \left( \frac{4}{M} C^{}_{BB}  \right)^2  \epsilon^{\mu \nu \rho \sigma}
 k_1^\rho k_2^\sigma \left(  \frac{i}{k^2 - \xi_B M_1^2}  \right) 
\epsilon^{\mu^\prime \nu^\prime \rho^\prime \sigma^\prime} k_1^{\rho^\prime} k_2^{\sigma^\prime}.
\eeqa
Using the anomaly equations for the AAA vertex we can evaluate the first diagram
\beqa
\mathcal{A}^{}_{\xi}&=&\frac{- \, i}{k^2 - \xi_B M_1^2} \frac{1}{M_1^2} \left(k^\lambda \Delta^{\lambda \mu \nu} \right)
\left(k^{\lambda^\prime} \Delta^{\lambda^\prime \mu^\prime \nu^\prime} \right)   \nonumber\\
 &=& \frac{- \, i}{k^2 - \xi_B M_1^2} \frac{1}{M_1^2} \left( - ( \,g^{}_{B})^3 \,  \frac{a_n}{3}  \epsilon^{\mu \nu \alpha \beta} 
k_1^\alpha k_2^\beta \right)   \left( ( \,g^{}_{B})^3 \, \frac{a_n}{3}  \epsilon^{\mu^\prime \nu^\prime \alpha^\prime \beta^\prime}
k_1^{\alpha^\prime} k_2^{\beta^\prime} \right)      \nonumber\\
&=& \frac{ \, i}{k^2 - \xi_B M_1^2} \frac{1}{M_1^2}  \left(\frac{a_n}{3}\,g^{\,3}_{B} \right)^2  \,  \epsilon^{\mu \nu \alpha \beta} \,
\epsilon^{\mu^\prime \nu^\prime \alpha^\prime \beta^\prime} \, k_1^\alpha k_2^\beta \, k_1^{\alpha^\prime} k_2^{\beta^\prime},
\eeqa 
while the axion exchange diagram gives
\beqa
\mathcal{B}^{}_{\xi}&=&4 \times \left(  \frac{4 C^{}_{BB}}{M} \right)^2   \left(   \frac{i}{k^2 - \xi_B M_1^2} \right) 
 \epsilon^{\mu \nu \alpha \beta} \,
\epsilon^{\mu^\prime \nu^\prime \alpha^\prime \beta^\prime} \, k_1^\alpha k_2^\beta \, 
k_1^{\alpha^\prime} k_2^{\beta^\prime}    \nonumber\\
&=&  \frac{64 \,  \,  {C^{\,2}_{BB}} }{M^2}   \frac{i}{k^2 - \xi_B M_1^2}  \,  \epsilon^{\mu \nu \alpha \beta} \,
\epsilon^{\mu^\prime \nu^\prime \alpha^\prime \beta^\prime} \, k_1^\alpha k_2^\beta \, 
k_1^{\alpha^\prime} k_2^{\beta^\prime}. 
\eeqa 
Adding the contributions from the two diagrams we obtain
\beqa
\mathcal{A}^{}_{\xi} + \mathcal{B}^{}_{\xi} = 0  \,\,\, \Longleftrightarrow \,\,\,
\frac{1}{M_1^2}  \left(\frac{a_n}{3}\,g^{\,3}_{B} \right)^2 + \frac{64 \,  \,  {C^{\,2}_{BB}}   }{M^2} = 0,
\eeqa
in fact substituting the proper value for the coefficient $C^{}_{BB}$ we obtain an identity 
\beqa
\frac{1}{M_1^2}  \frac{a_n^2}{9} \, g^{\,6}_{B} + \frac{64}{M^2}\left[  \,\frac{i g^{\,3}_{B} }{ 3! }   \, \frac{1}{4}  
\,a^{}_n \frac{M}{M_1}   \right]^2  = \frac{1}{M_1^2} \frac{a_n^2}{9} g^{\,6}_{B} - \frac{64}{M_1^2}
\frac{1}{64} \frac{a_n^2}{9}  g^{\,6}_{B} =0.
\eeqa
This pattern of cancellations holds for a massless fermion ($m_f=0$). 
\subsection{Gauge cancellations in the self-energy diagrams \label{app:self_brokenHiggs}}
In this case, following Fig.~\ref{bboxmass}, we isolate the following gauge-dependent amplitudes  
\beqn
{\mathcal A}^{}_{ \xi 0} &=&   \, \Delta^{\lambda \mu \nu} (-k^{}_{1}, -k^{}_{2})
\left[ \frac{- i }{k^{2} - \xi^{}_{B} M^{2}_{B}} \left( \frac{k^{\lambda} k^{\lambda'}}{ M^{2}_{B}} \right)  \right] 
\Delta^{\lambda' \mu' \nu'} (k^{}_{1}, k^{}_{2}) \, P^{\nu \nu'}_{o},    \nonumber\\
{\mathcal B}_{\xi 0} &=& 4 \times  \left( \frac{4}{M} \alpha^{}_{2} C^{}_{AA} \right)^{2} \epsilon^{\, \mu \nu \rho \sigma} 
k^{}_{1\rho} k^{}_{2\sigma} \frac{i}{k^{2} - \xi^{}_{B} M^{2}_{B}} \epsilon^{\, \mu' \nu' \rho' \sigma'} 
k^{}_{1\rho'} k^{}_{2\sigma'} \,  P^{\nu \nu'}_{o},   \nonumber\\
{\mathcal C}_{\xi 0} &=& \Delta^{\mu \nu}(-k^{}_{1}, -k^{}_{2}) \left(  2 i \frac{m^{}_{f}  }{ M^{}_{B} }\right) 
\frac{i}{k^{2} - \xi^{}_{B} M^{2}_{B} } \left(  2 i \frac{m^{}_{f}  }{ M^{}_{B} }\right) 
\Delta^{\mu' \nu'}(k^{}_{1}, k^{}_{2}) P^{\nu \nu'}_{o},     \nonumber\\
{\mathcal D}_{\xi 0} &=&  2 \times  \left( \frac{4}{M} \alpha^{}_{2} C^{}_{AA} \epsilon^{\, \mu \nu \rho \sigma} 
k^{}_{1\rho} k^{}_{2\sigma}   \right)  \frac{i}{k^{2} - \xi^{}_{B} M^{2}_{B} } 
\left(  2i \frac{m^{}_{f}  }{ M^{}_{B} }\right) 
\Delta^{\mu' \nu'}(k^{}_{1}, k^{}_{2}) P^{\nu \nu'}_{o},   \nonumber\\
{\mathcal E}_{\xi 0} &=& 
 2 \times  \Delta^{\mu \nu}(-k^{}_{1}, -k^{}_{2}) \left(  2i \frac{m^{}_{f} }{ M^{}_{B} }\right) 
\frac{i}{k^{2} - \xi^{}_{B} M^{2}_{B} }  \left( \frac{4}{M} \alpha^{}_{2} C^{}_{AA}  \epsilon^{\, \mu' \nu' \rho' \sigma'} 
k^{}_{1\rho'} k^{}_{2\sigma'} \right)  P^{\nu \nu'}_{o}, \nonumber\\
\eeqn
so that using the anomaly equations for the triangles
\beqn
k^{\lambda'} \Delta^{\lambda' \mu' \nu'}(k^{}_{1},k^{}_{2}) &=& a^{}_{3}(\beta) \epsilon^{\, \mu' \nu' \rho' \sigma'} 
k^{}_{1\rho'} k^{}_{2\sigma'}  + 2 m^{}_{f} \Delta^{\mu' \nu'},  \nonumber\\
k^{\lambda} \Delta^{\lambda \mu \nu}(-k^{}_{1},-k^{}_{2}) &=& - a^{}_{3}(\beta) \epsilon^{\, \mu \nu \rho \sigma} 
k^{}_{1\rho} k^{}_{2\sigma}  - 2 m^{}_{f} \Delta^{\mu \nu}, \nonumber
\eeqn
and substituting the appropriate value for the WZ coefficient, with the rotation coefficient of the axion $b$ 
to the Goldstone boson given by $\alpha^{}_{2} = M^{}_{1}/M^{}_{B}$, 
one obtains quite straightforwardly that the condition of gauge independence is satisfied
\beqn
{\mathcal A}^{}_{ \xi 0} + {\mathcal B}^{}_{ \xi 0} + {\mathcal C}^{}_{ \xi 0} + {\mathcal D}^{}_{ \xi 0} 
+ {\mathcal E}^{}_{ \xi 0}  = 0.   
\eeqn
\section{Appendix. Ward Identities on the tetragon} 
As we have seen in the previous sections, the shift dependence from the anomaly on each vertex, parameterized 
by  $\beta, \beta_1, \beta_2$, drops in the actual computation of the unitarity conditions on the $s$-channel amplitudes, which clearly signals the irrelevance of these shifts in the actual computation, as far as the Bose 
symmetry of the corresponding amplitudes that assign the anomaly on each vertex consistently, are respected.  
It is well known that all the contribution of the anomaly in correlators with more external legs 
is taken care of by the correct anomaly cancellation in three-point function. It is instructive to illustrate, for generic shifts, 
chosen so to respect the symmetries of the higher point functions, how a similar patterns holds. This takes place 
since anomalous Ward Identities for higher-order correlators are expressed in terms of standard triangle 
anomalies. This analysis and a similar analysis of other diagrams of 
this type, which we have included in an appendix, is useful for the 
investigation of some rare $Z$ decays (such as $Z \rightarrow \gamma \gamma \gamma$) which takes place 
with an on-shell $Z$ boson.  

\begin{figure}[t]
{\centering \resizebox*{5cm}{!}{\rotatebox{0}
{\includegraphics{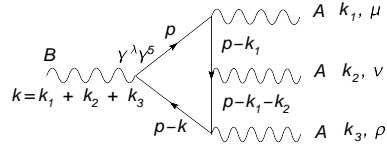}}}\par}
\caption{\small The tetragon contribution.}
\label{tetragon2}
\end{figure}

\begin{figure}[t]
{\centering \resizebox*{12.0cm}{!}{\rotatebox{0}
{\includegraphics{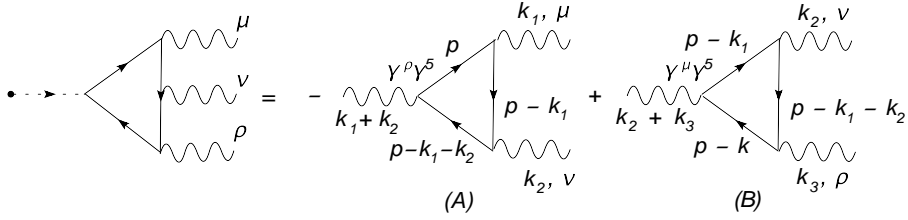}}}\par}
\caption{\small Momenta distribution for the external lines in a WI.}
\label{ward}
\end{figure}

Then let's consider the tetragon diagram BAAA shown in Fig.~\ref{tetragon2}, where $B$, being characterized 
by an axial-vector coupling, generates an anomaly in the related WI.  
We have the fermionic trace
\beqn
\Delta^{\lambda \mu \nu \rho}(k_{1}, k_{2}, k_{3}) = \overline{\Delta}^{\, \lambda \mu \nu \rho}(k_{1}, k_{2}, k_{3}) + \mbox{perm.}
\eeqn
where perm. means permutations of $\{(k_{1}, \mu),(k_{2}, \nu),(k_{3}, \rho) \}$.
One contribution to the axial WI comes for instance from 
\beqn
&&k^{\lambda} Tr \left[ \gamma^{\lambda} \gamma^{5} \frac{1}{\slash p - \slash k} \gamma^{\rho} 
\frac{1}{\slash p - \ds{ k}_{1} - \ds{k}_{2} } \gamma^{\nu} 
\frac{1}{\slash p -\ds{k}_{1} } \gamma^{\mu} \frac{1}{\slash p} \right]     \nonumber\\ 
&=& Tr \left[ \slash k \gamma^{5} \frac{1}{\slash p - \slash k} \gamma^{\rho} 
\frac{1}{\slash p - \ds{ k}_{1} - \ds{ k}_{2} } \gamma^{\nu} \frac{1}{\slash p -\ds{ k}_{1} } \gamma^{\mu} 
\frac{1}{\slash p} \right]     \nonumber\\
&=&  -  Tr \left[ \gamma^{5} \frac{1}{\slash p} \gamma^{\rho} 
\frac{1}{\slash p - \ds{k}_{1} - \ds{k}_{2}} \gamma^{\nu} \frac{1}{\slash p -\ds{ k}_{1} } \gamma^{\mu} \right]   \nonumber\\
&&+ Tr \left[ \gamma^{5} \frac{1}{\slash p - \slash k} \gamma^{\rho} 
\frac{1}{\slash p - \ds{ k}_{1} - \ds{ k}_{2} } \gamma^{\nu} \frac{1}{\slash p -\ds{ k}_{1} } \gamma^{\mu} \right],
\label{tetrag}
\eeqn 
which has been rearranged in terms of triangle anomalies using
\beqn
\frac{1}{\slash p} \slash k \gamma^{5}  \frac{1}{\slash p -\slash k} = \gamma^{5} \frac{1}{\slash p - \slash k} 
- \gamma^{5} \frac{1}{\slash p}.
\eeqn
Relation (\ref{tetrag}) is diagrammatically shown in Fig.~\ref{ward}.
Explicitly these diagrammatic equations become 
\beqn
k^{\lambda} \overline{\Delta}^{\, \lambda \mu \nu \rho} &=& -  \overline{\Delta}^{\, \rho \mu \nu}(k_{1}, k_{2}) 
+ \overline{\Delta}^{\, \mu \nu \rho}(\beta_1, k_{2}, k_{3}),   \nonumber\\
k^{\lambda} \overline{\Delta}^{\, \lambda \mu \rho \nu} &=& -  \overline{\Delta}^{ \, \nu \mu \rho}(k_{1}, k_{3}) 
+ \overline{\Delta}^{\, \mu \rho \nu}(\beta_2, k_{3}, k_{2}),   \nonumber\\
k^{\lambda} \overline{\Delta}^{\, \lambda \nu \rho \mu} &=& -  \overline{\Delta}^{\, \mu \nu \rho}(k_{2}, k_{3}) 
+ \overline{\Delta}^{\, \nu \rho \mu}(\beta_3, k_{3}, k_{1}),   \nonumber\\
k^{\lambda} \overline{\Delta}^{\, \lambda \nu \mu \rho} &=& -  \overline{\Delta}^{\, \rho \nu \mu}(k_{2}, k_{1}) 
+ \overline{\Delta}^{\,  \nu \mu \rho}(\beta_4, k_{1}, k_{3}),   \nonumber\\
k^{\lambda} \overline{\Delta}^{\, \lambda \rho \mu \nu} &=& -  \overline{\Delta}^{\, \nu \rho \mu}(k_{3}, k_{1}) 
+ \overline{\Delta}^{\,  \rho \mu \nu}(\beta_5, k_{1}, k_{2}),   \nonumber\\
k^{\lambda} \overline{\Delta}^{\, \lambda \rho \nu \mu} &=& -  \overline{\Delta}^{\, \mu \rho \nu}(k_{3}, k_{2}) 
+ \overline{\Delta}^{\,  \rho \nu \mu}(\beta_6, k_{2}, k_{1}),   \nonumber\\
\eeqn
where the usual (direct) triangle diagram is given for instance by
\beqn
\overline{\Delta}^{\, \mu \nu \rho} = \int \frac{d^4 p}{(2 \pi)^4} Tr \left[  \gamma^{\mu} \gamma^{5} \frac{1}{\slash p - \slash k} 
\gamma^{\rho} \frac{1}{\slash p - \ds{ k}_{1} - \ds{ k}_{2}} \gamma^{\nu} \frac{1}{ \slash p - \ds{ k}_{1}} \right].
\eeqn
Adding all the contributions we have 
\beqn
k^{\lambda} \Delta^{\lambda \mu \nu \rho}(k_1, k_2, k_3) &=& 
 \left[ \Delta^{\rho \mu \nu}(\beta_5, \beta_6,k_1, k_2) 
+ \Delta^{\nu \mu \rho}(\beta_3, \beta_4,k_1, k_3) 
+ \Delta^{\mu \nu \rho}(\beta_1, \beta_2, k_2, k_3) \right]  \nonumber\\
&&- \left[  \Delta^{\rho \mu \nu}(k_1, k_2) + \Delta^{\nu \mu \rho}(k_1, k_3) + \Delta^{\mu \nu \rho}(k_2, k_3) \right].
\label{WARD}
\eeqn
At this point, to show the validity of the WI independently of the chosen value of the CS shifts, 
we recall that under some shifts
\beqn
\Delta^{\mu \nu \rho}(\beta_1, \beta_2, k_2, k_3) &=&  \Delta^{\mu \nu \rho}(k_2, k_3) - \frac{i (\beta_1 + \beta_2)}{4 \pi^2} 
\epsilon^{\mu \nu \rho \sigma} (k_{2}^{\sigma} - k_{3}^{\sigma})   \nonumber\\
\Delta^{\nu \mu \rho}(\beta_3, \beta_4, k_2, k_3) &=&  \Delta^{\nu \mu \rho}(k_1, k_3) - \frac{i (\beta_3 + \beta_4)}{4 \pi^2} 
\epsilon^{\nu \mu \rho \sigma} (k_{1}^{\sigma} - k_{3}^{\sigma})   \nonumber\\
\Delta^{\rho \mu \nu}(\beta_5, \beta_6, k_1, k_2) &=&  \Delta^{\rho \mu \nu}(k_1, k_2) - \frac{i (\beta_5 + \beta_6)}{4 \pi^2} 
\epsilon^{\rho \mu \nu \sigma} (k_{1}^{\sigma} - k_{2}^{\sigma}),   \nonumber\\
\eeqn
and redefining the shifts by setting 
\beqn
\beta_5 + \beta_6 = \overline{\beta}_1 \qquad \beta_1+ \beta_2 = \overline{\beta}_3 \qquad \beta_3+ \beta_4 = \overline{\beta}_2 
\eeqn
we obtain
\beqn
k^{\lambda} \Delta^{\lambda \mu \nu \rho}(k_1, k_2, k_3) = - \frac{i}{ 4 \pi^2}  \left[ \,
\overline{\beta}_1 \epsilon^{\rho \mu \nu \sigma}(k_1^\sigma - k_2^\sigma) 
+ \overline{\beta}_2 \epsilon^{\nu \mu \rho \sigma}(k_1^\sigma - k_3^\sigma)  
+ \overline{\beta}_3  \epsilon^{\mu \nu \rho \sigma}(k_2^\sigma - k_3^\sigma)  \right].   
\eeqn
Finally, using the Bose symmetry on the r.h.s. (indices $\mu, \nu, \rho$) 
of the original diagram we obtain 
\beqn
\overline{\beta}_1 = \overline{\beta}_2 = \overline{\beta}_3, 
\eeqn
which is the correct WI: $k^{\lambda} \Delta^{\lambda \mu \nu \rho } = 0$.
\begin{figure}[t]
{\centering \resizebox*{4cm}{!}{\rotatebox{0}
{\includegraphics{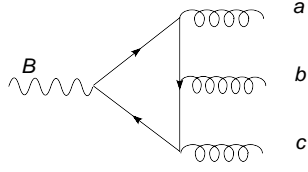}}}\par}
\caption{\small The tetragon diagram in the non-Abelian case.
\label{ward_non_ab}}
\end{figure}
We have shown that the correct choice of the CS shifts in tetragon 
diagrams, fixed by the requirements of Bose symmetries of the 
corresponding  amplitude and of the underlying three-point functions, gives the correct Ward Identities for these correlators. This is not unexpected, since 
the anomaly appears only at the level of three-point functions, 
but shows how one can work in full generality with these amplitudes and 
determine their correct structure. It is also interesting to underline the modifications that take place once this study is extended to the non-Abelian case.
In this case (shown in Fig.~\ref{ward_non_ab}) one obtains the same result already shown for the axial Abelian WI, but modified by color factors. We obtain
\beqn
&&Tr( \{ T^{a},T^{b}\} T^{c}) \, [ - \Delta^{\rho \mu \nu} +\Delta^{\rho \mu \nu}(\beta) ] 
+ \, Tr( \{ T^{c},T^{b}\} T^{a})  \, [- \Delta^{\mu \nu \rho} +\Delta^{\mu \nu \rho}(\beta) ]     \nonumber\\
&&+ Tr( \{ T^{a},T^{c}\} T^{b}) \, [- \Delta^{\nu \mu \rho} +\Delta^{\nu \mu \rho}(\beta)]    \nonumber\\
&=& d^{abc} \, [ - \Delta^{\rho \mu \nu} +\Delta^{\rho \mu \nu}(\beta) ] 
+ d^{cba} \, [- \Delta^{\mu \nu \rho} +\Delta^{\mu \nu \rho}(\beta) ] + d^{acb} \, [- \Delta^{\nu \mu \rho} +\Delta^{\nu \mu \rho}(\beta)],
\eeqn
where we have used the usual definition of the symmetric d-tensor
\beqn
d^{abc} = Tr ( \left\{ T^{a}, T^{b} \right\} T^{c}).
\eeqn
 Simple manipulations give a result which is proportional to the result of the 
Abelian case
\beqn
d^{abc} (  [ - \Delta^{\rho \mu \nu} +\Delta^{\rho \mu \nu}(\beta) ] 
+  [- \Delta^{\mu \nu \rho} +\Delta^{\mu \nu \rho}(\beta) ]+  [- \Delta^{\nu \mu \rho} +\Delta^{\nu \mu \rho}(\beta)]  ).
\eeqn
The vanishing of the shift-dependence is related to the Bose symmetry under exchange of the indices 
$$\{(a, \mu, k_{1}), (b, \nu, k_{2}), (c, \rho, k_{3}) \}.$$
This result is clearly expected, since the gauge current of $B$ is Abelian and behaves as a 
gauge-singlet current under the gauge interaction of $A$, the latter having been promoted to a 
non-Abelian current. 
\section{Appendix. DR-HVBM}
In this appendix we give some of the details for the computation of the direct plus exchanged diagrams in Dimensional Regularization using the HVBM scheme for a partially anticommuting $\gamma^{}_5$ 
\cite{'tHooft:1972fi, Breitenlohner:1977hr}. There are various results presented in the previous literature on the computation of these diagrams, most of them using a 
momentum shift without actually enforcing a regularization, shift that brings the anomaly contribution to 
the axial-vector vertex of the triangle diagram, keeping the vector Ward Identities satisfied, which takes to 
Rosenberg's parameterization (\ref{RRos}). We fill this gap by showing how the regularization works using an arbitrary tensor structure $T^{\la\mu\nu}$ rather than scalar amplitudes. We also keep 
the mass of the fermion arbitrary, so to obtain a general result concerning the mass dependence of the corrections to the anomaly contributions. We remind that momentum shift are allowed in DR-HVBM, once the integration measure is extended from 4 to $n= 4 -\epsilon$ dimensions and the Feynman parametrization can be used to reduce 
the integrals into symmetric forms. Symmetric integration can then be used exactly as in the standard DR 
case, but with some attention on how to treat the Lorentz indices in the two subspaces of dimensions 
4 and $n-4$, introduced by the regularization. These points are illustrated below.  
 
In the following we will use the notation $I_{x y}$ to denote the parametric integration after performing the loop 
integral 

\beq
I_{x y}\left[...\right]\equiv 2 \int_0^1\int_0^{1-x} dy \left[...\right].
\eeq
 
 There are various ways to implement $\gamma_5$ in D-dimensions, but the prescription 
that works best and is not so difficult to implement is the t'Hooft-Veltman-Breitenlohner-Maison (simply denoted as HVBM) prescription. In the HVBM prescription $\gamma_5$ is 
only partially anticommuting. The gamma algebra in this case is split into $n= 4 + (n-4)$, and the indices of the matrices are split accordingly: $\mu=(\tilde{\mu},\hat{\mu})$. There are now two subspaces, and the indices carrying a $\tilde{} $ are the 4-dimensional ones. The 4-dimensional part of the algebra is the same as usual, but now 
\beq
\left[\gamma^{\hat{\mu}},\gamma_5\right]_+=0, 
\eeq
where the commutators have been replaced by anticommutators. 
It is important to clarify some points regarding the use of symmetric integration. 
We recall that in DR the use of symmetric integration gives 
\beq
\int d^n q  \frac{q^{\hat{\alpha}},q^{\tilde{\alpha}}}{(q^2 - \Delta)^{L}}=0,
\eeq
and 
\beqa
\int d^n q  \frac{q^{\hat{\mu}} q^{\hat{\nu}}}{({q^2 - \Delta})^{L}}&=& 
g^{\hat{\mu} \hat{\nu}}\int d^n q  \frac{q^2/n}{{(q^2 - \Delta)}^{L}},  \nonumber \\
\int d^n q  \frac{q^{\tilde{\mu}} q^{\tilde{\nu}}}{({q^2 - \Delta})^{L}}&=& 
g^{\tilde{\mu} \tilde{\nu}}\int d^n q  \frac{q^2/n}{{(q^2 - \Delta)}^{L}}.
\eeqa
Integrals involving mixed indices are set to vanish. 
We now summarize other properties of this regularization. We denote by
\beq
g_{\mu\nu}, \qquad  \tilde{g}_{\mu\nu}, \qquad \hat{g}_{\mu\nu}  
\eeq
respectively the n -, 4 - and (n-4) -dimensional parts of the metric tensor. An equivalent notation is to set 
$\hat{g}_{\mu\nu}=g_{\hat{\mu} \hat{\nu}}$ and $\tilde{g}_{\mu\nu}=g_{\tilde{\mu} \tilde{\nu}}$, 
$\gamma^{\hat{\mu}}=\hat{\gamma}^\mu$, etc.
The contraction rules are 
\beq
g_\mu^\mu=n, \qquad g_{\mu\la}g^{\la}_{\nu}=g_{\mu\nu}, \qquad {\hat{g}}_{\mu}^{\mu}= n-4, \qquad 
\tilde{g}_\mu^\mu=4, \qquad \tilde{g}_{\mu \lambda}{\hat{g}}^{\lambda \nu}=0.
\label{rule1}
\eeq
Other properties of this regularization follow quite easily. For instance, from 
\beq
\tilde{\gamma}_\mu=\gamma^\sigma \tilde{g}_{\sigma \mu}, \qquad \hat{\gamma}^\mu=\gamma_\la \hat{g}^{\la\mu}, 
 \eeq
using (\ref{rule1}) it follows straightforwardly that  
\beqa
\tilde{\gamma}_\mu\gamma_{a_1}\gamma_{a_2}...\gamma_{a_D}\hat{\gamma}^\mu &=&0, \nonumber \\
{\gamma}_\mu\gamma_{a_1}\gamma_{a_2}...\gamma_{a_D}\hat{\gamma}^\mu &=& 
\hat{\gamma}_\mu\gamma_{a_1}\gamma_{a_2}...\gamma_{a_D}\hat{\gamma}^\mu.
\eeqa
The definition of $\gamma^{}_5$ involves an antisymmetrization over the basic gamma matrices 
\beq
\gamma_5 \equiv \frac{i}{4!} \epsilon_{\mu\nu\rho\sigma}\gamma^{\mu}\gamma^{\nu}\gamma^{\rho}\gamma^\sigma. 
\eeq
The definition is equivalent to the standard one 
$\gamma_5= i \gamma_0 \gamma_1 \gamma_2\gamma_3$. The $\epsilon$-tensor is a 4-dimensional 
projector that selects only the ${\bf \tilde{}}$ indices of a contraction,
\beq
 \epsilon_{\mu \nu \ro \si}\gamma^{\mu}\gamma^{\nu}\gamma^{\rho}\gamma^{\si}= 
 \epsilon_{\tilde{\mu}\tilde{ \nu} \tilde{\ro}\tilde{\si}}\gamma^{\tilde{\mu}}\gamma^{\tilde{\nu}}\gamma^{\tilde{\rho}}
\gamma^{\tilde{\si}}. 
\eeq
It is then easy to show that with this definition 
\beq
\left\{\gamma_5,\tilde{\gamma}^\mu\right\}=0, \qquad \left[\gamma_5,\hat{\gamma}^\mu\right]=0.
\eeq
These two relations can be summarized in the statement
\beq
\left\{\gamma_5,{\gamma}_\mu\right\}=2 \hat{\gamma}_\mu \gamma_5.
\eeq
We compute the traces and remove the hat-momenta of the two external vector currents. We illustrate some steps of the computation. 
We denote by $I[...]$ a typical momentum integral that appears in the computation 
\beq
I[...]\equiv \int \frac{d^n q}{(2 \pi)^n} \frac{[...]}{\left(q^2 - \Delta\right)^3},
\eeq
setting $n=4 - \epsilon$, for instance we get
\beqa
I\left[\epsilon[k_1, k_2, {\mu},{\nu}]\right] \hat{q}_{\la} &=&0,    \nonumber \\
I\left[\epsilon[k_2,{\la},{\mu},{\nu}] \right] \hat{q}\cdot\hat{q} &=& \epsilon[k_2, {\la}, {\mu}, {\nu}] (n - 4) I_2, \nonumber \\
I\left[\epsilon[k_2,q,{\mu},{\nu}]\right] \hat{q}_{\la}&=& 0, \nonumber \\
I \left[\epsilon[k_1,q,{\mu},{\nu}]\right] \tilde{q}_{\la} &=& \epsilon[k_1, {\la}, {\mu},{\nu}] I_2, \nonumber \\
I\left[\epsilon[k_2,q,{\mu},{\nu}]\right] q_{\la}&=& \epsilon[k_2,{\la},{\mu},{\nu}] I_2.
\eeqa
Denoting by D and E the direct and the exchanged diagram (before the integration over the Feynman parameters x,y), 
we obtain
\beq
D + E= - i I_{x y} \left[a_1 c_1 + a_2 c_2 + a_3 c_3 + a_4 c_4 + a_5 c_5 \right],
\eeq
where
\beqa
c_1 &=&-4i {I_2} [ n(-2 +x +y ) + 2(2 + x + y) ] \nonumber \\
&& +4 i {I_1} [ m_f^2 ( -2 + x + y) + sx(1 - x + xy -y + y^2)  ],  \nonumber \\
c_2 &=& - c_1, \nonumber \\
c_3 &=& 8i I_1 x(x-y-1)( {k_1}_{\la} + {k_2}_{\la}),   \nonumber \\
c_4 &=& 8i I_1( x + y -1)( y {k_1}_{\mu} - x {k_2}_{\mu}),    \nonumber \\
c_5 &=& 8i I_1 (x + y -1)(x {k_1}_{\nu} - y {k_2}_{\nu}),      \nonumber \\
a_1 &=& \epsilon[k_1,{\la},\mu,\nu],     \nonumber \\
a_2 &=&  \epsilon[k_2,{\la},\mu,\nu], \nonumber \\
a_3 &=&  \epsilon[k_1,k_2,\mu,\nu],   \nonumber \\
a_4 &=&  \epsilon[k_1,k_2,{\la},\nu],   \nonumber \\
a_5 &=&  \epsilon[k_1,k_2,{\la},\mu],     \nonumber \\
\eeqa
and introducing the dimensionally regulated expressions of $I_1$ and $I_2$ and expanding in $\epsilon$ we obtain
\beqa
c_1&= &\frac{1}{\epsilon}\frac{3 x+3 y-2}{4 \pi ^2} \nonumber \\
&& + \frac{x (x+y-1) (2 y-1) s+(3 x+3 y-2) \left(s x y  - m_f^2\right) 
\log
   \left(\frac{m_f^2 - s x y }{\mu ^2} \right)}{ 8 \pi ^2 \left( m_f^2 - s x y  \right)}, \nonumber \\
c_3 &=& - \frac{x(x-y-1)}{4 \pi^2 (s x y - m_f^2)}\left( k_{1\la} + k_{2\la}\right),
\nonumber \\
c_4 &=& 
\frac{(x+y-1) (k_{2\mu} x-k_{1 \mu} y)}{4 \pi ^2 \left(s x y-m_f^2 \right)},  \nonumber \\
c_5 &=& -\frac{(x+y-1) (k_{1 \nu} x- k_{2 \nu} y)}{4 \pi ^2 \left( s x y - m_f^2 \right)},
\eeqa
where $\mu$ is the renormalization scale in the $\overline{MS}$ scheme  
with $\overline{\mu}^{\, 2} = \mu^2 e^{\bf \gamma}/(4 \pi)$  and ${\bf \gamma}$ is the Euler-Mascheroni constant.
The pole singularity is related to tensor structures which have a lower mass dependence on $k_1$ and $k_2$ 
($a_1$ and $a_2$) which involve loop integrations with an additional powers of $q$ and are, therefore, UV divergent. 
However, the pole contributions vanish after integration over the Feynman parameter, since
\beq
I_{x y}\left[ 3 x + 3 y -2 \right]=0.
\eeq
Performing the integration over the Feynman parameters we obtain the result 
reported below in Eq.~(\ref{firstres}).

\subsection{The vanishing of a massive {\bf AAV/VVV}} 
The vanishing of {\bf AAV } in DR in the general case 
(with non-vanishing fermion masses) can be established by a direct 
computation, beside using C-invariance (Furry's theorem). The vanishing of this diagram is due to 
the specific form of all the Feynman parameters which multiply every covariant structure in the corresponding tensor amplitude. Denoting by $X^{\lambda\mu\nu} $ any of these generic structures, the parametric integral is of the form 

\beq
\Delta_{AAV}^{\lambda\mu\nu}= X^{\lambda\mu\nu}\int_{0}^{1} dx \int_0^{1-x} dy \frac{f(x,y)}{\Delta(x,y)} + \dots
\eeq
with $f(x,y)$ antisymmetric in $x,y$ and $\Delta(x,y)$ symmetric, giving a vanishing result. 
For {\bf VVV} the result is analogous.
\subsection{{\bf AVV} and shifts}  
If we decide to use a shift parameterization of the diagrams then the two values of the amplitudes 
$\ul{a_1}$ and $\ul{a_2}$, are arbitrary. This point has been discussed in the previous sections, 
although here we need to discuss with further detail and include in our analysis fermion mass effects as well. However, the use of Dimensional Regularization is such to determine an equal distribution of the anomaly among diagrams of the form {\bf AAA}, no matter which parameterization of the momentum we choose in the graph. Therefore, in this case, if a current is conserved, there is no need to add CS interactions or, equivalently, perform a shift in order to remove the anomaly from vertices which are vector-like.

The first significant parameterization of the anomalus diagram can be found in Rosenberg's paper, later used by Adler in his work on the axial anomaly. The shift is fixed by requiring CVC, which is practical matter rather than a fundamental 
issue. We will show that this method can be mapped into the DR-HVBM result using the Schouten identity. 
We start from Rosenberg's parameterization
\beqa
T^{\la\mu\nu} &=& A_1 \epsilon[k_1,\mu,\nu,\la] + A_2 \epsilon[k_2,\mu,\nu,\la] +
A_3 \epsilon[k_1,k_2,\mu,\la]{k_1}^{\nu}
 + A_4\epsilon[k_1,k_2,\mu,\la]k_2^{\nu}\nonumber \\
&& + A_5 \epsilon[k_1,k_2,\nu,\la]k_1^\mu 
+ A_6 \epsilon[k_1,k_2,\nu,\la]k_2^\mu
\label{Ros}
\eeqa
given in \cite{Rosenberg:1962pp}. By power-counting, two invariant amplitudes are divergent, $A_1$ and $A_2$, while the $A_i$ with $i\geq 3$ are finite\footnote{
We will be using the notation $\epsilon[a,b,\mu,\nu]\equiv \epsilon_{\alpha\beta\mu\nu}
a^\alpha b^\beta$ to denote the structures in the expansion of the anomalous triangle diagrams}. 
In general $A_1$ and $A_2$ are given by parametric integrals which are divergent and there are two free 
parameters in these integrals, amounting to momentum shifts, that can be chosen to render $A_1$ and $A_2$ finite. 
It is possible to redefine the momentum shifts so that the divergences are removed, and this can be 
obtained by imposing the defining Ward Identities (conservation of the two vector currents) in the diagrams
\beq
k_{1 \mu}T^{\la\mu\nu}=k_{2\nu}T^{\la\mu\nu}=0.
\eeq
This gives $A_1=s/2\,\, A_3$ and $A_2=s/2\,\, A_6$. The expressions of the invariant amplitudes 
$A_i$ are given in Rosenberg as implicit parametric integrals. 
They can be arranged in the form
\beqa
A_1 &=& -\frac{i}{4 \pi ^2}+ i C_0(m_f^2,s) \nonumber \\
A_2 &=& \frac{i}{4 \pi ^2}- i C_0(m_f^2,s) \nonumber \\
A_3 &=& - \frac{i}{2 s \pi^{2} } + \frac{2i}{s} C_0(m_f^2,s)\nonumber \\
A_4 &=& 
\frac{i}{s \pi ^2}  -  i f(m_f^2,s)\nonumber \\
A_5 &=&-A_4 \nonumber \\
A_6 &=& -A_3
\label{As}
\eeqa
where we have isolated the mass-independent contributions, which will appear in the anomaly, 
from the mass corrections dependent on the fermion mass ($m_f$), and we have defined 
\beq
C_0(m_f^2,s)= \frac{ {Li}_2\left(\frac{2 }{1 - \sqrt{ 1 - 4 m_f^2
   /s}} \right) m_f^2}{2 s \pi ^2} + \frac{ {Li}_2\left(\frac{2 }{1 + \sqrt{1 - 4 m_f^2/ s}}\right) m_f^2}{2 s \pi
   ^2},
\label{C0}
\eeq
\beq
f(m_f^2,s)=\frac{ \sqrt{1- 4 m_f^2/s}  \,\tanh ^{-1}\left(\frac{{1}}{\sqrt{1- 4
   m_f^2/s}}\right)}{s \pi ^2}.
\label{fm}
\eeq
Eqs.~(\ref{C0}) and (\ref{fm}) have been obtained integrating the parametric expressions of Rosenberg.

The axial vector WI is obtained from the contraction 
\beqa
&& \left( k_{1\la} + k_{2\la}\right) T^{\la\mu\nu}   \nonumber\\
&=&\Big(  - \frac{i}{2  \pi ^2}  
+ \frac{i {Li}_2 \left( \frac{2 }{ 1 - \sqrt{1 - 4 m_f^2 / s } } \right) m_f^2 }{ s \pi^2 } 
+ \frac{i {Li}_2 \left( \frac{2}{ 1 + \sqrt{1 - 4 m_f^2 /s } } \right) m_f^2}{ s \pi^2}  \Big)   
\epsilon[k_1,k_2,\mu,\nu]   \nonumber\\
&=& \left( - \frac{i}{2  \pi ^2} + 2 \,i\, C_0(m_f^2,s)
\right) \times \epsilon[k_1,k_2,\mu,\nu]
\label{finitemass}
\eeqa
where the first contribution is the correct value of the anomaly. The 
remaining term, expressed in terms of dilogarithmic functions, is related 
to the scalar three-point function, as shown below. 
\subsection{DR-HVBM scheme}
\subsubsection{ The {\bf AVV} diagram }
In this case if we use DR we obtain
\beqa
T_{\la\mu\nu} &=& -i \tau_1\,\,\Big( \epsilon[k_1,{\la},\mu,\nu] - \,\,\epsilon[k_2,{\la},\mu,\nu]\Big) 
- i {\tau_2}\left( k_{1\la} + k_{2 \la}\right)\,\, 
\epsilon[k_1,k_2,\mu,\nu] \nonumber \\
&& - i {\tau_3}\left(k_{1\mu} - k_{2 \mu}\right) \,\, \epsilon[k_1,k_2,{\la},\nu] 
- i {\tau_3}\left(k_{1\nu} - k_{2 \nu}\right)\,\, \epsilon[k_1,k_2,{\la},\mu] 
\eeqa
\beqa
\tau_1 &=& -  \frac{ {Li}_2 \left( \frac{2}{ 1 - \sqrt{1 - 4 m_f^2 /s }} \right) m_f^2 }{ 4 s \pi^2} 
 - \frac{{Li}_2 \left( \frac{2 }{ 1 + \sqrt{1 - 4 m_f^2 / s} } \right) m_f^2 }{4 s \pi ^2}   \nonumber\\
&& \,+ \frac{3}{8 \pi ^2}  - \frac{ \sqrt{4 m_f^2 /s - 1 }  \,\,  \tan ^{-1}  \left(  \frac{ 1 }{ \sqrt{ 4 m_f^2 /s - 1 } } \right) }
{4  \pi^{2} }   \nonumber\\
\tau_2 &=&
- \frac{{Li}_2 \left( \frac{2 }{ 1 - \sqrt{1 - 4 m_f^2 / s} } \right) m_f^2 }{ 2 s^{2} \pi^2} 
- \frac{ {Li}_2 \left( \frac{2}{ 1 + \sqrt{1 - 4 m_f^2 / s} } \right) m_f^2 }{ 2 s^{2} \pi^2 } \nonumber \\
&&
+ \frac{ \sqrt{ 4 m_f^2 /s - 1 } \,\,\tan^{-1} \left( \frac{ 1 }{ \sqrt{4 m_f^2 /s - 1 }} \right)}{ 2 s \pi^2} 
-\frac{1}{4 s \pi ^2} \nonumber\\
\tau_3 &=&
\frac{ {Li}_2 \left( \frac{2 }{ 1 - \sqrt{1 - 4 m_f^2 / s } } \right) m_f^2 }{2 s^{2} \pi^2 } 
+ \frac{ {Li}_2 \left( \frac{2}{ 1 + \sqrt{1 - 4 m_f^2 /s } } \right) m_f^2}{2 s^{2} \pi^2} \nonumber \\
&&
+ \frac{ \sqrt{ 4 m_f^2 /s - 1 } \,\, \tan ^{-1}\left( \frac{ 1 }{ \sqrt{4 m_f^2 /s - 1 }}  \right) }{ 2 s \pi^2 }
-\frac{3}{4 s \pi ^2}.
\label{firstres}
\eeqa
The expressions above require a suitable analytic continuation in order to cover all the kinematic 
range of the external invariant (virtuality) $s$. The position of the branch cut 
in the physical region is at $\sqrt{s}=2 m$, corresponding to an $s$-channel cut, where the virtual axial-vector line can produce two on-shell collinear massive fermions.

It is interesting to see how the vector and the axial-vector Ward Identities are satisfied 
for a generic fermion mass $m$. For the vector WI we get
\beqa
k_{1\mu}T^{\mu\nu\la} &=& \frac{i}{2} \left({\tau_3} s+2 {\tau_1}\right)
   {\epsilon[k_1,k_2,\la,\nu]} \nonumber \\
k_{2\nu}T^{\mu\nu\la} &=&- \frac{i}{2}  \left({\tau_3} s+2 {\tau_1}\right)
   {\epsilon[k_1,k_2,\la,\nu]}.
\eeqa
One can check directly that the combination $({\tau_3} s+2 {\tau_1})$ 
vanishes so that $k_{1\mu}T^{\mu\nu\la}=k_{2\nu}T^{\mu\nu\la}=0$.

The second and third term in (\ref{finitemass}) are related to the scalar three-point function 
\beq
C_{00}(k^2,k_1^2,k_2^2,m_f^2,m_f^2,m_f^2)=\int d^4 q \frac{1}{\left(q^2 - m_f^2\right) 
\left( (q + k_1)^2 - m_f^2\right)\left( (q + k_1 + k_2)^2 - m_f^2\right)} \nonumber \\
\eeq
\beqa
C_{00}(k^2,0,0,m_f^2,m_f^2,m_f^2) &=& -\frac{1}{k^2}\left(Li_2\left(\frac{1}{r_1}
\right) + 
Li_2\left(\frac{1}{r_2}\right)\right) \nonumber \\
r_{1,2}&=& \frac{1}{2}\left[ 1 \pm \sqrt{1 - 4\frac{m_f^2}{k^2}}\right]
\eeqa
giving the equivalent relation 
\beq
\left( k_{1\la} + k_{2\la}\right) T^{\la\mu\nu}= \left(  - \frac{i}{2  \pi ^2}  
+ \frac{i {Li}_2 \left( \frac{2 }{ 1 - \sqrt{1 - 4 m_f^2 / s } } \right) m_f^2 }{ s \pi^2 } 
+ \frac{i {Li}_2 \left( \frac{2}{ 1 + \sqrt{1 - 4 m_f^2 /s } } \right) m_f^2}{ s \pi^2}  \right)   
\epsilon[k_1,k_2,\mu,\nu].
\eeq
Our result for $T^{\la\mu\nu}$ can be easily matched to other 
parameterizations obtained by a shift of the momentum in the loop integral 
performed in 4 dimensions. We recall that in this case one needs 
to impose the defining Ward Identities on the amplitude, rather than obtaining them from a regularization, as in the case of the HVBM scheme. Before doing this, we present the analytically 
continued expressions of (\ref{firstres}) which 
are valid for $\sqrt{s} > 2 m^{}_{f}$ 
and are given by
\beqa
\tau_1 &=&  - \frac{1}{2} C_0(s,m_f^2) + \frac{3}{8 \pi^2} -
\frac{1}{4 \pi^2}\sqrt{1 - 4 m_f^2/s}\,\,\tanh^{-1}
\left(\frac{1}{\sqrt{1 - 4 m_f^2/s}}\right)    \nonumber \\
&=&  - \frac{1}{2} C_0(s,m_f^2) + \frac{3}{8 \pi^2} - \frac{s}{4} f(m_f^2,s),  \nonumber\\
\tau_2 &=&
- \frac{1}{s} C_0(s,m_f^2) - \frac{1}{4 s \pi^2 } +
\frac{1}{2 s \pi^2}\sqrt{1 - 4 m_f^2/s}\,\,\tanh^{-1}
\left(\frac{1}{\sqrt{1 - 4 m_f^2/s}}\right) \nonumber \\
&=& - \frac{1}{s} C_0(s,m_f^2) - \frac{1}{4 s \pi^2 } + \frac{1}{2}  f(m_f^2,s),    \nonumber\\
\tau_3 &=&
\frac{1}{s} C_0(s,m_f^2)  - \frac{3}{4 s\pi^2} + 
\frac{1}{2 s \pi^2}\sqrt{1 - 4 m_f^2/s}\,\,\tanh^{-1}
\left(\frac{1}{\sqrt{1 - 4 m_f^2/s}}\right) \nonumber \\
&=& \frac{1}{s} C_0(s,m_f^2)  - \frac{3}{4 s\pi^2} + \frac{1}{2}  f(m_f^2,s).
\label{ts}
\eeqa
\subsubsection{The {\bf AAA} diagram }
The second case that needs to be worked out in DR is that of a 
triangle diagram containing three axial-vector currents. We use the HVBM scheme for $\gamma_5$. The 
analysis is pretty similar to the case of a single $\gamma_5$. 
In this case we obtain 
\beqa
T_3^{\la\mu\nu} &=& - i \left( I_{xy}[c_1] \epsilon[k_1,\la,\mu,\nu] + I_{xy}[c_2] \epsilon[k_2,\la,\mu,\nu]  
+ I_{xy}[c_3] \epsilon[k_1,k_2,\mu,\nu]
\left( k_1^{\la} + k_2^{\la}\right)    \right. \nonumber \\
&&  \left.  + I_{xy}[c_4^{\mu}] \epsilon[k_1,k_2,\la,\nu] + I_{xy}[c_5^{\nu}]\epsilon[k_1,k_2,\la,\mu] \right)
\eeqa
where $I_ {xy}$ is the integration over the Feynman parameters. Also in this case the coefficients $c_1$ and $c_2$ are divergent 
and are regulated in DR. 
We obtain
\beqa
c_1 &=&
4 i \left({I_2} (n-6) (3 x+3 y-2)+{I_1} \left((-3 x-3
   y+2) m_f^2+ s x \left(y^2-y+x (y-1)+1\right)\right)\right)
\nonumber \\
c_2 &=&-c_1 \nonumber \\
c_3 &=& 8 i I_1 x( x-y-1) \nonumber \\
c_4 &=&-8 i {I_1} (x+y-1) (x {k_2}^{\mu}
   - y {k_1}^{\mu})
\nonumber \\
c_5 &=&8 i {I_1} (x+y-1) (x {k_1}^{\nu} - y {k_2}^{\nu}),
\eeqa
which in DR become 
\beqa
c_1 &=&
\frac{3 x+3 y-2}{4 \pi ^2 \epsilon} +
\frac{(3 x+3 y-2) \left(s x y-m_f^2\right) \log
   \left(\frac{m_f^2-s x y}{\mu ^2}\right)-s x
   (x+y-1) (2 y+1)}{8 \pi ^2 \left(m_f^2-s x y\right)} \nonumber \\
c_3 &=&\frac{x (x-y-1)}{4 \pi ^2 \left(m_f^2-s x y\right)} \nonumber \\
c_4 &=&-\frac{(x+y-1) ({{k_2}^{\mu}} x-{{k_1}^{\mu}} y)}{4 \pi
   ^2 \left(m_f^2-s x y\right)} \nonumber \\
c_5 &=&\frac{(x+y-1) ({{k_1}^{\nu}} x-{{k_2}^{\nu}} y)}{4 \pi
   ^2 \left(m_f^2-s x y\right)}.
\eeqa
After integration over x and y the pole contribution vanishes. We obtain 
\beqa
T^{(3)}_{\la\mu\nu} &=& -i \left(  \tau_1^{(3)}\,\,\left( \epsilon[k_1,{\la},\mu,\nu] - \,\,\epsilon[k_2,{\la},\mu,\nu]\right) 
+  {\tau_2^{(3)}}\left( k_{1\la} + k_{2 \la}\right)\,\, 
\epsilon[k_1,k_2,\mu,\nu]    \right. \nonumber \\
&&\left.   +  {\tau_3^{(3)}}\left(k_{1\mu} - k_{2 \mu}\right) \,\, \epsilon[k_1,k_2,{\la},\nu] 
+  {\tau_3^{(3)}}\left(k_{1\nu} - k_{2 \nu}\right)\,\, \epsilon[k_1,k_2,{\la},\mu]     \right),
\eeqa
\beqa
\tau_1^{(3)} &=& \frac{3 {Li}_2\left(\frac{2}{ 1 - \sqrt{1 - 4 m_f^2 /s}}
\right) m_f^2}{4 s \pi ^2}  + 
\frac{3 {Li}_2 \left(\frac{2 }{1 + \sqrt{1 - 4 m_f^2 / s}} \right) m_f^2}{4 s \pi ^2}    \nonumber\\
&& + \frac{\left( 64 \,m_f^4 / s^2 - 20\, m_f^2 /s  + 1  \right) \,\,\tan^{-1} \left( \frac{1}{ \sqrt{ 4 m_f^2 /s -1 } } 
\right) }{ 4  \pi^2 \sqrt{4 m_f^2 /s - 1 } } 
-\frac{4
   m_f^2}{s \pi ^2} +\frac{5}{24 \pi ^2},
\label{tau1}
\eeqa
\beqa
\tau_2^{(3)} &=& 
- \frac{{Li}_2 \left(\frac{2 }{1 - \sqrt{1 - 4 m_f^2 /s} } \right) m_f^2}{2 s^2 \pi^2} 
-\frac{ {Li}_2 \left( \frac{2}{1 + \sqrt{1  - 4 m_f^2 / s}} \right) m_f^2}{ 2 s^2 \pi ^2}  \nonumber \\
&& + \frac{ \sqrt{ 4 m_f^2 /s  - 1 } \tan ^{-1} \left( \frac{  1 }{\sqrt{4 m_f^2 /s - 1 }} 
\right)}{2 s \pi^2 }  -  \frac{1}{4 s \pi^2},
\label{tau2}
\eeqa
\beqa
\tau_3^{(3)} &=&
\frac{ {Li}_2 \left( \frac{2 }{1 - \sqrt{1 - 4 m_f^2 / s} } \right) m_f^2}{2 s^2 \pi^2 } 
+ \frac{ {Li}_2 \left( \frac{2}{ 1 + \sqrt{1 - 4 m_f^2 / s }}  \right) m_f^2}{ 2 s^2 \pi ^2} \nonumber \\
&& + \frac{ \sqrt{ 4 m_f^2 /s  - 1 } \tan^{-1} \left( \frac{  1  }{ \sqrt{4 m_f^2 / s - 1} } 
\right)}{2 s \pi ^2}-\frac{3}{4
   s \pi ^2}.
\label{tau3}
\eeqa
We present the analytically continued expressions of relations~(\ref{tau1}, \ref{tau2}, \ref{tau3}) valid for $\sqrt{s} > 2 m^{}_{f}$
\beqn
\tau_1^{(3)} &=& \frac{3 {Li}_2\left(\frac{2}{ 1 - \sqrt{1 - 4 m_f^2 /s}}
\right) m_f^2}{4 s \pi ^2}  + 
\frac{3 {Li}_2 \left(\frac{2 }{1 + \sqrt{1 - 4 m_f^2 / s}} \right) m_f^2}{4 s \pi ^2}    \nonumber\\
&& - \frac{\left( 64 \,m_f^4 / s^2 - 20\, m_f^2 /s  + 1  \right) \,\,\tanh^{-1} \left( \frac{1}{ \sqrt{1 - 4 m_f^2 /s} } 
\right) }{ 4  \pi^2 \sqrt{ 1 - 4 m_f^2 /s } } 
-\frac{4
   m_f^2}{s \pi ^2} +\frac{5}{24 \pi ^2},   
\eeqn
\beqn
\tau_2^{(3)} &=& 
- \frac{{Li}_2 \left(\frac{2 }{1 - \sqrt{1 - 4 m_f^2 /s} } \right) m_f^2}{2 s^2 \pi^2} 
-\frac{ {Li}_2 \left( \frac{2}{1 + \sqrt{1  - 4 m_f^2 / s}} \right) m_f^2}{ 2 s^2 \pi ^2}  \nonumber \\
&& + \frac{ \sqrt{1 - 4 m_f^2 /s} \tanh^{-1} \left( \frac{  1 }{\sqrt{1 - 4 m_f^2 /s }} 
\right)}{2 s \pi^2 }  -  \frac{1}{4 s \pi^2},  
\eeqn
\beqn
\tau_3^{(3)} &=&
\frac{ {Li}_2 \left( \frac{2 }{1 - \sqrt{1 - 4 m_f^2 / s} } \right) m_f^2}{2 s^2 \pi^2 } 
+ \frac{ {Li}_2 \left( \frac{2}{ 1 + \sqrt{1 - 4 m_f^2 / s }}  \right) m_f^2}{ 2 s^2 \pi ^2} \nonumber \\
&& + \frac{ \sqrt{  1 - 4 m_f^2 /s } \tanh^{-1} \left( \frac{  1  }{ \sqrt{ 1 - 4 m_f^2 / s} } 
\right)}{2 s \pi ^2}-\frac{3}{4 s \pi ^2}.
\eeqn
In the massless case, the contribution to the WI is given by 
\beqa
k_3^\lambda T^{AAA}_{\la\mu\nu} &=& -\frac{i}{6 \pi^2}\epsilon[k_1,k_2,\mu,\nu]\nonumber \\ 
k_1^\mu T^{AAA}_{\la\mu\nu}&=& -\frac{i}{6 \pi^2}\epsilon[k_2,k_3,\nu,\la] \nonumber \\
k_2^\nu T^{AAA}_{\la\mu\nu}&=& -\frac{i}{6 \pi^2}\epsilon[k_3,k_1,\la,\mu] 
\eeqa
where we have chosen a symmetric distribution of (outgoing) momenta $(k_1,k_2,k_3)$ attached to vertices 
$(\mu,\nu,\la)$, with $k_3=-k=-k_1 - k_2$.
\subsection{Equivalence of the shift-based (CVC) and of DR-HVBM schemes}
The equivalence between the HVBM result and the one obtained 
using the defining Ward Identities (\ref{As}) can be shown 
using the Schouten relation 
\beqa
&& k_i^{\mu_1} \epsilon[\mu_2,\mu_3,\mu_4,\mu_5] +k_i^{\mu_2} \epsilon[\mu_3,\mu_4,\mu_5,\mu_1] 
+ k_{i}^{\mu_3} \epsilon[\mu_4,\mu_5,\mu_1,\mu_2] \nonumber \\
&&+  k_i^{\mu_4} \epsilon[\mu_5,\mu_1,\mu_2,\mu_3] 
+ k_i^{\mu_5} \epsilon[\mu_1,\mu_2,\mu_3,\mu_4]=0, 
\eeqa
that allows to remove the $k_{1,2}^{\la}$ terms in terms of other contributions 
\beqa
k_1^\la \epsilon[k_1,k_2,\mu,\nu] &=& \frac{s}{2}\epsilon[k_1,\mu,\nu,\la] -
k_1^{\mu} \epsilon[k_1,k_2,\nu,\la] + k_1^\nu \epsilon[k_1,k_2,\mu,\la] \nonumber \\
k_2^\la \epsilon[k_1,k_2,\mu,\nu] &=& -\frac{s}{2}\epsilon[k_2,\mu,\nu,\la] -
k_2^{\mu} \epsilon[k_1,k_2,\nu,\la] + k_2^\nu \epsilon[k_1,k_2,\mu,\la].
\eeqa
The result in the HBVM scheme then becomes 
\beqa
T^{\la\mu\nu} &=& -i \left( \tau_1 + \frac{s}{2}\tau_2  \right) \epsilon[k_1,\mu,\nu,\la] -i 
\left( -\tau_1 - \frac{s}{2}\tau_2\right) \epsilon[k_2,\mu,\nu,\la] \nonumber \\
&& -i
\left(\tau_2 - \tau_3\right) \epsilon[k_1,k_2,\mu,\la]{k_1}^{\nu}
 -i \left(\tau_2 + \tau_3\right)\epsilon[k_1,k_2,\mu,\la]k_2^{\nu}\nonumber \\
&& -i \left(-\tau_2 -\tau_3\right) \epsilon[k_1,k_2,\nu,\la]k_1^\mu 
-i \left(\tau_3 -\tau_2\right) \epsilon[k_1,k_2,\nu,\la]k_2^\mu
\eeqa
and it is easy to check using (\ref{As}) and (\ref{ts}) that the invariant 
amplitudes given above coincide with those given  by Rosenberg. Therefore we have the correspondence 
\beqa
A_1 &=& -i (\tau_1 + \frac{s}{2}\tau_2 ) \nonumber \\
A_2 &=&-i(  - \tau_1 - \frac{s}{2}\tau_2 ) \nonumber \\
A_3 &=& -i ( \tau_2 -\tau_3 ) \nonumber \\
A_4 &=&-i ( \tau_2 + \tau_3  ) \nonumber \\
A_5 &=&-i(  - \tau_2 -\tau_3 ) \nonumber \\
A_6 &=& -i( \tau_3 - \tau_2 ). 
\label{idr}
\eeqa
A similar correspondence holds between the Rosenberg parameterization of 
{\bf AAA} and the corresponding DR-HVBM result
\beqa
A^{(3)}_1 &=&- i( \tau^{(3)}_1 + \frac{s}{2}\tau_2 ) \nonumber \\
A^{(3)}_2 &=&-i(  - \tau^{(3)}_1 - \frac{s}{2}\tau^{(3)}_2 ) \nonumber \\
A^{(3)}_3 &=&-i(  \tau^{(3)}_2 -\tau^{(3)}_3 ) \nonumber \\
A^{(3)}_4 &=&-i( \tau^{(3)}_2 + \tau^{(3)}_3 )\nonumber \\
A^{(3)}_5 &=& -i( - \tau^{(3)}_2 -\tau^{(3)}_3 )\nonumber \\
A^{(3)}_6 &=&-i( \tau^{(3)}_3 - \tau^{(3)}_2 ). 
\label{idr1}
\eeqa
\section{Appendix. The Chern-Simons and Wess-Zumino vertices \label{app:momentumCS_GS}}
The derivation of the CS vertex is straightforward and is given by
\beqn
&&\int dx \, dy \, dz\, T_{CS}^{\la\mu\nu}(z,x,y) \, B^{\lambda}(z) A^{\mu}(x) A^{\nu}(y)    \nonumber\\
&=&\int dx \, dy \, dz \, \int \frac{dk_1}{(2 \pi)^4} \frac{dk_2}{(2 \pi)^4} e^{- i k_1 (x - z) - i k_2 (y - z)} 
 \, \epsilon^{\lambda \mu \nu \alpha} \,( k^{\alpha}_1 -k_2^\alpha)\, B^{\lambda}(z) A^{\mu}(x) A^{\nu}(y)   \nonumber\\
&=& \int dx \, dy \, dz \,  i \left( \frac{\partial}{\partial x^{\alpha}}- \frac{\partial}{\partial y^{\alpha}} \right) 
\left( \int  \frac{dk_1}{(2 \pi)^4} \frac{dk_2}{(2 \pi)^4} e^{- i k_1 (x - z) - i k_2 (y - z)} \right) B^{\lambda}(z) 
A^{\mu}(x) A^{\nu}(y) \epsilon^{\lambda \mu \nu\alpha }     \nonumber\\
&=& (- i)  \int dx \, dy \, dz \,  \int \frac{dk_1 \, dk_2}{(2 \pi)^8} e^{- i k_1 (x - z) - i k_2 (y - z)} 
B^{\lambda}(z)\left(  \frac{\partial}{\partial x^{\alpha}} A^{\mu}(x)  A^{\nu}(y) - \frac{\partial}{\partial y^{\alpha}} 
A^{\nu}(y)  A^{\mu}(x) \right)  \epsilon^{\lambda \mu \nu \alpha} \nonumber\\
&=& (-i) \int dx \, dy \, dz \, \delta (x-z) \delta (y-z) B^{\lambda}(z)
\left(  \frac{\partial}{\partial x^{\alpha}}A^{\mu}(x)  A^{\nu}(y) - 
\frac{\partial}{\partial y^{\alpha}}A^{\nu}(y)  A^{\mu}(x)\right) \epsilon^{\lambda \mu \nu \alpha} \nonumber\\
&=& i \int dx A^{\lambda}(x) B^{\nu}(x) F^{A}_{\rho \sigma}(x) \epsilon^{\lambda \nu \rho \sigma}.
\eeqn
Proceeding in a similar way we obtain the expression of the Wess-Zumino vertex
\beqn
&& \int dx \, dy  \, dz  \,  \int \frac{d^{4}k_1}{(2 \pi)^{4}}  \frac{d^{4}k_2}{(2 \pi)^4} \epsilon^{\mu \nu \rho \sigma} 
k_{1}^{\rho} k_{2}^{\sigma} e^{- i k_{1} \cdot (x-z) - i k_{2} \cdot (y - z)} \,  b(z) B^{\mu}(x) B^{\nu}(y)   \nonumber\\
&=&  \int dx \, dy  \, dz  \,  \int \frac{d^{4}k_1}{(2 \pi)^{4}}  \frac{d^{4}k_2}{(2 \pi)^4} \epsilon^{\mu \nu \rho \sigma} 
 \left( \frac{1}{- i} \right)  \frac{\partial}{\partial x^{\rho}}   e^{- i k_{1} \cdot (x-z)}
 \left( \frac{1}{- i} \right)  \frac{\partial}{\partial y^{\sigma}} e^{ - i k_{2} \cdot (y - z)} 
\,  b(z) B^{\mu}(x) B^{\nu}(y)  \nonumber\\
&=&  (- 1) \int dx \, dy  \, dz  \, \delta^{(4)} (x-z) \delta^{(4)}(y -z) \, b(z) \,  
\frac{ \partial B^{\mu}}{\partial x^{\rho}}(x) \,  \frac{\partial B^{\nu}}{\partial y^{\, \sigma}}(y)  \,
 \epsilon^{\mu \nu \rho \sigma} \nonumber\\
&=&  - \frac{1}{4}  \int  dx \, b(x) \,  F^{B}_{\rho \mu}(x) \, F^{B}_{\sigma \nu}(x) \,  \epsilon^{\mu \nu \rho \sigma} 
=  \frac{1}{4}  \int  dx \, b \,  F^{B}_{\rho \mu} \, F^{B}_{\sigma \nu} \,  \epsilon^{\rho \mu \sigma \nu} 
\eeqn
so that we find the following correspondence between Minkowsky space and momentum space for the Green-Schwarz vertex
\beqn
4 \, \epsilon^{\mu \nu \rho \sigma} \, k^{\rho}_{1} k^{\sigma}_{2} \;\; \; \; \;
\leftrightarrow \;\; \; \; \; b F^{B} \wedge F^{B}.
\eeqn
\section{Appendix. Computation of the Effective Action}
In this appendix we illustrate the derivation of the variation of the effective action for typical anomalous contributions involving 
{\bf AVV} and {\bf AAA} diagrams. We consider the case of the A-B Model described in the first 
few sections. We recall that we have the relations 
\beq
\delta B^\mu\,=\, \partial_{\mu}\theta_B\nonumber\\
\qquad \delta A^\mu\,=\, \partial_{\mu}\theta_A.
\eeq
We obtain the following gauge variations
\beqn
\delta_B \mathcal{S}_{BAA}&=&  \delta_B  \int dx\,dy\,dz \, T_{\bf AVV}^{\lambda\mu\nu}(z,x,y)\,
B^\lambda(z)\,A^\mu(x)\,A^\nu(y)\nonumber\\
&=&- \int dx\,dy\,dz \, \partial_{z^\lambda}T_{\bf AVV}^{\lambda\mu\nu}(z,x,y)\,A^\mu(x)\,A^\nu(y)\,
\theta^{}_B(z)\nonumber\\
&=&- i a_3(\beta) \epsilon^{\mu\nu\alpha\beta}\int dx\,dy\,dz \, \partial_{x^\alpha}\partial_{y^\beta}[\delta(x-z)\,\delta(y-z)]\,
A^\mu(x)\,A^\nu(y)\theta_B(z)\nonumber\\
&=&- i a_3(\beta) \epsilon^{\mu\nu\alpha\beta}\int dx  \, \partial_{x^\alpha}A^\mu(x)\,
\partial_{x^\beta}A^\nu(x) \, \theta^{}_B(x)\nonumber\\
&=& i \frac{ a^{}_{3}(\beta) }{4}\int dx \, \theta^{}_B \, F_{\alpha\mu}^A F_{\beta\nu}^A \epsilon^{\alpha\mu\beta\nu},
\eeqn
\beqn
\delta_A \mathcal{S}_{BAA}&=&  \delta_A  \int dx\,dy\,dz \, T_{\bf AVV}^{\lambda\mu\nu}(z,x,y)\,
B^\lambda(z)\,A^\mu(x)\,A^\nu(y)\nonumber\\
&=&- \int dx\,dy\,dz \, \partial_{x^{\,\mu}} T_{\bf AVV}^{\lambda\mu\nu}(z,x,y)
\,B^\lambda(z)\,\theta^{}_A(x)\,A^{\nu}(y)\nonumber\\
&&- \int dx\,dy\,dz \, \partial_{y^{\,\nu}} T_{\bf AVV}^{\lambda\mu\nu}(z,x,y)
\,B^\lambda(z)\,A^{\mu}(x)\,\theta^{}_{A}(y)\nonumber\\
&=& i a_1(\beta) \epsilon^{\lambda\nu\alpha\beta}\int dx\,dy\,dz \, \partial_{x^{\,\alpha}} \partial_{y^{\,\beta}}
[\delta(x-z)\,\delta(y-z)]\,
B^\lambda(z)\,\theta^{}_{A}(x)\,A^\nu(y)\nonumber\\
&=& - i a_1(\beta) \epsilon^{\lambda\nu\alpha\beta}\int dx\, \, \partial_{x^\alpha}
B^\lambda(x)  \,\partial_{x^\beta} A^\nu(x) \,  \theta^{}_{A}(x)\nonumber\\
&&+  i a_1(\beta) \epsilon^{\lambda\mu\alpha\beta} \int dx\, \, 
\partial_{x^\beta} \,B^\lambda(x) \,\partial_{x^\alpha} \, A^{\mu} \,  \theta^{}_{A}(x)\nonumber\\
&=&  i \frac{ a^{}_{1}(\beta) }{4} \,2 \,\int dx \, \theta^{}_A \, F_{\alpha\lambda}^B F_{\beta\nu}^A \,
\epsilon^{\alpha\lambda\beta\nu}.
\eeqn
\section{Appendix. Decay of the axi-Higgs: the pseudoscalar triangle $\chi BB$}
The computation is standard and the result is finite. There are no problems 
with the handling of $\gamma_5$ and so we can stay in 4 dimensions.

We first compute the triangle diagram with the position of zero mass fermion $m^{}_f=0$
 \beq 
 \int \frac{ d^4 q }{ ( 2 \pi )^4 } \frac{ Tr \left[ \gamma^{5} ( \slash{q} - \slash{k}) \gamma^{\nu} ( \slash{q} - \slash{k_1})
 \gamma^{\mu} \slash{q} \right] }{q^2 ( q - k )^2  (q - k_1)^2 }  \mbox{+ exch.}
\eeq
which trivially vanishes because of the $\gamma$-algebra. 
Then the relevant contribution to the diagram comes to be proportional to the mass $m^{}_f \neq 0$, 
as we are now going to show.

We set $k=k_1 + k_2$ and set on-shell the $B$-bosons: $k_1^2 = k_2^2 = M_{B}^2$, so that 
$k^2 = 2 M_{B}^2 + 2 k_1 \cdot k_2 = m_{\chi}^2$

The diagram now becomes
\beq
\int \frac{ d^4 q }{ ( 2 \pi )^4 } \frac{ Tr \left[ \gamma^{5} ( \slash{q} - \slash{k} + m_f ) \gamma^{\nu} ( \slash{q} - 
\slash{k_1} + m_f)  \gamma^{\mu} ( \slash{q} + m_f ) \right] }{ \left[ q^2 - m^2_f \right] \left[ ( q - k )^2 - m^{2}_{f} \right]
\left[ (q - k_1)^2 - m^{2}_{f} \right] }  \mbox{+ exch.}
\label{diagr}
\eeq
Using a Feynman parameterization we obtain
\beqn
&=& 2 \int_{0}^{1}dx \int_{0}^{1-x} \frac{1}{ \left[ q^2 - 2 q [ k_2 y + k_1 (1-x) ] 
+ [\, y m_{\chi}^2 -m_{f}^2 + m_{B}^2(1-x-y)\,]   \right]^3}  \nonumber\\
&=&  2 \int_{0}^{1}dx \int_{0}^{1-x} dy \frac{1}{ \left[ q^2 - 2 q \Sigma +  D \right]^3} = 
2 \int_{0}^{1}dx \int_{0}^{1-x} dy \frac{1}{ \left[ (q - \Sigma)^2 -( \Sigma^2 - D) \,\right]^3}  \nonumber\\
&=&  2 \int_{0}^{1}dx \int_{0}^{1-x} dy \, \frac{1}{ \left[ (q - \Sigma)^2 - \Delta  \,\right]^3}.
\eeqn
We define
\beq 
\Sigma = y k_2 + k_1 ( 1 - x )   
\eeq
and 
\beq 
D =  y m_{\chi}^2 -m_{f}^2 + M_{B}^2(1-x-y),
\eeq
for the direct diagram and the function
\beq
\Delta = \Sigma^2 - D =  m_f^2 - x\,y \,m_{\chi}^2 + m_B^2 (x+y)^2 - x M_B^2 - y m_B^2 \equiv \Delta(x,y,m_f,m_\chi,M_B)
\eeq
and perform a shift of the loop momentum 
\beqn
q^\prime = q - \Sigma
\eeqn
obtaining
\beq
 2\int_{0}^{1}dx \int_{0}^{1-x}dy \int \frac{ d^D q }{ (2 \pi)^D  } \frac{ Tr [ \gamma^{5} ( \slash{q} + \slash{\Sigma}
- \slash{k} + m_{f}) \gamma^{\nu} ( \slash{q} + \slash{\Sigma} - \slash{k_1} + m_{f} ) \gamma^\mu ( \slash{q} + \slash{\Sigma} 
 + m_{f} ) ] }{ [ q^2 - \Delta ]^3 } 
\label{A}
\eeq
Using symmetric integration we can drop linear terms in $q$, together with
$q^\mu q^\nu = \frac{1}{D} q^2 g^{\mu\nu}$. Adding the exchanged diagram and after a routine calculation we obtain the amplitude for the decay
\beqn 
\Delta^{\mu\nu}=  \epsilon^{\alpha \beta \mu \nu }  k_1^{\alpha} k_2^{\beta}
m_f \left( \frac{1}{ 2 \pi^2}  \right) I(m_f,m_\chi,m_B)
\eeqn
with
\beq
I=\int^{1}_{0} dx \int^{1-x}_{0} dy  \frac{1}{\Delta(x,y,m_f,m_\chi,m_B)}. 
\eeq

\chapter{St\"{u}ckelberg Axions and the Effective Action of Anomalous Abelian Models 2.
\\A $SU(3)_C\times SU(2)_W\times U(1)_Y\times U(1)_B$ Model and its signature at the LHC  \label{chap:AbelianModels2}}
\fancyhead[LO]{\nouppercase{Chapter 2. An Abelian extension of the Standard Model}}
\section{Introduction to the chapter}
Among the possible extensions of the Standard Model (SM), those where the 
$SU(3)_C\times SU(2)_W\times U(1)_Y$ gauge group is enlarged by a number of
extra $U(1)$ symmetries are quite attractive for being modest enough departures 
from the SM so that they are computationally tractable, 
but at the same time predictive enough so that they are interesting and even 
perhaps testable at the LHC.
Of particular popularity among these have been models where at least one of the extra
$U(1)$'s is "anomalous", that is, some of the fermion triangle loops with
gauge boson external legs are non-vanishing. The existence of this possibility
was noticed in the context of the (compactified to four dimensions) 
heterotic superstring where the stability of the supersymmetric
vacuum \cite{Dine:1987xk, Atick:1987gy} can trigger in the four-dimensional low energy 
effective action a non-vanishing Fayet-Iliopoulos term proportional to the 
gravitational anomaly, i.e. proportional to the anomalous trace of the corresponding $U(1)$.  
The mechanism was recognized to be the low energy manifestation of the Green-Schwarz anomaly (GS)
cancellation mechanism of string theory.\footnote{Conventionally in this paper 
we will use both the term ``Green-Schwarz'' (GS) to denote the mechanism 
of cancellation of the anomalies, to conform to the string context, though 
the term ``Wess-Zumino'' (WZ) would probably be more adequate and sufficient for our analysis. The corresponding counterterm will be denoted, GS or WZ, with no distinction.} 
Most of the consequent developments were concentrated around exploiting this idea
in conjunction with supersymmetry and the Froggatt-Nielsen mechanism \cite{Froggatt:1978nt} in order to explain the 
mass hierarchies in the Yukawa sector of the SM \cite{Binetruy:1994ru}, supersymmetry breaking \cite{Binetruy:1996uv, Dvali:1996rj},
inflation \cite{Binetruy:1996xj} and axion physics \cite{Georgi:1998au}, in all of which the 
presence of the anomalous $U(1)$ is a crucial ingredient. In the context of theories with extra 
dimensions the analysis of anomaly localization and of anomaly inflow has also been at the center of interesting developments \cite{vonGersdorff:2003dt, Hill:2004uc, Hill:2006ei}. The recent explosion of string model building, in particular in the context
of orientifold constructions and intersecting branes \cite{Blumenhagen:2006ci, Bertolini:2005qh, Anastasopoulos:2006da}
but also in the context of the heterotic string \cite{Faraggi:2006bs},
have enhanced even more the interest in anomalous $U(1)$ models. 
There are a few universal characteristics that these 
vacua seem to possess. One is the presence of $U(1)$ gauge symmetries
that do not appear in the SM \cite{Ibanez:1998qp,Antoniadis:2002cs}. 
In realistic 4-dim heterotic string vacua the SM gauge group 
comes as a subgroup of the 10-dim $SO(32)$ or $E_8\times E_8$ 
symmetry \cite{Green:1987mn, Green:1987sp}, and in practice there is at least one anomalous $U(1)$ factor that appears
at low energies, tied to the SM sector in a particular way, which we will summarize
next. For simplicity and reasons of tractability we concentrate on the simplest non-trivial 
case of a model with gauge group $SU(3)_C\times SU(2)_W\times U(1)_Y\times U(1)_B$ 
where $Y$ is the hypercharge
and $B$ is the anomalous gauge boson and where the fermion spectrum 
is that of the SM. The mass term for the anomalous $U(1)_B$ appears through
a St\"{u}ckelberg coupling \cite{Antoniadis:2002cs, Ghilencea:2002da, Ghilencea:2002by, Anastasopoulos:2003aj, Anastasopoulos:2004ga} and the cancellation of its anomalies is due to axionic and Chern-Simons terms (in the open 
string context see the recent works \cite{Antoniadis:2002cs, Kiritsis:2002aj, Coriano:2005js, Anastasopoulos:2006cz, Anastasopoulos:2007qm}).

Despite of all this theoretical insight both from the top-down and bottom-up approaches,
the question that remains open is how to make concrete contact with experiment.
 However, as mentioned above,
in models with anomalous $U(1)$'s one should quite generally expect the presence
of a physical axion-like field $\chi$ and in fact in any decay that involves a non-vanishing fermion triangle like the 
decay $Z^*, Z^{'*}\longrightarrow \g\g$,  $Z, Z^\prime \longrightarrow Z\g$ etc., one should be able to see
traces of the anomalous structure \cite{Coriano:2005js, Anastasopoulos:2006cz, Anastasopoulos:2007qm, Coriano:2006xh, Coriano:2007fw}. 
In this chapter we will mostly concentrate on the gauge boson decays which, even though
hard to measure,  contain clear differences with respect to the SM - as is the case of
the $Z^*\longrightarrow \g\g$ decay -  and in addition with respect to anomaly free $U(1)$ extensions
- like the $Z^{'*}\longrightarrow \g\g$ decay - for example.
 
In \cite{Coriano:2005js} a theory which extends the SM with this minimal structure 
(for essentially an arbitrary number of extra $U(1)$ factors)
was called "Minimal Low Scale Orientifold Model" or mLSOM for short,
because in orientifold constructions one typically finds multiple anomalous $U(1)$'s.
Here, even though we discuss the case of a single anomalous $U(1)$ 
which could also originate from heterotic vacua or some field theory extension of the SM, 
we will keep on using the same terminology keeping in mind 
that the results can apply to more general cases.
We finally mention that other similar constructions with emphasis on other phenomenological 
signatures of such models have appeared before in \cite{Kiritsis:2002aj, Antoniadis:2002qm, Kors:2004dx, Kors:2004ri, Kors:2005uz, Feldman:2007fq, Feldman:2006wb, Feldman:2007wj, Kors:2004dx}. A perturbative study of the renormalization of these types of models is 
in \cite{McKeon:2006ym}. Other features of these models, in view of the recent activity 
connected to the claimed PVLAS result \cite{Abel:2006qt, Gies:2006ca, Mirizzi:2007hr, Masso:2006xg, DeAngelis:2007dy, DeAngelis:2007yu, Dupays:2006hz}, have been discussed in 
\cite{Coriano:2007fw}.

The chapter is organized as follows. In the first sections we will specialize the analysis of 
\cite{Coriano:2005js} to the case of an extension of the SM that contains one additional anomalous 
Abelian $U(1)$, with an Abelian structure of the form $U(1)_Y\times U(1)_B$. We will determine the structure of the entire Lagrangian and fix the 
counterterms in the one-loop anomalous effective action which are necessary 
to restore the gauge invariance of the model at quantum level. 
The analysis that we provide is the generalization of 
what is discussed in Chap.~\ref{chap:AbelianModels1} that was devoted primarily to the analysis of 
anomalous Abelian models and to the perturbative organization of the corresponding effective 
action. 
After determining the axion Lagrangian and after discussing Higgs-axion mixing 
in this extension of the SM, we will focus our attention on an analysis of the contributions  
to a simple process $(Z\to \gamma \gamma)$. Our analysis, in this case, aims to provide an 
example of how the new contributions included in the effective action - in the form of one-loop counterterms 
that restore unitarity of the effective action - modify the perturbative structure of the process. A detailed phenomenological analysis will be performed in the Chap.~\ref{chap:LHC}, since it requires, to be practically useful for searches at the LHC, a very accurate determination of the QCD and electroweak background around the Z/Z' resonance.  
\section{Effective models at low energy: the $SU(3)_C \times SU(2)_W \times U(1)_Y \times U(1)_B$ case} 
We start by briefly recalling the main features of the mLSOM starting from 
the expression of the Lagrangian which is given by
\begin{eqnarray}
{\cal L}\; =&-&\frac{1}{2} Tr\;[ F^{G}_{\mu\nu}F^{G\mu\nu} ] - \frac{1}{2} Tr[ \; F^{W}_{\mu\nu} F^{W\mu\nu}]
-\frac{1}{4} F^{B}_{\mu\nu} F^{B\mu\nu} -\frac{1}{4} F^{Y}_{\mu\nu} F^{Y\mu\nu}   \nonumber \\
&+&| ( \partial_{\mu} + i g^{}_{2} \frac{ \tau^j }{ 2 } W_{\mu}^j
+i g^{}_{Y} q^{Y}_{u} A_{\mu}^{Y} +i g^{}_{B} \frac{q^{B}_{u}}{2} B_{\mu} ) H_u|^2    \nonumber\\
&+& | ( \partial_{\mu} + i g^{}_{2} \frac{  \tau^j }{ 2 }  W_{\mu}^j
+i g^{}_{Y} q^{Y}_{d} A_{\mu}^{Y} +i g^{}_{B} \frac{q^{B}_{d}}{2} B_{\mu} ) H_d|^2
\nonumber \\
&+&\overline{Q}_{Li} \, i \gamma^{\mu} \left( \partial^{}_{\mu} 
+i g^{}_{3} \frac{\lambda^{a}}{2} G^{a}_{\mu}+ i g^{}_{2} \frac{\tau^{j}}{2} W^{j}_{\mu} 
+ i g^{}_{Y} q^{(Q_L)}_{Y} A^{Y}_{\mu} + i g^{}_{B} q^{(Q_L)}_{B} B_{\mu} \right) Q_{Li} \nonumber\\
&+& \overline{u}_{Ri}  \, i \gamma^{\mu}  \left( \partial_{\mu} + i g^{}_{Y} q^{(u_R)}_{Y}A^{Y}_{\mu} 
+ i g^{}_{B} q^{(u_R)}_{B}  B_{\mu}   \right) {u}_{Ri} 
+ \overline{d}_{Ri}  \, i \gamma^{\mu}  \left( \partial_{\mu} + i g^{}_{Y} q^{(d_R)}_{Y}A^{Y}_{\mu} 
+ i g^{}_{B} q^{(d_R)}_{B}   B_{\mu}  \right)    {d}_{Ri} \nonumber \\
&+& \overline{L}_{i} \, i \gamma^{\mu} \left( \partial^{}_{\mu} + i g^{}_{2} \frac{\tau^{j}}{2} W^{j}_{\mu} 
+ i g^{}_{Y} q^{(L)}_{Y} A^{Y}_{\mu} + i g^{}_{B} q^{(L)}_{B} B_{\mu} \right) L_{i} \nonumber\\
&+& \overline{e}^{}_{Ri} \, i \gamma^{\mu}  \left( \partial_{\mu} + i g^{}_{Y} q^{(e_R)}_{Y}A^{Y}_{\mu} 
+ i g^{}_{B} q^{(e_R)}_{B} B_\mu\right)  {e}_{Ri} +
\overline{\nu}_{Ri} \, i \gamma^{\mu} \left( \partial_{\mu} + i g^{}_{Y} q^{(\nu_R)}_{Y}A^{Y}_{\mu} 
+ i g^{}_{B} q^{(\nu_R)}_{B}  B_\mu   \right) {\nu}_{Ri}\nonumber \\
&-&  \Gamma^{d} \, \overline{Q}_{L} H_{d} d_{R} - \Gamma^{u} \, \overline{Q}_{L} (i \sigma_2 H^{*}_{u}) u_{R} 
+ c.c. \nonumber\\
&-&   \Gamma^{e} \, \overline{L} H_{d} {e}_{R} - \Gamma^{\nu} \, \overline{L} (i \sigma_2 H^{*}_{u}) \nu_{R} 
+ c.c.\nonumber\\
&+& \frac{1}{2}(\partial_{\mu}b + M^{}_{1} B_{\mu})^2 \nonumber\\ 
&+& \frac{C_{BB}}{M} b  F_{B} \wedge F_{B}  
+ \frac{C_{YY}}{M}  b F_{Y} \wedge F_{Y}   + \frac{C_{YB}}{M}  b F_{Y} \wedge F_{B}   \nonumber\\
&+& \frac{F}{M}  b Tr[F^W \wedge F^W]    +  \frac{D}{M}  b Tr[F^G \wedge F^G]   \nonumber\\ 
&+& d_{1} BY \wedge F_{Y} + d_{2}  YB \wedge F_{B}  + c_{1}  \epsilon^{\mu\nu\rho\sigma} B_{\mu} C^{SU(2)}_{\nu\rho\sigma} 
+ c_{2}  \epsilon^{\mu\nu\rho\sigma} B_{\mu} C^{SU(3)}_{\nu\rho\sigma}    \nonumber\\
&+& V(H_u,H_d,b),
\label{action}
\end{eqnarray}
where we have summed over the $SU(3)$ index $a=1,2,...,8$, over the $SU(2)$ index $j=1,2,3$ and over the fermion index $i=1,2,3$ denoting a given generation. 
We have denoted with $F^{G}_{\mu\nu}$ the field-strength for the
gluons and with $F^{W}_{\mu\nu}$ the field-strength of the weak gauge bosons $W_{\mu}$. 
$F^{Y}_{\mu\nu}$ and $F^{B}_{\mu\nu}$ are the field-strengths related to the Abelian hypercharge and the extra Abelian gauge boson B, which has anomalous interactions with a typical generation of 
the Standard Model.
The fermions in Eq.~($\ref{action}$) are either left- or right-handed Dirac spinors $f^{}_L$, 
$f^{}_R$ and they fall in the usual $SU(3)_C$ and 
$SU(2)_W$ representations of the Standard Model.
The additional anomalous $U(1)^{}_{B}$ is accompanied by a shifting St\"{u}ckelberg axion $b$. 
The $c_i$, $i=1,2$, are the coefficients of the Chern-Simons trilinear interactions \cite{Coriano:2005js, Anastasopoulos:2006cz, Anastasopoulos:2007qm} and we have also introduced a mass term 
$M^{}_{1} $ at tree level for the $B$ gauge boson, by means of 
the St\"{u}ckelberg term. As usual, the hypercharge is anomaly-free and its 
embedding in the so called ``$D$-brane basis'' has been discussed extensively  
in the previous literature \cite{Ibanez:1998qp, Antoniadis:2002qm, Ghilencea:2002da, Ghilencea:2002by}. Most of the features of the orientifold construction are preserved, 
but we don't work with the more general multiple $U(1)$ structure since our goal 
is to analyze as close as possible this model making contact with direct phenomenological 
applications, although our results and methods can 
be promptly generalized to more complex situations.  

Before moving to the specific analysis presented in this chapter, some comments are in order concerning the possible range of validity of effective actions of this type and the relation between the value of the cutoff 
parameter $\Lambda$ and the St\"uckelberg mass $M_1$. This point has been addressed before in 
great detail in \cite{Preskill:1990fr} and we omit any further elaboration, quoting the result. Lagrangians 
containing dimension-5 operators in the form of a Wess-Zumino term may have a range of validity constrained 
by  $M_1 \geq g_1 g^2/(64 \pi^3) a_n \Lambda$, where $g_1$ is the coupling at the chiral 
vertex where the anomaly $a_n$ is assigned and $g$ is the coupling constant of the other two 
vector-like currents in a typical AVV diagram. More quantitatively, this bound can be reasonably assumed 
to be of the order of $10^5$ GeV, by a power-counting analysis. Notice 
that the arguments of \cite{Preskill:1990fr}, though based on the picture of ``partial decoupling'' 
of the fermion spectrum, in which the pseudoscalar field is the phase of a heavier Higgs, 
remain fully valid in this context (see \cite{Preskill:1990fr} for more details).  
The actual value of $M_1$ is left undetermined, although in the context of string model building there are 
suggestions to relate them to specific properties of the compactified extra dimensions (see for instance 
\cite{Ibanez:1998qp, Ghilencea:2002da, Ghilencea:2002by}).
\section{The Effective Action of  the mLSOM  with a single anomalous $U(1)$}
\begin{figure}[t]
{\centering \resizebox*{9cm}{!}{\rotatebox{0}
{\includegraphics{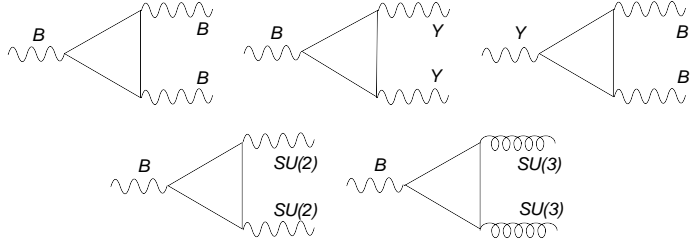}}}\par}
\caption{\small Anomalous triangle diagrams for the mLSOM.
\label{anomaly2}}
\end{figure}

Having derived the essential components of the classical Lagrangian of the model, 
now we try to extend our study to the quantum level, determining the 
anomalous effective action both for the Abelian and the non-Abelian sectors, fixing 
the $D$, $F$ and $C$ coefficients in front of the Green-Schwarz terms in Eq.~(\ref{action}). 
Notice that the only anomalous 
contributions to ${\cal S}_{an}$ in the $Y$-basis before symmetry breaking come from the triangle diagrams depicted in 
Fig.~\ref{anomaly2}.
Since hypercharge is anomaly-free, the only relevant non-Abelian 
anomalies to be canceled are those involving one boson $B$ with two $SU(2)_W$ bosons, 
or two $SU(3)_C$ bosons, while the Abelian anomalies are those containing three $U(1)$ bosons, 
with the $Y^{3}$ triangle excluded by the hypercharge assignment. These $(BSU(2)SU(2))$ and 
$(BSU(3)SU(3))$ anomalies must be canceled 
respectively by Green-Schwarz terms of the kind
$$F \,b \, Tr[F^W \wedge F^W], \qquad D \,b\, Tr[F^G \wedge F^G],$$
with $F$ and $D$ to be fixed by the conditions of gauge 
invariance. In the Abelian sector we have 
to focus on the $BBB, BYY$ and $YBB$ triangles which generate anomalous contributions that need to be 
canceled, respectively, by the Green-Schwarz terms $C_{BB}\,  b \, F^{B} 
\wedge F^{B}$, $C_{YY} \, b \, F^{Y} \wedge F^{Y}$ and $C_{YB} \, b \, F^{Y} \wedge F^{B}$. 
Denoting by ${\cal S}_{YM}$ the anomalous effective action involving 
the classical non-Abelian terms plus the non-Abelian anomalous diagrams, 
and with ${\mathcal S}_{ab}$ the analogous Abelian one, 
the complete anomalous effective action is given by
\beqn
 {\mathcal S}_{eff} &=& {\mathcal S}_0+ {\mathcal S}_{YM} + {\mathcal S}_{ab}   
\eeqn
with ${\mathcal S}_0$ being the classical Lagrangian and
\beqa
{\mathcal S}_{YM}&=& \int dx\, dy \, dz 
 \left(\frac{1}{2!} T^{\lambda \mu \nu, ij}_{BWW}(z,x,y) B^{\lambda}(z) W^{\mu}_{i}(x) 
W^{\nu}_{j}(y) + \frac{1 }{2!} T^{\lambda \mu \nu, ab}_{BGG}(z,x,y) B^{\lambda}(z) G^{\mu}_{a}(x) G_{b}^{\nu}(y) \right), \nonumber \\
\eeqa
\beqa
{\mathcal S}_{ab} &=& \int dx\, dy \, dz 
 \left( \frac{1}{ 3! } T^{\lambda \mu \nu}_{BBB}(z,x,y) B^{\lambda}(z) B^{\mu}(x) B^{\nu}(y) 
 + \frac{1}{2!} T^{\lambda \mu \nu}_{BYY}(z,x,y) B^{\lambda}(z) {Y}^{\mu}(x) {Y}^{\nu}(y) \right. 
\nonumber \\
&& \left. \qquad \qquad \qquad + 
 \frac{1}{2!} T^{\lambda \mu \nu}_{YBB}(z,x,y) {Y}^{\lambda}(z) B^{\mu}(x) B^{\nu}(y) \right).
\label{effeaction}
\eeqa
The corresponding three-point functions, for instance, are given by 
\beqn
 T^{\lambda \mu \nu,\,ij}_{BWW} B^{\lambda} W^{\mu}_{i} W^{\nu}_{j} 
&=&  \langle 0 | T( J^{\lambda, \, f}_{B} 
J^{\mu, \, f}_{W i} J^{\nu, \, f}_{W j})|0 \rangle  B^{\lambda} W^{\mu}_{i} W^{\nu}_{j}  \nonumber\\  
&\equiv&  \langle 0 | T( J^{\lambda, \, f_L}_{B} 
J^{\mu, \, f_L}_{W i} J^{\nu, \, f_L}_{W j}) |0 \rangle  B^{\lambda} W^{\mu}_{i} W^{\nu}_{j}, 
\eeqn
and similarly for the others. Here we have defined the chiral currents 
\beqn
J^{\lambda, f}_{B} = J^{\lambda, fR}_{B} + J^{\lambda, fL}_{B} 
= -  g^{}_{B} q^{\,fR}_{B} \, \overline{\psi}_{f} \gamma^{\lambda} P_{R} \psi_{f} 
-  g^{}_{B} q^{\,fL}_{B} \, \overline{\psi}_{f} \gamma^{\lambda} P_{L} \psi_{f}.
\eeqn
The non-Abelian $W$-current being chiral
\beqn
J^{\mu, f}_{Wi} \equiv J^{\mu, fL}_{Wi}   = -  g^{}_{2} \overline{\psi}_{f} \gamma^{\mu} \tau^{i} P_{L} \psi_{f},
\eeqn
it forces the other currents in the triangle diagram to be of the same chirality, as shown in Fig.~\ref{all_nonabelian}.
\section{Three gauge boson amplitudes and gauge-fixing}
\begin{figure}[t]
{\centering \resizebox*{13cm}{!}{\rotatebox{0}
{\includegraphics{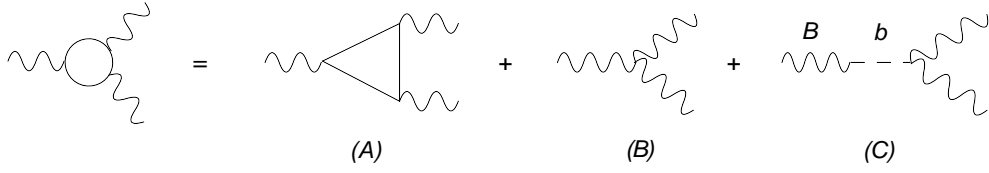}}}\par}
\caption{\small Contributions to a three Abelian gauge boson amplitude before the removal of the 
$B-\partial b$ gauge boson- St\"uckelberg mixing.
\label{non-abelian}}
\end{figure}
\subsection{The non-Abelian sector before symmetry breaking}
Before we get into the discussion of the gauge invariance of the model, it is convenient 
to elaborate on the cancellations of the spurious $s$-channel poles coming from the gauge-fixing conditions. These are imposed to remove the $\partial b-B$ mixing in the effective action. We will perform our analysis in the basis of the interaction eigenstates since in this basis recovering gauge 
independence is more 
straightforward, at least before we enforce 
symmetry breaking via the Higgs mechanism. The procedure that we follow 
is to gauge fix the $B$ gauge boson in the symmetric 
phase by removing the $B-\partial b$ mixing (see Fig.~\ref{non-abelian} (C)), so to derive simple Ward Identities 
involving only fermionic triangle diagrams and contact trilinear 
interactions with gauge bosons. For this purpose to the St\"uckelberg term
\beq
\frac{1}{2}(\partial_{\mu}b + M^{}_{1} B_{\mu})^2 \nonumber\\ 
\eeq
we add the gauge-fixing term  
\beq
\mathcal{L}_{gf}= -\frac{1}{2}\mathcal{G}_B^2 
\eeq
to remove the bilinear mixing, where
\beq
\mathcal{G}_B=\frac{1}{\sqrt{\xi_B}}\left( \partial\cdot B - \xi_B M_1 b\right), 
\eeq
with a propagator for the massive $B$ gauge boson separated in a gauge independent part 
$P_0$ and a gauge dependent one $P_\xi$:
\beqa
\frac{- \,i}{ k^2 - M_1^2} \left( g^{\,\lambda\, \lambda^\prime} - \frac{k^\lambda \, 
k^{\lambda^\prime}}{M_1^2} \right) 
+  \frac{- \,i}{ k^2 - \xi_B \,M_1^2}  \left( \frac{ k^{\lambda} k^{\lambda^\prime}}{M_1^2} 
\right)  
&=& P_0^{\lambda \, \lambda^\prime} +  P_{\xi}^{\lambda \, \lambda^\prime}.
 \eeqa
We will briefly illustrate here 
how the cancellation of the gauge dependence due to $b$ and $B$ exchanges in the 
$s$-channel goes in this (minimally) gauge-fixed theory. In the exact phase we have no mixing between all 
the $Y, B, W$ gauge bosons and the gauge dependence of the $B$ propagator is canceled by 
the St\"uckelberg axion. In the broken phase things get more involved, but essentially the pattern continues to hold. In that case the St\"uckelberg scalar has to be rotated into its physical 
component $\chi $ and the two Goldstones $G^{}_Z$ and $G^{}_{Z^\prime}$ which are linear combinations of $G^0_1$ and 
$G^0_2$. The cancellation of the spurious $s$-channel poles takes place, in this case, via the combined exchange 
of the $Z$ propagator and of the corresponding Goldstone mode $G^{}_Z $. Naturally 
the GS interaction will be essential for this to happen. 

\begin{figure}[t]
{\centering \resizebox*{8cm}{!}{\rotatebox{0}
{\includegraphics{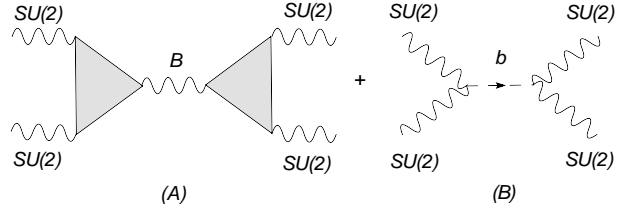}}}\par}
\caption{\small Unitarity check in the SU(2) sector for the mLSOM.
\label{non-abelian2}}
\end{figure}

For the moment we simply work in the exact symmetry phase and in the basis of the interaction eigenstates. 
We gauge fix the action to remove the $B-\partial b$ mixing, but for the rest we set the v.e.v. of the scalars 
to zero.
For definiteness let's consider the process $WW \rightarrow WW$ 
mediated by a $B$ boson as shown in Fig.~\ref{non-abelian2}. We denote by a bold-faced 
{\bf V} the $BWW$ vertex, constructed so to have gauge invariance on the $W$-lines. 
This vertex, as we are going to discuss next, requires a generalized CS 
counterterm to have such a property on the $W$-lines. Gauge invariance on the $B$-line, 
instead, which is clearly necessary 
to remove the gauge dependence in the gauge fixed action, is obtained at a diagrammatical 
level by the the axion exchange (Fig.~\ref{non-abelian2}).
The expressions of the two diagrams are
\beqn
A^{}_{\xi} + B^{}_{\xi} &=& 
\frac{- i}{k^2 - \xi_B M^2_1} \frac{1}{M^2_1} \Big( k^{\lambda} {\bf V}^{\lambda \mu \nu }_{BWW}(-k^{}_{1}, -k^{}_{2}) \Big) 
\left( k^{\lambda^{\prime}} {\bf V}^{\lambda' \mu' \nu'}_{BWW}( k^{}_{1},  k^{}_{2} ) \right)  
\left(g^{}_{B} g^{\,2}_{2} D^{(L)}_{B} \right)^2  \nonumber\\
&& + \, 4 \times \left( \frac{4 F}{M} \right)^2 
\left(  \frac{i}{ k^2 - \xi_B M^2_1} \right) \epsilon^{ \mu \nu \alpha \beta} k^{\alpha}_1 k^{\beta}_2  \,
\epsilon^{\mu' \nu' \alpha' \beta' }  k^{\alpha'}_1 k^{\beta'}_2. 
\eeqn
Using the equations for the anomalies and the correct value for the Green-Schwarz coefficient F given in 
Eq.~(\ref{coeff_GS}) (determined in the next section), we obtain
\beqn
A^{}_{\xi} + B^{}_{\xi} &=& 
\frac{- i}{k^2 - \xi_B M^2_1} \frac{1}{M^2_1} \Big( -  4 a_n \epsilon k_1 k_2 \Big) 
\Big(  4 a_n \epsilon' k'_1 k'_2 \Big)  \left( g^{}_{B} g^{2}_{2} D^{(L)}_{B} \right)^2   \nonumber\\
&&+ \frac{64}{M^2}  \frac{M^{2}}{M^{2}_1}
\left(  i g^{}_{B} g^{2}_{2} \frac{a_n}{2} D^{(L)}_{B} \right)^2 
\left(  \frac{i}{ k^2 - \xi_B M^2_1} \right) \epsilon k_1 k_2 \epsilon'  k'_1 k'_2 
\eeqn
so that the cancellation is easily satisfied. \footnote{For brevity we have adopted the notation $\epsilon k_1 k_2
= \epsilon^{\mu \mu \rho \sigma} k_{1 \rho} k_{2 \sigma}$.} The treatment of the $SU(3)$ sector is similar and we omit it.
\subsection{The Abelian sector before symmetry breaking}
\begin{figure}[t]
{\centering \resizebox*{7.5cm}{!}{\rotatebox{0}
{\includegraphics{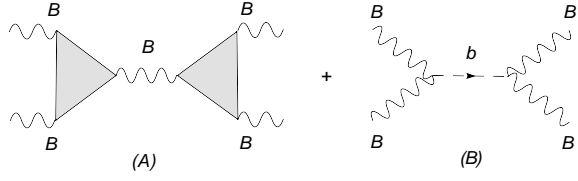}}}\par}
\caption{\small Unitarity check in the Abelian sector for the mLSOM.
\label{BBBanomaly}}
\end{figure}

\begin{figure}[t]
{\centering \resizebox*{7.5cm}{!}{\rotatebox{0}
{\includegraphics{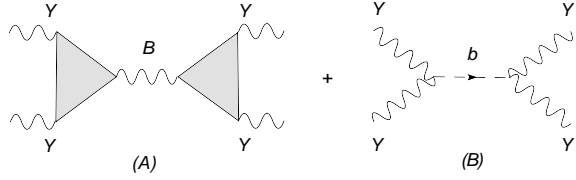}}}\par}
\caption{\small Unitarity check in Abelian sector for the mLSOM.
\label{BYYanomaly}}
\end{figure}

\begin{figure}[t]
{\centering \resizebox*{9cm}{!}{\rotatebox{0}
{\includegraphics{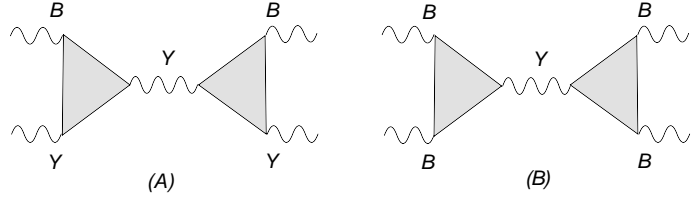}}}\par}
\caption{\small Unitarity check in the Abelian sector for the mLSOM.}
\label{Yunitarity}
\end{figure}

In the Abelian sector the procedure is similar. 
For instance, to test the cancellation of the gauge parameter $\xi_{B}$ in a process $BB \rightarrow BB$ mediated by a $B$ gauge boson we sum 
the two gauge dependent contributions coming from the diagrams in Fig.~\ref{BBBanomaly} (we consider only the gauge dependent 
part of the $s$-channel exchange diagrams)
\beqn
A^{}_{\xi} + B^{}_{\xi} &=& \frac{- i}{ k^2 - \xi_B M^{2}_{1}} \frac{1}{M^2_1} 
\Big(4 \,  k^{\lambda} { \Delta}^{\lambda \mu \nu}_{\bf AAA}(-k^{}_{1}, -k^{}_{2} ) \Big) 
 \left(4 \, k^{\lambda'} { \Delta}^{\lambda' \mu' \nu'}_{\bf AAA}(k^{}_{1}, k^{}_{2} ) \right) 
\left( g^{3}_{B} D^{}_{BBB} \right)^2    \nonumber\\
&&+ \, 4 \times \left( \frac{4}{M} C_{BB} \right)^{2} \frac{i}{k^2 - \xi_B M^{2}_{1}} 
 \, \epsilon k_1 k_2  \,   \epsilon' k'_1 k'_2, 
\eeqn
and cancellation of the gauge dependences implies that the following identity must hold
\beqn
 \frac{16}{M^2_1} \left( \frac{a_n}{3} \right)^2  \left( g^{\,3}_{B} D^{}_{BBB}  \right)^2 
+ 4 \times \left( \frac{4}{M} C_{BB}  \right)^2  = 0,
\eeqn
which can be easily shown to be true after substituting the value of the GS coefficient given in relation (\ref{abelian_BB}). 
In Fig.~\ref{BYYanomaly} we have depicted the anomalous triangle diagram $BYY$ (A) which 
has to be canceled by the Green-Schwarz 
term $\frac{C_{YY}}{M}bF^{Y} \wedge F^{Y}$, that generates diagram (B). 
In this case the two diagrams give
\beqn
A^{}_{\xi} + B^{}_{\xi}  &=&  \frac{- i}{ k^2 - \xi_B M^{2}_{1}} \frac{1}{M^2_1} \Big( k^{\lambda} 
{\bf V}^{\lambda \mu \nu}_{BYY}(-k^{}_{1}, -k^{}_{2} ) \Big) 
 \left( k^{\lambda'} {\bf V}^{\lambda' \mu' \nu'}_{BYY}(k^{}_{1}, k^{}_{2}) \right) \left( g^{}_{B} g^{2}_{Y} D^{}_{BYY} \right)^2
    \nonumber\\
&&+ \, 4 \times \left( \frac{4}{M} C_{YY} \right)^{2} \frac{i}{k^2 - \xi_B M^{2}_{1}} 
 \, \epsilon k_1 k_2  \,   \epsilon' k'_1 k'_2. 
\eeqn
The condition of unitarity of the amplitude requires the 
validity of the identity
\beqn
\frac{16}{M^2_1}  \, {a_n}^{2}   \left(  g^{}_{B} g^{\,2}_{Y}   D^{}_{BYY} \right)^2 
+ 4 \times \left( \frac{4}{M} C_{YY}  \right)^2  = 0,
\eeqn
which can be easily checked substituting the value of the GS coefficient $C^{}_{YY}$ given in relation (\ref{abelian_YY}). We will derive the expressions of these coefficients and the factors of all the other counterterms in the next section. 
The gauge dependences appearing in the diagrams shown 
in Fig.~\ref{Yunitarity} are analyzed in a similar way and we 
omit repeating the previous steps, but it should be obvious by now how the perturbative expansion is organized in terms of tree-level vertices and one-loop 
counterterms, and how gauge invariance is checked at higher orders when the propagators of the $B$ gauge boson and of the axion $b$ are both present. Notice that in the exact phase the axion $b$ is not coupled to the fermions and the pattern of cancellations to ensure gauge independence, in this specific case, is simplified.   
At this point we pause to make some comments.
The mixed anomalies analyzed above involve a non-anomalous Abelian 
gauge boson and the remaining gauge interactions (Abelian/non-Abelian). 
To be specific, in our model with a single non-anomalous $U(1)$, which is the hypercharge $U(1)_Y$ gauge group, these 
mixed anomalies are those involving triangle diagrams with the $Y$ 
and $B$ generators or the $B$ accompanied by the non-Abelian sector.    
Consider, for instance, the $BYY$  triangle, which appears in the $YB \rightarrow YB$ amplitude. There are two options that we can follow. Either we require that the 
corresponding traces of the generators over each generation vanish identically 
\beqn
Tr [ q^{2}_{Y} q_{B} ]   =- 2 \left( -\frac{1}{2}  \right)^{2} q^{(L)}_{B} + (-1)^{2} q^{(e_R)}_{B}  
+ 3 \left[ - 2 \left(\frac{1}{6} \right)^{2}  q^{(Q_L)}_{B} + \left( \frac{2}{3} \right)^{2}    q^{(u_R)}_{B}  
+  \left(  - \frac{1}{3}  \right)^{2} q^{(d_R)}_{B}  \right]   = 0,      \nonumber
\eeqn
which can be viewed as a specific condition on the charges of the model or, if this is not the case, we require that suitable one-loop counterterms balance the anomalous gauge variation. We are allowed, in other words, to fix the two divergent invariant 
amplitudes of the triangle diagram so that the corresponding Ward Identities 
for the $BYY$ vertex and 
similar anomalous vertices are satisfied. This is a condition on the parameterization of the Feynman vertex rather than on the charges and is, in principle, allowed. It is not necessary to have a specific determination of the charges for this to occur, as far as the counterterms are fixed accordingly.
For instance, in the Abelian sector the diagrams in question are
\beqn
 YB  \rightarrow YB \,\,\mbox{mediated by Y} \,\, &\propto& \,\,  Tr[ q^2_Y q_B]  \,\,   \nonumber\\
 YY  \rightarrow YY \mbox{mediated by B}\,\, &\propto& \,\,  Tr[ q^2_Y  q_B ]   \,\,   \nonumber\\
  BB  \rightarrow BB \,\,  \mbox{mediated by Y}    \,\,   &\propto&   \,\, Tr[ q_Y q^2_B]  \,\,    \nonumber\\
 YB  \rightarrow YB  \,\, \mbox{mediated by B}  \,\, &\propto&   \,\, Tr[ q_Y  q^2_B ].    \,\,   \nonumber
\eeqn
In the mLSOM these traces are, in general, 
non vanishing and therefore we need to introduce 
defining Ward Identities to render the effective action anomaly free. 
\section{Ward Identities, Green-Schwarz and Chern-Simons counterterms 
in the St\"uckelberg phase} \label{sec:WI_discuss}
Having discussed the structure of the theory in the basis of 
the interaction eigenstates, we come now to identify the coefficients 
needed to enforce cancellation of the anomalies in the one-loop effective action. 
In the basis of the physical gauge bosons we will be dropping, with this choice, 
a gauge dependent ($B\partial b$ mixing) term that is vanishing for 
physical polarizations. At the same time, for exchanges of virtual gauge bosons, the gauge 
dependence of the corresponding propagators is canceled by the associated Goldstone exchanges.  

\begin{figure}[t]
{\centering \resizebox*{13cm}{!}{\rotatebox{0}
{\includegraphics{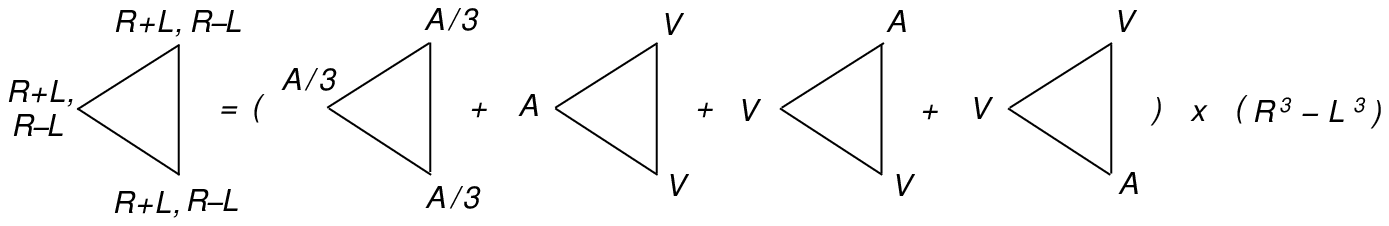}}}\par}
\caption{\small All the anomalous contributions to a triangle diagram in the Abelian sector for 
generic vector/axial-vector trilinear interactions in the massless fermion case.}
\label{all_abelian}
\end{figure}

Starting from the non-Abelian contributions, the $BWW$ amplitude, we separate the charge/coupling constant dependence of a 
given diagram from the rest of its 
parametric structure ${\bf T}$ using, in the $SU(2)$ case, the relations 
\beqn
 T^{\lambda \mu \nu,ij}_{BWW}  B^{\lambda} W^{\mu}_{i} W^{\nu}_{j}   &=&g^{}_{B} g^{\,2}_{2} \sum_{i,j} 
Tr[\tau^{}_{i} \tau^{}_{j}] \frac{1}{8} Tr[q^{L}_{B}] 
\,{\bf{T}}^{\,\lambda \mu \nu} B^{\lambda} W^{\mu}_{i} W^{\nu}_{j} \nonumber\\
  &=&  \frac{1}{2}  g^{}_{B} g^{\,2}_{2}  \sum_{i}  
\,D^{(L)}_{B} \,{\bf{T}}^{\,\lambda \mu \nu } B^{\lambda} W^{\mu}_{i} W^{\nu}_{i},   
\label{anom_coeff}
\eeqn
where $D^{(L)}_{B} =\frac{1}{8} Tr[q^{L}_{B}] = - \frac{1}{8} \sum_{f} q^{fL}_{B}$ and 
${\bf{T}}^{\lambda \mu \nu}$ is the three-point function in configuration space, with all the couplings 
and the charges factored out, symmetrized in $\mu \nu$. 
Similarly, for the coupling of $B$ to the gluons we obtain
\beqn
 T^{\lambda \mu \nu,ab}_{BGG} B^{\lambda} G^{\mu}_{a} G^{\nu}_{b} 
&=& g^{}_{B} g^{\,2}_{3} \sum_{a,b} \,Tr[T^{}_{a} T^{}_{b}]  \frac{1}{8} Tr[q^Q_B] 
\,{\bf{T}}^{\,\lambda \mu \nu } B^{\lambda} G^{\mu}_{a} G^{\nu}_{b} \nonumber\\
&=&\frac{1}{2}  g^{}_{B} g^{\,2}_{3}  \sum_{a} D^{(Q)}_{B} 
\,{\bf{T}}^{\,\lambda \mu \nu } B^{\lambda} G^{\mu}_{a} G^{\nu}_{a}, 
\eeqn
having defined $D^{(Q)}_B = \frac{1}{8} Tr[q^Q_B] =  \frac{1}{8} \sum_{Q} [ q^{Q_R}_B - q^{Q_L}_{B}]$, while the Abelian triangle diagrams are given by
\beqn
 T^{\lambda \mu \nu}_{BBB} B^{\lambda} B^{\mu} B^{\nu} 
&=& g^{\,3}_{B} \, \frac{1}{8} Tr[q^{\,3}_{B}] \, {\bf T}^{\,\lambda \mu \nu} B^{\lambda} B^{\mu} B^{\nu}  
= g^{\,3}_{B} \, D_{BBB} \,{\bf{T}}^{\,\lambda \mu \nu} B^{\lambda} B^{\mu} B^{\nu}, \label{BBBvertex}   \\
 {T}^{\lambda \mu \nu}_{BYY} B^{\lambda} Y^{\mu} Y^{\nu} &=& g^{}_{B} g^{\,2}_{Y} \,\frac{1}{8} Tr[q^{}_{B} q^{\,2}_{Y}] \, 
{\bf{T}}^{\,\lambda \mu \nu}  B^{\lambda} Y^{\mu} Y^{\nu}     \nonumber\\
& =& g^{}_{B} g^{\,2}_{Y} \, D_{BYY} \, {\bf{T}}^{\,\lambda \mu \nu}  B^{\lambda} Y^{\mu} Y^{\nu}, \\
T^{\lambda \mu \nu}_{YBB} Y^{\lambda} B^{\mu} B^{\nu} &=& g^{}_{Y} g^{\,2}_{B} \,\frac{1}{8} Tr[q^{}_{Y} q^{\,2}_{B}] \, 
{\bf{T}}^{\,\lambda \mu \nu} Y^{\lambda} B^{\mu} B^{\nu}   \nonumber\\
& =& g^{}_{Y} g^{\,2}_{B} \, D_{YBB} \, {\bf{T}}^{\,\lambda \mu \nu} Y^{\lambda} B^{\mu} B^{\nu},
\eeqn
with the following definitions for the traces 
\beqn
D^{}_{BBB}&=&   \frac{1}{8} Tr[q^{3}_{B}] = \frac{1}{8} \sum_{f} \left[ (q^{fR}_{B})^{3}
 - (q^{fL}_{B})^{3}  \right],    \\
D^{}_{BYY}&=&    \frac{1}{8} Tr[q^{}_{B} q^{2}_{Y}] =
                \frac{1}{8} \sum_{f} \left[ q^{fR}_{B} (q^{fR}_{Y})^{2}  - q^{fL}_{B} (q^{fL}_{Y})^{2}  \right],       \\
D^{}_{YBB}&=&    \frac{1}{8} Tr[q^{}_{Y} q^{2}_{B}] =
                \frac{1}{8} \sum_{f} \left[ q^{fR}_{Y} (q^{fR}_{B})^{2}  - q^{fL}_{Y} (q^{fL}_{B})^{2}   \right].      
\label{DDD}
\eeqn
The ${\bf T}$ vertex is given by the usual combination of vector and axial-vector components 
\beqn
  {\bf T}^{\lambda \mu \nu} = T^{\lambda \mu \nu}_{\bf AAA} + T^{\lambda \mu \nu}_{\bf AVV} 
+ T^{\lambda \mu \nu}_{\bf VAV} + T^{\lambda \mu \nu}_{\bf VVA},
\eeqn
and we denote by ${\bf \Delta}(k_1,k_2)$ its expression in momentum space 
\beqn
(2\pi)^4 \delta(k-k_1-k_2) {\bf \Delta}^{\lambda \mu \nu}(k_1, k_2) 
= \int dx \,dy \, dz \, e^{ik_1 \cdot x + i k_2 \cdot y - i k \cdot z}  \,
{\bf{T}}^{\lambda \mu \nu}(z,x,y). 
\label{configur}
\eeqn
We denote similarly with ${\bf \Delta}_{\bf AVV}^{\lambda \mu \nu},{\bf \Delta}_{\bf VAV}^{\lambda \mu \nu},{\bf \Delta}_{\bf VVA}^{\lambda \mu \nu}$ the momentum space expressions of the 
corresponding x-space vertices ${\bf T}_{\bf AVV}^{\lambda \mu \nu},{\bf T}_{\bf VVA}^{\lambda \mu \nu}, {\bf T}_{\bf VAV}^{\lambda \mu \nu}$ respectively.
\begin{figure}[t]
{\centering \resizebox*{11.5cm}{!}{\rotatebox{0}
{\includegraphics{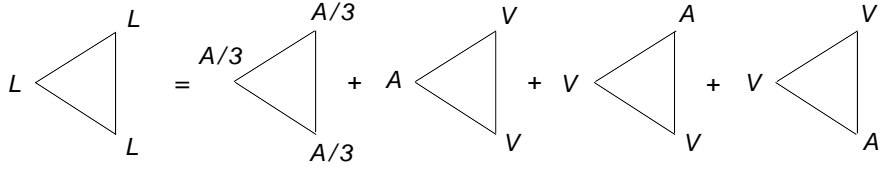}}}\par}
\caption{\small All the anomalous contributions to a triangle diagram in the non-Abelian sector in the massless fermion case.}
\label{all_nonabelian}
\end{figure}
As illustrated in Fig.~\ref{all_nonabelian} and Fig.~\ref{all_abelian}, the complete structure of 
${\bf T }$ is given by 
\beqn
{\bf \Delta}^{\lambda \mu \nu}(k_1, k_2)  &=& \frac{1}{3} \left[ {\Delta}^{\lambda \mu \nu}(- 1/2, k_1, k_2) 
+ { \Delta}^{ \mu \nu \lambda}(-1/2, k_2, - k)  + {\Delta}^{ \nu \lambda \mu}(-1/2, - k, k_1)  \right]   \nonumber\\
&&+ { \Delta}^{\lambda \mu \nu}(-1/2, k_1, k_2) 
+ { \Delta}^{ \mu \nu \lambda }(-1/2, k_2, -  k) + { \Delta}^{ \nu \lambda \mu}(-1/2, - k, k_1)    \nonumber\\
&=& \frac{4}{3} \left[   {\Delta}^{\lambda \mu \nu}(-1/2, k_1, k_2) 
+ { \Delta}^{ \mu \nu \lambda}(-1/2, k_2, - k)  + { \Delta}^{ \nu \lambda \mu}(-1/2, - k, k_1)    \right]   \nonumber\\
&=& 4 \Delta^{\lambda \mu \nu}_{\bf AAA},
\label{aaa}
\eeqn
where we have used the relation between the ${\bf \Delta}_{\bf AAA}$ (bold-faced) vertex and the usual 
$\Delta$ vertex, which is of the form ${\bf AVV}$. Notice that 
\beqa
{\bf \Delta}^{\lambda\mu\nu}_{\bf AVV}&=&{ \Delta}^{\lambda\mu\nu}(-1/2,k_1,k_2), \nonumber \\
{\bf \Delta}^{\mu\nu \lambda}_{\bf AVV}&=&{ \Delta}^{\mu\nu\lambda}(-1/2,k_2,-k), \nonumber \\
{\bf \Delta}^{\nu\lambda\mu}_{\bf AVV}&=&{ \Delta}^{\nu\lambda\mu}(-1/2,-k,k_1),
\eeqa
are the usual vertices with conserved vector current (CVC) on two lines and the anomaly on a single axial vertex. 

The {\bf AAA} vertex is 
constructed by symmetrizing the distribution of the anomaly on each of the 
three chiral currents, which is the content of (\ref{aaa}). The same vertex can be obtained from the basic {\bf AVV} vertex 
by a suitable shift, with $\beta= 1/6$, 
and then repeating the same procedure on the other 
indices and external momenta, with a cyclic permutation. We obtain
\beqa
\Delta^{\lambda \mu \nu}_{\bf AAA}(1/6, k_1, k_2) &=& 
\Delta^{\lambda \mu \nu}(-1/2,k_1, k_2) - \frac{i}{4 \pi^{2}} 
\frac{2}{3} \epsilon^{\lambda \mu \nu \sigma}(k_1 - k_2)_{\sigma} 
\eeqa
and its corresponding anomaly equations are given by
\beqn
 k_{ \lambda } \Delta^{  \lambda \mu \nu}_{\bf AAA}(1/6, k_1, k_2) &=&  \frac{a_n}{3} 
\epsilon^{\, \mu \nu \alpha \beta} k_{1 \alpha} k_{2 \beta} \nonumber\\
 k_{1 \mu } \Delta^{ \lambda \mu \nu }_{\bf AAA}(1/6, k_1, k_2) &=&  \frac{a_n}{3} 
\epsilon^{\,\lambda \nu \alpha \beta} k_{1\alpha} k_{2 \beta}\nonumber\\
  k_{2 \nu } \Delta^{  \lambda \mu \nu }_{\bf AAA}(1/6, k_1, k_2) &=& \frac{a_{n}}{3} 
\epsilon^{\,\lambda \mu \alpha \beta} k_{2\alpha} k_{1 \beta},
\eeqn
typical of a symmetric distribution of the anomaly.

These identities are obtained from the general shift-relation
\beqn
  \Delta^{\lambda \mu \nu}(\beta^\prime , k^{}_{1}, k^{}_{2}) =
 \Delta^{\lambda \mu \nu}(\beta, k^{}_{1}, k^{}_{2} ) 
+   \frac{i}{ 4 \pi^{2} }(\beta - \beta^{\prime} ) 
\epsilon^{\lambda \mu \nu \sigma}(k^{}_{1} - k^{}_{2})^{}_{\sigma}. 
\eeqn 
Vertices with conserved axial currents (CAC) can be related to the symmetric {\bf AAA} vertex in a similar way 
\beqn
\Delta^{ \lambda \mu \nu }_{\bf AAA}( + 1/6 , k^{}_{1}, k^{}_{2}) =
 \{ \Delta^{\lambda \mu \nu}(+ 1/2, k^{}_{1}, k^{}_{2} ) \}_{CAC} 
+ \frac{i}{ 4 \pi^{2} } \, \frac{1}{3} \,
\epsilon^{\lambda \mu \nu \sigma}(k^{}_{1} - k^{}_{2})^{}_{\sigma}.
\eeqn
At this point we are ready to introduce the complete vertices 
for this model, which are given 
by the amplitude ($\ref{configur}$) with the addition of 
the corresponding Chern-Simons 
counterterms, were required. These will be determined later in this section 
by imposing the conservation of the 
$SU(2)$, $SU(3)$ and $Y$ gauge currents. Following this definition for all the 
anomalous vertices, the amplitudes can then be written as 
\beqn
{\bf{\mathcal V}}^{\lambda \mu \nu, \, aa}_{BGG} B^{\lambda} G^{\mu}_{a} G^{\nu}_{a} &=& 
\frac{1}{2} g^{}_{B} g^{\,2}_{3} D^{(Q)}_B {\bf T}^{\lambda \mu \nu} B^{\lambda} G^{\mu}_{a} G^{\nu}_{a}
+ c^{}_{2} \epsilon^{ \mu \nu \rho \sigma} B^{}_{\mu} C^{SU(3)}_{\nu \rho \sigma}   \nonumber\\
{\bf {\mathcal V}}^{\lambda \mu \nu, \, ii}_{BWW} B^{\lambda} W^{\mu}_{i} W^{\nu}_{i} &=& 
\frac{1}{2} g^{}_{B} g^{\,2}_{2} D^{(L)}_{B} {\bf T}^{\lambda \mu \nu}  B^{\lambda} W^{\mu}_{i} W^{\nu}_{i}
+ c^{}_{1} \epsilon^{ \mu \nu \rho \sigma} B^{}_{\mu} C^{SU(2)}_{\nu \rho \sigma}   \nonumber\\
{\bf {\mathcal V}}^{\lambda \mu \nu}_{BYY} B^{\lambda} Y^{\mu} Y^{\nu} &=& 
 g^{}_{B} g^{\,2}_{Y} D^{}_{BYY} {\bf T}^{\lambda \mu \nu} B^{\lambda} Y^{\mu} Y^{\nu}
+ d^{}_{1} B Y \wedge F^{}_{Y}  \nonumber\\
{\bf{\mathcal V}}^{\lambda \mu \nu}_{YBB} Y^{\lambda} B^{\mu} B^{\nu}  &=&  
g^{}_{Y} g^{\,2}_{B} D^{}_{YBB} {\bf T}^{\lambda \mu \nu}  Y^{\lambda} B^{\mu} B^{\nu} 
+ d^{}_{2} Y B \wedge F^{}_{B},  \nonumber\\
{\bf{\mathcal V}}^{\lambda \mu \nu}_{BBB} B^{\lambda} B^{\mu} B^{\nu}  &=&  
g^{\,3}_{B} D^{}_{BBB} {\bf T}^{\lambda \mu \nu}  B^{\lambda} B^{\mu} B^{\nu} \nonumber\\
\label{definingv}
\eeqn
which are the anomalous vertices of the effective action, corrected when necessary by 
suitable CS interactions in order to conserve all the gauge currents at one-loop.

Before we proceed with our analysis, which has the goal to determine explicitly the 
counterterms in each of these vertices, we pause for some considerations. It is clear 
that the scheme that we have followed in order 
to determine the structure of the vertices of the effective action has been 
to assign the anomaly only to the chiral vertices and to impose conservation of the vector 
current. There are regularization schemes in the literature that enforce this principle, the most famous one 
being dimensional regularization with the t'Hooft Veltman prescription for $\gamma_5$ 
(see also the discussion in Chap.~\ref{chap:AbelianModels1} ). In this scheme the anomaly is equally distributed for 
vertices of the form {\bf AAA} and is assigned only to the axial-vector vertex in triangles of the form {\bf AVV} and similar. Diagrams of the form {\bf AAV } are zero by Furry's theorem, being equivalent to {\bf VVV}. 

We could also have proceeded in a different way, for instance 
by defining each $\mathcal{V}$, for instance ${\mathcal V}_{BYY}$, to have an anomaly only 
on the B vertex and not on the Y vertices, even if Y has both a vector and an axial-vector 
components at tree level and is, indeed, a chiral current. This implies 
that at one-loop the chiral projector has to be moved from the Y to to the B vertex ``by hand'', no matter if it  appears on the Y current or on the B current, rendering the Y current effectively vector-like at 
one-loop. This is also what a CS term does. In both cases we are anyhow bond 
to define separately the one-loop vertices as new entities, unrelated to 
the tree level currents. However, having explicit Chern-Simons counterterms renders the treatment compatible with dimensional regularization in the t'Hooft-Veltman prescription. 
It is clear, however, that one way or the other, the quantum action is not fixed at classical level since the counterterms are related to quantum effects 
and the corresponding Ward Identities, which force 
the cancellation of the anomaly to take place in a completely new way respect to the SM case, 
are indeed {\em defining conditions} on the theory.  

Having clarified this point, we return to the determination of the gauge invariance 
conditions for our anomalous vertices. 

Under $B$-gauge transformations we have the following variations (singlet anomalies) of the effective action 
\beqa
\frac{1}{2!} \delta_B < T_{ BWW} B W W> =  i \frac{ g^{}_{B} g^{\,2}_{2} }{2!}  \, \frac{4}{3}a_n \, \frac{1}{4} \langle \theta_B
 F^{W}_{i} \wedge F^{W}_{j}  \rangle \, Tr[ \tau^{i}  \tau^{j} ] \,  D^{(L)}_{B},   
\label{MLSOM_2}
\eeqn
\beqn
\frac{1}{2!} \delta_B < T_{ BGG } B G G> =  i \frac{g^{}_{B} g^{\,2}_{3} }{ 2!} \,   \frac{4}{3}a_n  \, \frac{1}{4} \langle   \theta_B
  F^{G}_{a} \wedge  F^{G}_{b}  \rangle \,  Tr[ T^{a} T^{b} ]  \, D^{(Q)}_B, 
\label{MLSOM_3}
\eeqn
and with the normalization given by 
\beqa
Tr [ \tau^{i} \tau^{j} ] = \frac{1}{2} \delta^{ij}    \qquad  Tr[T^{a}T^{b}] = \frac{1}{2} \delta^{ab}
\eeqa
we obtain 
\beqa
 \frac{1}{2!} \delta_B < T_{B SU(2) SU(2)} B W W > &=&  i \frac{g^{}_{B} g^{\,2}_{2}}{2!} \, \frac{a_{n}}{6}  \, 
\langle \theta_B  F^W_i \wedge F^W_i   \rangle   D^{(L)}_{B},  \\
\frac{1}{2!} \delta_B < T_{B SU(3) SU(3)} B G G>  &=&  i \frac{g^{}_{B} g^{\,2}_{3}}{2!} \,  \frac{a_{n}}{6}    
 \,  \langle  \theta_B F^G_a \wedge F^G_a \rangle D^{(Q)}_B.
\eeqa
Note, in particular, that the covariantization of the anomalous contributions requires the entire non-Abelian field-strengths 
$F^W_{i,\,\mu \nu}$ and $F^G_{a,\,\mu \nu}$
\beqn
F^W_{i, \,\mu \nu} &=& \partial_{\mu} W^{i}_{\nu} - \partial_{\nu} W^{i}_{\mu}  
-  g^{}_{2} \varepsilon_{ijk} W^{j}_{\mu} W^{k}_{\nu} 
=  \hat{F}^{W}_{i,\, \mu \nu}-  g^{}_{2} \varepsilon_{ijk} W^{j}_{\mu} W^{k}_{\nu} \\
F^G_{a,\,\mu \nu} &=& \partial_{\mu} G^{a}_{\nu} - \partial_{\nu} G^{a}_{\mu} 
 -  g^{}_{3} f_{abc} G^{b}_{\mu} G^{c}_{\nu} = \hat{F}^{G}_{a,\, \mu \nu} -  g^{}_{3} f_{abc} G^{b}_{\mu} G^{c}_{\nu}. 
\eeqn
The covariantization of the right-hand-side (r.h.s.) of the anomaly equations takes place 
via higher-order corrections, involving correlators with more external gauge lines. It is well known, though, that the cancellation of the anomalies in these higher-order 
non-Abelian diagrams (in d=4) is only related to the triangle diagram 
(see \cite{Coriano:2007fw}). 
Under the non-Abelian gauge transformations we have the following variations
\beqn
\frac{1}{2!} \delta_{SU(2)} \langle T_{BWW} BWW  \rangle &=& i  \frac{g^{}_{B} g^{\,2}_{2}}{2!} \frac{8}{3}
a^{}_{n} \frac{1}{4} \langle F^{B} \wedge Tr[ \theta   \hat{F}^{W}  ]  \rangle  D^{(L)}_{B}   \\
\frac{1}{2!} \delta_{SU(3)} \langle T_{BGG} BGG  \rangle &=& i  \frac{g^{}_{B} g^{\,2}_{3}}{ 2! } \frac{8}{3}
a^{}_{n} \frac{1}{4} \langle  F^{B} \wedge Tr[ \vartheta \hat{F}^{G} ]  \rangle   D^{(Q)}_B,
\eeqn
where the ``hat'' field-strengths $\hat{F}^{W}$ and $\hat{F}^{G}$ refer to the Abelian part of the non-Abelian field-strengths W and G. Introducing the notation
\beqn
Tr[\theta \hat{F}^{W}] &=&  Tr[\tau^{i} \tau^{j}] \theta^{}_{i} \hat{F}^{W}_{j}
 =  \frac{1}{2} \theta^{}_{i} \hat{F}^{W}_{i}   \;\;\,\,\,\,\,\,\,\, i,j = 1,2,3  \\
   Tr[\vartheta \hat{F}^{G}] &=& Tr[T^{a} T^{b}] \vartheta_{a} \hat{F}^{G}_{b} 
=  \frac{1}{2} \vartheta_{a} \hat{F}^{G}_{a}    \,\,\;\;\,\,\,\,\,\,  a,b = 1,2,..,8
\eeqn
the expressions of the variations become
\beqn
\frac{1}{2!} \delta_{SU(2)} \langle T_{BWW} BWW  \rangle &=& i  \frac{g^{}_{B} g^{\,2}_{2}}{2!} 
\frac{a^{}_{n}}{3}  \langle \theta^{}_{i} F^{B} \wedge  \hat{F}^{W}_{i}   \rangle  D^{(L)}_{B}    \\
\frac{1}{2!} \delta_{SU(3)} \langle T_{BGG} BGG  \rangle &=& i  \frac{g^{}_{B} g^{\,2}_{3}}{ 2! }
\frac{a^{}_{n}}{3} \langle \vartheta_{a}  F^{B} \wedge  \hat{F}^{G}_{a}   \rangle   D^{(Q)}_B.
\eeqn
We have now to introduce the Chern-Simons counterterms for the non-Abelian gauge variations 
\beqn
{\mathcal S}^{CS}_{non-ab} = {\mathcal S}^{CS}_{BWW} + {\mathcal S}^{CS}_{BGG} 
= c^{}_{1} \langle \epsilon^{\mu \nu \rho \sigma} B^{}_{\mu} 
C^{SU(2)}_{\nu \rho \sigma}   \rangle + c^{}_{2} \langle   \epsilon^{\mu \nu \rho \sigma} B^{}_{\mu} 
C^{SU(3)}_{\nu \rho \sigma}    \rangle,
\eeqn
with the non-Abelian CS forms given by
\beqn
C^{SU(2)}_{\mu \nu \rho} &=&  \frac{1}{6} \left[ W^{i}_{\mu} \left( F^W_{i,\,\nu \rho} + \frac{1  }{3} \, g^{}_{2}  
\, \varepsilon^{ijk} W^{j}_{\nu} W^{k}_{\rho}  \right) + cyclic   \right]              ,    \\
C^{SU(3)}_{\mu \nu \rho} &=&  \frac{1}{6} \left[ G^{a}_{\mu} \left( F^G_{a,\,\nu \rho} + \frac{1 }{3} \, g^{}_{3}
\, f^{abc} G^{b}_{\nu} G^{c}_{\rho}  \right) + cyclic  \right],      
\eeqn
whose variations under non-Abelian gauge transformations are 
\beqn
\delta_{SU(2)} C^{SU(2)}_{\mu \nu \rho} &=& \frac{1}{6}  \left[ \partial^{}_{\mu} \theta^{i} \,( \hat{F}^{W}_{i, \,\nu \rho}) 
+ cyclic \right],      \\
\delta_{SU(3)} C^{SU(3)}_{\mu \nu \rho} &=& \frac{1}{6} \left[ \partial^{}_{\mu} \vartheta^{a} \,( \hat{F}^{G}_{a,\,\nu \rho}) 
+ cyclic \right].      
\eeqn
The variations of the Chern-Simons counterterms then become 
\beqn
\delta_{SU(2)}  {\mathcal S}^{CS}_{BWW}  &=&  \frac{c^{}_{1}}{2} \, \frac{1}{2} 
\langle \theta^{i} F^{B} \wedge \hat{F}^{W}_{i}  \rangle    \\ 
\delta_{SU(3)}  {\mathcal S}^{CS}_{BGG}   &=& \frac{c^{}_{2}}{2} \,  \frac{1}{2}
\langle \vartheta^{a} F^{B} \wedge \hat{F}^{G}_{a}  \rangle,  
\eeqn
and we can choose the coefficients in front of the CS counterterms 
to obtain anomaly cancellations for the non-Abelian contributions
\beqn
c^{}_{1} =  - i g^{}_{B} g^{\,2}_{2}\frac{2}{3} a_n D^{(L)}_{B}  \qquad  
c^{}_{2} =  - i g^{}_{B} g^{\,2}_{3}\frac{2}{3} a_n D^{(Q)}_B. 
\label{fixx}
\eeqn 
The variations under $B$-gauge transformations for the related CS counterterms are then given by 
\beqn
\delta^{}_{B} {\mathcal S}^{CS}_{BWW} &=& - \frac{c^{}_{1}}{2} \, \frac{1}{2} \langle \theta^{}_{B}   F^W_i \wedge F^W_i  \rangle    \\
\delta^{}_{B} {\mathcal S}^{CS}_{BGG} &=& - \frac{c^{}_{2}}{2} \, \frac{1}{2} \langle \theta^{}_{B} F^G_{a} \wedge F^G_{a} \rangle,
\eeqn
where the coefficients $c_i$ are given in (\ref{fixx}). The variations under the $B$-gauge transformations for the $SU(2)$ and $SU(3)$ Green-Schwarz 
counterterms are respectively given by
\beqn
 \frac{F}{M} \, \delta_{B} \langle \, b \, Tr[F^W \wedge F^W]  \, \rangle &=& - F \frac{M_1}{M} \frac{1}{2}  \langle \theta_B 
 F^W_{i} \wedge F^W_{i} \rangle, \\
 \frac{D}{M} \, \delta_{B} \langle \, b \, Tr[ F^G \wedge F^G] \, \rangle &=& - D \frac{M_1}{M} \frac{1}{2}  \langle  \theta_B 
 F^G_{a} \wedge F^G_{a}  \rangle,  
\eeqn
and the cancellation of the anomalous contributions coming from the $B$-gauge transformations 
determines $F$ and $D$ as
\beqn
F =  \frac{M}{M_1} i   g^{}_{B} g^{\,2}_{2} \, \frac{a_n}{2} \, D^{(L)}_{B},    
\qquad D =  \frac{M}{M_1} i  g^{}_{B} g^{\,2}_{3} \,  \frac{a_n}{2} \,   D^{(Q)}_{B}.    
\label{coeff_GS}
\eeqn
There are some comments to be made concerning the generalized CS terms responsible for the cancellation of the mixed anomalies. These terms, in momentum space, generate standard trilinear CS interactions, whose momentum structure is exactly the same as that due to the Abelian ones (see the appendix of Chap.~\ref{chap:AbelianModels1} for more details), plus 
additional quadrilinear (contact) gauge interactions. These will be neglected in our analysis since we will 
be focusing in the next sections on the characterization of neutral tri-linear interactions. 
In processes such as $Z\to \gamma\gamma\gamma$ they re-distribute the anomaly appropriately 
in higher-point functions.
 
For the Abelian part ${\mathcal S}_{ab}$ of the effective action we first focus on gauge 
variations on $B$, obtaining 
\beqn
\frac{1}{3!} \delta_{B}  \langle T^{\lambda \mu \nu}_{ BBB } 
B^{\lambda}(z) B^{\mu}(x) B^{\nu}(y)  \rangle  =  i  \frac{g^{\,3}_{B}}{3!} \frac{4}{3} a_n \frac{3}{4}
\langle F^{B} \wedge F^{B} \theta_B \rangle  \, D^{}_{BBB},   
\eeqn 
\beqn
&&\frac{1}{2!} \delta_{B}  \langle T^{\lambda \mu \nu}_{ BYY  } 
B^{\lambda}(z) Y^{\mu}(x) Y^{\nu}(y)  \rangle =  i  \frac{g^{}_{B} g^{2}_{Y}}{2!} \frac{4}{3} a_n  \frac{1}{4}
\langle F^{Y} \wedge F^{Y} \theta_B \rangle  \,  D^{}_{BYY}, 
\eeqn 
\beqn
&&\frac{1}{2!} \delta_{B}  \langle T^{\lambda \mu \nu}_{ YBB  } 
Y^{\lambda}(z) B^{\mu}(x) B^{\nu}(y)  \rangle =  i  \frac{g^{}_{Y} g^{2}_{B}}{2!} \frac{4}{3} a_{n} \frac{2}{4} 
\langle F^{Y} \wedge F^{B} \theta_B \rangle  \,  D^{}_{YBB}, 
\eeqn 
and variations for $Y$ that give 
\beqn
&&\frac{1}{2!} \delta_{Y}  \langle T^{\lambda \mu \nu}_{ BYY } 
B^{\lambda}(z) Y^{\mu}(x) Y^{\nu}(y)  \rangle =  i  \frac{g^{}_{B} g^{2}_{Y}}{2!} \frac{4}{3} a_{n} \frac{2}{4}
\langle F^{Y} \wedge F^{B} \theta_Y \rangle  \,  D^{}_{BYY}, 
\label{var_Y2}
\eeqn 
\beqn
&&\frac{1}{2!} \delta_{Y}  \langle T^{\lambda \mu \nu}_{ YBB  } 
Y^{\lambda}(z) B^{\mu}(x) B^{\nu}(y)  \rangle =  i  \frac{g^{}_{Y} g^{2}_{B}}{2!}  \frac{4}{3} a_n  \frac{1}{4}
\langle F^{B} \wedge F^{B} \theta_Y \rangle  \,  D^{}_{YBB}. 
\label{var_Y3}
\eeqn 
Also in this case we introduce the corresponding Abelian Chern-Simons counterterms 
\beqn
{\mathcal S}^{CS}_{ab} = {\mathcal S}^{CS}_{BYY} + {\mathcal S}^{CS}_{YBB}  
=  d^{}_{1} \langle BY \wedge F^{}_{Y} \rangle +  d^{}_{2} \langle YB \wedge F^{}_{B} \rangle
\eeqn
whose variations are given by
\beqn
\delta_{Y}  {\mathcal S}^{CS}_{BYY}   &=&    \frac{ d^{}_{1} }{2} \langle \theta^{}_{Y} F^{B} \wedge F^{Y} \rangle,   \\
\delta_{Y}  {\mathcal S}^{CS}_{YBB}  &=& -  \frac{d^{}_{2} }{2} \langle \theta^{}_{Y} F^{B} \wedge F^{B}  \rangle,   
\eeqn
and we can fix their coefficients so to obtain the cancellation  of the Y-anomaly 
\beqn
d_{1} = - i g_{B} g^{2}_{Y}\frac{2}{3} a^{}_{n} D^{}_{BYY}  \qquad 
 d_{2} =  i g_{Y} g^{2}_{B}  \frac{ a^{}_{n} }{3}  D^{}_{YBB}.
\label{coefficiente_CS_abeliano}
\eeqn
Similarly, the gauge variation of $B$ in the corresponding Green-Schwarz terms gives
\beqn
  \frac{C_{BB}}{M}  \,   \delta_{B} \langle \, b \, F^{B} \wedge F^{B}  \rangle &=& - C_{BB} \frac{M_1}{M} \langle \theta_{B} 
F^{B}\wedge F^{B} \rangle      \\
 \frac{C_{YY}}{M} \, \delta_{B} \langle \, b \, F^{Y} \wedge F^{Y}  \rangle &=& - C_{YY} \frac{M_1}{M} \langle \theta_{B} 
F^{Y}\wedge F^{Y} \rangle      \\
 \frac{C_{YB}}{M} \,  \delta_{B} \langle \,  b \, F^{Y} \wedge F^{B}  \rangle &=& - C_{YB} \frac{M_1}{M} \langle \theta_{B} 
F^{Y}\wedge F^{B} \rangle 
\eeqn
and on the other hand the $B$-variations of the fixed CS counterterms are 
\beqn
\delta_{B}     {\mathcal S}^{CS}_{BYY}   &=& - \frac{ d^{}_{1} }{2} \langle \theta^{}_{B} F^{Y} \wedge F^{Y} \rangle,   \\
\delta_{B}   {\mathcal S}^{CS}_{YBB}   &=&   \frac{d^{}_{2} }{2} \langle \theta^{}_{B} F^{Y} \wedge F^{B}  \rangle.
\eeqn
Finally the cancellation of the anomalous contributions from the Abelian part of the effective action requires
following conditions 
\beqn
C_{BB} &=&  \frac{M}{M_1} \frac{ i g^{\,3}_{B} }{3!} a_n  D^{}_{BBB}, \label{abelian_BB}  \\
C_{YY} &=&  \frac{M}{M_1} i g^{}_{B} g^{\,2}_{Y} \frac{a^{}_{n}}{2} D^{}_{BYY}, \label{abelian_YY} \\
C_{YB} &=&  \frac{M}{M_1} i g^{}_{Y} g^{\,2}_{B}  \frac{a_{n}}{2} D^{}_{YBB}. \label{abelian_YB} 
\eeqn
Regarding the $Y$-variations $\propto Tr[q_B q^{2}_{Y}]$ and $\propto Tr[q^{2}_{B} q_{Y}]$, in general these traces are 
not identically vanishing and we introduce the CS and GS counterterms to cancel them. 
Having determined the factors in front of all the counterterms, 
we can summarize the structure of the one-loop anomalous 
effective action plus the counterterms as follows 
\beqn
{\mathcal S} &=&   {\mathcal S}_0 +{\mathcal S}_{an} + {\mathcal S}_{GS} + {\mathcal S}_{CS}  \nonumber\\
&=& {\mathcal S}_0 + \frac{1}{2!} \langle T_{BWW} BWW \rangle +  \frac{1}{2!} \langle T_{BGG} BGG \rangle
 + \frac{1}{3!} \langle T_{BBB} BBB \rangle     \nonumber\\
&&+ \frac{1}{2!} \langle T_{BYY} BYY \rangle + \frac{1}{2!} \langle T_{YBB} YBB \rangle  \nonumber\\
&&+ \frac{C_{BB}}{M} \langle b  F_{B} \wedge F_{B}  \rangle 
+ \frac{C_{YY}}{M} \langle b F_{Y} \wedge F_{Y}  \rangle + \frac{C_{YB}}{M} \langle b F_{Y} \wedge F_{B}  \rangle \nonumber\\
&&+ \frac{F}{M} \langle b Tr[F^W \wedge F^W]  \rangle  +  \frac{D}{M} \langle b Tr[F^G \wedge F^G] \rangle  \nonumber\\ 
&&+ d_{1} \langle BY \wedge F_{Y} \rangle + d_{2} \langle YB \wedge F_{B} \rangle  \nonumber\\
&&+ c_{1} \langle \epsilon^{\mu\nu\rho\sigma} B_{\mu} C^{SU(2)}_{\nu\rho\sigma} \rangle
+ c_{2} \langle \epsilon^{\mu\nu\rho\sigma} B_{\mu} C^{SU(3)}_{\nu\rho\sigma} \rangle,  
\eeqn
where ${\mathcal S}_0$ is the classical action. At this point we are ready to 
define the expressions in momentum space of the vertices introduced in 
Eq.~(\ref{definingv}), denoted by {\bf{V}}, obtaining
\beqn
{\bf V}^{\lambda \mu \nu}_{BGG} &=& 4 \, \frac{1}{2}  D^{(Q)}_{B} \, g^{}_{B} g^{\,2}_{3}
\, \Delta^{\lambda \mu \nu}_{\bf AAA} ( + 1/6 , k^{}_{1}, k^{}_{2} )  
+  D^{(Q)}_{B} \, g^{}_{B} g^{\,2}_{3} \frac{1}{2} \frac{i}{ \pi^{2} } \, \frac{2}{3} \,
\epsilon^{\lambda \mu \nu \sigma}(k^{}_{1} - k^{}_{2})^{}_{\sigma} 
\label{v1}    \\
{\bf V}^{\lambda \mu \nu}_{BWW} &=&4  \, \frac{1}{2} D^{(L)}_{B} \, g^{}_{B} g^{\,2}_{2}
\, \Delta^{\lambda \mu \nu}_{\bf AAA} ( + 1/6 , k^{}_{1}, k^{}_{2} )  
+  D^{(L)}_{B} \, g^{}_{B} g^{\,2}_{2} \frac{1}{2} \frac{i}{ \pi^{2} } \, \frac{2}{3} \,
\epsilon^{\lambda \mu \nu \sigma}(k^{}_{1} - k^{}_{2})^{}_{\sigma} 
\label{v1}    \\
{\bf V}^{\lambda \mu \nu}_{BYY} &=&4  D^{}_{BYY} \, g^{}_{B} g^{\,2}_{Y}
\, \Delta^{\lambda \mu \nu}_{\bf AAA} ( + 1/6 , k^{}_{1}, k^{}_{2} )  
+  D^{}_{BYY} \, g^{}_{B} g^{\,2}_{Y} \frac{i}{ \pi^{2} } \, \frac{2}{3} \,
\epsilon^{\lambda \mu \nu \sigma}(k^{}_{1} - k^{}_{2})^{}_{\sigma} 
\label{v1}    \\
{\bf V}^{\lambda \mu \nu}_{YBB} &=& 4 D^{}_{YBB} \, g^{}_{Y} g^{\,2}_{B}
 \, \Delta^{\lambda \mu \nu}_{\bf AAA} ( + 1/6 , k^{}_{1}, k^{}_{2} )  
-  D^{}_{YBB} \, g^{}_{Y} g^{\,2}_{B}  \frac{i}{  \pi^{2} } \, \frac{1}{3} \,
\epsilon^{\lambda \mu \nu \sigma}(k^{}_{1} - k^{}_{2})^{}_{\sigma}.
\label{v2} \\
{\bf V}^{\lambda \mu \nu}_{BBB} &=& 4 D^{}_{BBB} \, g^{\,3}_{B}
 \, \Delta^{\lambda \mu \nu}_{\bf AAA} ( + 1/6 , k^{}_{1}, k^{}_{2} ).
\label{v3}
\eeqn
where for the generalized CS terms we consider only the trilinear CS interactions whose momentum structure is the same as the Abelian ones as already discussed in Sec.~\ref{sec:WI_discuss}. The factor 1/2 overall in the non-Abelian vertices comes from the trace over the generators.
 These vertices satisfy standard Ward Identities on 
the external Standard Model lines, with an anomalous WI only on the $B$-line 
\beqn
k^{}_{1\mu} {\bf V}^{\lambda \mu \nu}_{BYY}( k^{}_{1}, k^{}_{2}) &=& 0   \\
k^{}_{2 \nu } {\bf V} ^{\lambda \mu \nu }_{BYY} (  k^{}_{1}, k^{}_{2}) &=& 0   \\
k^{}_{\lambda } {\bf V}^{ \lambda \mu \nu }_{BYY} ( k^{}_{1}, k^{}_{2}) &=& 4
 D^{}_{BYY} g^{}_{B} g^{\,2}_{Y} \, a^{}_{n} \,  \epsilon^{\mu \nu \alpha \beta} k_{1 \alpha} k_{2 \beta},
\eeqn
and obviously the $B$-currents contain the total anomaly $a^{}_{n}= - \frac{i}{2 \pi^{2}}$. The same anomaly equations given above for 
${\bf V}^{\lambda \mu \nu}_{BYY}$ hold for the ${\bf V}^{\lambda \mu \nu}_{BGG}$ and ${\bf V}^{\lambda \mu \nu}_{BWW}$ vertices but with a 1/2 factor overall. The anomaly equations for the $YBB$ vertex are 
\beqn
k^{}_{1\mu} {\bf V}^{\lambda \mu \nu}_{YBB}( k^{}_{1}, k^{}_{2}) &=& 4
  D^{}_{YBB} \,g^{}_{Y} g^{\,2}_{B} \, \frac{a^{}_{n}}{2} \,  \epsilon^{\lambda \nu \alpha \beta} 
k_{1 \alpha} k_{2 \beta} \\
k^{}_{2 \nu } {\bf V} ^{\lambda \mu \nu }_{YBB} \left(  k^{}_{1}, k^{}_{2}  \right) &=& 
4  D^{}_{YBB} \,g^{}_{Y} g^{\,2}_{B} \, \frac{a^{}_{n}}{2} \,  \epsilon^{\lambda \mu \alpha \beta} 
k_{2 \alpha} k_{1 \beta}     \\
k^{}_{\lambda } {\bf V}^{ \lambda \mu \nu }_{YBB} ( k^{}_{1}, k^{}_{2}) &=& 0, 
\eeqn
where the chiral current $Y$ has to be conserved so to render the one-loop effective action gauge invariant. 
Introducing a symmetric distribution of the anomaly, in the $BBB$ case the analogous equations are
\beqn
k^{}_{1\mu} {\bf V}^{\lambda \mu \nu}_{BBB}( k^{}_{1}, k^{}_{2}) &=& 4 \,
  D^{}_{BBB}  \,g^{\,3}_{B} \, \frac{a^{}_{n}}{3} \,   \epsilon^{\lambda \nu \alpha \beta} 
k_{1 \alpha} k_{2 \beta} \\
k^{}_{2 \nu } {\bf V} ^{\lambda \mu \nu }_{BBB} \left(  k^{}_{1}, k^{}_{2}  \right) &=& 
4 \, D^{}_{BBB} \, g^{\,3}_{B}  \, \frac{a^{}_{n}}{3} \,    \epsilon^{\lambda \mu \alpha \beta} 
k_{2 \alpha} k_{1 \beta}     \\
k^{}_{\lambda } {\bf V}^{ \lambda \mu \nu }_{BBB} ( k^{}_{1}, k^{}_{2}) &=& 4 \,
 D^{}_{BBB} \, g^{\,3}_{B}  \, \frac{a^{}_{n}}{3} \,    \epsilon^{\mu \nu \alpha \beta} k_{1 \alpha} k_{2 \beta} , 
\eeqn
A study of the issue of the gauge dependence in these types of models can be found in Chap.~\ref{chap:AbelianModels1}.
The cancellation of the 
gauge dependendent terms in specific classes of diagrams can be performed both in the 
exact phase and in the broken phase, similarly to the discussion presented in the previous chapter, 
having re-expressed the fields in the basis of the mass eigenstates. The approach that we follow is then clear: we worry about the cancellation of the anomalies in the exact phase, having performed a minimal gauge-fixing to remove the $B$ mixing with the axion $b$, then we rotate the fields and re-parameterize the Lagrangian around the non trivial vacuum of the potential. We will see in the next sections that with this simple procedure we can easily discuss simple basic processes involving neutral and charged currents exploiting the invariance of the effective action under re-parameterizations of the fields.
\section{ The neutral currents sector in the mLSOM}
In this section we move toward the phenomenological analysis of a typical process which exhibits 
the new trilinear gauge interactions at one-loop level. As we have mentioned in the introduction, our goal 
here is to characterize this analysis at a more formal level, leaving to the following chapters the numerical results with accurate studies of 
its predictions for applications at the LHC in the future.  

We proceed with our illustration starting from the definition of the neutral current in the model, which is given by 
\beqn
- {\mathcal L}_{NC} = \overline{\psi}^{}_{f} \gamma^\mu \left[ g^{}_{2} W^{3}_{\mu} T^{\,3} +  g^{}_{Y} Y A^{Y}_{\mu} 
+ g^{}_{B} Y^{}_{B} B^{}_{\mu}  \right]  \psi^{}_{f},
\eeqn
that we express in the two basis, the basis of the interaction eigenstates and of the mass eigenstates. 
Clearly in the  interaction basis the bosonic operator in the covariant derivative becomes
\beqn
\mathcal{F} &\equiv & g^{}_{2} W^{3}_{\mu} T^{3} +  g^{}_{Y} Y A^{Y}_{\mu} + g^{}_{B} Y^{}_{B} B^{}_{\mu} \nonumber \\
&=& g^{}_{Z} Q^{}_{Z} Z^{}_{\mu} 
+ g^{}_{Z^\prime} Q^{}_{Z^\prime} Z^{\prime}_{\mu} + e\, Q A^{\gamma}_{\mu}, 
 \eeqn
where $Q = T^{3} + Y$. The rotation in the photon basis gives
\beqn
W^{3}_{\mu} &=& O^{A}_{W_{3} \gamma} A^{\gamma}_{\mu} + O^{A}_{W_{3} Z} Z_{\mu} + O^{A}_{W_{3} Z^\prime} Z^{\prime}_{\mu}  \\
A^{Y}_{\mu} &=& O^{A}_{Y \gamma} A^{\gamma}_{\mu} + O^{A}_{Y Z} Z_{\mu} + O^{A}_{Y Z^\prime} Z^{\prime}_{\mu}  \\
B_{\mu} &=& O^{A}_{B Z} Z_{\mu} + O^{A}_{B Z^\prime} Z^{\prime}_{\mu}  
\eeqn
and performing the rotation on $\mathcal{F}$  we obtain
\beqn
\mathcal{F}&=&  A^{\gamma}_{\mu} \left[ g^{}_{2} O^{A}_{W_{3} \gamma} T^{3} + g^{}_{Y} O^{A}_{Y \gamma} Y  \right] 
+  Z_{\mu} \left[   g^{}_{2} O^{A}_{W_{3} Z} T^{3} + g^{}_{Y} O^{A}_{Y Z} Y + g^{}_{B} O^{A}_{BZ} Y^{}_{B} \right]   \nonumber\\
&&+  Z^{\prime}_{\mu} \left[  g^{}_{2} O^{A}_{W_{3} Z^\prime} T^{3} + g^{}_{Y} O^{A}_{Y Z^\prime} Y 
+ g^{}_{B} O^{A}_{BZ^\prime} Y^{}_{B} \right],
\eeqn
where the electromagnetic current can be written in the usual way
\beqn
 A^{\gamma}_{\mu} \left[ g^{}_{2} O^{A}_{W_{3} \gamma} T^{3} + g^{}_{Y} O^{A}_{Y \gamma} Y  \right]  = 
e A^{\gamma}_{\mu} Q,
\eeqn
with the definition of the electric charge as
\beqn
e= g_{2} O^{A}_{W_{3} \gamma} = g_{Y} O^{A}_{Y \gamma} = \frac{ g^{}_{Y} g^{}_{2} }{ \sqrt{ g^{2}_{Y} + g^{2}_{2}}}. 
\eeqn
Similarly for the neutral $Z$ current we obtain
\beqn
&& Z_{\mu} \left[  g^{}_{2} O^{A}_{W_{3} Z} T^{3} + g^{}_{Y} O^{A}_{Y Z} Y + g^{}_{B} O^{A}_{BZ} Y_{B} \right]  \nonumber\\
&=&  Z_{\mu} \left[ T^{3} ( g^{}_{2} O^{A}_{W_{3} Z} - g^{}_{Y} O^{A}_{YZ}) + g^{}_{Y} O^{A}_{YZ} Q 
+ g^{}_{B} O^{A}_{BZ} Y_{B} \right] \nonumber\\
&=&  Z_{\mu} g^{}_{Z}   \left[ T^{3} +   \frac{g^{}_{Y} O^{A}_{YZ}}{ g^{}_{2} O^{A}_{W_{3} Z} - g^{}_{Y} O^{A}_{YZ} } Q 
+ \frac{g^{}_{B} O^{A}_{BZ}}{ g^{}_{2} O^{A}_{W_{3} Z} - g^{}_{Y} O^{A}_{YZ} }  Y_{B}  \right],
\eeqn
where we have defined
\beqn
g_{Z} = g^{}_{2} O^{A}_{W_{3} Z} - g^{}_{Y} O^{A}_{YZ} \simeq  \, g = \frac{g^{}_{2}}{\cos \theta^{}_{W}}. 
\eeqn
We can easily work out the structure of the covariant derivative interaction 
applied on a left-handed or on a right-handed fermion.
For this reason it is convenient to introduce some notation. We define
\beqn
&&\mu^{Z}_{Q} = \frac{g^{}_{Y} O^{A}_{YZ}}{g_{Z}} \simeq - \sin^{2} \theta^{}_{W},  \\
&& \mu^{Z}_{B} = \frac{g^{}_{B} O^{A}_{BZ}}{g^{}_{Z}} \simeq  \frac{g^{}_{B}}{2} \epsilon^{}_{1} \mbox{\,\,\,\,\,so that\,\,} 
\lim_{M^{}_{1} \rightarrow \infty}  \mu^{Z}_{B}  = 0, 
\eeqn 
and similarly for the $Z^\prime$ neutral current 
\beqn
g^{}_{Z^\prime} = g^{}_{2} O^{A}_{W_{3} Z^\prime} - g^{}_{Y} O^{A}_{YZ^\prime}, \qquad
\mu^{Z^{\prime}}_{Q} = \frac{g^{}_{Y} O^{A}_{YZ^\prime}}{g^{}_{Z^\prime}},  \qquad \mu^{Z^\prime}_{B} 
= \frac{g^{}_{B} O^{A}_{B Z^\prime}}{g^{}_{Z^\prime}}. 
\eeqn
We can easily identify the generators in the (Z, $Z^\prime$, $A^{}_\gamma$) basis. These are given by
\beqn
\hat{Q}^{}_{Z} &=& \hat{Q}^{R}_{Z}   + \hat{Q}^{L}_{Z} = T^{3 L} + \mu^{Z}_{Q} Q^{L} 
+  \mu^{Z}_{B} Y^{L}_{B} +  \mu^{Z}_{Q} Q^{R} +  \mu^{Z}_{B} Y^{R}_{B}   \nonumber\\
\hat{Q}^{}_{Z^\prime} &=& \hat{Q}^{R}_{Z^\prime}  + \hat{Q}^{L}_{Z^\prime} = T^{3 L} + \mu^{Z^\prime}_{Q} Q^{L} 
+  \mu^{Z^\prime}_{B} Y^{L}_{B}  + \mu^{Z^\prime}_{Q} Q^{R} +  \mu^{Z^\prime}_{B} Y^{R}_{B}  \nonumber\\
\hat{Q}  &=& \hat{Q}_{L} + \hat{Q}_{R}
\eeqn
which will be denoted as $ Q^{}_{\, \overline p} = ( \hat{Q}, \hat{Q}_{Z}, \hat{Q}_{Z^\prime})$.
To express a given correlator, say $\langle Z A_{\gamma} A_{\gamma}  \rangle$ in the $(W_{3}, A_{Y}, B)$ basis we proceed as follows.
We denote with $ Q^{}_{\, \overline p} = ( \hat{Q}, \hat{Q}_{Z}, \hat{Q}_{Z^\prime})$ the generators in the photon basis 
$( A_{\gamma}, Z, Z^\prime)$ and with $g^{}_{\, \overline p}= (e, g^{}_{Z}, g^{}_{Z'})$ the corresponding couplings. Similarly, $ Q^{}_{p} = (T^{3}, Y, Y_{B})$ are 
the generators in the interaction basis 
$(W_{3}, A_{Y}, B)$ and $g^{}_{p} = (g^{}_{2}, g^{}_{Y}, g^{}_{B})$ the corresponding couplings, so that 
\beqn
- {\mathcal L}_{NC} &=& \overline{\psi} \gamma^{\mu} \left[  g^{}_{Z} \hat{Q}_{Z} Z_{\mu} 
+ g^{}_{Z^{\prime}} \hat{Q}_{Z^{\prime}} Z^{\prime}_{\mu} + e \, \hat{Q} A^{\gamma}_{\mu}  \right] \psi    \nonumber\\
&=&   \overline{\psi} \gamma^{\mu} \left[ g^{}_{2} T^{\,3} W^{\,3}_{\mu} 
+ g^{}_{Y} Y A^{Y}_{\mu}  + g^{}_{B} Y^{}_{B} B_{\mu}  \right] \psi.
\eeqn
\section{The $Z  \gamma \gamma$ vertex in the Standard Model}
Before coming to the computation of this vertex in the mLSOM we first start reviewing its structure in the SM. 

\begin{figure}[t]
{\centering \resizebox*{8cm}{!}{\rotatebox{0}
{\includegraphics{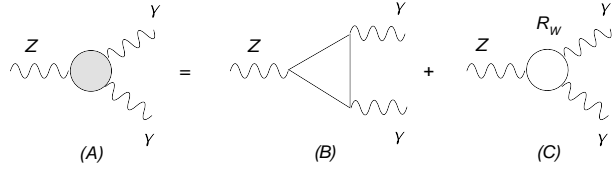}}}\par}
\caption{\small The $Z \gamma \gamma$ vertex to lowest order in the SM, with the anomalous contributions and the remaining weak corrections shown separately.}
\label{ZGGcontributions}
\end{figure}

We show in Fig.~\ref{ZGGcontributions} the $Z \gamma \gamma$ vertex in the SM, where we have separated
the QED contributions from the remaining corrections $R_{W}$. This vertex vanishes at all orders 
when all the three lines 
are on-shell, due to the Landau-Yang theorem. A direct proof of this property for the fermionic one-loop corrections will be shown in App.~\ref{app:LandauYang}.

\begin{figure}[t]
{\centering \resizebox*{12cm}{!}{\rotatebox{0}
{\includegraphics{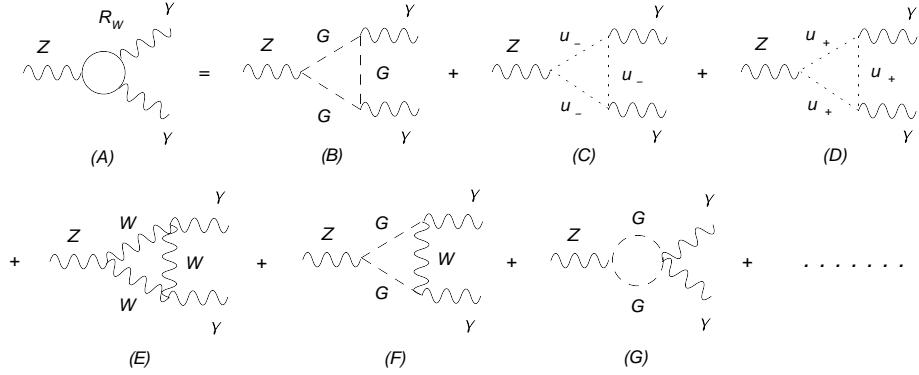}}}\par}
\caption{\small Some typical electroweak corrections, involving the charged Goldstones (here 
denoted by $G$, ghosts contributions ($u_\pm$) and $W$ exchanges.}
\label{electroweak}
\end{figure}

\begin{figure}[t]
{\centering \resizebox*{4.5cm}{!}{\rotatebox{0}
{\includegraphics{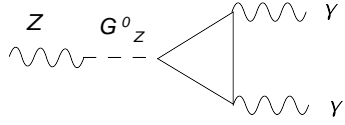}}}\par}
\caption{\small $Z-G^0_Z$ mixing in the broken phase in the SM.}
\label{gauge_mixing}
\end{figure}

\begin{figure}[t]
{\centering \resizebox*{9cm}{!}{\rotatebox{0}
{\includegraphics{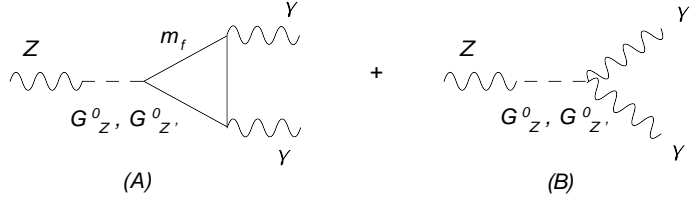}}}\par}
\caption{\small Same as in Fig.~\ref{gauge_mixing} but for the mLSOM.
\label{rotated_mixing_gauge}}
\end{figure}

\begin{figure}[t]
{\centering \resizebox*{12cm}{!}{\rotatebox{0}
{\includegraphics{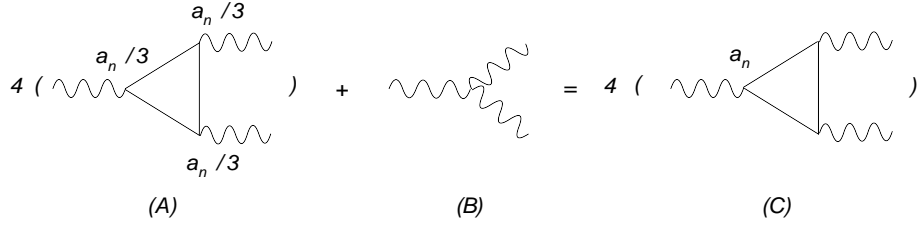}}}\par}
\caption{\small Re-distribution of the anomaly via the CS counterterm.
\label{anomaly_organization}}
\end{figure}

The QED contribution contains the fermionic triangle diagrams (direct plus exchanged) while $R^{}_W$ includes all the remaining contributions at one-loop level. Specifically, as shown in Fig.~\ref{electroweak}, $R_{W}$ contains ghosts, 
Goldstones and all other exchanges.
An exhaustive computation  of all these contributions 
is not needed for our discussion. We have omitted diagrams of the type shown in Figs.~\ref{gauge_mixing}, \ref{rotated_mixing_gauge}. These 
are removed by working in the $R^{}_\xi$ gauge for the $Z$ boson. 
Notice, however, that even without a gauge-fixing these decouple 
from the anomaly diagrams in the massless fermion limit since the Goldstone does not couple to massless fermions. 
In Fig.~\ref{anomaly_organization} we show how the anomaly is re-distributed in an $AAA$ diagram  
by a CS interaction, generating an $AVV$ vertex. 

To appreciate the role played by the anomaly in this vertex we perform a direct 
computation of the two anomaly diagrams and include the fermionic mass terms. 
A direct computation gives
\beqn
G^{\rho \nu \mu} (k, k_1, k_2) &=& - \frac{e^2 g}{\cos \theta_{W} } \sum_{f} g^{f}_{A} Q^{2}_{f} \int \frac{d^{4} p}{(2 \pi)^{4}}
\, Tr\left( \frac{1}{{\slashed p} - m_f} \gamma^{\rho} \gamma^{5} \frac{1}{ {\slashed p} - {\slashed k} - m_f } \gamma^{\nu}
\frac{1}{ { \slashed p} - {\slashed k_1} - m_f } \gamma^{\mu}  \right)    \nonumber\\
&&+ \, (k_1 \rightarrow k_2, \mu \rightarrow \nu).
\eeqn
which can be cast in the form 
\beqn
G^{\rho \nu \mu} (k, k_1, k_2) &=& - \frac{e^2 g}{ 2  \pi^2 \cos{ \theta_W} } \sum_f g^{f}_{A} Q^{2}_{f} \int^{1}_{0} 
d x_1  \int^{1 - x_1 }_{0} d x_2 \,   \nonumber\\
&& \cdot \, \frac{1}{\Delta} \, \Big[ \epsilon^{\rho \nu \mu \alpha } 
(1 - x_1 - x_2)(x_2 k_1 - x_1 k_2)_{\beta} ( k^{\beta}_{2} k_{1 \alpha} + k^{\beta}_{1} k_{2 \alpha } )    \nonumber\\
&&  + (1 - x_1 - x_2 ) (  \epsilon^{\alpha \rho \beta \nu } k_{1 \alpha} k_{2 \beta} ( x_2 k^{\mu}_{1} -  x_1 k^{\mu}_{2} )  
+ (\mu \rightarrow \nu ) )   \nonumber\\
&&  + \epsilon^{\alpha \nu \beta \mu } k_{1 \alpha} k_{2 \beta } 
( x_2 ( x_2 -  x_1 - 1 )  k^{\rho}_{1} - x_1 ( x_2 -  x_1 + 1 ) k^{\rho}_{2} )  \Big],
\label{cast}
\eeqn
where
\beqn
\Delta = m^{2}_{f} + x_2 (x_2 -1) k^{2}_{1} + x_1 (x_1 - 1 ) k^{2}_{2} - 2 x_1 x_2 \, k_1 \cdot k_2,
\eeqn
and 
we have introducing the $g_{Z,A}^f$ and $g_{Z,V}^f$ couplings of the $Z$ with 
\beq
g_{Z,A}^{f} = \frac{1}{2} T_3^f,  \qquad g_{Z,V}^{f} = \frac{1}{2} T_3^f - Q_f \sin^2\theta_W.
\eeq
This form of the amplitude is obtained 
if we use the standard Rosenberg definition of the anomalous diagrams and it agrees with \cite{Barroso:1984re}. In this case 
the Ward Identities on the photon lines are defining conditions for the vertex. 
Naturally, with the standard fermion multiplet assignment the anomaly vanishes since
\beqn
\sum_{f} g^{f}_{A} Q^{2}_{f} = 0. 
\eeqn 
Because of the anomaly cancellation, the fermionic vertex is zero 
also off-shell, 
if the masses of all the fermions in each generation are degenerate, in particular if they are massless.
Notice that this is not a consequence of the Landau-Yang theorem.
 
Let us now move to the WI on the $Z$-line. A direct computation gives
\beqn
k_{\rho} \, G^{\rho \nu \mu}  &=&  (k_1 + k_2 )_{\rho} \,  G^{\,  \rho \nu \mu }     \nonumber\\
&=& \frac{ e^2 g }{ \pi^{2} \cos \theta_{W}}  \sum_{f}  g^{f}_{A}  Q^{2}_{f} \,  \epsilon^{\, \nu \mu \alpha \beta } 
k_{1 \alpha} k_{2 \beta } \, \left[  \frac{1}{2} - m^{2}_{f} \int^{1}_{0} d x_1 \int^{1 - x_1 }_{0} 
d x_2  \,  \frac{1}{ \Delta }    \right]. 
\label{ABJ_anomaly}
\eeqn
The presence of a mass-dependent term on the r.h.s. of (\ref{ABJ_anomaly}) constitutes a break-down of axial 
current conservation for massive fermions, as expected.
\subsection{The $Z  \gamma \gamma$ vertex in anomalous Abelian models: the Higgs-St\"uckelberg phase} 
The presence of anomalous generators in a given vertex renders some trilinear interactions non-vanishing also for massless fermions. In fact, 
as we have shown in the previous section, in the SM the anomalous triangle diagrams vanish if we neglect the masses of all the fermions, and this occurs both on-shell and off-shell. The only left over corrections 
are related to the fermion mass and these will also vanish (off-shell) if all the 
fermions of a given generation are mass degenerate.  
The on-shell vanishing of the same vertices is a consequence of the 
structure of the amplitude, as we show in the appendix. 
The extraction of the contribution of the anomalous generators in the trilinear vertices can be obtained starting from the 1-particle-irreducible (1PI) effective action, written in the basis of the interaction eigenstates, and performing the rotation of the trilinear interaction that 
project onto the $Z \gamma \gamma$ vertex.

In order to appreciate the differences between the SM result and the analogous one in the anomalous extensions that we are considering, we start by observing that only in the St\"uckelberg phase ($M_1\neq 0$ and $v_u=v_d=0$) the anomaly-free 
traces vanish,   
\ba
\langle Y Y Y \rangle \,g_Y^3 \, Tr [Q_Y^3]&=& 0 \nonumber \\
\langle Y W_3 W_3 \rangle \, g_Y g_2^2 \,Tr [Q_Y T^3 T^3]&=& 0\,,
\ea
because of charge assignment. A similar result is valid also in the 
HS phase if the Yukawa couplings are neglected. Coming to extract the 
$Z \gamma \gamma$ vertex we rotate the anomalous diagrams of the effective action into the mass eigenstates, being careful to separate the massless from the massive fermion contributions. 

Hence, we split the $\langle YYY\rangle$ vertex into its chiral contributions
and performing the rotation of the fields we get the following contributions
\ba
&&\frac{1}{3!}\, \langle Y Y Y \rangle \,g_Y^3 \, Tr [Q_Y^3]=\nonumber\\
&&\hspace{2.5cm}\sum_f\left[g_Y^3\frac{1}{8}(Q_{Y,f}^{L})^3\langle LLL \rangle^{\lambda\mu\nu}
+g_Y^3\frac{1}{8}(Q_{Y,f}^{R})^3\langle RRR \rangle^{\lambda\mu\nu}
\right.\nonumber\\
&&\hspace{2.5cm}\left.+g_Y^3\frac{1}{8}Q_{Y,f}^{L}(Q_{Y,f}^{R})^2 \langle LRR \rangle^{\lambda\mu\nu}
+g_Y^3\frac{1}{8}Q_{Y,f}^{L}Q_{Y,f}^{R}Q_{Y,f}^{L}\langle LRL \rangle^{\lambda\mu\nu}
\right.\nonumber\\
&&\hspace{2.5cm}\left.+g_Y^3\frac{1}{8}(Q_{Y,f}^{L})^2 Q_{Y,f}^{R}\langle LLR \rangle^{\lambda\mu\nu}
+g_Y^3\frac{1}{8}Q_{Y,f}^{R}(Q_{Y,f}^{L})^2 \langle RLL \rangle^{\lambda\mu\nu}
\right.\nonumber\\
&&\hspace{2.5cm}\left.+g_Y^3\frac{1}{8}Q_{Y,f}^{R}Q_{Y,f}^{L}Q_{Y,f}^{R}\langle RLR \rangle^{\lambda\mu\nu}
+g_Y^3\frac{1}{8}(Q_{Y,f}^{R})^2 Q_{Y,f}^{L}\langle RRL \rangle^{\lambda\mu\nu}
\right]
Z^{\lambda} A_{\g}^{\mu}A_{\g}^{\nu}\frac{1}{3!} R^{YYY}
+\dots \nonumber\\
\ea
where the dots indicate all the other projections of the type $ZZ\g,Z^{\prime}\g\g$ etc. 
Here $\langle LLL\rangle$, $\langle RLR\rangle$ etc., indicate 
the (clockwise) insertion of $L/R$ chiral projectors on the $\lambda \mu\nu$ vertices of the anomaly diagrams.  

For the $\langle YWW\rangle$ vertex the structure is more simple because the generator
associated to $W_3$ is left-chiral
\ba
&&\frac{1}{2!}\, \langle Y W W \rangle \,g_Y g_2^2 \, Tr [Q_Y (T^{3})^2]=
\sum_f\left[g_Y g_2^2\frac{1}{8}Q_{Y,f}^{L}(T^{3}_{L,f})^2\langle LLL \rangle^{\lambda\mu\nu}
\right.\nonumber\\
&&\left.\hspace{2.5cm}
+g_Y g_2^2\frac{1}{8}Q_{Y,f}^{R}(T^{3}_{L,f})^2\langle RLL \rangle^{\lambda\mu\nu}
\right]Z^{\lambda} A_{\g}^{\mu}A_{\g}^{\nu}\frac{1}{2!} R^{YWW}
+\dots 
\ea
The $\langle BYY\rangle$ vertex works in same way of $\langle YYY\rangle$
\ba
&&\frac{1}{2!}\, \langle B Y Y \rangle \,g_B g_Y^2 \, Tr [Q_B Q_Y^2]=\nonumber\\
&&\hspace{1.5cm}\sum_f\left[g_B g_Y^2\frac{1}{8}Q_{B,f}^{L}(Q_{Y,f}^{L})^2\langle LLL \rangle^{\lambda\mu\nu}
+g_B g_Y^2\frac{1}{8}Q_{B,f}^{R}(Q_{Y,f}^{R})^2\langle RRR \rangle^{\lambda\mu\nu}
\right.\nonumber\\
&&\hspace{1.5cm}\left.+g_B g_Y^2\frac{1}{8}Q_{B,f}^{L}(Q_{Y,f}^{R})^2 \langle LRR \rangle^{\lambda\mu\nu}
+g_B g_Y^2\frac{1}{8}Q_{B,f}^{L}Q_{Y,f}^{R}Q_{Y,f}^{L}\langle LRL \rangle^{\lambda\mu\nu}
\right.\nonumber\\
&&\hspace{1.5cm}\left.+g_B g_Y^2\frac{1}{8}Q_{B,f}^{L}Q_{Y,f}^{L}Q_{Y,f}^{R}\langle LLR \rangle^{\lambda\mu\nu}
+g_B g_Y^2\frac{1}{8}Q_{Y,f}^{R}(Q_{Y,f}^{L})^2 \langle RLL \rangle^{\lambda\mu\nu}
\right.\nonumber\\
&&\hspace{1.5cm}\left.+g_B g_Y^2\frac{1}{8}Q_{B,f}^{R}Q_{Y,f}^{L}Q_{Y,f}^{R}\langle RLR \rangle^{\lambda\mu\nu}
+g_B g_Y^2\frac{1}{8}Q_{B,f}^{R} Q_{Y,f}^{R}Q_{Y,f}^{L}\langle RRL \rangle^{\lambda\mu\nu}
\right]Z^{\lambda} A_{\g}^{\mu}A_{\g}^{\nu}\frac{1}{2!} R^{BYY}
+\dots \nonumber\\
\ea
Finally, the $\langle BWW\rangle$ vertex is similar to $\langle YWW\rangle$
\ba
&&\frac{1}{2!}\, \langle B W W \rangle \,g_Y g_2^2 \, Tr [Q_B (T^{3})^2]=
\sum_f\left[g_B g_2^2\frac{1}{8}Q_{B,f}^{L}(T^{3}_{L,f})^2\langle LLL \rangle^{\lambda\mu\nu}
\right.\nonumber\\
&&\left.\hspace{5cm}
+g_B g_2^2\frac{1}{8}Q_{B,f}^{R}(T^{3}_{L,f})^2\langle RLL \rangle^{\lambda\mu\nu}
\right]Z^{\lambda} A_{\g}^{\mu}A_{\g}^{\nu}\frac{1}{2!} R^{BWW}
+\dots \nonumber\\
\ea
where we have defined
\ba
&&R^{YYY}=3\left[(O^{A\,T})_{22}(O^{A\,T})_{21}^{2}\right]\nonumber\\
&&R^{YWW}=\left[2(O^{A\,T})_{11}(O^{A\,T})_{12}(O^{A\,T})_{21}+(O^{A\,T})_{11}^{2}(O^{A\,T})_{22}\right]\nonumber\\
&&R^{BYY}=(O^{A\,T})_{21}^{2}(O^{A\,T})_{32}\nonumber\\
&&R^{BWW}=\left[(O^{A\,T})_{11}^{2}(O^{A\,T})_{32}\right]\,.
\ea
which are the product of rotation matrices that project the anomalous effective 
action from the interaction eigenstate basis over the $Z,\gamma$ gauge bosons.

We have expressed the generators in their chiral basis, and their mixing is due to mass insertions over each
fermion line in the loop. The ellypsis refers to additional contributions which do not project
on the vertex that we are interested in but which are present
in the analysis of the remaining neutral vertices, $ZZ\g,Z^{\prime}\g\g$ etc. The notation
$O^{AT} $ indicates the transposed of the rotation matrix from the interaction to the mass eigenstates.
To obtain the final expression of the amplitude in the interaction eigenstate basis one can easily observe that in the chiral amplitudes $\langle LLL \rangle$ and $\langle RRR\rangle$  
the mass dependence in the fermion loops  is all contained in the denominators of the propagators, not in the Dirac traces. The only diagrams that contain a mass dependence at the numerators are those involving chirality flips 
($\langle LLR\rangle,  \langle RRL \rangle$) which contribute with terms 
proportional to $m_f^2$. These terms contribute only to the invariant amplitudes 
$A_1$ and $A_2$ of the Rosenberg representation (see Chap.~\ref{chap:AbelianModels1}, Eq.~\ref{RRos}) and, although finite, they disappear once we impose a WI on the two photon lines, as requested by CVC for the two photons.   A similar result is 
valid for the SM, as one can easily figure out from Eq.~(\ref{cast}). Therefore, the amplitudes can be expressed just in terms of $LLL$ and $RRR$ correlators, and since the mass dependence is at the denominators of the propagators,  one can easily show the relation 
\beq
\langle LLL\rangle =-\langle RRR\rangle
\eeq
valid for any fermion mass $m_f$. Defining
$\langle LLL\rangle\equiv\Delta^{\lambda\mu\nu}_{LLL}(m_f\neq 0)$, we can express the only independent chiral graph as sum of two contributions
\ba
\Delta^{\lambda\mu\nu}_{LLL}(m_f\neq 0)=\Delta^{\lambda\mu\nu}_{LLL}(0)+\Delta^{\lambda\mu\nu}_{LLL}(m_f)
\ea
where we define
\ba
&&\Delta^{\lambda\mu\nu}_{LLL}(0)\equiv\Delta^{\lambda\mu\nu}_{LLL}(m_f=0)\nonumber\\
&&\Delta^{\lambda\mu\nu}_{LLL}(m_f)\equiv\Delta^{\lambda\mu\nu}_{LLL}(m_f\neq 0)-\Delta^{\lambda\mu\nu}_{LLL}(m_f=0).
\ea
Also, one can verify quite easily that
\ba
\Delta^{\lambda\mu\nu}_{LLL}( 0) &=&\Delta^{\lambda\mu\nu}_{AVV}(0)+\Delta^{\lambda\mu\nu}_{VAV}(0)+
\Delta^{\lambda\mu\nu}_{VVA}(0)+\Delta^{\lambda\mu\nu}_{AAA}(0)\nonumber \\
&&=4 \Delta^{\lambda\mu\nu}_{AAA}(0).
\ea
A second contribution to the effective action comes from the one-loop
counterterms containing generalized CS terms. There are two ways to express these counterterms:
either as separate trilinear interactions or as
modifications of the two invariant amplitudes of the Rosenberg
parameterization $A_1, A_2$. These amplitudes depend linearly on the momenta of the vertex (see App.~\ref{app:momentumCS_GS}).
For instance we use
\ba
\Delta_{AAA}(0)-\frac{a_n}{3}\epsilon^{\lambda\mu\nu\alpha}(k_{1\alpha}-k_{2\alpha})=\Delta_{AVV}(0),
\ea
which allows to absorb completely the CS term, giving conserved 
$Y/W_3$ currents in the interaction eigenstate basis. In this case we move 
from a symmetric distribution of the anomaly  in the $AAA$ diagram, to an 
$AVV$ diagram. 
These currents interpolate with the vector-like vertices (V) of the $AVV$ graph. 

Notice that once the anomaly is moved from any vertex involving a $Y/W_3$ current to a vertex with a $B$ current, it is then canceled by the GS interaction.
The extension of this analysis to the complete $m_f$-dependent case 
for $\Delta_{LLL}(m_f\neq 0)$ is  
quite straightforward.   
In fact,  after some re-arrangements of the $Z\g\g$ amplitude, we are left with the following contributions in the physical basis in the broken phase
\ba
&&\langle Z\g\g\rangle|_{m_f\neq 0}=
\frac{1}{4}\sum_f\Delta_{AVV}^{\lambda\mu\nu}(m_f\neq 0)
\left[g_Y^3\theta_f^{YYY}R^{YYY}+ g_Y g_2^2\theta_f^{YWW}R^{YWW}
\right.\nonumber\\
&&\hspace{2cm}\left.
+g_B g_Y^2\theta_f^{BYY}R^{BYY}+ g_B g_2^2\theta_f^{BWW}R^{BWW}\right]Z^{\lambda}A_{\g}^{\mu}A_{\g}^{\nu}
\label{final}
\ea
where we have defined the anomalous chiral asymmetries as
\ba
&&\theta^{BYY}_{f}=\left[Q_{B,f}^{L} (Q_{Y,f}^{L})^2-Q_{B,f}^{R} (Q_{Y,f}^{R})^2\right]
\nonumber\\
&&\theta^{BWW}_{f}=Q_{B,f}^{L}(T^{3}_{L,f})^2.
\ea
The conditions of gauge invariance force the coefficients in front of the CS terms to be
\ba
&&D_{BYY}=\frac{1}{8}\sum_f\theta^{BYY}_{f}
\nonumber\\
&&D_{BWW}=\frac{1}{8}\sum_f\theta^{BWW}_{f},\, 
\ea
which have been absorbed and do not appear explicitly, while the SM chiral asymmetries are defined as
\ba
&&\theta^{YYY}_{f}=\left[(Q_{Y,f}^{L})^3-(Q_{Y,f}^{R})^3\right]
\nonumber\\
&&\theta^{YWW}_{f}=Q_{Y,f}^{L}(T^{3}_{L,f})^2,
\ea
and the triangle $\Delta_{AVV}(m_f\neq 0)$ is given as in (\ref{cast}).
Notice that Eq.~(\ref{final}) is in complete agreement with the SM result 
shown in (\ref{cast}), obtained by removing the contributions proportional to the $B$ gauge bosons  and setting the chiral asymmetries of $Y$ and $W_3$ to zero. 
In particular, if the gauge bosons are not anomalous and in the 
chiral limit ($m_f=0$ or  $m_f=m$) 
this trilinear amplitude vanishes. 
As we have already pointed out, the amplitude for the $\langle Z\gamma\gamma\rangle$ process 
is espressed in terms of six invariant amplitudes that can be easily computed and take the form
\beqa
\Delta^{\lambda \mu\nu}_{AVV} &=& {A}_1(k_1,k_2) \epsilon[k_1,\mu,\nu,\la] +
{A}_2(k_1,k_2)\epsilon[k_2,\mu,\nu,\la]
+{A}_3(k_1,k_2) \epsilon[k_1,k_2,\mu,\la]{k_1}^{\nu}\nonumber \\
&&+ {A}_4(k_1,k_2) \epsilon[k_1,k_2,\mu,\la]{k_2}^{\nu} +
{A}_5(k_1,k_2) \epsilon[k_1,k_2,\nu,\la]k_1^\mu +
{A}_6(k_1,k_2)\epsilon[k_1,k_2,\nu,\la]k_2^\mu,   \nonumber\\
\label{Ros}
\eeqa
with
\beqa
A_1(k_1,k_2)&=&k_1\cdot k_2 A_3(k_1,k_2) + k_2^2 A_4(k_1,k_2) \nonumber \\ 
A_2(k_1,k_2) &=& -A_1(k_2,k_1) \nonumber \\
A_5(k_1,k_2) &=& -A_4(k_2,k_1) \nonumber \\
A_6(k_1,k_2) &=& -A_3(k_2,k_1). \nonumber \\
\eeqa
Also $A_1(k_1,k_2)=A_1(k_2,k_1)$ as one can easily check by a direct computation. 
We obtain 
\beqa
&&A_3(k_1,k_2)= -\frac{1}{2}\int_0^1 dx \int_0^{1-x} dy \frac{x y}{y (1-y) k_1^2 + 
x(1-x) k_2^2 + 2 x y \,k_1\cdot k_2 - m_f^2}  \nonumber \\
&& A_4(k_1,k_2)= \frac{1}{2}\int_0^1 dx \int_0^{1-x} dy \frac{x (1-x)}{y (1-y) k_1^2 + 
x(1-x) k_2^2 + 2 x y \,k_1\cdot k_2 - m_f^2}  \nonumber \\
\eeqa
The computation of these integrals can be done analytically and the various regions 
$0<  s< 4 m_f^2 $, $m_f>> \sqrt{s}/2$, and $m_f \to 0$  can be studied in detail.
In the case of both photons on-shell, for instance, and $s > 4 m_f^2 $ we obtain 
\beqa
&&A_3(k_1,k_2) = \frac{1}{2 s} - \frac{m_f^2}{s} \makebox{Li}_2\left(\frac{2}{1 - 
\sqrt{1- 4 m_f^2/s}}\right) - \frac{m_f^2}{s} \makebox{Li}_2\left(\frac{2}{1 + 
\sqrt{1- 4 m_f^2/s}}\right) \nonumber \\
&& A_4(k_1,k_2)= 
-\frac{1}{s} + \frac{\sqrt{1- 4 m_f^2/s}}{s}\makebox{ArcTanh}\left(\frac{1}
{\sqrt{1- 4 m_f^2/s}}\right)\eeqa
Notice that the case in which the two photons are on-shell and light fermions are running in the 
loop, then the evaluation of the integral requires particular care because 
of infrared effects which render the parameteric integrals ill-defined. 
The situation is similar to the case of the coupling of the axial anomaly to on-shell 
gluons in spin physics \cite{Carlitz:1988ab}, when the correct isolation of the massless quarks contributions 
is carried out by moving off-shell on the external lines and then performing the $m_f\to 0$ limit.
\subsection{ $q \bar{q} \to \gamma \gamma$ with an intermediate Z}
In this section we are going to describe the role played by the new anomaly cancellation 
mechanism in simple processes which can eventually be studied with accuracy at the LHC. A numerical analysis of processes involving neutral 
currents can be performed along the lines of \cite{Cafarella:2007tj} and we will return to this point in Chap.~\ref{chap:LHC} . Here we intend to discuss some of the phenomenological 
implications which might be of interest. 
Since the anomaly is canceled by a combination of Chern-Simons 
and Green-Schwarz contributions, the study of a specific process, such as 
$Z \to \gamma \gamma $, which differs from the SM prediction, requires, in general, 
a combined analysis both of the gauge sector and of the scalar sector.

We start from the case of a quark-antiquark annihilation mediated by a $Z$ that later undergoes a decay into two photons. At leading order this 
process is at parton level described by the annihilations of a valence quark $q$ 
and a sea antiquark $\bar{q}$ from the two incoming hadrons, both of them collinear and massless.
\begin{figure}[h]
{\centering \resizebox*{10cm}{!}{\rotatebox{0}
{\includegraphics{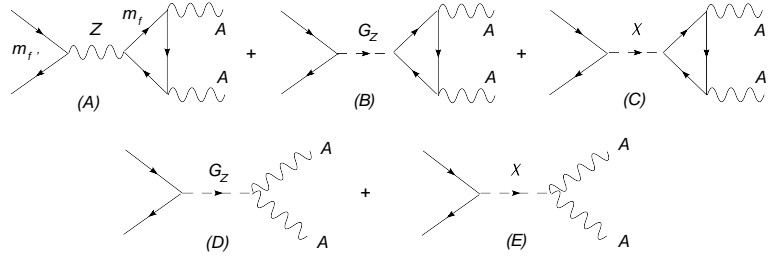}}}\par}
\caption{\small Two photon processes initiated by a $q \bar{q}$ annihilation with a $Z$ exchange.
\label{ampiezza3}}
\end{figure}
In Fig.~\ref{ampiezza3} we have depicted all the diagrams by which the process can take place to lowest 
order. Radiative corrections from the initial state are accurately known up to next-to-next-to-leading order, and are universal, being the same of the Drell-Yan cross section.
In this respect, precise QCD predictions for the rates are available, for instance around the $Z$ resonance \cite{Cafarella:2007tj}.  

In the SM, gauge invariance of the process requires both a $Z$ gauge boson exchange 
and the exchange of the corresponding Goldstone $G^{}_Z$, which involves 
diagrams (A) and (B). In the mLSOM a direct Green-Schwarz coupling to the photon (which is gauge dependent) 
is accompanied by a gauge independent axion exchange. If the incoming 
quark-antiquark pair is massless, then the Goldstone has no coupling to the incoming fermion pair, and 
therefore (B) is absent, while gauge invariance is trivially satisfied because of the massless condition 
on the fermion pair of the initial state. In this case only diagram (A) 
is relevant. Diagram (B) may also be set to vanish, for instance in suitable gauges, such as the unitary gauge. Notice also that the triangle diagrams have a dependence on $m^{}_f$, the mass of the fermion in the loop, 
and show two contributions: a first contribution which is proportional to the anomaly (mass independent) and a correction term which depends on $m^{}_f$. 

As we have shown above, the first contribution, which involves an off-shell vertex, is absent in the SM,
while it is non vanishing in the mLSOM. In both cases, on the other hand, we have $m^{}_f$ dependent contributions. 
It is then clear that in the SM the largest contribution to the process comes from the 
top quark circulating in the triangle diagram, the amplitude being essentially proportional only to the heavy top mass. On the $Z$ resonance and for on-shell photons, the cross section vanishes in both cases, as we have explained, in agreement with the Landau-Yang theorem. We have checked these properties explicitly, but they 
hold independently of the perturbative order at which they are analyzed, being based on the Bose 
symmetry of the two photons. The cross section, therefore, has a dip at $Q=M^{}_Z$, where it vanishes, and where $Q^2$ is the virtuality of the intermediate $s$-channel exchange.

\begin{figure}[h]
{\centering \resizebox*{12cm}{!}{\rotatebox{0}
{\includegraphics{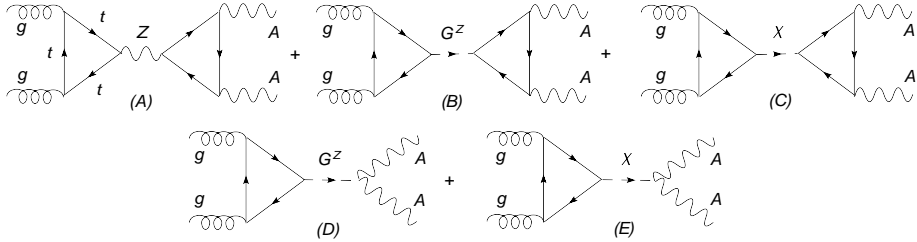}}}\par}
\caption{\small Gluon fusion contribution to double-photon production. Shown are also the scalar 
exchanges (B) and (D) that restore gauge invariance and the axi-Higgs exchange (E).}
\label{gluoncorretto}
\end{figure}

An alternative scenario is 
to search for neutral exchanges initiated by gluon-gluon fusion. In this case we replace the annihilation pair with a triangle loop (the process is similar to Higgs production via gluon fusion), as shown in Fig.~\ref{gluoncorretto}. 
As in the decay mechanism discussed above, the production 
mechanism in the SM and in the mLSOM are again different. In fact, 
in the mLSOM there is a massless contribution appearing already at the massless fermion level, which is absent in the SM. The production mechanism by gluon fusion has some special features as well.
In $ggZ$ production and $Z\gamma\gamma$ decay, the relevant diagrams are (A) and (B) since we need the 
exchange of a $G^{}_Z$ to obtain gauge invariance. 
As we probe smaller values of the Bjorken variable $x$,  the gluon density raises, and the process 
becomes sizable. 
On the other hand, in a $pp$ collider, although the quark annihilation channel 
is suppressed since the antiquark density 
is smaller than in a $p \bar{p}$ collision, 
this channel still remains rather significant. We have also shown in this figure one of the scalar channels, due to the exchange of an axi-Higgs. 

\begin{figure}[h]
{\centering \resizebox*{9cm}{!}{\rotatebox{0}
{\includegraphics{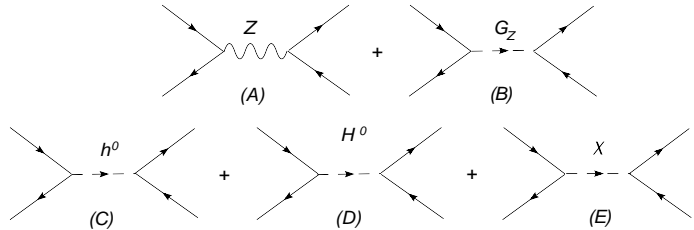}}}\par}
\caption{\small The $q \bar{q}$ annihilation channel (A,B). 
Scalar exchanges in the neutral sector 
involving the two Higsses and the axi-Higgs (C,D,E).}
\label{ampiezza2}
\end{figure}

Other channels such as those shown in Fig.~\ref{ampiezza2} can also be studied. 
These involve a 
lepton pair in the final state, and their radiative corrections also show the appearance of a triangle 
vertex. This is the classical Drell-Yan process, that we will briefly describe below. 
In this case, both the total cross section and the rapidity distributions of the lepton pair and/or an analysis 
of the charge asymmetry in $s$-channel exchanges of W's would be of major 
interest in order to disentangle the anomaly inflow. At the moment, errors on the parton 
distributions and scale dependences induce indeterminations which, just for the QCD 
background, are around $4 \% $ \cite{Cafarella:2007tj}, as shown in a high precision study. It is expected, 
however, that the statistical accuracy on the $Z$ resonance at the LHC is going to be a factor $10^2$ better. 
In fact this is a case in which the experiment can do better than the theory. 
\subsection{Isolation of the massless limit: the $Z^*\to \gamma^* \gamma^* $ amplitude}
The isolation of the massless from the massive contributions can be analyzed in the case of 
resolved photons in the final state. As we have already mentioned in the prompt photon case the amplitude, on the $Z$ resonance, vanishes because of Bose symmetry and angular momentum conservation. We can, 
however, be on the $Z$ resonance and produce one or two off-shell photons that undergo fragmentation. 
Needless to say, these contributions are small. However, the separation of the massless 
from the massive case is well defined. One can increase the rates by asking just for one single resolved photon and one prompt photon. Rates for this process in $pp$-collisions have been determined in \cite{Coriano:1996us}. We start from the case of off-shell external 
photons of virtuality $s_1$ and $s_2$  and an off-shell $Z$ $(Z^*)$.     
Following \cite{Kniehl:1989qu}, we introduce the total vertex $V^{\lambda\mu\nu}
(k^{}_1, k^{}_2, m^{}_f)$, which contains both the massive $m_f$ dependence (corresponding to the triangle amplitude $\Delta^{\lambda\mu\nu }$) and its massless counterpart ${\bf V}^{\lambda\mu\nu}(0)\equiv V(k^{}_1, k^{}_2, m^{}_f=0)$, obtained by sending the fermion mass to zero. 
The Rosenberg vertex and the V vertex are trivially related by a Schoutens 
transformation, moving the $\lambda$ index from the Levi-Civita tensor to the momenta 
of the photons
\beqn
&&V_{\lambda \mu \nu} (k_1, k_2, m_{f})    \nonumber\\
&&  = A(k_1, k_2,m_f) \epsilon[\lambda, \mu, \nu, k_2] s_1 
- A(k_2, k_1,m_f) \epsilon[\lambda, \mu , \nu, k_1] s_2 + A(k_1, k_2,m_f) \epsilon[\lambda, \nu, k_1, k_2] k_1^{\mu}  \nonumber\\
&&+ A(k_2, k_1,m_f) \epsilon[\lambda, \mu, k_2, k_1] k^{\nu}_{2} - B(k_1, k_2,m_f) \epsilon[\mu, \nu, k_1, k_2] k^{\lambda}
\eeqn
with $k - k_1 - k_2 = 0$ and $s_i = k_i^2 \,\, (i=1,2)$, and  
\beqn
A (k_1, k_2, m_f) &=& \frac{1}{\lambda} \left[ - \frac{1}{2}(s -s_1 + s_2) - \left( \frac{1}{2}(s + s_2) 
+ ( 6/{\lambda}) s s_1 s_2  \right) \Delta_{\#1} \right.   \nonumber\\
&& + s_2  \left[ \frac{1}{2} 
- ( 3/{\lambda}) s (s - s_1 - s_2) \right] \Delta_{\#2}   \nonumber\\
&&   \left. + \left[ s s_2 + (m_f^2 + (3/{\lambda})s s_1 s_2)(s 
- s_1 + s_2)   \right] C_{\#0} \right]
\label{Aeq}
\eeqn
\beqn
B(k_1, k_2, m_f) &=& \frac{1}{\lambda} \left[ \frac{1}{2} (s - s_1 - s_2) + s_1 \left[ \frac{1}{2} 
+ (3/{\lambda}) s_2 (s + s_1 - s_2)  \right]   \Delta_{\#1}   \right.   \nonumber\\
&&+ s_2 \left[ \frac{1}{2} 
+ (3/{\lambda}) s_1 (s - s_1 + s_2)  \right] \Delta_{\#2}     \nonumber\\
&&   \left. + \left[ s_1 s_2 
- (m_f^2 + (3/{\lambda}) s s_1 s_2)(s - s_1 - s_2)  \right] C_{\#0} \right]
\label{Beq}
\eeqn
with
\beq
\lambda = \lambda(s, s_1, s_2),  \nonumber\\
\eeq
being the usual Mandelstam function and
where the analytic expressions for $\Delta_{\#i}$ and $C_{\# 0}$ are given by
\beqn
\Delta_{\#i} &=& a_i \, \mbox{ln} \frac{a_i + 1}{a_i -1} - a_3 \, \mbox{ln} \frac{a_3 +1}{a_3 - 1} , \,\,\,\mbox{(i=1,2)}   \nonumber\\
C_{\#0} &=& \frac{1}{\sqrt{\lambda} } \sum^{3}_{i=1} \left[ \mbox{ Li}_{2} \left( \frac{b_i -1}{a_i + b_i} \right)
- \mbox{Li}_{2} \left( \frac{- b_i -1}{a_i - b_i} \right) + \mbox{Li}_{2} \left( \frac{- b_i +1}{a_i - b_i} \right) 
- \mbox{Li}_{2} \left( \frac{b_i + 1}{a_i + b_i} \right)  \right],     \nonumber\\
\eeqn
and
\beqn
&&t_i = - s_i - i \epsilon,     \qquad    a_i = \sqrt{1 + (2 m_f)^{2} /t_i}, \,\,\mbox{(i=1,2,3)},   \nonumber\\
&&\lambda = \lambda(t_1, t_2, t_3),  \qquad   b_1 = (t_1 - t_2 -t_3)/\sqrt{\lambda} \,\,\,\,\mbox{or cyclic}
\eeqn
For $m_f=0$ the two expressions above become 
\beqn
\Delta_{\#i} &=& \mbox{ln}(t_i/t_3), \,\,\,(i=1,2),  \nonumber\\
C_{\# 0} &=& (1/\sqrt{\lambda})  \Biggr[ 2  \Biggr( \zeta (2) - \mbox{Li}_{2}(x_1) - \mbox{Li}_{2}(x_2) 
+ \mbox{Li}_{2}  \left(  \frac{1}{x_3}  \right) \Biggr) + \mbox{ln} x_1 \mbox{ln}x_2 \Biggr]
\eeqn
with
\beqn 
x_i = \frac{(b_i + 1)}{(b_i - 1)}, \,\,\, (i = 1,2,3).
\eeqn
These can be inserted into (\ref{Aeq}) and (\ref{Beq}) together with $m_f=0$ to 
generate the corresponding ${\bf V}^{\lambda\mu\nu}(0)$ vertex needed for the computation of the massless contributions to the amplitude. 

With these notations we clearly have 
\beqa
\Delta^{\lambda\mu\nu} &=& V^{\lambda\mu\nu}(k_1,k_2,m_f) \nonumber \\
\Delta^{\lambda\mu\nu}(0) &=& {\bf V}^{\lambda\mu\nu}(k_1,k_2) \nonumber \\
\Delta^{\lambda\mu\nu}(m_f) &=& V^{\lambda\mu\nu}(k_1,k_2,m_f)
 - {\bf V}^{\lambda\mu\nu}(k_1,k_2). \nonumber 
\eeqa
\subsection{Extension to $Z\to \gamma^* \gamma$}
To isolate the contribution to the decay on the resonance, we keep one of the two photons off-shell 
(resolved). We choose $s_1=0$, and $s_2$ virtual. We denote by $\Gamma^{\lambda\mu\nu}$ the 
corresponding vertex in this special kinematical configuration. The $Z$ boson is on-shell.  
In this case at one-loop the result simplifies considerably \cite{Hagiwara:1991xy}
\beqn
\Gamma_{\lambda \mu \nu} &=& F_{2} (s_2 \epsilon[\lambda, \mu, \nu, k_1] + k^{\nu}_{2} \epsilon[\lambda, \mu, k_1, k_2] ),
\label{ampiezza_Zgg}
\eeqn
with $F_2$ expressed as a Feynman parametric integral 
\beqn
F_{2} =  \frac{1}{{2 \pi^{2}}}\int^{1}_{0} dz_{1} dz_{2} dz_{3} \delta(1 - z_{1} - z_{2} - z_{3}) 
\frac{- z_{2} z_{3}}{ m_f^{2} - z_{2}z_{3} s_2 - z_{1} z_{3} M_Z^2}.
\eeqn
Setting $F_2\equiv - F(z,r_f)$
where $f(z,r)$ is a dimensionless function of
\beqn
z= s_2/M^{2}_{Z}, \qquad r_f=m^{2}_{Z}/4m_{f}^2,
\eeqn
and for vanishing $m_f$ ($r_{f} = M^{2}_{Z}/4m^{2}_{f} \rightarrow \infty$), 
the corresponding massless contribution is expressed as $F(z,\infty)$
with, in general
\beqn
F(z,r_f) = \frac{1}{4(1 - z)^{2}} (I(r_f z,r_f) - I(r_f,r_f) + 1 - z ),
\eeqn
where
\beqn
I(x,r_f) &=& 2 \sqrt{\frac{x - 1}{x^{-}}} \mbox{ln}(\sqrt{-x} + \sqrt{1-x})
 - \frac{1}{r_f} (\mbox{ln}(\sqrt{-x} + \sqrt{1 -x}))^{2}  \,\,\,\,\,\,\,\,  \mbox{for \,x $<$ 0}  \nonumber\\
 &=& 2 \sqrt{\frac{1-x}{x}} \sin^{-1} \sqrt{x} + \frac{1}{r_f} (\sin^{-1} \sqrt{x})^{2} \,\,\,\,\,\,\,\mbox{for 0$<$\,x $<$ 1}  \nonumber\\
&=& 2 \sqrt{\frac{x-1}{x}} \left( \ln(\sqrt{x} + \sqrt{x-1}) - \frac{i \pi}{2}  \right) 
- \frac{1}{r_f} \left( \ln(\sqrt{x} + \sqrt{x-1} ) - \frac{i \pi}{2}  \right)^{2},    \nonumber\\
&&  \,\,  \mbox{for \,x $>$ 1}.
\eeqn
The $m_f=0$ contribution is obtained in the $r_f \rightarrow + \infty$ limit,
\beqn
F(z,\infty) &=& \frac{1}{4(1 - z)^{2}} (\ln z + 1 - z)  \,\,\,\,\,\,\,\,\,\,\,\,\,\,\,\,\,\,  \mbox{for \, z $>$ 0},  \nonumber\\
&=& \frac{1}{4(1 - z)^{2}} (\ln |z| + i \pi + 1 - z)  \,\,\,\,\,\,\,\mbox{for \, z $<$ 0}.
\eeqn
In these notations, the infinite fermion mass limit 
($m_{f} \rightarrow \infty$ or $r \rightarrow 0$), 
gives $F(z,0)= 0$ and we find 
\beqa
\Delta^{\lambda\mu\nu} &=& \Gamma^{\lambda\mu\nu}=F(z,r_f) \nonumber \\
\Delta^{\lambda\mu\nu}(0) &=& { \Gamma}^{\lambda\mu\nu}(0)=F(z,\infty) \nonumber \\
\Delta^{\lambda\mu\nu}(m_f) &=& \Gamma^{\lambda\mu\nu}- \Gamma^{\lambda\mu\nu}(m_f)=F(z,r_f) 
-F(z,\infty),
\eeqa
which can be used for a numerical evaluation. 
The decay rate for the process is given by 
\beq
\Gamma(Z\to \gamma^*\gamma)=
\frac{1}{4 M_Z}\int d^4 k_1 d^4 k_2 \delta(k_1^2)\,\delta(k_2^2 - Q_*^2) 
|{\cal M}_{Z\to \gamma\gamma^*}|^2 \, (2 \pi)^4 \delta( k - k_1 - k_2),
\eeq
where 
\beqa
|{\cal M}_{Z\to \gamma\gamma^*}|^2 &=& - A^{\lambda \mu\nu}_{Z\to \gamma\gamma^*}
\Pi_Z^{\lambda \lambda'}A^{\lambda' \mu\nu'}_{Z\to \gamma\gamma^*}\Pi_{Q^*}^{\nu \nu'}
\nonumber \\
\Pi_Z^{\lambda \lambda'} &=& - g^{\lambda \lambda'} + \frac{k^{\lambda}k^{\lambda'}}{M_Z^2} \nonumber \\
\Pi_{Q^*}^{\nu \nu' } &=& - g^{\lambda \lambda'} + \frac{k^{\lambda}k^{\lambda'}}{Q_*^2}. 
\eeqa
We have indicated with $Q_*$ the virtuality of the photon. A complete evaluation 
of this expression, to be of practical interest, would need the fragmentation functions of the photon 
(see \cite{Coriano:1996us} for an example).  We briefly summarize the main points 
involved in the analysis of this and similar processes at the LHC, where the decay rate 
is folded with the (NLO/NNLO) contribution from the initial state using QCD factorization.

Probably one of the best way to search for neutral current interactions in hadronic collisions at the LHC is in 
lepton pair production via the Drell-Yan mechanism (all the details of the numerical analysis will be presented in Chap.~\ref{chap:LHC}). 
 QCD corrections are known for this process up to O($\alpha_s^2$) (next-to-next-to-leading order, NNLO), 
which can be folded with the 
NNLO evolution of the parton distributions to provide accurate determinations of the hadronic $pp$ cross 
sections at the 4 $\%$ level of accuracy \cite{Cafarella:2007tj}. The same computation for Drell-Yan can be used 
to analyze the $pp\to Z\to \gamma \gamma^*$ process 
since the $W_V$ (hadronic) part of the process is universal, with $W_V $ defined below. An appropriate (and very useful) way to analyze this process would be to perform this study defining the invariant mass distribution

\beq
\frac{d \sigma}{d Q^2}= \tau \sigma_{Z\to \gamma^*\gamma}
(Q^2,M_V^2)\,W_V(\tau, Q^2) 
\eeq
where $\tau=Q^2/S$, 
which is separated into a point-like contribution $\sigma_{Z\to \gamma \gamma^*}$
\beq
\sigma_V(Q^2,M_V^2)=\frac{\pi \alpha}{4 M_Z \sin\theta_W^2 \cos\theta_W^2 N_c}
\frac{\Gamma(Z\to \gamma\gamma^*)}{(Q^2 - M_Z^2)^2 + M_Z^2 \Gamma_Z^2}.
\eeq
and a hadronic structure functions $W_Z$.
This is defined via the integral over parton distributions and coefficient functions 
$\Delta_{ij}$ 
\beq
W_Z(Q^2,M_Z^2)=\int_0^1 d x_1\, \int_0^1 d x_2 \int_0^1 d x \,  \delta( \tau - x x_1 x_2) 
P D^V_{ij}(x_1,x_2,\mu_f^2) \Delta_{ij}(x, Q^2,\mu_f^2)
\eeq
where $\mu_f$ is the factorization scale. The choice $\mu_f=Q$, with Q the invariant mass of the 
$\gamma\gamma^*$ pair , removes the $\log(Q/M)$ for the computation of the coefficient functions, which is, anyhow, 
arbitrary.
The non-singlet coefficient functions are given by 
\beqa
\Delta^{(0)}_{q \bar{q}} &=& \delta(1-x) \nonumber \\
\Delta^{(1)}_{q \bar{q}} &=& \frac{\alpha_S(M_V^2)}{4 \pi}C_F\left[ \delta(1-x)( 8 \zeta(2) - 16) + 
16 \left(\frac{\log(1-x)}{1-x}\right)_+ \right.\nonumber \\
&&\left. \qquad \qquad \quad \qquad -8 (1 + x)\log(1 -x) - 4\frac{1 + x^2}{1 -x} \log x\right]
\eeqa
with $C_F=(N_c^2-1)/(2 N_c)$ and the ``+'' distribution is defined by
\beq
 \left(\frac{\log(1-x)}{1-x}\right)_+= \theta( 1-x) \frac{\log(1-x)}{1-x} -
\delta(1-x)\int_0^{1-\delta} dx \, \frac{\log(1-x)}{1-x},
\eeq
while at NLO appears also a $qg$ sector 
\beq
\Delta^{(1)}_{q g} = \frac{\alpha_S(M_V^2)}{4 \pi}T_F\left[ 2(1 + 2 x^2 -2 x)\log 
\left( \frac{(1 -x)^2}{x}\right) + 1 - 7 x^2 + 6 x\right].
\eeq
Other sectors do not appear at this order.
Explicitly one gets 
\beqa
W_Z(Q^2,M_Z^2) &=&\sum_i \int_0^1 d x_1\, \int_0^1 d x_2 
\int_0^1 d x \delta( \tau - x x_1 x_2) \nonumber \\
 && \times \Big\{ \bigg(q_i(x_1,\mu_f^2)\bar{q_i}(x_2,\mu_f^2) +  \bar{q_i}(x_1,\mu_f^2)
{q_i}(x_2,\mu_f^2)\bigg) \Delta_{q\bar{q}}(x, Q^2,\mu_f^2) \nonumber \\
&&\qquad  + \bigg(q_i(x_1,\mu_f^2){g}(x_2,\mu_f^2) +  
{q_i}(x_2,\mu_f^2){g}(x_1,\mu_f^2)\bigg) \Delta_{q{g}}(x, Q^2,\mu_f^2)\Big\}
\eeqa
where the sum is over the quark flavours. The identification of the generalized mechanism of anomaly cancellation requires 
that this description be extended to NNLO. It involves a slight modification of the NNLO hard scatterings. An explicit computation will be performed in Chap.~\ref{chap:LHC} together with the numerical results. 
\section{Conclusions} 
We have presented a study of a model inspired by the structure encountered
in a typical string theory derivation of the Standard Model. In particular we have focused our investigation 
on the characterization of the effective action and worked out its expression 
in the context of an extension containing one additional anomalous $U(1)$. This analysis specializes and, at the same time, extends a previous study of models belonging to this class.
The results that we have presented are generic for models where the St\"{u}ckelberg and the Higgs mechanism are combined and where an effective Abelian anomalous interaction is present. 
Our analysis has then turned toward the study of simple processes mediated by neutral current exchanges, and we have focused, specifically, on one of them, the one involving the $Z \gamma \gamma$ vertex. In particular our findings clearly show that 
new massless contributions are presented at one-loop level when anomalous generators are involved in the fermionic triangle diagrams and the interplay between massless and massive fermion effects is modified respect to the SM case.
The typical processes considered in our analysis deserve a special attention, 
given the forthcoming experiments at the LHC, since they may provide a way to determine whether anomaly effects 
are present in some specific reactions. Other similar processes, involving the entire neutral sector should be considered, though the two-photon signal 
is probably the most interesting one phenomenologically. Given the high statistical precision 
($.05 \%$ and below on the $Z$ peak, for 10 $fb^{-1}$ of integrated luminosity) which can be easily obtained at the LHC, there are realistic 
chances to prove or disprove theories of these types.  
\section{Appendix. A Summary on the single anomalous $U(1)$ Model. \label{sec:Ochi}}
We summarize here some results concerning the model with a 
single anomalous $U(1)$ discussed in the main sections. These results 
specialize and simplify the general discussion of \cite{Coriano:2005js} to which we refer 
for further details. The hypercharge values used for our analysis are
\hskip 2cm
\begin{center}
\begin{tabular}{|c|c|c|c|c|c|c|}
\hline
$ f $ & $Q_{L}$ &  $ u_{R} $ &  $ d_{R} $ & $ L $ & $e_R$ & $\nu_R$ \\
\hline \hline
$q^{}_Y$  &  $1/6$  & $2/3$  &  $-1/3$ & $-1/2$ & $-1$ & $0$\\ \hline
\end{tabular}
\end{center}
and general $U(1)^{}_{B}$ charge assignments   \\
\hskip 2cm
\begin{center}
\begin{tabular}{|c|c|c|c|c|c|c|}
\hline
$ f $ & $Q_{L}$ &  $ u_{R} $ &  $ d_{R} $ & $ L $ & $e_R$ & $\nu_R$ \\
\hline \hline
$q^{}_B$  &  $q^{(Q_L)}_{B}$  & $q^{(u_R)}_{B}$  &  $q^{(d_R)}_{B}$ & $q^{(L)}_{B}$ & $q^{(e_R)}_{B}$ & $q^{(\nu_R)}_{B}$\\ \hline
\end{tabular}
\end{center}
The covariant derivatives act on the fermions $f_L,f_R$ as
\begin{eqnarray}
&& {\cal D}_{\mu}f^{}_{L} = \left(\partial_{\mu} +
i {\bf A}_{\mu} + i q_{l}^{(f_L)} g_{l}A_{l,\mu} \right)f^{}_{L}\nonumber \\
&& {\cal D}_{\mu}f^{}_{R} = \left(\partial_{\mu} +
i {\bf A}_{\mu} + i q_{l}^{(f_R)} g_{l}A_{l,\mu} \right)f^{}_{R}
\end{eqnarray}
with $l=Y,B$ Abelian index, 
where ${\bf A}_{\mu}$ is a non-Abelian Lie algebra element and write the lepton doublet as
\bea
L_{i} = \pmatrix { \nu^{}_{Li} \cr e^{}_{Li}}.
\eea
We will also use standard notations for the $SU(2)_W$ and $SU(3)_C$ gauge bosons 
\beqa
W_{\mu} &=& \frac{\sigma_i}{2} W^{i}_{\mu} = \tau_i W^{i}_{\mu}, \;\;\; \; \;  \mbox{with} \; \; \;\;   i = 1,2,3   \\
G_{\mu} &=& \frac{\lambda_a}{2} G^{\,a}_{\mu}  = T_a G^{\,a}_{\mu} \;\;\; \; \;  \mbox{with} \; \; \;\;   a = 1, 2,...,8
\eeqa
with the normalizations 
\beqn
Tr[ \tau^i \tau^j] = \frac{1}{2} \delta_{ij},    \qquad   Tr[ T^a T^b] = \frac{1}{2} \delta_{ab}.
\label{normalization}
\eeqn
The interaction Lagrangian for the leptons becomes 
\beqa
 {\cal L}_{int}^{lep} &= &   \pmatrix{ \overline{{\nu}}_{Li} & \overline{e}_{Li}} \gamma^{\mu}
\left[ -  g^{}_2 \frac{\tau^a}{2} W_{\mu}^a  - g^{}_Y \, q^{(L)}_Y  A^Y_{\mu} 
- g^{}_B  q^{(L)}_B  B_{\mu} \right]  \pmatrix{\nu_{Li} \cr e_{Li}}  + \nonumber\\
&& +  \; \overline{e}_{Ri} \; \gamma^{\mu} \left[
- g^{}_Y  q^{(e_{R})}_Y    A^Y_{\mu} -  g^{}_B  q^{(e_{R})}_B B_{\mu} \right]e_{Ri}\nonumber\\
&& +  \; \overline{\nu}_{Ri} \; {\gamma}^{\mu}\left[
- g^{}_Y  q^{({\nu}_{R})}_Y  A^Y_{\mu} - g^{}_B  q^{(\nu_{R})}_B   B_{\mu}\right]{\nu}_{Ri}.
\eeqa
As usual we define the left-handed and right-handed currents 
\beqn
J^{L}_{\mu} = \frac{1}{2}(J_{\mu} - J^{5}_{\mu}),  \qquad  J^{R}_{\mu} = \frac{1}{2}(J_{\mu} + J^{5}_{\mu}), 
\qquad J_{\mu} = J^{R}_{\mu} + J^{L}_{\mu}, \qquad  J^{5}_{\mu} = J^{R}_{\mu} - J^{L}_{\mu}.
\eeqn
Writing the quark doublet as
\bea
Q^{}_{Li} = \pmatrix {u^{}_{Li}\cr d^{}_{Li}},
\eea
we obtain the interaction Lagrangian
\beqa
 {\cal L}_{int}^{quarks} &= &   \pmatrix{ \overline{ u }^{}_{Li} & \overline{ d }^{}_{Li}} \gamma^{\mu}
\left[ - g^{}_3 \frac{\lambda^a}{2} G^{a}_{\mu}  -  g^{}_2 \frac{ \tau^i }{ 2 } W_{ \mu }^i  - g^{}_Y \, q^{(Q_L)}_Y  A^Y_{\mu} 
- g^{}_B  q^{(Q_L)}_B  B_{\mu} \right]  \pmatrix{ u^{}_{Li} \cr d^{}_{Li}}  + \nonumber\\
&& +  \; \overline{u}_{Ri} \; \gamma^{\mu} \left[
- g^{}_Y  q^{(u_{R})}_Y    A^Y_{\mu} - g^{}_B  q^{(u_{R})}_B B_{\mu} \right]  u^{}_{Ri}  \nonumber\\
&& +  \; \overline{d}^{}_{Ri} \; {\gamma}^{\mu}\left[
- g^{}_Y  q^{({d}_{R})}_Y  A^Y_{\mu} - g^{}_B  q^{(d_{R})}_B   B_{\mu}\right] d^{}_{Ri}.
\eeqa
As we have already mentioned in the introduction, we work with a two-Higgs doublet model, and therefore 
we parameterize the Higgs fields in terms of eight real degrees of freedom as
\beqa
H_u=\left(\begin{array}{c}
H_u^+\\
H_u^0
\end{array}\right) \qquad H_d = \left(\begin{array}{c}
H_d^+\\
H_d^0 \end{array}\right)
\eeqa
where $H_u^+$, $H_d^+$ and $H_u^0$, $H_d^0$ are complex fields. Specifically
\beq
H_u^+ =  \frac{H_{uR}^+ + i H_{uI}^+}{\sqrt{2}} ,\qquad
H_d^- =  \frac{H_{dR}^- + iH_{dI}^-}{\sqrt{2}} , \qquad
H_u^- = H_u^{+ *}, \qquad
H_d^+ = H_d^{- *}.
\eeq
Expanding around the vacuum we get for the neutral components
\beq
H_u^0 =  v_u + \frac{H_{uR}^0 + i H_{uI}^0}{\sqrt{2}} , \qquad
H_d^0 =  v_d + \frac{H_{dR}^0 + iH_{dI}^0}{\sqrt{2}}. \label{Higgsneut}
\eeq
The Weinberg angle is defined via
$\cos\theta_W= g_2/g, \sin\theta_W= g_Y/g$, with the following relation between the couplings
\bea
 g^2= g_Y^2 + g_2^2.
\eea
We also define $\cos \beta=v_d/v$, $ \sin \beta=v_u/v$, related by
\bea
v^2=v_d^2 + v_u^2.
\eea
\subsection{The neutral gauge sector \label{app:gauge}}
The mass-matrix of the neutral gauge bosons is given by
\bea
{\mathcal L}_{mass} =  \left( W_3 \,\,\,\, A^Y \,\,\,\, B  \right){\bf M}^2    \left(\begin{array}{c}
W_3\\
A^Y\\
B \\
\end{array}   \right),
\eea
where explicitly  
\bea
{\bf M}^2 = {1\over 4} \pmatrix{
{g^{}_2}^{2} v^2 & - {g^{}_2} \, {g^{}_Y} v^2 &  - {g^{}_2} \,  x^{}_B \cr
 - {g^{}_2} \,{g^{}_Y} v^2 &  {g^{}_Y}^{2} v^2 & {g^{}_Y}  x^{}_B \cr
 -{g^{}_2} \, x^{}_B &{g^{}_Y}  x^{}_B  & 2 M_1^2 + N^{}_{BB}}
\label{mmassmatrix}
\eea
with
\bea
N^{}_{BB}=  \left( q_u^{B\,2} \,{v^{\,2}_u} + q_d^{B\,2} \,{v^{\,2}_d} \right)\, g_B^{\,2},
\eea
\bea
x^{}_B=  \left(q_u^B {v^{\,2}_u} + q_d^B {v^{\,2}_d}  \right)\, g^{}_B.
\eea
The orthonormalized mass squared eigenstates corresponding to this matrix are given by
\beqa
\left(
\begin{array}{c}
 O_{11}^{A} \\
 O_{12}^A \\
 O_{13}^A
\end{array}
\right)
=\left(
\begin{array}{c}
  \frac{ g_Y }{ \sqrt{ g_2^2 + g_Y^2 }} \\
  \frac{ g_2}{\sqrt{  g_2^2 + g_Y^2 }} \\
  0 
\end{array}
\right),
\eeqa
\beqa
\left(
\begin{array}{c}
 O_{21}^{A} \\
 O_{22}^A \\
 O_{23}^A
\end{array}
\right)
=   \left(
\begin{array}{c}
   \frac{ g_2 \left( 2 M_1^2 - g^2 v^2 + N_{BB} 
+ \sqrt{  \left(  2 M_1^2 - g^2
   v^2 + N_{BB} \right)^2 + 4 g^2
   x_{B}^2}   \right)} 
{   g^2 x_B   \sqrt{  4  +  \frac{g^2}{g^4 x_B^2} 
\left( 2 M_{1}^2 - g^2 v^2 + N_{BB} 
+ \sqrt{ \left( 2 M_{1}^2 - g^2 v^2 + N_{BB} \right)^2 + 4 g^2 x_{B}^2 } \right)^2 } }   \\
    - \frac{ g_Y  \left( 2 M_1^2 - g^2 v^2 + N_{BB} 
+ \sqrt{  \left(  2 M_1^2 - g^2
   v^2 + N_{BB} \right)^2 + 4 g^2
   x_{B}^2}   \right)} 
{   g^2 x_B   \sqrt{  4  +  \frac{g^2}{g^4 x_B^2} 
\left( 2 M_{1}^2 - g^2 v^2 + N_{BB} 
+ \sqrt{ \left( 2 M_{1}^2 - g^2 v^2 + N_{BB} \right)^2 + 4 g^2 x_{B}^2 } \right)^2 } }    \\
  \frac{2}{  \sqrt{  4  +  \frac{g^2}{g^4 x_B^2} 
\left( 2 M_{1}^2 - g^2 v^2 + N_{BB} 
+ \sqrt{ \left( 2 M_{1}^2 - g^2 v^2 + N_{BB} \right)^2 + 4 g^2 x_{B}^2 } \right)^2 } }  
\end{array}
\right).
\eeqa
One can see that these results reproduce the analogous relations of the SM  in the limit of very large $M^{}_1$
\bea
\lim_{M_{1}\to\infty} O_{21}^{A} = \frac{g_2}{g}, \qquad  \lim_{M_{1}\to\infty} O_{22}^{A} = - \frac{g_Y}{g},  \nonumber\\
 O_{23}^{A}  \simeq \frac{g}{2} \frac{x_B}{M_{1}^2} \equiv  \frac{g}{2} \epsilon_1  \,\,\,
 \mbox{ so that} \,\,\, \lim_{M_{1}\to\infty} O_{23}^{A} = 0.   \nonumber
\eea
Similarly, for the other matrix elements of the rotation matrix $O^A $ we 
obtain
\beqa
\left(
\begin{array}{c}
 O_{31}^{A} \\
 O_{32}^A \\
 O_{33}^A
\end{array}
\right)
= \left(
\begin{array}{c}
 -  \frac{ g_2 \left( - 2 M_1^2 + g^2 v^2 - N_{BB} 
+ \sqrt{  \left(  2 M_1^2 - g^2
   v^2 + N_{BB} \right)^2 + 4 g^2
   x_{B}^2}   \right)} 
{   g^2 x_B   \sqrt{  4  +  \frac{g^2}{g^4 x_B^2} 
\left(- 2 M_{1}^2 + g^2 v^2 - N_{BB} 
+ \sqrt{ \left( 2 M_{1}^2 - g^2 v^2 + N_{BB} \right)^2 + 4 g^2 x_{B}^2 } \right)^2 } }   \\
     \frac{ g_Y  \left( - 2 M_1^2 + g^2 v^2 - N_{BB} 
+ \sqrt{  \left(  2 M_1^2 - g^2
   v^2 + N_{BB} \right)^2 + 4 g^2
   x_{B}^2}   \right)} 
{   g^2 x_B   \sqrt{  4  +  \frac{g^2}{g^4 x_B^2} 
\left( -2 M_{1}^2 + g^2 v^2 - N_{BB} 
+ \sqrt{ \left( 2 M_{1}^2 - g^2 v^2 + N_{BB} \right)^2 + 4 g^2 x_{B}^2 } \right)^2 } }    \\
  \frac{2}{  \sqrt{  4  +  \frac{g^2}{ g^4 x_B^2 } 
\left(- 2 M_{1}^2 + g^2 v^2 - N_{BB} 
+ \sqrt{ \left( 2 M_{1}^2 - g^2 v^2 + N_{BB} \right)^2 + 4 g^2 x_{B}^2 } \right)^2 } }  
\end{array}
\right),
\eeqa
whose asymptotic behavior is described by the limits 
\bea
O_{31}^{A}  \simeq -\frac{g^{}_2}{2} \frac{x^{}_B}{M^2_1} \equiv -\frac{g^{}_2}{2} {\epsilon_1} , 
\qquad  O_{32}^{A} \simeq \frac{g^{}_Y}{2} \frac{x^{}_B}{M^2_1} \equiv  \frac{g^{}_Y}{2}  \epsilon_1,
 \qquad   O_{33}^{A} \simeq  1,  
\eea
\bea
\lim_{M_{1}\to\infty} O_{31}^{A} = 0, \qquad  \lim_{M_{1}\to\infty} O_{32}^{A} = 0, \qquad   \lim_{M_{1}\to\infty} O_{33}^{A} = 1.
\eea
These mass-squared eigenstates correspond to one zero mass eigenvalue 
for the photon $A^{}_{\gamma}$, and two non-zero mass 
eigenvalues for the $Z$ and for the $Z^\prime$ vector bosons, corresponding to the mass values
\bea
m_{Z}^2 &=&  \frac{1}{4} \left( 2 M_1^2 + g^2 v^2 + N^{}_{BB} 
- \sqrt{\left(2  M_{1}^2 - g^2 v^2 + N^{}_{BB} \right)^2 + 4
   g^2 x_{B}^2} \right)    \nonumber\\
&\simeq&     \frac{g^2 v^2}{2} - \frac{1}{M_{1}^2} \frac{g^2 x_{B}^{2}}{4}
 + \frac{1}{M_{1}^4}\frac{g^2 x_{B}^2}{8 } (N^{}_{BB} - g^2 v^2) , 
\eea
\bea 
 m_{{Z}^\prime}^2 &=&   \frac{1}{4} \left( 2 M_1^2 + g^2 v^2 + N^{}_{BB} 
+ \sqrt{\left(2  M_{1}^2 - g^2 v^2 + N^{}_{BB} \right)^2 + 4   g^2 x_{B}^2} \right)   \nonumber\\
&\simeq&    M^{2}_{1} +  \frac{N^{}_{BB}}{2} . 
\label{ZZpmass} 
\eea
The mass of the $Z$ gauge boson gets corrected by terms 
of the order $v^{2}/M^{}_1$, converging to the SM value as $M_1\to \infty$, 
with $M^{}_1$ the St\"{u}ckelberg mass of the $B$ gauge boson, 
the mass of the $Z^\prime$ gauge boson can grow large with $M^{}_1$. 
The physical gauge fields can be obtained from the rotation matrix $O^A$ 
\ba
\pmatrix{A_\g \cr Z \cr {{Z^\prime}}} =
O^A\, \pmatrix{W_3 \cr A^Y \cr B}  \label{OA}
\ea
which can be approximated at the first order as
\bea
O^A  \simeq  \pmatrix{
\frac{g^{}_Y}{g}           &     \frac{g^{}_2}{g}         &      0   \cr
\frac{g^{}_2}{g} + O(\epsilon_1^2)          &     -\frac{g^{}_Y}{g} + O(\epsilon_1^2) &      \frac{g}{2} \epsilon_1    \cr
-\frac{g^{}_2}{2}\epsilon_1     &     \frac{g^{}_Y}{2}\epsilon_1  &   1 + O(\epsilon_1^2) }   .  
\label{matrixO}
\eea
The mass squared matrix (\ref{mmassmatrix}) can be diagonalized as
\bea
 \Big( A_\g \,\,\,\, Z \,\,\,\, Z^\prime   \Big) \, O^A   {\bf M}^2  (O^A)^T  \left(\begin{array}{c}
A_\g\\
Z\\
Z^\prime \\
\end{array}   \right)  =   \Big( A_{\g} \,\,\,\,  Z \,\,\,\,  Z^\prime  \Big)   \pmatrix{
0           &     0         &  0   \cr
0           &    m_{Z}^2    & 0    \cr
0           &     0         &   m_{{Z}\prime}^2 }      \left(\begin{array}{c}
A_{\g}  \\
Z   \\
Z^\prime   \\
\end{array}   \right).
\eea
It is straightforward to verify that the rotation matrix $O^A$ satisfies the proper orthogonality relation
\bea
O^A (O^A)^T = 1. 
\eea
\subsection{Rotation matrix $O^\chi$ on the axi-Higgs}
This matrix is needed in order to rotate into the mass eigenstates of the $CP$-odd sector, relating the axion $\chi$ and the two neutral Goldstones of this sector to the St\"uckelberg field $b$ and 
the $CP$-odd phases of the two Higgs doublets
\beq
\pmatrix{{\rm Im}H_u^0\cr {\rm Im}H_d^0\cr  b \cr }=\; O^{\chi} \;
\pmatrix{\chi \cr G_1^{\,0} \cr G_2^{\,0} \cr }.
\label{rotunit}
\eeq
We refer to \cite{Coriano:2005js} for a more detailed discussion of 
the scalar sector of the model, 
where, in the presence of explicit phases ($PQ$-breaking terms), 
the mass of the axion becomes massive from the massless case. 
The $PQ$ symmetric contribution is given by
\bea
V_{PQ}(H^{}_u, H^{}_d) = \sum_{a=u,d} \Bigl(  \mu_a^2  H_a^{\dagger} H_a + \lambda_{aa} (H_a^{\dagger} H_a)^2\Bigr)
-2\lambda_{ud}(H_u^{\dagger} H_u)(H_d^{\dagger} H_d)+2{\lambda^\prime_{ud}} |H_u^T\tau_2H_d|^2,
\eea
while the $PQ$-breaking terms are 
\begin{eqnarray}
V_{\ds{P} \ds{Q}}(H^{}_u, H^{}_d, b) &=&  b_{1} \, \left( H_u^{\dagger} H_d \, e^{-i  (q_u^B-q_d^B) \frac{b}{M_1}}  \right)
+ \lambda^{}_1 \left( H_u^{\dagger}H_d \,e^{-i  (q_u^B-q_d^B) \frac{b}{M_1}} \right)^2  \nonumber\\
&&+ \, \lambda^{}_2 \left( H_u^{\dagger}H_u \right) \left( H_u^{\dagger}H_d \,e^{-i  (q_u^B-q_d^B) \frac{b}{M_1}} \right)
+ \lambda^{}_3 \left( H_d^{\dagger}H_d \right) \left( H_u^{\dagger}H_d \,e^{-i (q_u^B-q_d^B)  \frac{b}{M_1}} \right) + c.c.
\nonumber\\ 
\label{PQbreak}
\end{eqnarray}
where $b^{}_{1}$ has mass squared dimension, while $\lambda^{}_{1}$, $\lambda^{}_{2}$, $\lambda^{}_{3}$ are dimensionless. 
From the scalar potential one can extract the mass 
eigenvalues of the model for the scalar sector. The mass matrix has two zero eigenvalues and one non-zero eigenvalue that corresponds to a physical
axion field, $ \chi$, with mass
\bea
m_{\chi}^2 =
-\frac{1}{2} \, c^{}_{ \chi} \, v^2  \left[ 1 + \left(  \frac{q_u^B-q_d^B}{M_1}\,
 \frac{v \, \sin{2\b} }{2} \right)^2 \right] = 
-\frac{1}{2} \, c^{}_{ \chi} \, v^2  \left[ 1 +  \frac{ ( q_u^B - q_d^B )^{2}}{M^{2}_{1}} \,
\frac{v^{2}_{u} v^{2}_{d} }{v^2}  \right],
\label{axionmass}
\eea
where
\bea
c^{}_{\chi} = 4 \left( 4 \lambda_1 +  \lambda_3 \cot \beta + \frac{ b_{1} }{ v^2 } \frac{ 2 }{ \sin 2\beta } 
+  \lambda_2  \tan\beta  \right),
\eea
and $v^{}_d = v \cos\beta, v^{}_u = v \sin\beta$ together with
\bea
 \cot\beta=\frac{\cos\beta}{\sin\beta} = \frac{v_{d}}{v_{u}}, \qquad
\tan\beta =\frac{\sin\beta}{\cos\beta} = \frac{v_{u}}{v_{d}}.
\eea
The mass of this state is positive if $c_{\chi} < 0$. Notice that the mass of the axi-Higgs is the result of two effects: 
the presence of the Higgs v.e.v.'s and the presence of a $PQ$-breaking 
potential whose parameters can be small enough to drive the mass of 
this particle to be very light. We refer to Sec.~\ref{sec:massaxion} for a simple 
illustration of this effect in an Abelian model. The rotation matrix $O^\chi$ is shown below.

Introducing $N$ given by
\bea
N = \frac{1}{ \sqrt{ 1+ \frac{  ( q_u^B - q_d^B )^2 }{ M^{\,2}_1 }  \frac{ v_d^2 v_u^2 }{ v^2 } } } 
  = \frac{1}{ \sqrt{ 1+  \frac{ ( q_u^B - q_d^B )^2 }{ M^{\,2}_1 }  \frac{ v^2 \sin^2{2\beta} }{4}  }}   
 \label{normcoeff}
\eea
and defining
\bea
Q_1 = - \frac{ ( q_u^B - q_d^B )}{M_1} v_u  = - \frac{  (q_u^B-q_d^B)}{M_1} v \sin{\b},
\eea
\bea
N_1 = \frac{1}{\sqrt{ 1 + Q_1^2 }},
\eea
$O^{\chi}$ is the following matrix
\beqa
O^{\chi} = \pmatrix{
-N\cos{\b} & \sin{\b} & {\overline N}_1 {\overline Q}_1 \cos{\b} \cr
 N\sin{\b} & \cos{\b} & -{\overline N}_1 {\overline Q}_1 \sin{\b} \cr
NQ_1 \cos{\b}& 0 & {\overline N}_1 },\nonumber\\
\eeqa
where we defined
\beq
{\overline Q}_1 = Q_1 \cos{\b}
\eeq
and
\beqa
{\overline N}_1 &=& \frac{1}{\sqrt{1+{\overline Q}_1^2}} = \frac{1}{\sqrt{ 1 +  Q_1^2 \cos^2\beta}}   \nonumber\\
&=&  \frac{1}{\sqrt{ 1 +  \frac{ (q_u^B  - q_d^B)^2 }{ M_1^{\,2} } \, v^2 \sin^2\beta \cos^2\beta}} 
=   \frac{1}{ \sqrt{ 1 + \frac{  (q_u^B  - q_d^B)^2 }{ M_1^{\,2} } \frac{ v_u^2 v_d^2}{v^2} }}.
\eeqa
One can see from (\ref{normcoeff}) that  ${\overline N}_1 = N$, and the 
explicit 
elements of the 3-by-3 rotation matrix $O^{\chi}$ can be written as
\bea
\left( O^{\chi}  \right)_{11} &=&  - \frac{1}{  \frac{ -  ( q_u^B - q_d^B ) }{M_1} v_u
   \sqrt{ \frac{M_1^{\,2} }{  ( q_u^B - q_d^B )^2 } \frac{ v^2 }{ v_u^2 v_d^2 } + 1 } }   \nonumber\\
   &=&     - \frac{1}{ v_u \, \frac{ v  }{v_u v_d}   } \, N      = -N\cos{\b}    \label{Ochi11} \\
\left( O^{\chi}   \right)_{21}&=&    \frac{1}{  \frac{ - ( q_u^B - q_d^B ) }{M_1} v^{}_d
   \sqrt{ \frac{M_1^{\,2} }{  ( q_u^B - q_d^B )^2 } \frac{ v^2 }{ v_u^2 v_d^2 }+1 } }    \nonumber\\
 &=&     \frac{1}{ v_d \, \frac{ v }{v_u v_d}   } N            =  N\sin{\b}   \label{Ochi21}   \\
\left( O^{\chi}  \right)_{31}  &=&  \frac{1}{ \sqrt{ \frac{ M_1^{\,2} }{  ( q_u^B - q_d^B)^2} 
\frac{ v^2 }{ v_u^2 v_d^2 } + 1 } }          \nonumber\\  
  &=&     \frac{1}{  \frac{M_1}{-  (q_u^B - q_d^B) \, v^{}_u} \,\, v^{}_u  
\sqrt{ \frac{  (q_u^B - q_d^B)^2}{M_1^{\,2} } + 
          \frac{ v^2 }{v_u^2 v_d^2} }   }     = NQ_1 \cos{\b}  \\
\nonumber\\
\left( O^{\chi} \right)_{12}&=&  \frac{v^{}_u}{\sqrt {v^{\,2}_u + v^{\,2}_d} }   = \sin{\b}     \\
\left( O^{\chi} \right)_{22}&=&  \frac{v^{}_d}{\sqrt {v^{\,2}_u + v^{\,2}_d} }    =   \cos{\b}  \\
\left( O^{\chi} \right)_{32}&=&    0         \label{coeffic1}    \\
\nonumber\\
\left( O^{\chi} \right)_{13}  &=&   \frac{1}{ \sqrt{ 1 + \frac{  ( q_u^B - q_d^B )^2  }{ M_1^{\,2} } 
    \frac{  v_u^{\,2} v_d^{\,2} }{  v^2  } } }   \left( 
 - \frac{  ( q_u^B - q_d^B ) }{ M_1 } \right)  \frac{v_u v_d^2}{  v^2 }  \nonumber\\
& =&   N   \left[  - \frac{ ( q_u^B - q_d^B ) }{ M_1 } v_u \cos\beta \right] \cos\beta  
 =  N {\overline Q}_1 \cos{\b}  \label{coeff_higgs_up}  \\
\left( O^{\chi}  \right)_{23} &=&  -   \frac{1}{ \sqrt{ 1 + \frac{  ( q_u^B - q_d^B )^2  }{ M_1^{\,2} } 
    \frac{  v_u^{\,2} v_d^{\,2} }{ v^2  } } }   \left( -
  \frac{  ( q_u^B - q_d^B ) }{ M_1 } \right) \frac{  v_u^2 v_d }{ v^2}       \nonumber\\
 &=&   -   N   \left[ 
  \frac{ -  ( q_u^B - q_d^B ) }{ M_1 } v_u \cos\beta  \right] \sin\beta 
 = -  N {\overline Q}_1  \sin\beta    \label{coeff_higgs_down}     \\
\left( O^{\chi}  \right)_{33}  &=&   \frac{1}{ \sqrt{ 1 + \frac{  ( q_u^B - q_d^B )^2  }{ M_1^{\,2} } 
    \frac{  v_u^{\,2} v_d^{\,2} }{  v^2 } }}  =   N.        \label{coeffic2}         
\eea
It can be easily checked that this is an orthogonal matrix
\bea
\left( O^{\chi}  \right)^{T}  O^{\chi} = {1}_{3\times 3}.
\eea 
\subsection{Appendix. Vanishing of $\Delta^{\la \mu \nu}$ for  on-shell external physical states \label{app:LandauYang}} 
An important property of the triangle amplitude is its vanishing for on-shell external physical states.  

The vanishing of the amplitude $\Delta$ for on-shell physical states can be verified once 
we have assumed conservation of the vector currents. This is a simple example 
of a result that, in general, goes under 
the name of the Landau-Yang theorem. In our case we use only the expression of the triangle in Rosenberg parametrization 
\cite{Rosenberg:1962pp} and its gauge invariance to 
obtain this result. 
We stress this point here since if we modify the WI on the correlator, as we are going to discuss next, 
additional interactions are needed in the analysis of processes mediated by this diagram in order to obtain consistency 
with the theorem. 

We introduce the three polarization four-vectors for the $\lambda$, $\mu$, and $\nu$ lines, denoted by ${\bf e}$,  
${\bf \epsilon_1}$ 
and ${\bf \epsilon_2}$ respectively, and we use the Sudakov parameterization of each of them, using the massless vectors 
$k_1$ and $k_2$ as a longitudinal basis on the light-cone, plus transversal $(\perp)$ components which are orthogonal to the 
longitudinal ones. We have 
\beqa
{\bf e} &=& \alpha (k_1 - k_2) + {\bf e_\perp} \qquad {\bf \varepsilon_1}= a k_1 + {\bf \varepsilon_{1\perp}}
 \qquad {\bf \varepsilon_2}= b k_2 + {\bf \varepsilon_{2\perp}}, 
\eeqa
where we have used the condition of transversality ${\bf e}\cdot k=0, {\bf \varepsilon_1}\cdot k_1=0, 
{\bf \varepsilon_2}\cdot k_2=0 $, the external lines being now physical.  Clearly ${\bf e_\perp}\cdot k_1={\bf e_\perp}\cdot k_2=0$, and similar relations hold also for ${\bf \varepsilon_{1 \perp}}$ and ${\bf \varepsilon_{2 \perp}}$, all 
the transverse polarization vectors being orthogonal to the light-cone spanned by $k_1$ and $k_2$. From gauge invariance 
on the ${\mu \nu}$ lines in the invariant amplitude, we are allowed to drop the light-cone components of the polarizators for 
these two lines 
\beq
\Delta^{\lambda \mu\nu}{\bf e}_\lambda {\bf \varepsilon_{1\mu}}{\bf \varepsilon_{2\nu}}= 
\Delta^{\lambda \mu\nu}{\bf e}_\lambda {\bf \varepsilon_{1\mu\perp}}{\bf \varepsilon_{2\nu\perp}}, 
\eeq
and a simple computation then gives (introducing ${\bf e}_\perp\equiv (0, \vec{{\bf e}})$ and similar)
\beqa
\Delta^{\lambda \mu\nu}{\bf e}_\lambda {\bf \varepsilon_{1\mu\perp}}{\bf \varepsilon_{2\nu\perp}} &=&
\underline{a}_1 \epsilon[k_1 - k_2,{\bf \varepsilon_{1\perp}},{\bf \varepsilon_{2\perp}},{\bf e}]
=\underline{a}_1 \epsilon[k_1 - k_2,{\bf \varepsilon_{1\perp}},{\bf \varepsilon_{2\perp}},
\alpha (k_1 - k_2) + {\bf e_\perp} ] \nonumber \\
&\propto & \left(\vec{\bf \varepsilon}_{1\perp}\times  \vec{\bf \varepsilon}_{2\perp}\right)\cdot \vec{\bf e}_\perp =0,   
\eeqa
since the three transverse polarizations are linearly dependent. Notice that 
this proof shows that $Z \to \gamma \gamma$ with all three particles on-shell 
does not occur.
\section{Appendix. Anomalous traces}
Here we briefly discuss the computation of the anomalous traces in front of the triangle amplitudes. We work with massless fermion amplitudes. 
The anomaly coefficient in Eq.~(\ref{anom_coeff}) can be obtained starting from the triangle diagram 
in momentum space. For instance we get
\beqn
&&\Delta^{\lambda \mu \nu, ij}_{BSU(2)SU(2)}   
=   g^{}_{B} g^{\,2}_{2} \,  Tr[\tau^{i} \tau^{j}]  \sum_{f} q^{fL}_{B} {\bf \Delta}^{L \lambda \mu \nu}  \nonumber\\
&=& g^{}_{B} g^{\,2}_{2} \,  Tr[\tau^{i} \tau^{j}]  \sum_{f} q^{fL}_{B} 
 (i)^{3} \int \frac{d^{4} q}{ (2 \pi)^{4} } 
\frac{Tr[ \gamma^{\lambda} P_{L} ( \ds{q} - \ds{k} ) \gamma^{\nu} P_{L} (\ds{q} - \ds{k}_{1} ) 
\gamma^{\mu} P_{L} \ds{q} ]}{ q^2 (q -k_1)^2 (q-k)^2 }     \nonumber\\
&&+ (k_1 \rightarrow k_2, \mu \rightarrow \nu)     \nonumber\\
&=&  g^{}_{B} g^{\,2}_{2} \,  Tr[\tau^{i} \tau^{j}] \frac{1}{8} \sum_{f} q^{fL}_{B} 
 (i)^{3} \int \frac{d^{4} q}{ (2 \pi)^{4} } 
\frac{Tr[ \gamma^{\lambda}  (1 - \gamma^{5}) ( \ds{q} - \ds{k} ) \gamma^{\nu}  (1-\gamma^{5}) 
 (\ds{q} - \ds{k}_{1} ) 
\gamma^{\mu} (1-\gamma^{5})  \ds{q} ]}{ q^2 (q -k_1)^2 (q-k)^2 }     \nonumber\\
&&+ (k_1 \rightarrow k_2, \mu \rightarrow \nu)     \nonumber\\
\eeqn  
and isolating the four anomalous contributions of the form $\bf AAA$, $\bf AVV$, $\bf VAV$ 
and $\bf VVA$ we obtain
\beq
D^{(L)}_{B} = \frac{1}{8} Tr[q^{fL}_{B}] \equiv - \frac{1}{8} \sum_{f} q^{fL}_{B}.
\eeq
Similarly in the Abelian case we obtain 
\beqn
&&\Delta^{\lambda \mu \nu}_{BBB}  
=   g^{\,3}_{B}  \, \sum_{f} (q^{fR}_{B})^{3} {\bf \Delta}^{R \lambda \mu \nu}  
+  g^{\,3}_{B}  \, \sum_{f} (q^{fL}_{B})^{3} {\bf \Delta}^{L \lambda \mu \nu} \nonumber\\
&=& g^{\,3}_{B}  \, \sum_{f} (q^{fR}_{B})^{3} 
 (i)^{3} \int \frac{d^{4} q}{ (2 \pi)^{4} } 
\frac{Tr[ \gamma^{\lambda} P_{R} ( \ds{q} - \ds{k} ) \gamma^{\nu} P_{R} (\ds{q} - \ds{k}_{1} ) 
\gamma^{\mu} P_{R} \ds{q} ]}{ q^2 (q -k_1)^2 (q-k)^2 }      \nonumber\\
&&+ \, g^{\,3}_{B}  \, \sum_{f} (q^{fL}_{B})^{3} 
 (i)^{3} \int \frac{d^{4} q}{ (2 \pi)^{4} } 
\frac{Tr[ \gamma^{\lambda} P_{L} ( \ds{q} - \ds{k} ) \gamma^{\nu} P_{L} (\ds{q} - \ds{k}_{1} ) 
\gamma^{\mu} P_{L} \ds{q} ]}{ q^2 (q -k_1)^2 (q-k)^2 }    \nonumber\\
&&\,+ \,\, (k_1 \rightarrow k_2, \mu \rightarrow \nu)     \nonumber\\
&=&  g^{\,3}_{B}  \,  \frac{1}{8} \sum_{f} (q^{fR}_{B})^{3} 
 (i)^{3} \int \frac{d^{4} q}{ (2 \pi)^{4} } 
\frac{Tr[ \gamma^{\lambda}  (1 + \gamma^{5}) ( \ds{q} - \ds{k} ) \gamma^{\nu}  (1+\gamma^{5}) 
 (\ds{q} - \ds{k}_{1} ) 
\gamma^{\mu} (1 + \gamma^{5})  \ds{q} ]}{ q^2 (q -k_1)^2 (q-k)^2 }     \nonumber\\
&&+ \,  g^{\,3}_{B}  \,  \frac{1}{8} \sum_{f} (q^{fL}_{B})^{3} 
 (i)^{3} \int \frac{d^{4} q}{ (2 \pi)^{4} } 
\frac{Tr[ \gamma^{\lambda}  (1 - \gamma^{5}) ( \ds{q} - \ds{k} ) \gamma^{\nu}  (1 - \gamma^{5}) 
 (\ds{q} - \ds{k}_{1} ) 
\gamma^{\mu} (1 - \gamma^{5})  \ds{q} ]}{ q^2 (q -k_1)^2 (q-k)^2 }     \nonumber\\
&&+ (k_1 \rightarrow k_2, \mu \rightarrow \nu)     \nonumber\\
\eeqn
with the related coefficient
\beqa
D^{}_{BBB}= \frac{1}{8} Tr[q^{3}_{B}] = \frac{1}{8} \sum_{f} \left[ (q^{fR}_{B})^{3}
 - (q^{fL}_{B})^{3}  \right]. \nonumber 
\eeqa
The other coefficients reported in Eq.~(\ref{DDD}) are obtained similarly.
\section{Appendix. CS and GS terms rotated}
The rotation of the CS and the GS terms into the physical fields and the Goldstone gives
\ba
&&V^{BYY}_{CS}=d_1 \langle B Y\wedge F_Y\rangle=(-i)d_1\epsilon^{\lambda\mu\nu\alpha}(k_{1\alpha}-k_{2\alpha})
\left[(O^{A\,T})_{21}^{2}(O^{A\,T})_{32}\right]Z^{\lambda} A_{\g}^{\mu}A_{\g}^{\nu}+\dots
\nonumber\\
&&V^{BWW}_{CS}=c_1 \langle \epsilon^{\mu\nu\rho\sigma} B_{\mu}C_{\nu\rho\sigma}^{Abelian}\rangle=
(-i)c_1\epsilon^{\lambda\mu\nu\alpha}(k_{1\alpha}-k_{2\alpha})
\left[(O^{A\,T})_{11}^{2}(O^{A\,T})_{32}\right]Z^{\lambda} A_{\g}^{\mu}A_{\g}^{\nu}+\dots
\nonumber\\
&&V^{bYY}_{GS}=\frac{C_{YY}}{M}b F_Y\wedge F_Y=4 \frac{C_{YY}}{M}b \epsilon^{\mu\nu\rho\sigma}k_{\mu}k_{\nu}Y_{\rho}Y_{\sigma}=
4 \frac{C_{YY}}{M}\epsilon^{\mu\nu\rho\sigma}k_{\mu}k_{\nu}\left[
O^{\chi}_{31} (O^{A\,T})_{21}^{2}\,\chi A_{\g}^{\mu}A_{\g}^{\nu}
\right.\nonumber\\
&&\hspace{5cm}\left.
+(O^{\chi}_{32}C_1+O^{\chi}_{33}C_1^{\prime}) (O^{A\,T})_{21}^{2}G_Z A_{\g}^{\mu}A_{\g}^{\nu}\right]
+\dots\nonumber\\
&&V^{bWW}_{GS}=\frac{F}{M}b Tr\left[F_W\wedge F_W\right]
=4 \frac{C_{YY}}{M}\frac{b}{2} \epsilon^{\mu\nu\rho\sigma}k_{\mu}k_{\nu}W^{i}_{\rho}W^{i}_{\sigma}=
4 \frac{F}{M}\epsilon^{\mu\nu\rho\sigma}k_{\mu}k_{\nu}\left[
O^{\chi}_{31} (O^{A\,T})_{11}^{2} \,\chi A_{\g}^{\mu}A_{\g}^{\nu}
\right.\nonumber\\
&&\hspace{5cm}\left.
+(O^{\chi}_{32}C_1+O^{\chi}_{33}C_1^{\prime}) (O^{A\,T})_{11}^{2}\right]G_Z A_{\g}^{\mu}A_{\g}^{\nu}
+\dots\nonumber\\
\ea
These vertices appear in the cancellation of the gauge dependence in $s$-channel exchanges of $Z$ gauge bosons in the $R^{}_\xi$ gauge. The dots refer to the additional contributions, proportional to interactions of $\chi$, the axi-Higgs, with
the neutral gauge bosons of the model.
                                           
  
\chapter{Unitarity Bounds for Gauged Axionic Interactions 
and the Green-Schwarz \\ Mechanism   \label{chap:UnitarityBound}}
\fancyhead[LO]{\nouppercase{Chapter 3. Unitarity Bound for Gauged Axionic Interactions}}
\section{Introduction to the chapter}
The cancellation of gauge anomalies in the Standard Model
is a landmark of modern particle theory that has contributed to
shape our knowledge on the fermion spectrum, its chiral charges 
and couplings. Other mechanisms of cancellation, based on the 
introduction of both local and non-local counterterms, have also 
received a lot of attention in the last two decades, from the 
introduction of the Wess-Zumino term in gauge theories \cite{Wess:1971yu} 
(which is local) to the Green-Schwarz mechanism of string theory 
\cite{Green:1984sg} (which is non-local). The field theory realization of 
this second mechanism is rather puzzling also on phenomenological 
grounds since it requires, in 4 dimensions, the non-local exchange
of a pseudoscalar to restore gauge invariance in the anomalous vertices.
In higher dimensions, for instance in 10 dimensions, the violation of
the Ward Identities due to the hexagon diagram is canceled
by the exchange of a 2-form \cite{Green:1984sg,Green:1987mn, Green:1987sp}.
In this chapter we are going to analyze the similarities between the
two approaches and emphasize the differences as well. We will try, 
along the way, to point out those unclear aspects of the field theory 
realization of this mechanism - in the absence of supersymmetry and 
gravitational interactions - which, apparently, suffers from the 
presence of an analytic structure in the energy plane that is in 
apparent disagreement with unitarity.
Moving to the WZ case, here we show that the restoration of gauge
invariance in the corresponding
one-loop effective Lagrangian via a local axion counterterm is not
able to guarantee unitarity beyond a certain scale, although this
deficiency is expected \cite{Preskill:1990fr, Coriano:2006xh}, given the local nature
of the counterterm. In the GS case, the restoration of the Ward Identities
suffers from the presence of unphysical massless poles in the trilinear gauge
vertices that, as we are going to show, are similar to those present in a
non-local version of axial electrodynamics, which has been studied extensively 
in the past \cite{Adam:1997gj} with negative conclusions concerning its unitarity
properties.  In particular, in the case of scalar potentials that include
Higgs-axion mixing, the phenomenological interpretation of the GS mechanism
remains problematic in the field theoretical construction.

We comment on the relation between the two mechanisms, when the axion 
is integrated out of the partition function of the anomalous theory, 
and on other issues of the gauge dependence of the
perturbative expansions, which emerge in the different formulations.
In the second part of the chapter we apply our analysis to a realistic
model characterizing numerically the bounds in effective actions of WZ type and discuss
the possibility to constrain brane and axion-like models at the LHC.
\subsection{WZ and GS counterterms}
Anomalous Abelian models are variations of the SM in which the gauge structure
of this is enlarged by one or more Abelian factors. The corresponding anomalies
are canceled by the introduction of a pseudoscalar, an axion ($b$),
that couples to 4-forms $F_I\wedge F_J$  (via $b/M F_I\wedge F_J$, the Wess-Zumino term)
of the gauge fields $(I,J)$. $M$ is a scale that 
is apparently unrelated to the rest of the theory and
simply  describes the range in which the anomalous model 
can be used as a good approximation to the underlying complete 
theory. The latter can be resolved at an energy $E > M$, 
by using either a renormalizable Lagrangian with an anomaly-free 
chiral fermion spectrum or a string theory. The motivations for 
introducing such models are several, ranging from the study of the 
flavor sector, where several attempts have been performed in the
last decade to reproduce the neutrino mixing matrix using theories
of this type, to effective string models, in which the extra $U(1)$ abound.
We also recall that in effective string models and in models characterized
by extra dimensions the axion ($b$) appears together with a mixing to
the anomalous gauge boson ($\partial{b}B$), which is, by coincidence,
natural in a (Higgs) theory in a broken phase. In a way, theories
of this type have several completions at higher energy \cite{Coriano:2006xh}.

Coming to the specific models that we analyze, these are complete
mLSOM-like \cite{Coriano:2005js, Irges:1998ax, Kiritsis:2003mc, Antoniadis:2002qm, Antoniadis:2000ena} models with three anomalous $U(1)$
\cite{Coriano:2007xg, Armillis:2007tb}, while most of the unitarity issues are easier to address
in simple models with two U(1) \cite{Coriano:2007fw}. In our phenomenological analysis,
which concerns only effective actions of WZ type, we will choose the charge assignments
and the construction of \cite{Ibanez:2001nd, Ibanez:1998qp}, but we will work in the region of parameter space
where only the lowest St\"uckelberg mass eigenvalue is taken into account, while the
remaining two extra $Z^{\prime}$ decouple. 
This configuration is not the most general but is enough to
clarify the key physical properties of these models.
\section{Anomaly cancellation and gauge dependences: the GS and the WZ mechanisms in field theory}
In this section we start our discussion of the unitarity properties
of the GS and WZ mechanisms, illustrating
the critical issues. We illustrate a pure diagrammatic 
construction of the WZ effective action using a set of basic 
local counterterms  and show how a certain class of amplitudes 
have an anomalous behavior that grows beyond their unitarity limit at high
energy. The arguments being rather subtle, we have decided to
illustrate the construction of the effective action for both
mechanisms in parallel. A re-arrangement of the same basic
counterterms of the WZ case generates the GS effective action,
which, however, is non-local.  The two mechanisms are different
even if they share a common origin. In a following
section we will integrate out the axion of the WZ formulation to
generate a non-local form of the same mechanism that resembles
more closely the GS counterterm. The two differ by a set of extra
non-local interactions in their respective effective actions.
One could go the other way and formulate the GS mechanism in a
local form using (two or more) extra auxiliary
fields. These points are relevant in order to understand the
connection between the two ways to cancel the anomaly.
\subsection{The Lagrangian}
Specifically, the toy model that we consider 
has a single fermion with a vector-like interaction with the gauge field $A$
and a purely axial-vector interaction with $B$. The Lagrangian is given by
\beqa
\mathcal{L}_0 &=& -\frac{1}{4} F_{A}^{2}
-\frac{1}{4} F_{B}^{2}   + \frac{1}{2}( \partial_{\mu} b +
M_1\ B_{\mu})^{2} + \overline{\psi} i \gamma^{\mu} ( \partial_{\mu} +i e A_{\mu}
+ i g^{}_{B} \gamma^{5} B_{\mu}  ) \psi,
\label{lagrangeBC}
\eeqa
where,  for simplicity, we have taken all the charges to be unitary, and we have allowed for
a St\"uckelberg term for $B$, with $M_1$ being the St\"uckelberg mass
\footnote{Even if (\ref{lagrangeBC}) is not the most general invariant Lagrangian
under the gauge group $U(1)_A\times U(1)_B$, our considerations are the same.
In fact, since $b$ shifts only under a gauge variation of the anomalous $U(1)$
gauge field $B$ (and not under $A$), the gauge invariance of the effective action under a gauge
transformation of the gauge field $A$ requires that there are no terms of the type $b F_A\wedge F_B$.}.
$A$ is massless and takes the role of a photon. The Lagrangian has a
St\"uckelberg-like symmetry with $b\to b -M_1\theta_B$ under a gauge
transformation of $B_\mu$, $\delta B_\mu=\partial_\mu \theta_B$.
The axion is a singlet under gauge transformations of $A$.  We call this simplified
theory the "A-B" Model. We are allowed not to perform any gauge-fixing on $B$
and keep the coupling of the longitudinal component of $B$ to the axion, $\partial B b$,
as an interaction vertex. If we remove $A$, we call the simplified model the "$B$ Model".
We will be interchanging between these two models for illustrative purposes and to
underline the essential features of theories of this type.

In the A-B Model, the $U(1)_A$ gauge freedom can be gauge-fixed in a generic Lorentz
gauge, with polarization vectors that carry a dependence on the
gauge parameter $\xi_A$, but $A$ being non-anomalous we will
assume trivially the validity of the Ward Identities on vector-like
currents. This will erase any dependence on $\xi_A$ both of the
polarization vectors of $A$ and of the propagators of the same gauge boson.
At the same time Chern-Simons (CS) interactions such as $A B\wedge F_B$ 
or $A B \wedge F_A$,  which are present if we define triangle diagrams 
with a symmetric distribution of the partial anomalies of each vertex 
both in the {\bf AVV} (axial-vector/vector/vector) and {\bf AAA} 
cases \cite{Coriano:2007fw, Armillis:2007tb}, can be absorbed by a re-distribution of the anomaly. 
For instance, if we assume vector Ward Identities on the $A$-current 
and move the whole anomaly to the axial-vector currents, then the CS
terms can be omitted. The anomalous corrections in the one-loop
effective action are due to triangle diagrams of the form $BAA$
({\bf AVV}, with conserved vector currents)
and  $BBB$ ({\bf AAA} with a symmetric distribution of the anomalies)
which require two WZ counterterms, given in $S_{WZ}$ below,
for anomaly cancellation. Since the analysis of anomalous gauge
theories containing WZ terms has been the subject of various 
analyses with radically different conclusions regarding the issue
of unitarity of these theories, we refer to the literature
for more details \cite{Adam:1997gj, Andrianov:1989by, Andrianov:1991ub, Andrianov:1993qy, Fosco:1993qx, Armillis:2008bg}.
Our goal here is to simply stress the relevance of these
analyses in order to understand the difference between the WZ and
GS cancellation mechanism and clarify that Higgs-axion
mixing does not find a suitable description within the standard
formulation of the GS mechanism. A detailed analysis of the unitarity issue will be presented in Chap.~\ref{chap:g-2}.
\section{Local and non-local formulations}
The GS mechanism is closely related to the WZ mechanism \cite{Wess:1971yu}.
The latter, in this case,  consists in restoring gauge invariance
of an anomalous theory by introducing a shifting pseudoscalar, an axion,
that couples to the divergence of an anomalous current. It can be formulated starting
from a massive Abelian theory and performing a
{\em field-enlarging transformation} \cite{Coriano:2007fw}
so as to generate a complete gauge invariant model in which
the usual Abelian symmetry is accompanied by a shifting axion.
The original Lagrangian of the massive gauge theory is interpreted
as the gauge-fixed case of the field-enlarged Lagrangian.
The gauge variation of the anomalous effective action is compensated
by the WZ term, so as to have a gauge invariant formulation of the model.
We will show next that a theory built in this way has a unitarity bound
that we will be able to quantify. The appearance of the axion in these
theories seems to be an artifact,
since the presence of a symmetry allows one to set the axion to vanish,
choosing a unitary gauge. In brane models, in the presence of a suitable
scalar potential, the axion ceases to be a gauge artifact and cannot be
gauged away, as shown in \cite{Coriano:2005js, Irges:1998ax}. This point is rather important,
since it shows that the GS counterterm is unable to describe Higgs-axion
mixing, which takes place when the
Peccei-Quinn \cite{Peccei:1977ur, Srednicki:2002ww} symmetry of the scalar potential is broken.
The reason is quite obvious: the GS virtual axion is a massless exchange
whose presence is just to guarantee the decoupling of the longitudinal
component of the gauge boson from the anomaly and that does not describe
a physical state. But before coming to a careful analysis of this point,
let us discuss the counterterms of the Lagrangian.

In the A-B Model the WZ counterterms are
\beq
\mathcal{L}_{WZ}= \frac{C_{AA}}{2! M_1} b F_A\wedge F_A +
\frac{C_{BB}}{2! M_1} b F_B \wedge F_B,
\label{WZ1}
\eeq
which are fixed by the condition of gauge invariance of the Lagrangian. 
The best way to proceed in the analysis of this theory is to work in the 
$R^{}_\xi$ gauge in order to remove the $B-b$ mixing \cite{Coriano:2007fw}. Alternatively, 
we are entitled to keep the mixing and perform a perturbative expansion of 
the model using the Proca propagator for the massive gauge boson, and treat 
the $b\partial B$ term as a bilinear
vertex. This second approach can be the source of some confusion,
since one could be misled and identify the perturbative expansion obtained 
by using the WZ theory with that of the GS mechanism, which involves, at a 
field theory level, only a re-definition of the trilinear fermionic vertex
with a pole-like counterterm. In the WZ effective action, treated with the
$b-B$ mixing (and not in the $R_\xi$ gauge), similar counterterms appear.
\begin{figure}[t]
{\centering \resizebox*{9.5cm}{!}{\rotatebox{0}
{\includegraphics{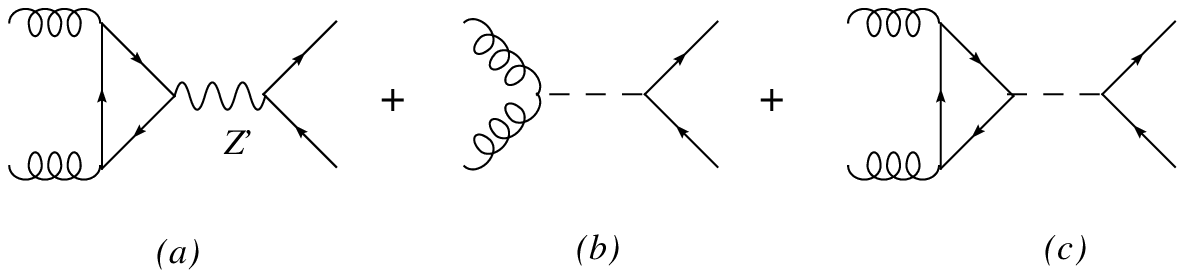}}}\par}
\caption{\small One-loop vertices and counterterms in 
the $R^{}_\xi$ gauge for the A-B Model for the WZ case.}
\label{onefigg}
\end{figure}
\begin{figure}[t]
{\centering \resizebox*{6.5cm}{!}{\rotatebox{0}
{\includegraphics{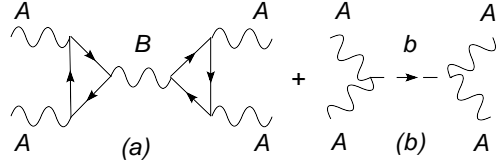}}}\par}
\caption{\small A typical Bouchiat-Iliopoulos-Meyer amplitude and 
the axion counterterm to restore gauge invariance in the $R^{}_\xi$ gauge in the WZ effective action. }
\label{twofig}
\end{figure}
We show in Fig.~\ref{onefigg} the vertices of the effective action
in the $R_\xi$ gauge approach, and we combine them to describe the process
$A A\to A A$, as shown in Fig.~\ref{twofig}.  Graph a) of Fig.~\ref{twofig} is a typical
BIM amplitude \cite{Bouchiat:1972iq}, first studied in '72 by Bouchiat, Iliopoulos and Meyer to
analyze the gauge independence of anomaly-mediated processes in the Standard Model.
The gauge independence of this process is a necessary condition in a gauge
theory in order to have a consistent $S$-matrix free of spurious singularities
\cite{Coriano:2007fw}, but is not sufficient to guarantee the absence of a unitarity bound.
Typically, gauge cancellations help to identify the correct power counting
(in $1/M_1$ and in the coupling constants) of the theory and are essential
to establish the overall correctness of the perturbative computations using
the vertices of the effective action.
\begin{figure}[t]
{\centering \resizebox*{12cm}{!}{\rotatebox{0}
{\includegraphics{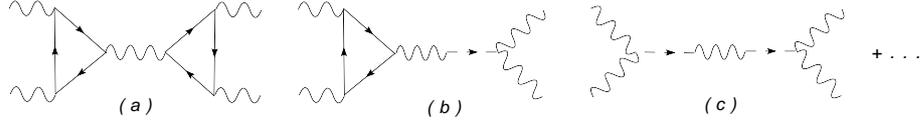}}}\par}
\caption{\small Perturbative expansion of the $A\,A\to A\,A$ amplitude in the presence of $B-b$ mixing.}
\label{fourfig}
\end{figure}
In our example, this can be established as follows:
diagram b) of Fig.~\ref{twofig} cancels the gauge dependence of diagram a)
but leaves an overall remnant, which is the contribution of diagram
a) computed in the unitary gauge ($\xi\to \infty$) in which the propagator takes the Proca form
\beq
D_P^{\lambda \lambda'} (k) = - \frac{i}{k^2 - M_1^2} \left( g^{\lambda \lambda'}
- \frac{k^{\lambda} k^{\lambda'}}{M_1^2}\right).
\eeq
Due to this cancellation, the total contribution of the two diagrams is
\beq
\Delta_A^{\lambda\mu\nu}(k,p_1,p_2)D_P^{\lambda \lambda'}
\Delta_A^{\lambda ' \mu' \nu '}(k,k_1,k_2),
\eeq
where $\Delta_A$ is given by the Dolgov-Zakharov parameterization
\ba
\Delta^{\lambda \mu \nu}_A(k, p_1, p_2) = A^{}_6(s)( k^{}_1 + k^{}_2 )^\lambda \epsilon \left[ k^{}_1, k^{}_2, \nu, \mu \right]\,,
\ea
where the coefficient $A^{}_6(s)$ in the massless case is $A^{}_6(s)=1/2(\pi^2 s)$.

We call $\Delta_B$ the triangle with a symmetric distribution
of the anomaly ($a_n/3$ for each vertex), which is obtained from $\Delta_A$
by the addition of suitable CS terms \cite{Coriano:2007xg, Armillis:2007tb}.
The bad behavior of this amplitude at high energy is then trivially given by
\footnote{
We will use the coincise notation $\epsilon[\lambda,p,k,\nu]
\equiv \epsilon^{\lambda\alpha\beta\nu}p_{\alpha}k_{\beta}$}
\beq
\frac{1}{M_1^2} \Delta_A^{\lambda \mu\nu}\, \frac{k_\lambda k_{\lambda^{\prime}}}{k^2-M_1^2}
\Delta_A^{\lambda^{\prime}\mu^{\prime}\nu^{\prime}}(k,k_1,k_2) = \frac{1}{M_1^2}
\frac{a_n^2}{k^2 - M_1^2}\epsilon[\mu, \nu, p_1, p_2] \epsilon[\mu^{\prime}, \nu^{\prime}, k_1, k_2],
\eeq
with $a_n= i/(2 \pi^2)$.
Squaring the amplitude, the corresponding cross section  grows linearly with $s=k^2$,
which signals the breaking of unitarity,
as expected in Proca theory, if the corresponding Ward Identities are violated.
A similar result holds for the $BBB$ case.
In the alternative  formulation, in which the $b-B$ term is treated as a vertex,
the perturbative expansion is formulated diagrammatically as in Fig.~\ref{fourfig}.
Though the expansion is less transparent in this case, it is still
expected to reproduce the results of the $R_\xi$
gauge and of the unitary gauge. Notice that the expansion seems to generate
the specific GS counterterms (Graph 3b)) that
limits the interaction of the gauge field with the anomaly to
its transverse component, together with some extra graphs, which
are clearly not absorbed by a re-definition of the gauge vertex.
\subsection{Integrating out the St\"uckelberg in the WZ case}
\begin{figure}[t]
{\centering \resizebox*{9cm}{!}{\rotatebox{0}
{\includegraphics{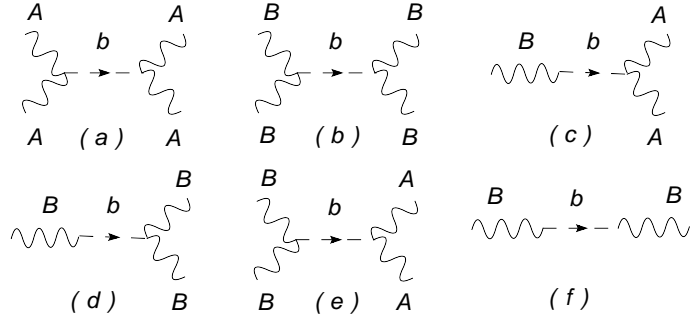}}}\par}
\caption{\small The structure of the WZ effective action having integrated out the axion.}
\label{sixfig}
\end{figure}

\begin{figure}[t]
{\centering \resizebox*{14cm}{!}{\rotatebox{0}
{\includegraphics{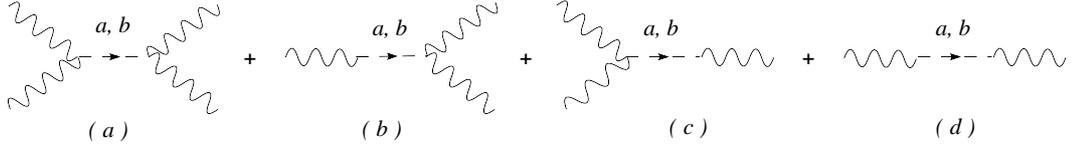}}}\par}
\caption{\small Effective action with two pseudoscalars, one of them ghost-like, $a$.}
\label{feder}
\end{figure}

We can make a forward step and try to integrate out the axion
from the partition function and obtain the non-local version of the WZ
effective action. Notice that this is straightforward only in the case in
which Higgs- axion mixing is absent. The partition function in this case is given by
\beq
Z=\int D\psi \, D\bar{\psi} \, D A \, DB \, Db \, \exp \left( i\langle \mathcal{L}(\psi,
\bar{\psi},A,B, b)\rangle\right),
\eeq
where $\langle\, \rangle$ denote integration over $x$ and
\beq
\mathcal{L}=\mathcal{L}_0 + \mathcal{L}_{WZ},
\eeq
with $\mathcal{L}_0$ and $\mathcal{L}_{WZ}$ given in (\ref{lagrangeBC}) and
(\ref{WZ1}), respectively. Indicating with $\mathcal{L}_b$ the $b$ sector of $\mathcal{L}$,
a partial integration on the axion gives
\footnote{We have re-defined
the coefficients in front of the counterterms absorbing the multiplicity factors.}
\beq
\mathcal{L}_b= -\frac{1}{2} b\,\square\, b + b\, J,
\eeq
where
\beq
J= M \partial B - \frac{\kappa_A}{M} F_A\wedge F_A  -
\frac{\kappa_B}{M} F_B\wedge F_B,
\eeq
and performing the path integration over $b$ we obtain
\beq
\int Db \exp\left( i \langle \mathcal{L}\rangle\right)=  \det \left( \square -
M_1^2\right)^{-1/2} \exp\left( \frac{i}{2} J\square^{-1} J\right),
\eeq
where
\beq
\langle J \square^{-1} J \rangle_{WZ} = \langle \left(M_1 \partial B 
-\frac{\kappa_A}{M_1} F_A\wedge F_A  -
\frac{\kappa_B}{M_1} F_B\wedge F_B\right)\square^{-1}
\left(M_1 \partial B - \frac{\kappa_A}{M_1} F_A\wedge F_A -
\frac{\kappa_B}{M_1} F_B\wedge F_B\right)
\rangle.
\eeq
The additional contributions to the effective action are
now non-local and are represented by the set of diagrams in Fig.~\ref{sixfig}.
Among these diagrams there are two GS counterterms, diagrams c) and d),
but there are also other contributions. To generate only the GS counterterms
one needs an additional pseudoscalar called $a$ in order to enforce the cancellation of the extra terms.
There are various ways of doing this \cite{Andrianov:1989by, Andrianov:1991ub, Andrianov:1993qy, Federbush:1996cp}.
In \cite{Federbush:1996cp} the non-local counterterm $\partial B \square^{-1} F\wedge F$ of
axial QED, which corresponds to the diagrams b) and c) of Fig.~\ref{feder},
is obtained by performing the functional integral over $a$ and $b$ of the following action \cite{Federbush:1996cp}
\beqa
\mathcal{L} &=& \overline{\psi} \left( i \slashed{\partial} + e \slashed{B} \gamma_5\right)\psi - \frac{1}{4} F_B^2 +
\frac{ e^3}{48 \pi^2 M_1} F_B\wedge F_B ( a + b) \nonumber \\
&& + \frac{1}{2}  \left( \partial_\mu b - M_1 B_\mu\right)^2 -
\frac{1}{2} \left( \partial_\mu a - M_1 B_\mu\right)^2.
\label{fedeqq}
\eeqa
The integral on $a$ and $b$ are gaussians and their contributions
to the effective action, after integrating them out, are shown in Fig.~\ref{feder}.
Notice that $b$ has a positive kinetic term and $a$ is ghost-like. The role of the
two pseudoscalars is to cancel the contributions in Fig.~\ref{feder} a) and d), leaving only
the contribution given by graphs b) and c), which has the pole structure typical
of the GS non-local counterterm.
Due to these cancellations, the effective action now reduces to
\beq
 \langle \partial B(x) \square^{-1}(x-y) F(y)\wedge F(y)\rangle
\eeq
besides the anomaly vertex and is represented by interactions of the form b) and c) of Fig.~\ref{feder}.
This shows that the WZ and the GS effective actions organize
the perturbative expansions in a rather different way.
It is also quite immediate that the cleanest way to analyze the
expansion is to use the $R_\xi$ gauge, as we have already stressed.
It is then also quite clear that in the WZ case we require the gauge
invariance of the Lagrangian {\em but not} of the trilinear gauge interactions,
while in the GS case, which is realized via Eq.~(\ref{fedeqq})  or, analogously,
by the Lagrangians proposed in \cite{Andrianov:1989by, Andrianov:1991ub, Andrianov:1993qy}, it is the trilinear
vertex that is rendered gauge invariant (together with the Lagrangian).
The presence of a ghost-like particle in the GS case renders the local
description quite unappealing and for sure the best way to define the mechanism
is just by adding the non-local counterterms. In the WZ case the local description
is quite satisfactory and allows one to treat the $bFF$ interaction
as  a real trilinear vertex, which takes an important role in the presence of
a broken phase.
The GS counterterms are, in practice, the same ones as appearing
in the analysis of axial QED with a non-local counterterm, as we are going to discuss next.
\begin{figure}[t]
{\centering \resizebox*{13cm}{!}{\rotatebox{0}
{\includegraphics{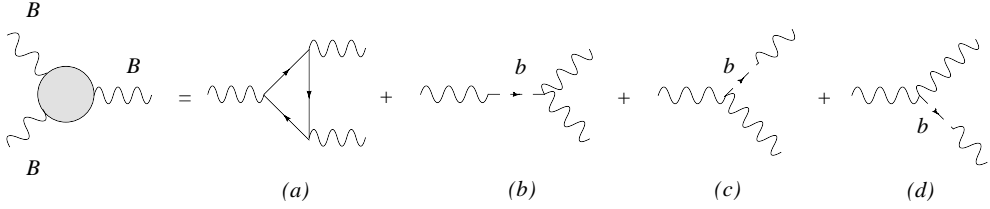}}}\par}
\caption{\small The GS vertex with the non-local 
contributions for the $BBB$ diagram symmetrized on each leg.}
\label{r1}
\end{figure}
\subsection{Non-local counterterms: axial QED}
The use of non-local counterterms to cancel the anomaly is for sure a 
debated issue in quantum field theory since most of the results 
concerning the BRS analysis of these
theories may not apply \cite{Andrianov:1991ub, Andrianov:1993qy}. In the GS case we may ignore all 
the previous constructions and just require ab initio that the
anomalous vertices are modified by the addition of a non-local counterterm
that cancels the anomaly on the axial lines.

\begin{figure}[t]
{\centering \resizebox*{12cm}{!}{\rotatebox{0} 
{\includegraphics{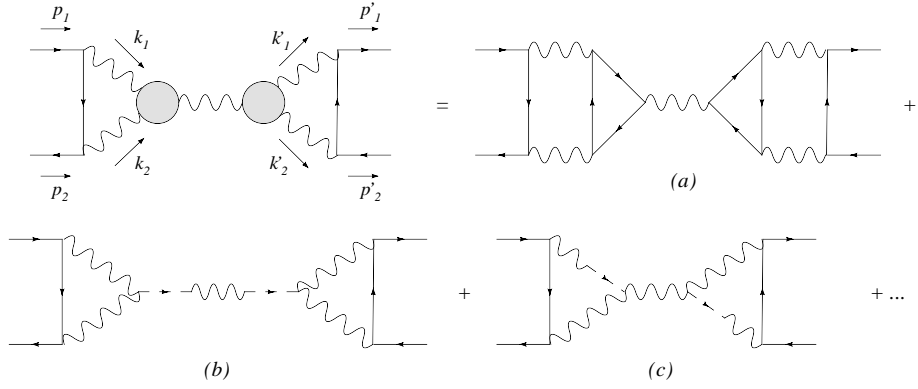}}}\par}
\caption{\small The structure of the fermionic scattering 
amplitudes with spurious massless poles (graphs b) and c)).}
\label{r2}
\end{figure}

Consider, for instance, the case of the $BBB$ vertex of Fig.~\ref{r1},
where the regularization of the anomalies has been obtained by adding
the three GS counterterms in a symmetric way
\cite{Anastasopoulos:2006cz}. In the $BAA$ case only a single countertem
is needed, but for the rest the discussion is quite similar to the 
$BBB$ case, with just a few differences. These concern the 
distribution of the partial anomalies on the $A$- and $B$-lines
in the case in which
also $BAA$ is treated symmetrically (equal partial anomalies).
In this particular case we need to compensate the vertex with CS
interactions, which are not, anyhow, observable if the $A$-lines correspond
to conserved gauge currents  such as in QED. In this situation the
Ward Identities would force the CS counterterms to vanish. We will
stick to the {\em consistent} definition of the anomaly in which only $B$
carries the total anomaly $a_n$ and $A$ is anomaly-free.
The counterterm used in the GS mechanism both for $BAA$ and $BBB$ is
nothing else but the
opposite of the Dolgov-Zakharov (DZ) expression \cite{Dolgov:1971ri}, which
in the $BAA$ case takes the form
\ba
C^{\lambda \mu \nu}_{AVV}(k,k_1,k_2) = -\frac{a_n}{k^2} k^\lambda \epsilon[\mu, \nu, k_1, k_2].
\ea
In the $BBB$ case a similar expression is obtained by creating a Bose symmetric combination of DZ poles,
\ba
C^{\lambda \mu \nu}_{AAA}(k,k_1,k_2) = -\left(
\frac{1}{3}\frac{a_n}{k^2} k^\lambda \epsilon[\mu, \nu, k_1, k_2]
+ \frac{1}{3}\frac{a_n}{k_1^2} k_1^\mu \epsilon[\nu, \lambda, k_2, k]
+ \frac{1}{3}\frac{a_n}{k_2^2} k_2^\nu \epsilon[\lambda, \mu, k, k_1]\right).
\ea
We have denoted by $k$ the incoming momenta of the axial-vector vertex and by
$k_1$ and $k_2$ the outgoing momenta of the vector vertices.
We keep this notation also in the $AAA$ case, since $k$ will denote
the momentum exchange in the $s$-channel when we glue together these 
amplitudes to obtain an amplitude of BIM type; 
this, we will analyze in the next sections.
These expressions are consistent with the following equations of the anomaly for the $BAA$ triangle
\ba
k_{1\mu}C^{\lambda \mu \nu}_{AVV}(k,k_1,k_2)= 0,   \nonumber\\
k_{2\nu}C^{\lambda \mu \nu}_{AVV}(k,k_1,k_2)= 0,   \nonumber\\
k_{\lambda}C^{\lambda \mu \nu}_{AVV}(k,k_1,k_2)= -a_n\epsilon[\mu, \nu, k_1, k_2],   \nonumber\\
\ea
and for the $BBB$ anomalous triangle
\ba
k_{1\mu}C^{\lambda \mu \nu}_{AAA}(k,k_1,k_2)= -\frac{a_n}{3}\epsilon[\lambda, \nu, k, k_2],   \nonumber\\
k_{2\nu}C^{\lambda \mu \nu}_{AAA}(k,k_1,k_2)= -\frac{a_n}{3}\epsilon[\lambda, \mu, k, k_1],   \nonumber\\
k_{\lambda}C^{\lambda \mu \nu}_{AAA}(k,k_1,k_2)=- \frac{a_n}{3}\epsilon[\mu, \nu, k_1, k_2]. \nonumber\\
\ea
So we can define a gauge invariant triangle amplitude, in both the $BBB$ and $BAA$ cases, by
\ba
\Delta^{\lambda \mu \nu\,GS }_{AAA}(k,k_1,k_2) &=& \Delta^{\lambda \mu \nu}_{AAA}(k,k_1,k_2) 
+ C^{\lambda \mu \nu}_{AAA}(k,k_1,k_2) \nonumber\\
\Delta^{\lambda \mu \nu \,GS }_{AVV}(k,k_1,k_2) &=& \Delta^{\lambda \mu \nu}_{AVV}(k,k_1,k_2) 
+ C^{\lambda \mu \nu}_{AVV}(k,k_1,k_2). \nonumber \\
\label{reg1}
\ea
Notice that the (fermionic) triangle diagrams, in the symmetric limit 
$k_1^2=k_2^2=k_3^2$, is exactly the opposite of the DZ counterterms, 
as we will discuss in the next section,
\beq
  \Delta_{AVV}(k_1^2=k_2^2=k^2)=-C_{AVV} \qquad  \Delta_{AAA}(k_1^2=k_2^2=k^2)=-C_{AAA},
\label{reg2}
\eeq
so the cancellation is identical at that point (only at that point), 
and the two vertices $\Delta^{GS}$ vanish. It is rather obvious that the 
cancellations of these poles in BIM amplitudes, corrected by the GS counterterms, 
is identical only for on-shell external gauge lines. It is also
quite straightforward to realize that
the massless $B$-model with the GS vertex correction
is equivalent to axial QED corrected by a non-local term \cite{Adam:1997gj} 
that is described by the Lagrangian
\beq
\mathcal{L}_{5\,QED}= \overline{\psi} \left( i \slashed{\partial} 
+ e \slashed{B} \gamma_5\right)\psi - \frac{1}{4} F_B^2
\eeq
plus the counterterm
\beq
\mathcal{S}_{ct}= \frac{1}{24\pi^2} \langle \partial B(x) \square^{-1}(x-y) F(y)\wedge F(y)
\rangle.
\eeq
This theory is equivalent to the (local) formulation given in 
Eq.~(\ref{feder}) and in \cite{Andrianov:1991ub, Andrianov:1993qy}, where the transversality
constraint ($\partial B=0$) is directly imposed on the Lagrangian via a multiplier.
\begin{figure}[t]
{\centering \resizebox*{3.0cm}{!}{\rotatebox{0}
{\includegraphics{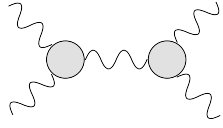}}}\par}
\caption{\small BIM amplitude with full GS vertices. 
For on-shell external lines the contributions from the extra poles disappear.}
\label{eightfig}
\end{figure}
\begin{figure}[h]
{\centering \resizebox*{5.5cm}{!}{\rotatebox{0}
{\includegraphics{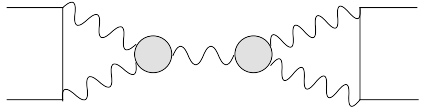}}}\par}
\caption{\small Embedding of the BIM amplitude with GS 
vertices in a fermion/antifermion scattering.}
\label{ninefigg}
\end{figure}
Unitarity requires these DZ poles in the C counterterms 
to disappear from a physical amplitude. In general, consider the 
diagrams depicted in Figs.~\ref{eightfig}
and \ref{ninefigg}. The structure of the GS vertex
is, for $BBB$, given by (\ref{reg1}) with the three massless poles
generated by the exchange of the pseudoscalar on the three legs, 
as shown in Fig.~\ref{r1}.
For on-shell external lines, in this diagram the 
contributions from the extra poles
cancel due to the transversality condition satisfied by the 
polarizators of the gauge bosons.
However, once these amplitudes are embedded into more 
general amplitudes, such as those shown
in Figs.~\ref{ninefigg}, \ref{r2}, the cancellation of the DZ extra poles introduced
by the counterterm requires a more detailed analysis that we postpone to Chap.~\ref{chap:g-2}, where this cancellation will be 
proven up to three-loop order.

We can summarize the issues that we have raised in the following points.

1) The GS and the WZ mechanism have different formulations in terms 
 of  auxiliary fields.
 
2) Previous analyses of axial gauge theories, though distinct
in their Lagrangian formulation, are all equivalent to axial 
QED plus a non-local counterterm. The regularization of the gauge
interactions, in these theories, coincides with that obtained by 
using the GS counterterm on the gauge vertex. In particular, the  
massless poles introduced by the regularization are not completely understood 
in the context of perturbative unitarity.
 
A special comment is needed when we move to the analysis of Higgs-axion mixing. 
This has been shown
to take place after electroweak symmetry breaking for a special class of potentials,
which are not supersymmetric. The axion, which in the St\"uckelberg phase is essentially
a Goldstone mode, develops a physical component and this component appears as a physical pole.
It is then clear, from the analysis presented above, that the regularization procedure introduced 
by the GS counterterm involves a virtual massless state  and not a physical pole. 
This is at variance with the WZ mechanism, in which  the $bFF$ vertex is introduced 
from the beginning as a vertex and not just as a virtual state. 
In this second case, $b$ can be decomposed in terms of a Goldstone mode and
a physical pseudoscalar, the {\em axi-Higgs},
which takes the role of a gauged Peccei-Quinn axion \cite{Coriano:2005js, Irges:1998ax}. This is
entitled to appear as a physical state (and a physical pole, massless or massive) 
in the spectrum. The re-formulation of the GS
counterterm in terms of a pseudoscalar comes also at a cost, due to the presence of
a ghost-like (phantom) particle in the spectrum, which is absent in the WZ case. There are some
advantages though, since the theory has, apparently, a nice UV behavior, 
given the gauge invariance condition on the vertex.  The presence of these spurious 
poles requires further investigations to see how they are really embedded into 
higher-order diagrams and we will return to this point in Chap.~\ref{chap:g-2}. 
 
In the next section we will move to the analysis of the unitarity bound
in the WZ  case. As we have shown, in this case it  is possible to
characterize it explicitly. We will work in a specific model, but the
implications of our analysis are general and may be used to constrain 
significantly entire classes of models  containing WZ interactions at
the LHC. Before coming to the specific phenomenological applications we
elaborate on the set of amplitudes that are instrumental in order to
spot the bad high energy behavior of the chiral anomaly
in $s$-channel processes: the so called BIM amplitudes.
\section{BIM amplitudes, unitarity and the resonance pole}
The uncontrolled growth of the cross section in the WZ case can be studied in a class of amplitudes that have two anomalous
({\bf AVV} or {\bf AAA} ) vertices connected by an $s$-channel exchange
as in Fig.~\ref{fourfig} a). We are interested in the expressions
of these amplitudes in the chiral limit, when all the fermions
are massless. Processes such as $AA\to AA$,  mediated by an
anomalous gauge boson $B$, with on-shell external $A$-lines and
massless fermions, can be expressed in a simplified form, which
is, also in this case, the DZ form. We therefore set
$k_1^2=k_2^2=0$  and $m_f=0$,
which are the correct kinematical conditions to obtain DZ poles.
We briefly elaborate on this point.

We start from the Rosenberg form of the $AVV$ amplitude, which is given by
\ba
&&T^{\lambda\mu\nu}=A_1\epsilon[k_1,\lambda,\mu,\nu]+A_2\epsilon[k_2,\lambda,\mu,\nu]
+A_3 k_1^{\mu}\epsilon[k_1,k_2,\nu,\lambda]
\nonumber\\
&&\hspace{1cm}+A_4 k_2^{\mu}\epsilon[k_1,k_2,\nu,\lambda]
+A_5 k_1^{\nu}\epsilon[k_1,k_2,\mu,\lambda]+A_6 k_2^{\nu}\epsilon[k_1,k_2,\mu,\lambda]\,,
\ea
and imposing the Ward Identities we obtain
\ba
&&A_1 = k_2^2 A_4 + k_1\cdot k_2 A_3
\nonumber\\
&&A_2 = k_1^2 A_5 + k_1\cdot k_2 A_6
\nonumber\\
&&A_3(k_1,k_2)=-A_6(k_1,k_2)
\nonumber\\
&&A_4(k_1,k_2)=-A_5(k_1,k_2),
\ea
where the invariant amplitudes $A_3,\dots,A_6$ are free from singularities.
In this specific kinematical limit we can use the following relations to
simplify our amplitude
\ba
&&\epsilon[k_2,\lambda,\mu,\nu]=\frac{2}{s}
\left(k_2^{\lambda}\epsilon[k_1,k_2,\nu,\mu]
+k_2^{\nu}\epsilon[k_1,k_2,\mu,\lambda]
+k_2^{\mu}\epsilon[k_1,k_2,\nu,\lambda] \right)
\nonumber\\
&&\epsilon[k_1,\lambda,\mu,\nu]=-\frac{2}{s}
\left(k_1^{\lambda}\epsilon[k_1,k_2,\nu,\mu]
+k_1^{\nu}\epsilon[k_1,k_2,\mu,\lambda]
+k_1^{\mu}\epsilon[k_1,k_2,\nu,\lambda] \right)
\ea
where $k^2=(k_1+k_2)^2=s$ is the center of mass energy.
These combinations allow us to re-write the expression 
of the trilinear amplitude as
\ba
T^{\mu\nu\lambda}=A_6 k^{\lambda}\epsilon[k_1,k_2,\nu,\mu]+
\left(A_4 + A_6\right)\left(k_2^{\nu}\epsilon[k_1,k_2,\mu,\lambda]
-k_1^{\mu}\epsilon[k_1,k_2,\nu,\lambda]\right)\,.
\ea
It is not difficult to see that the second piece drops off 
for physical external on-shell $A$-lines,
and we see that only one invariant amplitude
contributes to the result
\ba
&&T^{\mu\nu\lambda}=A_6^{f}(s)(k_{1}+k_{2})^{\lambda}
\epsilon\left[k_1,k_2,\nu,\mu\right].
\ea
There are some observations to be made concerning
this result. Notice that $A_6 $ multiplies a longitudinal 
momentum exchange and, as discussed in the literature on 
the chiral anomaly in QCD \cite{Dolgov:1971ri, Ioffe:2006ww, Achasov:1992bu}, 
brings about a {\em massless} pole in $s$. We just recall
that $A_6$ satisfies an unsubtracted dispersion relation 
in $s$ at a fixed invariant mass of the two photons,
$(k_1^2=k_2^2=p^2)$
\beq
A_6(s,p^2)=\frac{1}{\pi^2}\int_{4 m_f^2}^{\infty} dt \, \frac{{\it Im} A_6(t, p^2)}{t - s}
\eeq
and a sum rule 
\beq
\int_{4 m_f^2}^\infty Im A_6(t, p^2) \, d t = \frac{1}{2 \pi},
\eeq
while for on-shell external photons one can use the DZ relation \cite{Dolgov:1971ri}
\beq
Im A_6(k^2,0)= \frac{1}{\pi} \delta(k^2),
\eeq
to show that the only pole of the amplitude is actually at $s=k^2=0$. 
It can be simplified using the identity
\ba
Li_2(1- a)+Li_2(1- a^{-1})&=&-\frac{1}{2}\log^2(a),
\label{di_log}
\ea
with
\ba
a= \frac{\rho^{}_f + 1}{ \rho^{}_f - 1}  \qquad \rho^{}_f = \sqrt{1 - 4\frac{m_f^2}{s}}\,,
\ea
to give \cite{Achasov:1992bu}
\ba
&&A_6^{f}(s)=\frac{1}{2\pi^2 s}\left(1 -\frac{m_f^2}{s}\log^{2}\frac{\rho^{}_f + 1 }{\rho^{}_f - 1 }\right).
\ea
We can use this amplitude to discuss both the breaking of unitarity and the cancellation 
of the resonance pole in this simple model. The first point has already been addressed
in the previous section, where the computation of the diagrams in Fig.~\ref{twofig}
has shown that only a BIM amplitude survives in the WZ case in the scattering process
$AA\to AA$. If the sum of those two diagrams gives a gauge invariant result, 
with the exchange of the $Z^\prime$ described by a Proca propagator, there is a third 
contribution that should be added to this amplitude. This comes from the exchange 
of the physical axion $\chi$. We recall, in fact, that in the presence of Higgs-axion mixing,
when $b$ is the sum of a Goldstone mode and a physical axion $\chi$, each
anomalous $Z^\prime$ is accompanied by the exchange of the $\chi$ \cite{Coriano:2005js, Irges:1998ax}.
This is generated in the presence of electroweak symmetry breaking, 
after expansion of the Higgs scalar $\phi$ around a vacuum $v$
\beq
\phi = \frac{1}{\sqrt{2}}\left( v + \phi^{}_{1} + i \phi^{}_{2} \right),
\eeq
with the axion $b$ expressed as linear combination 
of the rotated fields $\chi$ and $G^{}_{B}$
\beqa
b = \alpha_1 \chi + \alpha_2 G^{}_{B} = 
\frac{q^{}_{B} g^{}_{B} v}{M_B} \chi + \frac{M_1}{M_B} G^{}_{B}.
\eeqa
We also recall that the gauge fields $B^{}_\mu$ get their masses
$M^{}_B$ through the combined Higgs-St\"{u}ckelberg
mechanism,
\beq
M_B=\sqrt{M_1^2 + (q^{}_{B} g^{}_{B} v)^2}.
\eeq
In the phenomenological analysis presented in the next sections the contribution due to $\chi$ has 
been included. Therefore, in the WZ case, the total contributions coming 
from the several BIM amplitudes related to the additional anomalous 
neutral currents should be accompanied not only by the set of Goldstone bosons,
to restore gauge invariance, but also by the exchange of the axi-Higgs.

The cancellation of the resonance pole for $s= M_B$ is an 
important characteristic of BIM amplitudes, which does not occur
in any other (anomaly-free) amplitude. This cancellation is the result 
of some trivial algebra already seen at work in the first two chapters. For instance, given a BIM amplitude and a Proca propagator, we have
\ba
A^{\mu \nu \mu^\prime \nu^\prime}_{BIM}&=&   \frac{a_n}{k^2} k^{\lambda} \epsilon[\mu,\nu,k_1,k_2]  \, \frac{- i}{k^2 - M_1^2}
\left( g^{\lambda \lambda^\prime}
- \frac{k^\lambda k^{\lambda^\prime} }{ M_1^2 }   \right)  \frac{a_n}{k^2}
( - k^{\lambda^\prime}) \epsilon[\mu^\prime,\nu^\prime,k_1^\prime,k_2^\prime]   \nonumber\\
&=&   \frac{a_n}{k^2}  \epsilon[\mu,\nu,k_1,k_2]  \, \frac{- i}{ k^2 - M_1^2 }   \frac{k^{ \lambda^\prime}( M_1^2 - k^2 ) }{ M_1^2 }  \,  \frac{a_n}{k^2}
( - k^{\lambda^\prime}) \epsilon[\mu^\prime,\nu^\prime,k_1^\prime,k_2^\prime]   \nonumber\\
&=&   \frac{a_n}{k^2}  \epsilon[ \mu,\nu,k_1,k_2]  \, \left(  \frac{- i k^2 }{  M_1^2 }  \right) \,  \frac{a_n}{k^2}
 \epsilon[\mu^\prime,\nu^\prime,k_1^\prime,k_2^\prime]   \nonumber\\
&=&  - \frac{a_n}{M_1}  \epsilon[ \mu,\nu,k_1,k_2]  \,\,  \frac{ i }{ k^2 }  \,\,  \frac{a_n}{M_1}
 \epsilon[\mu^\prime,\nu^\prime,k_1^\prime,k_2^\prime].
\label{resBIMM}
\ea
This result implies that the amplitude is described - in the chiral 
limit and for massless external states - by a diagram with the exchange 
of a pseudoscalar (see Fig.~\ref{twofig} b)) and that the resonance pole 
has disappeared. It is clear, from (\ref{resBIMM}), that these amplitudes 
break unitarity and give a contribution to the cross 
section that grows quadratically
in energy ($\sim s$).  Therefore, searching for
BIM amplitudes at the LHC can be a way to uncover the anomalous
behavior of extra neutral (or charged) gauge interactions.
There is one thing that might tame this growth, and this is
the exchange of the physical axion. We will show, working in
a complete brane model, that the exchange of the $\chi$
does lower the cross section, but insignificantly, independently
from the mass of the axion.
\section{A realistic model with WZ counterterms}
Clarified the relation between the WZ and GS 
mechanisms using a toy-model, we now move
towards the analysis of the issue of unitarity 
violation in the WZ Lagrangian at high energy.
Details on the structure of the effective action 
of the complete model that we
are going to analyze can be found in Chap.~\ref{chap:AbelianModels2}. 
We just mention that this is
characterized by a gauge structure of the form
$SU(3)\times SU(2)\times U(1)\times U(1)_B$, where the $U(1)_B$ is anomalous.
We work in the context of a two-Higgs doublet model with $H_u$ and $H_d$.
In our analysis the charge assignments are those of a realistic brane model with three
extra anomalous $U(1)$, but we will assume that only
the lowest mass eigenvalue taking part is significant, since the remaining
two additional gauge bosons are heavy and, essentially, decoupled.

We recall that the single anomalous gauge boson $B$, that we consider for
this analysis, is characterized by a generator $Y_B$ which is
anomalous ($Tr\left[ Y_B^3 \right] \neq 0$) but at the same time has mixed anomalies
with the remaining generators of SM and in particular 
with the hypercharge $Y$. In the presence of this
anomalous $U(1)_B$ both the $Z$ and the (extra) $Z^\prime$
gauge boson have an anomalous component, proportional to $B$.
We also recall that the effective action of the 
anomalous theory is rendered gauge
invariant using both CS and WZ counterterms,
while a given gauge invariant sector involves the exchange both of the
anomalous gauge boson and of the axion  in the $s$-, $t$- and $u$-channels.

In \cite{Armillis:2007tb} it has been shown how the trilinear vertices of
the effective WZ Lagrangian can be determined consistently for a generic
number of extra anomalous $U(1)$. Here, the goal is to identify and
quantify the contributions that cause a violation of unitarity in this
Lagrangian. For phenomenological reasons it is then convenient to select
those BIM amplitudes that have a better chance, at the experimental level,
to be measured at the LHC and for this reason we will focus on the
process $g\,g \to \gamma \gamma $. The gluon density grows at high
energy especially at smaller (Bjorken variable) $x$-values.  We choose to work with
{\em prompt} final state photons for obvious reasons, the signal being
particularly clean. To begin with, we will be needing the expressions of the
$Z \gamma \gamma$ and the $Z gg$ vertices. In the presence of three anomalous
$U(1)$'s, here denoted as $U(1)_B$, both the $Z$ and the (extra) $Z'$ gauge boson
have an anomalous component, which is proportional to the $B_{i \mu}$,
the anomalous gauge bosons of the interaction eigenstate basis (i=1,2,3).
The photon vertex is given by \cite{Armillis:2007tb}
\begin{eqnarray}
&&\langle Z_l\g\g\rangle|_{m_f\neq 0}=
-\frac{1}{2}Z^{\lambda}_l A_{\g}^{\mu}A_{\g}^{\nu}
\sum_f\left[
g_Y^3\theta_f^{YYY}\bar{R}^{YYY}_{Z_l\g\g}
+g_2^3\theta_f^{WWW}\bar{R}^{WWW}_{Z_l\g\g}
+g_Y g_2^2\theta_f^{YWW}R^{YWW}_{Z_l\g\g}
\right.\nonumber\\
&&\hspace{2cm}\left.
+ g_Y^2 g_2\theta_f^{YYW}R^{YYW}_{Z_l\g\g}
+\sum_i g_{B_i} g_Y g_2\theta_f^{B_iYW}R^{B_i Y W}_{Z_l\g\g}
\right.\nonumber\\
&&\hspace{2cm}\left.
+\sum_i g_{B_i} g_Y^2\theta_f^{B_iYY}R^{B_iYY}_{Z_l\g\g}
+ g_{B_i} g_2^2\theta_f^{B_iWW}R^{B_iWW}_{Z_l\g\g}\right]
\Delta_{AVV}^{\lambda\mu\nu}(m_f\neq 0)
\end{eqnarray}
with $l=1,2,3$ enumerating the extra anomalous neutral
currents. The explicit expressions of the rotation matrix $O^A$ can be found in App.~\ref{app:gauge}.
We have defined
\begin{equation}
\bar{R}^{YYY}_{Z_l\g\g}=(O^A)_{YZ_l}(O^{A})_{Y\g}^{2},
\hspace{1cm}
\bar{R}^{WWW}_{Z_l\g\g}=(O^A)_{W_3Z_l}(O^{A})_{W_3\g}^{2},
\end{equation}
and the triangle $\Delta_{AVV}(m_f\neq 0)$ is given by \cite{Armillis:2007tb}
\ba
&&\Delta_{AVV}^{\lambda\mu\nu}(m_f\neq 0,k_1,k_2)=\frac{1}{\pi^2}\int_0^1 dx \int_{0}^{1-x}dy
\frac{1}{\Delta(m_f)}\nonumber\\
&&\hspace{2cm}
\left\{\epsilon[k_1,\lambda,\mu,\nu]
\left[y(y-1)k_2^2 -x y k_1\cdot k_2\right]
\right.\nonumber\\
&&\hspace{2cm}\left.
+\epsilon[k_2,\lambda,\mu,\nu]
\left[x(1-x)k_1^2 +x y k_1\cdot k_2\right]
\right.\nonumber\\
&&\hspace{2cm}\left.
+\epsilon[k_1,k_2,\lambda,\nu]
\left[x(x-1)k_1^{\mu} -x y k_{2}^{\mu} \right]
\right.\nonumber\\
&&\hspace{2cm}\left.
+\epsilon[k_1,k_2,\lambda,\mu]
\left[x y k_1^{\nu} +(1-y)y k_{2}^{\nu} \right]
\right\}\,,
\nonumber\\
\nonumber\\
&&\Delta(m_f)=m_f^2+x(x-1)k_1^2+y(y-1)k_2^2-2 x y k_1\cdot k_2\,.
\ea
We have defined the following chiral asymmetries
\begin{eqnarray}
&&\theta^{B_{l}YY}_{f}=Q_{B_j,f}^{L}(Q_{Y,f}^{L})^2-Q_{B_l,f}^{R}(Q_{Y,f}^{R})^2,
\nonumber\\
&&\theta^{B_{l}WW}_{f}=Q_{B_l,f}^{L}(T^{3}_{L,f})^2,
\nonumber\\
&&\theta_{f}^{WWW}=(T^{3}_{L,f})^{3},
\nonumber\\
&&\theta_{f}^{YYW}=\left[(Q^{L}_{Y,f})^2 T^{3}_{L,f}\right],
\nonumber\\
&&\theta_{f}^{B_lYW}=\left[Q^{B_l,f}Q^{L}_{Y,f} T^{3}_{L,f}\right],
\end{eqnarray}
with $Q_B^{L/R}$ and $Q_Y^{L/R}$ denoting the charges of the
chiral fermions and $T^3_L$ is the generator of
the third component of the weak isospin, while the $R$ factors
are products of $O^A$ matrix elements.
The matrix $O^A$ relates the interaction eigenstate basis of the 
generators $(Y_B, Y, T_3)$ to those of the
mass eigenstate basis $(T_Z, T_{Z'}, Q)$, of the physical gauge bosons
of the neutral sector, $Z, Z'$ and $A_\gamma$. They are given by
\begin{eqnarray}
&&R^{YYY}_{Z_l\g\g}=3\left[(O^{A})_{Y Z_l}(O^{A})_{Y\g}^{2}\right]\nonumber\\
&&R^{YWW}_{Z_l\g\g}=\left[2(O^{A})_{W_3\g}(O^{A})_{YZ_l}(O^{A})_{Y\g}
+(O^{A})_{W_3\g}^{2}(O^{A})_{Y Z_l}\right]\nonumber\\
&&R^{WWW}_{Z_l\g\g}=\left[3(O^{A})_{B_i Z_l}(O^{A})^{2}_{W_3\g}\right]\nonumber\\
&&R^{YYW}_{Z_l\g\g}=\left[2(O^{A})_{Y Z_l}(O^{A})_{Y\g}(O^{A})_{W_3\g}
+(O^{A})_{W_3 Z_l}(O^{A})_{Y\g}^{2}\right]\nonumber\\
&&R^{B_i YY}_{Z_l\g\g}=(O^{A})_{Y\g}^{2}(O^{A})_{B_i Z_l}\nonumber\\
&&R^{B_i WW}_{Z_l\g\g}=\left[(O^{A})_{W_3\g}^{2}(O^{A})_{B_i Z_l}\right]\nonumber\\
&&R^{B_i YW}_{Z_l\g\g}=\left[2(O^{A})_{B_i Z_l}(O^{A})_{W_3\g}(O^{A})_{Y\g}\right]\,.\nonumber\\
\end{eqnarray}
These expressions will be used extensively in the next section
and computed numerically in a complete brane model.
\subsection{Prompt photons, the Landau-Yang theorem and the anomaly}
Since the analysis of the unitarity bound will involve the study of 
amplitudes with direct photons in the final state mediated by a $Z$ or
a $Z'$ in the $s$-channel, we briefly recall some facts concerning the 
structure of these amplitude and in particular the Landau-Yang theorem.
The condition of transversality of the $Z'$ boson ($e_{Z'}\cdot k=0$)
is essential for the vanishing of this
amplitude. 

The theorem states that a spin 1 particle cannot decay into two
on-shell spin 1 photons because of Bose symmetry and angular momentum conservation.
Angular momentum conservation tells us that the two photons must be
in a spin 1 state (which forces their angular momentum wave function 
to be antisymmetric), while their spatial part is symmetric.
The total wave function is therefore antisymmetric and violates 
the requirement of Bose statistics. For these reasons the amplitude has to vanish.
For a virtual exchange mediated by a $Z'$ the contribution is 
vanishing -after summing over the fermions in each generation of 
the SM-, the theory being anomaly-free, in the chiral limit.
The amplitude is non-vanishing only in the presence of chiral
symmetry breaking terms (fermion masses), which can be induced
both by the QCD vacuum and by the Yukawa couplings of SM in
the presence of electroweak symmetry breaking. For this reason 
it is strongly suppressed also in the SM.

However, the situation in the case of an anomalous vertex
is more subtle. The BIM amplitude is non-vanishing,
but at the same time, as we have explained, is {\em non-resonant}, 
which means that the particle pole due to the $Z^\prime$ has disappeared.
For the rest it will break unitarity at a certain stage.

In fact, a cursory look at the $AVV$ vertex shows that if the external 
photons are on-shell and transverse, the amplitude mediated by this
diagram is proportional to the momentum of the virtual $Z$, $k^{\mu}$. 
This longitudinal momentum
exchange does not set the amplitude to zero unless the production
mechanism is also anomalous. We will show first that in the SM these
processes are naturally suppressed, though not identically zero, since
they are proportional to the fermion masses, due to anomaly cancellation. 
We start our analysis by going back to the $AVV$ diagram, which summarizes
the kinematical behavior of the $Z\gamma \gamma$ amplitude.

Let $k_1$ and $k_2 $ denote the momenta of the two final state photons.
We contract the $AVV$ diagram with the polarization vectors of the photons,
$\varepsilon_{1\mu}$ and $\varepsilon_{2\nu}$ of
the $Z$ boson, $e_{\lambda}$, obtaining
\ba
&&e_{\lambda}\varepsilon_{1\mu}\varepsilon_{2\nu}
\Delta_{AVV}^{\lambda\mu\nu}(k_1,k_2,m_f\neq 0)=-\frac{1}{\pi^2}\int_0^1 dx \int_{0}^{1-x}dy
\frac{x y}{m_f^2 -2 x y k_1\cdot k_2}\nonumber\\
&&\hspace{2cm}e_{\lambda}\varepsilon_{1\mu}\varepsilon_{2\nu}
\left\{\epsilon_{\lambda\nu\mu\alpha}
\left( k_{2}^{\alpha}-k_{1}^{\alpha}\right)
+k_1^{\alpha}k_2^{\beta}
\left( \epsilon_{\alpha\lambda\beta\nu} k_2^{\mu}
-\epsilon_{\alpha\lambda\beta\mu}k_1^{\nu} \right)
\right\}\,,
\label{Davv}
\ea
where we have used the conditions
\ba
&&k_1^2=k_2^2=0\nonumber\\
&&\varepsilon_{1\mu}k_1^{\mu}=\varepsilon_{2\nu}k_2^{\nu}=0.
\ea
It is important to observe that if we apply Schouten's 
identity we can reduce this expression to the form
\ba
&&e_{\lambda}\varepsilon_{1\mu}\varepsilon_{2\nu}
\Delta_{AVV}^{\lambda\mu\nu}(k_1,k_2,m_f\neq 0)
=e_{\lambda}\left( k_1^{\lambda}+k_2^{\lambda}\right) 
\mathcal{F}_f(k^2,m_f)
\ea
with
\ba
\mathcal{F}_f(k^2, m_f)&=&\mathcal{J}_f(k^2) 
\epsilon[k_1,k_2,\varepsilon_1,\varepsilon_2] \nonumber \\
\mathcal{J}_f(k^2) &\equiv &-\frac{1}{\pi^2} 
\int_0^1 dx \int_{0}^{1-x}dy \frac{x y}{m_f^2 -2 x y k_1\cdot k_2},
\ea
which vanishes only if we impose the transversality condition on the polarization
vector of the $Z$ boson, $e_{\lambda}k^{\lambda}=0$.  
Alternatively, this amplitude vanishes if the anomalous 
$Z\gamma\gamma$ vertex is contracted with another gauge invariant vertex, as discussed above.
The amplitude has an anomalous behavior. In fact, contracting with the
$k_{\lambda}$ four-vector we obtain
\ba
&&k_{\lambda}\varepsilon_{1\mu}\varepsilon_{2\nu}\Delta_{AVV}^{\lambda\mu\nu}(k_1,k_2,m_f\neq 0)=
\left(-\frac{1}{\pi^2} \int_0^1 dx \int_{0}^{1-x}dy \frac{k_{\lambda}k^{\lambda}x y}
{m_f^2 -2 x y k_1\cdot k_2} \right)
\epsilon[k_1,k_2,\varepsilon_1,\varepsilon_2]
\nonumber\\
&&\hspace{1cm}=\left(\frac{1}{\pi^2} \int_0^1 dx \int_{0}^{1-x}dy \frac{-2 k_{1}\cdot k_{2}x y}
{m_f^2 -2 x y k_1\cdot k_2} \right)
\epsilon[k_1,k_2,\varepsilon_1,\varepsilon_2]
\nonumber\\
&&\hspace{1cm}=\left(\frac{1}{\pi^2} + \frac{m_f^2}{\pi^2} \int_0^1 dx \int_{0}^{1-x}dy \frac{1}
{m_f^2 -2 x y k_1\cdot k_2} \right)
\epsilon[k_1,k_2,\varepsilon_1,\varepsilon_2]
\ea
where the mass-independent and mass-dependent contributions have been separated. Summing over an anomaly-free generation, the first of these two contributions cancel. For this reason, it is also convenient to isolate the following quantity
\ba
{\mathcal{G}}_{f}(k^2,m_f)=\frac{m_f^2}{\pi^2} \int_0^1 dx \int_{0}^{1-x}dy \frac{1}
{m_f^2 -2 x y k_1\cdot k_2}
\epsilon[k_1,k_2,\varepsilon_1,\varepsilon_2]
\ea
which is the only contribution to the triangle amplitude in the SM for a given fermion flavor $f$.
\begin{figure}[t]
{\centering \resizebox*{11cm}{!}{\rotatebox{0}
{\includegraphics{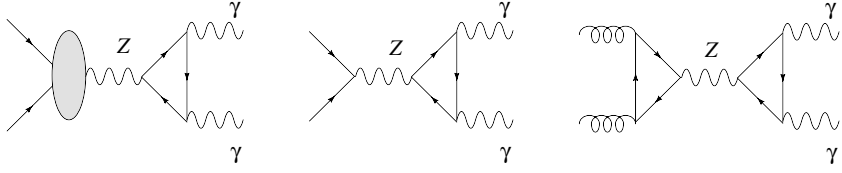}}}\par}
\caption{\small Vanishing and non-vanishing amplitudes mediated by an axial-vector spin 1.}
\label{VV}
\end{figure}

We illustrate in Fig.~\ref{VV} the contributions to the 
$q\bar{q}$ annihiliation channel at all orders (first graph), 
at leading order (second graph), and the BIM amplitude (third graph), 
all attached to an $AVV$ final state. In an anomaly-free theory all
these processes are only sensitive to the difference in masses among
the flavors, since degeneracy in the fermion 
mass sets these contributions
to zero. Only the second graph vanishes identically 
due to the WI on the $q\bar{q}$ channel also 
away from the chiral limit. For instance in the third diagram 
chiral symmetry breaking is sufficient to induce violations of 
the WI on the initial state vertex, due to the
different quark masses within a given fermion generation.

To illustrate, in more detail, how chiral symmetry breaking can induce the 
exchange of a scalar component in the process with two prompt photons, we
start from the SM case, where the $Z^* \gamma \gamma$ vertex, multiplied by 
external physical polarizations for the two photons becomes
\ba
V^\lambda_{Z^*\gamma\gamma} = k^\lambda \frac{g_2}{2\cos{\theta}_W} e^2 \sum_f(Q_f)^2
g_{A,f}^{Z}\mathcal{F}_f(k^2, m_f)
\ea
and consider the quark antiquark annihilation channel  
$q\bar{q}\rightarrow Z^{*}\rightarrow\g\g$.
We work in the parton model with massless light quarks. 
We can rewrite the amplitude as
\ba
\mathcal{ M} &=&V^\lambda_{ff Z^*}\Pi_{\lambda,\lambda^{\prime},\xi} V^{\lambda'}_{Z^*\gamma\gamma},
\ea
where we have introduced the $Z q\bar{q}$  vertex
\ba
V_{ff Z}=  \bar{v}(p_1)\Gamma^{\lambda'} u(p_2) \hspace {1cm} \Gamma^\lambda = i g_Z\g^{\lambda}(g_V-g_A\g^{5}),
\ea
and the expression of the propagator of the $Z$ in the $R^{}_\xi$ gauge
\ba
\Pi_{\lambda,\lambda^{\prime},\xi}=\frac{-i}{k^2-M_Z^2}\left[g^{\lambda\lambda^{\prime}}
-\frac{k^{\lambda}k^{\lambda^{\prime}}}{k^2-\xi M_Z^2}(1-\xi)\right].
\ea
To move to the unitary gauge, we split the propagator of the $Z$ boson as
\ba
\Pi_{\lambda,\lambda^{\prime},\xi}=\frac{-i}{k^2-M_Z^2}\left[g^{\lambda\lambda^{\prime}}
-\frac{k^{\lambda}k^{\lambda^{\prime}}}{M_Z^2}\right]
+\frac{-i}{k^2-\xi M_Z^2}\left(\frac{k^{\lambda}k^{\lambda^{\prime}}}{M_Z^2}\right)
\ea
and go to the unitary gauge by choosing $\xi\rightarrow \infty$.  The amplitude will then be written as
\ba
{\cal M} &=& \bar{v}(p_1)\Gamma^\lambda u(p_2)\frac{-i}{k^2-M_Z^2}\left[g^{\lambda\lambda^{\prime}}
-\frac{k^{\lambda}k^{\lambda^{\prime}}}{M_Z^2}\right]
k^{\lambda^{\prime}}\left( \frac{g_2}{2\cos{\theta}_W} e^2 \sum_f(Q_f)^2
g_{A,f}^{Z}\mathcal{F}_f(k^2, m_f) \right)
\nonumber\\
&=&\frac{i}{M_Z^2} \bar{v}(p_1)\Gamma^\lambda u(p_2) k^\lambda  \left( \frac{g_2}{2\cos{\theta}_W} e^2 \sum_f(Q_f)^2
g_{A,f}^{Z}\mathcal{F}_f(k^2, m_f) \right). 
\ea
Clearly, at the Born level, using the WI
on the left $V_{ff Z^*}$ vertex, we find that the
amplitude is zero. This result remains unchanged if we
include higher-order corrections (strong/electroweak),
since the structure of this vertex is just modified by a
Pauli (weak-electric) form factor and the additional
contribution vanishes after contraction with the momentum $k^\lambda $.
This amplitude is however non-vanishing if we replace
the $V_{ff Z^*}$ vertex with a $V_{gg Z^*}$ vertex, where now we assume that the
new vertex is computed for non-zero fermion masses (i.e. away from the chiral limit).
In this case we use the WI
\beqn
k_{\rho} \, V_{gg Z^*}^{\rho \nu \mu}  &=&  (p_1 + p_2 )_{\rho} \,  G^{\,  \rho \nu \mu }\nonumber\\
&=& -\frac{ e^2 g_2 }{2 \cos \theta_{W}}  \sum_{q}  g^{Z}_{A,q}  Q^{2}_{q} \,  \epsilon^{\, \nu \mu \alpha \beta }
p_{1 \alpha} p_{2 \beta } \, \left[  \frac{1}{\pi^{2}}  + \frac{m^{2}_{q}}{\pi^{2}} \int^{1}_{0} d x_1 \int^{1 - x_1 }_{0}
d x_2  \,  \frac{1}{ \Delta_q }    \right]
\label{ABJJ_anomaly}
\eeqn
with
\ba
\Delta_q= - x y k^2 +  m_q^2,
\ea
where the constant term $(1/\pi^2)$ vanishes in an anomaly-free theory.
It is convenient to define the function
\ba
\mathcal{G}_q(k^2,m_q)= \epsilon\left[\varepsilon_{1 g},
\varepsilon_{2 g},p_1,p_2\right] m^{2}_{q} \mathcal{I}_q(m_q)
\ea
with
\ba
\mathcal{I}_q(m_q) = \frac{1}{\pi^{2}}\int^{1}_{0} d x_1 \int^{1 - x_1 }_{0} d x_2  \,  \frac{1}{ \Delta_q},
 \ea
where $\varepsilon $ is the polarization of the gluon, which allows one to express the squared amplitude as
\ba
\langle |\mathcal{M}_{gg\to \gamma\gamma}|^2\rangle &=& \left( \frac{ e^2 g_2 }{2\cos \theta_{W}} \right)^4
 \frac{s^6}{4 M_Z^4} \left( \sum_{q}  g^{q}_{A}  Q^{2}_{q}\mathcal{G}_q(m_q)\right)^2
\left( \sum_{f}  g^{f}_{A}  Q^{2}_{f} \mathcal{G}_f(m_f)\right)^2, 
\ea
with $s\equiv k^2$. Notice that in the large energy limit  
$\mathcal{I}_q\sim \mathcal{J}_f\sim 1/{s^2}$
\cite{Achasov:1992bu}. This shows that double prompt photon 
production, in the SM, is non-resonant and is proportional
to the quark masses, neglecting the contributions coming 
from the masses of the leptons.
\section{Gauge parameter dependence in the physical basis}
When this analysis is extended to a complete anomalous
model such as the mLSOM
\cite{Coriano:2005js, Irges:1998ax, Coriano:2007xg, Armillis:2007tb} even the direct proof of the cancellation
of the gauge dependence
in the $Z^\prime$ exchange is quite complex and not obvious, 
although it is expected at a formal level.
We recall that we are working with a broken phase and the axion
has been decomposed into its physical component ($\chi$, which is the axi-Higgs)
and the Goldstone modes.
In \cite{Armillis:2007tb} it has been shown that the counterterms
of the theory can be fixed in the St\"uckelberg phase and then re-expressed,
in the Higgs-St\"uckelberg phase, in the physical base, which confirm
the correctness of the approach followed in Chap.~\ref{chap:AbelianModels2} for the
determination of the effective Lagrangian of the model after electroweak symmetry breaking.

\begin{figure}[t]
{\centering \resizebox*{8.5cm}{!}{\rotatebox{0}
{\includegraphics{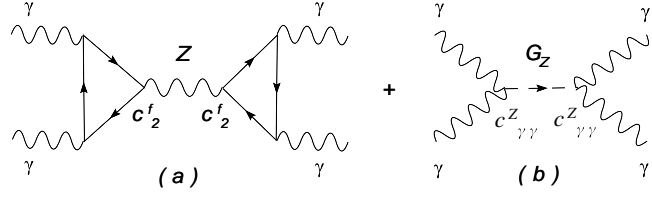}}}\par}
\caption{\small Cancellation of the spurious poles in the physical basis, $m^{}_{f}= 0$.}
\label{gauge_dependence}
\end{figure}

The matrix $O^{\chi}$, needed to rotate into the mass eigenstates
of the $CP$-odd sector, relating the axion $\chi$ and the
two neutral Goldstones of this sector to the
St\"uckelberg field $b$ and the $CP$-odd phases of the two Higgs doublets
satisfies the following relation
\begin{displaymath}
\left(
\begin{array}{c} ImH^{0}_{u} \\
                 ImH^{0}_{d} \\
					  b
\end{array} \right)=O^{\chi}
\left(
\begin{array}{c} \chi\\
                 G^{0}_1\\
					  G^{0}_2
\end{array} \right)\,,
\end{displaymath}
where the Goldstones in the physical basis are obtained by the following combination
\ba
\label{Goldstones}
G^{Z}&=&G^{0}_1\left[f_u \frac{v_u}{M_Z}O^{\chi}_{12}
+f_d \frac{v_d}{M_Z}O^{\chi}_{22} +g_B \frac{M_1}{M_Z}O^{A}_{ZB}O^{\chi}_{32}\right]
\nonumber\\
&+&G^{0}_2\left[f_u \frac{v_u}{M_Z}O^{\chi}_{13}
+f_d \frac{v_d}{M_Z}O^{\chi}_{23} +g_B \frac{M_1}{M_Z}O^{A}_{ZB}O^{\chi}_{33}\right]
\nonumber\\
&=&c_1 G^{0}_1 +c_2 G^{0}_2\nonumber\\
G^{Z^{\prime}}&=&G^{0}_1\left[f_{u,B} \frac{v_u}{M_Z^{\prime}}O^{\chi}_{12}
+f_{d,B} \frac{v_d}{M_Z^{\prime}}O^{\chi}_{22} +g_B \frac{M_1}{M_Z^{\prime}}
O^{A}_{Z^{\prime}B}O^{\chi}_{32}\right]
\nonumber\\
&+&G^{0}_2\left[f_{u,B} \frac{v_u}{M_Z^{\prime}}O^{\chi}_{13}
+f_{d,B} \frac{v_d}{M_Z^{\prime}}O^{\chi}_{23} +g_B \frac{M_1}{M_Z^{\prime}}
O^{A}_{Z^{\prime}B}O^{\chi}_{33}\right]
\nonumber\\
&=&c^{\prime}_1 G^{0}_1 +c^{\prime}_2 G^{0}_2\,.
\ea
Here we have defined the following coefficients
\ba
f_{u}=g_2 O^{A}_{ZW_3}-g_{Y}O^{A}_{ZY}-q^{B}_{u} g_B O^{A}_{ZB} &&
f_{d}=g_2 O^{A}_{ZW_3}-g_{Y}O^{A}_{ZY}-q^{B}_{d} g_B O^{A}_{ZB}
\nonumber\\
f_{u,B}=g_2 O^{A}_{Z^{\prime}W_3}-g_{Y}O^{A}_{Z^{\prime}Y}-q^{B}_{u} g_B O^{A}_{Z^{\prime}B} &&
f_{d,B}=g_2 O^{A}_{Z^{\prime}W_3}-g_{Y}O^{A}_{Z^{\prime}Y}-q^{B}_{d} g_B O^{A}_{Z^{\prime}B}\,,
\ea
and the $q^{B}_{u,d}$ charges are defined in Tab.~\ref{ccharge_higgs}.
The relations containing the physical Goldstones can be inverted so we obtain
\ba
&&G^{0}_{1}=C_1 G^Z + C_2 G^{Z^{\prime}}\nonumber\\
&&G^{0}_{2}=C_1^{\prime} G^Z +C_2^{\prime} G^{Z^{\prime}}\,,
\ea
where we give the explicit expression only for the coefficient $C^{\prime}_1$,
since this is the one relevant for our purposes.
Then, after the orthonormalization procedure, we obtain
\ba
C^{\prime}_1=\frac{c_2}{\sqrt{c_2^2+c^{\prime 2}_2} }\,.
\ea
We illustrate the proof of gauge independence from the $Z$ gauge parameter of
the amplitudes in Fig.~\ref{gauge_dependence}.
In this case the cancellation of the spurious poles takes place
via the combined exchange of the $Z$ propagator and of the corresponding
Goldstone mode $G_Z$. If we isolate the gauge-dependent
part in the $Z$ boson propagator we obtain
\ba
&&{\cal M}_{A\xi} + {\cal M}_{B\xi}    \nonumber\\
&=& \left( - \frac{1}{2}  c^{f}_{2} \Delta^{\lambda \mu \nu}_{AVV}(p_1, p_2) \right) \frac{- i }{k^2 - \xi_Z M^2_Z } \left( \frac{k^\lambda k^{\lambda^\prime} }{ M^{2}_{Z} } \right)   \left( \frac{1}{2} c^{f}_{2} \Delta^{\lambda^\prime \mu^\prime \nu^\prime}_{AVV}(k_1, k_2) \right)   \nonumber\\
&& + 4 \times \left( 4 c^{Z}_{\g \g} \epsilon[p_1, p_2, \mu, \nu] \right)  \frac{ i }{k^2 - \xi_Z M^2_Z }
 \left( 4 c^{Z}_{\g \g} \epsilon[k_1, k_2, \mu^{\prime}, \nu^{\prime}] \right)     \nonumber\\
&=&  \left( - \frac{a_n}{2}  c^{f}_{2}  \epsilon[p_1, p_2, \mu, \nu]    \right) \frac{- i }{k^2 - \xi_Z M^{2}_{Z} } \left( \frac{1}{ M^2_Z } \right)   \left( \frac{a_n}{2} c^{f}_{2}   \epsilon[k_1, k_2, \mu^{\prime}, \nu^{\prime}]     \right)   \nonumber\\
&& + 4 \times \left( 4 c^{Z}_{\g \g} \epsilon[p_1, p_2, \mu, \nu] \right)  \frac{ i }{k^2 - \xi_Z M^{2}_{Z} }
 \left( 4 c^{Z}_{\g \g} \epsilon[k_1, k_2, \mu^{\prime}, \nu^{\prime}] \right)\,,
\ea
where
\ba
\label{cZ_coeff}
&&c^{Z}_{\g \g}=\left[\frac{F}{M_1}(O^{A}_{W\g})^2+\frac{C_{YY}}{M_1}(O^{A}_{Y\g})^2\right]O^{\chi}_{33}C^{\prime}_1.
\ea
The coefficient $c^{Z}_{\g\g}$ in the ${\cal M}_{B\xi}$  amplitude ($G_Z$ exchange)
must be compared with the massless coefficients
$c^{f}_{2}$ of the ${\cal M}_{A\xi}$ amplitude (Z boson exchange) and
the explicit expressions of the coefficients $C_{YY}$ and $F$ have been already computed in Sec.~\ref{sec:WI_discuss}.
Adding the contributions of the two diagrams we obtain
\ba
&&{\cal M}_{A\xi} + {\cal M}_{B\xi}    \nonumber\\
&=&  \left\{  \frac{a_n}{2}  \left[ g_B g_Y^2\theta_f^{BYY}R^{BYY}_{Z\g\g}
+ g_B g_2^2\theta_f^{BWW}R^{BWW}_{Z\g\g} \right]  \epsilon[p_1, p_2, \mu, \nu]\right\}
\frac{ i }{k^2 - \xi_Z M^{2}_{Z} } \left( \frac{1}{ M^2_Z } \right)     \nonumber\\
&&\left\{ \frac{a_n}{2} \left[ g_B g_Y^2\theta_f^{BYY}R^{BYY}_{Z\g\g}
+ g_B g_2^2\theta_f^{BWW}R^{BWW}_{Z\g\g}\right]
\epsilon[k_1, k_2, \mu^{\prime}, \nu^{\prime}] \right\}   \nonumber\\
&& + 4 \times \left\{ 4 \left[\frac{F}{M_1}(O^{A}_{W\g})^2
+\frac{C_{YY}}{M_1}(O^{A}_{Y\g})^2 \right] O^{\chi}_{33}C^{\prime}_1
\epsilon[p_1, p_2, \mu, \nu] \right\}
\frac{ i }{k^2 - \xi_Z M^{2}_{Z} }    \nonumber\\
&& \left\{ 4  \left[\frac{F}{M_1}(O^{A}_{W\g})^2+\frac{C_{YY}}{M_1}(O^{A}_{Y\g})^2\right]
O^{\chi}_{33}C^{\prime}_1 \epsilon[k_1, k_2, \mu^{\prime}, \nu^{\prime}] \right\}.
\ea
At this point, the pattern of cancellation can be separated
in three different sectors, a pure $BYY$ sector, a pure $BWW$ and
mixed $BYY$-$BWW$ sectors, and it requires the validity of the relations
\ba
&& \left(  \frac{a_n}{2}  g_B g_Y^2\theta_f^{BYY}R^{BYY}_{Z\g\g}   \right)^2 \frac{1}{M^2_Z} +
4  \left( 4 \frac{C_{YY}}{M_1}(O^{A}_{Y\g})^2  O^{\chi}_{33}C^{\prime}_1 \right)^2     = 0, \\
&& \left(  \frac{a_n}{2}  g_B g_2^2 \theta_f^{BWW}R^{BWW}_{Z\g\g}   \right)^2 \frac{1}{M^2_Z} +
4  \left( 4 \frac{ F }{ M_1 }( O^{A}_{W\g})^2  O^{\chi}_{33}C^{\prime}_1 \right)^2  =  0  ,\\
&& \left( a_n \, g_B g_Y^2 \theta_f^{BYY} R^{BYY}_{Z\g\g}  a_n \, g_B g_2^2 \theta_f^{BWW} R^{BWW}_{Z\g\g}
\right)
\frac{1}{M^2_Z}  +
4 \left( 8 \frac{ F }{ M_1 }( O^{A}_{W\g})^2
\frac{C_{YY}}{M_1}(O^{A}_{Y\g})^2 O^{\chi}_{33}C^{\prime}_1  \right)  = 0.  \nonumber\\
\ea
We have been able to verify that these relations are automatically
satisfied because of the following identity, which connects the rotation
matrix of the interaction to the mass eigenstates $O^A$ to a component
of the matrix $O^{\chi}$. This matrix appears in the rotation from the basis
of St\"uckelberg axions to the basis of the
Goldstones $G^{}_Z$ and $G^{}_{Z^\prime}$ and of the axi-Higgs $\chi$.
The relation is
\ba
O^{A}_{BZ}  \frac{1}{M_Z} = 2 O^{\chi}_{33}C^{\prime}_1 \frac{1}{M_1}\,,
\label{magic_relation_GM}
\ea
with
\ba
O^{\chi}_{33} = \frac{1}{  \sqrt{ \frac{(q^B_u - q^B_d)^2}{M^2_1} \frac{v^2_u v^2_d}{v^2_u+v^2_d }  + 1} },
\ea
with $M_1$ the St\"uckelberg mass.
The origin of this connection has to be found in the 
Yukawa sector and the condition of gauge invariance of the Yukawa couplings.
\section{Unitarity bounds: the partonic contribution $gg\rightarrow \g\g$}
\begin{figure}[h]
{\centering \resizebox*{12cm}{!}{\rotatebox{0}
{\includegraphics{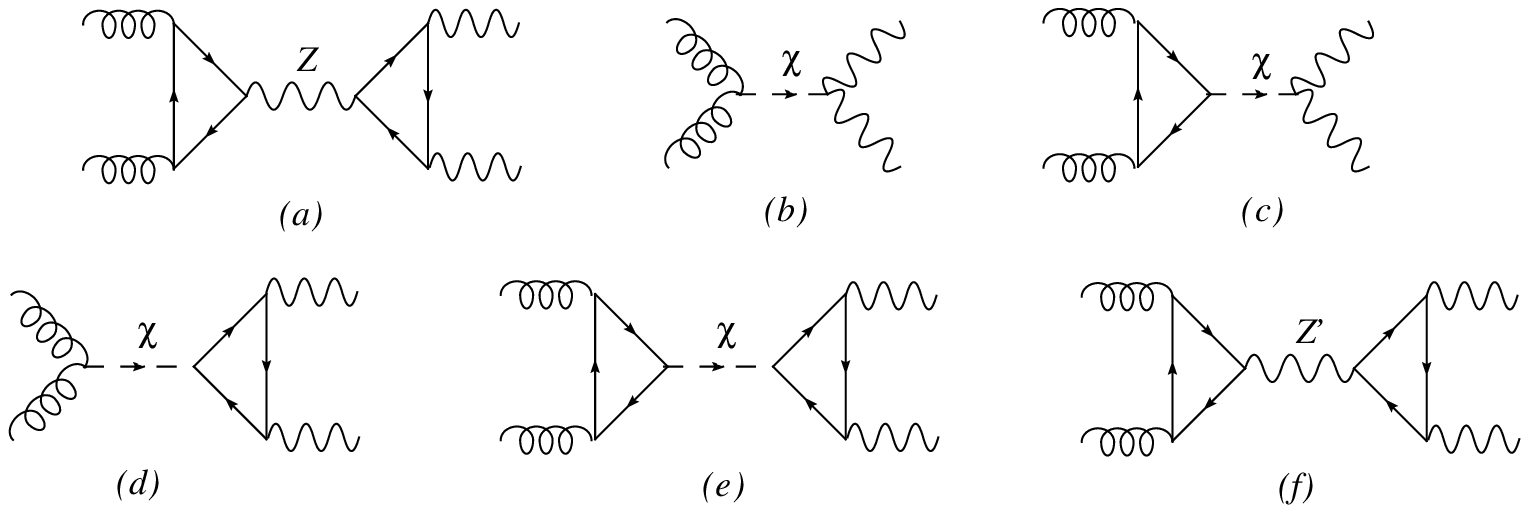}}}\par}
\caption{\small Anomalous contributions to $gg\to \gamma\gamma$ in the mLSOM.}
\label{GS}
\end{figure}

In this section we compute the $gg\rightarrow \g\g$ cross section
with two on-shell gluons and two on-shell photons in the final state.
The same computation is carried out both in the SM and in the mLSOM,
where the charge assignments have been determined as in \cite{Ibanez:2001nd, Ibanez:1998qp}, to determine
the different behavior of these amplitudes in the two cases. The list of contributions
that we have included are all shown in Fig.~\ref{GS}. 
We report some of the results of the
graphs in order to clarify the notation. For instance we obtain for diagram (a)
\ba
\sigma_A(s)=\frac{1}{2048 \pi} \left[\sum_q \frac{1}{2} c_1^{q}A_{6,q}\right]^2
\left[\sum_{f} \frac{1}{2} c_2^{f}A_{6,f}\right]^2
\frac{s^5}{M_Z^4},
\ea
where we have defined the coefficients
\begin{eqnarray}
&&c_1^{q} = g_3^2 \left[g_Y\theta^{Y}_{q}O^{A}_{YZ}+g_2\theta^{W}_{q}O^{A}_{WZ}
+g_B\theta^{B}_{q}O^{A}_{BZ}\right],
\nonumber\\
&&c_2^{f} = \left[g_Y^3\theta^{YYY}_{f}\bar{R}^{YYY}_{Z\g \g} + g_Y^3\theta^{WWW}_{f}\bar{R}^{WWW}_{Z\g \g}
+ g_Y g_2^2\theta^{YWW}_{f}R^{YWW}_{Z\g \g}+ g_Y^2 g_2\theta^{YYW}_{f}R^{YYW}_{Z\g \g}
\right.  \nonumber\\
 &&\left.+g_B g_Y^2\theta^{BYY}_{f}R^{BYY}_{Z\g \g}+g_B g_2^2\theta^{BWW}_{f}R^{BWW}_{Z\g \g}
+g_B g_Y g_2\theta^{BYW}_{f}R^{BYW}_{Z\g \g}\right]\,,
\end{eqnarray}
and the mass $M^{}_{Z^\prime}$ of the extra $Z^\prime$ is given by  Eq.~(\ref{ZZpmass}). 
We recall that $A_6^{f}(s)$ is approximately 
\ba
A_6^{f}(s)\approx \frac{1}{2\pi^2 s} -\frac{m_f^2}{2 s^2} 
+ O\left(\frac{m_f^2}{s^2}\log\left(\frac{m_f^2}{s^2}\right)\right),
\ea
at large values of $s$. We have seen that the SM contribution, in the
presence of a massive fermion circulating in the loop is suppressed
by a factor that is $O\left(m_f^2/s^2\right)$.
In the case of an anomalous model this contribution becomes
subleading, the dominant one coming from the anomalous parts,
proportional to the chiral asymmetries $\theta_f$
of the anomalous charge
assignments between left-handed and right-handed fermion modes.
The amplitude in diagram (b) is given by
\ba
\sigma_{B}(s)=\frac{2}{\pi}
\left(g^{\chi}_{gg}\right)^2 \left(g^{\chi}_{\g\g}\right)^2
\frac{ s^3}{(s-M_{\chi}^2)^2}\,.
\ea
where, for convenience, we have defined
\ba
\label{GS_coeffs}
&&g^{\chi}_{gg}=\frac{D}{M}O^{\chi}_{31},
\\
&&g^{\chi}_{\g\g}=\left[\frac{F}{M}(O^{A}_{W\g})^2
+\frac{C_{YY}}{M}(O^{A}_{Y\g})^2\right]O^{\chi}_{31}.
\ea
Here we have used the model-dependent WZ coefficients $D, F$ and $C^{}_{YY}$ already computed in Sec.~\ref{sec:WI_discuss}.
Proceeding in a similar way, graph c) gives
\ba
\sigma_C(s)= \frac{1}{32\pi}  \frac{ \,s^3}{(s-M_{\chi}^2)^2} (g^{\chi}_{gg})^{2}
\left[  \sum_{q} C_0(s,m_q)  c^{\chi, q}_{gg} \right]^2
\ea
where we have used Eq.~(\ref{di_log}) and we have set
\ba
C_0(s,m_f) = - \frac{m_f}{4\pi^2 s} \log^2{\frac{\rho^{}_f+1}{\rho^{}_f-1}}.
\ea
Starting from the Yukawa couplings shown in Lagrangian~(\ref{action}) it is convenient 
to define the following coefficients, which will be used below
\ba
&&c^{\chi,q}_{gg}= g^{2}_{3} c^{\chi,q},   \qquad q=u,d   \nonumber\\
&&c^{\chi,f}_{\gamma\gamma}= e^{2} c^{\chi,f},   \qquad q=u,d,\nu, e.
\label{chi_coupling}
\ea
We have used a condensed notation for the flavors, 
with u = \{u, c, t\}, d = \{d, s, b\}, $\nu$ = \{$\nu_{e}$, $\nu_{\mu}$, $\nu_{\tau}$\}
and e = \{e, $\mu$, $\tau$\}. The couplings of the physical axion to the fermions are given by
\ba
c^{\chi, u} &=& \Gamma^{u}  \frac{i}{\sqrt 2} O^{\chi}_{11} = \frac{m^{}_{u}}{v_u} i O^{\chi}_{11} ,  \qquad
c^{\chi, d} = - \Gamma^{d} \frac{i}{\sqrt 2} O^{\chi}_{21} = - \frac{m^{}_{d}}{v^{}_{d}} i O^{\chi}_{21},    \nonumber\\
c^{\chi, \nu} &=& \Gamma^{\nu}  \frac{i}{\sqrt 2} O^{\chi}_{11} = \frac{m^{}_{\nu}}{v^{}_u} i O^{\chi}_{11} ,  \qquad
c^{\chi, e} = - \Gamma^{e} \frac{i}{\sqrt 2} O^{\chi}_{21} = - \frac{m^{}_{e}}{v^{}_{d}} i O^{\chi}_{21}.
\ea
We have also relied on the $O^{\chi}$ matrix elements introduced in Sec.~\ref{sec:Ochi}.
The fermion masses have been expressed in terms of the Yukawa couplings by the relations
\ba
m_u = \frac{v_u \Gamma_u}{\sqrt 2},  \,\,\,\,\,\, m_d = \frac{v_d \Gamma_d}{\sqrt 2},
\,\,\,\,\,\, m_\nu = \frac{v_u \Gamma_\nu}{\sqrt 2}, \,\,\,\,\,\, m_e = \frac{v_d \Gamma_e}{\sqrt 2}.
\label{fermionmass}
\ea
\begin{figure}[t]
{\centering \resizebox*{9cm}{!}{\rotatebox{-90}
{\includegraphics{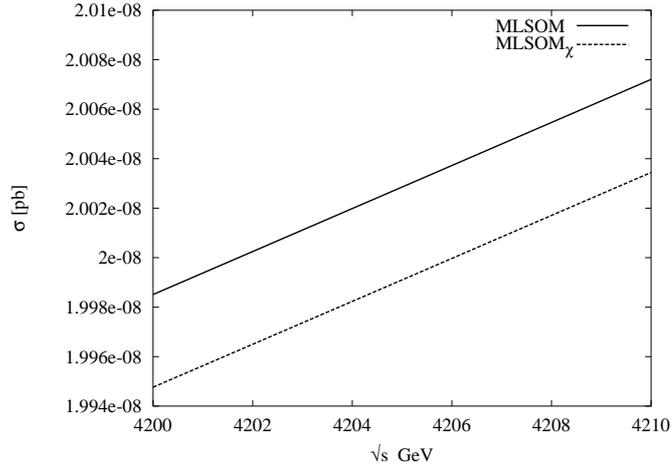}}}\par}
\caption{\small Partonic cross section for the anomalous process $gg\rightarrow \g\g$
with $m_f=0$, $\tan\beta=40$, $g_B=0.1$, $M_{\chi}=10$ GeV and $M_{1}=800$ GeV.
The solid line refers only to the exchange of the $Z$ and the $Z^\prime$, while
the dashed line refers to the complete cross section including the $\chi$ exchange.}
\label{MasslessGGgaga}
\end{figure}
\begin{figure}[h]
{\centering \resizebox*{9cm}{!}{\rotatebox{-90}
{\includegraphics{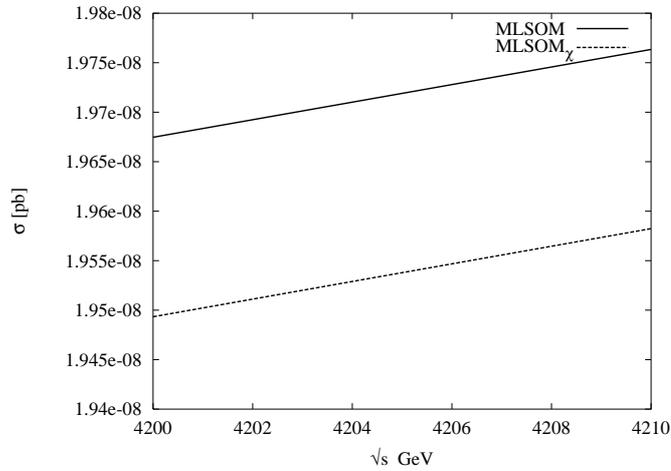}}}\par}
\caption{\small The same as in Fig.~\ref{MasslessGGgaga} but with $m_f\neq 0$.}
\label{MassiveGGgaga}
\end{figure}
\begin{figure}[t]
{\centering \resizebox*{9cm}{!}{\rotatebox{-90}
{\includegraphics{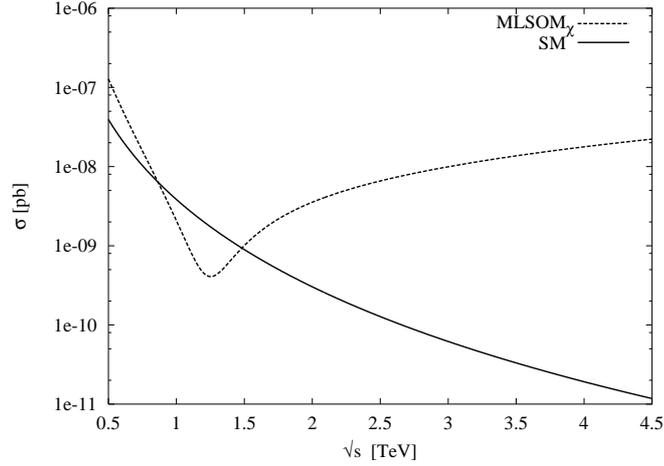}}}\par}
\caption{\small Partonic cross section for the anomalous process 
$gg\rightarrow \g\g$, with $\tan\beta=40$, $g_B=0.1$, $M_{\chi}=10$ GeV and $M_{1}=800$ GeV.
The solid line refers to the SM with the exchange of the $Z$ boson.
The dashed line refers to the mLSOM case. The point of minimum 
divides the anomaly-free region from the region where the anomalous contributions dominate.}
\label{SM_vs_MLSOm}
\end{figure}
The cross section for the amplitude (d) in Fig.~\ref{GS} is given by
\ba
\sigma_{D}(s) =  \frac{s^{3}}{ 32 \pi (s - M^{2}_{\chi})^{2}} (g^{\chi}_{\g\g})^{2}
 \left[  \sum_{f} C_{0}(s,m_{f})  c^{\chi, f}_{\g\g} \right]^2,
\ea
With these notations, we are now ready to express the cross section for graph e) as
\ba
\sigma_{E}(s) &=& \frac{s^{3}}{2048 \pi (s - M^{2}_{\chi})^{2}}
\left[  \sum_{q} C_{0}(s, m_q) c^{\chi, q}_{gg} \right]^2
\left[  \sum_{f} C_{0}(s, m_{f} ) c^{\chi, f}_{\g\g} \right]^2.
\ea
Finally, the cross section for the $Z^{\prime}$ exchange is given by
\ba
\sigma_F(s)=\frac{1}{2048 \pi} \left[\sum_q \frac{1}{2} {d}_1^{q}A_{6,q}\right]^2
\left[\sum_{f} \frac{1}{2} d_2^{f}A_{6,f}\right]^2
\frac{s^5}{M_{Z^\prime}^4}\,,
\ea
where, again, in order to simplify the notation we have defined the coefficients
\begin{eqnarray}
d_1^{q} &=& g_3^2 \left[g_Y\theta^{Y}_{q} O^{A}_{YZ^\prime}+g_2\theta^{W}_{q} O^{A}_{WZ^\prime}
+g_B\theta^{B}_{q} O^{A}_{BZ^\prime} \right],
\nonumber\\
d_2^{f} &=& \left[g_Y^3 \theta^{YYY}_{f} \bar{R}^{YYY}_{Z^\prime \g \g}
+g_2^3 \theta^{WWW}_{f} \bar{R}^{WWW}_{Z^\prime \g \g}
+ g^{}_Y g_2^2\theta^{YWW}_{f}R^{YWW}_{Z^\prime \g \g}+ g_Y^2 g^{}_2 \theta^{YYW}_{f}R^{YYW}_{Z^\prime \g \g}
\right.  
\nonumber\\ 
&&
\left. 
+ g^{}_B g_Y^2 \theta^{BYY}_{f} R^{BYY}_{Z^\prime \g \g } + g^{}_B  g_2^2 \theta^{BWW}_{f} R^{BWW}_{Z^\prime \g \g}
+ g^{}_B g^{}_Y g^{}_2 \theta^{BYW}_{f} R^{BYW}_{Z^\prime \g \g }  \right].
\end{eqnarray}
\begin{figure}[h]
{\centering \resizebox*{9cm}{!}{\rotatebox{-90}
{\includegraphics{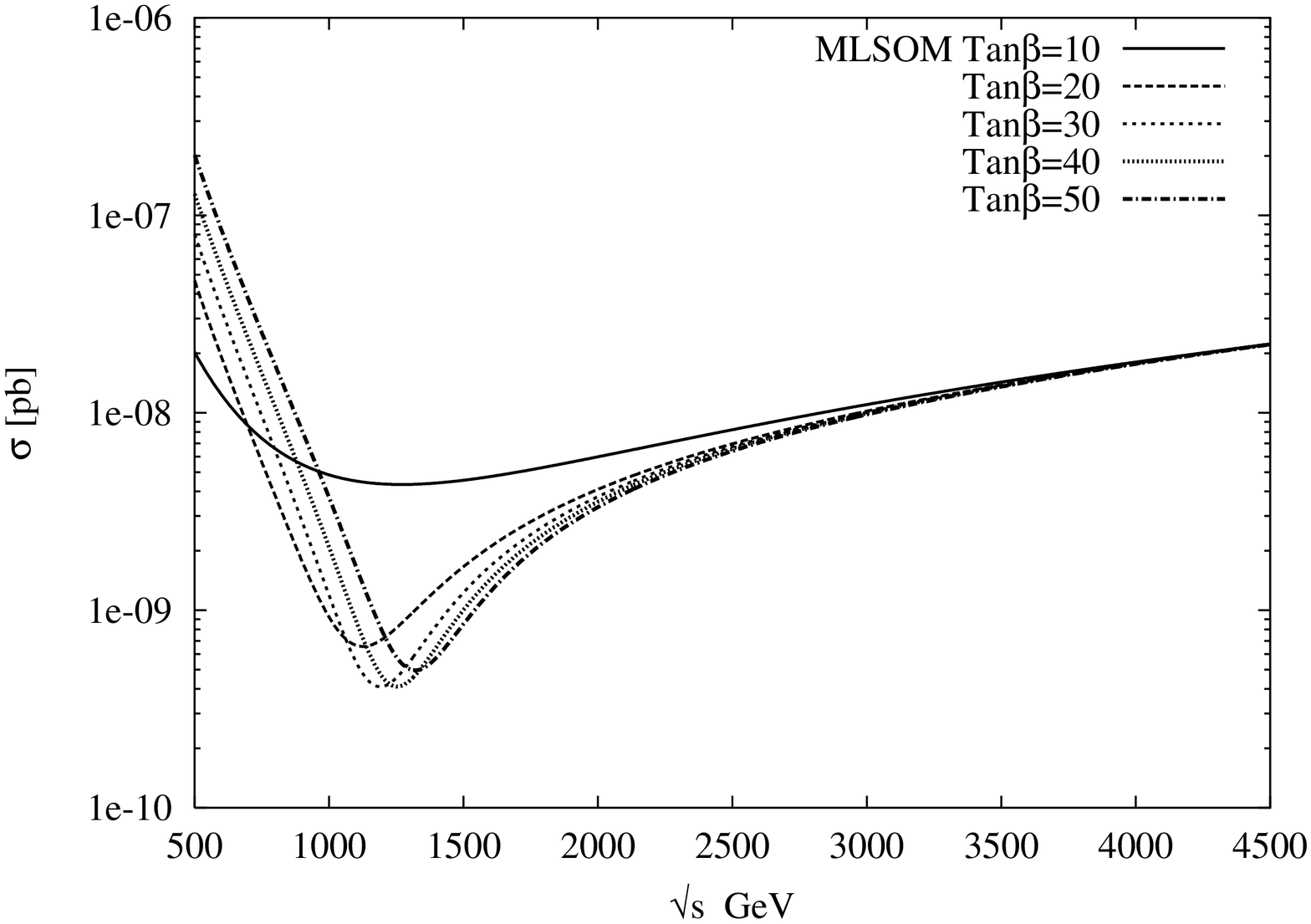}}}\par}
\caption{\small Partonic cross section for the 
anomalous process $gg\rightarrow \g\g$ with $g_B=0.1$, 
$M_{\chi}=10$ GeV and $M_{1}=800$ GeV.
The lines refer to the cross section evaluated 
for different values of $\tan\beta$.
\label{tanbeta_vs_MLSOm}}
\end{figure}
\begin{figure}[t]
{\centering \resizebox*{9cm}{!}{\rotatebox{-90}
{\includegraphics{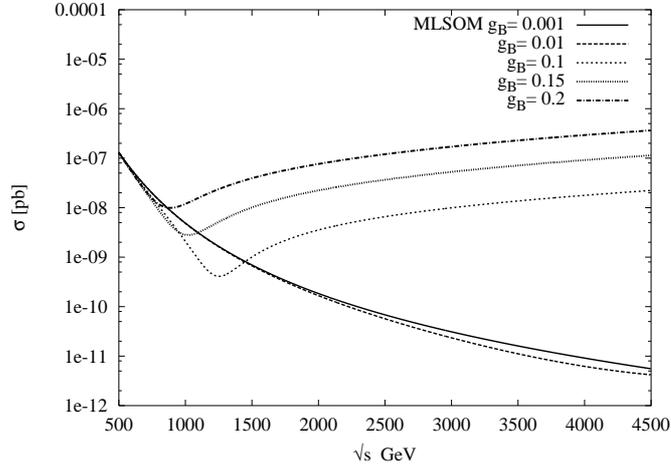}}}\par}
\caption{\small Partonic cross section for the anomalous 
process $gg\rightarrow \g\g$ with $\tan\beta=40$, $M_{\chi}=10$ GeV and $M_{1}=800$ GeV.
The lines refer to the cross section evaluated for different values of the
coupling constant $g_B$.
\label{gB_vs_MLSOm}}
\end{figure}
\begin{figure}[t]
{\centering \resizebox*{9cm}{!}{\rotatebox{-90}
{\includegraphics{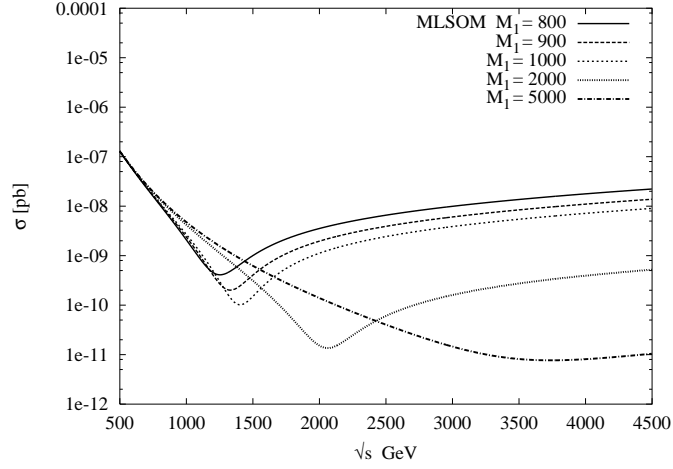}}}\par}
\caption{\small Partonic cross section for the anomalous process $gg\rightarrow \g\g$, with $\tan\beta=40$, $g_B=0.1$ and $M_{\chi}=10$ GeV.
The lines refer to the cross section evaluated for different values of the
St\"uckelberg mass $M^{}_1$.
}
\label{diciotto}
\end{figure}
In the SM case we work in the unitary gauge, 
being a tree-level $Z$ exchange, and we have only
one contribution whose explicit expression is given by
\ba
\sigma^{SM}(s)&=&\frac{1}{32768 \pi^9}\frac{1}{s^3 M_Z^4}
\left[\sum_{q}c_{1,q}^{SM} m^2_q \log^2\left(\frac{\rho_q+1}{\rho_q-1}\right) \right]^{2}
\nonumber\\
&&\times \left[\sum_{f}
c_{2,f}^{SM} m^2_{f} \log^2\left( \frac{\rho_{f} + 1 }{ \rho_{f} - 1 }\right) \right]^{2},
\ea
where we have defined the SM coefficients
\ba
&&c_{1,q}^{SM}= g^{2}_{3}\left[g_Y\theta^{Y}_{q} O^{A}_{YZ}+g_2\theta^{W}_{q}O^{A}_{WZ}\right]
\nonumber\\
&&c_{2,f}^{SM} = \left[ g_Y^3 \theta^{YYY}_{f }\bar{R}^{YYY} + g_2^3\theta^{WWW}_{f }\bar{R}^{WWW}
+ g_Y g_2^2\theta^{YWW}_{f }R^{YWW}+ g_Y^2 g_2\theta^{YYW}_{f }R^{YYW}
\right]\,.\nonumber\\
\ea
The partonic SM cross section is in agreement with unitarity 
in its asymptotic behavior, which is given by
\ba
\sigma^{SM}(s\rightarrow \infty)\approx\frac{\left[ \sum_{q} c_{1,q}^{SM} m^2_q \right]^2 \left[ \sum_{f}
c_{2,f}^{SM} m^2_{f}\right]^2}{s^3 M_Z^4} \, .
\ea
In our anomalous model the complete expression of the same cross section is given by
\ba
&&\sigma^{mLSOM}(s)  \nonumber\\
&&=  \left\{\frac{1}{2048\pi} \frac{s^5}{M_Z^4} 
\left[\frac{1}{2} \sum_q c_1^q \frac{\mathcal{A}(s,m_q)}{2\pi^2 s} \right]^{2}
\left[  \frac{1}{2} \sum_f c_2^f\frac{\mathcal{A}(s,m_f)}{2\pi^2 s}\right]^{2}
 +\frac{2}{ \pi} \frac{\left(g^{\chi}_{gg}\right)^2 \left(g^{\chi}_{\g\g}\right)^2 s^3}
{(s-M_{\chi}^2)^2}\right.  \nonumber \\
&&\left. + \frac{1}{32\pi}  \frac{ (g^{\chi}_{gg})^{2}\,s^3}{(s-M_{\chi}^2)^2}
\left[ \sum_q C_0(s,m_q)  c^{\chi, q}_{gg} \right]^2 +
 \frac{ (g^{\chi}_{\g\g})^{2}   s^{3}}{ 32 \pi (s - M^{2}_{\chi})^{2}}
 \left[ \sum_f C_{0}(s,m_{f})  c^{\chi, f}_{\g\g} \right]^2 
 \right.  \nonumber\\
&& \left.+\frac{s^{3}}{2048 \pi (s - M^{2}_{\chi})^{2}}
\left[ \sum_q  C_{0}(s, m_q) c^{\chi, q}_{gg} \right]^2  
\left[ \sum_f  C_{0}(s, m_{f} ) c^{\chi, f}_{\g\g} \right]^2 \right.  \nonumber\\
&&\left.+ \frac{1}{2048\pi} \frac{s^5}{M_{Z^\prime}^4}
\left[  \frac{1}{2} \sum_q d_1^{\,q}\frac{\mathcal{A}(s,m_q)}{2\pi^2 s} \right]^{2}
\left[  \frac{1}{2} \sum_f  d_2^f\frac{\mathcal{A}(s,m_f)}{2\pi^2 s}\right]^{2}   \right.  \nonumber\\
&&\left. +\frac{ g^{\chi}_{gg} g^{\chi}_{\g\g} s^4}{16\pi M_Z^2(s-M_{\chi}^2)}
\left[ \frac{1}{2} \sum_q c_1^q\frac{\mathcal{A}(s,m_q)}{2\pi^2 s}\right]
\left[ \frac{1}{2} \sum_f c_2^f\frac{\mathcal{A}(s,m_f)}{2\pi^2 s}\right]\right. \nonumber \\
&&+ \left.  \frac{s^{4}}{1024 M^{2}_{Z} \pi (s - M^{2}_{\chi}) }   
\left[ \frac{1}{2} \sum_q  c_{1}^{q} \frac{ \mathcal{A}(s,m_q)}{2\pi^2 s} \right]   
\left[ \frac{1}{2} \sum_f  c_{2}^{f}  \frac{\mathcal{A}(s,m_f)}{2\pi^2 s} \right] \times \right.  \nonumber\\
 && \left. \times \left[ \sum_{f^\prime} C_{0}(s, m_{ f^\prime}) c^{\chi,  f^\prime}_{\g\g}   \right]
\left[  \sum_{q^\prime}   C_{0}(s, m_{ q^\prime}) c^{\chi, q^\prime}_{gg}   \right]   \right.  \nonumber\\
&& \left.+\frac{1}{1024 \pi} \frac{s^5}{M_{Z}^{2}\, M_{Z^\prime}^2}   
\left[ \frac{1}{2} \sum_q {d}_1^{\,q}  \frac{\mathcal{A}(s,m_q)}{2\pi^2 s}   \right]
\left[  \frac{1}{2}  \sum_f d_2^{f}  \frac{\mathcal{A}(s,m_f)}{2\pi^2 s}   \right]  \times   \right.  \nonumber\\
&& \left. \times \left[  \frac{1}{2} \sum_{q^\prime} {c}_1^{q^\prime} 
 \frac{\mathcal{A}(s,m_{q^\prime})}{2\pi^2 s} \right] \left[  \frac{1}{2} \sum_{f^\prime} c_2^{f^\prime}  
 \frac{\mathcal{A}(s,m_{f^\prime})}{2\pi^2 s} \right]  \right.   \nonumber\\
&&  \left. +\frac{s^{4}}{1024\, M^{2}_{Z^\prime}\, \pi \, (s - M^{2}_{\chi}) }   
\left[ \frac{1}{2} \sum_q d_{1}^{\,q}    \frac{\mathcal{A}(s,m_q)}{2\pi^2 s}         \right]   
\left[ \frac{1}{2} \sum_f d_{2}^{f}    \frac{\mathcal{A}(s,m_f)}{2\pi^2 s}  \right]   \times  \right. \nonumber\\
&& \left. \times \left[ \sum_{f^\prime} C_{0}(s, m_{ f^\prime}) c^{\chi,  f^\prime}_{\g\g}   \right]
\left[ \sum_{q^\prime} C_{0}(s, m_{ q^\prime}) c^{\chi, q^\prime}_{gg}   \right]  \right.  \nonumber\\
&&  \left.  +\frac{1}{16\pi}   \frac{  g^{\chi}_{gg}g^{\chi}_{\g\g} \,s^4}{M_{Z^\prime}^2(s-M_{\chi}^2)}
\left[ \frac{1}{2} \sum_q  d_1^{\,q} \frac{\mathcal{A}(s,m_q)}{2\pi^2 s} \right]
\left[ \frac{1}{2} \sum_f d_2^{f} \frac{\mathcal{A}(s,m_f)}{2\pi^2 s} \right]  \right\},
\label{Mchi}
\ea
where we have introduced the notation
\beq
\mathcal{A}(s,m_f)\equiv\left[1-\frac{m^2_f}{s} \log^2\left(\frac{\rho_f+1}{\rho_f-1}\right) \right].
\eeq
At high energy we can neglect the mass of the axion  ($s-M_{\chi}^2\approx s$) 
 and from the limits
\ba
\mathcal{A}(s\rightarrow \infty,m_f)=1,  \qquad C_{0}(s \rightarrow \infty, m_{ f})=0,
\ea
we obtain that the total cross section reduces to
\beq
\sigma^{mLSOM}(s \rightarrow \infty)= \frac{\mathcal{K}^2}{\pi} s,
\eeq
which is linearly divergent and has a unitarity bound.
$\mathcal{K}\equiv \mathcal{K}(s_b,g_B,\alpha_S(s),\tan \beta )$ is defined by
\ba
\mathcal{K}^2=&&   \frac{1}{ 2048 M_Z^4}  \left( \sum_q  \frac{c_{1,q}}{4\pi^2}\right)^2
\left( \sum_f \frac{c_{2,f}}{4\pi^2}\right)^2
+ 2 (g^{\chi}_{gg})^2 (g^{\chi}_{\g\g})^2     \nonumber\\
&&  +\frac{1}{16 M_Z^2} \left( \sum_q \frac{c_{1,q}}{4\pi^2}\right)
\left( \sum_f \frac{c_{2,f}}{4\pi^2}\right)
g^{\chi}_{gg} g^{\chi}_{\g\g} 
+\frac{1}{16 M_{Z^\prime}^2} \left( \sum_q \frac{d_{1,q}}{4\pi^2}\right)
\left( \sum_f \frac{d_{2,f}}{4\pi^2}\right)
g^{\chi}_{gg} g^{\chi}_{\g\g}      \nonumber\\
&& +  \frac{1}{ 1024 M_Z^2 M_{Z^\prime}^2 } \left( \sum_q \frac{c_{1,q}}{4\pi^2}\right)
\left( \sum_f \frac{c_{2,f}}{4\pi^2}\right)
\left( \sum_{q^\prime}  \frac{d_{1,q^\prime}}{4\pi^2}\right) 
\left( \sum_{f^\prime} \frac{d_{2,f^\prime}}{4\pi^2}  \right)    \nonumber\\
&&  + \frac{1}{ 2048 M_{Z^\prime}^4} \left( \sum_q \frac{d_{1,q}}{4\pi^2}\right)^2
\left(  \sum_f \frac{d_{2,f}}{4\pi^2}\right)^2. 
\ea
The derivation of the unitarity bound for this cross section is based, 
in analogy with Fermi theory, on the partial wave expansion
\ba
\frac{d\sigma}{d\Omega}=|f(\theta)|^2=|\frac{1}{2ik}\sum_{l=0}^{\infty}(2l+1)f_l P_l(\cos{\theta})|^2,
\ea
with an $s$-wave contribution given by
\ba
\frac{d\sigma}{d\Omega}=\frac{1}{s}|f_{0}|^2 +\, ...
\ea
Since unitarity requires that  $|f_{l}|\leq 1$ we obtain the bound 
\ba
\frac{d\sigma}{d\Omega}\leq\frac{1}{s}, 
\ea
or, equivalently,
\ba
\sigma\leq\frac{4\pi}{s}
\ea
\ba
\sqrt{s}\geq \sqrt{\frac{2\pi}{\mathcal{K}}}.
\ea
The bound is computed numerically by looking for values $s_b$ at which 
\beq
s_b^2={\frac{2\pi}{\mathcal{K}(s_b,g_B,\alpha_S(s_b),\tan \beta )}}
\eeq
where in the total parametric dependence of the factor ${\mathcal K}$, 
$\mathcal{K}(s_b,g_B,\alpha_S(s_b),\tan \beta )$, we have included the whole energy
dependence, including that coming from running of the coupling (up to three-loop level).
We will analyze below the bound numerically in the context of the specific brane model
of \cite{Ibanez:2001nd, Ibanez:1998qp, Ghilencea:2002da}.
\section{Couplings and Parameters in the Madrid Model \label{sec:Madrid}}
\begin{figure}[t]
{\centering \resizebox*{9cm}{!}{\rotatebox{-90}
{\includegraphics{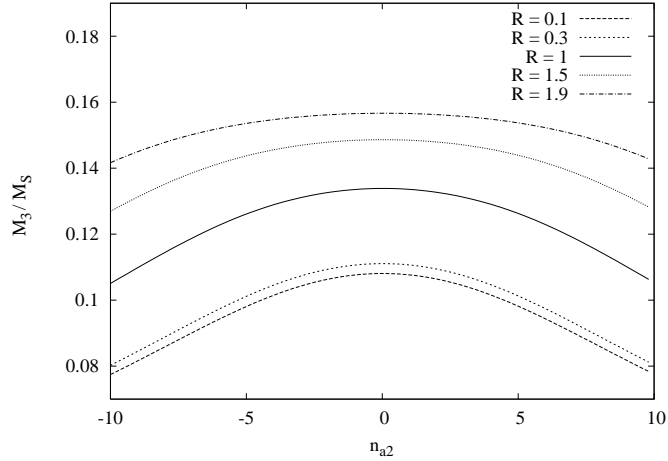}}}\par}
\caption{\small Lightest eigenvalue related to the generator 
$X_B$ for a model of Class A. We have chosen the couplings $g_c$ 
and $g_d$ such that their ratio ${\mathcal{R}}=g_d / g_c$.}
\label{ratio_plot}
\end{figure}

We now turn to a brief illustration of the specific charge assignments of
the class of models that we have implemented in our numerical analysis.
These are defined by a set of free parameters, which can be useful in order
to discern between different scenarios. In our implementation we rotate the
fields from the $D$-brane basis to the hypercharge basis and at the same time
we redefine the Abelian charges and couplings.
The four $U(1)$ in the hypercharge basis are denoted $U(1)_{X_i}$ with
$i=A,B,C$ and $U_Y$, where the last is the hypercharge $U(1)$, which is
demanded to be anomaly-free. This fixes the hypercharge generator
in the hypercharge basis in terms of the generator $q_{\alpha}$ $(\alpha=a,b,c,d)$
in the $D$-brane basis. The $U(1)_{a}$ and $U(1)_{d}$ symmetries can be identified
with (three times) the baryon number and (minus) the lepton number respectively.
The $U(1)_{c}$ symmetry can be identified with the third component of
right-handed weak isospin and finally the $U(1)_{b}$ is a $PQ$-like symmetry.
Specifically the hypercharge generator is given by
\beq
 Y = \frac{1}{6} (q_a  + 3 q_d )   - \frac{1}{2} q_c,
 \label{constraint}
\eeq
which in fact is a linear combination of the two anomaly-free 
generators $(q_a + 3q_d)$ and $q_c$, while the orthogonal combinations
\beq
X_A =3q_a - q_d ,   \qquad    X_B = q_b,
\eeq
represent anomalous generators in the hypercharge basis. 
Note that relation (\ref{constraint}) must be imposed in 
these models in order to obtain a correct massless hypercharge 
generator as in the SM. The set $(Y,A,B)$ does not depend on 
the model, while the fourth generator $X_C$ is model-dependent and is given by
\beq
X_{C} = \left(  \frac{ 3 \, \beta_2 \, n_{a2} }{ \beta_1 } \, q_a\ +\ 6 \, \rho \, n_{b1} \, q_{b} \ +\
2 \, n_{c1} \, q_c \ +\ \frac{ 3 \, \rho \, \beta_2 \,  n_{d2} }{ \beta_1 } \, q_d  \right).
\eeq
As can be seen in the detailed analysis performed in \cite{Ibanez:2001nd, Ibanez:1998qp, Ghilencea:2002da}, the general solutions are parametrized by a phase
$\epsilon =\pm1$, the Neveu-Schwarz background
on the first two tori $\beta_i=1-b_i=1,1/2$, the four integers
$n_{a2},n_{b1},n_{c1}$ and $n_{d2}$, which are the wrapping numbers
of the branes around the extra (toroidal) manifolds 
of the compactification, and a parameter $\rho=1,1/3$,
with an additional constraint in order to obtain the correct massless hypercharge
\beq
n_{c1} \ =\ {{\beta_2}\over {2\beta_1} } (n_{a2} + 3 \, \rho \, n_{d2}).
\label{Y_massless}
\eeq
This choice of the parameters identifies a particular class of models 
which are called Class A Models \cite{Ghilencea:2002da} and all the parameters are
listed in Tabs.~\ref{pparameters}, \ref{ccharge_higgs}, \ref{ttabpssm}.
Whether anomalous or not, the Abelian fields have mass terms induced 
by the St\"uckelberg mechanism, the mass matrix of the $U(1)$ gauge
bosons in the $D$-brane basis is given by the following expression
\beq
({\mathcal{M}}^2)_{\alpha \beta} = g_{\alpha} g_{\beta} M^{2}_{S} 
\sum^{3}_{i=1} c^{\alpha}_{i} c^{\beta}_{i},
\eeq
where $M_{S}$ is some string scale to be tuned. 
Greek indices run over the $D$-brane basis $\{ a,b,c,d\}$, the
Latin index $i$ runs over the three additional Abelian gauge groups,
while the $g^{}_{\alpha}$ and $g^{}_{\beta}$ are the couplings of the four $U(1)$.
The eigenvectors $w_{i=Y, A, B, C}$ and their eigenvalues $\lambda_i$ 
for the matrix $({\mathcal{M}}^2)_{\alpha \beta}$ have been computed in terms of 
the various classes of models in reference \cite{Ibanez:2001nd, Ibanez:1998qp}
\beqa
w^{}_Y &=&   \frac{1}{|w^{}_Y|}  \left\{ \frac{g_d}{3 g_a}, 0, -\frac{g_d}{g_c}, 1  \right\} _{\alpha}    \\
w^{}_{i} &=&  \frac{1}{|w^{}_i|}  \left\{ w_{ia}  , w_{ib} , w_{ic} , 1  \right\}
\eeqa
where the $w_{i=A,B,C}$ are the components of the eigenvectors. 
On the basis of the analysis of the mass matrix one can derive a plot 
of the lightest eigenvalue $M_3$, which corresponds to the $X_B$
generator in the hypercharge basis \cite{Ghilencea:2002da}. We have reproduced 
independently the result for the lightest eigenvalue $M_3$, which 
is in agreement with the predictions of
\cite{Ghilencea:2002da}. We have implemented numerically the diagonalization 
of the mass matrix and shown in
Fig.~\ref{ratio_plot} the behavior of the ratio $M_3/M_S$ as a 
function of the wrapping number $n_{a_2}$ and for several 
values of the ratio $\mathcal{R}=g_d/g_c$. This ratio, which 
characterizes the couplings of $U(1)_c$ and $U(1)_d$, appears 
as a free parameter in the gauge boson mass matrix. $M_S$, the 
string scale, is arbitrary and can be tuned at low values in the region of a few TeV.
We have selected a St\"uckelberg mass $M_3$ 
(which is essentially the mass of the extra $Z'$) 
of $800$ GeV. From Fig.~\ref{ratio_plot}, if we choose the curve
with $\mathcal{R}=1$ at the peak value, then this value 
is $13\%$ of the string scale, which in this case is 
lowered to approximately 6.1 TeV. It is quite obvious that 
the mass of the extra $Z^\prime$ can be reasonably assumed to be a
free parameter for all practical purposes.
\begin{table}[t]
\begin{center}
\begin{tabular}{|c|c|c|c|c|c|c|c|c|}
\hline
    $\nu$  & $\beta_1$ & $\beta_2 $ & $n_{a2}$  &  $n_{b1}$ & $n_{c1}$ & $n_{d2}$ \\
\hline  1/3 & 1/2  & $  1 $ &  $n_{a2}$ &  -1 & 1 & 1 - $n_{a2}$\\
\hline \end{tabular}
\end{center}
 \caption{\small Parameters for a Class A Model with a $D6$-brane .}
\label{pparameters}
\end{table}
The matrix $E_{i \alpha}\equiv (w_i)_\alpha$ constructed with the 
eigenvectors of the mass matrix defines the rotation matrix
$U_{i \alpha}$ for the $U(1)$ charges from the $D$-brane basis $\{ a,b,c,d\}$ into the
hypercharge basis $\{ Y,A,B,C\}$ as follows
\beq
q_{i} = \sum_{\alpha=a,b,c,d} \, U_{i \alpha} q_{\alpha} \,, \,\,\,\,\,  U_{i \alpha} = \frac{g_\alpha}{g_i} E_{i \alpha}, \,\,\,\,\,\,\,\,\,\,\,  (i=Y,A,B,C).
\eeq
So for the hypercharge we find that 
\beq
q_Y = \frac{ g_d }{ 3 g_Y  |w_Y|} (q_a - 3 q_c + 3 q_d)\equiv  \frac{1}{6} (q_a  + 3 q_d )   - \frac{1}{2} q_c,
\eeq 
to be identified with the correct hypercharge assignment given in expression (\ref{constraint}). This identification gives a relation between the gauge couplings
\beq
\frac{1}{g_Y^2} = \frac{|w^{}_Y|^2}{4 g^2_d} = \frac{1}{36 g_a^2} + \frac{1}{4g^2_c} + \frac{1}{4 g^2_d}.
\label{vincolo_gY}
\eeq
For the third generator a similar argument gives
\beq
q^{}_B =   \frac{g_a}{g_B} w^{}_{Ba} \, q_a + \frac{g_b}{g_B} w^{}_{Bb}\, q_b + \frac{g_c}{g_B} w^{}_{Bc}\, q_c + \frac{g_d}{g_B}w^{}_{Bd}\, q_d \equiv q^{}_b,
\eeq
where we identify the gauge symmetry $U(1)^{}_B$ corresponding to the 
lightest mass eigenvalue as an anomalous generator $q_B$. 
The charges $q_B$ of the SM spectrum are given in Tab.~\ref{ccharges}.
We recall that given a particular non-Abelian $SU(N)$
gauge group, with coupling $g_N$, arising from a stack of $N$ 
parallel branes and the corresponding $U(1)$ field living in the same stack,
the two coupling constants are related by $g_1 = g_N / \sqrt{2 N}$. 
Therefore, in particular the couplings $g_a$ and $g_b$ are determined
using the SM values of the couplings of the non-Abelian gauge groups
\beq
g^{2}_{a} = \frac{g^{2}_{QCD}}{6},  \qquad  g^{2}_{b} = \frac{g^{2}_{L}}{4},
\eeq
where $g_L$ is the $SU(2)_L$ gauge coupling, while $g_c$ and $g_d$ 
are constrained by relation (\ref{vincolo_gY}).
Imposing gauge invariance for the Yukawa couplings \cite{Coriano:2007xg}
we obtain the assignments for the Higgs doublets 
shown in Tab.~\ref{ccharge_higgs}.
\begin{table}[t]
\begin{center}
\begin{tabular}{|c|c|c|c|c|}
\hline
   &  Y &$ X^{}_A$  & $X^{}_{B} $    \\
\hline $H^{}_{u}$ & 1/2 & 0  & 2     \\
\hline  $H^{}_{d}$   & 1/2  &0  & -2     \\
\hline \end{tabular}
\end{center}
\caption{\small Higgs charges in the Madrid Model.
\label{ccharge_higgs}}
\end{table}
\begin{table}[t]
\begin{center}
\begin{tabular}{|c|c|c|c|c|c|}
\hline
   &   $q_a$  & $q_b $ & $q_c $ & $q_d$   \\
\hline $Q_L$ & 1  & -1 & 0 & 0   \\
\hline  $u_R$   &  -1  & 0  & 1  & 0   \\
\hline   $d_R$  &  -1  & 0  & -1  & 0   \\
\hline  $ L$    &  0   & -1   & 0  & -1    \\
\hline   $e_R$  &  0  & 0  & -1  & 1      \\
\hline   $N_R$  &  0  & 0  & 1  & 1    \\
\hline \end{tabular}
\end{center}
 \caption{\small SM spectrum charges in the $D$-brane basis for the Madrid Model.}
\label{ttabpssm}
\end{table}
\begin{table}[t]
\begin{center}
\begin{tabular}{|c|c|c|c|c|c|c|}
\hline
   &   $Q_L$  & $u_R $ & $d_R $ & $L$  &  $e_R$ & $N_R$ \\
\hline  $q_{Y}$  &  1/6    & - 2/3  & 1/3   &  -1/2   & 1 &  0  \\
\hline   $q_{B}$  & -1    & 0  & 0   & -1   & 0  & 0 \\
\hline \end{tabular}
\end{center}
 \caption{\small Fermion spectrum charges in the $Y$-basis for the Madrid Model.}
\label{ccharges}
\end{table}
\subsection{Anomalous and anomaly-free regions: numerical results}
\begin{figure}[t]
{\centering \resizebox*{9cm}{!}{\rotatebox{-90}
{\includegraphics{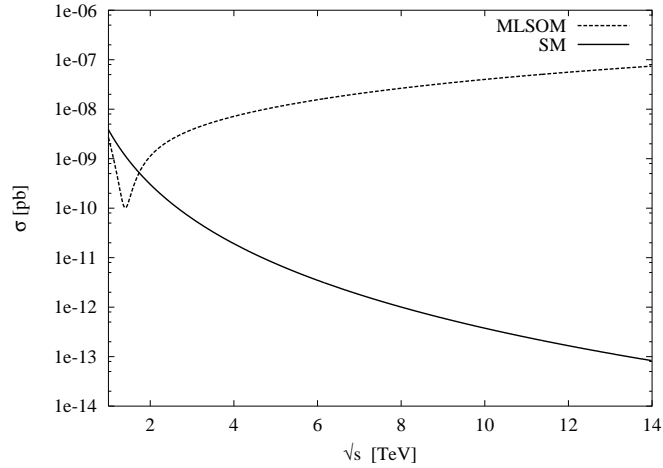}}}\par}
\caption{\small Partonic cross section for the anomalous 
process $gg\rightarrow \g\g$ with $\tan\beta=40$, $g_B=0.1$, $M_{\chi}=10$ GeV and $M_{1}=1$ TeV.
The lines refer to the cross section evaluated for different values of the
center of mass energy $\sqrt{s}$.}
\label{MLSOm_s}
\end{figure}
\begin{figure}[t]
{\centering \resizebox*{9cm}{!}{\rotatebox{-90}
{\includegraphics{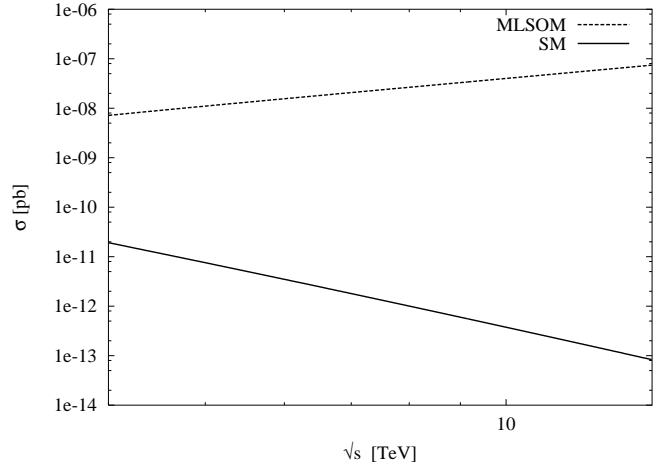}}}\par}
\caption{\small The partonic cross section with a parameter choice as in Fig.~\ref{MLSOm_s}.
The dashed upper lines refer to mLSOM cross sections evaluated with
and without the $\chi$ exchange, while the decreasing solid line refers to the SM.
\label{mlsom_sm_14T}}
\end{figure}
\begin{figure}[t]
{\centering \resizebox*{9cm}{!}{\rotatebox{-90}
{\includegraphics{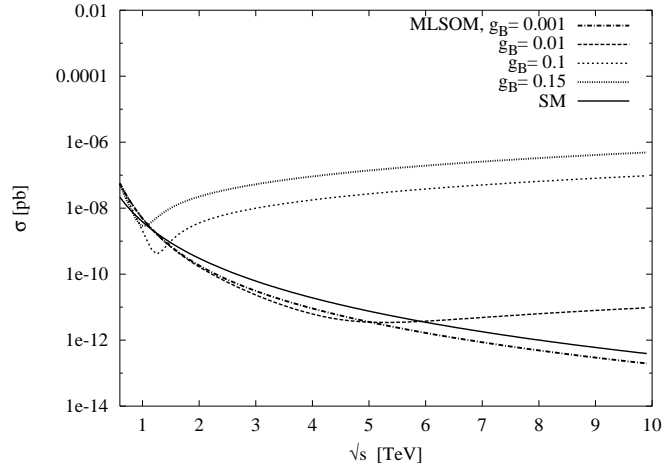}}}\par}
\caption{\small Partonic cross section
plotted for different values of the coupling constant $g_B$. The parameters are chosen as before.
\label{mlsom_gB_14T}}
\end{figure}
\begin{figure}[t]
{\centering \resizebox*{9cm}{!}{\rotatebox{-90}
{\includegraphics{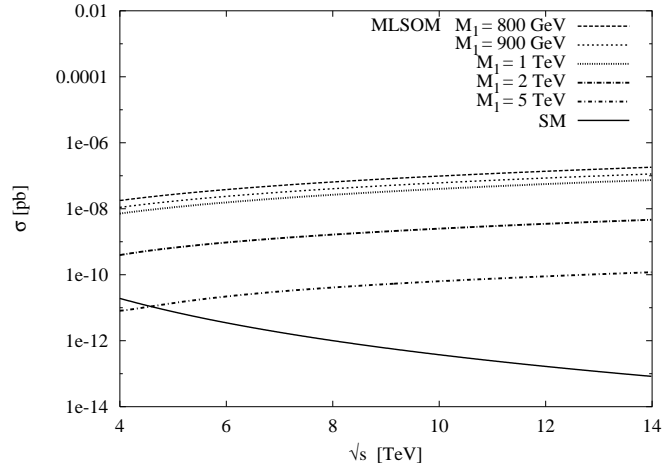}}}\par}
\caption{\small Partonic cross section for the anomalous 
process $gg\rightarrow \g\g$, $\tan\beta=40$, $g_B=0.1$, and $M_{\chi}=10$ GeV
plotted for different values of the St\"uckelberg mass $M^{}_1$.
\label{mlsom_M1_14T}}
\end{figure}
\begin{figure}[t]
{\centering \resizebox*{9cm}{!}{\rotatebox{-90}
{\includegraphics{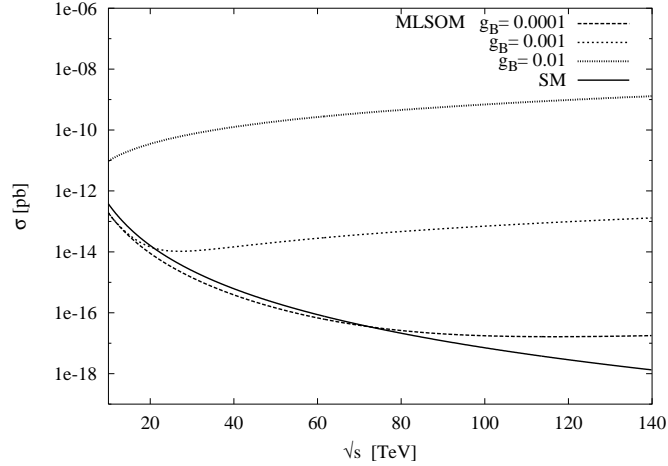}}}\par}
\caption{\small Partonic cross section for the anomalous 
process $gg\rightarrow \g\g$, $\tan\beta=40$ $M_1=800$ 
GeV and $M_{\chi}=10$ GeV. The different plots show a 
comparison between the SM and the mLSOM cross sections,
at very high energies, for small value of the coupling constant $g_B$.
\label{mlsom_M1_140T}}
\end{figure}
\begin{figure}[t]
{\centering \resizebox*{9cm}{!}{\rotatebox{-90}
{\includegraphics{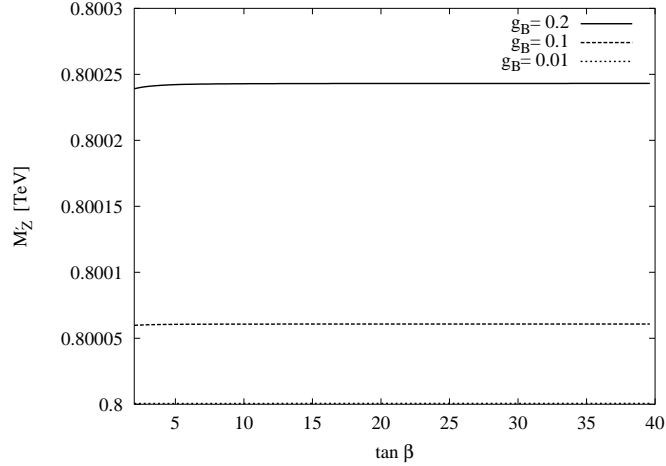}}}\par}
\caption{\small Behavior of the mass of the additional anomalous $Z^{\prime}$
as a function of $\tan{\beta}$ for different values of $g_B$. The variations are very small.
\label{Mzp_tanbeta}}
\end{figure}
\begin{figure}[t]
{\centering \resizebox*{9cm}{!}{\rotatebox{-90}
{\includegraphics{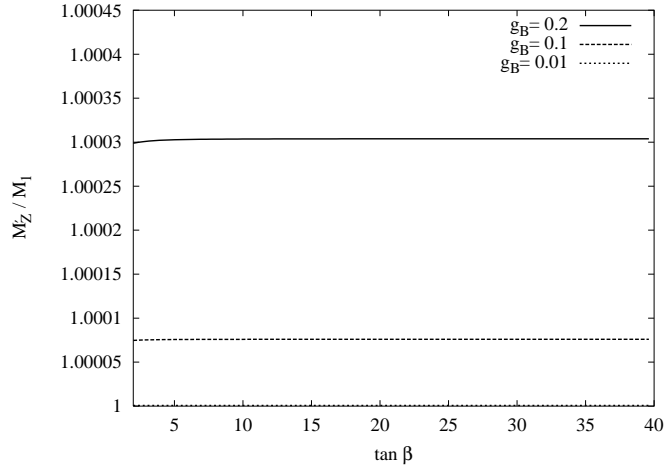}}}\par}
\caption{\small Mass of the extra $Z^\prime$ gauge boson 
for different values of the St\"uckelberg mass.
\label{M1Mzp}}
\end{figure}
\begin{figure}[t]
{\centering \resizebox*{9cm}{!}{\rotatebox{-90}
{\includegraphics{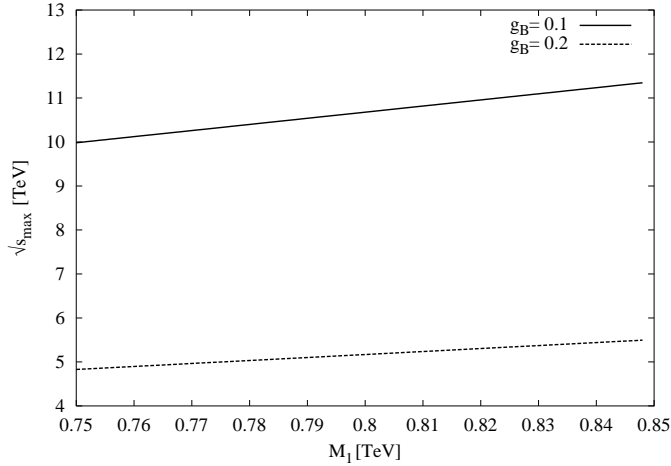}}}\par}
\caption{\small Bound for different values of the St\"uckelberg mass and $\tan{\beta}=40$.
The bound grows as we reduce the anomalous coupling,
and approaches the Standard Model behavior.
For $g_B$ = 0.01 the model exhibits a bound around $\sqrt{s}_{max}$ = 100 TeV (not shown).}
\label{newbound_M1}
\end{figure}
\begin{figure}[t]
{\centering \resizebox*{9cm}{!}{\rotatebox{-90}
{\includegraphics{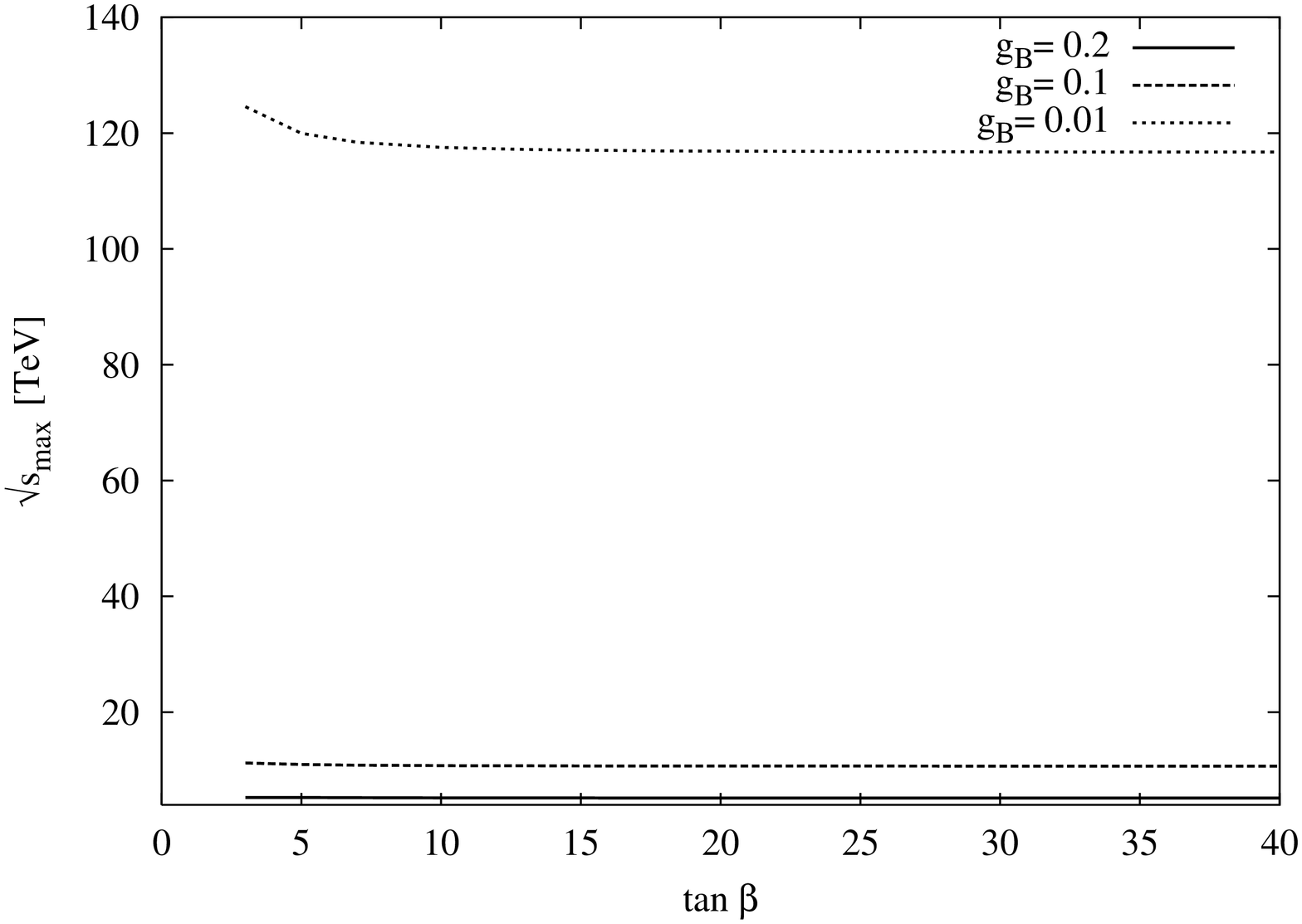}}}\par}
\caption{\small Bound for different values of $\tan\beta$, for $M_1=800$ GeV.
\label{newbound_tanbeta}}
\end{figure}
We have implemented in \textsc{Candia 1.0} \cite{Cafarella:2008du}
a numerical program that will provide full support for the
experimental collaborations for their analysis of the main 
signals in this class of models at the LHC,
with the charge assignments discussed in the previous section.
\textsc{Candia 1.0} has been planned to deal with the analysis of extra
neutral interactions at hadron colliders in specific channels,
such as Drell-Yan processes and double prompt photon processes 
with the highest precision. 
The program is entirely based on
the theory developed in \cite{Coriano:2005js, Irges:1998ax, Coriano:2007fw, Coriano:2007xg, Coriano:2006xh, Coriano:2008pg} and is tailored
to determine the basic processes, which provide the signal
for the anomalous and anomaly-free extra $Z^\prime$ both from string and GUT models.
The QCD corrections, including the parton evolution, is treated with extreme
accuracy using the theory of the logarithmic expansions, developed in the
last several years. Here we provide only parton level results for
the anomalous processes that we have discussed in the previous sections,
which clarify the role played by the WZ Lagrangian in the restoration of unitarity at high energy.
A complete analysis for the LHC is presented in Chap.~\ref{chap:LHC}.

\begin{itemize}
\item{ \bf $\sigma$ reduction by the exchange of $\chi$}

We show in Fig.~\ref{MasslessGGgaga} a plot of the (small) 
but increasing partonic cross section
for double prompt photon production from anomalous gluon fusion. 
We have chosen a typical
SM-like value for the coupling constant of the extra $Z^\prime$ included 
in the analysis and varied the center
of mass energy of a few GeV around $4.2$ TeV.
We show two plots, both in the brane model, one with the inclusion of
the axi-Higgs $\chi$ and one without it, with only the exchange of the $Z$ and $Z^\prime$.
Notice that the exchange of the $\chi $ is a separate gauge invariant contribution.
We have chosen a St\"uckelberg mass of 800 GeV.
The plots show the theoretically expected reduction of the
linear growth of the cross section, but the improvement is tiny,
for these values of the external parameters.
In these two plots, the fermion masses have been removed,
as we worked in the chiral limit. The inclusion of all the mass
effects in the amplitude has an irrelevant effect on the growth
of the cross section. This is shown in Fig.~\ref{MassiveGGgaga} where,
again, the inclusion of the axi-Higgs lowers the growth, but not in a significantly way.
We have analyzed the behavior of the cross sections in the
presence of a light axi-Higgs (ten or more GeV), 
but also in this case the effects are negligible.
This feature can be easily checked from Eq.~(\ref{Mchi}); in fact,
the mass term $M_{\chi}$ is contained in the denominator of the propagator
for the scalar, and in the TeV region we have $(s-M_{\chi}^2)\approx s$.
The numerical value of the unitarity bound remains essentially unchanged.

\item{\bf  Anomaly-free and anomalous regions}

An interesting behavior shows up in Fig.~\ref{SM_vs_MLSOm},
where we compare the results in the SM
and in the mLSOM for the same cross section, starting
at a lower energy. It is clear that the SM result, being anomaly-free,
is characterized by a fast-falling cross section, while
the mLSOM result is very different. In particular, one
finds a region of lower energy, where essentially the Model
follows the SM behavior (below 1 TeV) - but  smaller by a
factor of 10 - the growth of the anomalous contributions
still being not large, and a region of higher energy, where the
anomalous contributions take over (at about 2 TeV)
and which drive the growth of the cross section, as in the previous
two plots. There is a minimum at about 1.2 TeV, which is the point
at which the anomalous subcomponent becomes sizeable.

\item{\bf $\tan \beta$,  $g_b$ and $M_1$ variations}

The variation of the same behavior shown in the previous
plot with $\tan \beta$ is shown in Fig.~\ref{tanbeta_vs_MLSOm},
where we have varied this parameter from small to larger
values (10$\div$50). The depth of the minimum increases as we increase
this value. At the same time, the cross section tends to
fall much steeper, starting from larger values in the anomaly-free region.

In Fig.~\ref{gB_vs_MLSOm} we study the variation of $\sigma$
as we tune the coupling $g_B$ of the anomalous $U(1)_B$.
A very small value of the coupling tends to erase the
anomalous behavior, rendering the anomalous components subleading.
The cross section then falls quite fast before increasing reaching the bound.
The falling region is quite visible for the two values of $g_B=0.001$ and $0.01$,
showing that the set of minimum points, or the anomaly-free region,
is pushed up to several TeV,
in this case above $4.5$ TeV. The unitarity bound is weaker,
being pushed up significantly. The situation is opposite for stronger values of $g^{}_B$.

A similar study is performed in Fig.~\ref{diciotto}, but
for different St\"uckelberg masses $M^{}_1$. As this mass parameters
increase, the anomaly-free region tends to grow wider and
the cross section stabilizes. For instance, for a value of
the St\"uckelberg mass around 5 TeV, the region in which $\sigma$
has a normal behavior moves up to 3.5 TeV. The explanation of this
result has to be found in the fact that the anomalous growth is controlled by the
mass of the anomalous $Z'$ in the $s$-channel, appearing in
the denominator of the cross section. This suppression is seen both
in the direct diagram and in the counterterm diagram, which describes the exchange
of the axi-Higgs. Obviously, it is expected that as we reduce the
coupling of the anomalous gauge boson, the anomalous behavior is reduced as
well.  The different behavior of the cross section in the SM and
mLSOM cases can easily be inferred from
Fig.~\ref{MLSOm_s} having chosen a St\"uckelberg
mass of the order of 1 TeV.
A similar behavior is quite evident also from Fig.~\ref{mlsom_sm_14T},
from which it appears that in the mLSOM the deviations compared to the SM
partonic predictions get sizeable at parton level already at an energy of 4-6 TeV.
Notice also from Fig.~\ref{MassiveGGgaga} that the presence of the axi-Higgs seems
to be irrelevant for the chosen values of the couplings and parameters of the model.
We show in Fig.~\ref{mlsom_gB_14T} a plot of the dependence of the predictions on $g_B$ at
larger energy values, which appears to be quite significant.

Furthermore, in the TeV region the mLSOM predictions for small values of
the coupling constant ($g_B=0.001$) go below the SM prediction and
this is due to the axi-Higgs exchange, which is negative in this kinematical domain.
Moving below to 1 TeV the axi-Higgs interference has an opposite sign and
the mLSOM predictions are above the SM.
A similar analysis, this time for a varying St\"uckelberg mass,
is shown in Fig.~\ref{mlsom_M1_14T}, and also in this case, as in the previous one,
the results confirm that this dependence is very relevant.

Finally, in Fig.~\ref{mlsom_M1_140T} we plot the SM and mLSOM 
results on a larger interval, from 10 to 140 TeV,
from which the drastically different behavior of the two cross 
sections are quite clear. Notice that as we lower $g_B$, for 
instance down to $10^{-4}$, the anomaly-free region extends up 
to energy values that are of the order of 200 TeV or so. 
We conclude that the enhancement of the anomalous contributions with respect to the SM
prediction are in general quite large and very sensitive to the mass
of the extra $Z'$ and to the strength of the anomalous coupling.
Interestingly, a very weakly coupled $Z'$ gives a cross section that has
a faster fall-off compared to the SM case in the anomaly-free region. 
The mLSOM and SM predictions
intersect at a very large energy scale (140 TeV), when the 
anomalous contribution starts to increase.

Before drawing conclusions, it is necessary to comment on the other dependence,
that on $\tan\beta$, which appears to be far less significant compared 
to that on $M^{}_1$ (or $M_Z'$) and $g_B$. This third parameter essentially has a
(very limited) influence on the mass of the extra $Z^\prime$ and on the overall
predictions. This is clearly shown in
Figs.~\ref{Mzp_tanbeta} and \ref{M1Mzp}, where we have varied both 
$\tan\beta$ and $g^{}_B$. Therefore, the mass of the extra $Z^\prime$ and the
St\"uckelberg mass may be taken to be essentially coincident, to a first approximation.

\item{\bf The bounds}

We conclude our analysis with two plots, depicted 
in Figs.~\ref{newbound_M1} and \ref{newbound_tanbeta}, 
which show the variations of the bounds with the parameters
$g^{}_B$ and $M^{}_1$. In the first plots, shown in Fig.~\ref{newbound_M1}, 
we choose a large value of $\tan\beta$ and we have varied both the 
St\"uckelberg mass and the strength of the anomalous coupling. 
For a St\"uckelberg mass around 1 TeV, the bound is around 5 TeV, 
for an anomalous coupling $g_B=0.2$. For a smaller value of $g_B=0.1$ 
the bound grows to 10 TeV. For a smaller value of $g_B=10^{-2}$, 
the bound is around 100$\div$120 TeV. This result is particularly interesting, 
because it should allow one to set limits on the St\"uckelberg mass and the
value of the anomalous couplings at the LHC in the near future. 
Smaller values of $g^{}_B$ are tested in Fig.~\ref{newbound_tanbeta}, 
where the bound is shown to increase significantly as $g^{}_B$ gets smaller.

\end{itemize}
\section{Conclusions}
We have analyzed the connection between the WZ term and the GS 
mechanism in the context first of simple models and then 
in a complete brane model, containing three extra anomalous
$U(1)$. We have shown that the WZ method of cancellation 
of the anomaly does not protect
the theory from an excessive growth, which is bound to
violate unitarity beyond a certain
scale. We have also studied the connection between the two 
mechanisms, illustrating the corresponding differences.

We have quantified the unitarity bound for several choices of the parameters
of the theory. The significant dependences are those on the St\"uckelberg
mass $M^{}_1$ and the coupling constant $g^{}_B$ of the anomalous generator. 
We have also shown that the exchange of a physical axion $\chi$ lowers
the cross section, but not significantly, whose growth remains
essentially untamed at high energy. We have shown that in these
models one can identify a region that is SM-like, where some 
anomalous processes have a fast fall-off, from a
second region, where the anomaly dominates. 

The analysis in this chapter has been confined to the parton level,
but we postpone more definitive predictions 
for the detection of anomalous interactions at the LHC to Chap.~\ref{chap:LHC}. 

The approach that we
have suggested, the use of BIM amplitudes to search for unitarity 
violations at future colliders, can be a way in the
search to differentiate between non-anomalous 
\cite{Langacker:2008yv} and anomalous extra $Z^\prime$ \cite{Coriano:2008wf}.
Our objective here has been to
show that there is a systematic way to analyze the two mechanisms
for canceling the
anomalies at the phenomenological level and that unitarity issues
are important in order to characterize the region in which a
certain theory starts to be dominated by the chiral anomaly.
\section{Appendix. The cross section }
The total cross section is given by the sum of all the contributions shown in Fig.~\ref{GS}
\ba
\sigma^{mLSOM}(s)&=&\sigma_A(s) + \sigma_{B}(s) + \sigma_{C}(s)+  \sigma_{D}(s) +  \sigma_{E}(s) +  \sigma_{F}(s)  \nonumber\\
&& + \, \sigma_{AB}(s) + \sigma_{AE}(s) +  \sigma_{AF}(s)+  \sigma_{BF}(s) +  \sigma_{EF}(s).
\label{totalcross}
\ea
The interference term between $Z$ exchange and $\chi$ exchange, diagram (b) in Fig.~\ref{GS}, is given by
\ba
\sigma_{AB}(s)= \frac{1}{16 \pi}
\left[\sum_q \frac{1}{2} c_1^{q}A_{6,q}\right]\left[\sum_{f} \frac{1}{2} c_2^{f}A_{6,f}\right]
g^{\chi}_{gg}g^{\chi}_{\g\g}\frac{s^4}{M_Z^2(s-M_{\chi}^2)}\,.
\ea
The following interference terms are vanishing 
\ba
&&{\cal M}_A {\cal M}_C^\dagger + {\cal M}_C {\cal M}_A^\dagger = 0,  \qquad 
{\cal M}_A {\cal M}_D^\dagger + {\cal M}_D {\cal M}_A^\dagger = 0,   \nonumber\\
&&{\cal M}_B {\cal M}_C^{\dagger}  + {\cal M}_C {\cal M}_B^\dagger = 0,   \qquad
{\cal M}_B {\cal M}_D^{\dagger} + {\cal M}_D {\cal M}_B^{\dagger} = 0.
\ea
The interference term between the exchange of the $Z$ boson and $\chi$ exchange, 
diagram (e) in Fig.~\ref{GS}, gives
\ba
\sigma_{AE}(s)& =&   \frac{s^{4}}{1024 M^{2}_{Z} \pi (s - M^{2}_{\chi}) }
\left[ \sum_q \frac{1}{2} c_{1}^{q}  A_{6,q}  \right]
\left[  \sum_{f} \frac{1}{2} c_{2}^{f}   A_{6,f}  \right]   \nonumber\\
&& \times \left[\sum_{ f^\prime } C_{0}(s, m_{ f^\prime}) c^{\chi,  f^\prime}_{\g\g}   \right]
\left[ \sum_{ q^\prime} C_{0}(s, m_{ q^\prime}) c^{\chi, q^\prime}_{gg}   \right],
\ea
while the interference term between $\chi$ exchange and $Z^\prime$ boson exchange contributes with
\ba
\sigma_{BE}(s) =- \frac{ s^3}{16 \pi (s - M_{\chi}^{2})^{2}} g^{\chi}_{\gamma \gamma}  g^{\chi}_{gg}
\left[ \sum_{ f} C_{0}(s, m_{ f}) c^{\chi, f}_{\g\g} \right]
\left[\sum_ q C_{0}(s, m_q) c^{\chi, q}_{gg}  \right].
\ea
The other interference term in the cross section is given by
\ba
\sigma_{CD}(s) = \frac{ s^3}{16 \pi (s - M_{\chi}^{2})^{2}}
g^{\chi}_{\gamma \gamma} g^{\chi}_{gg}     \left[ \sum_{q} C_{0}(s, m_{q} ) c^{\chi, q}_{gg}  \right]
 \left[ \sum_f C_{0}(s, m_f) c^{\chi, f}_{\g\g}    \right],
\ea
 so we obtain
\ba
\sigma_{CD}(s) + \sigma_{BE}(s) = 0.
\ea
Other interference terms also vanish; in fact, we get
\ba
{\cal M}_C {\cal M}_E^{\dagger} + {\cal M}_E {\cal M}_C^\dagger = 0, \qquad
{\cal M}_D {\cal M}_E^{\dagger} + {\cal M}_E {\cal M}_D^{\dagger} = 0,
\ea
\ba
{\cal M}_C {\cal M}_F^{\dagger} + {\cal M}_F {\cal M}_C^\dagger = 0, \qquad
{\cal M}_D {\cal M}_F^{\dagger} + {\cal M}_F {\cal M}_D^{\dagger} = 0.
\ea
The interference term between $Z$ exchange and $Z^\prime$ exchange takes the form
\ba
\sigma_{AF} = \frac{1}{1024 \pi} \left[\sum_q \frac{1}{2} {d}_1^{\,q}A_{6,q}\right]
\left[\sum_{f} \frac{1}{2} d_2^{f}A_{6,f}\right]  \left[\sum_{q^\prime} \frac{1}{2} c_1^{q^\prime}A_{6,q^\prime}\right]
\left[\sum_{f^\prime} \frac{1}{2} c_2^{f^\prime}A_{6,f^\prime}\right]
\frac{s^5}{M_{Z}^{2}\, M_{Z^\prime}^2}\,.
\ea
The interference term between $Z^\prime$ exchange and $\chi$ exchange 
of diagram (b) in Fig.~\ref{GS} is given by
\ba
\sigma_{BF}(s)=\frac{1}{16\pi}
\left[\sum_q \frac{1}{2} d_1^{\,q}A_{6,q}\right]\left[\sum_{f} \frac{1}{2} d_2^{f}A_{6,f}\right]
g^{\chi}_{gg}g^{\chi}_{\g\g}\frac{s^4}{M_{Z^\prime}^2(s-M_{\chi}^2)}\,.
\ea
Finaly, the interference between $Z^\prime$ exchange and $\chi$ exchange, 
diagram (e) in Fig.~\ref{GS}, gives
\ba
\sigma_{EF}(s)& =&   \frac{s^{4}}{1024\, M^{2}_{Z^\prime}\, \pi \, (s - M^{2}_{\chi}) }
\left[ \sum_q \frac{1}{2} d_{1}^{\,q}  A_{6,q}  \right]
\left[  \sum_{f} \frac{1}{2} d_{2}^{f}   A_{6,f}  \right]   \nonumber\\
&&  \times \left[\sum_{ f^\prime } C_{0}(s, m_{ f^\prime}) c^{\chi,  f^\prime}_{\g\g}   \right]
\left[ \sum_{ q^\prime} C_{0}(s, m_{ q^\prime}) c^{\chi, q^\prime}_{gg}   \right].
\ea
                            
\chapter{Axions and Anomaly-Mediated Interactions: 
The Green-Schwarz and Wess-Zumino Vertices at Higher Orders and  ${g-2}$ of the muon\label{chap:g-2}}
\fancyhead[LO]{\nouppercase{Chapter 4. The Green-Schwarz and Wess-Zumino Vertices at Higher Orders}}
\section{Introduction to the chapter}
Understanding the role played by the Green-Schwarz (GS) and Wess-Zumino (WZ) mechanisms in quantum field theory is important in order to grasp the implications of the chiral gauge anomaly at the level of model building, especially in the search of extra trilinear gauge interactions at future colliders. In recent years several proposals coming either from string theory or from theories with extra dimensions have introduced new perspectives in regard to the various mechanisms of cancellation of the gauge anomalies in effective low energy Lagrangians, which require further investigation in order to be fully understood. These effective models are characterized by the presence of higher dimensional operators
and interactions of axion-like particles. In a class of vacua of string theory this enterprise has some justification, for instance in orientifold models (see \cite{Kiritsis:2003mc, Antoniadis:2000ena, Ibanez:2001nd}), where deviations from the Standard Model may appear in the form of higher dimensional corrections which are not heavily suppressed and which could be accessible at the LHC.

In anomaly-free realizations of chiral gauge theories the trilinear anomalous gauge interactions vanish (identically) in the chiral limit, by a suitable distribution of charges among the fermions of each generation (or inter-generational), showing that residual interactions are proportional to the mass differences of the various fermions. In the GS realization this request is far more relaxed and the mechanism requires only the cancellation, in the presence of anomalous contributions, of the longitudinal component of the anomaly vertex rather than that of the entire triangle diagram. In the WZ case, the cancellation of the anomaly takes place at Lagrangian level, rather than at the vertex level, and requires an axion as an asymptotic state, which is a generalization of the Peccei-Quinn interaction.

The effective field theory of the WZ mechanism has been analyzed in \cite{Coriano:2005js, Coriano:2007fw, Coriano:2007xg, Coriano:2006xh, Coriano:2008pg, Armillis:2007tb}, together with  its supersymmetric extensions \cite{Anastasopoulos:2008jt} while a string derivation of the GS constructions has been outlined in \cite{Anastasopoulos:2006cz}. Pseudoscalar fields (axion-like particles) - with a mass and a coupling to gauge fields which are left unrelated - have been the subject of several investigations and proposals for their detection either in ground-based experiments \cite{Jaeckel:2006xm} or to explain some puzzling results on gamma ray propagation \cite{DeAngelis:2008sk, DeAngelis:2007dy},
while new solutions of the strong $CP$-problem in more general scenarios have also received attention \cite{Berezhiani:2000gh}. At the same time the search for extra $Z^\prime$  at the LHC
from string models and extra dimensions, together with precision studies on the  resonance to uncover new effects,
has also received a new strength (\cite{Adam:2008ge, Lee:2007qx, Langacker:2008yv, Fuks:2008yp, Feldman:2006wb, Fuks:2007gk}).

If the GS and the WZ mechanisms are bound to play any role at future experiments (see for instance \cite{Kumar:2007zza}) remains to be seen, given the very small numerical impact of the anomaly corrections in the cleanest processes that can be studied, for instance, at the LHC; nevertheless more analysis is needed in order to understand the theoretical implications of ``anomaly mediation'' and of its various realizations, in the form of GS and WZ interactions, in effective models.

Both mechanisms are quite tricky, since they show some unusual features which are not common to the rest of anomaly-free field theories and it is not hard to find in the literature several issues which have been debated for a long time, concerning the consistency of these approaches \cite{Andrianov:1989by, Andrianov:1993qy, Fosco:1993qx}.

For instance, in the GS case, one of them concerns unitarity, due to the claimed presence of extra ``double poles''  \cite{Adam:1997gj} in a certain class of interactions which would render completely invalid a perturbative prescription; another one concerns the physical interpretation of the longitudinal subtraction, realized within the same mechanism, which is usually interpreted as due to the exchange of an axion, which, however, as pointed out in \cite{Coriano:2008pg} is not an asymptotic state.

The first part of this chapter is a study of the organization of the perturbative expansion for anomalous theories in the presence of counterterms containing double poles in virtual corrections and, in principle, in $s/t/u-$channel exchanges. Our point of view and conclusions are in contradiction with those of
\cite{Adam:1997gj}, formulated within axial QED, where the analysis of the anomaly pole counterterm was not taking into account the fact that the subtracted term is an intrinsic part of the triangle (anomaly) diagram, corresponding to one of its invariant amplitudes, in a specific formulation.

The picture that emerges from our analysis is that of consistency -rather than of inconsistency- of the GS mechanism at the level of effective field theory. In other words, it should be possible to subtract the longitudinal pole of the anomaly diagram with no further consequences at perturbative level.
The structure of the perturbative expansion in the presence of explicit GS counterterms is worked out and we detail the methods for the computations of graphs containing extra poles in the propagators and compare the general features of this expansion to an ordinary expansion.

In any case, in the absence of a direct check of the unitarity equations -which is hard to perform given the rather high order at which these anomalous corrections appear- the problems in perturbation theory can potentially appear in the form of double poles in some ({\em external} ) propagators. A re-examination of several diagrams brings us to conclude that this situation is avoided.

Coming to a direct phenomenological application, we investigate the role of these vertices in the study of the anomalous magnetic moment of the muon. We stress that if the physical mechanism introduced for the cancellation of the anomaly is of WZ type, then a physical axion appears in the spectrum. This is the case if the anomalous extra $Z^\prime$ receives its mass both by the Higgs and the St\"uckelberg mechanisms. Both for the GS and the WZ case we outline the role of the anomalous extra $Z^\prime$ and of the pseudoscalar exchange up to two-loop level. A previous analysis of the leading contribution to $g-2$ for intersecting brane models can be found in \cite{Kiritsis:2002aj}.

More recently, the GS vertex has been used in the study of the coupling of the Kaluza-Klein (KK) \cite{Kumar:2007zza, Djouadi:2007eg} excitations of gauge bosons to fermions, where it has been pointed out the possibility to detect these coupling at the LHC, for instance in $t\bar{t}$ production. We find that several of these results are based on a still unsatisfactory understanding of the GS mechanism at theoretical level, and our work is an attempt to clarify some of these points. From our analysis will emerge the correct structure of the broken Ward Identities for the GS vertex, which are specific of a non-local theory.
These points will be carefully analyzed in the final section of this chapter.
\section{The GS and WZ vertices}
The field theory version of the GS mechanism, deprived of all its stringy features, appears in an attempt to cancel the anomaly by introducing a specific non-local counterterm added to the anomalous theory. This attempt had been the cause of serious debates which have questioned the consistency of the approach. The mechanism uses a ghost-like particle, which in string theory is generically identified as an axion - although there it does not appear as an asymptotic state - to restore the broken Ward Identities due to the anomaly.
A paradigm for the GS mechanism in field theory is an anomalous version of axial QED in 4-dimensions defined by the Lagrangian
\beq
\mathcal{L}_{5\,QED}= \overline{\psi} \left( i \slashed{\partial} + e \slashed{B} \gamma_5\right)\psi - \frac{1}{4} F_B^2
\label{count0}
\eeq
plus the counterterm
\beq
\mathcal{S}_{ct}= \frac{1}{24\pi^2} \langle \partial B(x) \square^{-1}(x-y) F(y)\wedge F(y)
\rangle.
\label{count}
\eeq
Federbush \cite{Federbush:1996cp} proposed to reformulate this  Lagrangian in terms of one axion and one ghost-like particle interacting via a Wess-Zumino (WZ) counterterm (see the discussion in Chap.~\ref{chap:UnitarityBound}). An equivalent formulation of the same subtraction counterterm is given in \cite{Andrianov:1989by}, where a transversality constraint ($\partial B=0$) is directly imposed on the Lagrangian
via a multiplier. Eq.~(\ref{count}) can be obtained by performing the functional integral over $a$ and $b$ of the following action
\beqa
\mathcal{L} &=& \overline{\psi} \left( i \not{\partial} + e \not{B} \gamma_5\right)\psi - \frac{1}{4} F_B^2 +
\frac{ e^3}{48 \pi^2 M} F_B\wedge F_B ( a + b) \nonumber \\
&& + \frac{1}{2}  \left( \partial_\mu b - M B_\mu\right)^2 -
\frac{1}{2} \left( \partial_\mu a - M B_\mu\right)^2.
\label{fedeq}
\eeqa
The integral on $a$ and $b$ are gaussians and one recovers the non-local contribution in (\ref{count}) after partial integration.
Notice that $b$ has a positive kinetic term and $a$ is ghost-like. Both $a$ and $b$ shift by the same amount under a gauge transformation of $B$
\beq
a\rightarrow a + M \theta,\,\,\,\,\, b\rightarrow b + M \theta
\eeq
where $\theta$ is the gauge parameter.
This second (local) formulation of the pole counterterm contained in (\ref{count}) shows the connection between this action and the WZ mechanism. Both actions, in fact, share some similarities, but describe different theories. In particular, the WZ action is obtained by removing the ghost term ($b$) and keeping only the
axion. This second theory is characterized by a unitarity bound, as shown in Chap.~\ref{chap:UnitarityBound}. The bound is due to the fact that in effective models containing Wess-Zumino interactions, gauge invariance of the effective action requires a cancellation  between different trilinear vertices: the anomalous vertices and the axion counterterm $\phi F\wedge F$, while for Green-Schwarz vertices the subtraction of the longitudinal component of the anomaly is sufficient to make the effective vertex  gauge-invariant to all orders. In both cases the physical amplitudes are gauge-independent.

In the WZ case the proof of gauge
  independence is rather involved and has been discussed before in Chaps.~\ref{chap:AbelianModels1}, \ref{chap:AbelianModels2}. In the GS case, instead, this is trivial since the vertex is gauge-invariant by construction. Notice that a local counterterm in the form of a Peccei-Quinn term is not sufficient to remove the power-like growth with energy of  a class of amplitude (BIM amplitudes) that are characterized by anomalous production and anomalous decay of massless gauge bosons in the initial and final states, mediated by the exchange of an anomalous $Z^\prime$ in the $s$-channel (see Chap.~\ref{chap:UnitarityBound} for more details). These amplitudes are quite interesting since they evade the Landau-Yang theorem, triggering a $Z \gamma \gamma $ vertex. The phenomenological implications of these amplitudes are discussed in Chap.~\ref{chap:LHC}.
\begin{figure}[t]
\begin{center}
\includegraphics[scale=0.8]{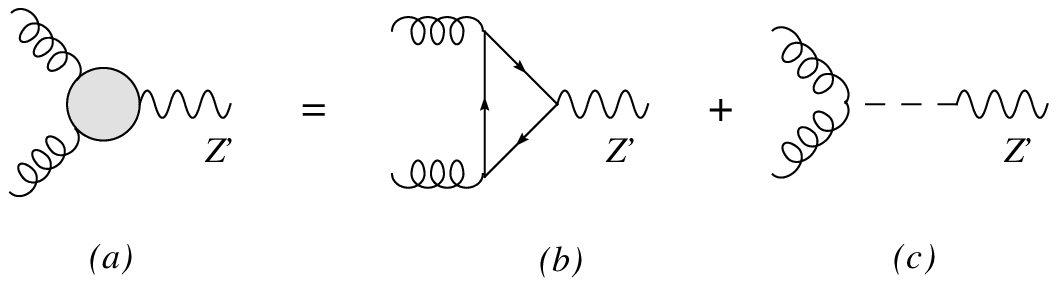}
\caption{\small A  gauge invariant GS vertex of the AVV type, composed of an AVV triangle  and a single counterterm of Dolgov-Zakharov form. Each term is denoted by  $\Delta_{AVV}^{\lambda \mu \nu \, GS}$ (a), $\Delta_{AVV}^{\lambda \mu \nu}$ (b) and $C^{\lambda \mu \nu}_{AVV}$ (c).}
\label{GSS_AVV}
\end{center}
\end{figure}
\begin{figure}[t]
\begin{center}
\includegraphics[scale=0.8]{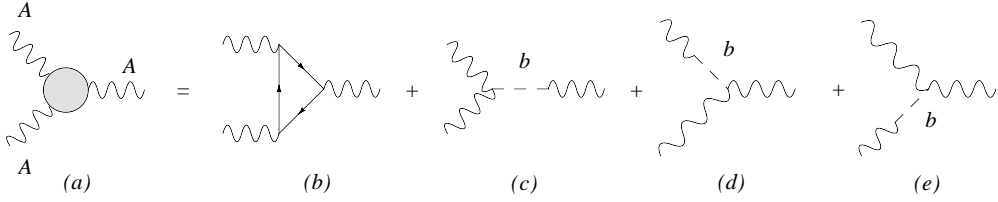}
\caption{\small All the contributions to the GS gauge invariant vertex, for an AAA triangle. The single terms are denoted by $\Delta_{AAA}^{\lambda \mu \nu \, GS}$ (a), $\Delta_{AAA}^{\lambda \mu \nu}$ (b), $C^{\lambda \mu \nu}_{AVV}$ (c), $C^{\mu \nu \lambda}_{AVV}$ (d) and $C^{\nu \mu \lambda}_{AVV}$ (e).}
\label{GS_AAA}
\end{center}
\end{figure}
\subsection{The GS vertex in the $AAA$ and $AVV$ cases}
In our analysis, we denote with $T_{\mu\nu\lambda}$ the three-point function in momentum space, obtained from the Lagrangians (\ref{count0}) and (\ref{count}). In the case of three axial-vector currents we define the correlator
\begin{eqnarray}
(2\pi)^4 \delta (k_1+k_2-k)\Delta_{AAA}^{\lambda\mu\nu}(k, k_1, k_2)
=\int dx_1 \, dx_2 \, dx_3 \, e^{i(k_1 x_1 + k_2 x_2 - k x_3)}
\langle J_{\mu}^5(x_1)J_{\nu}^5(x_2)J_{\lambda}^5(x_3)\rangle.
\end{eqnarray}
and a symmetric distribution of the anomaly for the $AAA$ vertex
\footnote{We have used the following notation
$a_n=-\frac{i}{2\pi^2}$ and
$\epsilon[\mu,\nu,k_1,k_2]=\epsilon^{\mu\nu\alpha\beta}k_{1\alpha}k_{2\beta}$}
\begin{eqnarray}
\label{anomaly}
&&k_{\lambda} \Delta_{AAA}^{\lambda\mu\nu}(k,k_1,k_2)=
\frac{a_n}{3}\epsilon[\mu, \nu, k_1, k_2]
\nonumber\\
&&k_{1\mu} \Delta_{AAA}^{\lambda\mu\nu}(k,k_1,k_2)=
\frac{a_n}{3}\epsilon[\lambda, \nu, k, k_2]
\nonumber\\
&&k_{2\nu} \Delta_{AAA}^{\lambda\mu\nu}(k,k_1,k_2)=
\frac{a_n}{3}\epsilon[\lambda, \mu, k, k_1].
\end{eqnarray}
In the $AVV$ case, a second vector-like gauge interaction ($A_\mu$) is introduced in Eq.~(\ref{fedeq}) for more generality and we have
\begin{eqnarray}
(2\pi)^4 \delta (k_1+k_2-k)\Delta_{AVV}^{\lambda\mu\nu}(k,k_1,k_2)
=\int dx_1 \, dx_2 \, dx_3 \, e^{i(k_1 x_1 + k_2 x_2 - k x_3)}
\langle J_{\mu}(x_1)J_{\nu}(x_2)J_{\lambda}^5(x_3)\rangle,
\end{eqnarray}
where the anomaly equations are
\begin{eqnarray}
&&k_{\lambda} \Delta_{AVV}^{\lambda\mu\nu}(k,k_1,k_2)=
a_n \epsilon^{\mu\nu\alpha\beta}k_{1\alpha}k_{2\beta}
\nonumber\\
&&k_{1\mu} \Delta_{AVV}^{\lambda\mu\nu}(k,k_1,k_2)=0
\nonumber\\
&&k_{2\nu} \Delta_{AVV}^{\lambda\mu\nu}(k,k_1,k_2)=0.
\end{eqnarray}
Below we will consider both the $AVV$ and $AAA$ cases.
The GS counterterm that corresponds to the exchange of the massless pole of Eq.~(\ref{count}) takes the following form in momentum space in the $AVV$ case
\bea
C^{\lambda \mu \nu}_{AVV}(k,k_1,k_2) =
C^{\mu \nu}(k_1,k_2) k^\lambda =
-\frac{a_n}{k^2} k^\lambda \epsilon[\mu,\nu,k_1,k_2].
\eea
Similarly, a GS counterterm in the $AAA$ case, with incoming momentum $k$  and outgoing momenta $k_1, k_2$,  is defined as
\beqa
C^{\lambda \mu \nu}_{AAA}(k,k_1,k_2) &=&
\frac{1}{3} \left( C^{\lambda \mu \nu}_{AVV}(k,k_1,k_2)
+ C^{\mu \nu \lambda}_{AVV}(-k_1,k_2,-k)
+ C^{\nu \lambda \mu}_{AVV}(-k_2, -k,k_1)\right)
\nonumber\\
&=&  \frac{1}{3} \Bigg( C^{\mu \nu}(k_1,k_2)k^\lambda
- C^{\nu \lambda}(k_2,-k)k_1^\mu -  C^{\lambda \mu}(-k,k_1)k_2^\nu   \Bigg) \nonumber\\
&=&- \frac{1}{3} \left(
\frac{a_n}{k^2} k^\lambda \epsilon[\mu, \nu, k_1, k_2]
+ \frac{a_n}{k_1^2} k_1^\mu \epsilon[\lambda, \nu, k, k_2]
+ \frac{a_n}{k_2^2} k_2^\nu \epsilon[\lambda, \mu, k, k_1]\right),
\eeqa
and corresponds to the Dolgov-Zakharov form (DZ) of the anomaly diagram \cite{Dolgov:1971ri} (modulo a minus sign).
The re-defined vertex shown in Fig.~\ref{GSS_AVV} is written as
\begin{eqnarray}
\label{cance}
\Delta_{AVV}^{\lambda\mu\nu \, GS}(k,k_1,k_2)=
\Delta_{AVV}^{\lambda\mu\nu}(k,k_1,k_2)+C_{AVV}^{\lambda\mu\nu}(k,k_1,k_2)
\end{eqnarray}
and we obtain a similar expression for the $AAA$ vertex (Fig.~\ref{GS_AAA})
just by replacing $AVV$ with $AAA$ in Eq.~(\ref{cance}) and taking into account the different form of the counterterms.
These gauge invariant vertices trivially satisfy the Ward Identities
\begin{eqnarray}
\label{WI}
k_{\lambda}\Delta^{\lambda\mu\nu\,  GS}(k,k_1,k_2)
=k_{1\mu} \Delta^{\lambda\mu\nu\,  GS}(k,k_1,k_2)
=k_{2\nu} \Delta^{\lambda\mu\nu\,  GS}(k,k_1,k_2)
=0,
\end{eqnarray}
where again $\Delta^{\lambda\mu\nu\,  GS}$  refers either to an $AVV$ or to an $AAA$ correlator.
\subsection{Implications of the GS vertex: vanishing of (real) light-by-light  scattering at two-loop}
\begin{figure}[t]
\begin{center}
\includegraphics[scale=1.1]{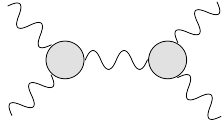}
\caption{\small Amplitude with two full GS vertices and the exchange of an axial-vector current in the $s$-channel. For on-shell external lines the contributions from the extra poles disappear.}
\label{GSGS}
\end{center}
\end{figure}
To illustrate some of the properties of the GS vertex and its implications, we consider a two-loop process in which we have two massless vector bosons in the initial and in the final state with the exchange in the $s$-channel of an axial-vector current (Fig.~\ref{GSGS}). The example that we provide here comes from anomalous axial QED,  but it can be extended to more realistic models with no major variants. In Fig.~\ref{GSGS} both the initial and the final state contain anomalous subdiagrams, but the two GS vertices are defined in such a way to absorb all the longitudinal subtractions terms inside each of the blobs.
The amplitude of such a process is given by
\ba
\label{GS_BIMeq}
{\mathcal M}^{\, \mu \nu \mu^\prime \nu^\prime}=
\Delta_{AVV}^{\lambda \mu \nu \, GS}(- k, - k_1, - k_2) \left(- \frac{i g_{\lambda \lambda^\prime}}{k^2} \right)
\Delta_{AVV}^{\lambda^\prime \mu^\prime \nu^\prime \,  GS}(k,k^\prime_1,k^\prime_2).
\ea
\begin{figure}[t]
\begin{center}
\includegraphics[scale=0.7]{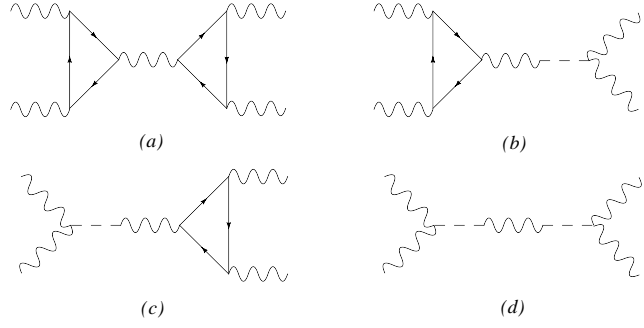}
\caption{\small All the contributions from the GS gauge invariant vertex, in the AVV case, to the amplitude $VV \rightarrow VV$ via an axial-vector current.}
\label{GS_BIM}
\end{center}
\end{figure}
In the expression above, the propagator is deprived of its longitudinal momentum dependence due a WI. The amplitude in Eq.~(\ref{GS_BIMeq}) can be decomposed into the four sub-amplitudes shown in Fig.~\ref{GS_BIM} after expanding the two GS vertices with Eq.~(\ref{cance})
\bea
{\mathcal M}^{\, \mu \nu \mu^\prime \nu^\prime} =
- \left( \Delta_{AVV}^{\lambda \mu \nu}(- k,- k_1,- k_2)-
k^\la C^{\mu \nu}(-k_1, -k_2)\right) \frac{i g_{\lambda \lambda^\prime}}{k^2}
\left( \Delta_{AVV}^{\lambda^\prime \mu^\prime \nu^\prime}(k, k^\prime_1,k^\prime_2)+
C^{\la^\prime \mu^\prime \nu^\prime}(k^\prime_1, k^\prime_2)\right), \nn\\
\eea
but only two sub-amplitudes survive (Fig.~\ref{GS_BIM}a) and \ref{GS_BIM}b)) because of the Ward Identities in Eq.~(\ref{WI}).
\begin{figure}[t]
\begin{center}
\includegraphics[scale=0.7]{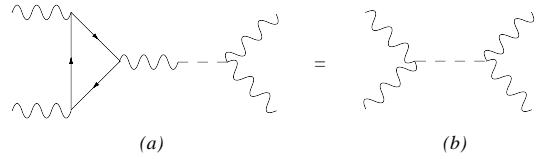}
\caption{\small The sub-amplitude in Fig.~\ref{GS_BIM}b after the contraction $k^\la \Delta_{AVV}^{\lambda \mu \nu}$ which gives the anomaly equation.}
\label{GS_BIM_WI}
\end{center}
\end{figure}
We are left with two contributions which cancel, for on-shell matrix elements. In fact, while off-shell the graph in Fig.~\ref{GS_BIM_WI} spoils unitarity, when instead the four external lines are on-shell  the triangle contribution (the first term) reduces to the DZ form and the cancellation between the two terms is identical. In view of the structure of the anomaly vertex and of the GS vertex given before, this cancellation implies that the anomaly diagram, for on-shell (axial-vector) photons and in the chiral limit, is purely longitudinal (DZ form). A similar result holds also for the $AAA$ case.

It is then natural to look at more general situations when these types of diagrams appear into higher-order contributions. In these more general  cases, the anomaly diagram does  not coincide with its DZ form, except for specific kinematical points ($k_1^2=k_2^2=k^2$, off-shell), and there is no identical cancellation of the anomalous trilinear gauge interactions.  We conclude that compared to the identical cancellation of the anomaly by charge assignment on each generation, which eliminates all-together all the trilinear gauge interactions, the GS vertex can be either transversal or vanishing (in the chiral limit) for each given flavour.

As we have previously mentioned, an anomaly vertex with the addition of the pole counterterms
(i.e. deprived of the anomaly pole) has been criticized in previous works in \cite{Adam:1997gj}. There the author brings in as an example a class of amplitudes which are affected by double poles, claiming a unitarity failure of the model. We will argue against this interpretation.
\begin{figure}[t]
\begin{center}
\includegraphics[scale=1.0]{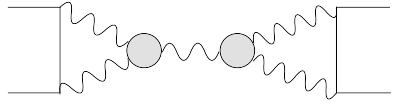}
\caption{\small The embedding of the BIM amplitude with GS vertices in a fermion/antifermion scattering.}
\label{ninefig}
\end{center}
\end{figure}
\begin{figure}[t]
\begin{center}
\includegraphics[scale=0.8]{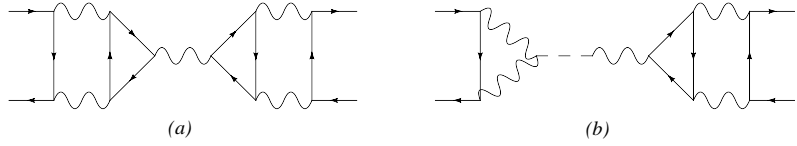}
\caption{\small Contributions to the fermion-antifermion scattering with the GS mechanism. 
\label{antifermion_fermion_GS}}
\end{center}
\end{figure}
\subsection{Embedding the GS vertex into higher-order diagrams \label{sec:GSembedding}}
When we embed the amplitude into higher-order diagrams (see Fig.~\ref{ninefig}), and consider an on-shell fermion-antifermion scattering, according to \cite{Adam:1997gj}, we are forced to move away from a symmetric configurations of the loop momenta and the identical vanishing of the anomaly is not ensured any longer, for the reasons that we have
just  raised above. In particular, according to \cite{Adam:1997gj}, the $s$-channel exchange is affected
by a double pole.

There is no better way to check the correctness of these conclusions than going through an explicit computation of this amplitude.

Expanding the two GS vertices, which are clearly non-zero in this case, we end up with several contributions,  such as a graph with two triangle diagrams, and a specific set of counterterms.
The two contributions involved in the study of fermion-antifermion scattering are shown
in Fig.~\ref{antifermion_fermion_GS}. Notice the presence of the box-triangle diagram ($\mathcal{BT}$) in the first graph, which remains non-trivial to compute even in the chiral limit. The second graph is the only contribution which does not disappear, according to \cite{Adam:1997gj}, and therefore describes a spurious $s$-channel exchange characterized by a double pole.

The conclusions of \cite{Adam:1997gj} are erroneous for two reasons:
1) in this specific case the double pole cancels in the explicit computation, so it is not a good counter example;
2) the cancellation or the presence of double poles should be analyzed together with the anomalous vertex and not separately. This second point will be addressed in the next sections.

To prove point 1) we need the two-loop structure of the $\mathcal{BT}$ graph, which
is just proportional to its tree-level axial-vector form times a form factor ${\bf G}(s)$,
function of $s=(p_1 + p_2)^2$,
\ba
\bar{v}(p_2) \gamma^\lambda \gamma^5  u(p_1) {\bf G}(s).
\ea
Explicit expressions of the ${\bf G}(s)$ coefficient in the massless case are obtained from \cite{Bernreuther:2005rw} and are given by
\ba
&& {\mathcal Re}{\bf G}(s,m_i=0,m_q=0) =3\log\left(\frac{s}{\mu^2}\right)-9+2\zeta(2)
\nonumber\\
&& {\mathcal Im}{\bf G}(s,m_i=0,m_q=0) =-3,
\ea
where $m_i$ is the mass of the internal fermion with flavor $i$ circulating in the triangle diagram
while $m_q$ is the mass of the fermionic external leg with flavor $q$.

We have for the two sub-amplitudes in Fig.~\ref{antifermion_fermion_GS}
\bea
&&{\mathcal M}_a= - {\bf G}^2(s)\bar{v}(p_2) \gamma^\lambda \gamma^5u(p_1) \frac{i}{k^2} \bar{u}(p^\prime_1)
\gamma_\lambda \gamma^5 v(p^\prime_2)
\nonumber\\
&&{\mathcal M}_b= i \, \int \frac{d^4 k_1}{(2 \pi)^4} \Bigg[ \bar{v}(p_2)
\gamma^\nu  \frac{ \ds p_1 -\ds k_1} {(p_1 -k_1)^2} \gamma^\mu u(p_1)
\frac{1}{k^2_1}  \frac{1}{k^2_2} a_n \epsilon[\mu, \nu, k_1, k_2]  \Bigg]
\frac{1}{(k^2)^2} \bar{u}(p^\prime_1) \ds k \gamma^5 {\bf G}(s) v(p^\prime_2)=0\,,
\nonumber\\
\eea
where ${\mathcal M}_b$ is identically zero, because of the equations of motion satisfied by the external fermion lines ($k = p^\prime_1 + p^\prime_2$).

\begin{figure}[t]
\begin{center}
\includegraphics[scale=0.6]{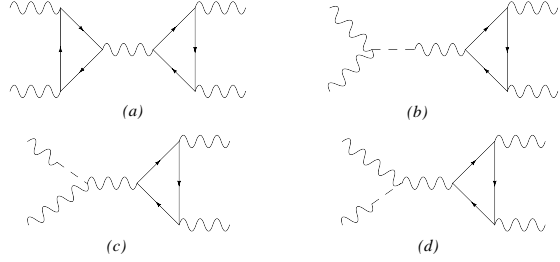}
\caption{\small All the contributions in the symmetric GS vertex. 
\label{four_terms}}
\end{center}
\end{figure}

We can easily generalize our analysis to an $AAA$ case. When an $AAA$ triangle is embedded in the two-loop fermion-antifermion scattering process
we can formally write the amplitude as follows
\ba
{\mathcal S}=\int\frac{d^4 k_1}{(2\pi)^4}\frac{d^4 k^{\prime}_1}{(2\pi)^4}
\bar{v}(p_2)\gamma^{\nu}\frac{1}{\ds p_1-\ds k_1}\gamma^{\mu}u(p_1)
\frac{1}{k_1^2}\frac{1}{k_2^2}{\mathcal S}^{\, \mu \nu \mu^\prime \nu^\prime}
\bar{u}(p^{\prime}_1)\gamma^{\mu^{\prime}}\frac{1}{\ds k_1^{\prime}-\ds p_1^{\prime}}
\gamma^{\nu^{\prime}}v(p^{\prime}_2)\frac{1}{{k^{\prime}_1}^2}\frac{1}{{k^{\prime}_2}^2}
\ea
where the tensor sub-amplitude is defined as
\bea
{\mathcal S}^{\, \mu \nu \mu^\prime \nu^\prime}&=&
- \, \Delta_{AAA}^{\lambda \mu \nu \, GS}(-k,-k_1,-k_2)
\frac{i g^{\lambda \lambda^\prime}}{k^2}
\Delta_{AAA}^{\lambda^\prime \mu^\prime \nu^\prime \,  GS}(k,k^\prime_1,k^\prime_2)
\nonumber\\
&=&
- [ \Delta_{AAA}^{\lambda \mu \nu}(-k,-k_1,-k_2)
+ C^{\lambda \mu \nu}_{AAA}(-k,-k_1,-k_2) ]   \frac{i}{k^2}
  [ \Delta_{AAA}^{\lambda \mu^\prime \nu^\prime}(k,k^\prime_1,k^\prime_2)
+ C^{\lambda \mu^\prime \nu^\prime}_{AAA}(k,k^\prime_1,k^\prime_2)].
\nonumber\\
\label{S_AAA}
\eea
Using the WI $k^{\lambda} \Delta_{AAA}^{\lambda \mu \nu \, GS}(-k,-k_1,-k_2) = 0$,
we can drop the GS counterterm $C^{\lambda \mu^\prime \nu^\prime}_{AAA}(k,k^\prime_1,k^\prime_2)$.
In this way the sixteen terms which contribute to the sub-amplitude
${\mathcal S}^{\, \mu \nu \mu^\prime \nu^\prime}$ given in
Eq.~(\ref{S_AAA}) reduce to the four terms shown in Fig.~\ref{four_terms}.
Therefore the total femion-antifermion scattering amplitude is given by the sum of four terms
${\mathcal S}_a$, ${\mathcal S}_b$, ${\mathcal S}_c$ and ${\mathcal S}_d$ which are defined below
\begin{figure}[t]
\begin{center}
\includegraphics[scale=0.4]{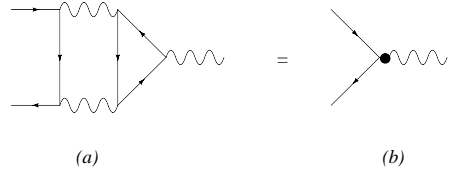}
\caption{\small The axial-vector form factor for the $\mathcal{BT}$ diagram. }
\label{remiddi}
\end{center}
\end{figure}
\begin{figure}[t]
\begin{center}
\includegraphics[scale=0.8]{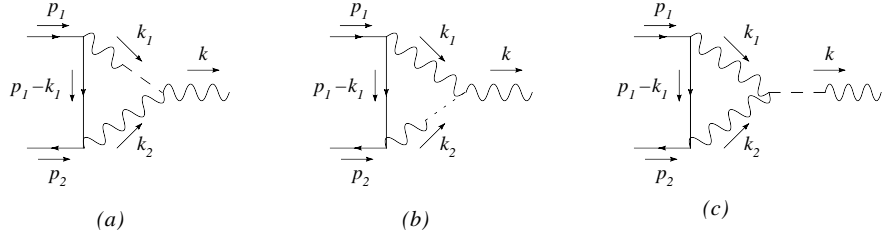}
\caption{\small The one-loop GS counterterms included in the $\mathcal{BT}$ as its longitudinal part. }
\label{null2}
\end{center}
\end{figure}
\ba
&&{\mathcal S}_a= - \, \mathcal{ BT}^{ \lambda }_{AAA}  \frac{i}{k^2} \mathcal{ BT}^{ \lambda}_{AAA}
\nonumber\\
&&{\mathcal S}_b= i \, \int  \frac{d^4 k_1}{(2 \pi)^4}   \Bigg( \bar v (p_2)
\gamma_\nu \frac{1}{\ds p_1 - \ds k_1} \gamma_\mu u(p_1) \frac{1}{k_1^2}
\frac{1}{k_2^2}  \Bigg)  \frac{a_n}{3} \frac{k^\lambda}{k^2}
\epsilon[\mu, \nu, k_1, k_2]   \frac{1}{k^2}
\mathcal{ BT}^{ \lambda}_{AAA}
\nonumber\\
&&{\mathcal S}_c= i \, \int\frac{d^4 k_1}{(2 \pi)^4}
\Bigg( \bar v (p_2) \gamma_\nu \frac{1}{\ds p_1 - \ds k_1} \gamma_\mu u(p_1) \frac{1}{k_2^2}
\frac{1}{k_1^2}  \frac{a_n}{3}\frac{k^\mu_1}{k_1^2}
\epsilon[\nu, \lambda, k_2, k]\Bigg) \frac{1}{k^2}
\mathcal{ BT}^{ \lambda}_{AAA}
\nonumber\\
&&{\mathcal S}_d=i \, \int  \frac{d^4 k_1}{(2 \pi)^4} \Bigg( \bar v (p_2) \gamma_\nu \frac{1}{\ds p_1 - \ds k_1} \gamma_\mu u(p_1) \frac{1}{k_1^2}
\frac{1}{k_2^2}  \frac{a_n}{3} \frac{k^\nu_2}{k_2^2}  \epsilon[\lambda, \mu, k, k_1]   \Bigg)   \frac{1}{k^2}
\mathcal{ BT}^{ \lambda}_{AAA}\,,
\ea
where we have defined
\ba
\mathcal{ BT}^{ \lambda}_{AAA}(k,p^\prime_1,p^\prime_2)= - \int\frac{d^4 k^{\prime}_1}{(2\pi^4)}
\Delta_{AAA}^{\lambda \mu^\prime \nu^\prime}(k,k^\prime_1,k^\prime_2)
\bar{u}(p^{\prime}_1)\gamma^{\mu^{\prime}}\frac{1}{\ds k_1^{\prime}-\ds p_1^{\prime}}
\gamma^{\nu^{\prime}}v(p^{\prime}_2)\frac{1}{{k^{\prime}_1}^2}\frac{1}{{k^{\prime}_2}^2}
\ea
and the total amplitude is given by
\bea
{\mathcal S}={\mathcal S}_a +{\mathcal S}_b+{\mathcal S}_c+{\mathcal S}_d.
\eea
In the ${\mathcal S}_b$ sub-amplitude we distribute the anomaly symmetrically on each vertex and using the following Ward Identities on the vector currents we obtain
\bea
k^{\lambda}  \mathcal{ BT}^{ \lambda}_{VAV}  =  k^{\lambda}  \mathcal{ BT}^{ \lambda}_{VVA} = 0.
\eea
This allows us to simplify the ${\mathcal S}_b$ expression as follows
\bea
{\mathcal S}_b = i \, \int  \frac{d^4 k_1}{(2 \pi)^4}
\Bigg( \bar v (p_2) \gamma_\nu \frac{1}{\ds p_1 - \ds k_1} \gamma_\mu u(p_1) \frac{1}{k_1^2}
\frac{1}{k_2^2}    \Bigg)  \frac{a_n}{3} \frac{1}{(k^2)^2}
\epsilon[\mu, \nu, k_1, k_2] \bar u(p^\prime_1) \ds k \gamma^5
G(s) v(p^\prime_2)=0\,,   \nonumber\\
\eea
where we have used the result shown in \cite{Bernreuther:2005rw}.

Also the third amplitude ${\mathcal S}_c$ does not contribute to ${\mathcal S}$,
in fact we have
\ba
{\mathcal S}_c = i \, \bar v (p_2) \gamma_\nu u(p_1)
\frac{a_n}{3}   \epsilon[\nu, \lambda, \rho, \sigma] k^\sigma \int  \frac{d^4 k_1}{(2 \pi)^4}   \Bigg(  \frac{k_1^\rho}{(k-k_1)^2 k_1^4 }
\Bigg) \frac{1}{k^2}\mathcal{ BT}^{ \lambda}_{AAA}=0,
\ea
which vanishes by symmetry due to the structure of the tensor integral, which is proportional to $k^{\rho}$.
In the same way also the fourth contribution vanishes, ${\mathcal S}_d =0$. This explicit computations contradicts the
conclusions of \cite{Adam:1997gj} where the same amplitude was conjectured to be affected by double poles in the $s$-channel.

There are some conclusions to be drawn. The first is that
the replacement of the two anomaly vertices in the amplitude with two GS vertices, in this case,
is irrelevant as far as the fermions are massless.
Equivalently, the longitudinal components of the two anomaly vertices decouple in the graph and the
only invariant amplitudes coming from the anomaly vertices that survive - after the integration on the second loop momentum - are the transverse ones. This is one special case in which the anomaly diagram
is transverse in the virtual corrections just by itself. In other cases this does not happen and the extra poles introduced by the counterterm are sufficient to cancel those generated by the anomaly. For this reason the presence of extra poles in a partial amplitude is not necessarily the sign of an inconsistency.
\begin{figure}[h]
\begin{center}
\includegraphics[scale=0.4]{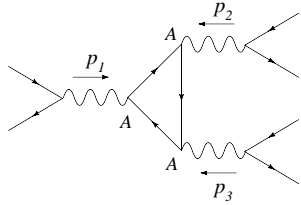}
\caption{\small Typical graph in which the longitudinal anomalous component vanishes.}
\label{fftriangle}
\end{center}
\end{figure}
\subsection{The vertex in the longitudinal/transverse (L/T) formulation \label{sec:LTformulation}}
The analysis presented above becomes more transparent if we use a special parameterization of the anomaly diagram in which the longitudinal part of the vertex is separated from the transverse one, as done in recent studies of radiative corrections to the anomalous magnetic moment of the gluon \cite{Knecht:2003xy}. This parameterization is more convenient than the usual Rosenberg form \cite{Rosenberg:1962pp}.

While the longitudinal component of the anomaly diagram is given by its DZ form, once this component is subtracted from the general triangle diagram, it leaves behind an anomaly-free vertex which is purely transverse and corresponds to the GS trilinear interaction.
The Ward Identities restrict the general covariant decomposition of ${{\Delta^{GS}}}_{\lambda\mu\nu}(k_3,k_1,k_2)$ into invariant functions to three terms (with all incoming momenta)
\beqa
\label{calw}
{{ \Delta^{GS}}}_{\lambda\mu\nu}(k_1,k_2) &=&
 -\,\frac{1}{8\pi^2}\,\left(
w_T^{(+)}\left(k_1^2,k_2^2,k_3^2\right)\,t^{(+)}_{\lambda\mu\nu}(k_1,k_2)
 +\,w_T^{(-)}\left(k_1^2,k_2^2,k_3^2\right)\,t^{(-)}_{\mu\nu\rho}(k_1,k_2) \right. \nonumber \\
 && \left. + {\widetilde{w}}_T^{(-)}\left(k_1^2,k_2^2,k_3^2\right)\,{\widetilde{t}}^{(-)}_{\mu\nu\rho}(k_1,k_2)
\right),
 \eeqa
with the transverse tensors given by
\beqa
t^{(+)}_{\mu\nu\rho}(k_1,k_2) &=&
k_{1\nu}\,\epsilon_{\mu\rho\alpha\beta}\ k_1^\alpha k_2^\beta \,-\,
k_{2\mu}\,\epsilon_{\nu\rho\alpha\beta}\ k_1^\alpha k_2^\beta \,-\, (k_{1}\cdot
k_2)\,\epsilon_{\mu\nu\rho\alpha}\ (k_1 - k_2)^\alpha
\nonumber\\
&& \quad\quad+\ \frac{k_1^2 + k_2^2 - k_3^2}{k_3^2}\
\epsilon_{\mu\nu\alpha\beta}\ k_1^\alpha k_2^\beta(k_1 + k_2)_\rho
\nonumber \ , \\
t^{(-)}_{\mu\nu\rho}(k_1,k_2) &=& \left[ (k_1 - k_2)_\rho \,-\, \frac{k_1^2 - k_2^2}{(k_1
+ k_2)^2}\,(k_1 + k_2)_\rho \right] \,\veps_{\mu\nu\alpha\beta}\ k_1^\alpha k_2^\beta
\nonumber\\
{\widetilde{t}}^{(-)}_{\mu\nu\rho}(k_1,k_2) &=& k_{1\nu}\,\epsilon_{\mu\rho\alpha\beta}\
k_1^\alpha k_2^\beta \,+\, k_{2\mu}\,\epsilon_{\nu\rho\alpha\beta}\ k_1^\alpha k_2^\beta
\,-\, (k_{1}\cdot k_2)\,\epsilon_{\mu\nu\rho\alpha}\ (k_1 + k_2)^\alpha \,,
\lbl{tensors}
\eeqa
where, due to Bose symmetry ($k_1,\mu \leftrightarrow k_2,\nu$) we have
\beqa
w_T^{(+)}\left(k_2^2,k_1^2,k_3^2\right)
&=&+w_T^{(+)}\left(k_1^2,k_2^2,k_3^2\right),
\nonumber \\
w_T^{(-)}\left(k_2^2,k_1^2,k_3^2\right) &=& -
w_T^{(-)}\left(k_1^2,k_2^2,k_3^2\right),
~~~
{\widetilde{w}}_T^{(-)}\left(k_2^2,k_1^2,k_3^2\right) = -
{\widetilde{w}}_T^{(-)}\left(k_1^2,k_2^2,k_3^2\right) \,.
\eea
This version of the GS-corrected AVV vertex satisfies the Ward Identities on all the three external lines. The explicit expression of these invariant amplitudes can be obtained from
\cite{Jegerlehner:2005fs}
\begin{eqnarray}
{\widetilde{w}}_{T}^{(-)}(k_1^2,k_2^2,k_3^2) &=& -  w_{T}^{(-)}(k_1^2,k_2^2,k_3^2),
\\
k_3^2 \Delta^2  w^{(-)}_{T}(k_1^2,k_2^2,k_3^2)
&=& 8 (x-y)\Delta
+8(x-y)(6xy + \Delta)\Phi^{(1)}(x,y)
\nonumber
\\
&-&4 [18 x y + 6 x^2-6 x + (1+x+y)\Delta)] L_x
\nonumber
\\
&+& 4[18 x y + 6 y^2-6 y + (1+x+y)\Delta)] L_y \, ,
\\
\nonumber
\\
k_3^2 \Delta^2  w^{(+)}_{T}(k_1^2,k_2^2,k_3^2)&=&8[6xy + (x+y)\Delta]\Phi^{(1)}(x,y)
+8\Delta
\nonumber
\\
&-&4 [6 x+ \Delta ](x-y-1) L_x
\nonumber \\
&+&4 [6 y+ \Delta] (x-y+1) L_y
\end{eqnarray}
with
\begin{equation}
L_x = \ln x,~~~~
L_y = \ln y,~~x=\frac{k_1^2}{k_3^2}~~y=\frac{k_2^2}{k_3^2}
\end{equation}
which involves the scalar triangle diagram for general off-shell lines and determines the function $\Phi^{(1)}$ \cite{Usyukina:1992jd} as
\be
\label{Phi1}
\Phi^{(1)} (x,y) = \frac{1}{\lambda} \left\{ \frac{}{}
2 \left( \Li{2}{-\rho x} + \Li{2}{-\rho y} \right)
+ \ln\frac{y}{x} \ln{\frac{1+\rho y}{1+\rho x}}
+ \ln(\rho x) \ln(\rho y) + \frac{\pi^2}{3}
\right\} ,
\ee
where
\be
\label{lambda}
\lambda(x,y) \equiv \sqrt{\Delta} \; \; \; ,
\; \; \; \rho(x,y) \equiv 2 \; (1-x-y+\lambda)^{-1}, ~~\Delta\equiv(1-x-y)^2 - 4 x y \;.
\ee
The full anomaly amplitude is simply obtained by adding the anomaly pole to this expression

\beq
\Delta^{\lambda\mu\nu}= w_L k^\lambda \epsilon[\mu,\nu,k_1,k_2] +\Delta^{GS\, \lambda\mu\nu}
\label{long}
\eeq
with $w_L=1/(8 \pi^2 k^2)$.

In a non-anomalous theory  a specific charge assignment -in the chiral limit- sets to zero the entire trilinear gauge interaction (identically), while in theories characterized by the GS vertex we require the vanishing of the anomalous part (the anomaly pole). The pole is part of the expression of the triangle diagram, which may or may not contribute in certain graphs.
A typical example is shown in Fig.~\ref{fftriangle} which is not sensitive (in the massless fermion limit) to the longitudinal component of the
anomaly, due to the Ward Identities satisfied by the fermion antifermion currents on each external photon line (we are considering axial-vector interactions for each photon). In fact, this is a case in which a GS vertex or a complete anomaly vertex give the same contributions.  In this sense, the anomaly, for this graph, is harmless since the external fermion current is conserved, but this situation is not general.  In fact, we will analyze cases in which a similar situation occurs, and others in which the decoupling of the longitudinal part requires the subtraction of $w_L$ from the anomaly diagram. As we have seen above, there are other cases in which the GS vertex is identically vanishing, and this happens in graphs in which the transverse part of the anomaly diagram is zero, as in light-by-light scattering. The anomaly diagrams are purely longitudinal, and their replacement with the GS vertex has to give necessarily zero. We come therefore to discuss point 2) which has been raised in the previous section. We cannot address the issue of double poles {\em only} in the DZ counterterms and forget that the same poles are also present in the triangle anomaly. In other words: the cancellation of the counterterms in specific graphs takes place if and only if the anomaly diagram is harmless.
\subsection{Examples of explicit GS counterterms: anomaly in muon decay}
\begin{figure}[t]
\begin{center}
\includegraphics[scale=0.8]{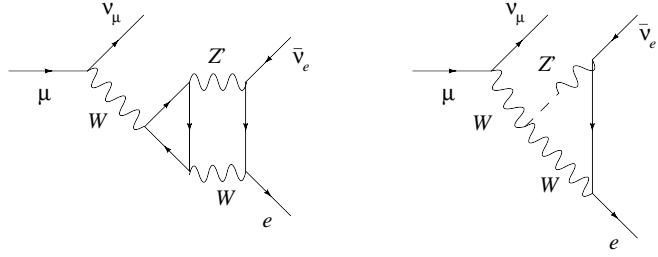}
\caption{\small The muon decay process via a $BT$ diagram. }
\label{muon_decay}
\end{center}
\end{figure}
In general, a given amplitude containing an anomalous extra $Z^{\prime}$ can be harmless if we neglect all the fermion masses and harmful in the opposite case.
An interesting example is shown in Fig.~\ref{muon_decay} which describes a special decay of the muon, mediated by a $WW Z$ vertex. In general, in anomalous extensions of the SM, this amplitude requires a longitudinal subtractions either with the inclusion of a GS or a WZ counterterm. For massless fermions, for instance, the process is anomaly-free. To show this point consider the amplitude for the second diagram in Fig.~\ref{muon_decay} which is given by
\bea
\mathcal{P}&=& i \, \int \frac{d^4 k_2}{(2 \pi)^4} \bar u (p^\prime_1) \g^\mu \frac{1}{\ds p^\prime_1+ \ds k_1} \g^\nu v(p^\prime_2)\, \frac{1}{k_1^2} \left( g^{ \b \nu} - \frac{k_2^\b k_2^\nu}{M_W^2} \right) \frac{1}{k_2^2-M_W^2} \nn \\
&& \frac{a_n}{3} \frac{k_2^\b}{k_2^2} \epsilon[\alpha, \mu, k, k_1] \frac{1}{k^2}\bar u (p_2) \g^\alpha u(p_1) \nn \\
&=& i \, \bar u (p^\prime_1) \g^\mu v(p^\prime_2)\frac{1}{k^2}\bar u (p_2) \g^\alpha u(p_1)\epsilon[\alpha, \mu, p^\prime_{12}, \tau] \,\int \frac{d^4 k_2}{(2 \pi)^4} \frac{k_2^\tau}{k_2^2\,(-p^\prime_{12} - k_2)^2}.
\eea
After using the equations of motion for on-shell spinors with $p^\prime_1+p^\prime_2+k_1+k_2=0$ and the tensor integral decomposition in terms of the only momentum in the loop, $p^\prime_{12}=p^\prime_1+p^\prime_2$, it is trivial to verify that the expression vanishes. If we switch-on the external fermion masses, violation of the Ward Identities will induce a longitudinal coupling of the anomaly pole on the neutral current, which need an explicit GS subtraction. We will come back
to address the structure of the anomalous contributions away from the chiral limit below, when we will analyze the contribution of similar diagrams to $g-2$ of the muon.
\subsection{Self-energy and gauge invariance}
\begin{figure}[t]
\begin{center}
\includegraphics[scale=0.8]{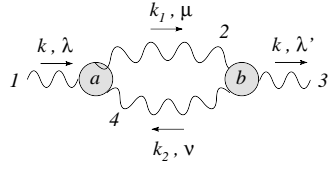}
\caption{\small Self-energy  amplitude. }
\label{self}
\end{center}
\end{figure}
Anomalous contributions, in these models, appear also in the running of the coupling, though at a rather large order.
Also in this case transversality of the self-energy is ensured by construction, being the GS vertex transverse
by definition, however, the separation of the vertex into anomaly graph and GS counterterms illustrates how the cancellation of all the double poles takes place.

The corrections appear at three-loop level and are shown in Fig.~\ref{self}. We denote with $a$, $b$ the GS vertices and assign a number on each line of all the vertices. For instance, in
Fig.~\ref{self} vertex $a$ shares lines 2 and 4 with vertex $b$. The anomaly diagrams are denoted by
$\Delta_a$ and $\Delta_b$, respectively, and are separated into the linear combinations $AVV, VAV$ and $VVA$  as in the previous example, carrying partial anomalies. The pole counterterm can be ``emitted'' by the
vertex either toward the initial or the final state of the diagram along the numbered line. For instance $C_{a2}$ denotes the DZ counterterm that is generated by vertex $a$ with a mixing/ double pole term generated on line $2$. Other trivial cancellations are obtained due to the orthogonality relations between DZ counterterms associated to different lines when they are contracted together.

The expression of the integrand in the amplitude is given by
\bea
{\mathcal M}_{self}^{\lambda \lambda^\prime} &=& - \Delta_{AAA, a}^{\lambda \mu \nu \, GS} (k, k_1, -k_2)
\frac{ g^{\mu \mu^\prime} }{ k_1^2 }  \frac{ g^{ \nu \nu^\prime} }{(k_1-k)^2} \Delta_{AAA, b}^{\lambda^\prime \mu^\prime \nu^\prime  \, GS}( - k, - k_1, k_2)    \nonumber\\
&=&  - \left[  \Delta_{AAA,a} + C_{AAA,a}  \right] ^{\lambda \mu \nu } \frac{1}{k_1^2 (k_1-k)^2}
  \left[  \Delta_{AAA,b} + C_{AAA,b}  \right]^{\lambda^\prime \mu \nu },
\eea
which is given explicitly by
\beqa
C^{\lambda^\prime \mu \nu}_{AAA, b}(-k,-k_1,k_2) &=& \frac{1}{3} \left(C^{\lambda^\prime \mu \nu}_{b3}(-k,-k_1,k_2)
+ C^{\mu \nu \lambda^\prime}_{b2}(k_1,k_2,k)  +  C^{\nu \lambda^\prime \mu}_{b4}(-k_2,k,- k_1)   \right)    \nonumber\\
&=& C^{\mu \nu}_{b3}(- k_1,k_2)k^{\lambda^\prime} + C^{\nu \lambda^\prime}_{b2}(k_2,k)k_1^\mu
 + C^{\lambda^\prime \mu}_{b4}(k,- k_1) k_2^\nu .
\eeqa
\beqa
C^{\lambda \mu \nu}_{AAA, a}(k,k_1,-k_2) &=&
\frac{1}{3} \left( C^{\lambda \mu \nu}_{a1}(k,k_1,-k_2) + C^{\mu \nu \lambda}_{a2}(-k_1,-k_2,-k)
+  C^{\nu \lambda \mu}_{a4}(k_2, -k,k_1)   \right)    \nonumber\\
&=&   C^{\mu \nu}_{a1}(k_1,-k_2)k^\lambda + C^{\nu \lambda}_{a2}(-k_2,-k)k_1^\mu
 +  C^{\lambda \mu}_{a4}(-k,k_1)k_2^\nu
\eeqa
so using the Ward Identities
\bea
k^{\mu}_1 \Delta_{AAA, b}^{\lambda^\prime \mu \nu \, GS}( - k, - k_1, k_2)=
k^{\nu}_2 \Delta_{AAA, b}^{\lambda^\prime \mu \nu \, GS}( - k, - k_1, k_2)=0
\eea
we can reduce the sixteen contributions of the amplitude ${\mathcal M}_{self}^{\lambda \lambda^\prime}$ to eight terms.

\begin{figure}[t]
\begin{center}
\includegraphics[scale=0.8]{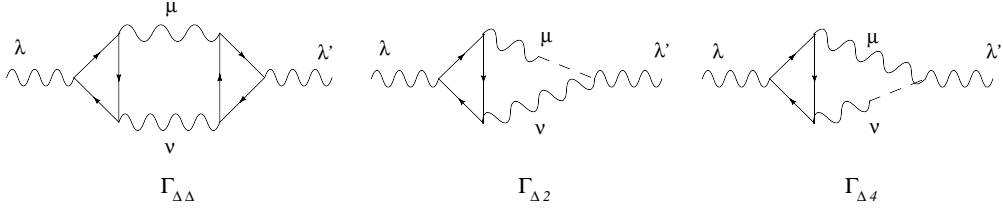}
\caption{\small Contributions to the self-energy amplitude. }
\label{self_terms}
\end{center}
\end{figure}

The GS counterterms $C_{a1}$ and $C_{b3}$  are non-zero for off-shell external photons and are needed to remove the longitudinal poles from the anomaly diagram, while the remaining counterterm contributions are those shown in Fig.~\ref{self_terms}. The latter are transverse just by themselves, as we are going to show. They are given by
\bea
{\mathcal M}_{self}^{\lambda \lambda^\prime} &=&
 -  \Delta_{AAA,a} ^{\lambda \mu \nu } \frac{1}{k_1^2 (k_1-k)^2}
  \left[  \Delta_{AAA,b} + C_{b2} + C_{b4}  \right]^{\lambda^\prime \mu \nu }     \nonumber\\
  &=& -\left(  \Delta_{AAA,a}^{\lambda \mu \nu}  \frac{1}{k_1^2 (k_1-k)^2}   \Delta_{AAA,b}^{\lambda^\prime \mu \nu}
  +   \Delta_{AAA,a}^{\lambda \mu \nu}  \frac{1}{k_1^2 (k_1-k)^2} C_{b2}^{ \mu \nu \lambda^\prime}
 +  \Delta_{AAA,a}^{\lambda \mu \nu}  \frac{1}{k_1^2 (k_1-k)^2}    C_{b4}^{\nu \lambda^\prime \mu } \right) \nonumber\\
 &=&  \Gamma_{\Delta \Delta}^{\lambda \lambda^\prime} + \Gamma_{\Delta 2}^{\lambda \lambda^\prime}
 + \Gamma_{\Delta 4}^{\lambda \lambda^\prime}.
\eea
The amplitude $\Gamma_{\Delta2}$ can be cast in this form using dimensional regularization in $D$ dimensions
\bea
\Gamma^{\lambda \lambda^\prime}_{\Delta 2} &=&
- \int \frac{d^4 k_1}{(2 \pi)^4} \Delta^{\lambda \mu \nu}_{AAA,a}(k, k_1, -k_2)
\frac{1}{k^2_1(k_1-k)^2} C^{\mu \nu \lambda^\prime}_{b2}(k_1, k_2,k)  \nonumber\\
&=&   \frac{a_n}{3} \epsilon[\lambda, \nu, \alpha, k]   \frac{a_n}{3} \epsilon[\nu, \lambda^\prime, \beta, k] \int \frac{d^4 k_1}{(2 \pi)^4}
\frac{k_1^\alpha k_1^\beta}{k_1^4 (k_1 -k)^2}       \nonumber\\
&=& - \frac{1}{2} \left( \frac{a_n}{3} \right)^2 \epsilon[\lambda, \nu, \alpha, k] \epsilon[\nu, \lambda^\prime, \beta, k]
g^{\alpha \beta} \frac{1-D}{s} Bub^{D+2}(s)    \nonumber\\
 &=& \left( \frac{a_n}{3} \right)^2 (k^\lambda k^{\lambda^\prime} - k^2 g^{\lambda \lambda^\prime})
 (1-D) \frac {Bub^D(s)}{8 \pi (3-2 \eps )} \nn \\
 &=&  C\, (k^\lambda k^{\lambda^\prime} - k^2 g^{\lambda \lambda^\prime})  Bub^D(s),
\eea
 where the explicit expressions of the two master integrals $Bub^D(s)$ and $Bub^{D+2}(s)$ can be found in \cite{Armillis:2008bg}  and
 \bea
 C= \left( \frac{a_n}{3} \right)^2 \frac{1-D}{8 \pi (3-2 \eps )}.
 \eea
\begin{figure}[t]
\begin{center}
\includegraphics[scale=0.8]{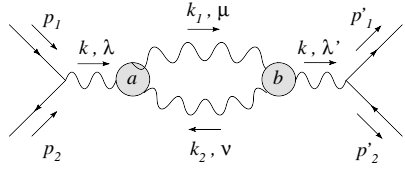}
\caption{\small A self-energy amplitude embedded in a physical process of fermion-antifermion scattering with on-shell external lines. }
\label{self_ffbar}
\end{center}
\end{figure}
If we include the same amplitude in a fermion-antifermion scattering, see Fig.~\ref{self_ffbar}, we obtain
\bea
\mathcal{S}_{\Delta2} &=& - \bar v(p_2) \gamma^\lambda u(p_1) \frac{1}{k^2}
 \left( \frac{k^\lambda k^{\lambda^\prime}}{k^2} - g^{\lambda \lambda^\prime} \right) k^2 C \,Bub^D(s)\,  \frac{1}{k^2} \bar u(p^\prime_1)
\gamma^{\lambda^\prime}  v(p^\prime_2)    \nonumber\\
&=& \bar v(p_2) \gamma^\lambda u(p_1) \frac{1}{k^2}
C \, Bub^D(s) \, \bar u(p^\prime_1)
\gamma^{\lambda}  v(p^\prime_2),
\eea
with
\beq
Bub^{D}(s) = \frac{i \pi^{D/2}}{(2 \pi)^D}\, \mu^{2 \eps} \, \biggl(\frac{e^{\g}}{4 \pi}\biggr)^\eps\frac{c_\Gamma}{ \eps (1-2 \eps)} (s)^{- \eps} (-1)^{\eps},
\label{BubDs0}\\
\eeq
where we have used the equations of motion for the on-shell spinors ($k=p_1+p_2=p^\prime_1 + p^\prime_2$).
The transversality of the pole counterterm comes as a surprise, since while the total amplitude with GS vertices is transverse by construction, the anomalous contribution, in principle, is not expected separately to be so. The computation shows that internal double poles, those due to the GS counterterms, give contributions which are also transversal. This shows once more that there are no apparent inconsistencies in the perturbative expansion of the theory.
\section{Higher-order diagrams}
Having worked out several examples in which either the extra poles appear explicitly or cancel by themselves, signalling a harmless anomaly, we now
move to discuss more complex cases, where these techniques will be systematized.

We have two ways to apply the GS vertex at higher-order. We could use its explicit form -in terms of its transverse invariant amplitudes- or we could use it in the form "anomaly diagrams plus counterterms". This second form is the most useful one.
The presence of higher poles in the counterterms, which balance those -not explicit- in the anomaly
diagrams, can be treated perturbatively as a field theory of a higher perturbative order. We will illustrate below one case from which we can easily infer the general features of the perturbative expansion with these types of graphs. It should be clear that the cancellation of all the poles from the external lines takes place only on-shell, but this is not a problem since we are interested in $S$-matrix elements.
\subsection{Three-point functions}
\begin{figure}[th]
\begin{center}
\includegraphics[scale=0.75]{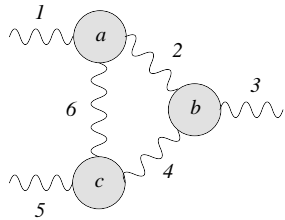}
\caption{\small A four-loop amplitude given by three GS vertices with on-shell external lines. }
\label{threeblobs}
\end{center}
\end{figure}

For this reason we consider the four-loop diagram shown in Fig.~\ref{threeblobs} with three symmetric GS vertices of the $AAA$ type connected together, which is given by
\bea
\mathcal{M}^{\la \rho \tau}= i\, (\Delta + C_1 + C_2 + C_6)_a^{\la \mu \nu} \frac{1}{k_2^2}\,
(\Delta + C_2 + C_3 + C_4)_b^{\mu \rho \sigma} \, \frac{1}{k_4^2}
(\Delta + C_4 + C_5 + C_6)_c^{\sigma \tau \nu} \frac{1}{k_6^2},
\eea
where, as done in the previous sections, $\Delta$ denotes an $AAA$ triangle amplitude with a symmetric anomaly distribution on each vertex and  $C_i$ a single GS counterterm with the derivative coupling on the i-th line.
At this stage we start simplifying the term $(\Delta + C_2 + C_3 + C_4)_b$ as $(GS)_b$ and in a similar way the $c$ blob using the Ward Identities
\bea
C_{a2} (GS)_b = C_{a6} (GS)_c= C_{b4} (GS)_c =0,
\eea
and then omit the GS counterterms $C_{a1}$, $C_{b3}$, $C_{c5}$ in which the transversality conditions
\bea
\eps_{1\la} k_1^{\la} = \eps_{5\tau} k_5^{\tau} = \eps_{3\rho} k_3^{\rho} =0
\eea
 act on the derivative coupling, which allow to reduce the $\mathcal{M}$ amplitude to the six contributions
\bea
\mathcal{M}^{\la \rho \tau} &=& i\, (\Delta)_a^{\la \mu \nu} \frac{1}{k_2^2}\,
(\Delta  + C_2)_b^{\mu \rho \sigma} \frac{1}{k_4^2}\,
(\Delta + C_4 + C_6)_c^{\sigma \tau \nu} \frac{1}{k_6^2}\,  \nn \\
&=& \frac{i}{k_2^2 \, k_4^2 \, k_6^2}
\left(\Delta_a \Delta_b \Delta_c + \Delta_a \Delta_b C_{c4} + \Delta_a \Delta_b C_{c6}
+ \Delta_a C_{b2} \Delta_c + \Delta_a C_{b2} C_{c4} + \Delta_a C_{b2} C_{c6} \right)^{\la \rho \tau} \nn \\
&=& \left( \Delta_a \Delta_b\Delta_c + \Gamma_4 + \Gamma_6 + \Gamma_2 + \Gamma_{24} + \Gamma_{26}\right)^{\la \rho \tau},
\label{Mthree}
\eea
where the notation $\Gamma_i$ and $\Gamma_{ij}$ refers to the line corresponding to the counterterm in Fig.~\ref{threeblobs}.
\begin{figure}[ht]
\begin{center}
\includegraphics[scale=0.7]{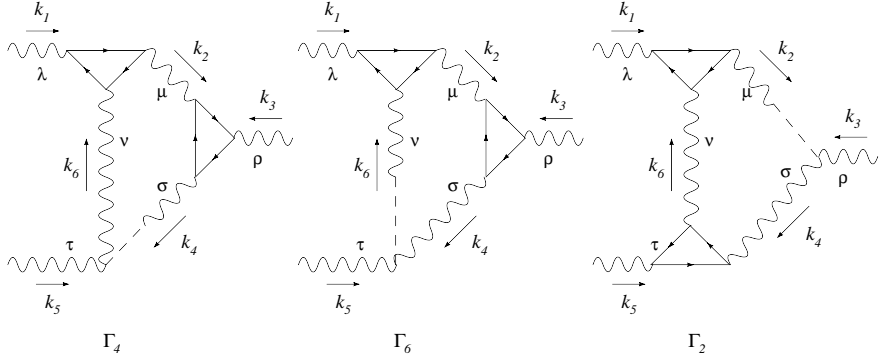}
\caption{\small The $\Gamma_4$,  $\Gamma_6$ and  $\Gamma_2 $ contributions taken from $\mathcal{M}$ at four-loop level.}
\label{DDC3}
\end{center}
\end{figure}
At this point we consider the $\Delta_a \Delta_b C_{c4}$ contribution represented in Fig.~\ref{DDC3} with a counterterm on the line $4$  denoted by $\Gamma_4^{\la \rho \tau}$
\bea
{\Gamma}_4^{\la \rho \tau} &=& i  \int \frac{d^4k_4}{(2 \pi)^4}
\bigg[ \Delta_a^{\lambda \mu \nu } (k_1, k_2, -k_6) \frac{1}{ k_2^2 }
\Delta_b^{\mu \rho \sigma} (k_2, -k_3, k_4)
\frac{ 1 }{k_4^2} \frac{a_n}{3 \,k_4^2}
k_4^{\sigma} \epsilon[\tau, \nu , k_5, k_6] \frac{1}{ k_6^2 }\bigg],
\eea
in which we substitute the Rosenberg parametrization for the triangle amplitude $\Delta_a^{\lambda \mu \nu } (k_1, k_2, -k_6)$ given by
\bea
\Delta_a^{\lambda \mu \nu } (k_1, k_2, -k_6) &=&
A_1 \, \epsilon[k_2,\mu, \nu, \la] -
A_2 \, \eps[k_6,\mu, \nu, \la] -
A_3 \, k_2^\nu \, \eps[k_2, k_6, \mu, \la] \nn \\
&+& A_4 \, k_6^\nu \, \eps[k_2, k_6, \mu,\la] -
A_5 \, k_2^\mu \, \eps[k_2, k_6, \nu, \la] +
A_6 \, k_6^\mu \, \eps[k_2, k_6, \nu, \la],
\label{Rosen1}
\eea
and  the anomaly equation
\bea
k_4^{\sigma} \Delta_b^{\mu \rho \sigma} (k_2, -k_3, k_4) = - \frac{a_n}{3} \eps [\mu, \rho, k_2, k_3].
\label{an_eq}
\eea
We choose $k_1$ and $k_5$ as independent momenta, so we have
\bea
k_3=-(k_1+k_5), \qquad \qquad k_2=k_1+k_4+k_5, \qquad \qquad k_6=k_4+k_5,
\eea
with  the on-shell conditions  $k_1^2=k_5^2=k_3^2=0$.
A direct computation of the $\Gamma_4^{\la \rho \tau}$ amplitude shows the complete cancellation of  the spurious double pole relative to the  $k_4$ momentum.

In an analogous way we can consider the $\Delta_a \Delta_b C_{c6}$ term or $\Gamma_6$ in Eq.~(\ref{Mthree}), that is
\bea
{\Gamma}_6^{\la \rho \tau} &=& i \int \frac{d^4k_6}{(2 \pi)^4}
\bigg[ \Delta_a^{\lambda \mu \nu } (k_1, k_2, -k_6) \frac{1}{ k_2^2 }
\Delta_b^{\mu \rho \sigma} (k_2, -k_3, k_4)
\frac{1}{k_4^2} \frac{a_n}{3 \,k_6^2}
\, k_6^\nu \epsilon[\sigma, \tau, k_4, k_5] \frac{1}{ k_6^2 }\bigg]
\eea
and the $\Delta_a C_{b2} \Delta_c $ term or $\Gamma_2$
\bea
{\Gamma}_2^{\la \rho \tau} &=& i \int \frac{d^4 k_2}{(2 \pi)^4}
\bigg[ \Delta_a^{\lambda \mu \nu } (k_1, k_2, -k_6) \frac{1}{ k_2^2 }
\frac{a_n}{3 \,k_2^2}  k_2^\mu \epsilon[\rho, \sigma, k_3, k_4] \frac{ 1}{k_4^2}
\Delta_c^{\sigma \tau \nu} (k_4, -k_5, k_6) \frac{1}{ k_6^2 }\bigg],
\eea
for which the conditions in Eqs.~(\ref{Rosen1}) and (\ref{an_eq})  have to be modified in a suitable form.   After the expansion of the tensor integrals in terms of the two external momenta $k_1$ and $k_5$ we can conclude that also in this case the double poles don't contribute to the physical on-shell amplitude.\\
\begin{figure}[ht]
\begin{center}
\includegraphics[scale=0.75]{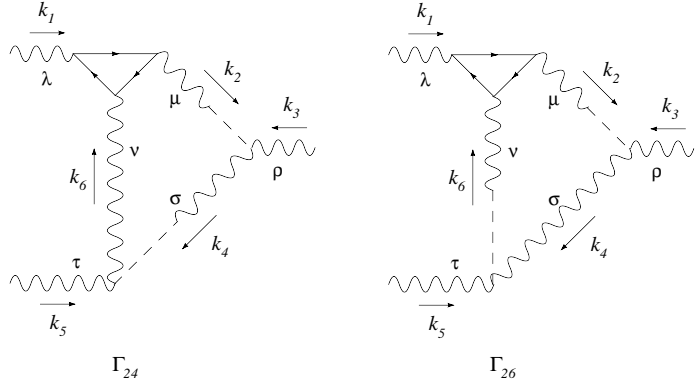}
\caption{\small Representation of $\Gamma_{24}$ and  $\Gamma_{26}$, the two terms with double poles on the internal lines. }
\label{DCC}
\end{center}
\end{figure}
In a similar way we can show the vanishing of the last contributions $\Delta_a\, C_{b2}\, C_{c4}$ ($\Gamma_{24}$) and
 $\Delta_a\, C_{b2}\, C_{c6}$ ($\Gamma_{26}$) shown in Fig.~\ref{DCC} due to antisymmetry
 \bea
{\Gamma}_{24}^{\la \rho \tau} = i \int \frac{d^4k_2}{(2 \pi)^4}
\bigg[ \Delta_a^{\lambda \mu \nu } (k_1, k_2, -k_6) \frac{1}{ k_2^2 }
\frac{a_n}{3 \,k_2^2}  k_2^{\mu} \eps[\rho, \sigma, k_3, k_4]
\frac{1}{k_4^2} \frac{a_n}{3 \,k_4^2}
k_4^{\sigma} \eps[\tau, \nu, k_5, k_6]
\frac{1}{ k_6^2 }\bigg] =0.
 \eea

In the $\Delta_a C_{b2} C_{c6}$ case one obtains the same result after using the anomaly equation, so that
 \bea
{\Gamma}_{26}^{\la \rho \tau} = i \int \frac{d^4k_2}{(2 \pi)^4}
\bigg[ \Delta_a^{\lambda \mu \nu } (k_1, k_2, -k_6) \frac{1 }{ k_2^2 }
\frac{a_n}{3 \,k_2^2}  k_2^{\mu} \eps[\rho, \sigma, k_3, k_4]
\frac{1}{k_4^2} \frac{a_n}{3 \,k_6^2}
\, k_6^\nu \, \eps[\sigma, \tau, k_4, k_5]
\frac{1}{ k_6^2 }\bigg],
 \eea
where the contraction
\bea
k_2^{\mu} \Delta_a^{\lambda \mu \nu } (k_1, k_2, -k_6)= -\frac{a_n}{3} \eps[\nu, \la, k_6, k_1]
\eea
gives $\Gamma_{26}=0$ for antisymmetry.
In conclusion, the amplitude $\mathcal{M}$ at four-loop level, composed by three GS symmetric vertices, is not affected by unphysical massless poles arising from the derivative coupling present on some internal lines.
As a result of this analysis it is clear that there are far more cancellations than expected in some of these complex diagrams, due to the structure of the pole counterterms. In fact each DZ counterterm induces  a
WI on an attached triangle diagram and brings in antisymmetric $\epsilon$-tensors into the integrand. This is enough, in many cases, to cause a diagram to vanish by symmetry/antisymmetry of the integrand.
\section{WZ and GS interactions, anomalous magnetic moment of the muon and muonium}
In this section we move to discuss the role played by the GS and the WZ mechanism in $g-2$ of the muon and in
muonium. This is the case where the L/T decomposition of the anomaly amplitude shows its direct relevance and the role of the GS and WZ 
vertices can be easily worked out.

Our aim is not to proceed with a complete study of these corrections, some of which require a separate study, but to highlight the role played by the two mechanisms in the context of specific
processes which can be accurately quantified in future studies. The possibility of searching for anomalous extra $Z^\prime$ and axions in precision measurements of several observables is challenging but realistic.
\subsection{The GS case}
\begin{figure}[t]
\begin{center}
\includegraphics[scale=0.8]{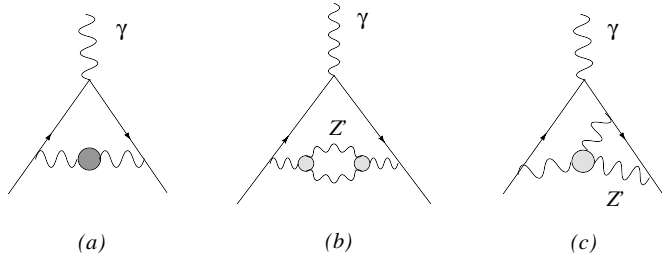}
\caption{\small Higher-order contributions to the muon magnetic moment in the GS case.}
\label{magn1}
\end{center}
\end{figure}

\begin{figure}[t]
\begin{center}
\includegraphics[scale=0.8]{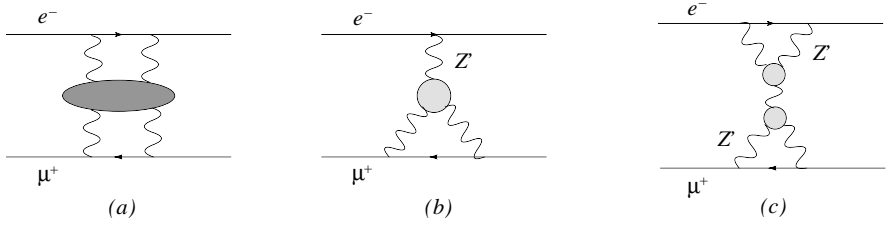}
\caption{\small Hadronic contributions (a), higher-order anomalous contributions (b) and light-by-light contributions (c) to the hyperfine splitting in muonium.}
\label{magn2}
\end{center}
\end{figure}

We show in Figs.~\ref{magn1} and \ref{magn2} some of the lowest order GS contributions to the anomalous magnetic moment of the muon and to the hyperfine splitting of muonium.  Some of the recent theoretical attention to $a_\mu\equiv g-2$ has been focused on the study of effects at two-loop level and higher, such as those shown in Fig.~\ref{magn1}a and
~\ref{magn1}d. The first indicates generically the hadronic contributions coming from self-energy insertions in the lowest order vertex. Of these types are also the corrections coming
from the self-energy graphs involving GS vertices. The corrections are tiny, being of order $g^8$
and their computation involves a 4-loop graph with ordinary propagators (the two-triangle diagram of Fig.~\ref{self_terms}) and two-loop graphs related to the pole counterterms that we have studied in the analysis of
the self-energy. Clearly, the underlying Lagrangian should allow an anomalous extra $Z^\prime$ in the spectrum. Working models of this type have been studied previously, and include several anomalous $U(1)'s$, such as in the case of intersecting
branes. The presence of a physical axion that mixes with the Higgs sector (the axi-Higgs) via a kinetic St\"uckelberg term (and eventually a Peccei-Quinn breaking term) makes these models quite attractive. The axi-Higgs is massless in the first case and massive in the second case. Models with an axi-Higgs are constructed using only WZ interactions and not GS interactions.
\begin{figure}[t]
\begin{center}
\includegraphics[scale=0.75]{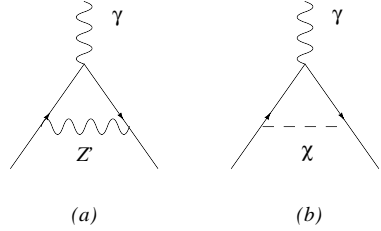}
\caption{\small Leading order corrections to the anomalous magnetic moment of the muon.}
\label{magn_LO}
\end{center}
\end{figure}
\begin{figure}[t]
\begin{center}
\includegraphics[scale=0.8]{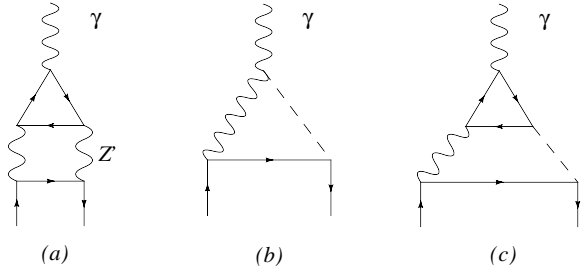}
\caption{\small Higher-order corrections to the anomalous magnetic moment of the muon with a WZ vertex.}
\label{magn_WZ}
\end{center}
\end{figure}
\begin{figure}[t]
\begin{center}
\includegraphics[scale=0.8]{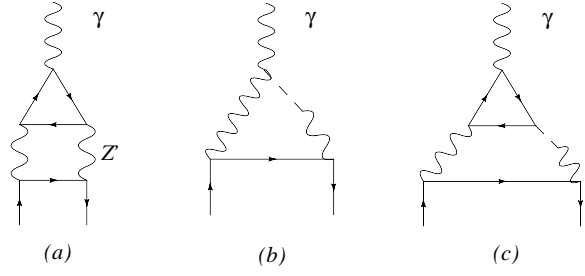}
\caption{\small As in Fig.~\ref{magn_WZ} for a GS vertex.}
\label{magn_GS}
\end{center}
\end{figure}
\subsection{Anomalous corrections to $a_\mu$}
The evaluation of the anomalous corrections to $a_\mu$ in the GS and WZ cases are quite different. In the WZ case there is a larger set of contributing graphs due to the interaction of the axion with the fermions and involve the exchange of an axi-Higgs (massless or massive), which is proportional to the fermion mass. The simplest corrections due to the presence of an anomalous extra $Z^\prime$ are shown in Fig.~\ref{magn_LO}. These do not involve the anomaly diagram and are the leading ones. They have been computed in \cite{Kiritsis:2002aj}. Higher-order corrections are those shown in Fig.~\ref{magn_WZ}, also involving  a physical axi-Higgs.

It is convenient to describe in some detail the structure of the perturbative expansion at higher orders to emphasize the differences between the two mechanisms.

The structure of the expansion can be grasped more easily if we work in the chiral limit
(all the fermions are taken to be massless) and focus our attention, for example, on graph (b) in Fig.~\ref{magn1}
since in this case there is no direct point-like interaction of the axion with the fermion.
If we decide to cancel the anomaly with WZ counterterms, we know that we can draw
a counterterm diagram in which the axion is emitted and absorbed by the gauge line.
In this case it is clear that the anomaly is potentially harmful and only a direct
computation is able to show if the counterterm is zero or not. In this specific diagrams we
know that explicit pole counterterms are needed, as we have shown in the previous sections.
If we consider diagram (c), however, the application of this argument shows immediately that the anomaly,
in this case, is harmless, since there is no axion counterterm of WZ type that we can draw.
A similar result is obtained for diagram (a) in Fig.~\ref{magn_WZ}. Also in this case we are unable to draw a WZ counterterm in which the axion is attached only to gauge lines. Therefore this diagram is also well defined even in the presence of an anomaly diagram, since its longitudinal part cancels automatically due to the topology of the graph. In these last two diagrams the gauge lines have to be attached in all possible ways to the muon lines for this to happen. Diagram (c) appears in the massive case, but it is not a counterterm.

Coming to the GS case in the massive fermion case, the anomaly diagram developes a mass-dependence in
the residue of the anomaly pole, shown in graphs (c) of Fig.~\ref{magn_GS}. As we are going to show in the next section,  this is not an anomaly  counterterm. The only counterterm is still given only by diagram (b).
More details will be given below and in the final section.
\begin{figure}[t]
\begin{center}
\includegraphics[scale=0.9]{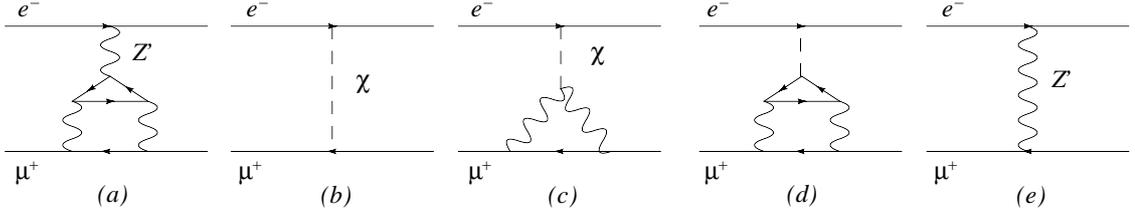}
\caption{\small Leading contributions to the hyperfine splitting in muonium and denoted as $\mathcal N_i$, with i=a,b,c,d in the WZ case.}
\label{muonium_others}
\end{center}
\end{figure}
\begin{figure}[h]
\begin{center}
\includegraphics[scale=0.9]{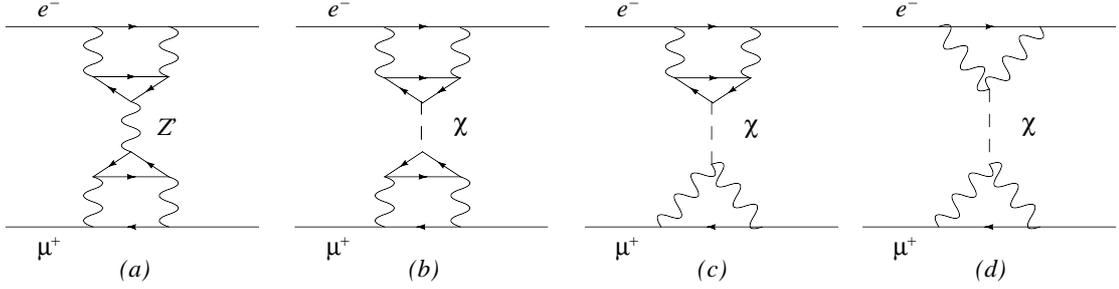}
\caption{\small Explicit (higher-order) expansion of the diagram in Fig.~\ref{magn2}c  for a trilinear WZ vertex. Each amplitude is denoted as $\mathcal{M}_i$ with i=a,b,c,d in the text with $p_1$, $p_1^\prime$ as incoming momenta and $p_2$, $p_2^\prime$ as outcoming ones.}
\label{LBL_WZ}
\end{center}
\end{figure}
\section{Two-loop Contributions to $g-2$: Anomalous Diagrams}
In this section we briefly analyze the general structure of these corrections for both mechanisms when anomaly diagrams are present. The analysis that we follow is close to the discussions for $g-2$ presented in \cite{Czarnecki:2002nt, Knecht:2002hr, Kukhto:1992qv}, adapted to our case. Most of the physical discussion carried out in these papers, in the case of the muon anomalous magnetic moment, has to do with the identification of the effects due to chiral symmetry breaking in the computation of the anomaly diagrams, which are related both to perturbative and to non-perturbative effects, treated within the operator product expansion. In our case we will be interested only in the perturbative contributions with the GS and WZ vertices.  We will point out the differences
compared to those previous studies while reviewing their derivation in order to be self-contained.  In the GS case the anomaly diagram, corrected by the pole subtraction, does not satisfy any longer the Vainshstein relation \cite{Vainshtein:2002nv} between the longitudinal ($w_L(q^2)$) and transverse ($w_T(q^2)$) component of the anomaly vertex
\beq
w_L(q^2)=2 w_T(q^2),
\eeq
which is obtained in a specific kinematical limit of the anomaly diagram. In particular, in the chiral limit, the
longitudinal component $w_L$ of the GS vertex is zero. Away from the chiral limit a pole $O(1/q^4)$ reappears, multiplied by additional contributions proportional to the fermion mass squared $(m_f^2)$, but it is not an anomaly pole.  The separation between $L$ and $T$ components, away from the chiral limit, for $m_f\neq 0$,
can be done in several ways. In \cite{Knecht:2002hr} this is obtained by isolating the anomalous pole contribution from the rest. After the subtraction of the pole term, the new anomaly-free vertex is still not transverse and satisfies a broken WI. The truly transverse component ($\tilde{w}^T$) is isolated by acting with a specific projection on the vertex, as we shall see below. This assumes a special form in the limit
in which one of the photons is on-shell ($k^2\to 0$) and soft ($k\to 0$). It can be expressed in terms of a set of scalar diagrams which come from the rank-2 tensor decomposition of the fermionic triangle ($C_{ij}$), and which are well-known in the literature. The explicit expressions of these integrals, which are for instance given in  \cite{Kniehl:1989qu}, are singular in the soft/on-shell photon limit that is needed in order to extract their contribution to $g-2$. This is the reason
for  the re-analysis of these contributions using the operator product expansion (OPE), which in this case follows the approach of  \cite{Czarnecki:2002nt, Knecht:2002hr}. In our case, both for the GS and WZ vertices, the OPE analysis would be similar, and can be performed on the two currents carrying large momentum ($q^2\to \infty$), therefore we omit it. In the WZ case the pseudoscalar exchanges involve a Goldstone and of a physical axion, this second one being not present in the SM.

We start our analysis by stating our  conventions.
 The coupling of the extra neutral current to the fermions is given by
\beq
-\frac{ig_2}{4\cos\theta_W}\bar{\psi_i}\left(g_V^{Z,Z'}\gamma^{\mu}
+g_A^{Z,Z'}\gamma^{\mu}\gamma^{5}\right)\psi_i V_{\mu}
\eeq
where the vector boson $V_{\mu}$ stays for the $Z$ or the $Z^\prime$ and
the vector and the axial-vector couplings can be written as
\ba
&&\frac{-i g_2}{4 c_w}\gamma^{\mu} {g_V}^{Z^{\prime},j}=\frac{-i g_2}{c_w}
\frac{1}{2}\left[ -\varepsilon c_w^2 T_3^{L,j}
+\varepsilon s_w^2(\frac{\hat{Y}^{j}_L}{2}+\frac{\hat{Y}^{j}_R}{2})
+\frac{g_z}{g_2}c_w(\frac{\hat{z}_{L,j}}{2}+
\frac{\hat{z}_{R,j}}{2})\right]\gamma^{\mu}
\nonumber\\
&&\frac{-i g_2}{4 c_w}\gamma^{\mu}\gamma^{5} {g_A}^{Z^{\prime},j}=\frac{-i
g_2}{c_w} \frac{1}{2}\left[ \varepsilon c_w^2 T_3^{L,j}
+\varepsilon s_w^2(\frac{\hat{Y}^{j}_R}{2}-\frac{\hat{Y}^{j}_L}{2})
+\frac{g_z}{g_2}c_w(\frac{\hat{z}_{R,j}}{2}-
\frac{\hat{z}_{L,j}}{2})\right]\gamma^{\mu}\gamma^{5}.
\ea
Here $j$ is an index which represents the quark or the lepton and we have
set $\sin\theta_W=s_w,\cos\theta_W=c_w$ for brevity.
We denote with $\hat{z}_{R,L}$ the charges of the fermions
under the extra anomalous $U(1)$ and with
$g_z$ the coupling constant of the anomalous gauge interaction.

The electroweak vertex that we need to compute in order to take into account the corrections to the anomalous magnetic moment of the muon, due to
the exchange of an extra anomalous $Z^\prime$, in analogy with the discussion presented in \cite{Knecht:2002hr},  is given by
\ba
&&\langle\bar{\mu}(p')\vert
V_{\rho}^{em}(0)\vert\mu(p)\rangle=
\bar{u}(p')\Gamma_{\rho}(p',p)u(p)=
\nonumber\\
&&\int\frac{d^4q}{(2\pi)^4}\,
\frac{-i}{q^2}\,\frac{-i}{(p'-p-q)^2-M_{Z}^2}(-ie)(-ie)
\left(\frac{ig_2}{4\cos\theta_{W}}\right)
\left(\frac{-ig_2}{4\cos\theta_{W}}\right)\times
\nonumber\\
&&\bar{u}(p')\left[\gamma^{\mu}\frac{i}{\pslsh'\ -\qsls-m_{\mu}}g_{A,\mu}^{Z'}\gamma^{\nu}\gamma_{5}+
g_{A,\mu}^{Z'}\gamma^{\nu}\gamma_{5}\frac{i}{\pslsh\ +\qsls-m_{\mu}}
\gamma^{\mu}\right]u(p)\times
\nonumber\\
&&\int dx \, e^{iq \cdot x}\int dy \, e^{i(p'-p-q) \cdot y}
\langle 0\vert
T\{A_{\mu}^{\gamma}(x)Z_{\nu}^{\prime}(y)A_{\rho}^{\gamma }(0)\}\vert
0\rangle\,,
\ea
where
\beq
A_{\mu}^{\gamma}(x)=\bar{q}(x)\gamma_{\mu}\, Q_f\, q(x)\,,\qquad
Z_{\nu}^{\prime}(y)=\bar{q}(y)\gamma_{\nu}\gamma_{5}\, g_{A,f}^{Z'}\,q(y)
\eeq
are the fermion currents of quarks, and $g_{A,f}^{Z^\prime}$ refers to a quark of flavor $f$.
The most general $CP$ invariant expression for a vertex function
satisfying the current conservation is defined by
\ba
\Gamma^{\mu}=-i e\bar{u}(p_2)\left[F_1(q^2) \gamma^{\mu}
+ F_2(q^2)\frac{q_{\alpha}}{4 m_{\mu}}\sigma^{\alpha\mu}
+F_3(q^2)\frac{\left(q^{\mu}\ds{q}-\gamma^{\mu}q^2\right)\gamma^5}{4 M_W^2}
\right]u(p_1)
\ea
where the coefficients $F_i(q^2)$ are the form-factors and $q=p_2-p_1$, $M_W$ and $m_\mu$ denote the mass of the $W$ and of the muon, respectively.
Taking the limit  $q^2\rightarrow 0$ in the Pauli form-factor we obtain the value
of the anomalous magnetic moment
\beq
a=\frac{g-2}{2}=F_2(0),
\eeq
and using the equation of motion we obtain \cite{Kukhto:1992qv}
\ba
\Gamma_{\mu}=a a_{\mu} && a_{\mu}=i e \bar{u}(p_2)\frac{1}{2 m_{\mu}}(p_1+p_2)_{\mu}u(p_1).
\ea
We can use a project operator to extract $F_{2}(0)$
\beq
F_{2}(0)=\lim_{k^2\rightarrow 0}\tr\left\{(\pslsh+m_{\mu})\Lambda_{2}^{\rho}(p',p)(\pslsh'+m_{\mu})
\Gamma_{\rho}(p',p)
\right\}\,,
\eeq
where $(p'=p+k)$
\beq
\Lambda_{2}^{\rho}(p',p)=\frac{m_{\mu}^2}{k^2}\frac{1}
{4m_{\mu}^2-k^2}\gamma^{\rho}
-\frac{m_{\mu}}{k^2}\frac{2m_{\mu}^2+k^2}{(4m_{\mu}^2-k^2)^2}
(p+p')^{\rho}
\eeq
is the projector on the Pauli form factor.
The triangle contribution is obtained from the one-loop correlator of the electroweak currents
\beq
(2\pi)^4 \delta( p' - p - q ) \Delta^{\mu\nu\rho}_{VAV}(q, k)=\int dx \,e^{iq \cdot x}\int dy \, e^{i(p'-p-q) \cdot y}
\langle 0\vert T\{A_{\mu}^{\gamma}(x)Z_{\nu}^{\prime}(y)A_{\rho}^{\gamma }(0)\}\vert
0\rangle
\eeq
with $p'$ the incoming photon four-momentum.
The corresponding tensor structure of the triangle in the $k^2\rightarrow 0$ limit for
a fermion of flavor $f$ is given by \cite{Kukhto:1992qv}, obtained from the Rosenberg representation
\cite{Rosenberg:1962pp}
\ba
&&\Delta^{\mu\nu\rho}_{VAV}(q,k)\longrightarrow \frac{g}{\pi^2\cos\theta_W} g_{A,f}^{Z'}~e^2 Q_f^2~
q^{\alpha}q^{\beta} S^{\mu\nu\rho}_{\alpha\beta}(k)\int_0^{1}d x\frac{x(1-x)}{x(1-x)q^2-m_{f}^2}
\nonumber\\
&&S^{\mu\nu\rho}_{\alpha\beta}(k)=-2 k_{\tau} \epsilon^{\tau\lambda\mu\rho}
\left(g_{\alpha\lambda}g^{\nu}_{\beta} - g_{\alpha\beta}g^{\nu}_{\lambda}\right)
+g_{\alpha\lambda} \epsilon^{\lambda\tau\nu\rho}\left(k_{\beta}g^{\mu}_{\tau}-g^{\mu}_{\beta}k_{\tau}\right),
\ea
where $k=p'-p$. This expression, in the GS case, is simply modified by the subtraction of the longitudinal pole due to the anomaly.
The tensor $\Delta^{\mu\nu\rho}_{VAV}(q,k)$
in momentum space is affected by the longitudinal (anomaly) pole, similarly to the case of axial QED discussed above, in the form of a longitudinal $w_L(q^2)$ contribution \cite{Jegerlehner:2005fs}.
In fact the asymptotic behavior at large $Q^2=-q^2$ is given by \cite{Czarnecki:2002nt}
\ba
w_L^{f}(Q^2)=\frac{g_2}{\cos\theta_W} g_{A,f}^{Z'}e^2 Q_f^2
\left[ \frac{1}{2 \pi^2 Q^2}-2\frac{m_f^2}{\pi^2 Q^4}\log{\frac{Q^2}{m_f^2}}+O\left(\frac{1}{Q^6}\right)\right].
\label{carn}
\ea
and in the GS case it becomes
\ba
{w_{L}}^{f}(Q^2)\vert_{GS}=-\frac{g_2}{\pi^2\cos\theta_W} g_{A,f}^{Z'}e^2 Q_f^2
\left[ 2\frac{m_f^2}{Q^4}\log{\frac{Q^2}{m_f^2}}+O\left(\frac{1}{Q^6}\right)\right].
\ea
Following \cite{Knecht:2002hr}
 we can always write
\beq
\Delta^{f}_{\mu\nu\rho}(q,k)=\Delta_{\mu\nu\rho}(q,k)_{\mbox{\rm\tiny
anomaly}}+\tilde{\Delta}^{f}_{\mu\nu\rho}(q,k) \,,
\eeq
where
\beq
\Delta_{\mu\nu\rho}(q,k)_{\mbox{\rm\tiny anomaly}}=
\sum_f g_{A,f}^{Z'}e^2 Q_f^2 a_n\frac{(q-k)_{\nu}}{(q-k)^2}
\epsilon_{\mu\rho\alpha\beta}q^{\alpha}k^{\beta}\,,
\eeq
with $a_n=-i/(2\pi^2)$.
The function $\tilde{\Delta}^f_{\mu\nu\rho}(q,k)$ is transverse with respect to
the momenta $q^{\mu}$ and $k^{\rho}$
\beq
q^{\mu}\tilde{\Delta}^f_{\mu\nu\rho}(q,k)=0\,,\nonumber\\ \;
k^{\rho}\tilde{\Delta}^f_{\mu\nu\rho}(q,k)=0\,,
\label{wi}
\eeq
but in the presence of massive fermions we isolate the longitudinal components of the
corresponding broken WI
\beq
\tilde{\Delta}^f_{\mu\nu\rho}(q,k)=
\tilde{\Delta}_{\mu\nu\rho}^{f,\mbox{\tiny long}}(q,k)
+\tilde{\Delta}_{\mu\nu\rho}^{f,\mbox{\tiny trans}}(q,k)\,.
\eeq
Differentiating the 2nd expression in Eq.~(\ref{wi}) with respect to $k_{\rho}$ we obtain
\ba
\tilde{\Delta}_{\mu\nu\rho}(q,k)=-k^{\sigma}\frac{\partial}{\partial k^{\rho}}\tilde{\Delta}_{\mu\nu\sigma}(q,k)
\ea
where we have suppressed the flavor index $f$ for simplicity.
Since we are interested in the soft photon limit, the relevant contributions are those linear in
$k$.  In \cite{Knecht:2002hr} these are extracted in the form
\ba
\tilde{\Delta}_{\mu\nu\rho}^{\mbox{\tiny trans}}(q,k)=k^{\sigma}\Delta_{\mu\nu\rho\sigma}(q) + ...
\label{rosi}
\ea
where the tensor $\Delta_{\mu\nu\rho\sigma}(q)$ is obtained by using the projection operator
$\Pi^{\nu\nu'}$ as follows
\ba
&&\Pi^{\nu\nu'}(q,k)=\left( g^{\nu\nu'}-\frac{(q-k)^{\nu}}{(q-k)^2}(q-k)^{\nu'}\right),
\nonumber\\
&&\Delta_{\mu\nu\rho\sigma}(q)=-\frac{\partial}{\partial k^{\rho}}\left(\Pi^{\nu\nu'}\Delta_{\mu\nu'\rho}\right)\vert_{\lim k\rightarrow 0}.
\label{proj}
\ea
It is not difficult to notice that
\ba
\Pi^{\nu\nu'}\Delta_{\mu\nu'\rho}=\Pi^{\nu\nu'}\tilde{\Delta}_{\mu\nu'\rho}.
\ea
As we have already mentioned, the action of $\Pi^{\nu\nu'}$ is to remove all the longitudinal parts from
the $\Delta_{\mu\nu\rho}$ tensor, including the anomalous term.
 $\Delta_{\mu\nu\rho\sigma}(q)$ in (\ref{rosi}) has the form
\ba
\Delta_{\mu\nu\rho\sigma}(q)=i \Delta(Q^2)\left[ q_{\rho} \epsilon_{\mu\nu\alpha\sigma}q^{\alpha}
- q_{\sigma} \epsilon_{\mu\nu\alpha\rho} q^{\alpha}\right],
\ea
where $q^2=-Q^2$.
We can now try to apply this formalism to the anomalous
triangle diagrams. We use the generic parametrization of a $AVV$ triangle given in \cite{Kniehl:1989qu},
\ba
&&\Delta^f_{\mu\nu\rho}(q,k)=-\frac{g_2}{\pi^2 \cos\theta_W} g_{A,f}^{Z'}e^2Q_f^2
\left[
A(k,-q,m_f)(k_{\rho}\epsilon_{\nu\mu\beta\sigma}k^{\beta}
-k^2\epsilon_{\nu\mu\rho\sigma})(-q)^{\sigma}
\right.\nonumber\\
&&\hspace{1.5cm}\left.+A(-q,k,m_f)
(q_{\mu}\epsilon_{\nu\rho\beta\sigma}q^{\beta}
-q^2\epsilon_{\nu\rho\mu\sigma})k^{\sigma}
\right.\nonumber\\
&&\hspace{1.5cm}\left.
-B(k,-q,m_f)(q-k)_{\nu}\epsilon_{\rho\mu\alpha\beta}k^{\alpha}(-q)^{\beta}
\right],
\ea
where the functions $A(k,-q,m_f)$ and $B(k,-q,m_f)$ are given by in terms of the
tensor-reduction coefficients $C^{}_{ij}$ as follows
\ba
&&A(k,-q,m_f)=(C_{11} -C_{12}+C_{21}-C_{23})(k,-q,m_f)
\nonumber\\
&&B(k,-q,m_f)=(C_{12}+C_{23})(k,-q,m_f),
\ea
and are defined in Eqs.~(A.2), (A.3) of Ref.~\cite{Kniehl:1989qu}.
The WI on the axial-vector current is given by
\ba
(q-k)^{\nu}\Delta^f_{\mu\nu\rho}(q,k)=
-\frac{g_2}{\pi^2 \cos\theta_W} g_{A,f}^{Z'}e^2Q_f^2\left[\frac{1}{2}-2m_f^2 C_0\right] \epsilon_{\rho\mu\alpha\beta} \, k^{\alpha}(-q)^{\beta}
\label{gen}
\ea
and the most general expression of the coefficient $C_0$ is given
in Eq.~(A.8) of ref.~\cite{Kniehl:1989qu}. $C_0$ is the scalar three-point function with a fermion of
mass $m_f$ circulating in the loop. In the soft photon limit the invariant amplitude defined by the right-hand-side of (\ref{gen}) reduces to (\ref{carn}).
The purely transverse part (for $m_f\neq 0$) is obtained by applying the projection operator given in
(\ref{proj})
\ba
\Delta^{trans}_{\mu\nu\rho}(q,k)&=&-k^{\sigma}\frac{\partial}{\partial k^{\rho}}\left(\Pi^{\nu\nu'}(q,k)\Delta^f_{\mu\nu\sigma}(q,k)\right)\vert_{\lim k\rightarrow 0}
\nonumber\\
&=&-k^{\sigma}\frac{\partial}{\partial k^{\rho}}
\Delta^T_{\mu\nu\sigma}(q,k)\vert_{\lim k\rightarrow 0},
\ea
where
\ba
&&\Delta^T_{\mu\nu\sigma}(q,k)=A(k,-q,m_f)\left[
q^{\alpha}\epsilon_{\alpha\mu\nu\sigma}k^2 -\frac{k_{\nu}\,k^2}{(q-k)^2} \epsilon_{\mu\sigma\alpha\beta}k^{\alpha}q^{\beta}
+\frac{q_{\nu}\,k^2}{(q-k)^2} \epsilon_{\mu\sigma\alpha\beta}k^{\alpha}q^{\beta}
+k_{\sigma} \epsilon_{\mu\nu\alpha\beta}k^{\alpha}q^{\beta}
\right]
\nonumber\\
&&\hspace{1.5cm}+A(-q,k,m_f)\left[
k^{\alpha} \epsilon_{\alpha\mu\nu\sigma}q^2
-\frac{k_{\nu}\,q^2}{(q-k)^2} \epsilon_{\mu\sigma\alpha\beta}k^{\alpha}q^{\beta}
+\frac{q_{\nu}\,q^2}{(q-k)^2} \epsilon_{\mu\sigma\alpha\beta}k^{\alpha}q^{\beta}
-q_{\mu} \epsilon_{\nu\sigma\alpha\beta}k^{\alpha}q^{\beta}\right],
\nonumber\\
\ea
Differentiating with respect to $k_{\rho}$ and taking the $\lim k \rightarrow 0$ we obtain
\ba
\frac{\partial}{\partial k^{\rho}}\Delta^T_{\mu\nu\sigma}(q,k)\vert_{\lim k \rightarrow 0}&=&
A(Q^2)\left[ \epsilon_{\rho\mu\nu\sigma}q^2
+q_{\nu}\epsilon_{\mu\sigma\rho\beta}q^{\beta}
-q_{\mu}\epsilon_{\nu\sigma\rho\beta}q^{\beta}\right]
\nonumber\\
&=&A(Q^2)\left[q_{\rho}\epsilon_{\mu\nu\alpha\sigma}q^{\alpha}
- q_{\sigma}\epsilon_{\mu\nu\alpha\rho}q^{\alpha}\right].
\ea
where $A(Q^2)$ denotes the soft limit of the $A(-q,k,m_f)$ amplitude.
The intermediate steps to simplify the contribution to $a_\mu$  are those of \cite{deRafael:1993za}.
In our case, with the modifications discussed above, the Pauli form-factor for a circulating fermion of flavor $f$
we obtain
\ba
&&F_{2}(0)|_{Z'}=(-e^2)\frac{g_2^2}{16\cos^2\theta_{W}}
\frac{1}{M_{Z'}^2}
\lim_{k^2\rightarrow 0}\int\frac{d^4q}{(2\pi)^4}\frac{1}{q^2}
\left(\frac{M_{Z'}^2}{q^2 -M_{Z'}^2}\right)\times
\nonumber\\
&&\hspace{2cm}\frac{1}{4k^2}\tr\left\{(\pslsh+m_{\mu})
\left[\gamma^{\rho}\ksls-\left(k^{\rho}+
\frac{p^{\rho}}{m_{\mu}}\ksls\right)\right]
\right.\nonumber\\
&&\left.\hspace{2cm}\left[\gamma^{\mu}\frac{(\pslsh\ -\qsls+m_{\mu})}
{q^2-2q\dd p}g_{A,\mu}^{Z'}
\gamma^{\nu}\gamma_{5}+g_{A,\mu}^{Z'}\gamma^{\nu}\gamma_{5}
\frac{(\pslsh\ +\qsls+m_{\mu})}{q^2+2q\dd p}\gamma^{\mu}
\right]
\right\}\times
\nonumber\\
&&\hspace{2cm}g_{A,f}^{Z'} Q_f \left[
-i\tilde{\Delta}_{\mu\nu\rho}^{f,\mbox{\rm\tiny long  }}(q,k)+
k^{\sigma}\left[q_{\rho} \epsilon_{\mu\nu\alpha\sigma}q^{\alpha}-q_{\sigma}
\epsilon_{\mu\nu\alpha\rho}q^{\alpha}\right]A(Q^2)\right]\,,
\label{TOT}
\ea
where
$\tilde{\Delta}_{\mu\nu\rho}^{\mbox{\rm\tiny f,long}}$ is not anomalous
and in the soft photon limit it is given by
\ba
\tilde{\Delta}_{\mu\nu\rho}^{\mbox{\rm\tiny f,long}}(q,k)= \frac{q_{\nu}}{q^2}\tilde{w}_L^f \epsilon_{\mu\rho\alpha\sigma}q^{\alpha}k^{\sigma}
\ea
where
\ba
\tilde{w}_L^f= -\left[-\frac{2}{\pi^2}\frac{m_f^2}{q^4}
\log{\frac{(-q)^2}{m_f^2}}+O\left(\frac{1}{q^6}\right)\right].
\ea
It is obvious from this analysis that in presence of the GS mechanism there is a two-loop counterterm which removes the pure anomalous contribution in Eq.~(\ref{TOT}) and is given by see Fig.~\ref{magn_GS}b. Diagram c) in the same figure is the longitudinal part of the diagram and appears in the broken WI that we will discuss in the last section. Finally, after some manipulations, similar to those performed in \cite{Knecht:2003xy, Knecht:2002hr, Peris:1995bb, Czarnecki:2002nt}, the final result for the anomalous contributions to $a_\mu$ takes the form
\ba
F_{2}(0)|_{Z'}=(-e^2 Q_f)\frac{g_2^2 g_{A,\mu}^{Z'}
g_{A,f}^{Z'}}{16\cos^2\theta_{W}}\left(\frac{m_{\mu}^2}
{M_{Z'}^2}\right)\frac{1}{4\pi^2}\int_{m_{\mu}}^{\infty}dQ^2
\left(\tilde{w}_L^{f}(Q^2)+\frac{M_{Z'}^2}{Q^2+M_{Z'}^2}A(Q^2)\right)\,,
\ea
where $-q^2=Q^2$ and $\tilde{w}_L^{f}(Q^2)$ vanishes in the chiral limit.
\subsubsection{The Wess-Zumino Counterterm}
A similar analysis can be performed in the case of the WZ mechanism.
The leading non anomalous one-loop contributions Fig.~\ref{magn_LO} have been calculated in \cite{Kiritsis:2002aj} for a specific $D$-brane model. These are due to the coupling of the axi-Higgs to the fermions. The organization of the perturbative expansion for a theory with an axion-like particle has been discussed in \cite{Coriano:2007fw, Coriano:2006xh}, where the explicit cancellation of the gauge dependence has been discussed on general grounds. We show in Fig.~\ref{magn_WZ} the contribution coming from the $Z^\prime$ propagator (graph a) in the anomalous exchange, the additional graphs b) and c) represent the axion counterterm due to the WZ interaction (b) and the correction due to the coupling of the axion to the massive fermions (c). We have omitted a graph similar to (c) in which the exchanged pseudoscalar is a Goldstone and cancels the gauge dependence of the $Z^\prime$ propagator.

The computation of graph a) follows exactly the analysis of \cite{Knecht:2002hr} and can be performed in dimensional regularization in the unitary gauge, to give
\ba
&&\lambda_{\mbox{\tiny{$\overline{\mbox{MS}}$}}}\,\equiv\,
\frac{\mu^{d-4}}{16\pi^2}\,\Big[ \,\frac{1}{d-4}\,-\,
\frac{1}{2}\,\big(\log 4\pi +2+ \Gamma '(1)\big)\Big]\,,
\nonumber\\
&&F_2(0)\Big\vert^{(f)}_{\mbox{\tiny anom}} = \frac{g_2^2}{16\pi^2\cos^2\theta_W}
\frac{m_{\mu}^2}{M_{Z'}^2}\frac{1}{4\pi^2}Q_f^2 g_A^{Z',f}\times
\left[ \log\left(\frac{\mu_R^2}{M_{Z'}^2}\right)\,
-\,32\pi^2\lambda_{\mbox{\tiny{$\overline{\mbox{MS}}$}}}\,+
\log\left(\frac{M_{Z'}^2}{m_{\mu}^2}\right)\,+\, \frac{1}{2}
\right]\,.
\nonumber\\
\ea
The expression of the extra contributions when a physical axion is exchanged are given by
\ba
&&F_{2}(k^2)\big\vert_{(c)\mbox{\rm\tiny long}}=
e Q_{\mu}c^{\chi}_{\mu} \lim_{k^2\rightarrow 0}\int\frac{d^4q}{(2\pi)^4}\,
\frac{1}{q^2}\,\frac{1}{(p'-p-q)^2-M_{Z'}^2}\times
\nonumber\\
&&\bar{u}(p')\left[\gamma^{\mu}\frac{1}{\pslsh'\ -\qsls-m_{\mu}}\gamma_{5}+
\gamma_{5}\frac{1}{\pslsh\ +\qsls-m_{\mu}}\gamma^{\mu}\right]u(p)\times
\nonumber\\
&&\left[2 \frac{m_{\mu}}{M_{Z'}^2} \sum_f e Q_f c^{\chi}_f \frac{1}{(q-k)^2-M_{\chi}^2}\right]
\Delta_{\mu \rho}(m_f,q,k,q-k),
\ea
where $c^{\chi}_f$ is coupling of the axi-Higgs to the fermions and
\ba
&&\Delta_{\mu \rho}(m_f,q,k,q-k)=\epsilon_{\mu\rho\alpha\beta}q^{\alpha}k^{\beta}
\left(-\frac{1}{2\pi^2}\right)I(m_f)
\nonumber\\\nonumber\\
&&I(m_f)\equiv -\int_0^1\int_0^{1-x} dx \, dy \, \frac{1}{m_f^2+(x-1)x q^2+(y-1)y k^2-2 x y q\cdot k}\,.
\ea
Using the projection operator we get
\ba
&&F_{2}(0)\big\vert_{(c)\mbox{\rm\tiny long}}=
e Q_{\mu}c^{\chi}_{\mu} \lim_{k^2\rightarrow 0}\int\frac{d^4q}{(2\pi)^4}\,
\frac{1}{q^2}\,\frac{1}{(p'-p-q)^2-M_{Z'}^2}\times
\nonumber\\
&&\frac{1}{4k^2}\tr\left\{(\pslsh+m_{\mu})
\left[\gamma^{\rho}\ksls-\left(k^{\rho}+
\frac{p^{\rho}}{m_{\mu}}\ksls\right)\right]\times
\right.\nonumber\\
&&\left.\left[\gamma^{\mu}\frac{1}{\pslsh'\ -\qsls-m_{\mu}}\gamma_{5}+
\gamma_{5}\frac{1}{\pslsh\ +\qsls-m_{\mu}}\gamma^{\mu}\right]\right\}\times
\nonumber\\
&&\left[2 \frac{m_{\mu}}{M_{Z'}^2} \sum_f e Q_f c^{\chi}_f \frac{1}{(q-k)^2-M_{\chi}^2}\right]
\Delta_{\mu \rho}(m_f,q,k,q-k).
\ea
The contribution coming from the diagram in Fig.~\ref{magn_WZ}b is similar to Fig.~\ref{magn_WZ}c and we obtain
\ba
&&F_{2}(0)\big\vert_{(b)\mbox{\rm\tiny long}}=
e Q_{\mu}c^{\chi}_{\mu} \lim_{k^2\rightarrow 0}\int\frac{d^4q}{(2\pi)^4}\,
\frac{1}{q^2}\,\frac{1}{(p'-p-q)^2-M_{Z'}^2}\times
\nonumber\\
&&\frac{1}{4k^2}\tr\left\{(\pslsh+m_{\mu})
\left[\gamma^{\rho}\ksls-\left(k^{\rho}+
\frac{p^{\rho}}{m_{\mu}}\ksls\right)\right]\times
\right.\nonumber\\
&&\left.\left[\gamma^{\mu}\frac{1}{\pslsh'\ -\qsls-m_{\mu}}\gamma_{5}+
\gamma_{5}\frac{1}{\pslsh\ +\qsls-m_{\mu}}\gamma^{\mu}\right]\right\}\times
\nonumber\\
&&\left[-2 \frac{m_{\mu}}{M_{Z'}^2} g^{\chi}_{\gamma\gamma} \frac{1}{(q-k)^2-M_{\chi}^2}\right]
\epsilon_{\mu\rho\alpha\beta}q^{\alpha}k^{\beta}.
\ea
where the coefficient $g^{\chi}_{\gamma\gamma}$ is the coupling of the axi-Higgs to the photons
and it will be given explicitly in the next section together with the coefficient $c^{\chi}$.
\subsection{Corrections to muonium}
\begin{figure}[t]
\begin{center}
\includegraphics[scale=0.7]{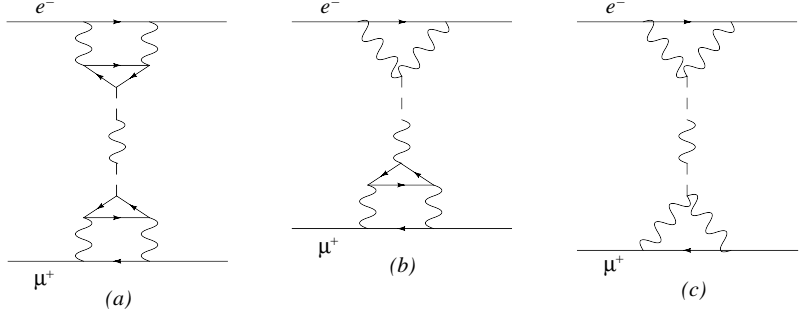}
\caption{\small As in Fig.~\ref{LBL_WZ} for a GS vertex.}
\label{LBL_GS}
\end{center}
\end{figure}

A similar analysis of the role played by both mechanisms in anomalous processes at higher orders can be
done in the case of muonium. A recent analysis of the hadronic effects in this type of systems can be found
in \cite{Vainshtein:2007zz}. One of the typical contributions is given by virtual light-by-light scattering, shown in Fig.~\ref{magn2}. In the presence of anomalous gauge interactions a dominant contribution for the GS case is given by diagram (b). Diagram (c) is subdominant. This is expanded in terms single and double counterterms, typically given in
Fig.~\ref{LBL_GS}. In the WZ case we report some of the corresponding contributions in Figs.~\ref{muonium_others} and \ref{LBL_WZ}, where we allow a coupling of the axi-Higgs to the fermions.  The leading contribution are diagrams
(b) and (e) of Fig.~\ref{muonium_others}, which are the analogue of (a) and (b) of Fig.~\ref{magn_LO}.
The diagrams involving light-by-light scattering in the presence of a WZ vertex with a physical axi-Higgs $\chi$ coupled to fermions are shown in Fig.~\ref{LBL_WZ}; their expression can be easily obtained by taking into account some recent results  on two-loop QCD corrections \cite{Bernreuther:2005rw} and a specific choice of parameters for an anomalous model developed and fully described in Chaps.~\ref{chap:AbelianModels1}, \ref{chap:AbelianModels2}. So we have
\ba
{\mathcal M}_a &=&
\bar u (p_2)  \Bigg[ e^4 \sum_f g_{Z^\prime} a^f_{Z^\prime} Q^2_f \, \Lambda_\mu(s,m_f,m_e) \Bigg]
u(p_1) \frac{-i}{k^2-M_{Z^\prime}^2}  \left( g^{\mu \nu}-\frac{k^\mu k^\nu}{M^2_{Z^\prime}}  \right)    \nonumber\\
&&\bar v( p^\prime_1)
 \Bigg[ e^4 \sum_{f^\prime} g_{Z^\prime} a^{f^\prime}_{Z^\prime} Q^2_{f^\prime} \Lambda_\nu(s,m_{f^\prime},m_\mu) \Bigg]
 v(p^\prime_2),
\ea
where the axial-vector vertex function $\Lambda_\mu(s,m_f,m_{ext})$ is given \cite{Bernreuther:2005rw} in terms of some coefficients named $G_1$ and $G_2$ as
\ba
\Lambda_\mu(s,m_f,m_{ext})  = \gamma_\mu \gamma_5 G_1(s,m_f,m_{ext})
+ \frac{1}{2 m_{ext}} k_\mu \gamma_5 G_2(s,m_f,m_{ext})
\ea
and $m_{ext}$ refers to the electron or the muon. Their explicit expression can be found in \cite{Bernreuther:2005rw}.
For $\mathcal M_b$, with an axi-Higgs exchanged in the $t$-channel we obtain
\ba
{\mathcal M}_b =
\bar u (p_2) \Bigg[ e^2  \sum_f c^{\chi, f}_{\g\g}  \Lambda(s,m_f,m_e) \Bigg]
u(p_1) \frac{i}{k^2-m_\chi^2}
\bar v( p^\prime_1)
 \Bigg[ e^2 \sum_{f^\prime} c^{\chi, f^\prime}_{\g\g}  \Lambda(s,m_{f^\prime},m_\mu)  \Bigg]    v(p^\prime_2) 
\ea
with the general coupling of the physical axion
\ba
c^{\chi,f}_{\gamma\gamma}= e^{2} Q^2_f c^{\chi,f},   \qquad (f=u,d,\nu, e)
\label{chi_coupling}
\ea
and  the pseudoscalar vertex function $\Lambda(s,m_f,m_{ext})$ (see \cite{Bernreuther:2005rw})
\ba
\Lambda(s,m_f,m_{ext}) = \gamma_5 A(s,m_f ,m_{ext}).
\ea
We have used a condensed notation for the flavors in Eq.~\ref{chi_coupling}
with u = \{u, c, t\}, d = \{d, s, b\}, $\nu$ = \{$\nu_{e}$, $\nu_{\mu}$, $\nu_{\tau}$\}
and e = \{e, $\mu$, $\tau$\}, whose expansion yields
\ba
c^{\chi, u} &=& \Gamma^{u}  \frac{i}{\sqrt 2} O^{\chi}_{11} = \frac{m^{}_{u}}{v_u} i O^{\chi}_{11} ,  \qquad
c^{\chi, d} = - \Gamma^{d} \frac{i}{\sqrt 2} O^{\chi}_{21} = - \frac{m^{}_{d}}{v^{}_{d}} i O^{\chi}_{21},    \nonumber\\
c^{\chi, \nu} &=& \Gamma^{\nu}  \frac{i}{\sqrt 2} O^{\chi}_{11} = \frac{m^{}_{\nu}}{v^{}_u} i O^{\chi}_{11} ,  \qquad
c^{\chi, e} = - \Gamma^{e} \frac{i}{\sqrt 2} O^{\chi}_{21} = - \frac{m^{}_{e}}{v^{}_{d}} i O^{\chi}_{21},
\ea
where the elements of the $O^{\chi}$ rotation matrix from the interaction to the mass eigenstate basis are given in App.~\ref{sec:Ochi}.\\
The most difficult to analyze are those of higher order, shown in Fig.~\ref{LBL_WZ}
\ba
{\mathcal M}_c &=&
\bar u (p_2) \Bigg[ e^2  \sum_f c^{\chi, f}_{\g\g}  \Lambda(s,m_f,m_e) \Bigg]
u(p_1) \frac{i}{k^2-m_\chi^2} \bar v( p^\prime_1)  F(s,m_\mu) g^\chi_{\g\g}  e^2  v(p^\prime_2) \\
{\mathcal M}_d &=& \bar u(p_2) F(s,m_e) g^\chi_{\g\g} e^2
u(p_1) \frac{i}{k^2-m_\chi^2} \bar v( p^\prime_1)  F(s,m_\mu) g^\chi_{\g\g}  e^2  v(p^\prime_2),
\ea
with the one-loop anomalous vertex function $F(s, m_{ext})$  \cite{Bernreuther:2005rw}
\ba
F(s,m_{ext})= 2 i \, m_{ext} \,  f(s,m_{ext})  \,\gamma_5
\ea
where $m_{ext}$ is the mass of the external fermion, in this case the muon mass,
and the specific choice of coupling given by
\ba
\label{GS_coeffs}
g^{\chi}_{\g\g}=\left[\frac{F}{M_1}(O^{A}_{W\g})^2
+\frac{C_{YY}}{M_1}(O^{A}_{Y\g})^2\right]O^{\chi}_{31}
\ea
in terms of model dependent parameters defined in Chap.~\ref{chap:AbelianModels2}. \\
The simplest corrections are those of Fig.~\ref{muonium_others} and can be written as
\ba
{\mathcal N}_a &=&
\bar u(p_2) ( g_{Z^\prime} a^e_{Z^\prime} ) \g^\mu \gamma^5 u(p_1)  \frac{-i}{k^2-M_{Z^\prime}^2}
\left( g^{\mu \nu} - \frac{k^\mu k^\nu}{M^2_{Z^\prime}}  \right)    
\bar v( p^\prime_1) \Bigg[ e^4 \sum_f g_{Z^\prime} a^f_{Z^\prime} Q^2_f \Lambda_\nu(s,m_f,m_\mu) \Bigg] v(p^\prime_2), \nn \\ \\
{\mathcal N}_b &=&
\bar u(p_2)c^{\chi, e} \gamma^5 u(p_1) \frac{i}{k^2-m_\chi^2}\bar v( p^\prime_1) c^{\chi, \mu} \gamma^5 v(p^\prime_2), \\
{\mathcal N}_c &=&
\bar u(p_2) c^{\chi, e} \gamma^5 u(p_1) \frac{i}{k^2-m_\chi^2}
\bar v( p^\prime_1) F(s,m_\mu) g^{\chi}_{\g\g} e^2  v(p^\prime_2), \\
{\mathcal N}_d &=&
\bar u(p_2) c^{\chi, e} \gamma^5 u(p_1) \frac{i}{k^2-m_\chi^2}
\bar v( p^\prime_1)  \Bigg[ e^2  \sum_f c^{\chi,f}_{\g\g} \Lambda(s,m_f,m_\mu) \Bigg] v(p^\prime_2),
\ea
where the one-loop functions  $\Lambda_\nu$, $F$ and $\Lambda$ have been given above and can be found in the literature.
\section{The longitudinal subtraction and the broken Ward Identities of the GS vertex}
There is one last important point that we will address in this final section which concerns the correct interpretation of the anomaly counterterm in both (chiral) phases of theory. The GS vertex satisfies a broken WI, which is easy to derive diagrammatically. The identity is similar to that of the ordinary triangle diagram, but with a subtraction of the (massless) anomaly pole. In the massless fermion case, the GS counterterm restores the WI on the anomaly vertex; in the massive case the mass-dependent terms are a signal of chiral symmetry breaking, but are not counterterms. The only counterterm is the anomaly pole. We briefly clarify this point.

We recall that, for on-shell photons (analogously for gluons) in an anomalous theory,
 the pole contribution to the $AVV$ triangle is given by
\ba
&&T^{\lambda\mu\nu}=g_{V_{KK}}\theta_f^{KK} g_s^2 Tr[t^a t^b] k^{\lambda}\epsilon[k_1,k_2,\mu,\nu]
\left[\frac{1}{2\pi^2 s}-\frac{m_f^2}{ 2\pi^2 s^2}\log^2 \left(\frac{\rho_f+1}{\rho_f-1}\right) \right]
\nonumber\\
&&\rho_f=\sqrt{1-\frac{4 m_f^2}{s}}
\ea
and the anomalous WI on the axial-vector line gives
\ba
&&k_{\lambda}T^{\lambda\mu\nu}=g_{V_{KK}}\theta_f^{KK} g_s^2 Tr[t^a t^b]\left(\epsilon[k_1,k_2,\mu,\nu]\frac{1}{2\pi^2}
+2 m_f T^{\mu\nu}\right)
\nonumber\\
&&T^{\mu\nu}=\frac{m_f}{2\pi^2}
\int_0^1\int_0^{1-x} d x \, d y \, \frac{\epsilon[k_1,k_2,\mu,\nu]}{m_f^2 -2 x y k_1\cdot k_2}.
\label{reT}
\ea
The second term in the WI above, or $T_{\mu\nu}$, in a local gauge theory with  spontaneous symmetry breaking (in an anomaly-free theory), is determined by the BRS invariance of the correlator. In an anomalous theory the first term is the anomaly, while the second term comes from chiral symmetry breaking. If we use  a WZ counterterm to restore the gauge symmetry, the WI is modified with the addition of the $bF\tilde{F}$ graph, and the analysis can be found  in \cite{Armillis:2007tb}. Eq.~(\ref{reT}) takes a more general form for off-shell gauge lines.
The general corrections to the anomaly pole are of the form
\ba
&&\Delta^{\lambda\mu\nu}=g_{V_{KK}}\theta_f^{KK} g_s^2 Tr[t^a t^b] k^{\lambda} \epsilon[k_1,k_2,\mu,\nu]
\left[\frac{1}{2\pi^2 s}-2 {m_f^2} C_0(t, k_1^2,k_2^2,m_f)  \right] 
\label{TT}
\ea
where $C_0(t, k_1^2,k_2^2,m_f) $ is the scalar triangle diagram. Also in this case the $C_0$ terms are not counterterms. We don't need to add any mass-dependent term to the GS vertex to restore the WI of the non-local theory. These longitudinal contributions, following the analysis of $g-2$ in \cite{Knecht:2002hr} and the discussion of the previous sections,
are easily interpreted as the longitudinal parts of the non-anomalous components of the vertex, generated by the breaking of the chiral symmetry.
There are two ways to write the broken WI in GS case. The first form is given by
\ba
k_{\lambda} \left(\Delta^{\lambda\mu\nu} + \Gamma_{GS}^{\lambda\mu\nu}\right) + {{T}}^{\mu\nu}=0,
\ea
in which the $T_{\mu\nu}$ term, which is of the form~(\ref{TT}), is derived simply by acting with the WI on the GS vertex (anomaly plus massless pole term) and bringing the result to the first member. The chiral symmetry breaking corrections to the pole term are then obtained from the decomposition
\beq
\Delta^{GS\,\lambda\mu\nu}=
\tilde{\Delta}_{long}^{\lambda\mu\nu} + \tilde{\Delta}_{trans}^{\lambda\mu\nu}
\eeq
where
\beq
k_{\lambda}\tilde{\Delta}_{long}^{\lambda\mu\nu}=\frac{1}{4 \pi^2} {m_f^2} C_0(t, k_1^2,k_2^2,m_f)
\epsilon[\mu,\nu,k_1,k_2]\equiv -T^{\mu\nu} \eeq
and
\beq
k_{\lambda}\tilde{\Delta}_{trans}^{\lambda\mu\nu}=0.
\eeq
A second form of the same equation is obtained by extracting a massless pole from $T_{\mu\nu}$
\ba
k_{\lambda} \left(\Delta^{\lambda\mu\nu} + \Gamma_{GS}^{\lambda\mu\nu} + \frac{k^{\lambda}}{k^2} {{T}}^{\mu\nu}\right) =0,
\ea
\begin{figure}[t]
\begin{center}
\includegraphics[scale=0.7]{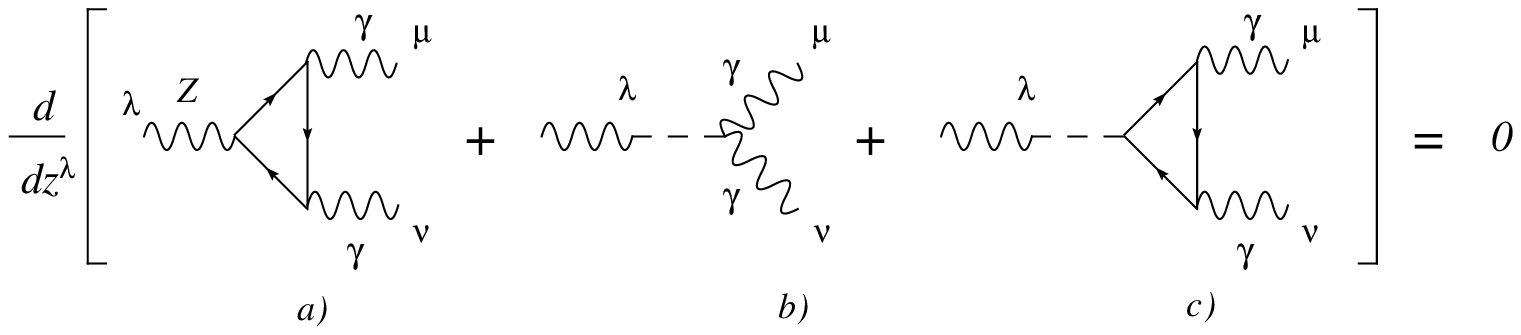}
\caption{\small Broken WI in the presence of a GS interaction.}
\label{GS1}
\end{center}
\end{figure}
whose explicit form is shown in Fig.~\ref{GS1}. This result is in disagreement with \cite{Djouadi:2007eg}, where the authors write down an {\em exact} WI for the GS vertex in the chirally broken phase, identity which clearly does not exist, since the pole counterterm and the effects due to chiral symmetry breaking should be kept separate.

There are other issues concerning the use of this vertex to describe the mixing of the Kaluza-Klein excitations of gauge bosons to an axion, claimed to be relevant in $t\bar{t}$ production,
which also point toward an inconsistency of these types of formulations in theories with extra dimensions and chiral delocalization on the brane. These quietly assume that the GS vertex is generated by sewing together local-interactions ($b F\tilde F$ vertex and $B\,b$ mixing), which are claimed to be obtained from extra dimensional theories \cite{Djouadi:2007eg}. The bilinear mixing is assumed to be physical (i.e. no gauge-fixing condition can remove it). If these constructions were consistent, this would imply that the anomaly can be removed by adding a finite number of local interactions. Instead, the anomaly pole can be removed, but at the expense of building a non-local theory.
This result does not contradict the use of the WZ mechanism for the "cancellation" of the anomaly, since the WZ theory, being local, generates an effective theory which is unitary only below a certain  scale, while remaining gauge invariant at all scales. The theory, in fact, needs to be amended by higher dimensional operators for the restoration of unitarity, and therefore the description of axion-like particle, involving St\"uckelberg axions and $PQ$ interactions, are not unitary at all scales. In fact, the number of local interactions needed to obtain an anomaly-free theory is infinite, which is the price to pay for not having a pole counterterm as in the GS case. The presence of BIM amplitudes for the WZ mechanism provides a clear example of processes with a non-unitary growth at high energy, as clearly shown in Chap.~\ref{chap:UnitarityBound}.
\section{Conclusions}
We have investigated the consistency of the subtraction of pole counterterm  in an anomalous theory,
re-analysing the problem of the generation of double poles in the perturbative expansion due to the extra subtractions, and in particular, in some $s$-channel exchanges. Having the anomaly diagram a natural separation into longitudinal and transverse contributions, the subtraction of the longitudinal component can be viewed simply as the remotion of one of its {\em independent} invariant amplitudes. If the structure of a given graph does not render the anomaly vertex harmless, the longitudinal subtraction is explicit, otherwise
the subtraction vanishes by itself, as does the longitudinal component of the anomaly in that case.

In principle, the perturbative expansion for the GS vertex can be formulated directly in terms of its transverse components. Away from the chiral limit there is still no anomaly pole, and the decoupling of the anomaly should hold at all orders and also in the broken chiral phase.

We have argued by explicit examples that the organization of the perturbative expansion in terms of anomaly diagrams and pole counterterms
(or DZ counterterms) is just  a matter of convenience, especially if a given computation has to be carried out to
higher orders. In this case, the double poles due to the counterterms have to be interpreted as genuine contributions which are embedded in two-loop graphs.  We have pointed out that
the emergence of double poles is not an isolated case, but a standard
result, common to a specific way to address the tensor decomposition of a Feynman graph.

In this approach the computation of tensor integrals is performed using scalar integrals with higher power of the denominators and then re-formulated in terms of suitable sets of master integrals. Therefore, an ordinary
perturbative expansion at two-loop level - after integration on one of the loop momenta -  gives -with no surprise-  a theory with propagators of second order and higher.

A final comment goes to the high energy behavior od the GS vertex.
The good high energy behavior of the vertex is related to its gauge invariance, with BIM amplitudes which are identically vanishing in the chiral limit. A similar feature is absent in the WZ case, which violates unitarity at high energy.
Finally, we have investigated the emergence of GS and of WZ vertices in $g-2$ of the muon and in muonium, describing the differences between the SM case and its anomalous extensions, involving one axion-like particle and an extra anomalous $Z^\prime$, concentrating our attention, in particular, on the anomalous contributions, which can be studied accurately in the future in view of the planned experiments on $g-2$  at BNL.  In general, in the WZ case, the leading contributions to $g-2$ come from the exchange of an axi-Higgs and an anomalous $Z^\prime$, while in the GS case they involve directly the transverse components of the GS vertex. 
\section{Appendix. Some features of the GS and WZ vertices} 
We comment on the relation between the WZ and GS formulation. 

There are several ways to parameterize an anomaly vertex ($AVV$), the most well known being the one due to Rosenberg \cite{Rosenberg:1962pp} which involves six invariant amplitudes $(A_1,A_2,...A_6)$, two of which are ill-defined and determined by the Ward Identities of the theory in terms of the finite ones. The presence of an anomaly pole is not obvious in this formulation, although its structure was clearly established by Dolgov and Zakharov in their work \cite{Dolgov:1971ri} using dispersion relations. The basic interpretation of this result is that the anomaly is not just an ultraviolet but also an infrared effect.

The extraction of the anomaly pole from the rest of the amplitude is not so evident from the Rosenberg parameterization, but is quite obvious from the L/T formulation of this vertex, discussed in Sec.~\ref{sec:LTformulation}.
As we have discussed in the same section, the GS mechanism corresponds to a redefinition of the anomaly vertex. It means that whenever we encounter an anomaly diagram we replace it with 
another vertex in which the DZ pole has been explicitly removed. In a Lagrangian formulation this operation is equivalent to the addition of the counterterm shown in diagram c) of 
Fig.~\ref{GSS_AVV}. We stress once more that there is no direct coupling of the axion to the fermion, since in this approach the axion is not an asymptotic state. As we have extensively discussed in the previous 
sections this subtraction can be understood in a local version of the effective action by using Federbush's formulation of the GS mechanism with two pseudoscalars, Eq.~(\ref{fedeq}), one of them being actually a ghost, with negative kinetic energy. This formulation could, in principle, be extended so to describe a coupling of one of these two axions to the fermions. 

In the WZ case the local counterterm  $b F \wedge F$ introduces the axion as an asymptotic state of the corresponding $S$-matrix. Therefore, the axion takes an important role in the mechanisms of symmetry breaking, being this due either to a Higgs sector or to the St\"uckelberg mechanism, or to both. 
 For this reason, in the presence of electroweak symmetry breaking, there is a direct coupling of the axion to the fermions via the corresponding Yukawa couplings. One point which is worth to stress is that the WZ mechanism guarantees the gauge invariance of the one-loop effective action but not of the trilinear gauge vertex. 
The differences between the two mechanisms can be seen rather clearly, for instance, by comparing Fig.~\ref{LBL_WZ} and \ref{LBL_GS}. Notice that fermion mass effects, in the WZ case, induced either by chiral symmetry breaking and/or electroweak symmetry breaking cause a direct interaction of the axion to the fermion. If they are both absent, then those diagrams in which the axion couples to the fermions are trivially vanishing. 
\subsection{Gauge choices} 
The cancellation of the gauge dependence in the perturbative expansion is rather trivial in the GS case while it is less straightforward in the WZ case. In the first case, the redefinition of the trilinear gauge vertex is sufficient to obtain from the beginning a gauge invariant result. For this purpose we may work directly in the 
$R_\xi$ gauge, denoting with $\xi_B$ the gauge-fixing parameter. The gauge dependent propagator for the gauge field is given by
\beqa
\frac{- i}{k^2}\left[  g^{\, \lambda\, \lambda^{\prime}} - \frac{k^\lambda \, k^{\lambda^\prime}}{k^2 }
(1 - \xi_B) \right] 
\eeqa 
and the longitudinal components disappear whenever they are attached to a GS vertex, due to the Ward Identities satisfied on all the gauge lines. In the WZ case the cancellation of the gauge dependence is more subtle and has been discussed extensively in Chap.~\ref{chap:AbelianModels1}. 
\section{Appendix. Simplifications in some of the integrands on higher-point functions}
\subsection{Computation of the diagrams in Fig.~\ref{null2}}
We show the vanishing of the counterterms in Fig.~\ref{null2}.
We have
\bea
C_1^\la &=& \int  \frac{d^4 k_1}{(2 \pi)^4} \bar v (p_2) \gamma_\nu \frac{1}{\ds p_1 - \ds k_1} \gamma_\mu u(p_1) \frac{1}{k_1^2} \frac{1}{k_2^2} \, C^{\mu\nu\la}_{AVV}(k_1, -k_2, k) \nn \\
&=& \int  \frac{d^4 k_1}{(2 \pi)^4} \bar v (p_2) \gamma_\nu \frac{1}{\ds p_1 - \ds k_1} \gamma_\mu u(p_1)
\frac{1}{k_2^2} \frac{k_1^\mu}{k_1^4}\frac{a_n}{3}\epsilon[\la,\nu,k,k_2],
\eea
with $k_2=k-k_1$ so that
\bea
C_1^\la&=& \bar v (p_2) \gamma_\nu u(p_1)
 \frac{a_n}{3}   \epsilon[ \lambda, \nu, k, \rho]
 \int  \frac{d^4 k_1}{(2 \pi)^4} \frac{k_1^\rho}{(k-k_1)^2 k_1^4 },
\eea
where the expansion of the integrand function yields a result proportional to the independent momentum $k^\rho$ and finally $C_1^\la=0$.
The $C_2$ counterterm vanishes in an analogous way as $C_1^\la$, so we take into account the last diagram in Fig.~\ref{null2}
\bea
C_3^{\la} &=&
\int  \frac{d^4 k_1}{(2 \pi)^4}
\bar v (p_2) \gamma_\nu \frac{\ds p_1 - \ds k_1}{(p_1 -k_1)^2} \gamma_\mu u(p_1) \frac{1}{k_1^2}
\frac{1}{k_2^2} C^{\la\mu\nu}_{AVV}(-k,-k_1,-k_2),
\label{C3}
\eea
with $k_2=k-k_1$ and $C^{\la\mu\nu}_{AVV}(-k,-k_1,-k_2)=\frac{a_n}{3} \frac{k^\la}{k^2} \epsilon[\mu, \nu, k_1, k_2]$.
$C^\la_3$ is given by the sum of a first-rank and a second-rank tensor integral which can be further reduced with the well-known tensor-reduction technique. The general expansion for the two integrands is
\bea
\int \frac{d^4 k_1}{(2 \pi)^4} \frac{k_1^\alpha}{(p_1 -k_1)^2 k_1^2 (k-k_1)^2} &=& C^{}_1\, p_1^\alpha + C^{}_2 \, k^\alpha , \\
\int \frac{d^4 k_1}{(2 \pi)^4} \frac{k_1^\alpha k_1^\beta}{(p_1 -k_1)^2 k_1^2 (k-k_1)^2} &=&
C^{}_{00} \, g^{\alpha \beta} + C^{}_{12} \, (p_1^\alpha k^\beta + p_1^\beta k^\alpha)+ C^{}_{11} \, p_1^\alpha p_1^\beta 
+ C^{}_{22} \, k^\alpha k^\beta,
\eea
first we notice that all the terms proportional to $k^\alpha$ trivially vanish after the contraction with the antisymmetric Levi-Civita tensor in Eq.~(\ref{C3}) and then we conclude $C^\la_3=0$ by using the following relations in Eq.~(\ref{C3})
\bea
\bar v(p_2) \gamma_\nu \ds p_1 \gamma_\mu \epsilon[\mu, \nu, p_1, k] u(p_1)
&=& 2 i \, \bar v(p_2) (p_1^2 \ds k  - k \cdot p_1\ds p_1 ) \, \gamma^5 u(p_1) = 0,\\
\bar v(p_2) \gamma_\nu \gamma_\beta \gamma_\mu  \epsilon[\mu, \nu, \beta, k] u(p_1)
&=& 6 i\,  \bar v(p_2) \ds k \, \gamma^5 u(p_1)=0.
\label{AM}
\eea
for massless external fermions with momenta $p_1$ and $p_2$ and $k=p_1+p_2$.
\subsection{Simplifications of the integrands in Sec.~\ref{sec:GSembedding}}
The third amplitude ${\mathcal S}_c$ does not contribute to ${\mathcal S}$,
in fact we have
\bea
{\mathcal S}_c &=&  \int  \frac{d^4 k_1}{(2 \pi)^4} \Bigg( \bar v (p_2) \gamma_\nu \frac{1}{\ds p_1 - \ds k_1} \ds k_1 u(p_1) \frac{1}{k_2^2}
 \frac{1}{k_1^2}  \frac{a_n}{3} \frac{1}{k_1^2}  \epsilon[\nu, \lambda, k_2, k]   \Bigg) \frac{1}{k^2}
 \mathcal{ BT}^{ \lambda}_{AAA}    \nonumber\\
 &=&-   \bar v (p_2) \gamma_\nu u(p_1)
 \frac{a_n}{3}  \int  \frac{d^4 k_1}{(2 \pi)^4}   \Bigg(  \frac{1}{(k - k_1)^2}
 \frac{1}{k_1^4}  \epsilon[\nu, \lambda, k - k_1, k]   \Bigg) \frac{1}{k^2}\mathcal{ BT}^{ \lambda}_{AAA} \nonumber\\
&=& \bar v (p_2) \gamma_\nu u(p_1)
 \frac{a_n}{3}   \epsilon[\nu, \lambda, \rho, \sigma] k^\sigma \int  \frac{d^4 k_1}{(2 \pi)^4}   \Bigg(  \frac{k_1^\rho}{(k-k_1)^2 k_1^4 }
  \Bigg) \frac{1}{k^2}\mathcal{ BT}^{ \lambda}_{AAA}   \nonumber\\
 &\propto& \bar v (p_2) \gamma_\nu u(p_1)
 \frac{a_n}{3}   \epsilon[\nu, \lambda, k, k] \frac{1}{k^2}  \mathcal{ BT}^{ \lambda}_{AAA}  = 0,
\eea
where by the tensor integral decomposition we obtain the following result
\ba
\epsilon[\nu, \lambda, \rho, \sigma] k^\sigma \int  \frac{d^4 k_1}{(2 \pi)^4}
\Bigg(  \frac{k_1^\rho}{(k-k_1)^2 k_1^4 }\Bigg)
=\epsilon[\nu, \lambda, \rho, \sigma] k^\sigma B k_{\rho}=0.
\ea
Here we omit the explicit form of the coefficient of the rank-1 tensor decomposition $B$,
since it is not essential for the calculation.
We can apply the same arguments to prove that ${\mathcal S}_d =0$.

\chapter{An Anomalous Extra $Z^\prime$ from Intersecting Branes
with Drell-Yan and Direct Photons at the LHC  \label{chap:LHC}}
\fancyhead[LO]{\nouppercase{Chapter 5. An Anomalous Extra $Z^\prime$ from Intersecting Branes}}
\section{Introduction to the chapter}
The study of anomalous gauge interactions at the
LHC and at future linear colliders is for sure a difficult topic,
but also an open possibility that deserves close theoretical and experimental attention.
Hopefully, these studies will be able to establish if an
additional anomalous extra $Z^\prime$ is present in the spectrum, introduced by an
Abelian extension of the gauge structure of the Standard Model (SM), assuming that
extra neutral currents will be found in the next several years of
running of the LHC \cite{Langacker:2008yv}. The interactions that we discuss are characterized by
 {\em anomalous vertices}  in which gauge anomalies
cancel in some non trivial way, not by a suitable (anomaly-free)
charge assignment of the chiral fermion spectrum for each generation.

The phenomenological investigation of this topic is rather new, while
various mechanisms of cancellation of the gauge anomalies involving
axions have been around for quite some time. Global anomalies,
for instance, introduced for the solution of the strong $CP$-problem,
such as the Peccei-Quinn solution \cite{Peccei:1977hh, Peccei:1977ur, Weinberg:1977ma, Wilczek:1977pj, Kim:1979if, Shifman:1979if, Dine:1981rt, Zhitnitsky:1980tq}
(reviewed in \cite{Peccei:2006as}) require an axion, while local anomalies, cancelled by a Wess-Zumino counterterm,
allow an axion-like particle in the spectrum, whose mass and gauge coupling
- differently from $PQ$ axions - are independent. Similar constructions
hold also in the supersymmetric case and a generalization
of the WZ mechanism is the Green-Schwarz mechanism (GS) of string theory.
The two mechanisms  are related but not identical,
the first of them being characterized by a unitarity bound \cite{Coriano:2008pg}.
Details on the relation between the two at the level of effective field theory can be found in
\cite{Coriano:2008pg, Armillis:2008bg}.

Intersecting brane models, in which several anomalous $U(1)$'s
and St\"uckelberg mass terms are present, may offer a realization of these constructions
\cite{Antoniadis:2001np, Ibanez:2001nd, Kiritsis:2003mc, Blumenhagen:2006ci},
which can also be investigated in a bottom-up approach by using effective
Lagrangians built out of the requirements of gauge invariance of the
one-loop effective action \cite{Coriano:2005js}.
In our analysis we will consider the simplest extension of these anomalous Abelian gauge factors, which
involves a single anomalous $U(1)$, denoted as $U(1)_B$. The corresponding anomalous gauge boson
$B$ gets its mass via a combination of the Higgs and of St\"uckelberg mechanisms. Axions play a key role
in the cancellation of the anomalies in these theories although they may appear in other constructions
as well, due to the decoupling of a chiral fermion  in anomaly-free theories \cite{Coriano:2006xh}.

The presence of an anomalous $U(1)$ in effective models derived from string theory is quite common, although in all the previous literature before \cite{Coriano:2005js} and \cite{Coriano:2006xh,Coriano:2007fw,Coriano:2007xg} the phenomenological relevance of the anomalous $U(1)$ had not been worked out in any detail.
In particular, the dynamics of the anomalous extra gauge interaction had been neglected, by invoking a decoupling of the anomalous sector on the assumption of a large mass of the extra gauge boson.
In \cite{Coriano:2005js} it was shown that only one physical axion appears in the spectrum of these models, independently of the number of Abelian factors, which is the most important feature of these realizations.
In our case, the axion can be massless or massive, depending on the structure of the scalar potential.
Recent developements in the study of these models include their supersymmetric extensions \cite{Anastasopoulos:2008jt} and their derivations as symplectic forms of supergravity \cite{DeRydt:2007vg,
DeRydt:2008hw}. Other interesting variants include the St\"uckelberg extensions considered in
\cite{Feldman:2006wb,Feldman:2006ce,Feldman:2007wj} which depart significantly from the Minimal Low Scale Orientifold Model (mLSOM) introduced in \cite{Coriano:2005js} and discussed below. Specifically these models are also characterized by the presence of two mechanisms of symmetry breaking (Higgs and
St\"uckelberg) but do not share the anomalous structure.  As such they do not describe the anomalous
$U(1)$'s of these special vacua of string theory.

Axion-like particles, beside being a natural candidate for dark matter, may play a role in explaining some puzzling results regarding the propagation of high energy gamma rays  \cite{DeAngelis:2007dy, DeAngelis:2008sk} due to the oscillations of photons into axions in the presence of intergalactic magnetic field. In general, the presence of independent mass/coupling relations for these particles allows to evade most of the experimental bounds coming from \textsc{CAST} and other experiments on the detection of $PQ$ axions, characterized by a suppression of both mass and gauge couplings of this particle by the same large scale ($10^{10} \div 10^{12}$ GeV), (see \cite{Jaeckel:2006xm, Ahlers:2007qf}).
Here we focus our attention on the gauge sector, quantifying the rates for the detection of anomalous neutral currents at the LHC in some specific and very important channels.

\begin{itemize}
\item{\bf Drell-Yan}
\end{itemize}
Being leptoproduction the best way to search for extra neutral
interactions, it is then obvious that the study of the anomalous vertices and
of possible anomalous extra $Z^\prime$ should seriously consider the investigation of this process.
We describe the modifications induced on Drell-Yan computed in the Standard Model (SM)
starting from the description of some of the properties of the new anomalous vertices and of the corresponding one-loop counterterms, before moving to the analysis of the corrections. These appear - both in the WZ and GS cases in the relevant partonic channels at NNLO in the strong coupling constant
($O(\alpha_s^2)$). We perform several comparisons between anomalous
and non-anomalous extra $Z^\prime$ models
and quantify the differences with high accuracy.
\begin{itemize}
\item{\bf Direct Photons (Di-photon, DP)}
\end{itemize}
Double prompt (direct) photons offer an interesting signal which
is deprived of the fragmentation contributions especially at large
values of their invariant mass $Q$, due to the steep falling of the
photon fragmentation functions. In addition, photon isolation may
provide an additional help in selecting those events coming from channels in which
the contribution of the anomaly is more sizeable, such as gluon fusion.
Also in this case we perform a detailed investigation of this sector.
For direct photons, the anomaly appears in gluon fusion - at parton level
- in a class of amplitudes which are characterized by two-triangles graphs
- or BIM amplitudes - using the definitions of Chap.~\ref{chap:UnitarityBound}.

In both cases the quantification of the background needs extreme care,
due to the small signal, and the investigation of the renormalization/factorization
scale dependence of the predictions is of outmost importance. In particular,
we consider all the sources of scale-dependence in the analysis, including
those coming from the evolution of the parton densities (\textsc{Pdf}'s)
which are just by themeselves enough to overshadow the anomalous corrections.
For this reason we have used the program \textsc{Candia 1.0} in the evolution
of the \textsc{Pdf}'s,  which has been documented in \cite{Cafarella:2008du}.
The implementation of DY and DP processes is part of two programs $\textsc{Candia}_{DY}$
and $\textsc{Candia}_{Axion}$ for the study of the QCD background with the
modifications induced by the anomalous signal. The QCD background in DP
is computed using \textsc{Diphox} \cite{Binoth:1999qq} and \textsc{Gamma2MC}
based on Ref.~\cite{Bern:2002jx}. The NLO corrections to DP before the
implementation of \textsc{Diphox} have been computed
by Gordon and Corian\`o back in 1995 \cite{Coriano:1996us}
and implemented in a Monte Carlo based on the phase space slicing method.

In the numerical analysis that we present we have included all the
contributions coming from the two mechanisms as separate cases,
corrections that are implemented in DY and DP processes. We will start
analyzing the contributions to these processes in more detail in the next
sections, discussing the specific features of the anomalous contributions and of the corresponding
counterterms at a phenomenological level.

The chapter is organized as follows.
After a brief description of the anomalous interactions
and the counterterms that appear either at Lagrangian level
(WZ case) or at the level of the trilinear gauge vertex (GS case),
we discuss the main properties of these vertices and we address the
structure of the corrections in DY and in DP. Our study is mainly focused
on the invariant mass distributions in the two cases. The need for performing
these types of analysis in parallel will be explained below, and there is the
hope that it may be extended to other processes and observables  in the future,
such as rapidity distributions and rapidity correlations \cite{Chang:1997sn}.
We present high precision estimates of the QCD background at NNLO, which is
the order where, in these processes, the anomalous corrections start
to appear. Other analysis, of course, are also possible, such as those
involving four-fermion decays in trilinear gauge interactions which could,
in principle, be sensitive to Chern-Simons terms
\cite{Coriano:2005js, Anastasopoulos:2006cz} if at least two anomalous $U(1)$'s
are present in the spectrum. These additional interactions are allowed, as discussed in~\cite{Armillis:2007tb}, 
whenever the distribution of the partial anomalies
on a diagram is not fixed by symmetry requirements.
A complete description of these vertices has been carried out in
\cite{Armillis:2007tb}, useful for direct phenomenological studies.

As we are going to show, the search for effects due to anomalous $U(1)$ at the LHC in
$pp$ collisions cannot avoid an analysis of the QCD background at NNLO.
DY and DP are the only two cases where this level of precision has been
obtained in perturbation theory. As we are going to show, the anomalous
effects at the LHC in these two key processes are tiny, since the invariant
mass distributions are down by a factor of $10^{3} \div 10^{4}$ compared to the
NNLO (QCD) background. The accuracy required at the LHC to identify these
effects on these observables should be of a fraction of a percent
($0.1 \%$ and below), which is beyond reach at a hadron collider due
to the larger indetermination intrinsic in QCD factorization and the parton model.
\section{Anomaly-free versus an anomalous extra $Z^\prime$ in Drell-Yan}
As we have already mentioned, the best mode to search for extra $Z^\prime$
at the LHC is in the production of a lepton pair via the Drell-Yan mechanism
($q\bar{q}$ annihilation) mediated by neutral currents. The final state is
easily tagged and resonant due to the $s$-channel exchange of the extra gauge boson.
In particular, a new heavier gauge boson modifies the invariant mass distribution
also on the $Z$ peak due to the small modifications induced on the couplings and
to the $Z-Z^\prime$ interference. In the anomalous model
that we have investigated, though based on a specific charge assignment, we find
larger rates for these distributions both on the peak of the $Z$ and of the
$Z^{\prime}$ compared to the other models investigated, if the extra resonance
is around 1 TeV. This correlation is expected to drop as the mass of the extra
$Z^\prime$ increases. In our case, as we will specify below, the mass of the extra resonance is
given by the St\"uckelberg ($M_1$) mass, which appears also (as a suppression scale)
in the interaction of the physical axion to the gluons and it is essentially a free parameter.

In DY, the investigation of the NNLO hard scatterings goes
back to \cite{Hamberg:1990np}, with a complete computation of the invariant
mass distributions, made before that the NNLO corrections to the DGLAP evolution
had been fully completed. In our analysis we will compare three anomaly-free
models against a model of intersecting brane with a single anomalous $U(1)$.
The anomaly-free charge assignments come from a gauged $B-L$ Abelian symmetry,
a ``$q+u$'' Model -both described in \cite{Carena:2004xs} - and the Free Fermionic Model analyzed in
\cite{Coriano:2008wf}. We start by summarizing our definitions and conventions.

In the anomaly-free case we address Abelian extensions of the gauge structure
of the form $SU(3)\times SU(2) \times U(1)_Y \times U(1)_z$,
with a covariant derivative in the $W_{\mu}^{3},B_{Y}^{\mu},B_{z}^{\mu}$
(interaction) basis defined as
\ba
\hat{D}_{\mu}=\left[\partial_{\mu} -i g_2 \left( W_{\mu}^{1}T^{1} +
W_{\mu}^{2}T^{2} + W_{\mu}^{3}T^{3} \right) -i\frac{g_{Y}}{2}\hat{Y}
B_{Y}^{\mu}-i\frac{g_{z}}{2}\hat{z} B_{z}^{\mu} \right]
\ea
where we denote with $g_2,g_Y, g_z$ the couplings of $SU(2)$, $U(1)_Y$ and
$U(1)_z$, with  $\tan\theta_W=g_Y/g_2$. After the diagonalization of the
mass matrix we have
\ba
\left( \begin{array}{c}
A_{\mu} \\
Z_{\mu}  \\
Z^{\prime}_{\mu}
\end{array} \right)
=
\left( \begin{array}{ccc}
\sin\theta_W & \cos\theta_W & 0\\
\cos\theta_W & -\sin\theta_W & \varepsilon \\
-\varepsilon\sin\theta_W& \varepsilon\sin\theta_W & 1
\end{array} \right)
\left( \begin{array}{c}
W^{3}_{\mu} \\
B^{Y}_{\mu}  \\
B^{z}_{\mu}
\end{array} \right)
\label{massmatrixnoanomaly}
\ea
where $\varepsilon$ is a perturbative parameter which is
around $10^{-3}$ for the models analyzed, introduced in
\cite{Carena:2004xs} and \cite{Coriano:2008wf}. It  is defined as
\ba
\varepsilon=\frac{\delta M^2_{Z
Z^{\prime}}}{M^2_{Z^{\prime}}-M^2_{Z}}
\ea
while the mass of the $Z$ boson and of the extra $Z^{\prime}$ are
\ba
&&M_Z^2=\frac{g_2^2}{4
\cos^2\theta_W}(v_{H_1}^2+v_{H_2}^2)\left[1+O(\varepsilon^2)\right]
\nonumber\\
&&M_{Z^{\prime}}^2=\frac{g_z^2}{4}(z_{H_1}^2
v_{H_1}^2+z_{H_2}^2v_{H_2}^2+z_{\phi}^2
v_{\phi}^2)\left[1+O(\varepsilon^2)\right]
\nonumber\\
&&\delta M^2_{Z Z^{\prime}}=-\frac{g_2 g_z}{4\cos\theta_W}(z_{H_1}^2
v_{H_1}^2+z_{H_2}^2 v_{H_2}^2).
\ea
In this class of models we have two Higgs doublets $H_1$ and $H_2$,
whose v.e.v.'s are $v^{}_{H_1}$ and $v^{}_{H_2}$ and an extra $SU(2)_{W}$ singlet $\phi$
whose v.e.v. is $v^{}_\phi$. The extra $U(1)_z$ charges of the Higgs doublets
are respectively $z_{H_1}$ and $z_{H_2}$, while for the singlet this is denoted as $z_\phi$.
Taking the value of $v_{H_2}$ of the order of the electroweak scale ($\approx 246$ GeV),
we fix $v_{H_1}$ with $\tan\beta = v_{H_2}/v_{H_1}$, and we still have one free parameter, $v_{\phi}$, which enters in the calculation of the mass of the extra $Z^{\prime}$.
Then it is obvious that we can take the mass $M_{Z^{\prime}}$ and the coupling constant
$g^{}_z$ as free parameters. We choose $\tan\beta\approx 40$ in order to reproduce the
mass of the $Z$ boson at $91.187$ GeV, choice that is performed, for consistency, also in the anomalous model. In this last case the Higgs sector is characterized only by 2 Higgs doublets, with the v.e.v. of the extra singlet being replaced by the St\"uckelberg mass. We define $g_2 \sin\theta_W  =g_Y \cos\theta_W =e$
and construct the $W^{\pm}$ charge eigenstates and the corresponding
generators $T^{\pm}$ as usual
\ba
W^\pm=\frac{W_1\mp iW_2}{\sqrt{2}} \qquad  T^\pm=\frac{T_1\pm iT_2}{\sqrt{2}},  \nonumber
\ea
while in the neutral sector we introduce the rotation matrix
\ba
\left( \begin{array}{c}
W^{3}_{\mu} \\
B^{Y}_{\mu}  \\
B^{z}_{\mu}
\end{array} \right)
=
\left( \begin{array}{ccc}
\frac{\sin\theta_W (1+\varepsilon^2)}{1+\varepsilon^2} &
\frac{\cos\theta_W}{1+\varepsilon^2} & \varepsilon
\frac{\cos\theta_W}{1+\varepsilon^2}\\
\frac{\cos\theta_W(1+\varepsilon^2)}{1+\varepsilon^2} &
-\frac{\sin\theta_W}{1+\varepsilon^2} & \varepsilon
\frac{\sin\theta_W}{1+\varepsilon^2}\\
0 & \frac{\varepsilon}{1+\varepsilon^2}& \frac{1}{1+\varepsilon^2}
\end{array} \right)
\left( \begin{array}{c}
A_{\mu} \\
Z_{\mu}  \\
Z^{\prime}_{\mu}
\end{array} \right)
\ea
which relates the interaction and the mass eigenstates.
Substituting these expression in the covariant derivative we obtain
\ba
&&\hat{D}_{\mu}=\left[\partial_{\mu} -i A_{\mu} \left(g_2 T_3\sin\theta_W+
g_Y\cos\theta_W \frac{\hat{Y}}{2}\right)
-ig_2\left(W^{-}_{\mu}T^{-}+ W^{+}_{\mu}T^{+}\right)\right.
\nonumber\\
&&\hspace{1cm}\left.-iZ_{\mu}\left( g_2\cos\theta_W T_{3} -g_Y \sin\theta_W
\frac{\hat{Y}}{2}
+g_z \varepsilon\frac{\hat{z}}{2}\right)\right.
\nonumber\\
&&\hspace{1cm}\left.-iZ^{\prime}_{\mu}\left(-g_2\cos\theta_W
T_{3}\varepsilon +g_Y\sin\theta_W \frac{\hat{Y}}{2}\varepsilon
+g_z\frac{\hat{z}}{2}\right)\right]
\ea
where we have neglected all the $O(\varepsilon^2)$ terms.
In the limit $g_z\rightarrow 0$ and $\varepsilon\rightarrow 0$
we obtain the SM expression.
The vector and the
axial-vector couplings of the $Z$ and $Z^\prime$ to the fermions
are expressed equivalently in terms of the left - ($z_L$) and right
- ($z_R$) $U(1)_z$ chiral charges  and hypercharges ($Y_R$, $Y_L$)
of the  models that we have implemented. These can be found in
\cite{Coriano:2008wf} for the free fermionic case and in
\cite{Carena:2004xs} for the remaining models with a V-A structure given by
\ba
&&\frac{-i g_2}{4 c_w}\gamma^{\mu} {g_V}^{Z,j}=\frac{-i g_2}{c_w}
\frac{1}{2}\left[c_w^2
T_3^{L,j}-s_w^2(\frac{\hat{Y}^{j}_L}{2}+\frac{\hat{Y}^{j}_R}{2})
+\varepsilon \frac{g_z}{g_2} c_w
(\frac{\hat{z}_{L,j}}{2}+\frac{\hat{z}_{R,j}}{2})\right]\gamma^{\mu}
\nonumber\\
&&\frac{-i g_2}{4 c_w}\gamma^{\mu}\gamma^{5} {g_A}^{Z,j}=\frac{-i g_2}
{c_w}\frac{1}{2}\left[-c_w^2 T_3^{L,j}
-s_w^2(\frac{\hat{Y}^{j}_R}{2}-\frac{\hat{Y}^{j}_L}{2})
+\varepsilon \frac{g_z}{g_2} c_w
(\frac{\hat{z}_{R,j}}{2}-\frac{\hat{z}_{L,j}}{2})\right]\gamma^{\mu}\gamma^{5}
\nonumber\\
&&\frac{-i g_2}{4 c_w}\gamma^{\mu} {g_V}^{Z^{\prime},j}=\frac{-i g_2}{c_w}
\frac{1}{2}\left[ -\varepsilon c_w^2 T_3^{L,j}
+\varepsilon s_w^2(\frac{\hat{Y}^{j}_L}{2}+\frac{\hat{Y}^{j}_R}{2})
+\frac{g_z}{g_2}c_w(\frac{\hat{z}_{L,j}}{2}+
\frac{\hat{z}_{R,j}}{2})\right]\gamma^{\mu}
\nonumber\\
&&\frac{-i g_2}{4 c_w}\gamma^{\mu}\gamma^{5} {g_A}^{Z^{\prime},j}=\frac{-i
g_2}{c_w} \frac{1}{2}\left[ \varepsilon c_w^2 T_3^{L,j}
+\varepsilon s_w^2(\frac{\hat{Y}^{j}_R}{2}-\frac{\hat{Y}^{j}_L}{2})
+\frac{g_z}{g_2}c_w(\frac{\hat{z}_{R,j}}{2}-
\frac{\hat{z}_{L,j}}{2})\right]\gamma^{\mu}\gamma^{5},
\nonumber\\
\ea
where $j$ is an index which represents the quark or the lepton and we have
set $\sin\theta_W=s_w,\cos\theta_W=c_w$ for brevity.
\subsection{An anomalous extra $Z^\prime$}
In the presence of anomalous interactions we can use the same formalism developed so far
for anomaly-free models with some appropriate changes.
Since the effective Lagrangian of the class of the anomalous
models that we are investigating includes both a St\"uckelberg and a two-Higgs doublet sector,
the masses of the neutral gauge bosons are provided by a combination
of these two mechanisms, as can be clearly understood from the expressions of $M^{}_Z$ and $M^{}_{Z^\prime}$ in Eqs.~(\ref{ZZpmass}). 
In this case we have as free parameters the St\"ueckelberg mass $M_1$ and the anomalous coupling
constant $g_B$, with $\tan\beta$ as in the remaining anomaly-free models. As we have
already stressed, the analysis does not depend significantly on the choice of this parameter.
The value of the St\"uckelberg mass $M_1$
is loosely constrained by the $D$-brane model in terms of suitable wrapping numbers ($n$) of the 4-branes
which define the charge embedding \cite{Ghilencea:2002da, Ibanez:2001nd}
reported in Tabs.~\ref{pparameters}, \ref{ccharge_higgs}, \ref{ttabpssm} of Sec.~\ref{sec:Madrid} and Tab.~\ref{charges}.
As shown in Sec.~\ref{sec:Ochi}, the physical gauge fields are obtained from the hypercharge basis by means of the rotation matrix $O^A$
which can be approximated at the first order as
\bea
O^A  \simeq  \pmatrix{
\frac{g^{}_Y}{g}           &     \frac{g^{}_2}{g}         &      0   \cr
\frac{g^{}_2}{g} + O(\epsilon_1^2)          &     -\frac{g^{}_Y}{g} + O(\epsilon_1^2) &      \frac{g}{2} \epsilon_1 \cr
-\frac{g^{}_2}{2}\epsilon_1     &     \frac{g^{}_Y}{2}\epsilon_1  &   1 + O(\epsilon_1^2) }
\eea
which is the analogue of the matrix in Eq.~(\ref{massmatrixnoanomaly}) for the anomaly-free models, but here the role of the mixing parameter $\epsilon_1$ is taken by  the expression
\ba
\epsilon_1=\frac{x_B}{M_1^2}.
\label{alogue}
\ea
A relation between the two expansion parameters can be easily obtained in an approximate way by a direct comparison. For simplicity we take all the charges to be O(1) in all the models obtaining
\ba
&&M_Z^2\sim {g_2^2} v^2
\nonumber\\
&&M_{Z^{\prime}}^2 - M_Z^2 \sim g_z^2 v_{\phi }^2 \nonumber \\
&& \delta M^2_{Z Z^{\prime}}\sim{g_2 g_z} v^2
\ea
giving
\beq
\epsilon_1\sim \frac{v^2}{M_1^2},
\eeq
which is the analogue of Eq.~(\ref{alogue}), having identified the St\"uckelberg mass with the v.e.v. of the extra singlet Higgs, $M_1\sim g_z v_{\phi}$. This is natural since the St\"uckelberg mechanism can be thought of as the low energy remnant of an extra Higgs whose radial fluctuations have been frozen and with the imaginary phase surviving at low energy as a $CP$-odd scalar \cite{Coriano:2006xh}.
Concerning the charge assignments, the corresponding model is obtained form the
intersection of four branes $(a, b, c, d)$ with generators $(q_a, q_b, q_c, q_d)$ which are rotated to the hypercharge basis, with an anomaly free hypercharge. The identification of the generators involve the solution of some constraint
equations, solutions which for a $T^6$ compactification are parametrized by a phase $\epsilon =\pm1$; the Neveu-Schwarz background
on the first two tori $\beta_i=1-b_i=1,1/2$, four integers
$n_{a2}, n_{b1}, n_{c1}, n_{d2}$, which are the wrapping numbers of the branes around the extra (toroidal) manifolds of the compactification, and a parameter $\rho=1,1/3$, with an additional constraint in order to obtain the correct massless hypercharge.  One of the choice for these parameters is reported in Tab.~\ref{pparameters} of Sec.~\ref{sec:Madrid}.
\begin{figure}[t]
\begin{center}
\includegraphics[width=6.5cm,angle=-90]{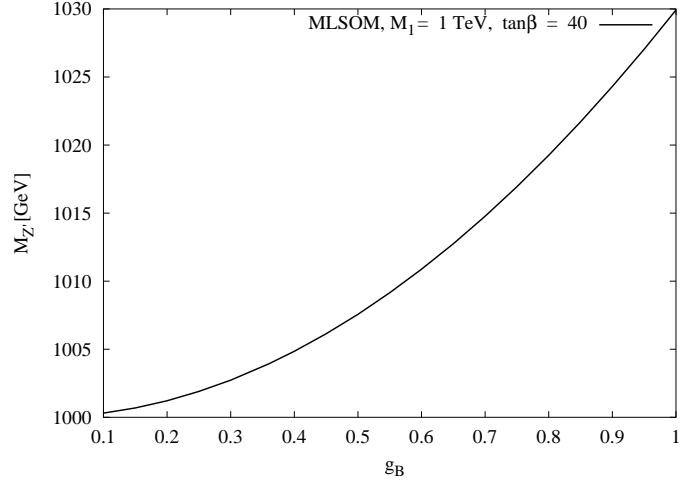}
\caption{\small Anomalous $M_{Z^\prime}$ as a function of the coupling $g^{}_B$.}
\label{MZp_mass_gB}
\end{center}
\end{figure}
\begin{table}[t]
\begin{center}
\begin{tabular}{|c|c|c|c|c|c|c|}
\hline
   &   $Q_L$  & $u_R $ & $d_R $ & $L$  &  $e_R$ & $N_R$ \\
\hline  $q_{Y}$  &  1/6    & - 2/3  & 1/3   &  -1/2   & 1 &  0  \\
\hline   $q_{B}$  & -1    & 0  & 0   & -1   & 0  & 0 \\
\hline \end{tabular}
\end{center}
 \caption{\small Fermion spectrum charges in the $Y$-basis for the Madrid Model \cite{Ghilencea:2002da}.}
\label{charges}
\end{table}
\begin{figure}[t]
\begin{center}
\includegraphics[scale=0.8]{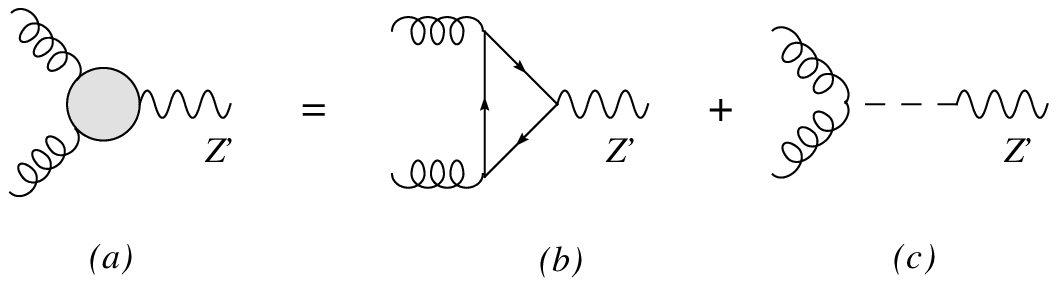}
\caption{\small A  gauge invariant GS vertex of the AVV type, composed of an AVV triangle  and a single counterterm of the Dolgov-Zakharov form.}
\label{GS_AVV}
\end{center}
\end{figure}
\section{ The GS and WZ vertices and gluon fusion}
As we have mentioned above in the previous sections, the two available
mechanisms that enforce at the level of the effective Lagrangian the cancellation
of the anomalies involve either $PQ$-like (axion-like) interactions
- in the WZ case -  or the subtraction of the anomaly pole (for the GS case).
In the GS case, the anomaly of a given diagram is removed
by subtracting the longitudinal pole of the triangle amplitude in the chiral limit.
We have stressed that the counterterm (the pole subtraction) amounts to the removal of one of the
invariant amplitudes of the anomaly vertex (the longitudinal component) and
corresponds to a vertex re-definition.

The procedure is exemplified in Fig.~\ref{GS_AVV} where we show the triangle
anomaly and the pole counterterm which is subtracted from the first amplitude.
The combination of the two contributions defines the GS vertex, which is made
of purely transverse components in the chiral limit \cite{Armillis:2008bg} and
satisfies an ordinary WI. Notice that the vertex does not require
an axion as an asymptotic state in the related
$S$-matrix; for a non-zero fermion mass in the triangle diagram, the vertex
satisfies a broken WI. We now proceed and summarize some of
these properties, working in the chiral limit.

Processes such as $g g\to \gamma\gamma$,  mediated by an
anomalous gauge boson $Z^{\prime}$, can be expressed in a simplified
form in which only the longitudinal component of the anomaly appears. We therefore set
$k_1^2=k_2^2=0$  and $m_f=0$,
which are the correct kinematical conditions to obtain the anomaly pole,
necessary for a parton model (factorized) description of the cross section
in a $pp$ collision at the LHC, where the initial state of the partonic hard-scatterings are on-shell.

We start from the Rosenberg form of the $AVV$ amplitude, which is given by
\ba
&&T^{\lambda\mu\nu}=A_1\epsilon[k_1,\lambda,\mu,\nu]+A_2\epsilon[k_2,\lambda,\mu,\nu]
+A_3 k_1^{\mu}\epsilon[k_1,k_2,\nu,\lambda]
\nonumber\\
&&\hspace{1cm}+A_4 k_2^{\mu}\epsilon[k_1,k_2,\nu,\lambda]
+A_5 k_1^{\nu}\epsilon[k_1,k_2,\mu,\lambda]+A_6 k_2^{\nu}\epsilon[k_1,k_2,\mu,\lambda]\,,
\ea
and imposing the Ward Identities to bring all the anomaly on the axial-vector vertex, we obtain the usual conditions
\ba
&&A_1 = k_2^2 A_4 + k_1\cdot k_2 A_3
\nonumber\\
&&A_2 = k_1^2 A_5 + k_1\cdot k_2 A_6
\nonumber\\
&&A_3(k_1,k_2)=-A_6(k_1,k_2)
\nonumber\\
&&A_4(k_1,k_2)=-A_5(k_1,k_2),
\ea
where the invariant amplitudes $A_3,\dots,A_6$ are free from kinematical singularities for off-shell external lines. We set $k^2=(k_1+k_2)^2=s$.
As we have mentioned, in the parton model we take the initial gluons to be on-shell, while the hadronic cross section is obtained by convoluting the hard scattering given above (corrected by a color factor)  with the \textsc{Pdf}'s. The amplitude simplifies drastically in this case and takes the form
\ba
T^{\mu\nu\lambda}=A_6 k^{\lambda}\epsilon[k_1,k_2,\nu,\mu]+
\left(A_4 + A_6\right)\left(k_2^{\nu}\epsilon[k_1,k_2,\mu,\lambda]
-k_1^{\mu}\epsilon[k_1,k_2,\nu,\lambda]\right),\,
\ea
in which the second piece drops off
for physical on-shell photon/gluon lines, leaving only a single invariant
amplitude to contribute to the final result
\ba
&&T^{\mu\nu\lambda}=A_6^{f}(s)(k_{1}+k_{2})^{\lambda}
\epsilon\left[k_1,k_2,\nu,\mu\right]
\label{massiveT}
\ea
where
\ba
&&A_6^{f}(s)=\frac{1}{2\pi^2 s}\left(1 +\frac{m_f^2}{s}\log^{2}\frac{\rho^{}_f + 1 }{\rho^{}_f - 1 }\right),
\hspace{1cm} \rho^{}_f = \sqrt{1 - 4\frac{m_f^2}{s}}, \,\,\,\,\, s<0.
\label{a6}
\ea
The anomaly pole is given by the first term of  Eq.~(\ref{a6})
\ba
&& T_c^{\mu\nu\lambda}\equiv
\frac{1}{2\pi^2 s} (k_{1}+k_{2})^{\lambda}
\epsilon\left[k_1,k_2,\nu,\mu\right].
\label{tc}
\ea
The logarithmic functions in the expression above are continued
in the following way in the various region
\ba
&&0 < s < 4 m_f^2:
\nonumber\\
&&\rho_f\rightarrow i \sqrt{-\rho_f^2}; \hspace{0.5cm}
\frac{1}{2}\log\left(\frac{\rho_f +1}{\rho_f - 1}\right)\rightarrow
-i\arctan\frac{\sqrt{s}}{\sqrt{4 m_f^2 - s}},
\nonumber\\
&& s > 4 m_f^2 > 0:
\nonumber\\
&&\sqrt{-\rho_f^2} \rightarrow -i\rho_f; \hspace{0.5cm}
\arctan\frac{1}{\sqrt {- \rho_f^2}}\rightarrow \frac{\pi}{2}
+\frac{i}{2}\log\left(\frac{\sqrt{s - 4 m_f^2}+\sqrt{s}}{\sqrt{s}-\sqrt{s - 4 m_f^2}} \right).
\ea
\begin{figure}[t]
\begin{center}
\includegraphics[scale=0.8]{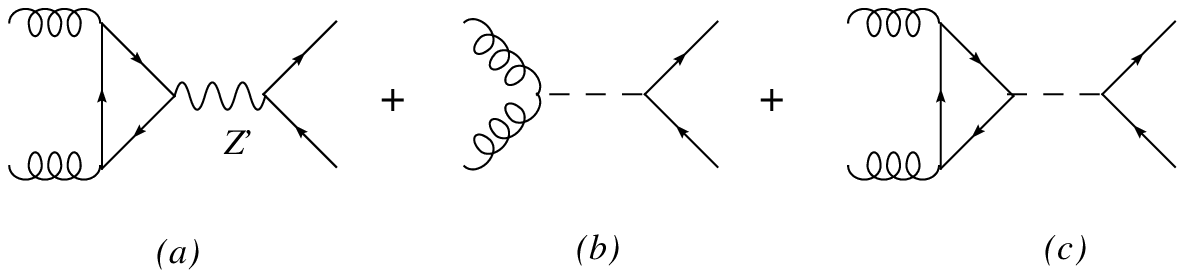}
\caption{\small One loop vertices and counterterms for the WZ mechanism.}
\label{onefig}
\end{center}
\end{figure}
\begin{figure}[t]
\begin{center}
\includegraphics[scale=0.85]{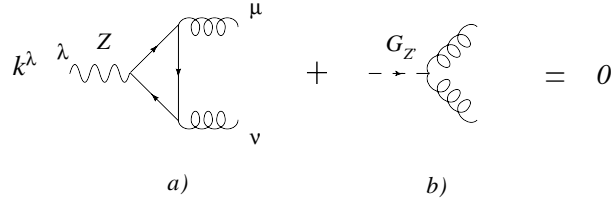}
\caption{\small  WI in the WZ case in the chiral limit.}
\label{STIggch}
\end{center}
\end{figure}

Notice that the surviving amplitude $A_6$ multiplies a longitudinal
momentum exchange and, as discussed in the literature on
the chiral anomaly in QCD \cite{Achasov:1992bu, Dolgov:1971ri},
is characterized by a massless pole in $s$, which is the anomaly pole,
as one can clearly conclude from Eq.~(\ref{a6}). This equation shows also how
chiral symmetry breaking effects appear in this amplitude at this special
kinematical point by the $m_f$ terms.

The subtraction of the anomaly pole is shown in
Fig.~\ref{GS_AVV} and is represented by diagram c).
The combination of diagrams b) and c) defines the GS vertex
of the theory \cite{Armillis:2008bg}, with diagram c) described by Eq.~(\ref{tc}) ($- T_c$).
It is easily verified that in the massless fermion limit and for on-shell gluon lines,
the GS vertex is trivially vanishing by construction. In general, for any asymmetric
configuration of the external lines in the vertex, even in the massless limit,
the vertex has non-zero transverse components \cite{Knecht:2003xy,Jegerlehner:2005fs}.
The expression is well known \cite{Knecht:2002hr, Jegerlehner:2005fs} in the chiral
limit and has been shown to satisfy the Adler-Bardeen theorem \cite{Jegerlehner:2005fs}.

For a non-vanishing $m_f$ the GS vertex, for generic virtualities,
can be defined to be the general $AVV$ vertex, for instance extracted
from \cite{Kniehl:1989qu} or, in the longitudinal/transverse formulation,
by the amplitudes given in \cite{Jegerlehner:2005fs}, with the subtraction of the anomaly pole, as given in
\cite{Armillis:2008bg}. We will refer to the anomaly (subtraction)
counterterm of diagram b) as to the Dolgov-Zakharov \cite{Dolgov:1971ri}
(DZ) counterterm. The anomaly diagram reduces to its DZ form for two on-shell
gauge lines (photons/gluons) and in this case the transverse components
completely disappear. There are other cases in which, instead,
the longitudinal components cancel. This occurs if, for instance,
a conserved current is attached to the anomalous line, rendering the anomaly
"harmless", as explained in \cite{Armillis:2008bg}.

The analogous interaction in the WZ case is shown in Fig.~\ref{onefig},
where we have attached a fermion pair in the final state to better
identify the contributions. In this case, beside the anomalous contribution
of diagram a), the mechanism will require the  exchange of a physical axion,
shown in diagram b) and c). Diagram b) is the usual WZ counterterm
(or generalized $PQ$ interaction)  while the third diagram is non-vanishing
only in the presence of fermions of non-zero mass. This third contribution
is numerically irrelevant and in DY is usually omitted. The WZ mechanism
re-establish gauge invariance of the effective Lagrangian but is not based
on a vertex re-definition and, furthermore, involves an asymptotic axion state.
As shown in Chap.~\ref{chap:UnitarityBound} the presence of a unitarity bound in this
mechanism is a signal of its limitation as an effective theory
(see also the discussion in \cite{Armillis:2008bg}). We have summarized
in an appendix the discussion of this point
in a simple case.
\subsection{Ward Identities}
\begin{figure}[t]
\begin{center}
\includegraphics[scale=0.85]{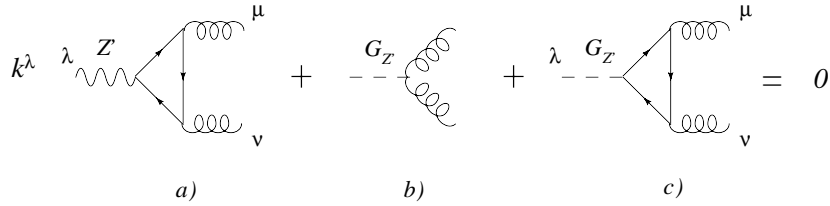}
\caption{\small Generalized WI in the WZ case. }
\label{STInochiral}
\end{center}
\end{figure}
Both vertices satisfy ordinary Ward Identities in the chiral limit and
generalized Ward Identities away from it. In the chiral limit, for instance,
the WZ mechanism adds to the effective action of the anomalous theory an
interaction of the St\"uckelberg field ($b$) with the gluons $(b\, G\wedge G)$,
shown in diagram b) of Fig.~\ref{STIggch}. In this figure we have shown a diagrammatic realization
of the WI for this case.

In WZ, being the cancellation based on a local field theory, the derivation of the generalized WI can be formally obtained from the requirement of BRST invariance of the gauge-fixed effective
action, as shown in~\cite{Armillis:2007tb}. This is illustrated in Fig.~\ref{STInochiral},  in the case of an anomalous $Z^{\prime}$, where we show the coupling of the Goldstone - in the broken Higgs phase -  to the gluons (diagram b)) and to
the massive fermion (diagram c)) \cite{Armillis:2007tb}. The normalization of the counterterm in b) can be chosen to remove
the anomaly of diagram a) when a single fermion runs inside the anomaly loop. Alternatively, the same graphical representation holds if in the first and the last diagram we sum over the entire generation. In this case the counterterm is normalized to cancel the entire anomaly of the complete vertex.
\begin{figure}[t]
\begin{center}
\includegraphics[scale=0.8]{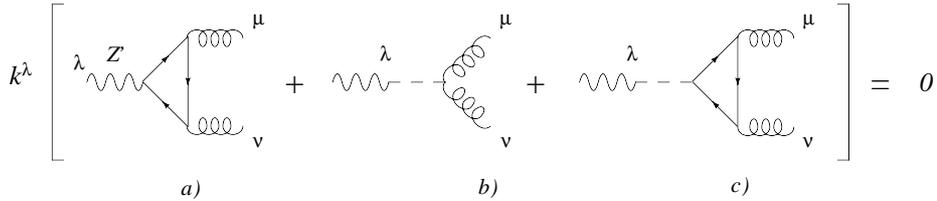}
\caption{\small Generalized WI in the GS case.}
\label{STIgg}
\end{center}
\end{figure}
\begin{figure}[t]
\begin{center}
\includegraphics[scale=0.75]{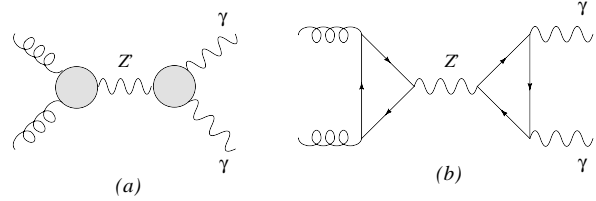}
\caption{\small  BIM amplitude with an anomalous $Z^\prime$ exchange.}
\label{ggBIM}
\end{center}
\end{figure}
The analysis in the GS case is slightly more subtle. We show in Fig.~\ref{STIgg}
the generalized WI satisfied by the vertex in the massive case.
In the massless case only diagrams a) and b) survive, while  the contribution
of diagram c) comes from a direct computation. It is obtained by multiplying
typical pseudoscalar interaction - such as the one shown in Fig.~\ref{STInochiral},
diagram c) - by a massless pole. If we denote by
$T^{\mu\nu} $  the diagram describing the decay of a pseudoscalar
into two gluons, diagram c) takes the form
$k^{\lambda}/k^2 T_{\mu\nu}$, with a factorized pole on the anomalous external line.
We refer to \cite{Armillis:2008bg} for a detailed discussion of these points.

In our analysis we encounter a class of \textsc{BIM} amplitudes which are characterized
by two anomaly vertices connected by an $s$-channel exchange
of the anomalous gauge boson. These amplitudes grow quadratically with
the energy and are not eliminated by fine tuning Fig.~\ref{ggBIM}.
The true \textsc{BIM} amplitude is the one shown in diagram b) and
appears in the gluon fusion sector in the WZ case.  In the SM a similar
graph contributes only if heavy fermions run in the loop. They are comparable
in size to the anomalous BIM amplitude. Obviously, this contribution would be
identically vanishing if all the fermions of a given generation would be mass-degenerate.

Diagram a)
shown in the same figure, instead, is the GS version of the \textsc{BIM} amplitude and is identically zero in the chiral limit for on-shell gluon lines, as is the case in the parton model. For this reason the gluon fusion sector
disappears completely  for DP (in the GS mechanism), since the BIM amplitude in this case is obtained by replacing diagram b) of this figure with diagram a) which is indeed vanishing.

We can summarize the basic features of the anomalous sectors in anomaly free-models after QCD factorization, for generic virtualities of the external gauge lines, according to the following points:

1) In the SM the residual contributions coming from anomalous diagrams, such as in the $V V Z, V V Z^{\prime}$ vertices, where $V$ is a gauge field, are proportional to the mass of the heavy quarks in the anomaly loop. In the chiral limit, instead, both the anomaly pole contribution
and the transverse component of the anomaly cancel by charge assignment.

2) In the GS case, as we have just discussed, the anomaly pole is absent by definition, while the transverse
contributions are allowed. This separation between longitudinal and transverse components is less transparent for a heavy fermion mass, which induce a longitudinal component, proportional to
$m_f^2/s^2 $ times a small logarithmic correction of the ratio of the same variables,
away from the chiral limit. This longitudinal component, however, should not be confused
with the anomaly pole and {\em is not shifted or corrected} perturbatively in any way. It can couple, for instance, to a $t\bar t $ (top) quark current because of a broken WI and can be interpreted as a manifestation of the GS mechanism at the LHC, but can be easily overshadowed by SM contributions.
This point will be re-addressed more formally below in Eq.~(\ref{translong}).

At some special kinematical points (two massless gauge lines, or three massless
gauge lines of the same virtualities) where the anomalous vertex takes its DZ form,
the GS vertex is identically vanishing in the massless case. In the presence of a
heavy fermion the logarithmic correction shown in
Eq.~(\ref{a6}) reappears.

3) In the WZ case the anomaly pole is not cancelled. A second sector (the exchange of the axion) is
needed to restore the gauge invariance of the effective action.
In a hadronic collision the BIM amplitudes induce very small deviations
from the SM behavior after the convolution with the gluon density.
They are absent in DY at NNLO. In DP they affect the invariant mass distributions
- at large $Q$ - of the photon pair, for a given center of mass energy of the two
colliding protons. As such they are sensitive to large (Bjorken) $x$-values of the
gluon \textsc{Pdf}'s, region where the gluon density is rapidly decreasing.
In particular, in DP their contribution becomes more sizeable via intereference
with some box-like amplitudes ($gg \to \gamma \gamma$). In previous NLO study of this process
\cite{Coriano:1996us} they had been included even though they exceed the NLO accuracy, being truly NNLO
contributions. These amplitudes and vertices are the
basic building blocks of our numerical analysis and are responsible
for all the anomalous signal both in DY and in DP. We will quantify
their impact in the invariant mass distributions in both cases.
\section{ Invariant mass distributions in Drell-Yan }
Our NNLO analysis of the invariant mass distributions for lepton pair production,
for the computation of the QCD sectors, is based on the hard scatterings of Ref.~\cite{Hamberg:1990np}, and the NNLO evolution 
of the parton distributions
(\textsc{Pdf}'s) has been obtained with \textsc{Candia 1.0} \cite{Cafarella:2008du}.
The anomalous corrections to the invariant mass distributions have been evaluated
separately, since at NNLO they appear in DY in the
interference with the lowest order graph, and added to the standard QCD background.
It is important to recall that lepton pair production at low $Q$ via Drell-Yan is
sensitive to the \textsc{Pdf}'s at small-$x$ values, while in the high mass region
this process is essential in the search of
additional neutral currents. In our analysis we have selected a mass of 1 TeV for
the extra gauge boson and analyzed the signal and the background both on the peaks
of the $Z$ and of the of the new resonance.

At hadron level the colour-averaged inclusive differential cross section
for the reaction $H_1 +H_2 \rightarrow l_1 +l_2 +X $, is given by the expression \cite{Hamberg:1990np}
\ba
\frac{d\sigma}{dQ^2}=\tau \sigma_{\cal Z}(Q^2,M_{\cal Z}^2) W_{\cal Z}(\tau,Q^2)\hspace{1cm} \tau=\frac{Q^2}{S},
\label{factor}
\ea
where ${\cal Z}\equiv Z, Z^\prime$ is the point-like cross section and
all the information from the hadronic initial state is contained in the \textsc{Pdf}'s.
The hadronic structure function $W_{\cal Z}(\tau, Q^2)$ is given by a convolution product
between the parton luminosities $\Phi_{i j}(x, \mu_R^2, \mu_F^2)$ and the Wilson coefficients
$\Delta_{i j}(x,Q^2,\mu_R^2,\mu_F^2)$
\ba
W_{\cal Z}(\tau,Q^2, \mu_R^2,\mu_F^2)&=&\sum_{i, j}\int_{\tau}^{1}\frac{d x}{x}\Phi_{i j}(x,\mu_R^2,\mu_F^2)
\Delta_{i j}(\frac{\tau}{x},Q^2,\mu_F^2),
\ea
where the luminosities are given by
\ba
\Phi_{i j}(x,\mu_R^2,\mu_F^2)=\int_{x}^{1}\frac{d y}{y}f_{i}(y,\mu_R^2,\mu_F^2) f_{j}\left(\frac{x}{y},\mu_R^2,\mu_F^2\right)
\equiv  \left[f_{i}\otimes f_{j}\right](x,\mu_R^2,\mu_F^2)
\ea
and the Wilson coefficients (hard scatterings) depend on both the factorization ($\mu_F$) and renormalization scales $(\mu_R)$, formally expanded in the strong coupling $\alpha_s$ as
\ba
\Delta_{i j}(x,Q^2,\mu_F^2)=
\sum_{n=0}^{\infty}\alpha_s^n(\mu_R^2)\Delta^{(n)}_{i j}(x,Q^2,\mu_F^2,\mu_R^2).
\ea
We will vary $\mu_F$ and $\mu_R$ independently in order to determine the sensitivity of the prediciton on their variations and their optimal choice.

The anomalous corrections to the hard scatterings computed in the SM will be
discussed below. We just recall that the relevant point-like cross sections
appearing in the factorization formula (\ref{factor}) and which are part
of our analysis include, beside the Z and the $Z^{\prime}$ resonance, also
the contributions due to the photon and the $\gamma-Z,\, \gamma -Z^{\prime}$
interferences. For instance in the $Z^{\prime}$ case we have

\ba
&&\sigma_{\gamma}(Q^2)=\frac{4\pi\alpha_{em}^2}{3 Q^4}\frac{1}{N_c}
\nonumber\\
&&\sigma_{{Z^{\prime}}}(Q^2)=\frac{\pi\alpha_{em}}{4
M_{{Z^{\prime}}}\sin^2\theta_W \cos^2\theta_W N_c}
\frac{\Gamma_{{Z^{\prime}}\rightarrow \bar{l}
l}}{(Q^2-M_{Z^{\prime}}^2)^2 + M_{Z^{\prime}}^2 \Gamma_{Z^{\prime}}^2}
\nonumber\\
&&\sigma_{{Z^{\prime}},\gamma}(Q^2)=\frac{\pi\alpha_{em}^2}{6 N_c}
\frac{g_V^{Z^{\prime},l}g_V^{\gamma,l}}{\sin^2{\theta_W}\cos^2{\theta_W}}
\frac{(Q^2-M_{Z^{\prime}}^2)}{Q^2(Q^2-M_{Z^{\prime}}^2)^2+
M_{Z^{\prime}}^2\Gamma_{Z^{\prime}}^2},\nonumber\\
&&\sigma_{{Z^{\prime}},Z}(Q^2)=\frac{\pi\alpha_{em}^2}{96}
\frac{\left[g_V^{Z^{\prime},l}g_V^{Z,l}+
g_A^{Z^{\prime},l}g_A^{Z,l}\right]}{\sin^4{\theta_W}\cos^4{\theta_W}N_c}
\frac{(Q^2-M_Z^2)(Q^2-M^2_{Z^{\prime}})
+M_Z\Gamma_{Z}M_{Z^{\prime}}\Gamma_{Z^{\prime}}}
{\left[(Q^2-M_{Z^{\prime}}^2)^2 + M_{Z^{\prime}}^2\Gamma_{Z^{\prime}}^2\right]
\left[(Q^2-M_Z^2)^2 + M_Z^2\Gamma_{Z}^2\right]}.\nonumber\\
\ea
where $N_C$ is the number of colours, $\Gamma_{{Z^{\prime}}\rightarrow \bar{l}l}$
is the partial decay width of the gauge boson and the total hadronic widths are defined by
\ba
\Gamma_Z\equiv\Gamma(Z\rightarrow hadrons)=\sum_{i}\Gamma(Z\rightarrow \psi_i\bar{\psi_i}),
\nonumber\\
\Gamma_{Z^\prime}\equiv\Gamma(Z^{\prime}\rightarrow
hadrons)=\sum_{i}\Gamma(Z^{\prime}\rightarrow \psi_i\bar{\psi_i}),
\ea
where we refer to hadrons not containing bottom and top quarks (i.e.
$i=u,d,c,s$). We also ignore electroweak corrections of higher-order and we have included the top-quark mass and QCD corrections. We have included only tree level decays into SM fermions, with a total decay rate for the $Z$ and $Z^\prime$ which is given by
\ba
&&\Gamma_Z=\sum_{i=u,d,c,s}\Gamma(Z\rightarrow
\psi_i\bar{\psi_i})+\Gamma(Z\rightarrow b\bar{b})
+3\Gamma(Z\rightarrow l\bar{l})+3\Gamma(Z\rightarrow \nu_l\bar{\nu_l}),\\
&&\Gamma_{Z^{\prime}}=\sum_{i=u,d,c,s}\Gamma(Z^{\prime}\rightarrow
\psi_i\bar{\psi_i})+\Gamma(Z^{\prime}\rightarrow b\bar{b})
+\Gamma(Z^{\prime}\rightarrow t\bar{t})+3\Gamma(Z^{\prime}\rightarrow l\bar{l})
+3\Gamma(Z^{\prime}\rightarrow \nu_l\bar{\nu_l}).
\ea
\begin{figure}[t]
\begin{center}
\includegraphics[scale=0.8]{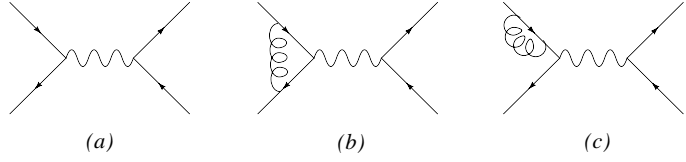}
\caption{\small $q\bar q \rightarrow Z, Z^{\prime}$  at LO and NLO (virtual corrections).}
\label{DYLO}
\end{center}
\end{figure}
\begin{figure}[t]
\begin{center}
\includegraphics[scale=0.75]{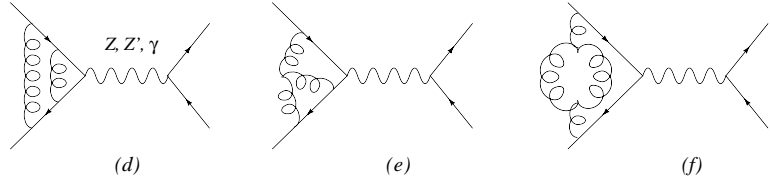}
\caption{\small $q\bar q \rightarrow Z, Z^{\prime}$  at NNLO (virtual corrections).}
\label{DYNLOv}
\end{center}
\end{figure}
Coming to illustrate the contributions included in our analysis, these are shown in some representative
graphs. The complete NNLO expressions of the hard scatterings and the corresponding Feynman diagrams can be found in \cite{Hamberg:1990np}.

\begin{itemize}
\item{\bf SM QCD contributions}
\end{itemize}

\begin{figure}[t]
\begin{center}
\includegraphics[scale=0.8]{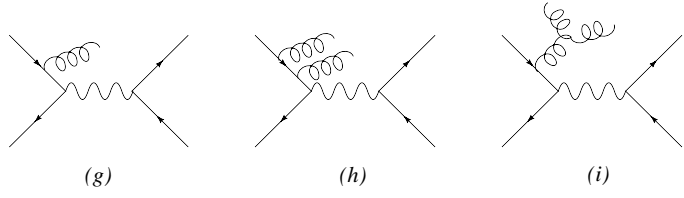}
\caption{\small $q\bar q \rightarrow Z, Z^{\prime}$  with real corrections at NLO $(g)$ and at NNLO $(h)$, $(i)$. }
\label{DYNLOr}
\end{center}
\end{figure}

We show in Fig.~\ref{DYLO} the leading $O(\alpha_w)$ and some typical
next-to-leading order $O(\alpha_w\alpha_s)$ (LO, NLO) contributions to the process
in the annihilation channel (virtual corrections). Examples of higher-order
virtual corrections included in the hard scatterings are shown in Fig.~\ref{DYNLOv},
which are of $O(\alpha_s^2 \alpha_w)$, while the corresponding real emissions,
integrated over the final state gluons, are shown in Fig.~\ref{DYNLOr} at NLO (graph g)) and
NNLO (graphs h) and i)).
\begin{figure}[t]
\begin{center}
\includegraphics[scale=0.8]{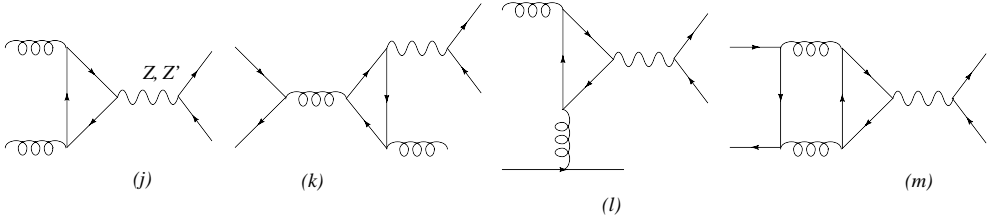}
\caption{\small Anomalous contributions for  $Z^{\prime}$
production in the $gg$, $q \bar q $ and $q g$ sectors at lower orders. }
\label{DYanom}
\end{center}
\end{figure}
\begin{itemize}
\item{\bf Anomalous  corrections}
\end{itemize}
We show in Fig.~\ref{DYanom} the leading anomalous corrections to leptoproduction.
At $O(\alpha_s\alpha_w)$ there is a first contribution coming from the interference
between graph j) and the leading order
$q\bar{q}$ annihilation vertex (graph a) of Fig.~\ref{DYLO}. The square of the same
graph appears in the anomalous corrections at $O(\alpha_s^2 \alpha_w)$.
Other contributions that we have included are those due to the exchange
of a physical axion and Goldstone modes, which can be removed in the
unitary gauge. Of higher order are the contributions
shown in diagram k), l) and n), which contribute via their interference with NLO tree
level graphs. For instance k) interferes with diagram g) of Fig.~\ref{DYNLOr}, while m)
interferes with the LO annihilation graph. The analogous contributions in the WZ and GS cases are
obtained by replacing the triangle graph with the GS vertex,
as in Fig.~\ref{GS_AVV}, or, for the WZ case, with
Fig.~\ref{onefig}.
Notice that in Fig.~\ref{onefig}, in the WZ case the anomaly pole is
automatically cancelled by the WI on the lepton pair of the final
state, if the two leptons are taken to be massless at high energy, as is the case.
Then, the only new contributions from the anomaly vertex that survive are those
related to the transverse component of this vertex.  This is an example, as we
have discussed in \cite{Armillis:2008bg}, of a "harmless" anomaly vertex.
A similar situation occurs whenever there is no coupling of the longitudinal
component of the anomaly to the (transverse) external leptonic current.
This property continues to hold also away from the chiral limit, since the
corrections due to the fermion mass in the anomaly have the typical structure
\beq
\Delta_{\mu\nu\rho}(q,k)_{\mbox{\rm\tiny anomaly}}=
\sum_f g_{A,f}^{Z'}e^2 Q_f^2 a_n\frac{(q-k)_{\nu}}{(q-k)^2}\left( \frac{1}{2} - 2 m_f^2 C_0\right)
\epsilon_{\mu\rho\alpha\beta}q^{\alpha}k^{\beta} + \tilde{\Delta}^{trans}\,,
\label{translong}
\eeq
where $\tilde{\Delta}^{trans}$ is the truly transversal component
away from the chiral limit. The most general expression of the coefficient $C_0$ is given
in Eq.~(A.8) of ref.~\cite{Kniehl:1989qu}. $C_0$ is the scalar three-point function with a fermion of
mass $m_f$ circulating in the loop.
 In both mechanisms anomaly (strictly massless) effects are comparable
 with the corresponding contributions
coming from the SM for massive fermions.  It should be clear by now that
in the WZ case the anomaly pole is not cancelled, rather an additional
exchange is necessary to re-establish the gauge independence of the $S$-matrix (the axion).
In DY
this sector does not play a significant role due to the small mass of the lepton pair.
As we have discussed above, the cancellation of the anomaly is due, in this case,
to the WI of the leptonic current and there is no axion exchanged in the $s$-channel.
\subsection{Precision studies on the $Z$ resonance}
\begin{figure}[t]
\begin{center}
\includegraphics[scale=0.8]{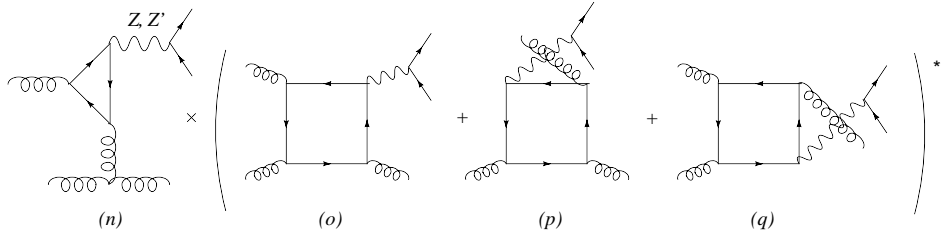}
\caption{\small Anomalous contributions  for the $gg \rightarrow gg$ process mediated by an anomalous $Z^\prime$
at higher perturbative orders. }
\label{DYgg}
\end{center}
\end{figure}
The quantification of the corrections due to anomalous Abelian gauge structures in
DY requires very high precision, being these of a rather high order. For this reason
we have to identify all the sources of indeterminations in QCD which come from the
factorization/renormalization scale dependence of the cross section, keeping into
account the dependence on $\mu_F$  and $\mu_R$ {\em both} in the DGLAP evolution
{\em and} in the hard scatterings. The set-up of our analysis is similar to that
used for a study of the NNLO DGLAP evolution in Refs.~\cite{Cafarella:2007tj,Cafarella:2005zj}, where the study has covered
every source of theoretical error, including the one related to the various
possible resummations of the DGLAP solution, which is about $2-3\%$ in DY and
would be sufficient to swamp away any measurable deviation due to new physics at the LHC.

These previous studies have been focused on the DY distributions on the resonance
peaks, in particular on the peak of the $Z$, where the accuracy at the LHC is of
outmost importance for QCD partonometry. The presence
  of anomalous corrections on the $Z$ peak is due both to the anomalous components of the
$Z$ in the anomalous models and to the interference between the $Z^\prime$ and the
$Z$, that we have taken into account. Notice that in DY the treatment of the anomalous
corrections to the $Z$
is drastically simplified if we neglect the (small) mass of the lepton pair, as usual.
In fact, these are due to trilinear (anomaly) vertices which involve the $BBB$,
$BYY $, $B W_3 W_3$ and $B G G$ gauge fields - in the interaction basis - all
of them involving  interactions of the St\"uckelberg field with the corresponding
field-strengths of the gauge fields, such as $b F_B\wedge F_B$,
$b F_Y \wedge F_Y$, $b F_W\wedge F_W$ and $b F_G\wedge F_G$, where $G$ denotes
the gluon field. The only contribution that is relevant for the LHC is then one
obtained by projecting the $b F_G\wedge F_G$ vertex on the physical axion
$\chi$, whose mass is, in principle, a free parameter of the anomalous models.
It is then clear that the axion channel plays a more important role in the
production of the $top$, due to its large mass, than in leptoproduction.
We will now briefly summarize the results for the new contributions in DY,
starting from the non-anomalous ones.

In the $q\bar{q}$ sector we have two contributions involving triangle fermion loops
see Fig.~\ref{DYanom} k,m.
The one depicted in Fig.~\ref{DYanom}m is a two-loop virtual correction with a
$Z$ or a $Z^{\prime}$ boson in the final state, while in
Fig.~\ref{DYanom}k we have a real emission of a gluon in the final state which is integrated out.
The first contribution has been calculated in \cite{Bernreuther:2005rw, Larin:1993tq, Gonsalves:1991qn, Rijken:1995gi},
\ba
&&\Delta^{V}_{q\bar{q}}(x,Q^2,\mu_F^2,m^2) =\delta(1-x)a^{Z'}_q a^{Z'}_Q C_F T_f\frac{1}{2}
\left(\frac{\alpha_s}{\pi}\right)^2\times\nonumber\\
&&\hspace{2cm}\left[\theta(Q^2-4 m^2) G_1(m^2/Q^2)
+\theta(4 m^2-Q^2)G_2(m^2/Q^2)\right)
\ea
where $C_F$ and $ T_f$ are the color factors, $q=u,d,c,s$, $Q=t,b$ and $m$ the mass of the
heavy flavors, while in the massless limit the functions $G_1$ and $G_2$ are given by
\ba
&&G_1(m=0)=3\log\left(\frac{Q^2}{\mu_R^2}\right)-9 +2 \zeta(2)
\nonumber\\
&&G_2(m=0)=0
\ea
and $Q$ represents the invariant mass of the system.
\begin{figure}[h]
\begin{center}
\includegraphics[scale=0.8]{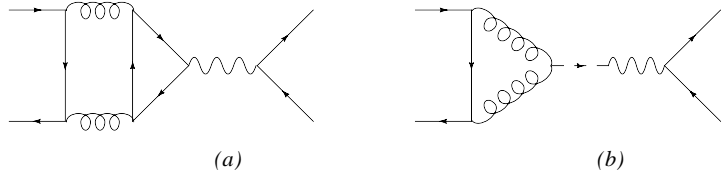}
\caption{\small GS mechanism: anomalous contribution
and counterterm for the $q\bar{q}$ scattering sector.}
\label{qq_AVV_WZ_ff}
\end{center}
\end{figure}
The contribution of Fig.~\ref{DYanom}k in the massless limit is given by
\ba
\Delta^{R}_{q\bar{q}}(x,Q^2,\mu_F^2,m=0)=a^{Z'}_q a^{Z'}_Q C_F T_f\frac{1}{2}
\left(\frac{\alpha_s}{\pi}\right)^2\times
\left\{\frac{(1+x)}{(1-x)_{+}}\left[-2 + 2 x (1-\log(x))\right]\right\},
\ea
while in the $qg$ sector we have the contribution shown in Fig.~\ref{DYanom}l
which is given by
\ba
&&\Delta_{qg}(x,Q^2,\mu_F^2,m^2)     \nonumber\\
&&=a^{Z'}_q a^{Z'}_Q T_f^2\frac{1}{2}
\left(\frac{\alpha_s}{\pi}\right)^2\times
\left[\theta(Q^2-4 m^2) H_1(x,Q^2,m^2)+\theta(4 m^2-Q^2)H_2(x,Q^2,m^2)\right]
\ea
with the massless limit of  $H_1(x,Q^2,m^2)$ given by
\ba
&&H_1(x,Q^2,m=0)    \nonumber\\
&&=2 x \left[\log\left(\frac{1}{x}\right)\log\left(\frac{1}{x}-1\right)
+Li_2\left(1-\frac{1}{x}\right)\right] + 2 \left(1-\frac{1}{x}\right)
\left[1-2 x \log\left(\frac{1}{x}\right)\right].
\ea
Separating the anomaly-free from the anomalous contributions,
the factorization formula for the invariant mass distribution in DY is given by
\ba
&&\frac{d\sigma}{dQ^2}=\tau \sigma_{\cal Z}(Q^2,M_{\cal Z}^2)\left\{
W_{\cal Z}(\tau,Q^2) + W_{\cal Z}^{anom}(\tau,Q^2)\right\}
\\
&&W_{\cal Z}^{anom}(\tau,Q^2)=\sum_{i, j}\int_{\tau}^{1}\frac{d x}{x}\Phi_{i j}(x,\mu_R^2,\mu_F^2)
\Delta_{i j}^{anom}(\frac{\tau}{x},Q^2,\mu_F^2)
\\
&&\Delta_{i j}^{anom}(x,Q^2,\mu_F^2)  \nonumber\\
&&=\Delta^{V}_{q\bar{q}}(x,Q^2,\mu_F^2,m=0)
+\Delta^{R}_{q\bar{q}}(x,Q^2,\mu_F^2,m=0)+\Delta_{qg}(x,Q^2,\mu_F^2,m=0)
\ea
that we will be using in our numerical analysis below.
\subsection{Di-lepton production: numerical results}
\begin{figure}[t]
\subfigure[SM vs mLSOM at NLO]{\includegraphics[%
 width=5.6cm,
 angle=-90]{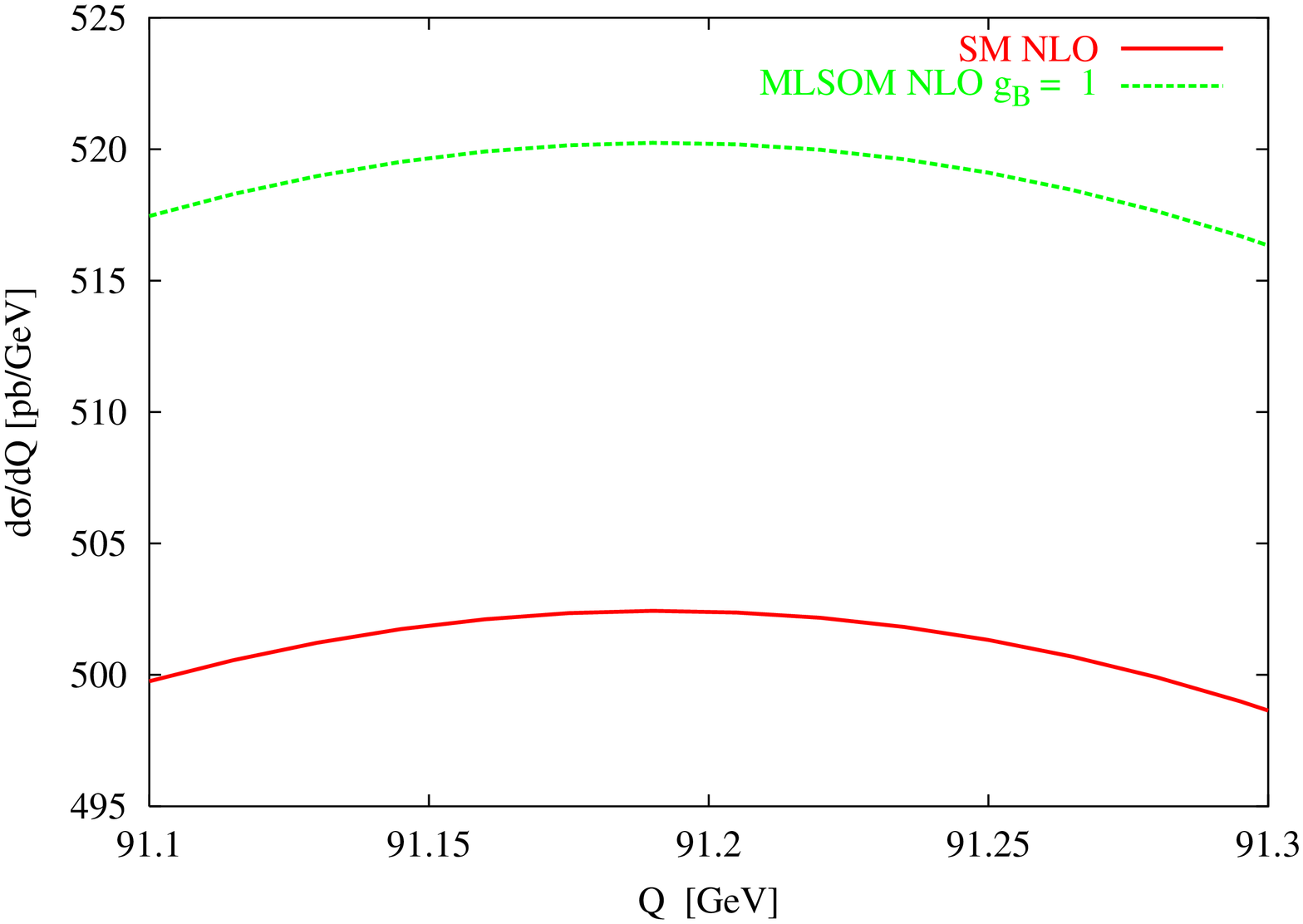}}
\subfigure[SM vs Anomaly free models at NLO]{\includegraphics[%
 width=5.6cm,
 angle=-90]{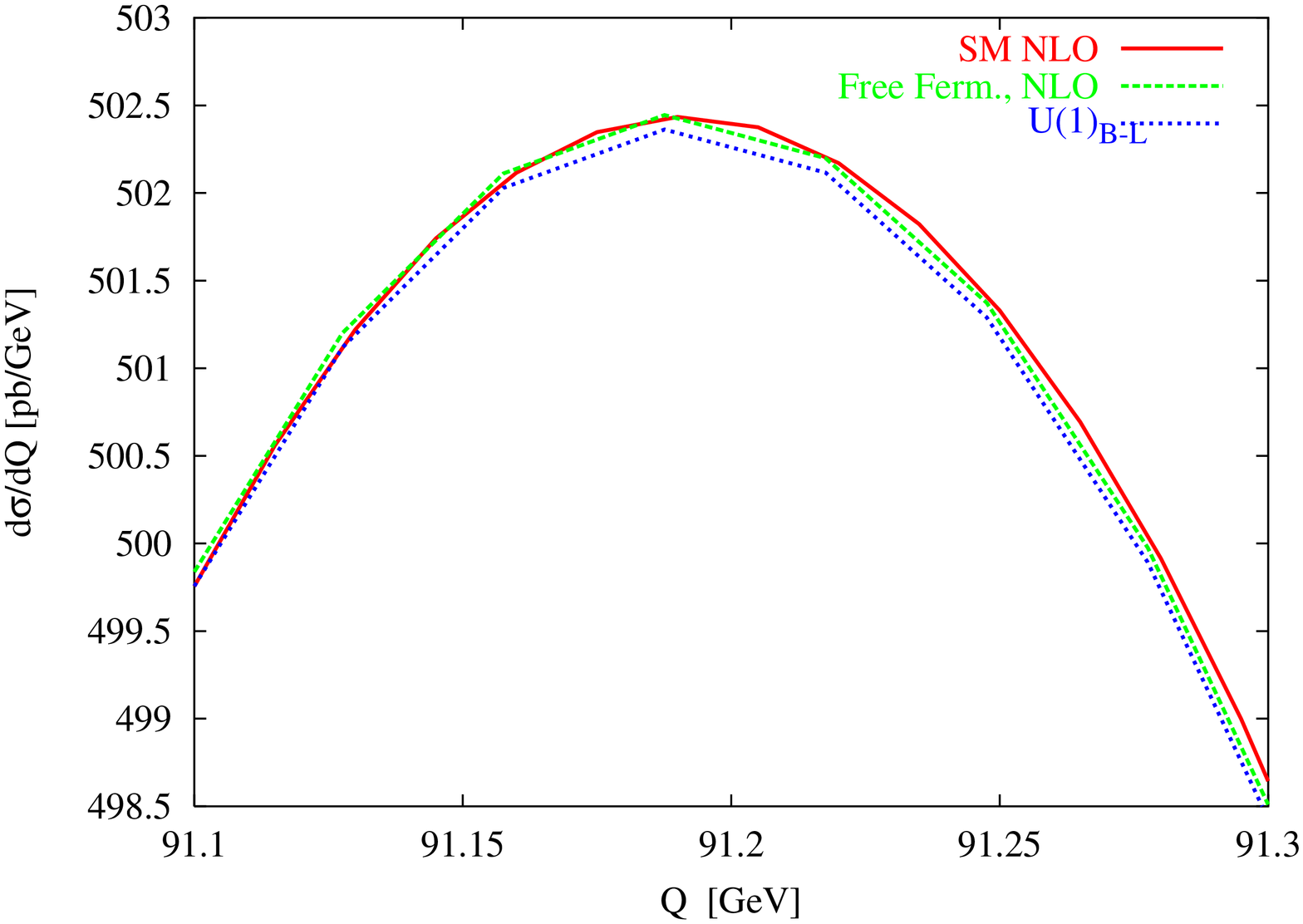}}
\subfigure[SM vs mLSOM at NNLO]{\includegraphics[%
 width=5.6cm,
 angle=-90]{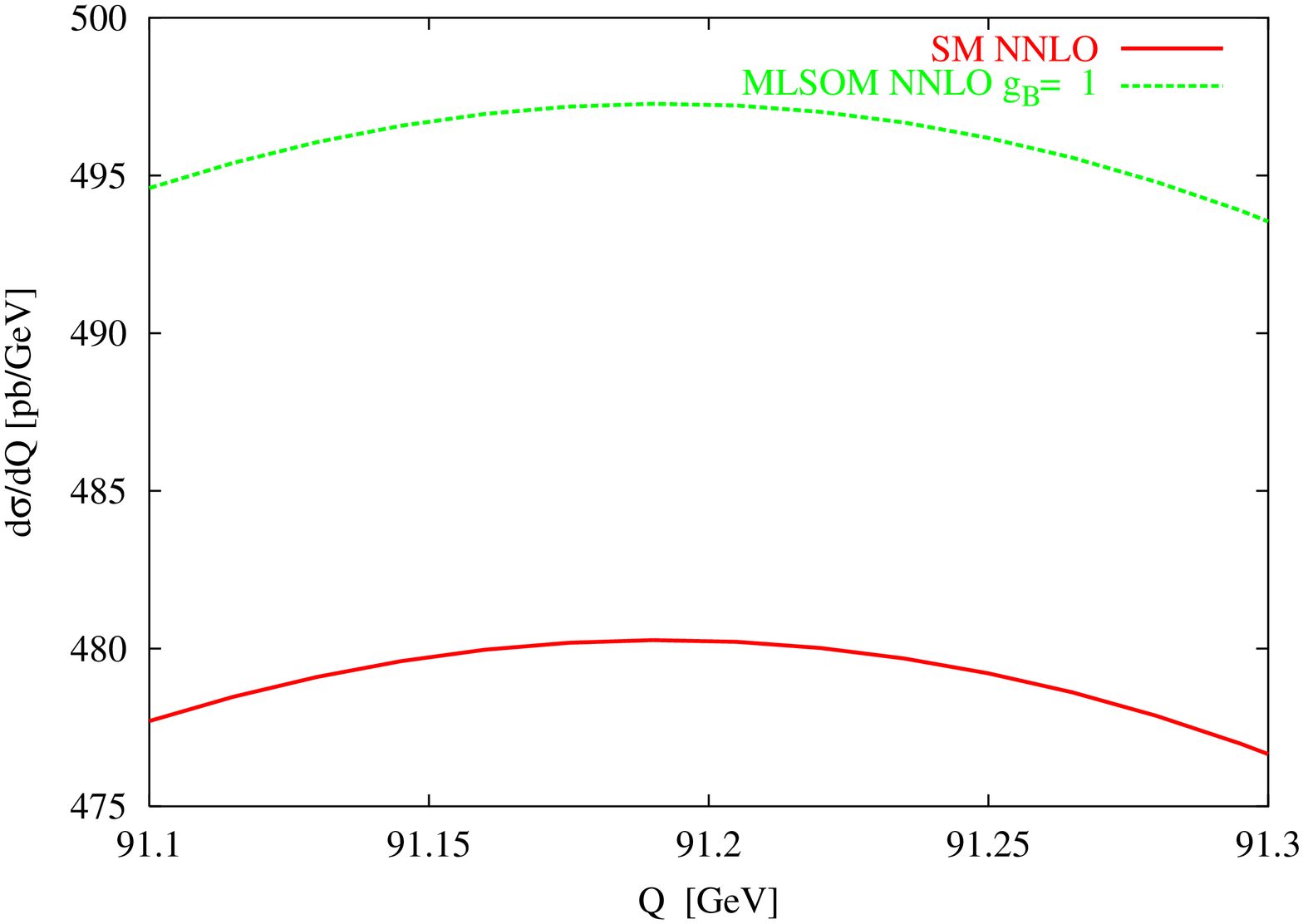}}
\subfigure[SM vs Anomaly free models at NNLO]{\includegraphics[%
 width=5.6cm,
 angle=-90]{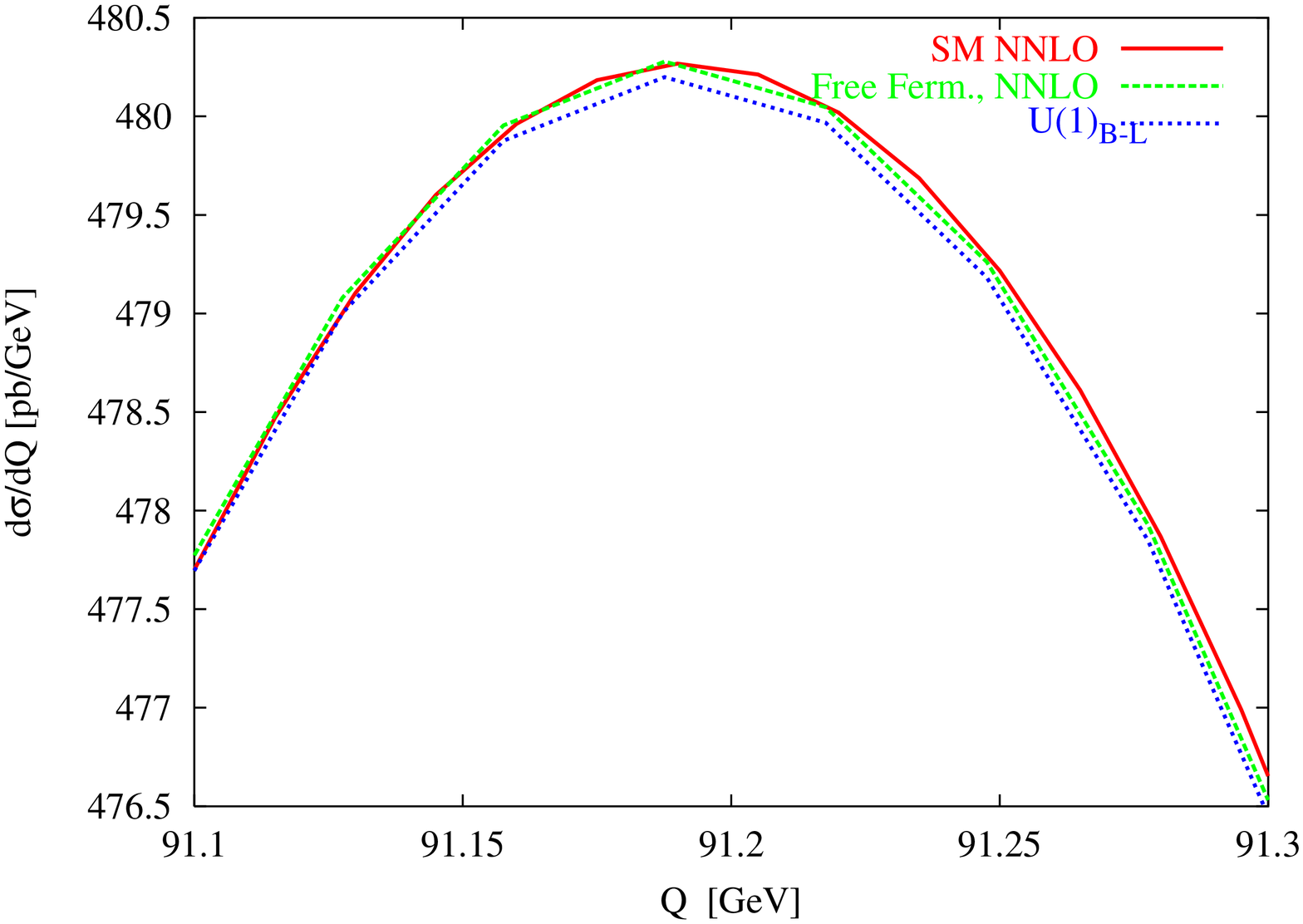}}
\caption{\small Zoom on the $Z$ resonance for anomalous Drell-Yan in
the $\mu_F=\mu_R=Q$ at NLO/NNLO for all the models.}
\label{zpeaknnloall}
\end{figure}
We have used the MRST-2001 set of \textsc{Pdf}'s given in Refs.~\cite{Martin:2001es} and \cite{Martin:2002dr}. We start by showing in
Fig.~\ref{zpeaknnloall} various zooms of the differential cross
section on the peak of the $Z$ - for all the models -  both at NLO and at NNLO.
We have kept the factorization and renormalization scales coincident and
equal to $Q$, while the mass of the extra $Z^\prime$ has been chosen around 1 TeV.
The anomaly-free models, from the SM to the three Abelian extensions that we
have considered (Free Fermionic \cite{Coriano:2008wf} and $U(1)_{B-L}$ \cite{Carena:2004xs}
in Fig.~\ref{zpeaknnloall}, while $U(1)_{q+u}$ appears in Tab.~\ref{tablethird} )
show that the cross section is more enhanced for the mLSOM, illustrated in
Fig.~\ref{zpeaknnloall}a, c. The plots show a sizeable difference
(at a 3.5  $\%$ level) between
the anomalous and all the remaining anomaly-free models.
A comparison between (a) and (c) indicates, however, that this difference
has to be attributed to the specific
charge assignment of the anomalous model and not to the anomalous partonic sector,
which is present in (c) but not in (a). The anomalous corrections in DY appear at
NNLO and not at NLO, while in both figures the difference between the SM and the
mLSOM remains almost unchanged.

\begin{figure}
\subfigure[      ]{\includegraphics[%
 width=5.6cm,
 angle=-90]{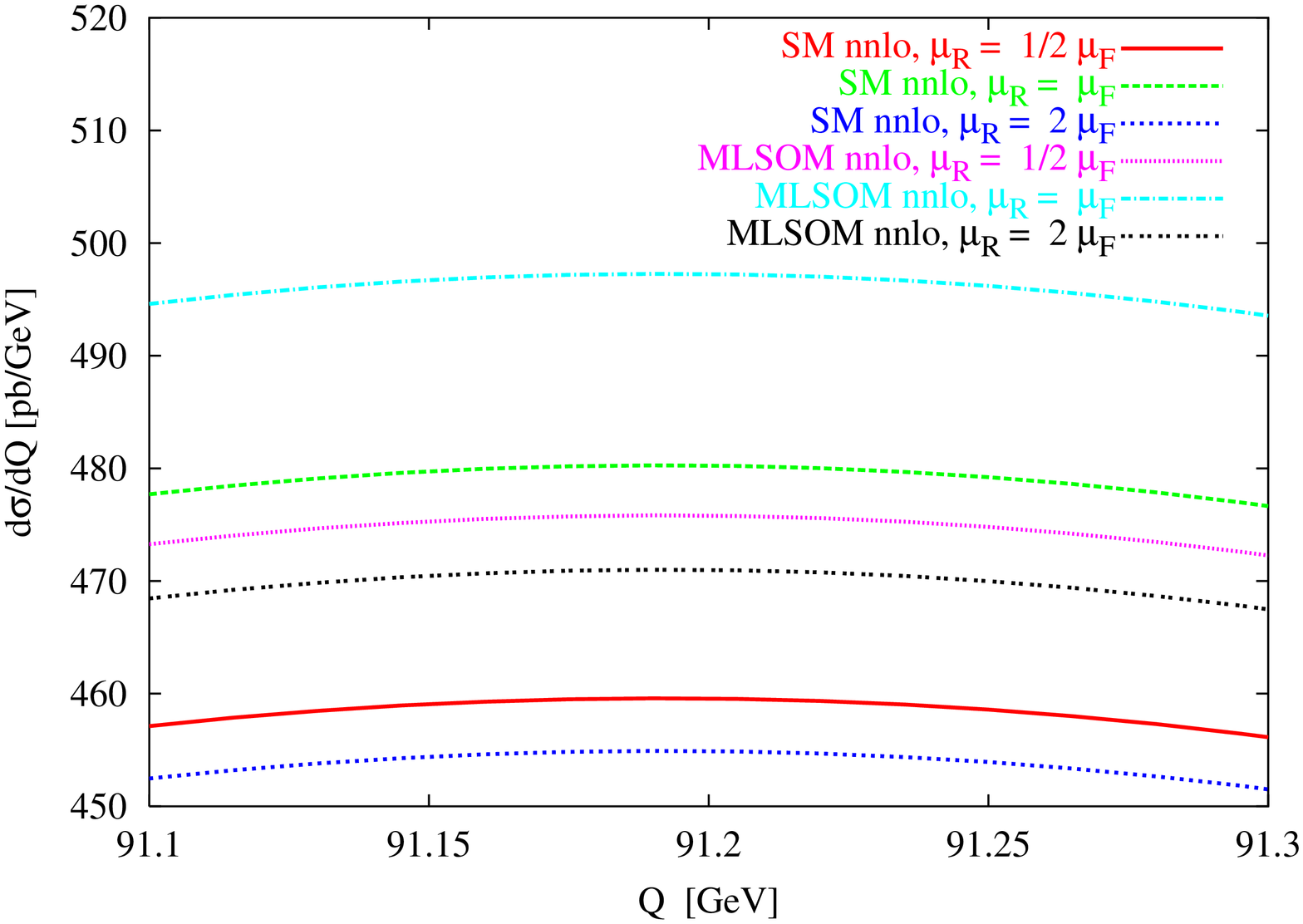}}
\subfigure[        ]{\includegraphics[%
 width=5.6cm,
 angle=-90]{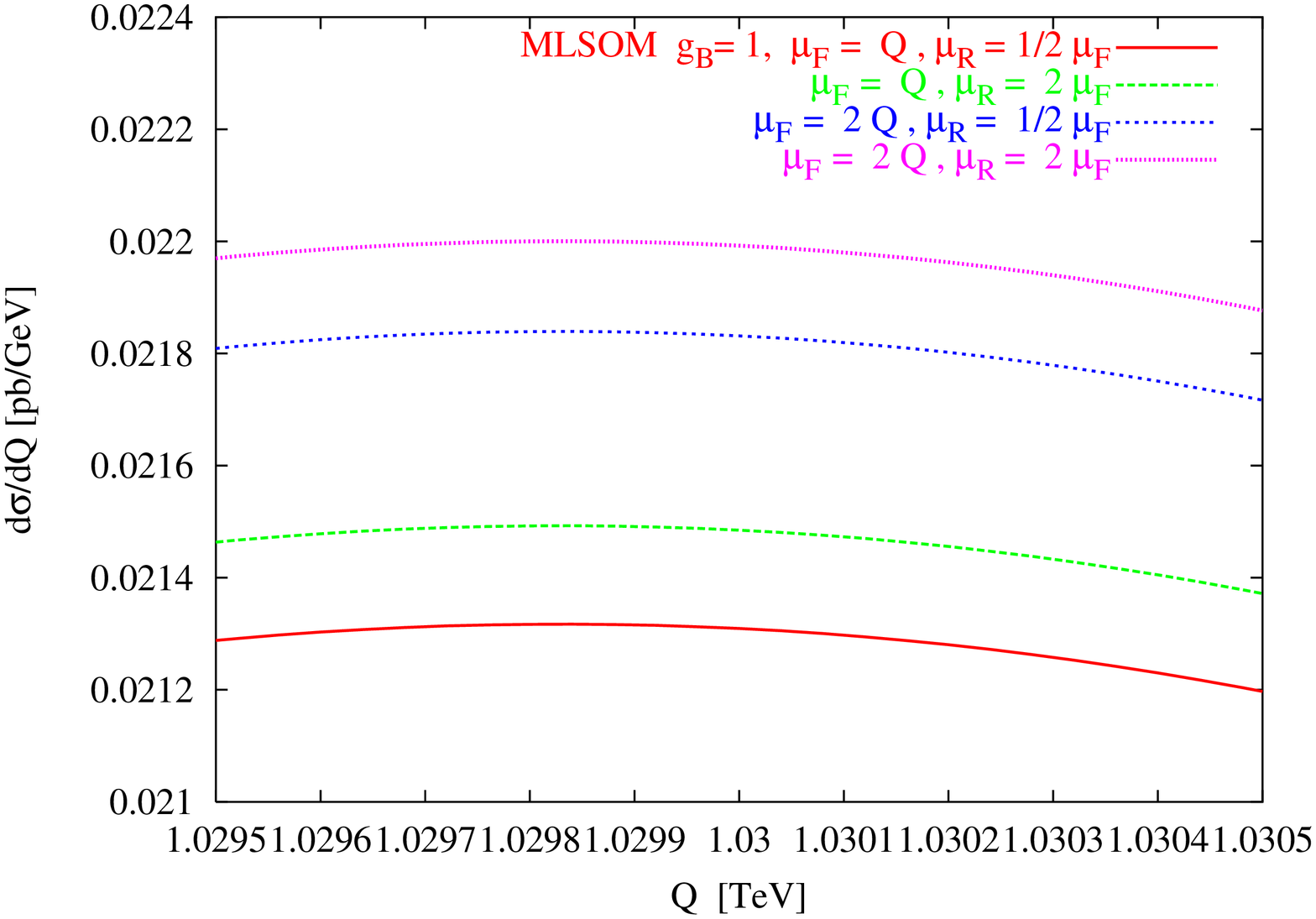}}
\caption{\small Zoom on the $Z$ resonance for anomalous Drell-Yan for varying
factorization and renormalization scales at NNLO for the SM and the mLSOM.
Results are shown for $Q\sim 91$ GeV (a) and 1 TeV  (b) both for $\sqrt{S}= 14$ TeV.}
\label{factren}
\end{figure}

Moving from NLO to NNLO the cross section is reduced. Defining the $K$-factor
\beq
\frac{\sigma_{NNLO} - \sigma_{NLO}}{\sigma_{NLO}}\equiv K_{NLO}
\label{var}
\eeq
in the case of the mLSOM this factor indicates a reduction of about
$4 \%$ on the peak and can be attributed to the NNLO terms in the DGLAP
evolution, rather than to the NNLO corrections to the hard scatterings.
This point can be explored numerically by the (order) variation \cite{Cafarella:2007wd, Cafarella:2007tj}
\ba
\Delta\sigma &\sim& \Delta \hat{\sigma}\otimes \phi + \hat{\sigma}\otimes \Delta \phi \nonumber \\
\Delta\sigma &\equiv& |\sigma_{NNLO} - \sigma_{NLO}|
\ea
which measures the ``error'' change in the hadronic cross section $\sigma$
going from NLO to NNLO ($\Delta \sigma$) in terms of the analogous changes
in the hard scatterings ($\Delta \hat{\sigma}$) and parton luminosities $\Delta{\phi}$).
The dominance of the first or the second term on the rhs of Eq.~(\ref{var}) is an indication
of the dominance of the hard scatterings or of the evolution in moving from lower to higher order.
The same differences emerge also from Tabs.~\ref{table_nlo}, \ref{table_nnlo}. Differences in the resonance region of this size can be considered
marginally relevant for the identification of anomalous components in this observables.
In fact, in \cite{Cafarella:2007tj} a high precision study of this distributions on the
same peak (in the SM case) shows that the total theoretical error is reasonably below
the 4 $\%$ level and can decrease at 1.5-2 $\% $ level when enough statistics will allow
to reduce the experimental errors on the \textsc{Pdf}'s. It is then obvious that the
isolation/identification of a specific model
- whether anomalous or not - appears to be rather difficult from the measurement of
a single observable even with very high statistics, such as the $Z$ resonance.
\begin{table}
\begin{center}
\begin{footnotesize}
\begin{tabular}{|c||c|c|c|c|c|}
\hline
\multicolumn{6}{|c|}{$d\sigma^{nlo}/dQ$ [pb/GeV] for the mLSOM with $M_{1}=1$ TeV, $\tan\beta=40$, \textsc{Candia 1.0} evol.}
\tabularnewline
\hline
$Q ~[\textrm{GeV}]$      &
$g_B=0.1$                 &
$g_B=0.36$                 &
$g_B=0.65$                 &
$g_B=1$                   &
$\sigma_{nlo}^{SM}(Q)$\tabularnewline
\hline
\hline
$90.50$&
$3.8551\cdot10^{+2}$&
$3.8711\cdot10^{+2}$&
$3.9106\cdot10^{+2}$&
$3.9902\cdot10^{+2}$&
$3.8543\cdot10^{+2}$
\tabularnewline
\hline
$90.54$&
$3.9712\cdot10^{+2}$&
$3.9877\cdot10^{+2}$&
$4.0284\cdot10^{+2}$&
$4.1105\cdot10^{+2}$&
$3.9704\cdot10^{+2}$
\tabularnewline
\hline
$90.59$&
$4.0861\cdot10^{+2}$&
$4.1030\cdot10^{+2}$&
$4.1449\cdot10^{+2}$&
$4.2294\cdot10^{+2}$&
$4.0852\cdot10^{+2}$
\tabularnewline
\hline
$90.63$&
$4.1988\cdot10^{+2}$&
$4.2162\cdot10^{+2}$&
$4.2592\cdot10^{+2}$&
$4.3461\cdot10^{+2}$&
$4.1979\cdot10^{+2}$
\tabularnewline
\hline
$90.68$&
$4.3084\cdot10^{+2}$&
$4.3263\cdot10^{+2}$&
$4.3705\cdot10^{+2}$&
$4.4596\cdot10^{+2}$&
$4.3075\cdot10^{+2}$
\tabularnewline
\hline
$90.99$&
$4.9041\cdot10^{+2}$&
$4.9245\cdot10^{+2}$&
$4.9749\cdot10^{+2}$&
$5.0766\cdot10^{+2}$&
$4.9031\cdot10^{+2}$
\tabularnewline
\hline
$91.187$&
$5.0254\cdot10^{+2}$&
$5.0463\cdot10^{+2}$&
$5.0981\cdot10^{+2}$&
$5.2024\cdot10^{+2}$&
$5.0243\cdot10^{+2}$
\tabularnewline
\hline
$91.25$&
$5.0143\cdot10^{+2}$&
$5.0352\cdot10^{+2}$&
$5.0869\cdot10^{+2}$&
$5.1911\cdot10^{+2}$&
$5.0133\cdot10^{+2}$
\tabularnewline
\hline
$91.56$&
$4.6103\cdot10^{+2}$&
$4.6296\cdot10^{+2}$&
$4.6772\cdot10^{+2}$&
$4.7732\cdot10^{+2}$&
$4.6094\cdot10^{+2}$
\tabularnewline
\hline
$91.77$&
$4.1178\cdot10^{+2}$&
$4.1350\cdot10^{+2}$&
$4.1776\cdot10^{+2}$&
$4.2635\cdot10^{+2}$&
$4.1170\cdot10^{+2}$
\tabularnewline
\hline
$92.0$&
$3.5297\cdot10^{+2}$&
$3.5444\cdot10^{+2}$&
$3.5810\cdot10^{+2}$&
$3.6547\cdot10^{+2}$&
$3.5289\cdot10^{+2}$
\tabularnewline
\hline
\end{tabular}
\end{footnotesize}
\end{center}
\caption{\small Invariant mass distributions at NLO for the mLSOM and the SM around the peak of the $Z$.
The mass of the anomalous extra $Z^\prime$ is taken to be 1 TeV with $\mu_F=\mu_R= Q$.}
\label{table_nlo}
\end{table}
\begin{table}
\begin{center}
\begin{footnotesize}
\begin{tabular}{|c||c|c|c|c|c|}
\hline
\multicolumn{6}{|c|}{$d\sigma^{nnlo}/dQ$ [pb/GeV] for the mLSOM with $M_{1}=1$ TeV, $\tan\beta=40$, \textsc{Candia 1.0} evol.}
\tabularnewline
\hline
$Q ~[\textrm{GeV}]$      &
$g_B=0.1$                 &
$g_B=0.36$                 &
$g_B=0.65$                 &
$g_B=1$                   &
$\sigma_{nnlo}^{SM}(Q)$\tabularnewline
\hline
\hline
$90.50$&
$3.6845\cdot10^{+2}$&
$3.6997\cdot10^{+2}$&
$3.7374\cdot10^{+2}$&
$3.8132\cdot10^{+2}$&
$3.6835\cdot10^{+2}$
\tabularnewline
\hline
$90.54$&
$3.7956\cdot10^{+2}$&
$3.8112\cdot10^{+2}$&
$3.8500\cdot10^{+2}$&
$3.9282\cdot10^{+2}$&
$3.7945\cdot10^{+2}$
\tabularnewline
\hline
$90.59$&
$3.9054\cdot10^{+2}$&
$3.9215\cdot10^{+2}$&
$3.9615\cdot10^{+2}$&
$4.0419\cdot10^{+2}$&
$3.9043\cdot10^{+2}$
\tabularnewline
\hline
$90.63$&
$4.0132\cdot10^{+2}$&
$4.0298\cdot10^{+2}$&
$4.0708\cdot10^{+2}$&
$4.1535\cdot10^{+2}$&
$4.0121\cdot10^{+2}$
\tabularnewline
\hline
$90.68$&
$4.1180\cdot10^{+2}$&
$4.1351\cdot10^{+2}$&
$4.1772\cdot10^{+2}$&
$4.2621\cdot10^{+2}$&
$4.1169\cdot10^{+2}$
\tabularnewline
\hline
$90.99$&
$4.6879\cdot10^{+2}$&
$4.7073\cdot10^{+2}$&
$4.7554\cdot10^{+2}$&
$4.8523\cdot10^{+2}$&
$4.6866\cdot10^{+2}$
\tabularnewline
\hline
$91.187$&
$4.8040\cdot10^{+2}$&
$4.8239\cdot10^{+2}$&
$4.8733\cdot10^{+2}$&
$4.9727\cdot10^{+2}$&
$4.8027\cdot10^{+2}$
\tabularnewline
\hline
$91.25$&
$4.7935\cdot10^{+2}$&
$4.8134\cdot10^{+2}$&
$4.8627\cdot10^{+2}$&
$4.9619\cdot10^{+2}$&
$4.7922\cdot10^{+2}$
\tabularnewline
\hline
$91.56$&
$4.4076\cdot10^{+2}$&
$4.4259\cdot10^{+2}$&
$4.4713\cdot10^{+2}$&
$4.5628\cdot10^{+2}$&
$4.4064\cdot10^{+2}$
\tabularnewline
\hline
$91.77$&
$3.9371\cdot10^{+2}$&
$3.9535\cdot10^{+2}$&
$3.9941\cdot10^{+2}$&
$4.0759\cdot10^{+2}$&
$3.9360\cdot10^{+2}$
\tabularnewline
\hline
$92.0$&
$3.3750\cdot10^{+2}$&
$3.3891\cdot10^{+2}$&
$3.4239\cdot10^{+2}$&
$3.4942\cdot10^{+2}$&
$3.3741\cdot10^{+2}$
\tabularnewline
\hline
\end{tabular}
\end{footnotesize}
\end{center}
\caption{\small Invariant mass distributions at NNLO for the mLSOM and the SM around the peak of the $Z$.
The mass of the anomalous extra $Z^{\prime}$ is taken to be 1 TeV with $\mu_F=\mu_R = Q$.}
\label{table_nnlo}
\end{table}
The evolution of the \textsc{Pdf}'s has been performed with \textsc{Candia 1.0}
\cite{Cafarella:2008du} which allows independent variations of  $\mu_F$ and
$\mu_R$ in the initial state. This analysis is shown in
Fig.~\ref{factren}, where we vary $\mu_F$ up to $2 Q$, while we have taken
$1/2 \mu_F \leq \mu_R \leq 2 \mu_F$. We observe that by increasing
both scales there is an enhancement
in the result and this is due to the logarithms $\ln{\mu_R^2/\mu_F^2}$ and
$\ln{Q^2/\mu_F^2}$, contained in the hard scatterings.
The scale variations induce changes of about $4\%$ in the SM
case at NNLO and about $3.5 \%$ in the mLSOM on the peak of the
$Z$. Notice that the variations are not symmetric as we vary the
scales and the percentual changes refer to the maximum variability.
This typical scale dependence is universal for all the studies presented
so far on the peak of the $Z$ and is a limitation of the parton model prediction.
After a large data taking, optimal choices for the \textsc{Pdf}'s and for $\mu_R$
and $\mu_F$ will allow a considerable reduction of this indetermination.
In Fig.~\ref{factren}b we repeat the same analysis, for the same c.m. energy,
this time for $Q\sim 1$ TeV, on the $Z^\prime$ resonance in the mLSOM, for a sizeable
coupling of the anomalous gauge boson,  $g^{}_B=1$. Compared to the value on
the $Z$ peak, the reduction of the cross section is by a factor of
$2\times 10^4$. Also in this interval the variation of the differential
cross section with the two scales is around $3 \%$.
\begin{table}
\begin{center}
\begin{footnotesize}
\begin{tabular}{|c||c|c|c|c|}
\hline
\multicolumn{5}{|c|}{$\sigma_{tot}^{nnlo}$ [fb], $\sqrt{S}=14$ TeV, $M_{1}=1$ TeV, $\tan\beta=40$}
\tabularnewline
\hline
$g_z$&
mLSOM&
$U(1)_{B-L}$&
$U(1)_{q+u}$&
$Free Ferm.$
\tabularnewline
\hline
\hline
$0.1$ & $5.982$& $3.575$& $2.701$& $1.274$  \\
       & $0.173$& $0.133$& $0.177$& $0.122$  \\
       & $0.277$& $0.445$& $0.252$& $0.017$
\tabularnewline
\hline
$0.36$ & $106.674$& $105.567$& $53.410$& $42.872$  \\
       & $2.248$& $1.733$& $2.308$& $1.583$  \\
       & $4.937$& $13.138$& $4.991$& $0.586$
\tabularnewline
\hline
$0.65$ & $240.484$& $143.455$& $108.344$& $51.155$  \\
       & $7.396$& $5.700$& $7.592$& $5.205$  \\
       & $11.127$& $17.853$& $10.124$& $0.699$
\tabularnewline
\hline
$1$ & $532.719$& $317.328$& $239.401$& $113.453$  \\
       & $17.810$& $13.720$& $18.274$& $12.530$  \\
       & $24.639$& $39.491$& $22.370$& $1.550$
\tabularnewline
\hline
\end{tabular}
\end{footnotesize}
\end{center}
\caption{\small Total cross sections,  widths and
$\sigma_{tot}\times BR(Z\rightarrow l\bar{l})$, where
$BR(Z\rightarrow l\bar{l})=\Gamma_{Z^{\prime}\rightarrow l\bar{l}}/\Gamma_{Z^{\prime}}$,  for the mLSOM and three anomaly-free extensions of the SM; they are all shown as functions of the coupling constant.}
\label{tablethird}
\end{table}
We have added Tab.~\ref{tablethird} in which we show results
for the total cross sections for the various models at the $Z$ peak. In the first
line of each column we show the results for
the total cross section in $[fb]$, in the 2nd line the total width
$\Gamma^{}_{Z^\prime}$, expressed in GeV and
in the 3rd line the observable $\sigma_{tot}\times BR(Z\rightarrow l\bar{l})$, where
$BR(Z\rightarrow l\bar{l})=\Gamma_{Z^{\prime}\rightarrow l\bar{l}}/\Gamma_{Z^{\prime}}$.
These quantities refer to the value of the coupling constant $g_z$ listed in the first column.
\footnote{Notice that we have chosen $g_z=g_B$ for the mLSOM.}

\begin{figure}[t]
\subfigure[  ]{\includegraphics[%
 width=5.6cm,
 angle=-90]{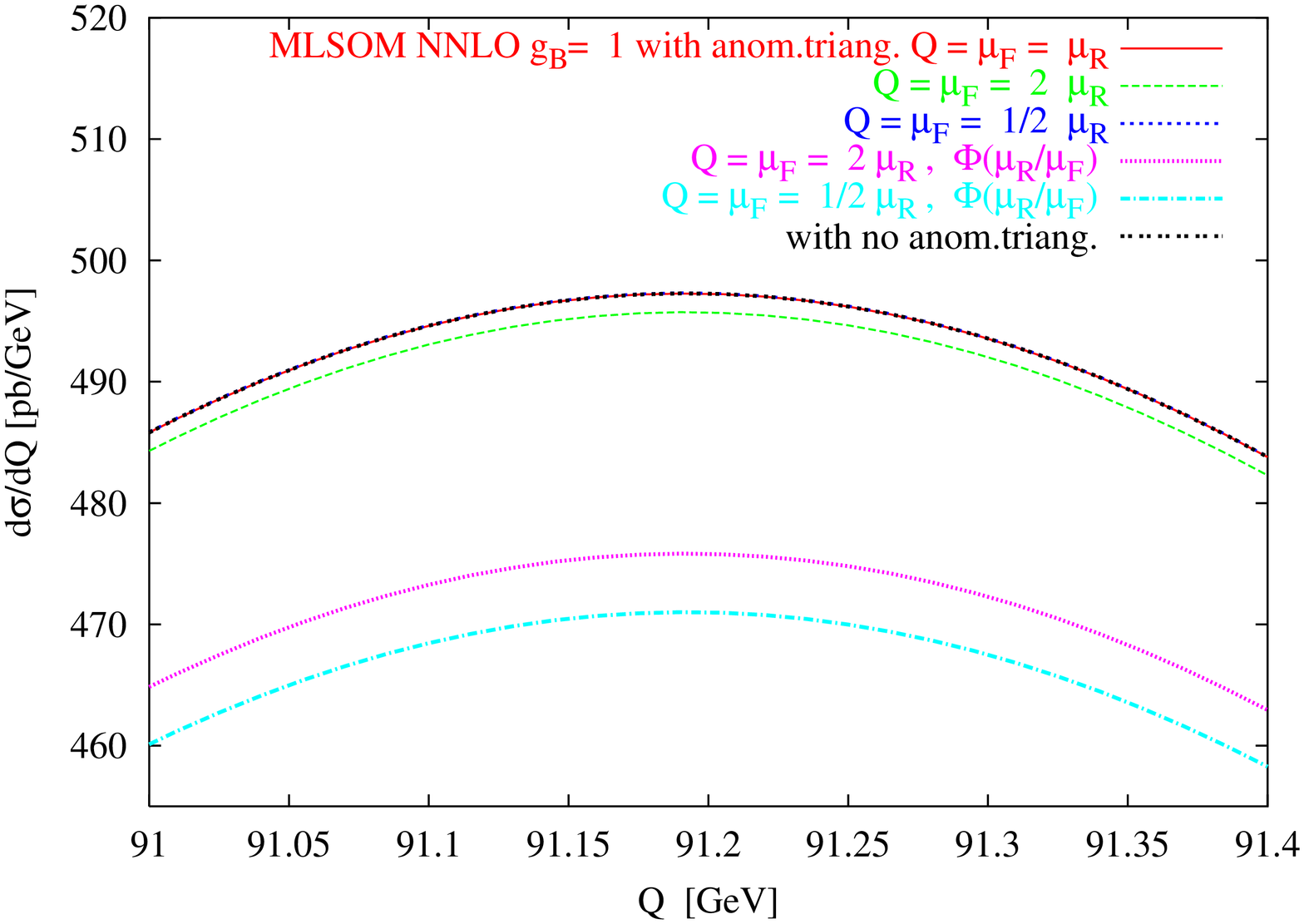}}
\subfigure[  ]{\includegraphics[%
 width=5.6cm,
 angle=-90]{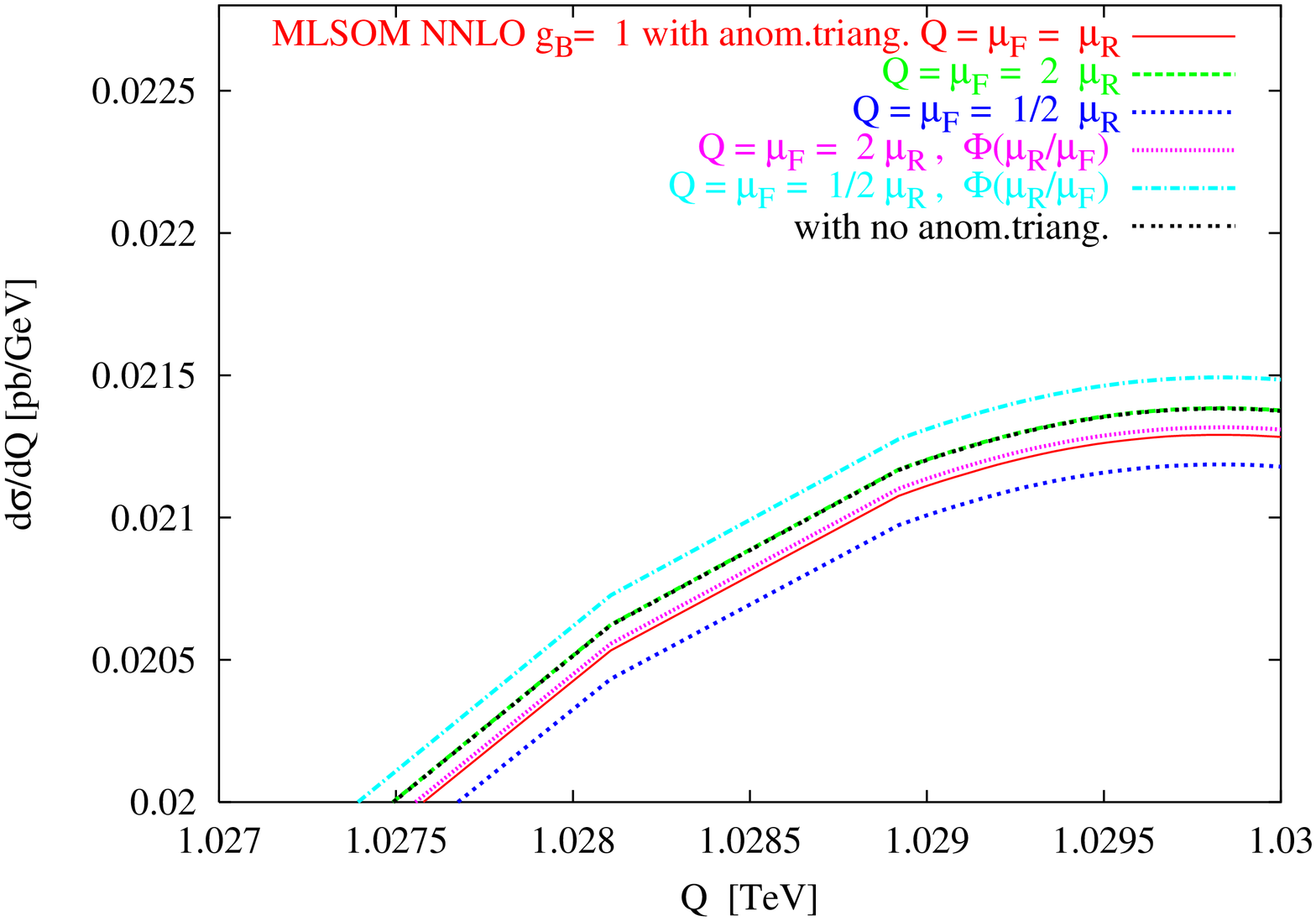}}
\caption{\small  Plot of the DY invariant mass distributions on the
peak of the $Z$ (a) and of the $Z^\prime$ (b). Shown are the total
contributions of the mLSOM and those in which the anomalous terms have been
removed. The variation of the result on $\mu_F$ and $\mu_R$ is included
both in the hard scatterings and in the luminosities ($\Phi(\mu_F/\mu_R)$).}
\label{mlsomef0}
\end{figure}

\begin{figure}[t]
\subfigure[  ]{\includegraphics[%
 width=5.6cm,
 angle=-90]{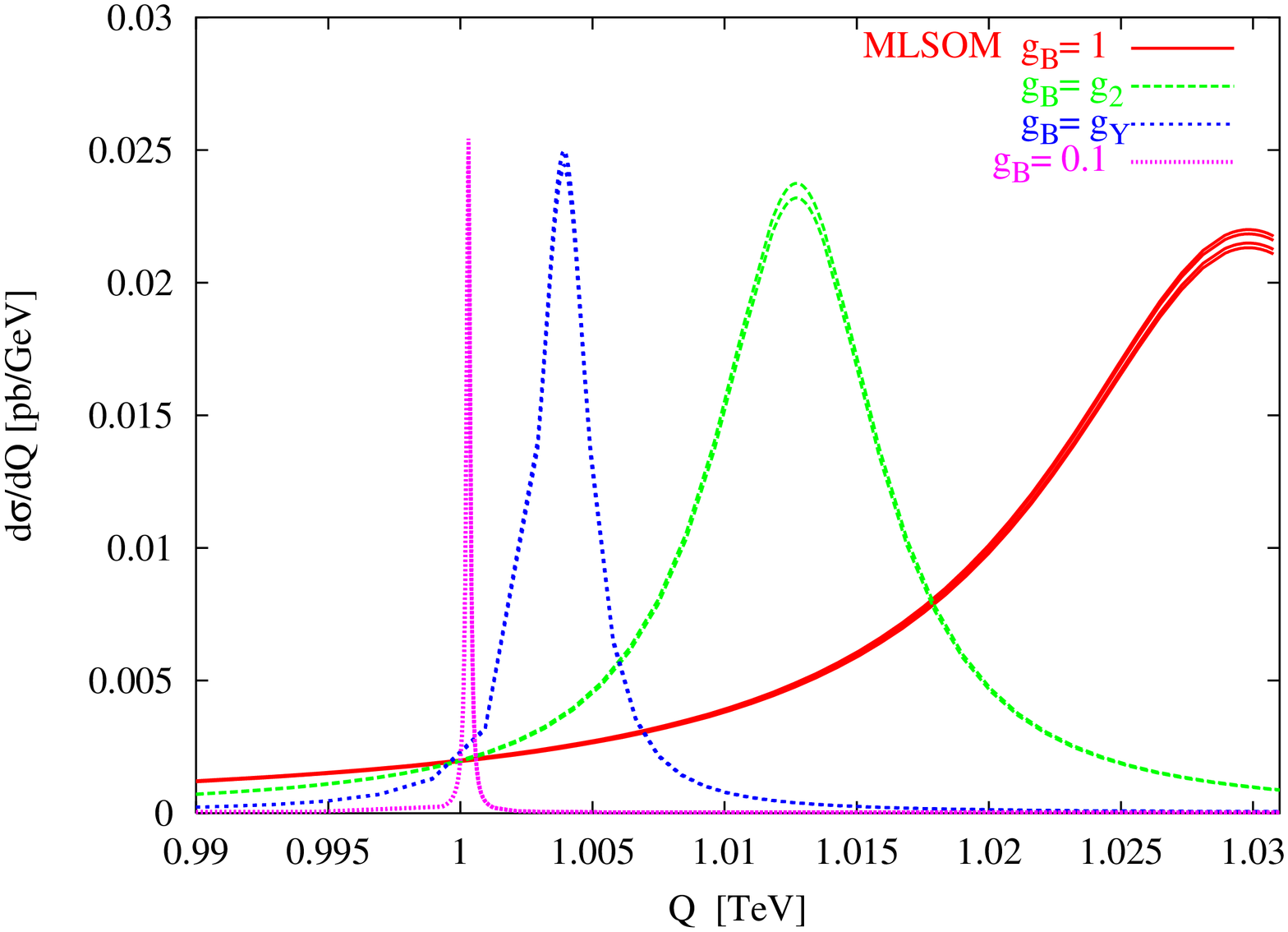}}
\subfigure[  ]{\includegraphics[%
 width=5.6cm,
 angle=-90]{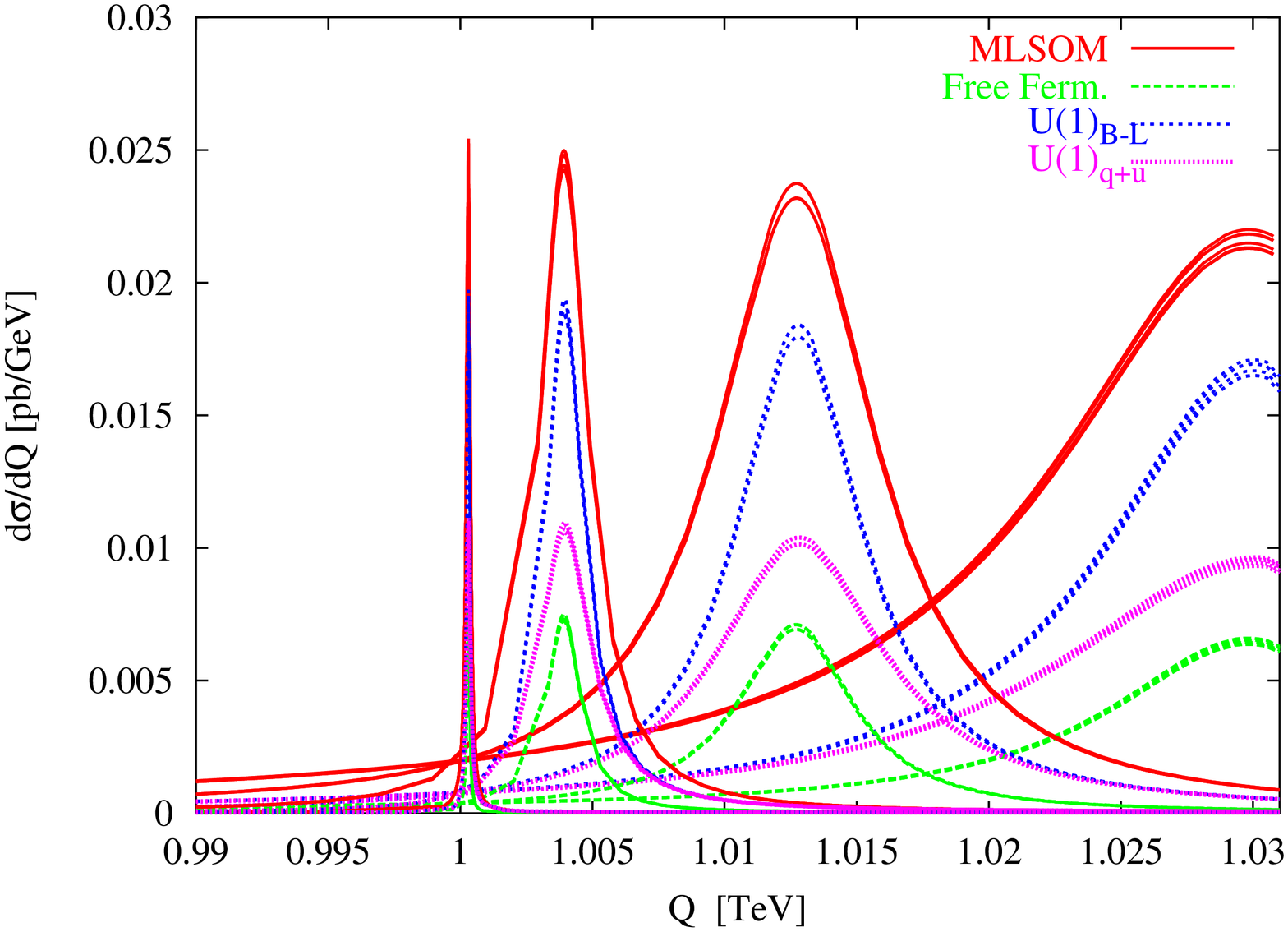}}
\caption{\small  (a) Anomalous $Z^\prime$ resonances obtained by varying $g^{}_B$. (b) Comparisons among anomalous Drell-Yan in the mLSOM versus several anomaly-free models.}
\label{mlsomef}
\end{figure}

We show in Fig.~\ref{mlsomef0}a,b two plots of the results
for the mLSOM of the DY cross section on the peak of the Z and
of the extra $Z^\prime$, where we vary the scales both in the
hard scatterings and in the parton luminosities. We have added and
subtracted the anomalous sector in order to estimate their size respect
to the remaining contributions. As we have already pointed out, the
scale variability at NNLO is larger than the
changes induced on the result by the anomalous graphs.
The anomalous effects are more visible at large
$Q$ (subfig. (b)), on the resonance of the extra $Z^\prime$,
and are due to the behavior of the anomalous components at large-$x$ due to a growing $Q$.
In Fig.~\ref{mlsomef}a we show a plot of the mLSOM for
different values of $g_B$ and for different values of $\mu_R$ and $\mu_F$.
The first peak (purple line) corresponds to $g_B=0.1$ the 2nd (blue line)
to $g_B=g_Y$ and so on. As $g_B$ grows the width of each peak
gets larger but the peak-value of the cross section decreases.
Different choices of $g_B$ correspond to slightly different values of
the mass of the extra $Z^\prime$ because of the relation between the
St\"uckelberg mass $M_1$ and $M_{Z^\prime}$ given in Eq.~(\ref{ZZpmass}).
For a fixed value of the coupling, the effects due to the variations
of the scales become visible only
for $g_B=1$ and in this case they are around 2-3\%.
In the case $g_B=1$  (red line), the uppermost lines correspond
to the choice $\mu_F=2 Q$, $\mu_R=1/2 \mu_F$ and $\mu_R= 2 \mu_F$,
while the lowermost lines correspond to the choice
$\mu_F=Q$, $\mu_R=1/2 \mu_F$ and $\mu_R= 2 \mu_F$.
Again, we notice that if we increase $\mu_F$ and $\mu_R$ the cross section
grows.

In Fig.~\ref{mlsomef}b we show the result of a comparison between
the mLSOM and the anomaly-free extensions. We have also included the $\mu_R/\mu_F$ scale dependence,
which appears as a band, and the variations with respect to $g_B$.
As shown in this figure, the red lines correspond to the mLSOM, the
blue lines to the $U(1)_{B-L}$ model, the green lines to the free fermionic model
and the purple lines to $U(1)_{q+u}$. Right as before, the first peak
corresponds to $g_B=0.1$, the 2nd to $g_B=g_Y$ etc.
The peak-value of the anomalous model is the largest of all, with a cross section which is
around $0.022$ [pb/GeV], the free fermionic appears to be the smallest with
a value around $0.006$ [pb/GeV].
\section{Direct Photons with GS and WZ interactions}
The analysis of $pp\to \gamma \gamma $ proceeds similarly to the DY
case, with a numerical investigation of the background and of the anomalous signal at parton level.

We start classifying the strong/weak interference
effects that control the various sectors of the process and then identify
the leading contributions due to the presence of anomaly diagrams.
\begin{figure}[ht]
\begin{center}
\includegraphics[scale=0.7]{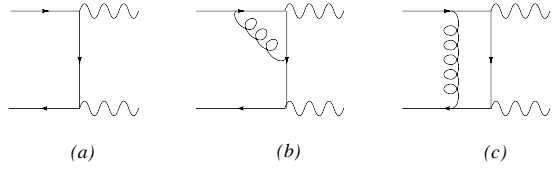}
\caption{\small $q\bar q$ sector for the process $q \bar q \rightarrow \g \g$ including virtual corrections at LO (a) and NLO (b,c).
\label{qqbarvirtual}}
\end{center}
\end{figure}
\begin{figure}[ht]
\begin{center}
\includegraphics[scale=0.7]{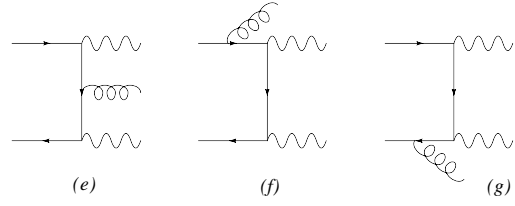}
\caption{\small Real emissions for $q \bar q \rightarrow \g \g$ at NLO.
\label{qqbarreal}}
\end{center}
\end{figure}
\begin{figure}[ht]
\begin{center}
\includegraphics[scale=0.75]{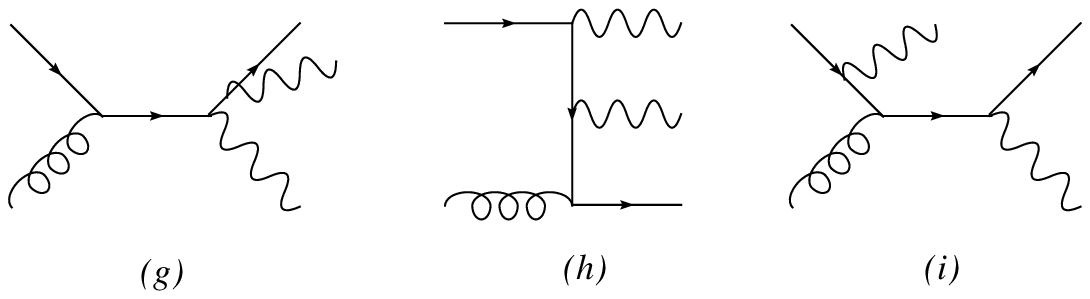}
\caption{\small $q g $ sector for the process $q g \rightarrow \g \g$. }
\label{qgall}
\end{center}
\end{figure}
\begin{figure}[ht]
\begin{center}
\includegraphics[scale=0.8]{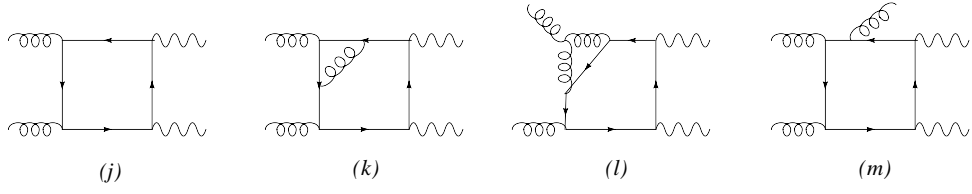}
\caption{\small $gg $ sector for the process $gg \rightarrow \g \g$ with virtual and real radiative corrections. }
\label{ggall}
\end{center}
\end{figure}
\begin{figure}[h]
\begin{center}
\includegraphics[scale=0.9]{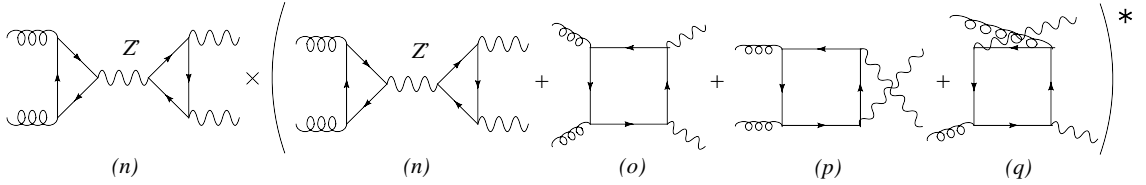}
\caption{\small Anomalous contributions for $ gg \rightarrow \g \g$ involving the BIM amplitude and its interference with the box graphs. }
\label{nopq}
\end{center}
\end{figure}
\begin{figure}[h]
\begin{center}
\includegraphics[scale=0.8]{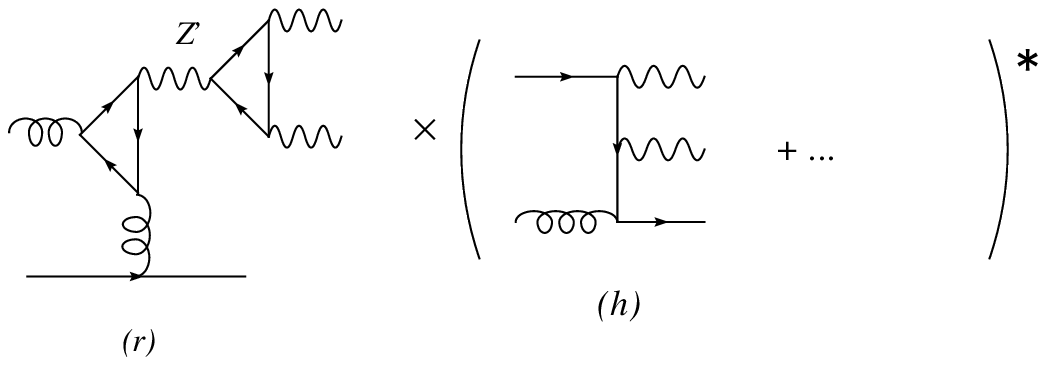}
\caption{\small Total amplitude for $ qg \rightarrow \g \g$. }
\label{qgtwodelta}
\end{center}
\end{figure}
\begin{figure}[h]
\begin{center}
\includegraphics[scale=0.85]{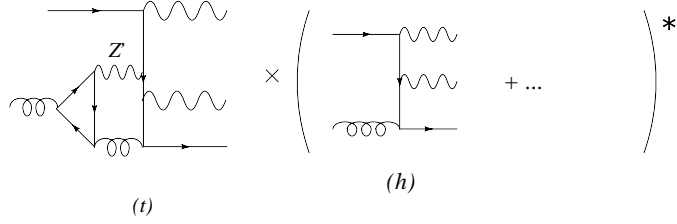}
\caption{\small Another configuration for the total amplitude of the  $ qg \rightarrow \g \g$ process. }
\label{delta2}
\end{center}
\end{figure}
\begin{figure}[t]
\begin{center}
\includegraphics[scale=0.8]{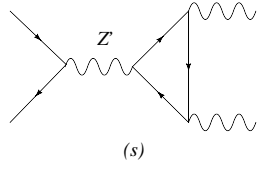}
\caption{\small Single diagram with an exchanged $Z^{\prime}$ boson in the $s$-channel. }
\label{s}
\end{center}
\end{figure}
\begin{figure}[t]
\begin{center}
\includegraphics[scale=0.85]{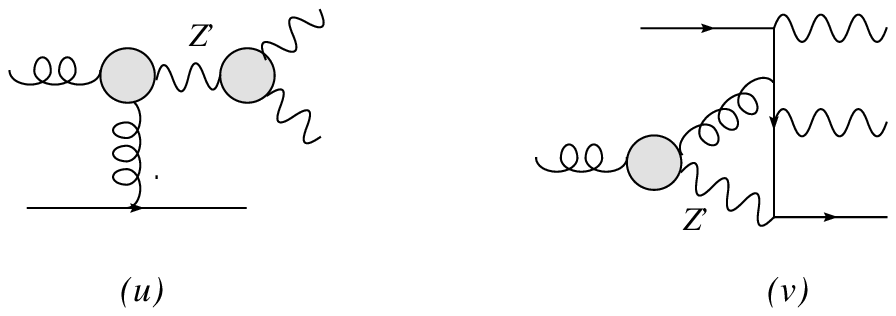}
\caption{\small Generic representation of the $qg\rightarrow
\g\g$ process in the presence of a GS vertex of the $AVV$ type.}
\label{qgGS}
\end{center}
\end{figure}
Direct photons are one of the possible channels to detect anomalous gauge
interactions, although, as we are going to see, also in this case the
anomalous signal remains rather small.
Direct photons are produced by partonic interactions rather than as a
result of the electromagnetic decay of hadronic states. At leading order
(LO) they carry the transverse momentum of the hard scatterers,
offering a direct probe of the underlying quark-gluon dynamics.
The two main channels in $pp$ collisions are the annihilation $q \bar{q}$
and Compton ($qg$), the second one being roughly $80 \%$ of  the entire
signal at large $p_T$  (from $p_T=4$ GeV on). The annihilation channel is
subleading, due to the small antiquark densities in the proton. The cross
section is also strongly suppressed (by a factor of approximately $10^{-3}$)
compared to the jet cross section. For this reason the electromagnetic decay
of the produced hadrons is a significant source of background, coming mostly
from the $\pi^0\to \gamma \gamma$ decay. In this case the angular opening of
the two photons is in general rather small, due to the small pion mass, and
proportional to their energy asymmetry
$((E_1-E_2)/(E_1 + E_2))$. A limitation in the granularity of the detector,
therefore,  may cause the two photons to be unresolved experimentally, giving
a spurious signal for direct photons. Even in the presence
of enough granularity in the detector, very asymmetric decays can also cause
the failure of the experimental
apparatus to resolve the low energy photon. A second source of background is
due to $\eta\to \gamma\gamma$ decay, which is about $20 \%$ of the pion contribution.
These two contributions account for almost all the background to direct photons.
The overall signal to background, though small, is supposed to raise with an
increasing $p_T$. This is due to the steepening of the
$\pi$ and $\eta$ spectra -at higher $Q^2$- respect to the $p_T$ spectra of
the parent jet, so there are smaller fractions of these particles that can
fragment into photons. The reduction of the background can be performed either
using reconstruction of the contributions due to the pions by measurements of
the invariant mass of the pair in the final state, or by imposing isolation cuts.
In the isolation procedure one can eliminate events with more than 2 particle in
the final state, considering that emissions from fragmentation are usually accompanied
by a large multiparticle background.
The selection of appropriate isolation cuts are one of the way to render the anomalous
signal more significant, considering that the tagged photon signal, although being of
higher-order in $\alpha_s$ (NNLO in QCD), is characterized by a two-photon-only final
state. As we are going to see this signal is non-resonant, even in the presence of
an $s$-channel exchange, due to the anomaly.

We show in Fig.~\ref{qqbarvirtual} a partial list of the various background
contributions to the DP channel in
$pp $ collisions.
We show the leading order (LO) contribution in diagram (a)
with some of the typical virtual corrections included in (b)
and (c). These involve the $q\bar{q}$ sector giving a cross section of the form
\beq
\sigma_{q\bar{q}}= \alpha_{em}^2 ( c_1 + c_2 \alpha_s).
\eeq
These corrections are the NLO ones in this channel.
The infrared safety of the process is guaranteed at the same perturbative order
by the real emissions in Fig.~\ref{qqbarreal} with an integrated gluon in the final
state, which are also of $O(\alpha_{em}^2 \alpha_s)$.

A second sector is the $qg$ one, which is shown in Fig.~\ref{qgall}, also of the same
order $(O(\alpha_s \alpha_{em}))$. These corrections are diagrammatically
the NLO ones. In general, the NLO prediction for this process
are improved by adding a part of the NNLO
(or $O(\alpha_{em}^2\alpha_s^2)$) contributions, such as
the box contribution (j) of the $gg$ sector which is of
higher order ($O(\alpha_{em}^2 \alpha_s^2)$) in $\alpha_s$,
the reason being that these contributions have been shown to be
sizeable and comparable with the genuine NLO ones. All these corrections have
been computed long ago \cite{Coriano:1996us} and implemented independently
in a complete Monte Carlo in \cite{Binoth:1999qq,Binoth:2001vm} with a more
general inclusion of the fragmentation.
More recently, other NNLO contributions have been added
to the process, such as those involving the gg sector through $O(\alpha_{em}^2 \alpha_s^3)$,
\beq
\sigma_{gg}= \alpha_{em}^2( d_1 \alpha_s^2 + d_2 \alpha_s^3),
\eeq
shown in graphs $(k),(l),(m)$.
The other sectors have not yet been computed with the same
accuracy, for instance in the $qq$ and $q\bar{q}$ channels
they involve 2 to 4 emission amplitudes which need to be
integrated over 2 gluons. For instance, graph (m) is a real
emission in $\sigma_{gg}$ which is needed to cancel the
infrared/collinear singularities of the virtual ones at the same order.

\begin{figure}[t]
\begin{center}
\includegraphics[scale=0.8]{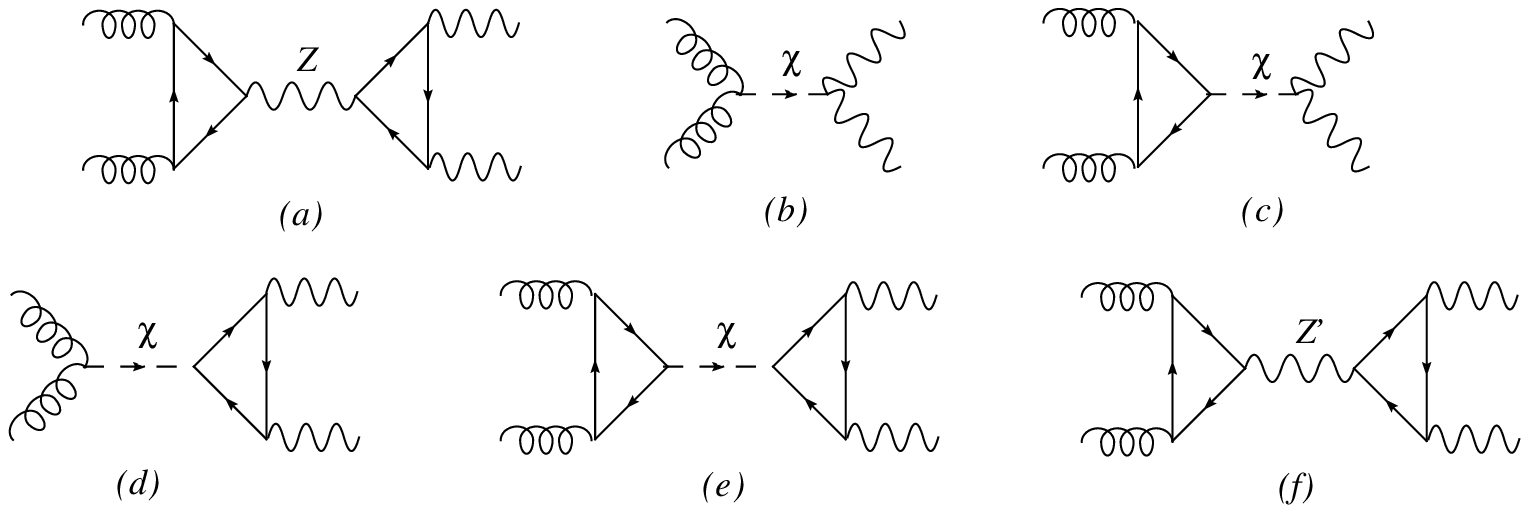}
\caption{\small Complete list of amplitudes included in the type of graphs shown in Fig.~\ref{nopq}. They also have the exchange of a physical axion and contributions proportional to the mass of the internal fermion.}
\label{qgonedelta}
\end{center}
\end{figure}

The anomalous contributions are shown in Figs.~\ref{nopq}, \ref{qgtwodelta}, \ref{delta2}, \ref{s}
and \ref{qgGS}.\footnote{As stated in the previous sections, the diagram in
Fig.~\ref{s} vanish because of a WI.}
One of the most important partonic
process, in this case, is the BIM amplitude shown in diagram (n)
which violates unitarity at very high energy. In the WZ case, where the axion $b$
appears in trilinear interactions not as a virtual state (such as in $(b F\wedge F)$)
this amplitude has two main properties;
1) it is well defined and finite in the chiral limit even
for on-shell physical gluons/photons and
2) it is non-resonant and can grow beyond the unitarity limit.
Working in the chiral limit, its expression is given by the Dolgov-Zakharov
limit of the anomaly amplitude (Eqs. (\ref{massiveT}) and (\ref{a6}), with $m_f=0$),
which appears both in the production mechanism of the extra $Z^{\prime}$,
$(gg\to Z^{\prime}) $, where the $Z^\prime$ in the $s$-channel is virtual, and in
its decay into two photons.
We recall from Chap.~\ref{chap:UnitarityBound} that the presence of an anomaly in the initial and in the final state
cancels the resonance of a given channel.
For simplicity we consider amplitude Fig.~\ref{nopq}n, assuming to have photons both in the
initial and in the final state. The amplitude is given by
\ba
A^{\mu \nu \mu^\prime \nu^\prime}_{BIM}&=&   \frac{a_n}{k^2} k^{\lambda} \epsilon[\mu,\nu,k_1,k_2]  \, \frac{- i}{k^2 - M^2}
\left( g^{\lambda \lambda^\prime}
- \frac{k^\lambda k^{\lambda^\prime} }{ M^2 }   \right)  \frac{a_n}{k^2}
( - k^{\lambda^\prime}) \epsilon[\mu^\prime,\nu^\prime,k_1^\prime,k_2^\prime]   \nonumber\\
&=&   \frac{a_n}{k^2}  \epsilon[\mu,\nu,k_1,k_2]  \, \frac{- i}{ k^2 - M^2 }   \frac{k^{ \lambda^\prime}( M^2 - k^2 ) }{ M^2 }  \,  \frac{a_n}{k^2}
( - k^{\lambda^\prime}) \epsilon[\mu^\prime,\nu^\prime,k_1^\prime,k_2^\prime]   \nonumber\\
&=&   \frac{a_n}{k^2}  \epsilon[ \mu,\nu,k_1,k_2]  \, \left(  \frac{- i k^2 }{  M^2 }  \right) \,  \frac{a_n}{k^2}
 \epsilon[\mu^\prime,\nu^\prime,k_1^\prime,k_2^\prime]   \nonumber\\
&=&  - \frac{a_n}{M}  \epsilon[ \mu,\nu,k_1,k_2]  \,\,  \frac{ i }{ k^2 }  \,\,  \frac{a_n}{M}
 \epsilon[\mu^\prime,\nu^\prime,k_1^\prime,k_2^\prime].
\label{resBIM}
\ea
where $M$ denotes, generically, the mass of the anomalous gauge boson
in the $s$-channel. If we multiply this amplitude by the external polarizators of
the photons, square it and perform the usual averages, one finds that it grows
quadratically with energy.  The additional contributions in the $s$-channel for this amplitude are shown in Fig.~\ref{qgonedelta}.
The exchange of a massive axion (Fig.~\ref{qgonedelta}b),
due to a mismatch between the coupling and the parameteric dependence
between Fig.~\ref{qgonedelta}a and b, does not erase the growth
(see the discussion in the appendix).
This mismatch is at the origin of the unitarity bound for this
theory analyzed.
The identification of this
scale in the context of QCD is quite subtle, since the lack of unitarity in a partonic
process implies a violation of unitarity also at hadron level, but at
a different scale compared to the partonic one, which needs to be
determined numerically directly from the total hadronic cross
section $\sigma_{pp}$. Overall, the convolution of a BIM amplitude
with the parton distributions will cause a suppression of the rising partonic contributions,
due to the small gluon density at large Bjorken $x$. Therefore,
the graphs do not generate a large anomalous signal in this channel.
However, the problem of unitarizing the theory by the inclusion of
higher dimensional operators beyond the minimal dimension-5 operator $b F\wedge F$ remains.

The anomalous terms, beside the $(\textrm{n}){(\textrm{n})}^* $ contribution with the
exchange of an extra $Z/Z^\prime$ which carry an anomalous component,
which is $O(\alpha_{em}^2 \alpha_s \alpha_w^2)$, include the
interference between the $s/t/u$ box diagrams of $gg\to\gamma\gamma$
with the same BIM amplitude (n). In the $gg$ sector
the anomalous terms give, generically, an expression of the form
\beq
\sigma_{gg}^{an}= \alpha_{em}^2(a_1\alpha_s^2\alpha_w + a_2 \alpha_s^2 \alpha_w^2),
\eeq
with the first contribution coming from $(\textrm{n}){(\textrm{o})}^* $
and from the interference with the ($gg\to\gamma \gamma$) box diagram,
while the second from $(\textrm{n}){(\textrm{n})}^* $.
Other contributions which appears at $O(\alpha_{em}^2\alpha_s \alpha_w)$
are those shown in Fig.~\ref{qgtwodelta} which involve 2 anomaly diagrams (r) and their
interference with the NLO real emission diagram of type (m). These contributions are phase-space
suppressed. If we impose isolation cuts on the amplitude, we can limit our analysis, for the anomalous signal, only to 2-to-2 processes.
\subsection{Helicity amplitudes: massless box diagrams and anomalous interferences}
Moving to the computation of the anomalous contributions to
$g(p_1,\pm)+g(p_2,\pm)\rightarrow \gamma(k_1,\pm)+\gamma(k_2,\pm)$,
coming from the 2-to-2 sector we identify the
following non-vanishing helicity amplitudes for the diagrams shown in
Fig.~\ref{nopq}, with the usual conventions
\ba
&&s=(p_1+p_2)^2
\nonumber\\
&&t=(p_1-k_1)^2=-s/2(1-\cos{\theta})
\nonumber\\
&&u=(p_1-k_2)^2=-s/2(1+\cos{\theta}),
\ea
where $\theta$ is the angle between $\vec{p}_1$ and $\vec{k}_1$, and we obtain
\ba
M^{{\cal Z}}_{++++}&=&\frac{s}{(2\pi)^4 M_{\cal Z}^2},
\nonumber\\
M^{{\cal Z}}_{----}&=&\frac{s}{(2\pi)^4 M_{\cal Z}^2},
\nonumber\\
M^{{\cal Z}}_{++--}&=& i\,\epsilon[k_1,k_2,p_1,p_2]
\frac{{\left( t^2 - u^2 \right) }^3 - s^2\,\left( t^4 - u^4 \right)}{8 M_{\cal Z}^2\,{\pi }^4\,s^3\,t^2\,u^2}
\nonumber\\
&-&\frac{ {\left( t^2 - u^2 \right) }^4 -2\,s^2\,{\left( t^2 - u^2 \right) }^2\,
\left( t^2 + u^2 \right) + s^4\,\left( t^4 + u^4 \right)}{32\,M_{\cal Z}^2\,{\pi }^4\,s^3\,t^2\,u^2},
\nonumber\\
M^{{\cal Z}}_{--++}&=& -i\,\epsilon[k_1,k_2,p_1,p_2]
\frac{{\left( t^2 - u^2 \right) }^3 - s^2\,\left( t^4 - u^4 \right)}{8 M_{\cal Z}^2\,{\pi }^4\,s^3\,t^2\,u^2}
\nonumber\\
&-&\frac{ {\left( t^2 - u^2 \right) }^4 -2\,s^2\,{\left( t^2 - u^2 \right) }^2\,
\left( t^2 + u^2 \right) + s^4\,\left( t^4 + u^4 \right)}{32\,M_{\cal Z}^2\,{\pi }^4\,s^3\,t^2\,u^2},
\nonumber\\
M^{\chi}_{++++}&=&\frac{16 s^2}{(s-M_{\chi}^2)},
\nonumber\\
M^{\chi}_{----}&=&\frac{16 s^2}{(s-M_{\chi}^2)},
\nonumber\\
M^{\chi}_{++--}&=&-32\, i \epsilon[k_1,k_2,p_1,p_2]
\frac{s^2 \left( t^4 - u^4 \right)-\left( t^2 - u^2 \right)^3 }
{s^2 \left( s -M_{\chi}^2 \right) \,t^2\,u^2}
\nonumber\\
&-&8\,\frac{\left( t^2 - u^2 \right)^4 - 2\,s^2\,\left( t^2 - u^2 \right)^2\,
\left( t^2 + u^2 \right) + s^4\,\left( t^4 + u^4 \right)}{s^2\,
\left( s-M_{\chi}^2 \right) \,t^2\,u^2},
\nonumber\\
M^{\chi}_{--++}&=& 32\, i \epsilon[k_1,k_2,p_1,p_2]
\frac{s^2 \left( t^4 - u^4 \right)-\left( t^2 - u^2 \right)^3 }
{s^2 \left( s -M_{\chi}^2 \right) \,t^2\,u^2}
\nonumber\\
&-&8\,\frac{\left( t^2 - u^2 \right)^4 - 2\,s^2\,\left( t^2 - u^2 \right)^2\,
\left( t^2 + u^2 \right) + s^4\,\left( t^4 + u^4 \right)}{s^2\,
\left( s-M_{\chi}^2 \right) \,t^2\,u^2},
\ea
where again ${\cal Z}$ indicates either $Z$ or $Z^\prime$
and the $M_\chi$ refers to the contributions with the exchange of a axi-Higgs.
In the above formulas we have omitted all the coupling constants to obtain
more compact results. The helicity amplitudes for the massless box contribution
have been computed in \cite{Bern:2001df} and are given by
\ba
&&M^{box}_{--++}=1,
\nonumber\\
&&M^{box}_{-+++}=1,
\nonumber\\
&&M^{box}_{++++}=-\frac{1}{2} {t^2+u^2\over s^2}
\Bigl[ \ln^2\Bigl({t\over u}\Bigr) + \pi^2 \Bigr]
- {t-u\over s} \ln\Bigl({t\over u}\Bigr) - 1,
\nonumber \\
&&M_{+--+}^{box} =- {1\over2} {t^2+s^2\over u^2}
\ln^2\Bigl(-{t\over s}\Bigr)
- {t-s\over u} \ln\Bigl(-{t\over s}\Bigr) - 1
\nonumber \\
&& \null \hskip 2 cm
- i \pi \biggl[ {t^2+s^2\over u^2} \ln\Bigl(-{t\over s}\Bigr)
+ {t-s\over u} \biggr],
\nonumber \\
&&M_{+-+-}^{box}(s,t,u)=M_{+--+}^{box}(s,u,t),
\ea
giving a differential cross section
\beq
{d\sigma^{box} \over d\cos\theta} = {\alpha_{em}^2 \alpha_{s}^2 N_c^2\over 64 \pi \, s}
\left[\sum_f Q_f^2\right]^2   \Bigl\{
     | M_{--++}^{box} |^2 + 4 \, | M_{-+++}^{box} |^2
   + | M_{++++}^{box} |^2 + | M_{+--+}^{box} |^2 + | M_{+-+-}^{box} |^2
   \Bigr\}.
\label{Box_cross}
\eeq
The interference terms are listed in Fig.~\ref{nopq} and the interference differential cross
section is given by
\ba
{d\sigma^{int} \over d\cos\theta} = \sum_{{\cal Z}=Z,Z^{\prime}} {d  \sigma^{ {\cal Z}, box} \over d\cos\theta}  +
{d\sigma^{\chi, box} \over d\cos\theta},
\ea
where
\ba
 {d  \sigma^{ {\cal Z}, box} \over d\cos\theta} &=& \frac{1}{256 \pi s}   \sum_{q}
  \frac{1}{2}    c_1^q     \sum_{q^\prime} \frac{1}{2}  c_2^{ q^\prime }
 \sum_f Q_f^2 \alpha_{em}\alpha_{s} N_c
{\mbox {Re}}\left[2 M_{++++}^{\cal Z} M_{++++}^{* box} + (M_{++--}^{\cal Z} + M_{--++}^{\cal Z})M_{++--}^{* box}
\right]\nonumber\\
&=& - \frac{1}{8192 \pi^5 M^2_{\cal Z} } \sum_{q}
  \frac{1}{2}    c_1^q     \sum_{q^\prime} \frac{1}{2}  c_2^{ q^\prime }   \sum_f Q_f^2 \alpha_{em}\alpha_{s} N_c
 \left[   \left(       \cos^2 \theta + 1  \right)
   \log^2  \left(  \frac{ 1 -   \cos\theta  }{  1+   \cos\theta    }  \right)    \right.    \nonumber\\
 &&    +   4  \cos\theta  \log \left( \frac{ 1-   \cos\theta }{ 1 +  \cos\theta   } \right)
 + \left(  \cos^2\theta + 1 \right)  \pi^2 + 8   \Big],
\ea
\ba
 {d  \sigma^{ \chi, box} \over d\cos\theta} &=& \frac{ g^{\chi}_{gg} g^{\chi}_{\g\g} }{256 \pi s}
\sum_f Q_f^2 \alpha_{em}\alpha_{s} N_c   \,{\mbox Re}
\left[2 M_{++++}^{\chi} M_{++++}^{* box} + (M_{++--}^{\chi} + M_{--++}^{\chi})M_{++--}^{* box}
\right]
\nonumber\\
&=&  - \frac{s}{32 \pi ( s- M_\chi^2) } g^{\chi}_{gg} g^{\chi}_{\g \g}
\sum_f Q_f^2 \alpha_{em}\alpha_{s} N_c
    \left[  \left(    \cos^2 \theta + 1 \right)
   \log^2  \left(  \frac{ 1 -  \cos\theta   }{ 1 +  \cos\theta  } \right)     \right.   \nonumber\\
&&  +4  \cos\theta \log \left( \frac{ 1 -    \cos\theta  }{ 1+ \cos\theta  }  \right)
    + \left( 1+  \cos^2\theta  \right)
   \pi^2 + 8  \Big] ,
\ea
where $g^{}_3$ is the strong coupling constant while   $g_B$ is the anomalous coupling
constant of the model. Here $M_{++--}^{box}=M_{--++}^{box}$ and we have used the parameters
of the mLSOM, explicitly given in Chaps.~\ref{chap:AbelianModels2}, \ref{chap:UnitarityBound}. The rotation matrix
elements from the hypercharge basis to the physical basis after
electroweak symmetry breaking of the neutral gauge sector ($O^A$) and rotation matrix elements of the $CP$-odd sector
($O^{\chi}$) are defined in App.~\ref{sec:Ochi}, while the couplings of the physical axion to the gluons and the photons, $g^{\chi}_{gg}$ and 
$g^{\chi}_{\g\g}$, are given in Eqs.~\ref{GS_coeffs}. The axial-vector coupling $g_{A,q}^{\cal Z}$ of the neutral gauge bosons
to a quark flavour $q$
\begin{eqnarray}
g_{A,q}^{\cal Z} = \frac{1}{2}(Q_{\cal Z}^{R,q}-Q_{\cal Z}^{L,q}),   \qquad (\mathcal{Z}=Z, Z^\prime)
\end{eqnarray}
\subsection{ Numerical analysis for direct photons}
\begin{figure}[t]
\begin{center}
\includegraphics[width=7cm,angle=-90]{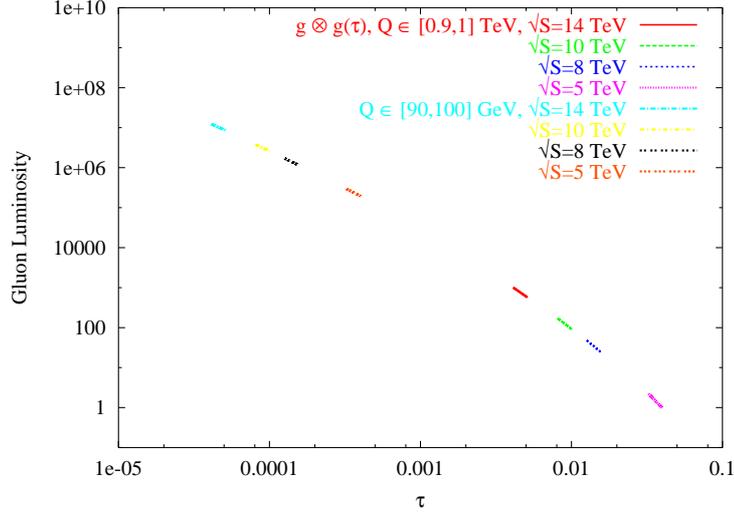}
\caption{\small The NNLO gluon luminosity as a function of $\tau=Q^2/S$  for various value of the invariant mass of the photon pair ($Q$) and energy of the hadronic beam ($S$) at the LHC evolved with \textsc{Candia 1.0}.}
\label{Gluons}
\end{center}
\end{figure}
In our numerical implementation of double prompt photon production
we compare the size of the anomalous
corrections respect to the SM background evaluated by a Monte Carlo
\cite{Binoth:1999qq, Bern:2002jx}.
Since the anomalous signal is small compared to that of the SM, we have extracted
both for the SM case and the anomalous case the $gg$ sector and compared them at
hadron level by convoluting the partonic contributions with
the \textsc{Pdf}'s (see Fig.~\ref{Gluons}).
In this comparison, the SM sector is given by the graphs shown in
Fig.~\ref{ggall} plus the interference graphs shown in Fig.~\ref{nopq}. In the
SM case this second set of graphs contributes proportionally to the mass of the
heavy quarks in the anomaly loop. At high energy the hard scatterings coming from
this interference are essentially due to the mass of the top quark running inside
a BIM amplitude and are, therefore, related to heavy quark effects.
In the anomalous case the same set of graphs is considered, but now the
anomaly contributions are explicitly included. The hadronic differential
cross section due to the anomalous interactions for massless quarks
is given by
\ba
&&\frac{d\sigma}{d Q}=\int_{0}^{2\pi}d\phi \int_{-1}^{1} d\cos{\theta}\,\,
\frac{\tau}{4 Q}\int_{\tau}^{1}\frac{d x}{x}\Phi_{gg}(\frac{\tau}{x})\Delta(x,\theta),
\nonumber\\
&&\Phi_{gg}(y)=\int_{y}^{1}\frac{d z}{z}g(y/z)g(z),
\nonumber\\
&&\Delta(x)=\delta(1-x)\left[\frac{d\sigma_{Z}}{d\cos{\theta}}
+\frac{d\sigma_{Z'}}{d\cos{\theta}}+\frac{d\sigma_{\chi}}{d\cos{\theta}}
+\frac{d\sigma_{int}}{d\cos{\theta}}\right]
\nonumber\\
&&\frac{d\sigma_{int}}{d\cos{\theta}}=\frac{d\sigma^{Z,box}}{d\cos{\theta}}+
\frac{d\sigma^{Z',box}}{d\cos{\theta}}+\frac{d\sigma^{\chi,box}}{d\cos{\theta}}.
\ea
The contributions which are part of this sector due to exchange
of a $Z$ or a $Z^{\prime}$ and a $\chi$ (see (a), (b) and (f) of the BIM set
in Fig.~\ref{qgonedelta}) are those labelled above, while $\sigma_{int}$
refers to the interferences shown in Fig.~\ref{nopq}, with the inclusion
of a $Z^{\prime}$ and a physical axion
(such as Fig.~\ref{qgonedelta}b).

Defining
\ba
&&\sigma_{gg\to \gamma \gamma} \equiv  \int_{0}^{2\pi}d\phi \int d\cos{\theta} \,\, \Delta(x,\theta)
\ea
the hadronic cross section takes the form of a product of the gluon luminosity and the partonic $gg\to\gamma \gamma$ cross section
\ba
\frac{d\sigma}{dQ}=\frac{Q}{4 S}\sigma_{gg\to \gamma\gamma} \Phi(\tau).
\label{factor2}
\ea
\subsection{The $gg$ sector}
\begin{figure}[h]
\begin{center}
\includegraphics[width=7cm, angle=-90]{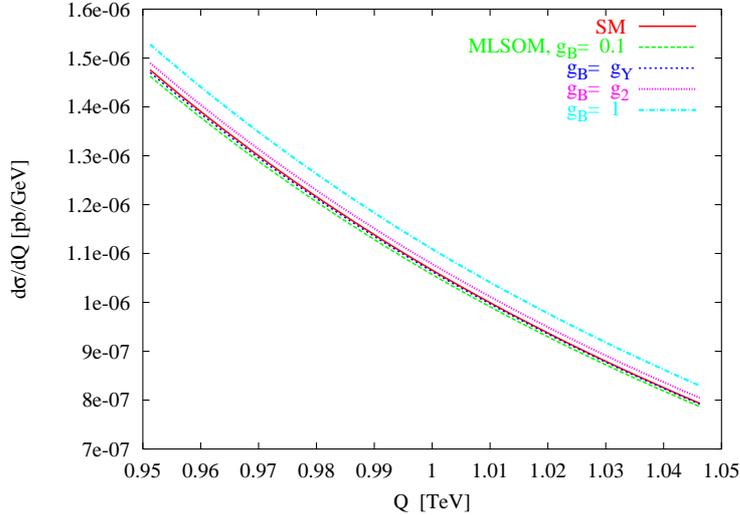}
\caption{\small Comparison plots for the gluon sector in the SM and in the anomalous model for a resonance of
1 TeV. The box-like contributions are not included,
while they appear in the interference with the BIM amplitudes.}
\label{new2}
\end{center}
\end{figure}

Coming to the analysis of the gluon fusion sector, the result of this study
is shown in Fig.~\ref{new2}
where we plot the gluon contribution to the hadronic cross section
for both the SM and the mLSOM, having chosen $M_1=1$ TeV.
We have used the MRST99 set of parton distributions
to generate the NNLO gluon luminosity
with $\alpha_s(M_Z)=0.1175$, $Q=\frac{1}{2}\mu_R$ and $\sqrt{S}=14$ TeV.
We have chosen $\tan\beta=40$ and different values of $g_B$. The size of the
cross section is around $10^{-6}$ [pb/GeV] - right on
the mass of the resonance - for both models, with a difference that grows as
we rise the coupling constant for the anomalous $U(1)$ ($g_B$).
We have chosen four possible values for $g_B$: a small parametric value
($g_B=0.1$); equal to the coupling of the hypercharge $g_Y$ at the same scale
($g_B=g_Y$) or to
the $SU(2)_w$ coupling $g_2$  ($g_B=g_2$) or, finally, parameterically
sizeable, with $g_B=1$. In the interference graphs used for this comparison
between the anomalous signal of the mLSOM and the SM  (in this second case
the BIM amplitudes contribute via the heavy quark mass in the loops) we have
included, beside the BIM amplitude, the entire set of contributions shown in
Fig.~\ref{qgonedelta}, with the exchange of a $Z$, a
$Z^{\prime}$ and the axi-Higgs $\chi $.
We have chosen a light St\"uckelberg axion with $m_\chi=30$ GeV.
For $Q$ around $m_\chi$ the
anomalous signal grows quite substantially, as we are going to show next.
The overall flatness of the result in
this region - the width of the interval is just 5 GeV in Fig.~\ref{new2}
- shows that the corrections are non-resonant and rather small.
They are also overlapping for both models and the extraction of additional
information concerning the anomalous sector appears to be very difficult.
Before we get into a more detailed analysis of the various contributions
to this sector, we mention that we have performed isolation
cuts on the cross section of the SM background in DP with a choice
of $R=0.4$  for the radius of the cone of isolation of the photons.
This is defined in terms of an azimuthal
angle $\phi$ and pseudorapidity $\eta=\ln{\tan{\theta/2}}$, with a
maximal value of transverse energy $E_{Tmax}=15$ GeV in the cone,
as implemented in \cite{Bern:2002jx}.
\begin{figure}
\subfigure[  ]{\includegraphics[%
 width=5.6cm,
 angle=-90]{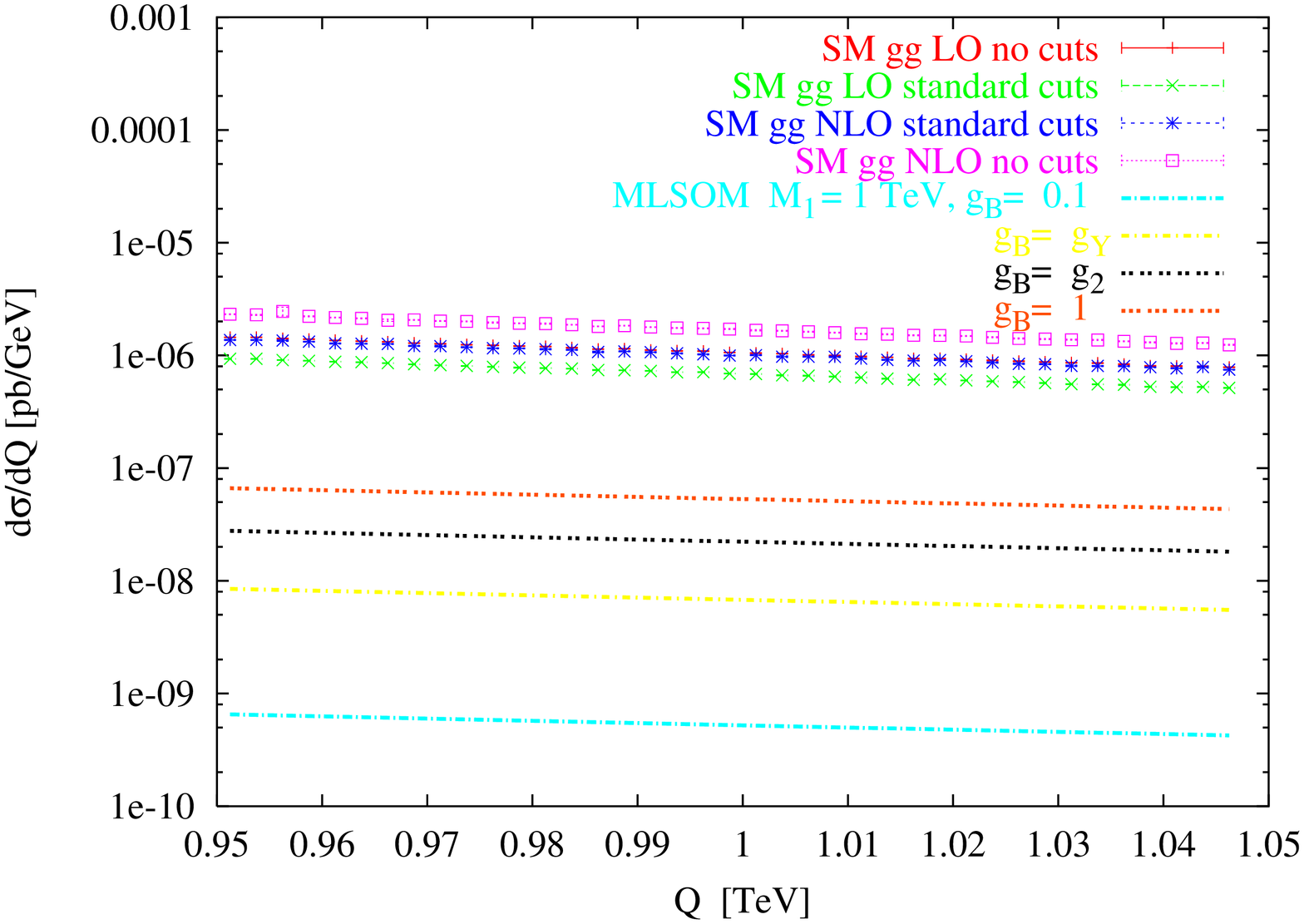}}
\subfigure[  ]{\includegraphics[%
 width=5.6cm,
 angle=-90]{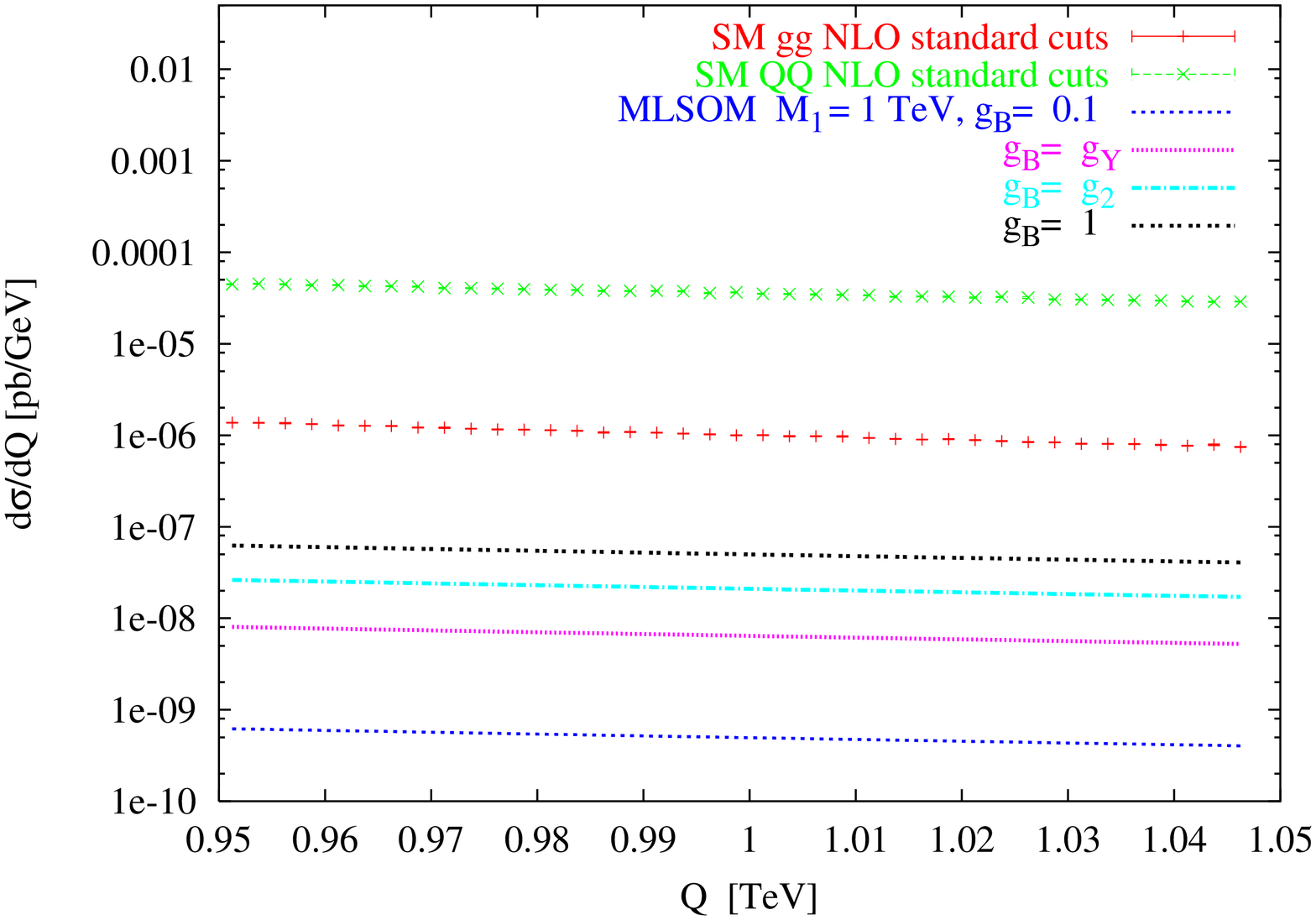}}
\caption{\small (a): SM contributions for the gluon-gluon channel obtained with the Monte
Carlo \textsc{Gamma2MC}. These are indicated by dotted
lines and include all the interferences and the box graphs.
Shown are also the anomalous contributions of the mLSOM (no box).
(b): as in (a) but we have included the Monte Carlo results for the SM qq channel at NLO.}
\label{ab}
\end{figure}
We show a more detailed investigation of the results for the various contributions
in the $gg$ sector in Fig.~\ref{ab}a,b. The dotted lines are the results obtained by the Monte Carlo and include
both 2-to-2 and 2-to-3 contributions (pure QCD) with and without cuts, computed at LO and at NLO. The size of these contributions is around $2\times 10^{-6}$ pb/GeV in the SM case. We show in the same subfigure the
anomalous corrections in the mLSOM, which vary between $10^{-9}$ and $10^{-7}$ pb/GeV. Therefore, for
$g_B\sim1$, the anomalous sector of the mLSOM (the square of the box terms here are not included for the MSLOM) is suppressed by a factor of 10 respect to the signal from the same sector coming from the SM.
In subfig.~(b) we show the same contributions but we include in the SM also the quark channel (shown separately from the gluon channel), which is around $10^{-4}$ pb/GeV. Therefore, the quark sector overshadows the anomalous corrections by a factor of approximately $10^{3}$, which are difficult to extract at this value of the invariant mass.

\begin{figure}
\subfigure[  ]{\includegraphics[%
width=5.6cm,
 angle=-90]{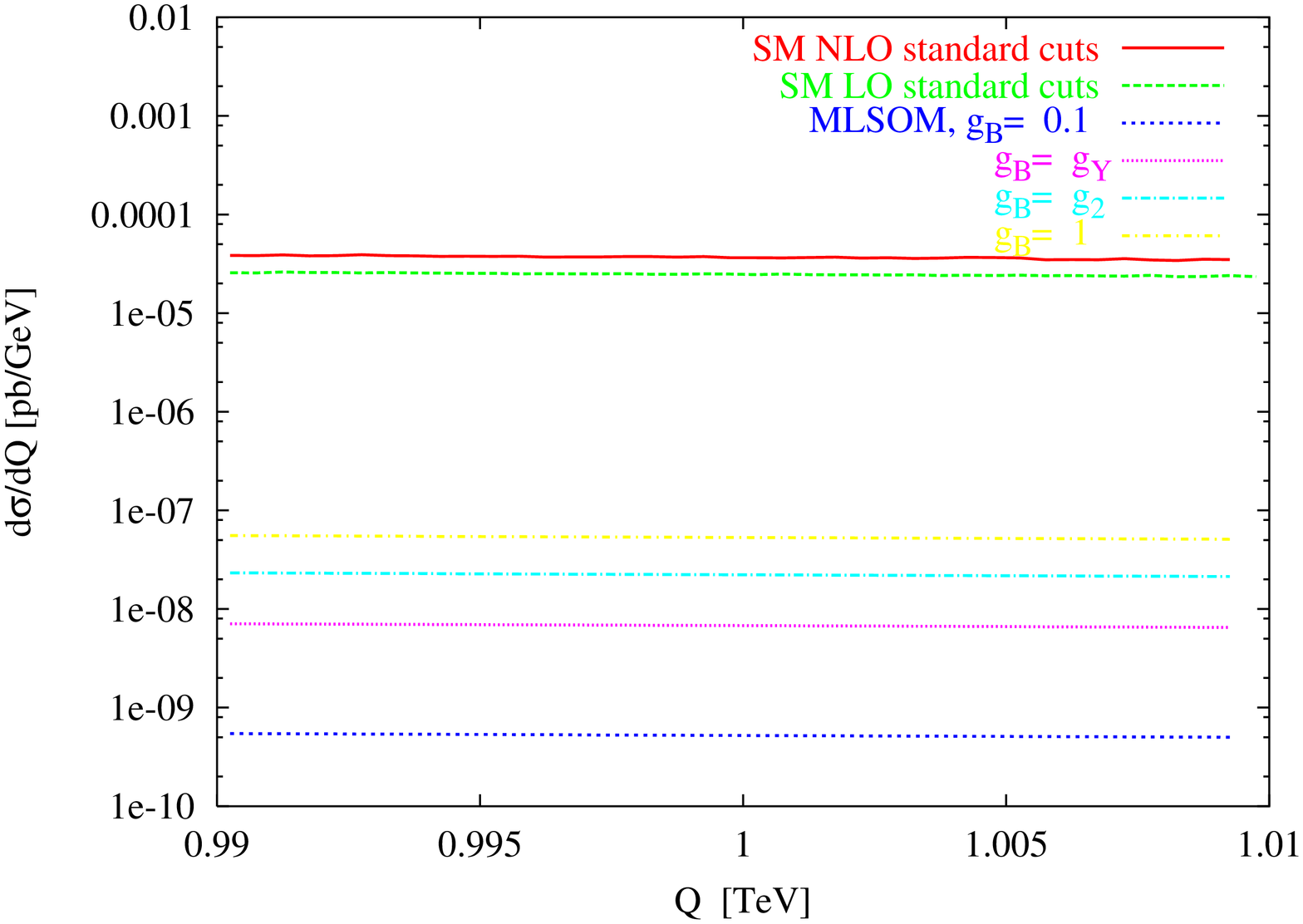}}
\subfigure[  ]{\includegraphics[%
 width=5.6cm,
 angle=-90]{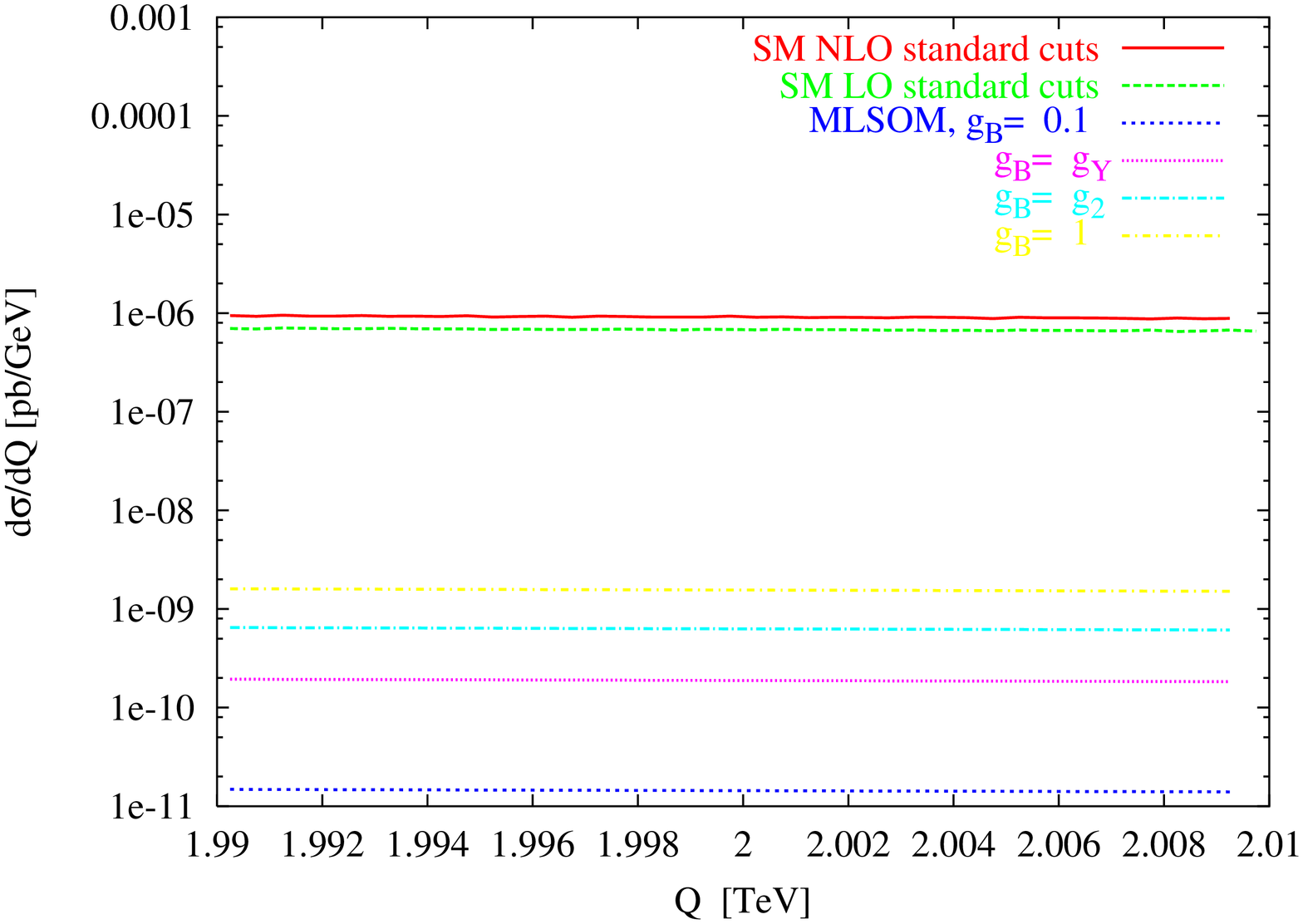}}
\caption{\small (a) Plot of the invariant mass distribution for $Q\sim 1$TeV and for a  St\"uckelberg mass of 1 TeV
showing a comparison between the anomalous contributions (anomalous massless and massive BIM +
interference with the box) and the SM contributions. All the sectors at one-loop and two-loop have been obtained with the Monte Carlo \textsc{Gamma2MC}.
(b) As in the previous figure but the invariant mass distribution has been chosen around 2 TeV.}
\label{new678}
\end{figure}

A comparison between the differential cross section  obtained by
the Monte Carlo and the anomalous contributions is shown in Fig.~\ref{new678}a,
from which one can see that the anomalous components are down by a factor of $10^3-10^7$
respect to the background, depending on the value of the anomalous coupling
$g_B$.

\begin{figure}
\subfigure[  ]{\includegraphics[%
width=5.6cm,
 angle=-90]{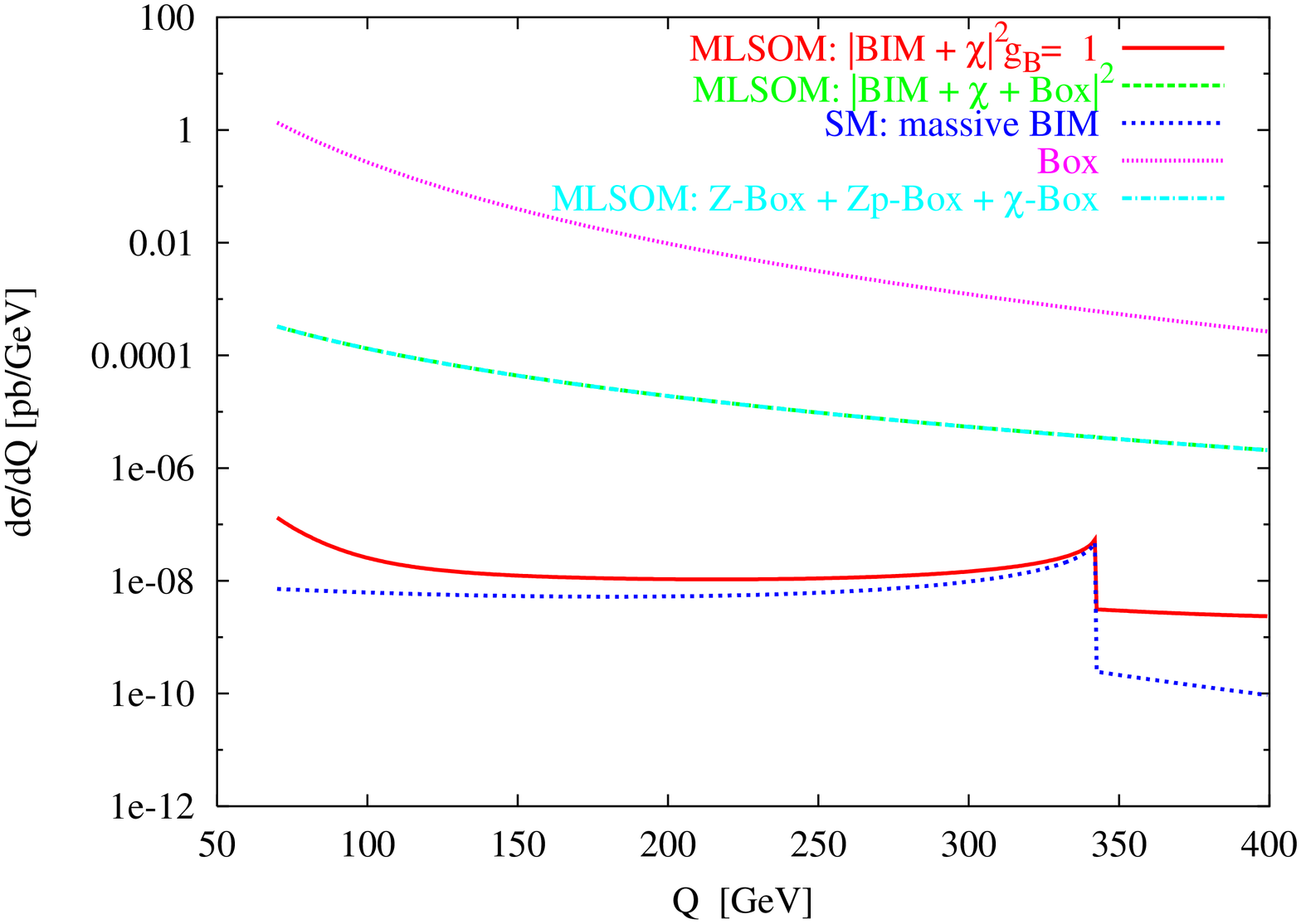}}
\subfigure[  ]{\includegraphics[%
 width=5.6cm,
 angle=-90]{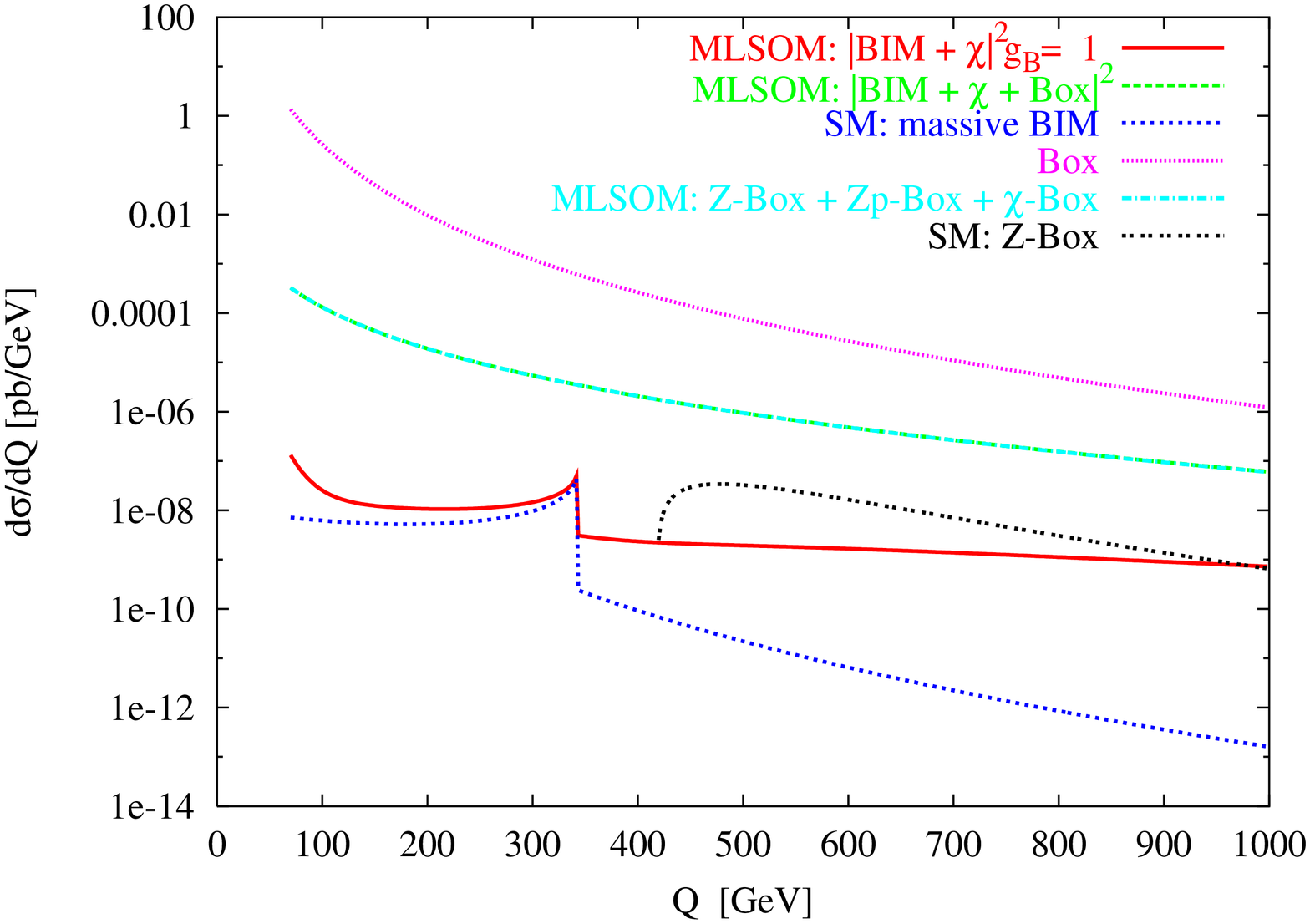}}
\caption{\small Plots of the various anomalous and box-like
components in the SM and mLSOM sector  at (a) lower $Q$ and (b) higher $Q$. }
\label{ex12}
\end{figure}

A similar comparison between anomalous signal and background
for a 1 TeV extra gauge boson but at larger invariant mass
(2 TeV) of the di-photon is shown in Fig.~\ref{new678}b. In both cases the
ratio between the size of the anomalous signal and the background
is $10^{-3}$, showing the large suppression as in the other regions.
\begin{itemize}
\item{\bf Anatomy of the gluon sector}
\end{itemize}
We show in Fig.~\ref{ex12}a and b) two plots that illustrate the size
of the various contributions to the
$gg$ sector in the SM and in the anomalous model. We have separated
these contributions into several
components in the mLSOM and SM cases.

In the anomalous model we have ``pure BIM-like'' amplitudes:

1) the square of the ``BIM set''  shown in Fig.~\ref{qgonedelta},
which contains the $s$-channel exchanges of the $Z$, of the
$Z^{\prime}$ and of the $\chi$. In Figs.~\ref{ex12}a and b) these contributions are indicated as
``BIM + $\chi$''.

Analogously, in the SM case we have that

1$'$) the BIM amplitudes contribute away from the chiral limit due to
the exchange of the $Z$ gauge boson and with top/bottom  quarks
running inside each of the two loops. In Figs.~\ref{ex12}a and b
the contributions in the SM are denoted by ``SM: massive BIM'', and are
just obtained by squaring the single BIM amplitude of Fig.~\ref{qgonedelta}a.
In the SM this component is sizeable around the
threshold $s=4m_t^2$, with $m_t$ being the mass of the top quark,
which explains the cusp in the figure around this energy value.

The interferences sets of the mLSOM  (``BIM-Box'') in which the
BIM amplitudes interfere with the box graphs:

2) these are denoted as ``Z-Box + Zp-Box + $\chi$-Box''.
The three box diagrams are those shown in Fig.~\ref{nopq},
corresponding to graphs (o), (p) and (q) in this figure.

In the SM we have similar contributions. These are generated by

2$'$) interfering the same box graphs mentioned above with the graph  in
Fig.~\ref{qgonedelta}a. In the SM case, as we have already mentioned, these
contributions are due to heavy quarks, having neglected the mass of the leptons
and of the light quarks. This interference is denoted as ``Z-Box''.

Finally,

3)
In both cases we have the squared ``Box'' contributions, which are
just made from the o), p) and q) diagrams of Fig.~\ref{nopq}.

From a look at Figs.~\ref{ex12}a and b) it is quite evident
that the contributions obtained by squaring the three box diagrams are
by far the most important at $\sqrt{S}=14$ TeV. We have tried to
cover both the region $Q < 400$ GeV and the region $400$ GeV < Q < $1$ TeV.
The second largest contributions, in these two plots, are those due to the
interference between the BIM amplitudes and the box. These contributions
are enhanced because of the presence
of the box amplitude. Notice that in the SM case this interference is
small and negative for $Q < 400$ GeV
(see Fig.~\ref{ex12}a) and is not reported. In the second region
($400$ GeV < Q < $1$ TeV) this contribution gets sizeable
around the two-particle cut of the scalar triangle diagram due
to the top quark in the loop and shows up as a steepening in the
line labeled as ``Z-box'' in Fig.~\ref{ex12}b.

The ``pure anomalous'' contributions, due to the squared BIM amplitudes
are down by a factor of about $10^{5}$ respect to the dominant box contributions.
A similar trend shows up also in the SM case. Around the 2-particle cut the
BIM amplitudes both in the SM and in the mLSOM have a similar behavior,
but differ substantially away from this point in the larger-$Q$ and smaller-$Q$ region.
These are the two regions
where the effects of the anomaly are more apparent.
For instance, for $Q< 200$ GeV the steep rise of
the anomalous contributions of the
``BIM + $\chi$'' line is due to the fact that $Q$ is getting closer to the
resonance of the axion $\chi$. In the SM the same region is characterized
by a suppression by the ratio ${m_f}^2/Q^2$. At larger $Q$ values,
the growth of the anomalous contributions are due to the bad behavior of the
anomaly, which grows quadratically with energy, as we have discussed above.
This trend is more visible in Fig.~\ref{ex12}b (red line) in the region
$Q> 800$ GeV.
The possibility of increasing the di-photon (gluon-fusion)
signal is related to the possibility of imposing suitable
kinematical constraints in the analysis of the final state
- or phase-space cuts - in the experimental analysis.
To achieve this goal there are several features of the process
that should be kept into account. The first is the sharp decrease
of the cross section at large invariant mass $Q$ of the di-photon final state;
the second is its increase with $\sqrt{S}$, the collision energy of the two colliding protons.
In the first case we have an increase of the relevant parameter
$\tau=Q^2/S$ characterizing the gluon density; in the second
case this is reduced considerably, giving a fast enhancement of the gluon luminosity.
\begin{figure}[t]
\begin{center}
\includegraphics[width=7cm,angle=-90]{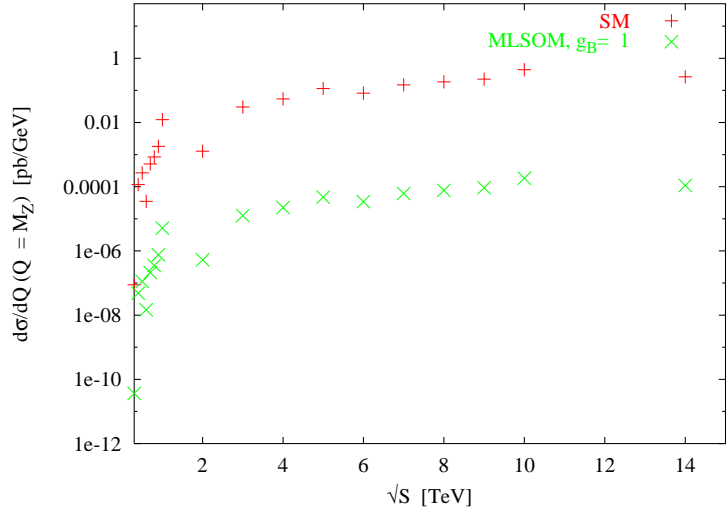}
\caption{\small Double photon invariant mass cross section at $Q=M_Z$ plotted as a function of the energy.
Here the SM contributions include the massive BIM + Box + interference, while the
mLSOM contributions include the massless and massive BIM + interference.}
\label{sqrts}
\end{center}
\end{figure}

\begin{figure}[t]
\begin{center}
\includegraphics[width=7cm,angle=-90]{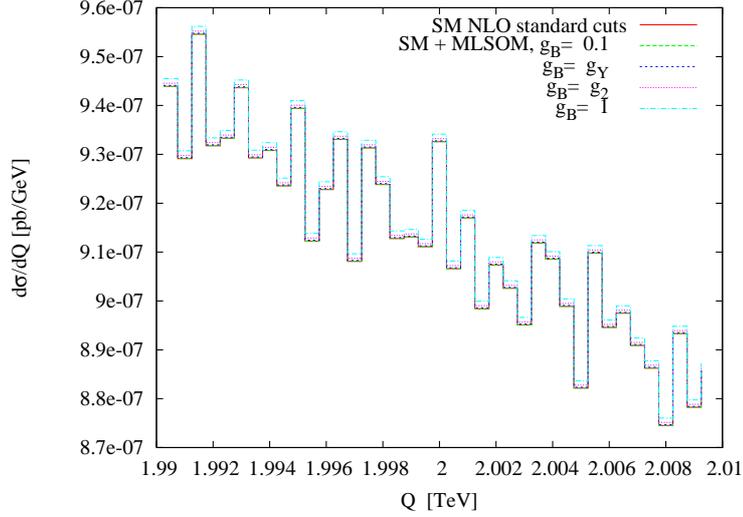}
\caption{\small Binned cross sections for the SM and the mLSOM. The small anomalous corrections are the thick lines at the top of each bin.}
\label{merge}
\end{center}
\end{figure}
 Since bounds on the coupling of the anomalous models may
come both from a combined analysis of both DY and of the
di-photon cross section, we have chosen $Q$ around the
$Z$ peak (small $Q$ option) and the same luminosity evaluated
on the peak of the extra $Z^{\prime}$ (large $Q$ option), both as
a function of the collisional energy $S$.  This is shown in Fig.~\ref{sqrts}.
The anomalous signal grows with $\sqrt{S}$ but the non-unitary growth is not apparent at LHC
energies after convolution with the parton densities; on the other end,
the growths in the SM and in the anomalous sector of the mLSOM appear to be,
at LHC energies, quite similar.

If an extra $Z^\prime$
is found in DY, then the analysis of di-photon both on the peak
of the $Z$ and on the peak of the
$Z^\prime$ could be used to set reasonable bounds on the coupling
of the underlying model, trying to uncover the possible presence of
an anomalous signal. A $\tau\sim 10^{-4}$ can be reached at the
$Z$ peak for a collision energy of 10 TeV, which causes an enhancement
of the anomalous cross section by a factor of approximately $10^4$
respect to the same anomalous contribution measured for $Q\sim 1$ TeV.

In a final figure we show the shapes of the distributions for the SM and the mLSOM using the Monte Carlo
with a binning of the cross section in both cases. This is provided in Fig.~\ref{merge},
where the colored lines on top of each bin indicate the small anomalous signal in the cross section.
\section{Conclusions}
Both DY and DP have some special features, being characterized by a clean final state. In DY the identification of a new resonance in the neutral current sector would bring to the immediate conclusion that an extra $Z^\prime$ is present in the spectrum, but would give not specific indication concerning its true nature. Current experimental bounds constrain the mixing of a possible extra neutral component with the $Z$ gauge boson, with a mass which should be larger than 800 GeV, rendering the future search of extra neutral interactions, at least for DY, quite delicate, being the allowed mass range at the tail of the invariant mass distribution of this process. For this reason, the identification of a restricted mass range for the extra gauge boson would be crucial for its discovery, but unfortunately, there is no extension of the SM that comes with a definite prediction for it.

In intersecting brane models one encounters a similar indetermination and for this reason in our construction the St\"uckelberg mass has been assumed to be a free parameter.
If an extra $s$-channel resonance is found, then the investigation of its specific properties would require much more effort and several years of data collection at the LHC, especially for larger mass values (above 1.5 TeV or so) of the extra $Z^\prime$.  A sizeable width of the resonance could then allow to study the V-A structure of the coupling, though this study is quite complex. In most of the models studied so far the corresponding
widths are expected to be quite small ($\ll 30 \div 40$ GeV)
even for sizeable couplings ($g_B\sim O(1)$) \cite{Coriano:2008wf}, probably below or barely close to the limits of the effective resolution of the detector.

Under these conditions, deciding over the true nature of the extra $Z^\prime$, whether anomalous or not, would then
be far more challenging and would require a parallel study of several independent channels. For this reason we have analyzed two processes which are both affected by anomalous contributions
and could be used for correlated studies of the same interaction.

 We have seen that changes in the factorization/renormalization scales both in the hard scatterings and in the evolution of the PDF's can easily overshadow the anomalous corrections, making a NLO/NNLO analysis
 truly necessary.
 We have focused our investigation on an extra
$Z^{\prime}$ of mass 1 TeV and searched for anomalous
effects in the invariant mass distributions  on the $Z$ peak,
at 1 TeV and for large $Q$ values (up to 2 TeV). From our analysis the large
suppression of the anomalous signal compared to the QCD background
is rather evident, although there are ways to improve on our results.
In fact, our analysis has been based on the characterization of inclusive observables, which are not sensitive
to the geometric structure of the final state. In principle, the shape of the final state event could be
resolved at a finer level of detail, by analyzing, for instance, the rapidity correlations between the
di-photon and a jet, or other similar less inclusive cross sections, as in other cases \cite{Chang:1997sn}.   Obviously, these types of studies are theoretically challenging and require an excellent knowledge of the QCD background at NNLO for these observables.
\section{Appendix}
The decay rates into leptons for the $Z$ and the $Z^\prime$ are
universal and are given by
\ba
&&\Gamma({\cal Z}\rightarrow l\bar{l})=\frac{g^2}{192\pi c_w^2}
M_{{\cal Z}}\left[(g_{V}^{{\cal Z},l})^2+(g_{A}^{{\cal Z},l})^2\right]=
\frac{\alpha_{em}}{48 s_w^2 c_w^2}M_{{\cal Z}}\left[(g_{V}^{{\cal Z},l})^2+
(g_{A}^{{\cal Z},l})^2\right]\,,
\nonumber\\
&&\Gamma({\cal Z}\rightarrow \psi_i\bar{\psi_i})=\frac{N_c\alpha_{em}}{48
s_w^2 c_w^2}
M_{{\cal Z}}\left[(g_{V}^{{\cal Z},\psi_i})^2+(g_{A}^{{\cal
Z},\psi_i})^2\right]\times\nonumber\\
&&\hspace{3cm}\left[1+ \frac{\alpha_s(M_{{\cal Z}})}{\pi}
+1.409\frac{\alpha_s^2(M_{{\cal
Z}})}{\pi^2}-12.77\frac{\alpha_s^3(M_{\cal Z})}{\pi^3}\right],\,
\ea
where $i=u,d,c,s$ and ${\cal Z}=Z,Z^{\prime}$.

For the $Z^{\prime}$ and $Z$ decays into heavy quarks we obtain
\ba
&&\Gamma({\cal Z}\rightarrow b\bar{b})=\frac{N_c\alpha_{em}}{48 s_w^2 c_w^2}
M_{{\cal Z}}\left[(g_{V}^{{\cal Z},b})^2+(g_{A}^{{\cal
Z},b})^2\right]\times\nonumber\\
&&\hspace{3cm}\left[1+ \frac{\alpha_s(M_{{\cal
Z}})}{\pi}+1.409\frac{\alpha_s^2(M_{{\cal Z}})}{\pi^2}
-12.77\frac{\alpha_s^3(M_{\cal Z})}{\pi^3}\right]\,,
\nonumber\\
&&\Gamma({\cal Z}\rightarrow t\bar{t})=\frac{N_c\alpha_{em}}{48 s_w^2 c_w^2}
M_{{\cal Z}}\sqrt{1 - 4 \frac{m_t^2}{M_{{\cal Z}}^{2}} }\times\nonumber\\
&&\hspace{3cm}\left[(g_{V}^{{\cal Z},t})^2\left(1 + 2
\frac{m_t^2}{M_{{\cal Z}}^{2}}\right)
+(g_{A}^{{\cal Z},t})^2\left(1 - 4 \frac{m_t^2}{M_{{\cal
Z}}^{2}}\right)\right]\times\nonumber\\
&&\hspace{3cm}\left[1+ \frac{\alpha_s(M_{{\cal
Z}})}{\pi}+1.409\frac{\alpha_s^2(M_{{\cal Z}})}{\pi^2}
-12.77\frac{\alpha_s^3(M_{\cal Z})}{\pi^3}\right]\,.\nonumber\\
\ea
\section{Appendix. A comment on the unitarity breaking in WZ Lagrangians}
\begin{figure}[t]
\begin{center}
\includegraphics[scale=0.65]{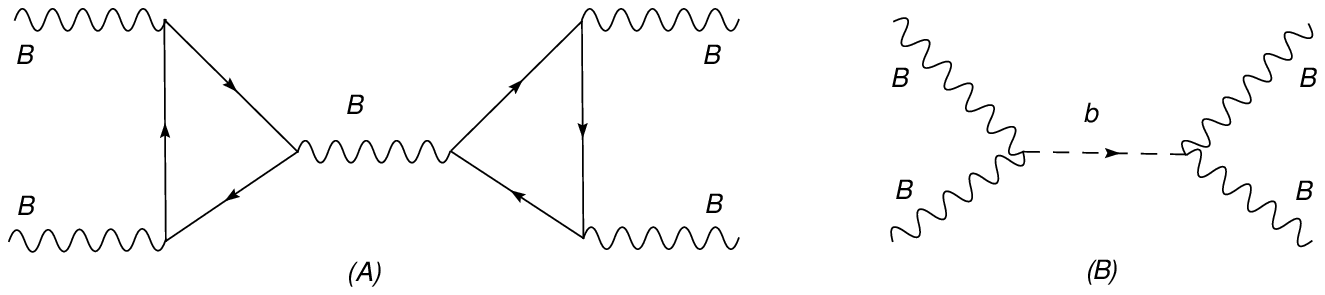}
\caption{\small BIM amplitude for a toy model with a $BBB$ anomaly and $\chi$ exchange diagram.}
\label{unitarity}
\end{center}
\end{figure}

\begin{figure}[t]
\begin{center}
\includegraphics[scale=0.7]{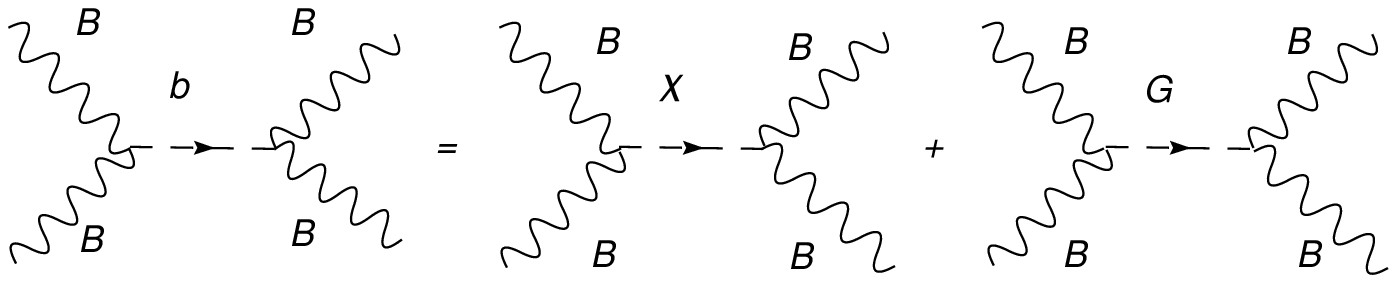}
\caption{\small Decomposition of the St\"uckelberg axion in a Goldstone boson and a physical axion.}
\label{rotationb}
\end{center}
\end{figure}

In the framework of the Abelian A-B Model analyzed in Chap.~\ref{chap:AbelianModels1},
the presence of an untamed growth of the amplitude for a 2-to-2
process can be simplified as follows. In this
simple model $A$ is vector-like and $B$ is axial-vector like.
We also have a single chiral fermion, with an uncancelled $BBB$
anomaly that requires an axion-like interaction $b F_B\wedge F_B$ of the St\"uckelberg
$(b)$ with the $B$ gauge field for the restoration of gauge invariance.
The two contributions to $BB\to BB$ are the BIM amplitude see Figs.~\ref{unitarity}, \ref{rotationb}
for the process mediated by a $B$ boson and a similar one mediated by a physical
axion $\chi$. If we neglect the Yukawa couplings
(we take the fermion to be massless) the only contributions
involved are those already shown in Fig.~\ref{qgonedelta}, diagrams (a) and (b),
but with different couplings.
\ba
A_{BIM}^{\mu \nu \mu^\prime \nu^\prime}= (g_B)^3 \Delta^{\lambda \mu \nu}(-k_1, -k_2) \frac{- i}{k^2-M_B^2} \left( g^{\lambda \lambda^\prime} - \frac{k^\lambda k^{\lambda^\prime}}{M_B^2} \right) (g_B)^3  \Delta^{\lambda^\prime \mu^\prime \nu^\prime}(k_1, k_2)
\ea
where $M_B=\sqrt{M_1^2 + (2 g_B v)^2}$ is the mass of the gauge boson in the $s$-channel after symmetry breaking.
The exchange of the physical axion gives
\ba
B^{\mu \nu \mu^\prime \nu^\prime}= 4 \times \left( \frac{4}{M} \alpha_1 C_{BB} \right)^2   \epsilon[\mu, \nu, k_1, k_2]
\frac{i}{k^2-m_\chi^2}  \epsilon[\mu^\prime, \nu^\prime, k^\prime_1, k^\prime_2]
\label{axi_exchange}
\ea
where the overall factor of 4 in front is a symmetry factor,
the coefficient $\alpha_1=\frac{2 g^{}_B v}{M_B}$ comes from the rotation of
the $b$ axion over the axi-Higgs $\chi$, and the coefficient
$C_{BB}$ has been determined from the condition of gauge invariance
of the anomalous effective action before symmetry breaking
\ba
C_{BB}= \frac{i g_B^3}{3!} \frac{1}{4} a_n \frac{M}{M_1}.
\ea
 The anomaly diagrams are longitudinal, taking the DZ form
\ba
A_{BIM}^{\mu \nu \mu^\prime \nu^\prime}(a_n)= (g_B)^3 \frac{a_n}{k^2} (-k^\lambda) \epsilon[\mu, \nu, k_1,k_2] \frac{- i}{k^2-M_B^2} \left( g^{\lambda \lambda^\prime} - \frac{k^\lambda k^{\lambda^\prime}}{M_B^2} \right) (g_B)^3 \frac{a_n}{k^2} (k^{\lambda^\prime}) \epsilon[\mu^\prime, \nu^\prime, k^\prime_1,k^\prime_2].
\ea
It is easy to recognize in this anomalous amplitude the same
structure of the amplitude for the axi-Higgs exchange in Eq.~(\ref{axi_exchange}).
If we add up these two amplitudes and impose the cancellation
of the two amplitudes (which is a fine tuning) we obtain the condition
\ba
\frac{1}{M_B^2} + \frac{(2 g_B v)^2}{M_B^2 M_1^2} = 0
\ea
in terms of the mass of the $B$ and the St\"uckelberg mass $M_1$,
which does not have a real solution. This is a simple example that conveys the
issue of the presence of unitarity limit for (local) WZ interactions.
                                                          

\addcontentsline{toc}{chapter}{Bibliography}\markboth{}{Bibliography}\fancyhead[LO]{\nouppercase{Bibliography}}

\bibliographystyle{h-elsevier3}
\bibliography{PhDThesis_MorelliFINALE}

\end{document}